ORGANISATION EUROPÉENNE POUR LA RECHERCHE NUCLÉAIRE

**CERN** EUROPEAN ORGANIZATION FOR NUCLEAR RESEARCH

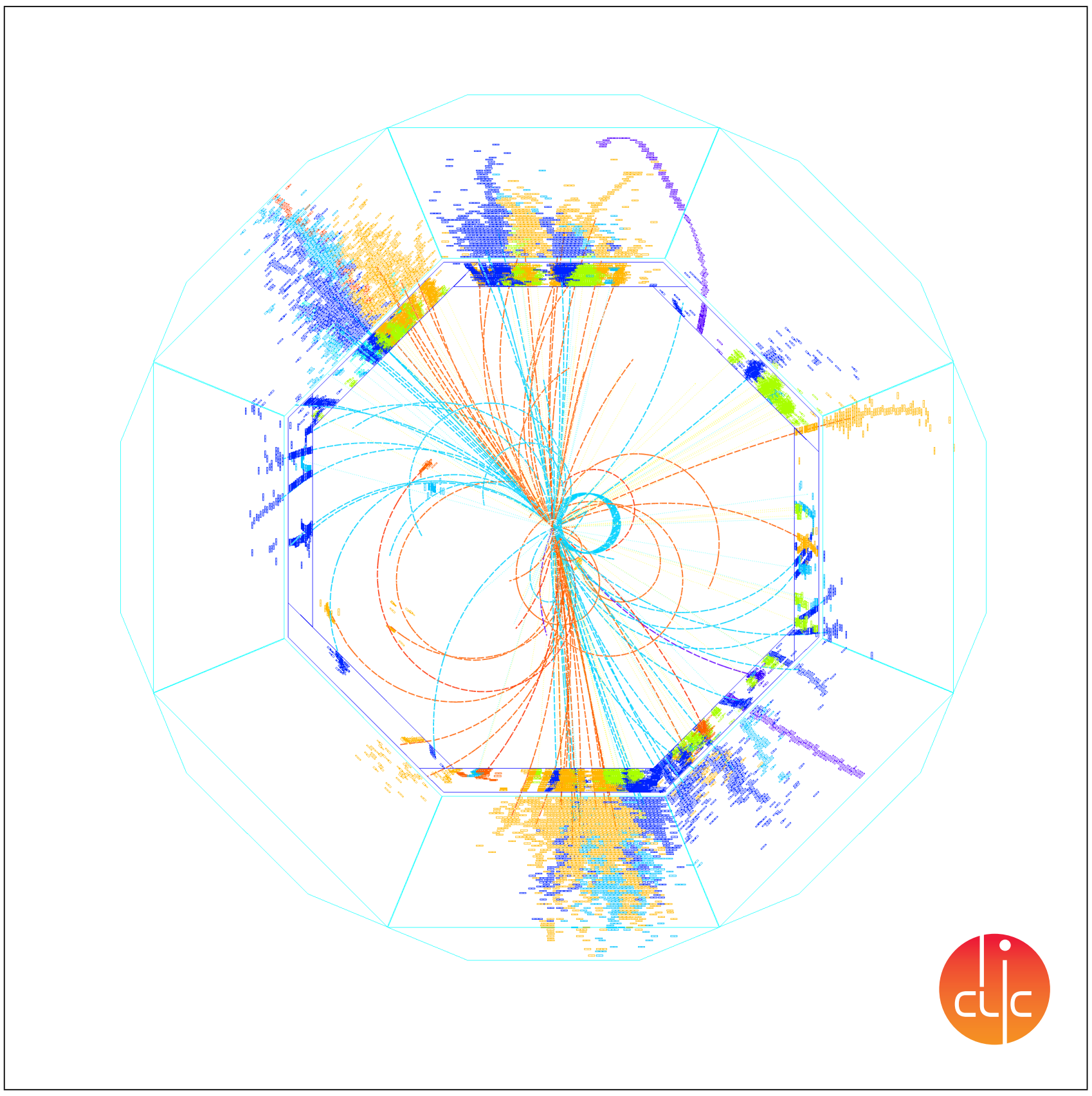

# Physics and Detectors at CLIC

## CLIC Conceptual Design Report

**Abstract**
This report describes the physics potential and experiments at a future multi-TeV $e^+e^-$ collider based on the Compact Linear Collider (CLIC) technology. The physics scenarios considered include precision measurements of known quantities as well as the discovery potential of physics beyond the Standard Model. The report describes the detector performance required at CLIC, taking into account the interaction point environment and especially beam-induced backgrounds. Two detector concepts, designed around highly granular calorimeters and based on concepts studied for the International Linear Collider (ILC), are described and used to study the physics reach and potential of such a collider. Detector subsystems and the principal engineering challenges are illustrated. The overall performance of these CLIC detector concepts is demonstrated by studies of the performance of individual subdetector systems as well as complete simulation studies of six benchmark physics processes. These full detector simulation and reconstruction studies include beam-induced backgrounds and physics background processes. After optimisation of the detector concepts and adopting the reconstruction algorithms the results show very efficient background rejection and clearly demonstrate the physics potential at CLIC in terms of precision mass and cross section measurements. Finally, an overview of future plans of the CLIC detector and physics study is given and a list of key detector R&D topics needed for detectors at CLIC is presented.

CORRESPONDING EDITORS

Lucie Linssen (CERN), Akiya Miyamoto (KEK), Marcel Stanitzki (DESY), Harry Weerts (ANL)

# List of Signatories

## Introductory Remarks

Starting in September 2011, an invitation has been addressed to the particle physics and accelerator community to sign up for a common signatories list for the CLIC Accelerator and Physics & Detectors CDR. In the invitation, the following text was used:

*You are cordially invited to subscribe to the CDR Signatories List*

- *If you have made contributions to the CLIC accelerator or the Linear Colliders Physics and Detector studies, or intend to contribute in the future,*

*OR / AND*

- *If you wish to express support to the physics case and the study of a multi-TeV Linear Collider based on the CLIC technology, and its detector concepts. (Note that signing the CDR does not imply an expression of exclusive support for CLIC versus other major collider options under development.)*

The list of signatories printed below reflects the status of February $13^{th}$, 2012. The call for signatories in support for CLIC remains open[a], and the list is regulary updated[b].

A. Ioannisian

*Yerevan Physics Institute, Yerevan, Armenia*

M. Boland[1]

*Australian Synchrotron, Clayton, Australia*

D. Kim

*Monash University, Melbourne, Australia*

R. Foot, A. Medina[2], R. Rassool[3], B. von Harling, K. Wootton

*University of Melbourne, Melbourne, Australia*

T. Bergauer, M. Dragicevic, S. Frank, M. Krammer, W. Mitaroff, C. Schwanda, M. Valentan

*Austrian Academy of Sciences, Vienna, Austria*

S.D. Bass, P. Jussel

*University of Innsbruck, Innsbruck, Austria*

A. Hoang

*University of Vienna, Vienna, Austria*

J. van Hoorne[4]

*Vienna University of Technology, Vienna, Austria*

[a] https://indico.cern.ch/conferenceDisplay.py?confId=136364

[b] https://edms.cern.ch/document/1183227/

K. Afanaciev[5]
*Belarusian State University, Minsk, Belarus*

V. Gilewsky, A. Patapenka, A. Safronava, I. Zhuk
*Joint Institute for Power and Nuclear Research - Sosny, Minsk, Belarus*

A. Pankov, I. Serenkova[6], A. Tsytrinov
*P.O. Sukhoi State Technical University of Gomel, Gomel, Belarus*

M. Grunewald, M. Tytgat, E. Yazgan
*Ghent University, Ghent, Belgium*

K. Mawatari
*Vrije Universiteit Brussel, Brussels, Belgium*

A. Maciel
*Centro Brasileiro de Pesquisas Físicas - CBPF, Rio de Janeiro, Brazil*

C. Lagana[4]
*Universidade Estadual Paulista Júlio de Mesquita Filho, Sao Paulo, Brazil*

D. Stamenov, I. Tsakov
*Institute for Nuclear Research and Nuclear Energy, Bulgarian Academy of Sciences, Sofia, Bulgaria*

A. Bellerive, F. Tarrade
*Carleton University, Ottowa, Canada*

J. Cline, P. Scott
*McGill University, Montreal, Canada*

N. Toro
*Perimeter Institute for Theoretical Physics, Waterloo, Canada*

I. Trigger
*TRIUMF, Vancouver, Canada*

M. Barbi
*University of Regina, Regina, Canada*

R. Teuscher
*University of Toronto, Toronto, Canada*

R. McPherson[7]
*University of Victoria, Victoria Canada*

Q. Li
*Beijing University, Beijing, China*

S.L. Chen
*Central China Normal University, Wuhan, China*

G.L. Wang, T. Wang
*Harbin Institute of Technology, Harbin, China*

G. Chen, J. Gao, W. Li, C.D. Lu, D. Wang, Y. Wang[4], Z.Z. Xing, C. Xu[8]
*Institute of High Energy Physics, Beijing, China*

Y. Gao, Y.P. Kuang, B. Li[4], Y. Li
*Tsinghua University, Beijing, China*

Y. Li
*Yantai University, Yantai, China*

N. Bilic
*Ruder Boskovic Institute, Zagreb, Croatia*

A. Ilakovac, I. Picek
*University of Zagreb, Zagreb, Croatia*

J. Chyla, P. Gallus, T. Jakoubek[9], J. Kvasnicka, T. Lastovicka, M. Marcisovsky, I. Polak, J. Ridky, P. Sicho, P. Travnicek, V. Vrba[9], J. Zalesak
*Institute of Physics, Academy of Sciences, Prague, Czech Republic*

J. Esberg[4], U.I. Uggerhøj
*Aarhus University, Aarhus, Denmark*

E. Del Nobile, T. Hapola
*CP3-Origins, University of Southern Denmark, Odense, Denmark*

P. Damgaard, J.B. Hansen, J.D. Hansen[4], P. Hansen[10], M.D. Joergensen
*Niels Bohr Institute, Copenhagen, Denmark*

E. Gabrielli, M. Kadastik, A. Racioppi, M. Raidal
*National Institute of Chemical Physics and Biophysics, Tallinn, Estonia*

S. Groote[11]
*University of Tartu, Tartu, Estonia*

E. Brücken, F. Djurabekova, P. Eerola[12], F. Garcia, J. Hahkala, K. Huitu, J. Huopana, V. Karimaki[4], S. Lehti, T. Niinikoski, K. Nordlund, R. Nousiainen[13], K. Österberg[12], J. Paro[13], S. Parviainen, A. Pohjonen, R. Raatikainen, A. Ruzibaev, N.A. Törnqvist, J. Turunen[4], J. Väinölä[13]
*Helsinki Institute of Physics, Helsinki, Finland*

S. Lavignac, G. Soyez
*CEA, IPhT, Saclay, France*

M.N. Bakirci[14], M. Besancon, S. Choudhury, P. Colas, B. Dalena[4], F. Deliot, W. Farabolini, P. Girardot, Y. Guler[14], C. Guy, P. Jarry, F. Kircher, E.C. Lancon, F. Peauger, C. Royon, I. Shreyber[15], M. Titov
*CEA, Irfu, Saclay, France*

E. Aslanides, O. Leroy, G. Mancinelli, S. Muanza, M. Perrin-Terrin, L. Vacavant
*Centre de Physique des Particules de Marseille (CPPM), Aix-Marseille Université, CNRS/IN2P3, Marseille, France*

R. Coquereaux[4], M. Knecht
*Centre de Physique Théorique (CPT), Aix-Marseille Université, CNRS/IN2P3, Marseille, France*

D. Tapia Takaki
*Institut de Physique Nucléaire d'Orsay (IPN), IN2P3/CNRS, Orsay, France*

A. Arbey[4], X. Artru, R. Chehab, M. Chevallier, A. Deandrea
*Institut de Physique Nucléaire de Lyon (IPNL), IN2P3/CNRS, Villeurbanne, France*

B. Fuks
*Institut Pluridisciplinaire Hubert Curien (IPHC), IN2P3/CNRS, Strasbourg, France*

C. Adloff, J. Allibe, G. Balik, J. Blaha, J.J. Blaising, L. Brunetti, M. Chefdeville, G. Coignet, G. Deleglise, A. Espargilière, R. Gaglione, N. Geffroy, A. Jeremie, Y. Karyotakis, N. Massol, J.M. Nappa, S. Rosier Lees, J. Tassan Viol, S. Vilalte, G. Vouters
*Laboratoire d'Annecy le Vieux de Physique des Particules (LAPP), Université de Savoie, IN2P3/CNRS, Annecy, France*

P. Bambade, J. Brossard, S. Callier, O. Dadoun, N. Dinu, F. Dulucq, J.F. Grivaz, J. He, G. Martin-Chassard, F. Poirier, R. Roux, N. Seguin-Moreau, A. Variola
*Laboratoire de l'Accélérateur Linéaire (LAL), Université de Paris-Sud XI, IN2P3/CNRS, Orsay, France*

D.E. Boumediene, P. Gay, N. Ghodbane
*Laboratoire de Physique Corpusculaire (LPC), IN2P3/CNRS, Clermont-Ferrand, France*

W. Da Silva, S. de Cecco, F. Kapusta
*Laboratoire de Physique Nucléaire et des Hautes Energies (LPNHE), IN2P3/CNRS, Paris, France*

S. Kraml, G. Sajot
*Laboratoire de Physique Subatomique et de Cosmologie (LPSC), Université Joseph Fourier Grenoble 1, IN2P3/CNRS, Grenoble, France*

A. Djouadi, D. Guadagnoli, Y. Mambrini
*Laboratoire de Physique Théorique d'Orsay (LPT), IN2P3/CNRS, Orsay, France*

S. Rychkov
*Laboratoire de Physique Théorique ENS, Paris, France*

C. Clerc, M. Ruan, Y. Sirois, H. Videau
*Laboratoire Leprince-Ringuet (LLR), Ecole Polytechnique, IN2P3/CNRS, Palaiseau, France*

A. Badel, B. Caron
*Laboratoire Systèmes et Matériaux pour la Mécatronique (SYMME)-Polytech Annecy Chambéry, Université de Savoie, Annecy, France*

L. Basso, V. Bertone, O. Brein, M. Schumacher
*Albert-Ludwigs-Universität Freiburg, Freiburg, Germany*

G. Müller, S. Pataraia[4], S. Weber, M. Worek
*Bergische Universität Wuppertal, Wuppertal, Germany*

S. Aplin, M. Berggren, K. Buesser, S.S. Caiazza[16], A. Cakir, J. Dietrich, W. Ehrenfeld, G. Eigen[17], B. Foster[18], F. Gaede, E. Garutti[16], A. Grohsjean, C. Günter, O. Hartbrich[19], A. Hartin, B. Hermberg[4], A. Ignatenko[20], S. Karstensen, A. Kaukher, J. List, S. Lu, B. Lutz, M. Medinnis, I. Melzer-Pellmann, V. Morgunov[15], S. Morozov[16], J. Reuter, A. Rosca, I. Rubinskiy[21], S. Schreiber, F. Sefkow[4], M. Stanitzki, N. Styles, M. Terwort, A. Weiler, P. Zerwas
*DESY, Hamburg, Germany*

I. Bloch, J. Bluemlein, C. Grah, H. Henschel, W. Lange[4], W. Lohmann[22], K. Mönig[4], O. Novgorodova[22], S. Riemann, T. Riemann, H.J. Schreiber, M. Stanescu Bellu
*DESY, Zeuthen, Germany*

O. Arnaez, V. Buescher, T. Hurth, H. Spiesberger, S. Tapprogge, S. Weinzierl
*Johannes Gutenberg-Universität Mainz, Mainz, Germany*

H. Stenzel
*Justus Liebig Universität Giessen, Giessen, Germany*

A. Bernhard, A.S. Brogna, S. Casalbuoni, M. Fitterer[4], S. Gieseke, R. Grober, N. Hiller, V. Judin, A.S. Mueller, M. Mühlleitner, M. Nasse, U. Nierste, P. Peiffer, G. Quast, R. Rossmanith, F.P. Schilling, M. Schuh, M. Steinhauser, M. Weber
*Karlsruhe Institute of Technology, Karlsruhe, Germany*

G. Buchalla
*Ludwig-Maximilians-Universität München, Munich, Germany*

M. Schmelling, H. Zhang
*Max Planck Institute for Nuclear Physics, Heidelberg, Germany*

A. Caldwell, G. Grindhammer, K. Seidel[23], R. Settles[4], F. Simon[23], C. Soldner[23], L. Weuste[23], S. Zhou[24]
*Max Planck Institute for Physics, Munich, Germany*

W. Bernreuther, M. Czakon, M. Krämer, A. Kulesza, A. Meyer, A. Perieanu
*RWTH Aachen University, Aachen, Germany*

H. Henke
*Technische Universität Berlin, Berlin, Germany*

W.F. Mader, J.H. Park, G.M. Pruna, A. Straessner
*Technische Universität Dresden, Dresden, Germany*

M. Ratz, J. Torrado Cacho, R. Ziegler
*Technische Universität Munich, Munich, Germany*

Y. Bilevych[25], I. Brock, J. Conley, M. Cristinziani[4], K. Desch, M. Drees, H. Dreiner, S. Hillert, J. Kaminski, M. Lupberger, C. Marinas[26], J. Tattersall, N. Wermes, P. Wienemann
*Universität Bonn, Bonn, Germany*

S. Albino, N. Feege[27], J. Kersten, B. Kniehl, G. Moortgat-Pick[27], S. Schuwalow[5]
*Universität Hamburg, Hamburg, Germany*

P. Eckert, I. Peric, H.C. Schultz-Coulon
*Universität Heidelberg, Heidelberg, Germany*

I. Fleck, C. Grupen, K. Ikematsu
*Universität Siegen, Siegen, Germany*

A. Denner, T. Ohl, W. Porod, R. Rückl
*University of Würzburg, Würzburg, Germany*

G.J. Gounaris, C. Petridou[4], D. Sampsonidis
*Aristotle University of Thessaloniki, Thessaloniki, Greece*

P. Zisopoulos
*National and Kapodistrian University of Athens, Athens, Greece*

E. Gazis[4], N. Gazis[4], E. Ikarios[4]
*National Technical University of Athens, Athens, Greece*

S. Czellar[4], D. Horvath
*Wigner Research Centre for Physics, Hungarian Academy of Sciences, Budapest, Hungary*

B. Ananthanarayan, S.K. Garg, R. Godbole, K. Mohan, M. Patra, R. Varma[28], S. Vempati
*Centre for High Energy Physics, Indian Institute of Science, Bangalore, India*

A. Nyffeler
*Harish-Chandra Research Institute, Allahabad, India*

D.K. Ghosh
*Indian Association For The Cultivation Of Science, Calcutta, India*

R. Singh
*Indian Institute of Science Education and Research, Calcutta, India*

K. Rao
*Indian Institute of Technology, Bombay, India*

S. Rakshit
*Indian Institute of Technology, Indore, India*

S. Gopalakrishna
*Institute of Mathematical Sciences, Chennai, India*

R. Gupta
*Panjab University, Chandigarh, India*

S. Rindani, P. Sharma
*Physical Research Laboratory, Ahmedabad, India*

P.D. Gupta, P. Shrivastava
*Raja Ramanna Centre for Advanced Technology, DAE, Indore, India*

S. Banerjee, G. Bhattacharyya, N. Majumdar, P. Mathews, S. Mukhopadhyay
*Saha Institute of Nuclear Physics, Calcutta, India*

G. Majumder
*Tata Institute of Fundamental Research, Mumbai, India*

A. Kundu
*University of Calcutta, Calcutta, India*

B. Choudhary, D. Choudhury, P. Saxena
*University of Delhi, Delhi, India*

S. Paktinat Mehdiabadi
*Institute for Research in Fundamental Sciences, Tehran, Iran*

H. Abramowicz[29], G. Alexander, G. Bella, S. Kananov, A. Levy, I. Sadeh, R. Schwartz
*Tel Aviv University, Tel Aviv, Israel*

E. Duchovni
*Weizmann Institute of Science, Rehovot, Israel*

M. Maggi, A. Ranieri
*INFN, Bari, Italy*

M. Caffo[30], D. Hatzifotiadou[30]
*INFN, Bologna, Italy*

A. Tricomi
*INFN, Catania, Italy*

S. Catani[31], S. de Curtis[31], S. Paoletti, G. Sguazzoni[4], G. Valenti
*INFN, Florence, Italy*

D. Alesini, M.E. Biagini, C. Biscari, A. Ghigo, G. Isidori[4], F. Marcellini, G. Pancheri, P. Raimondi[32], M. Serio, A. Stella
*INFN, Frascati, Italy*

C. Troncon[33]
*INFN, Milan, Italy*

C. Gatto
*INFN, Naples, Italy*

O. Nicrosini, F. Piccinini
*INFN, Pavia, Italy*

O. Panella
*INFN, Perugia, Italy*

F. Bedeschi, T. Boccali[34], F. Palla, P. Spagnolo[34], R. Tenchini[34]
*INFN, Pisa, Italy*

B. Mele, L. Silvestrini
*INFN, Rome, Italy*

A. Ballestrero, C. Mariotti[4]
*INFN, Turin, Italy*

M. Nemevsek
*International Centre for Theoretical Physics, Trieste, Italy*

K. Kannike, A. Lusiani[35]
*Scuola Normale Superiore, Pisa, Italy*

D. Marzocca[36], M. Monaco, A. Romanino, M. Spinrath
*SISSA, Trieste, Italy*

R. Bonomi, A. Degiovanni[37], R. Kieffer
*TERA Foundation, Novara, Italy*

M. Caccia[38]
*Università degli Studi dell'Insubria, Como, Italy*

F.L. Navarria[39]
*Università degli Studi di Bologna, Bologna, Italy*

I. Masina, M. Moretti
*Università degli Studi di Ferrara, Ferrara, Italy*

M. Calvetti, R. Casalbuoni[40], V. Ciulli[40], F. Coradeschi[40], D. Dominici[40]
*Università degli Studi di Firenze, Florence, Italy*

C.M. Becchi[41]
*Università degli Studi di Genova, Genoa, Italy*

U. Amaldi[42], C. Oleari
*Università degli Studi di Milano-Bicocca, Milan, Italy*

M. Levchenko[4], F. Ragusa
*Università degli Studi di Milano, Milan, Italy*

G. Montagna
*Università degli Studi di Pavia, Pavia, Italy*

G. Bellettini[43], A. Strumia[44]
*Università degli Studi di Pisa, Pisa, Italy*

R. Contino[45], S. Gentile[45], C. Luci[45], G. Organtini[45]
*Università degli Studi di Roma La Sapienza, Rome, Italy*

L. Magnea[46], G. Passarino
*Università degli Studi di Torino, Turin, Italy*

G. Della Ricca[36], A. Schizzi[36]
*Università degli Studi di Trieste, Trieste, Italy*

G. Blankenburg
*Università degli Studi Roma Tre, Rome, Italy*

T. Abe, A. Aryshev, K. Fujii, T. Higo, Y. Makida, T. Matsuda, A. Miyamoto, T. Shidara, T. Takatomi, Y. Takubo, T. Tauchi, N. Toge, K. Ueno, J. Urakawa, A. Yamamoto, M. Yamanaka
*High Energy Accelerator Research Organization, KEK, Tsukuba, Japan*

T. Tanabe, S. Yamashita
*International Center for Elementary Particle Phyiscs, The University of Tokyo, Tokyo, Japan*

M. Ishino
*Kyoto University, Kyoto, Japan*

K. Kawagoe, K.I. Okumura
*Kyushu University, Fukuoka, Japan*

J. Hisano
*Nagoya University Nagoya, Japan*

H. Ono
*Nippon Dental University, Tokyo, Japan*

K. Kotera, T. Takeshita
*Shinshu University, Nagano, Japan*

T. Nagamine, H. Yamamoto
*Tohoku University, Sendai, Japan*

S.Y. Choi, E.J. Kim
*Chonbuk National University, Jeonju, Korea*

P. Ko, S.C. Park[47]
*Korea Institute for Advanced Study, Seoul, Korea*

A. Aranda
*Universidad de Colima, Colima, Mexico*

H. Beijers, S. Brandenburg
*Kernfysisch Versneller Instituut, University of Groningen, Groningen, the Netherlands*

S. Bentvelsen, P. de Jong, M. Merk, D.B. Ta, J. Timmermans[27], N. van Bakel, H. van der Graaf
*Nikhef, Amsterdam, the Netherlands*

S. Caron[25], N. de Groot[25], S. de Jong[25], F. Filthaut[25], W. Metzger[25]
*Radboud University of Nijmegen, Nijmegen, the Netherlands*

P. Osland
*University of Bergen, Bergen, Norway*

E. Adli[32], A. Raklev, K. Sjobak[4], A. Strandlie
*University of Oslo, Oslo, Norway*

I. Ahmed
*COMSATS Institute of Information Technology, Islamabad, Pakistan*

R. Khalid
*National University of Sciences and Technology, Islamabad, Pakistan*

H. Hoorani
*Quaid-i-Azam University, Islamabad, Pakistan*

J. Aguilar, M. Idzik, S. Kulis
*AGH University of Science and Technology, Krakow, Poland*

T. Lesiak, B. Pawlik, M. Skrzypek, Z. Was[4], W. Wierba[27], L. Zawiejski
*Institute of Nuclear Physics PAN, Krakow, Poland*

W. Placzek
*Jagiellonian University, Krakow, Poland*

J. Kalinowski, M. Krawczyk, M. Misiak
*University of Warsaw, Warsaw, Poland*

G.C. Branco, F. Joaquim, M.N. Rebelo, J. Romão
*Instituto Superior Técnico, Lisbon, Portugal*

A. David, J. Varela
*Laboratório de Instrumentação e Física Experimental de Partículas, Lisbon, Portugal*

C. Coca, M.O. Dima, O. Marius Ciprian, E. Teodorescu
*Horia Hulubei National Institute of Physics and Nuclear Engineering, Bucharest, Romania*

F. Elena, M. Marian, G. Mogildea, A.T. Neagu, P.M. Potlog, T. Preda, G. Veta
*Institute of Space Science, Bucharest, Romania*

A. Bragin, E. Levichev, P. Piminov, S. Sinyatkin, V.M. Strakhovenko, V. Telnov[48], K. Zolotarev
*Budker Institute of Nuclear Physics, Novosibirsk, Russia*

S. Gninenko
*Institute for Nuclear Research, Moscow, Russia*

M. Chadeeva, M. Danilov, O. Markin[27], E. Tarkovsky, M. Vysotsky
*Institute for Theoretical and Experimental Physics, Moscow, Russia*

M. Chizhov[49], M. Filippova, A. Gongadze[50], S. Grigoryan[51], D. Gudkov[4], A. Olyunin, A. Samochkine[4], V. Samoylov, A. Sapronov[27], V. Soldatov, A. Solodko, E. Solodko, I. Tyapkin, V. Uzhinsky[4], A. Vorozhtsov[4]
*Joint Institute for Nuclear Research, Dubna, Russia*

V. Grichine[4]
*Lebedev Physical Institute of the Russian Academy of Sciences, Moscow, Russia*

V. Saveliev[27]
*National Research Nuclear University, Moscow, Russia*

E. Boos, L. Gladilin, L. Smirnova
*Skobeltsyn Institute of Nuclear Physics, Moscow State University, Moscow, Russia*

P. Michael
*The Institute of Applied Physics of the Russian Academy of Sciences, Nizhny Novgorod, Russia*

A. Gurtu[52]
*King Abdulaziz University, Jeddah, Saudi Arabia*

P. Adzic[53], I. Bozovic-Jelisavcic, S. Lukic, M. Pandurovic, I. Smiljanic
*Vinca Institute of Nuclear Sciences, Belgrade, Serbia*

J. Kamenik
*J. Stefan Institute, Ljubljana, Slovenia*

A. Hamilton
*University of Cape Town, Cape Town, South Africa*

M. Aguilar-Benitez, J. Alcaraz Maestre, F.M. de Aragón, L. Garcia-Tabares, D. Iglesias, C. Oliver, I. Podadera, E. Rodríguez García, L. Sanchez Garcia, F. Toral, C. Vazquez
*CIEMAT, Madrid, Spain*

S. Heikkinen
*Fusion for Energy, The European Joint Undertaking for ITER and the Development of Fusion Energy, Barcelona, Spain*

J.R. Espinosa
*Institut de Física d'Altes Energies, Barcelona, Spain*

C. Belver-Aguilar, C. Blanch Gutierrez, A. Celis, A. Faus-Golfe, J. Fuster[54], I. García García, J.J. Garcia-Garrigos, E. Higon Rodriguez[4], C. Lacasta, A. Pich, J. Resta Lopez, J.W.F. Valle[55], M. Vos
*Instituto de Física Corpuscular, Valencia, Spain*

M.R. Celso, M. Felcini[56], C. Martinez Rivero, A. Ruiz Jimeno, I. Vila Alvarez
*Instituto de Física de Cantabria, (CSIC-Univ. Cantabria), Santander, Spain*

M. Grefe, M. Herrero, L. Ibanez, C. Munoz, R. Torre
*Instituto de Física Teórica UAM/CSIC, Madrid, Spain*

J.J. Fernandez-Melgarejo, E. Torrente-Lujan
*TH-FISPAC Murcia University, Murcia, Spain*

F. Cornet, O. Kittel
*Universidad de Granada, Granada, Spain*

N. Armesto, H. Lin
*Universidade de Santiago de Compostela, Santiago de Compostela, Spain*

A. Dieguez, S. Penaranda-Rivas[57], B. Pie Valls, J. Sola, J. Trenado
*Universitat de Barcelona, Barcelona, Spain*

R. Morón Ballester

*Universitat Jaume I, Castelló de la Plana, Spain*

Y. Kubyshin[58], E. Marin[4], G. Montoro

*Universitat Politècnica de Catalunya, BarcelonaTech, Barcelona, Spain*

P. Christiansen, V. Hedberg, L. Jönsson, U. Mjörnmark, A. Oskarsson, L. Österman, T. Sjöstrand, E. Stenlund

*Lund University, Lund, Sweden*

C. Ohm

*Stockholm University, Stockholm, Sweden*

G. Angelova Hamberg, R. Brenner, T. Ekelöf, A. Ferrari, M. Jacewicz, T. Muranaka, A. Palaia, R. Ruber, R. Santiago Kern[4], V. Ziemann

*Uppsala University, Uppsala, Sweden*

D. Abbaneo, A. Adiguzel, M. Aicheler[59], G. Altarelli[60], M. Anastasopoulos, A. Andersson, G. Anelli, I. Antoniadis, F. Antoniou[61], J. Apostolakis, A. Apyan[62], G. Arnau Izquierdo, K. Artoos, P. Aspell, S. Atieh, E. Auffray, J. Baechler, R. Ballabriga, G. Ballesteros, D. Banfi, M.J. Barnes, J. Barranco Garcia, A. Bartalesi, A. Behrens, M. Benoit, A. Benot-Morell[26], L. Benucci, D. Bergesio[42], S. Bettoni, G. Blanchot, B. Bolzon, J. Boyd, D. Bozzini, C. Bracco, J. Bremer, H. Breuker, M. Brugger, P. Buncic, H. Burkhardt, F. Butin, M. Buzio, S. Calatroni, E. Calvo, M. Campbell, T. Camporesi, G. Cancio, O. Capatina, N. Catalan Lasheras, A. Cattai, E. Chevallay, P. Chiggiato, J. Christiansen, E. Ciapala, M. Cirilli, C. Collette[63], B. Constance, R. Corsini, G. Cosmo, B. Cure, D. D'Enterria, A. Dabrowski[64], A. Dallocchio, D. Dannheim, M. De Gaspari, G. de Rijk, A. De Roeck, L. Deacon[65], S. Deghaye, J.P. Delahaye, A. Descoeudres, B. Di Girolamo, T. Dobers, S. Doebert, A. Dotti, M. Draper, F. Duarte Ramos, A. Dubrovskiy, K. Elsener, M. Esposito, P. Fassnacht, V. Fedosseev, P. Fernandez Carmona, P. Ferracin, K. Foraz, T. Fowler, H. Franca[66], M. Frank, D. Froidevaux, J.F. Fuchs, C. Fuentes, E. Fullana[67], A. Gaddi, D. Gamba[33], G. Ganis, E. Garcia Garcia[68], H. Garcia Morales[69], J. Garcia Perez, C. Garion, M. Gastal, L. Gatignon, P. Gavillet, J.C. Gayde, A. Gerbershagen[70], M. Gersabeck, H. Gerwig, G. Geschonke, S. Gibson[70], M. Girone, G.F. Giudice, B. Goddard, M. Goulette[71], S. Gowdy, C. Grefe[72], S. Griffet, C. Grojean[73], J.F. Grosse-Oetringhaus, A. Grudiev, E. Gschwendtner, M. Guinchard, K. Hamilton, M. Hauschild, C. Hauviller, R. Hawkings, A. Hektor[44], A. Henriques, M. Herdzina, C. Hessler, J. Holma[74], E.B. Holzer, A. Honma, Y. Inntjore Levinsen[75], V. Ivanchenko, S. Janssens[76], B. Jeanneret, E. Jensen, J.M. Jimenez, M. Jones, O.R. Jones, M. Jonker, C. Joram, J.M. Jowett, M. Kaya[77], M. Keil, J. Kemppinen, K. Kershaw, V. Khan, M. Killenberg, W. Klempt, A. Kluge, J. Knobloch, R. Knoops[78], K. Koeneke, O. Kononenko, E. Koukovini Platia[37], J.W. Kovermann[79], A. Kuzmin, F. Lackner, T. Lagouri[80], C.B. Lam[81], S. Langeslag[81], A. Latina, J.M. Le Goff, P. Lebrun, P. Lecoq, H.M. Lee, T. Lefevre, A. Lenz, B. Lenzi, R. Leuxe, T. Lienart[82], R.L. Lillestol[75], M. Limper, L. Linssen, X. Llopart Cudie, J.J. Lopez-Villarejo[83], C. Lopez, M. Losasso[84], A.I. Lucaci Timoce, D. Luckey, M. Ludwig, R. Maccaferri[42], C. Maglioni, R. Mahbubani, F. Mahmoudi[51], C.O. Maidana[85], H. Mainaud Durand, L. Malgeri, M. Mangano, D. Manglunki[76], M. Mannelli, C. Marrelli, Z. Marshall, P. Mato, G. Mavromanolakis, G. McMonagle, A. Mereghetti, D. Mergelkuhl, O. Mete, D. Missiaen, M. Modena, M. Moll, L. Moneta, A. Muennich, D. Muenstermann, M. Mulders, L. Musa, J. Nardulli, A. Naumann, A. Newborough, B. Nicquevert, D. Nisbet, M. Nonis, M. Nordberg, P. Nouvel[86], M. Olvegård[87], A. Onnela, T. Orimoto, J. Osborne, R. Ostojic, T. Otto, S. Palestini, Y. Papaphilippou, S. Pasinelli, C. Pasquino, G. Pasztor, M. Pepe Altarelli, A.T. Perez Fontenla[88], G. Perez, T. Persson, P. Petagna, C. Petrone, A. Pfeiffer, J. Pfingstner,

S. Pittet, S. Poss, D. Prieur, M. Pullia, S. Pütz, L. Quertenmont, A. Raimondo, A.U. Rehman[89], Y. Renier, J.P. Revol, A. Ribon, E. Richards, G. Riddone, P. Riedler, L. Rinolfi, S. Roesler, J. Rojo, P. Roloff, L. Ropelewski, F. Rossi, G. Roy, V. Rude, G. Rumolo, S. Russenschuck, A. Sailer[90], G. Salam[91], J. Salicio-Diez, E. Salvioni[92], M. Sapinski, P. Schade[27], K.M. Schirm, D. Schlatter, H. Schmickler, B. Schmidt, D. Schoerling[93], D. Schulte, S. Sekmen, G. Servant[73], A. Sfyrla, S. Sgobba, S.H. Shaker[94], A. Sharma, J. Shi, P. Sievers, M. Silari, P. Skands, P. Skowronski, J. Snuverink, L. Soby, P. Speckmayer, P. Sphicas[95], J. Stafford-Haworth[64], S. Stapnes, J. Steinberger, G. Sterbini, J. Strube, M. Struik, F. Stulle, I. Syratchev, A. Sznajder, A. Tauro, F. Tecker, H. ten Kate, F. Teubert, A. Thamm[37], P.A. Thonet, L. Thorndahl, L. Timeo, H. Timko[96], R. Tomas Garcia, D. Tommasini, F. Tramontano[97], P. Tropea, J. Uythoven, P. Valerio[98], E. van der Kraaij, E. van Herwijnen, R. van Weelderen, G. Vandoni, R. Veness, M. Vesterinen, H. Vincke, A. Vivoli[8], V. Vlachoudis, J. Vollaire, R. Wegner, J.D. Wells[99], P.S. Wells, T. Wengler, P. Wertelaers, U.A. Wiedemann, H. Wilkens, I. Wilson, S. Worm[100], W. Wuensch, M. Zakaria, J. Zalieckas[17], C. Zamantzas, T. Zickler, F. Zimmermann

*CERN, Geneva, Switzerland*

A. Bay, N. Charitonidis[4], T. Montaruli[71], D. Pappadopulo, R. Rattazzi

*Ecole Polytechnique Fédérale de Lausanne, Lausanne, Switzerland*

G. Dissertori, P. Le Coultre, F. Moortgat, L. Pape, F.J. Ronga, D. Treille[4], R. Wallny

*ETH Zurich, Zurich, Switzerland*

H.H. Braun[4], G. Chachamis, M. Csatari Divall[4], G. De Michele[37], M. Dehler, T. Garvey[4], T. Pieloni[4], J.Y. Raguin, L. Rivkin[37], R. Zennaro

*Paul Scherrer Institut, Villigen, Switzerland*

H.P. Beck, G. Colangelo, A. Ereditato, C. Greub, P. Minkowski[4], A. Oeftiger[4]

*University of Bern, Bern, Switzerland*

R. Durrer, D. Haas, M. Pohl

*University of Geneva, Geneva, Switzerland*

M. Grazzini[40]

*University of Zurich, Zurich, Switzerland*

C.H. Lin

*Academia Sinica, Taipei, Taiwan*

G.W.S. Hou

*National Taiwan University, Taipei, Taiwan*

B. Tali

*Adiyaman University, Adiyaman, Turkey*

H. Aksakal[101], V. Ari, O. Cakir, A.K. Ciftci, R. Ciftci, H. Duran Yildiz, S. Kuday, I. Turk Cakir[4]

*Ankara University, Ankara, Turkey*

B. Antmen, S. Cerci, Z.S. Demiroglu, I. Dumanoglu, E. Eskut, F.H. Geçit, E. Gurpinar, I. Hos, O. Kara, T. Karaman, A. Kayís Topaksu, G. Onengut, M. Özcan, K. Ozdemir, A. Polatoz, D. Sunar Cerci, H. Topakli, S. Türkçapar

*Cukurova University, Adana, Turkey*

N. Sonmez

*Ege University, Izmir, Turkey*

B. Belma Sirvanli

*Gazi Universitesi Rektorlugu, Ankara, Turkey*

B. Demirkoz, A.M. Guler

*Middle East Technical University, Ankara, Turkey*

S. Sultansoy

*TOBB Economics and Technology University, Ankara, Turkey*

L. Teodorescu

*Brunel University, Uxbridge, United Kingdom*

F. Krauss, A. Signer

*Durham University, Durham, United Kingdom*

P.D. Dauncey, K. Stelle, M. Vazquez Acosta

*Imperial College, London, United Kingdom*

J. Ellis[4]

*King's College London, London, United Kingdom*

P.K. Ambattu[102], G. Burt[102], A. Dexter[102], B. Woolley[102]

*Lancaster University, Lancaster, United Kingdom*

A. Bevan

*Queen Mary, University of London, London, United Kingdom*

T. Aumeyr, G. Blair, L.M. Bobb[4], V. Boisvert, S. Boogert[65], G. Boorman, N. Joshi[4], K. Lekomtsev, A. Lyapin, S. Molloy, W. Shields, P. Teixeira-Dias, S. West[100]

*Royal Holloway, University of London, Egham, United Kingdom*

J.A. Clarke[102], N.A. Collomb, S.P. Jamison, B.J.A. Shepherd

*STFC Daresbury Laboratory, Warrington, United Kingdom*

C. Damerell, K. Harder, S. Moretti[103], G.N. Patrick, F. Wilson

*STFC Rutherford Appleton Laboratory, Didcot, United Kingdom*

G. Shore

*Swansea University, Swansea, United Kingdom*

C. Welsch[104]
*The Cockcroft Institute, Daresbury, United Kingdom*

R. Apsimon, R. Bartolini, D. Bett, K. Peach[70], A. Reichold[70], A. Seryi
*The John Adams Institute for Accelerator Science, Oxford University, Oxford, United Kingdom*

A. Bosco, F. Cullinan[64], P. Karataev
*The John Adams Institute for Accelerator Science, Royal Holloway, University of London, Egham, United Kingdom*

F. Deppisch, T. Gonzalo Velasco, O. Grachov, D. Miller, M. Wing
*University College London, London, United Kingdom*

C. Hawkes, N.K. Watson
*University of Birmingham, Birmingham, United Kingdom*

J. Goldstein, D. Newbold[100], J. Velthuis
*University of Bristol, Bristol, United Kingdom*

B. Allanach, F. Brochu, B. Gripaios[4], C. Lester, J. Marshall, D.J. Munday, M.A. Parker, J. Stirling, M. Thomson, D. Ward
*University of Cambridge, Cambridge, United Kingdom*

A. Gillespie
*University of Dundee, Dundee, United Kingdom*

V.J. Martin, H. Tabassam[105]
*University of Edinburgh, Edinburgh, United Kingdom*

B. Colquhoun, A. Moraes, A. Robson
*University of Glasgow, Glasgow, United Kingdom*

R. Barlow
*University of Huddersfield, Huddersfield, United Kingdom*

M. Korostelev[102], S. Mallows[4], T. Teubner
*University of Liverpool, Liverpool, United Kingdom*

R.B. Appleby[102], J. Forshaw, C. Glasman[102], R. Jones[102], G. Lafferty, J. Masik, H. Owen[102], C. Parkes, A. Pilaftsis, C. Schwanenberger, N. Shipman[4], A. Siodmok[4], G. Xia[102], U.K. Yang
*University of Manchester, Manchester, United Kingdom*

N. Blaskovic Kraljevic[18], P. Burrows, G. Christian, L. Corner[18], M. Davis, N. Harnew, C. Hays, B.T. Huffman, A. Nomerotski, S. Sarkar, C. Swinson[106], R. Walczak, T. Weidberg
*University of Oxford, Oxford, United Kingdom*

E. Berger, B. Bilki[107], M. Demarteau, G. Drake, K. Francis, W. Gai, J. Gainer[62], T. Le Compte, J. Repond, J.R. Smith[108], D. Trojand[109], H. Weerts, L. Xia, G. Yang, J. Zhang
*Argonne National Laboratory, Argonne, USA*

B. Ward
*Baylor University, Waco, USA*

J. Butler, A. Heister
*Boston University, Boston, USA*

C. Amelung
*Brandeis University, Waltham, USA*

M. Harrison, S. Mtingwa, M.-A. Pleier
*Brookhaven National Laboratory, Upton, USA*

B. Barish, J. Schwarz
*California Institute of Technology, Pasadena, USA*

M. Paulini
*Carnegie Mellon University, Pittsburgh, USA*

G. Brooijmans, E. Ponton
*Columbia University, New York, USA*

M. Blanke, J. Crittenden, A. Mikhailichenko, M. Palmer, D. Peterson, J. Serra
*Cornell University, Ithaca, USA*

J. Alwall[110], S. Cihangir, W. Cooper, L. Elementi, H.E. Fisk, D. Green, C.T. Hill, Y.K. Kim[2], A.S. Kronfeld, P. Levchenko[4], A. Para, R. Roser, N. Shah, L. Taylor, M. Wendt, H. Wenzel
*Fermi National Accelerator Laboratory, Batavia, USA*

D. Krohn
*Harvard University, Cambridge, USA*

W. Hoelzl
*IEEE, New York, USA*

R. Dermisek, R. van Kooten
*Indiana University, Bloomington, USA*

J. Hauptman
*Iowa State University, Ames, USA*

J. Bagger, B. Barnett, B. Blumenfeld, A. Gritsan, K. Grizzard, M. Swartz, Y. Zhou
*Johns Hopkins University, Baltimore, USA*

Y. Maravin
*Kansas State University, Manhattan, USA*

S. Alioli, J.F. Arguin, S. Lidia
*Lawrence Berkeley National Laboratory, Berkeley, USA*

M. Graesser
*Los Alamos National Laboratory, Los Alamos, USA*

P. Fileviez Perez, G. Gabadadze
*New York University, New York, USA*

G. Blazey, D. Chakraborty, D. Hedin, G. Lima, S.P. Martin[43]
*Northern Illinois University, DeKalb, USA*

M.M. Velasco
*Northwestern University, Illinois, USA*

C. Tully
*Princeton University, Princeton, USA*

D. Bortoletto, I. Shipsey
*Purdue University, West Lafayette, USA*

C. Adolphsen, K. Bane, T. Barklow, M. Breidenbach, S.J. Brodsky, R. Cassell, S. Dimopoulos[4], S. Gessner, N. Graf, J. Hewett, M.P. Le, T. Markiewicz, T. Maruyama, J. McCormick, K. Moffeit, T. Nelson, D. Onoprienko, M. Oriunno, R. Partridge, N. Phinney, M. Pivi, T. Raubenheimer, T. Rizzo, J.C. Sheppard, S. Smith[4], S. Tantawi[111], F. Wang, J. Wang, F. Zhou
*SLAC National Accelerator Laboratory, Menlo Park, USA*

R. Rios
*Southern Methodist University, Dallas, USA*

D. Curtin, K. Dehmelt[27], M. Douglas, R. Essig, P. Grannis
*Stony Brook University, New York, USA*

Y. Pakhotin
*Texas A&M University, College Station, USA*

K. Sliwa
*Tufts University, Medford, USA*

L. Clavelli, A. Das, N. Okada
*University of Alabama, Tuscaloosa, USA*

G. Chambers III
*University of Arkansas at Little Rock, Little Rock, USA*

J. Gunion
*University of California at Davis, Davis, USA*

Y. Nomura
*University of California, Berkeley, USA*

J.L. Feng
*University of California, Irvine, USA*

V. Andreev
*University of California, Los Angeles, USA*

B. Grinstein, D. Stone, P. Uttayarat
*University of California, San Diego, La Jolla, USA*

J. Incandela[4]
*University of California, Santa Barbara, USA*

T. Banks[112], M. Battaglia[4], M. Dine, P. Draper, H. Haber, B. Schumm
*University of California, Santa Cruz, USA*

M. Oreglia, C. Wagner[67], L.T. Wang
*University of Chicago, Chicago, USA*

J. Zupan
*University of Cincinnati, Cincinnati, USA*

S. de Alwis, U. Nauenberg, J.G. Smith, S. Wagner
*University of Colorado, Boulder, USA*

S. Hussain
*University of Delaware, Newark, USA*

D. Bourilkov
*University of Florida, Gainesville, USA*

G. Varner
*University of Hawaii, Manoa, USA*

Y.O. Günaydin[113], G. Halladjian, U. Mallik, E. Norbeck, Y. Onel, S. Ozturk, R. Zaidan
*University of Iowa, Iowa City, USA*

K.C. Kong, D. Marfatia, G. Wilson
*University of Kansas, Lawrence, USA*

D. Brown
*University of Louisville, Louisville, USA*

R. Franceschini, N. Hadley
*University of Maryland, College Park, USA*

B. Brau, D. Ventura, S. Willocq
*University of Massachusetts, Amherst, USA*

D. Feldman, D. Gerdes, S. Goldfarb, R.S. Gupta, K. Riles
*University of Michigan, Ann Arbor, USA*

R. Kriske[114]
*University of Minnesota, Minneapolis, USA*

S. Seidel
*University of New Mexico, Albuquerque, USA*

Y.J. Ng
*University of North Carolina at Chapel Hill, Chapel Hill, USA*

I. Bigi, A. Delgado, M. Hildreth, C. Kolda
*University of Notre Dame, Notre Dame, USA*

J. Brau, R. Frey, G. Kribs, N. Sinev[32], D.M. Strom, E. Torrence
*University of Oregon, Eugene, USA*

C. Ainsley[115]
*University of Pennsylvania, Philadelphia, USA*

T. Han, V. Savinov
*University of Pittsburgh, Pittsburgh, USA*

I. Bars
*University of Southern California, Los Angeles, USA*

C. Jackson, E. Sarkisyan-Grinbaum[4], A. White, J. Yu
*University of Texas at Arlington, Arlington, USA*

G. Watts
*University of Washington, Seattle, USA*

S. Dasu, A. Hervé, P. Huang, G. Shaughnessy
*University of Wisconsin, Madison, USA*

T. Weiler
*Vanderbilt University, Nashville, USA*

P. Karchin
*Wayne State University, Detroit, USA*

T. Appelquist, S. Dhawan, J. Shelton
*Yale University, New Haven, USA*

A. Font
*Universidad Central de Venezuela, Caracas, Venezuela*

[1] *Melbourne University, Melbourne, Australia*
[2] *University of Chicago, Chicago, USA*
[3] *Australian Synchrotron, Clayton, Australia*
[4] *CERN, Geneva, Switzerland*
[5] *DESY, Zeuthen, Germany*
[6] *International Centre for Theoretical Physics, Trieste, Italy*
[7] *TRIUMF, Vancouver, Canada*
[8] *Laboratoire de l'Accélérateur Linéaire (LAL), Université de Paris-Sud XI, IN2P3/CNRS, Orsay, France*
[9] *Czech Technical University in Prague, Prague, Czech Republic*
[10] *University of Copenhagen, Copenhagen, Denmark*
[11] *Johannes Gutenberg-Universität Mainz, Mainz, Germany*
[12] *University of Helsinki, Helsinki, Finland*
[13] *VTT Technical Research Centre of Finland, Finland*
[14] *Cukurova University, Adana, Turkey*
[15] *Institute for Theoretical and Experimental Physics, Moscow, Russia*
[16] *Universität Hamburg, Hamburg, Germany*
[17] *University of Bergen, Bergen, Norway*
[18] *The John Adams Institute for Accelerator Science, Oxford University, Oxford, United Kingdom*
[19] *Bergische Universität Wuppertal, Wuppertal, Germany*
[20] *Belarusian State University, Minsk, Belarus*
[21] *National Research Nuclear University, Moscow, Russia*
[22] *Brandenburgische Technische Universität, Cottbus, Germany*
[23] *Excellence Cluster 'Universe', TU Munich, Garching, Germany*
[24] *Institute of High Energy Physics, Beijing, China*
[25] *Nikhef, Amsterdam, the Netherlands*
[26] *Instituto de Física Corpuscular, Valencia, Spain*
[27] *DESY, Hamburg, Germany*
[28] *Indian Institute of Technology, Bombay, India*
[29] *Max Planck Institute for Physics, Munich, Germany*
[30] *Università degli Studi di Bologna, Bologna, Italy*
[31] *Università degli Studi di Firenze, Florence, Italy*
[32] *SLAC National Accelerator Laboratory, Menlo Park, USA*
[33] *Università degli Studi di Milano, Milan, Italy*
[34] *Università degli Studi di Pisa, Pisa, Italy*
[35] *INFN, Pisa, Italy*

[36]INFN, Trieste, Italy
[37]Ecole Polytechnique Fédérale de Lausanne, Lausanne, Switzerland
[38]INFN, Milan, Italy
[39]INFN, Bologna, Italy
[40]INFN, Florence, Italy
[41]INFN, Genoa, Italy
[42]TERA Foundation, Novara, Italy
[43]Fermi National Accelerator Laboratory, Batavia, USA
[44]National Institute of Chemical Physics and Biophysics, Tallinn, Estonia
[45]INFN, Rome, Italy
[46]INFN, Turin, Italy
[47]Chonnam National University, Gwangju, Korea
[48]Novosibirsk State University, Novosibirsk, Russia
[49]Sofia University, Sofia, Bulgaria
[50]Andronikashvili Institute of Physics, Georgian Academy of Sciences, Tbilisi, Georgia
[51]Laboratoire de Physique Corpusculaire (LPC), IN2P3/CNRS, Clermont-Ferrand, France
[52]University of Delhi, Delhi, India
[53]University of Belgrade, Belgrade, Serbia
[54]University of Valencia and CSIC, Valencia, Spain
[55]Joint Institute of University of Valencia and Spanish Research Council (CSIC)
[56]University of California, Los Angeles, USA
[57]Universidad de Zaragoza, Zaragoza, Spain
[58]Skobeltsyn Institute of Nuclear Physics, Moscow State University, Moscow, Russia
[59]Ruhr-Universität Bochum, Bochum, Germany
[60]Università degli Studi Roma Tre, Rome, Italy
[61]National Technical University of Athens, Athens, Greece
[62]Northwestern University, Illinois, USA
[63]University of Brussels, Brussels, Belgium
[64]Royal Holloway, University of London, Egham, United Kingdom
[65]The John Adams Institute for Accelerator Science, Royal Holloway, University of London, Egham, United Kingdom
[66]Instituto Superior Técnico, Lisbon, Portugal
[67]Argonne National Laboratory, Argonne, USA
[68]Universitat Politècnica de València, Valencia, Spain
[69]Universitat Politècnica de Catalunya, BarcelonaTech, Barcelona, Spain
[70]University of Oxford, Oxford, United Kingdom
[71]University of Geneva, Geneva, Switzerland
[72]Universität Bonn, Bonn, Germany
[73]CEA, IPhT, Saclay, France
[74]Aalto University School of Science and Technology, Espoo, Finland
[75]University of Oslo, Oslo, Norway

[76]*Université Libre de Bruxelles, Brussels, Belgium*
[77]*Bogazici University, Istanbul, Turkey*
[78]*KU Leuven, Leuven, Belgium*
[79]*RWTH Aachen University, Aachen, Germany*
[80]*Yale University, New Haven, USA*
[81]*University of Twente, Enschede, the Netherlands*
[82]*Université catholique de Louvain, Louvain-la-Neuve, Belgium*
[83]*Universidad Autónoma de Madrid, Madrid, Spain*
[84]*Fusion for Energy, The European Joint Undertaking for ITER and the Development of Fusion Energy, Barcelona, Spain*
[85]*Idaho State University, Pocatello, USA*
[86]*Université de Toulouse, Toulouse, France*
[87]*Uppsala University, Uppsala, Sweden*
[88]*Universidade de Vigo, Vigo, Spain*
[89]*GSI Helmholtzzentrum für Schwerionenforschung, Darmstadt, Germany*
[90]*Humboldt-Universität zu Berlin, Berlin, Germany*
[91]*Princeton University, Princeton, USA*
[92]*Università degli Studi di Padova, Padua, Italy*
[93]*Technische Universität Bergakademie Freiberg, Freiberg, Germany*
[94]*Institute for Research in Fundamental Sciences, Tehran, Iran*
[95]*National and Kapodistrian University of Athens, Athens, Greece*
[96]*Helsinki Institute of Physics, Helsinki, Finland*
[97]*Università degli Studi di Napoli Federico II, Naples, Italy*
[98]*Università degli Studi di Roma La Sapienza, Rome, Italy*
[99]*University of Michigan, Ann Arbor, USA*
[100]*STFC Rutherford Appleton Laboratory, Didcot, United Kingdom*
[101]*Nigde University, Nigde, Turkey*
[102]*The Cockcroft Institute, Daresbury, United Kingdom*
[103]*University of Southampton, Southampton, United Kingdom*
[104]*University of Liverpool, Liverpool, United Kingdom*
[105]*Quaid-i-Azam University, Islamabad, Pakistan*
[106]*High Energy Accelerator Research Organization, KEK, Tsukuba, Japan*
[107]*University of Iowa, Iowa City, USA*
[108]*University of Texas at Arlington, Arlington, USA*
[109]*McGill University, Montreal, Canada*
[110]*National Taiwan University*
[111]*Stanford University, Palo Alto, USA*
[112]*Rutgers, the State University of New Jersey, Piscataway, USA*
[113]*Kahramanmaras Sutcu Imam University, Kahramanmaras, Turkey*
[114]*California Institute of Technology, Pasadena, USA*
[115]*University of Cambridge, Cambridge, United Kingdom*

# Contents

# Executive Summary

This report forms part of the Conceptual Design Report (CDR) of the Compact LInear Collider (CLIC). The CLIC accelerator complex is described in a separate CDR volume [1]. A third document, to appear later, will assess strategic scenarios for building and operating CLIC in successive centre-of-mass energy stages. It is anticipated that CLIC will commence with operation at a few hundred GeV, giving access to precision standard-model physics like Higgs and top-quark physics. Then, depending on the physics landscape, CLIC operation would be staged in a few steps ultimately reaching the maximum 3 TeV centre-of-mass energy. Such a scenario would maximise the physics potential of CLIC providing new physics discovery potential over a wide range of energies and the ability to make precision measurements of possible new states previously discovered at the Large Hadron Collider (LHC).

The main purpose of this document is to address the physics potential of a future multi-TeV $e^+e^-$ collider based on CLIC technology and to describe the essential features of a detector that are required to deliver the full physics potential of this machine. The experimental conditions at CLIC are significantly more challenging than those at previous electron-positron colliders due to the much higher levels of beam-induced backgrounds and the 0.5 ns bunch-spacing. Consequently, a large part of this report is devoted to understanding the impact of the machine environment on the detector with the aim of demonstrating, with the example of realistic detector concepts, that high precision physics measurements can be made at CLIC. Since the impact of background increases with energy, this document concentrates on the detector requirements and physics measurements at the highest CLIC centre-of-mass energy of 3 TeV. One essential output of this report is the clear demonstration that a wide range of high precision physics measurements can be made at CLIC with detectors which are challenging, but considered feasible following a realistic future R&D programme.

A pre-release of this report was reviewed by an international review committee in autumn 2011 [2]. Comments from the review have been taken into account in the current document.

## Overview of this Report

This volume of the CLIC CDR commences with an overview of the physics potential at CLIC. It then describes the machine backgrounds and the corresponding detector requirements which must be met to fully exploit the CLIC physics potential. Two detector concepts are presented. They combine high precision measurement capability with the ability to operate in the CLIC background environment. These two detector concepts, which are designed around highly granular calorimeters, build heavily on those under study for the International Linear Collider (ILC) [3]. The detector subsystems are then described in more detail, concentrating on the aspects relevant to operation at CLIC. Results from detailed simulation studies are presented and, in many cases, these results are corroborated by test beam experiments. The principal engineering challenges, such as the magnet systems, the electronics, the design of the interaction region and detector integration are also addressed. The overall performance of the CLIC detector concepts are presented in the context of detector benchmark processes. These detailed physics studies, which use full detector simulation and reconstruction including the main machine backgrounds, demonstrate the effectiveness of the background rejection strategy developed in this report. They also provide a clear demonstration of the physics potential at CLIC in terms of precision mass and cross section measurements. Finally an overview of future plans of the CLIC detector and physics study is given, highlighting the R&D prospects.

## Physics Potential

At present (end 2011), the LHC has accumulated approximately 5 $fb^{-1}$ of integrated luminosity per experiment at a centre-of-mass energy of 7 TeV. With the subsequent increase in energy to 14 TeV and significant increases in integrated luminosity, the LHC provides a large discovery potential in proton-proton interactions. A high-energy $e^+e^-$ collider is the best option to complement and to extend the LHC physics programme. A lepton collider gives access to additional physics processes, beyond those observable at the LHC, and therefore provides new discovery potential. It can also provide complementary and/or more precise information about new physics uncovered at the LHC. This information would provide powerful discrimination between models of new physics and their fundamental parameters. These aspects of CLIC physics are illustrated in this report using examples which include: Higgs physics within and beyond the Standard Model (SM); supersymmetry and alternative models such as Higgs strong interactions; a possible new $Z'$ sector; contact interactions and extra dimensions. Examples of where the physics potential can be enhanced with polarised electron and positron beams are provided. Whilst the CLIC physics potential is generally viewed in the light of complementarity with LHC, this report also highlights the predictive power of precision measurements at CLIC to distinguish between models of new physics and to provide insight into deeper questions like dark matter and grand unification.

The optimal choice of the CLIC centre-of-mass energy and the possible energy stages will be driven by LHC results and is, therefore, not currently known. This report principally addresses the 3 TeV CLIC physics case. This corresponds to the highest energy currently assumed for the CLIC accelerator complex. The choice to study primarily 3 TeV concurrently addresses the ultimate physics reach of CLIC and corresponds to the most challenging experimental conditions.

## Experimental Environment

Many of the interesting physics processes at CLIC are likely to have small cross sections, typically in the few femtobarn range. Therefore, together with the quest for high acceleration gradients to reach multi-TeV collision energies, the need for high luminosities is a driving factor in the CLIC accelerator design. One consequence of the small bunch sizes required to achieve high luminosities at CLIC, is the phenomenon of the strong electromagnetic radiation (beamstrahlung) from the electron and positron bunches in the high field of the opposite beam. These beamstrahlung effects, which are largest at the highest centre-of-mass energy, have a major impact on the effective luminosity spectrum of a 3 TeV machine which has a peak around 3 TeV and a long tail towards lower energies. For 3 TeV operation at a total luminosity of $5.9 \cdot 10^{34}$ cm$^{-2}$s$^{-1}$, the luminosity in the most energetic 1% fraction of the spectrum is $2.0 \cdot 10^{34}$ cm$^{-2}$s$^{-1}$. Most physics measurements at CLIC will be significantly above production threshold and will therefore profit from the majority of the total luminosity.

The CLIC bunch structure corresponds to 50 bunch trains per second, occurring at 20 ms time intervals. Each 156 ns long bunch train consists of 312 distinct bunch crossings separated by 0.5 ns. Beamstrahlung results in the creation of a large background of $e^+e^-$ pairs which are predominantly produced in forward directions and with low transverse momenta. Whereas numerous background particles from beamstrahlung will be created in every single bunch crossing, only a small fraction of the bunch trains will produce a hard $e^+e^-$ physics interaction. The presence of the pair background mainly impacts the design of the very forward region of the detector and results in potentially high detector occupancies in the inner layers of the vertex detector and in the forward tracking detectors. The main source of background particles with higher transverse momenta is multi-peripheral hadronic two-photon interactions (where the photons can either be virtual or originate from beamstrahlung). For 3 TeV operation, on average there are 3.2 $\gamma\gamma \rightarrow$ hadrons interactions for each bunch crossing. The pile-up of this background over the entire 156 ns bunch-train deposits 19 TeV of energy in the calorimeters, of which approximately 90% occurs in the endcap and 10% in the barrel regions of the calorimeters. The presence of the $\gamma\gamma \rightarrow$ hadrons background is a major consideration for the design of a CLIC detector and its readout.

## CLIC Detector Concepts

As is the case for the ILC, it is assumed that CLIC will have a single interaction point and that two detectors share the interaction point, moving in and out of the beam a few times per year using a so-called push-pull system. The detector performance requirements at CLIC are determined by the precision physics aims at an $e^+e^-$ collider, and imply challenging goals for jet energy resolution, track momentum resolution, impact parameter resolution, flavour tagging performance, lepton identification as well as detection of electrons down to small angles. The principal factors driving the overall design of a detector at CLIC are the requirement of excellent jet energy resolution and the need to efficiently identify and reject calorimeter energy depositions from beam-induced background. Highly segmented calorimetry, optimised for particle flow techniques which make optimal use of tracking and calorimeter information, meets these design requirements [4, 5].

In high granularity particle flow calorimetry, individual particles formed from charged particle tracks and/or calorimeter information are reconstructed. Jets are formed from these reconstructed particles. The use of tracking information reduces the dependence on hadronic calorimetry and results in the required excellent jet energy and di-jet mass resolution. At the same time, timing information in the calorimeters and tracking detectors allows the individual reconstructed particles to be associated with a relatively small range of bunch crossings within a 156 ns bunch train. Background particles from $\gamma\gamma \rightarrow$ hadrons, which have relatively low transverse momenta and are predominantly produced in the forward direction, are distributed uniformly in time within the bunch train. Therefore, by combining timing and transverse momentum information of fully reconstructed particles, the background from $\gamma\gamma \rightarrow$ hadrons can be separated effectively from the particles in the physics event. One of the main conclusions of this CDR is that a detector based on high granularity particle flow calorimetry can meet the jet energy resolution goals at CLIC and provides a robust way of mitigating the impact of background.

The general purpose ILD [4] and SiD [5] detector concepts, developed in the context of the 500 GeV ILC, form an excellent starting point for a detector design for CLIC as demonstrated in an earlier study [6]. These concepts, which are based on high granularity particle flow calorimetry, are the result of many years of physics simulation and detector optimisation studies, combined with a wealth of worldwide hardware development. In 2009 the ILD and SiD concepts were reviewed and validated by the IDAG panel of international experts [7]. Modified versions of the ILD and SiD detector concepts, CLIC_ILD and CLIC_SiD, form the basis for the studies in this report. The main changes with respect to the ILC case are modifications of vertex detectors and very forward detector regions, to mitigate the impact of backgrounds, and increased hadron calorimeter depth. The detailed detector simulation and sophisticated reconstruction software developed for the ILD and SiD concepts underlies the studies presented in this report.

The overall structures and calorimeters of the CLIC_ILD and CLIC_SiD detector concepts are similar. Both concepts have a barrel and endcap geometry with the barrel calorimeters and tracking detectors located inside a superconducting solenoid which provides an axial magnetic field of 4 T in CLIC_ILD and 5 T in CLIC_SiD. The diameter and length are about 14 m and 13 m respectively for both detectors, including the return yoke.

In both cases the electromagnetic calorimeters are silicon-tungsten sampling calorimeters with 30 samplings in depth and have silicon cells of 25 $mm^2$ and 13 $mm^2$ for CLIC_ILD and CLIC_SiD respectively. The hadron calorimeters are also highly granular sampling calorimeters, with either analog ($3\times3$ $cm^2$) or digital ($1\times1$ $cm^2$) readout, and have 60 to 75 sampling layers in depth. The passive material in the HCAL is taken to be steel in the endcap region and, for reasons of compactness, tungsten in the barrel region. For both concepts, the overall depth of the calorimeter system is chosen to be approximately 8.5 interaction lengths to avoid loss of energy resolution due to leakage. In both cases the superconducting solenoids are surrounded by thick iron yokes which have instrumented gaps allowing to measure punch through from high-energy hadronic showers and to identify muons.

Charged particle track reconstruction in the CLIC_ILD concept is based on a Time Projection Chamber (TPC) with a large radius of 1.8 m for increased spatial separation of calorimeter deposits to support particle flow. The TPC allows for highly redundant continuous tracking with relatively little material in the tracking volume itself. Tracking in CLIC_ILD is supplemented by silicon pixel and micro-strip detectors and forward tracking disks. The CLIC_SiD concept has a compact all-silicon tracking system with an outer radius of approximately 1.3 m in a stronger magnetic field, providing relatively few but highly accurate space points. Unlike a TPC which necessarily integrates over the entire bunch train, the CLIC_SiD tracker has the advantage of fast charge collection. The vertex detectors of both concepts are based on silicon technology with pixels of approximately 20 µm × 20 µm. To minimise the effect of multiple scattering a very low material budget is assumed for the vertex and tracking detectors.

The required track momentum and impact parameter resolutions of $\sigma_{p_T}/p_T^2 \approx 2 \cdot 10^{-5}$ GeV$^{-1}$ and $\approx 5$ µm respectively are achievable for high-momentum tracks. The CLIC_ILD and CLIC_SiD tracking and calorimeter systems give a jet energy resolution of approximately 3.5% for high-energy jets using PFA reconstruction.

The detailed design of the forward region of a detector at CLIC must accommodate final-focus machine elements, which require excellent mechanical stability. In both the CLIC_ILD and CLIC_SiD the forward region is instrumented with two low angle electromagnetic calorimeters, one providing the absolute luminosity and the other for tagging very low angle electrons.

## Timing Requirements

To operate with the 0.5 ns bunch spacing at CLIC, the subdetector systems must provide precise hit timing information in order to suppress the relatively high levels of beam-induced background. This report presents a scheme for background suppression based on time-stamping capabilities assumed to be 10 ns for all silicon tracking elements and a hit time resolution of 1 ns for all calorimeter hits. Even if practically achievable, simply imposing tight timing cuts at the hit level does not provide a viable solution; it can neither account for the finite hadronic shower development time nor for precise time-of-flight corrections for lower momentum particles. A two-stage approach is adopted. In the first stage, all raw detector hits in a 10 ns time window[1] are used as input to the offline reconstruction. Here the combination of the high-granularity calorimeters and particle flow reconstruction allows hits from a single particle to be clustered together. The combined timing information of the calorimeter hits in a cluster allows a precise time to be assigned to each reconstructed particle. Cuts on the reconstructed time and $p_T$ of the fully reconstructed particles enable the background from $\gamma\gamma \rightarrow$ hadrons to be reduced from 19 TeV per bunch train to approximately 100 GeV per reconstructed physics event. This impressive level of background rejection is achieved without significantly impacting the detector performance.

## Physics Benchmarks

Six detector benchmark processes are studied using full event simulation and reconstruction in the CLIC_ILD and CLIC_SiD detectors. The benchmark processes are chosen to address different aspects of the detector performance and collectively they cover a wide range of measurements of Standard Model and new physics signatures at a high energy $e^+e^-$ collider. The prime purpose of the benchmark studies is to demonstrate overall capability of the detector concepts to deliver precision physics measurements in the CLIC environment. Consequently, an essential part of the studies is the inclusion of machine background from $\gamma\gamma \rightarrow$ hadrons and the adoption of the treatment of timing information described above. Five benchmark processes are studied at 3 TeV in order to assess the impact of background in the most extreme case. These are: Standard Model Higgs production and subsequent decay into b-quarks, c-quarks and muon pairs; heavy neutral and charged Higgs production in SUSY; production of squarks; chargino and neutralino pair production; and slepton production. In addition, the production and decay of top pairs is studied with the 500 GeV CLIC machine parameters to allow a comparison with the sensitivity of the top mass measurement for the ILC operating at 500 GeV. The studies at 3 TeV assume an integrated luminosity of 2 ab$^{-1}$ based on four years of operation of a fully commissioned machine with 200 days running per year with an effective up-time of 50%.

[1] A 100 ns time window is used for the HCAL barrel to account for the slower shower development time in tungsten.

The results of the six benchmark studies confirm that the two detector concepts considered here provide the excellent performance required at CLIC and demonstrate that the impact of beam-induced background can be effectively mitigated. As a result, in all six studies considered, the production cross sections and particle masses can be measured very precisely at CLIC. The Higgs coupling to b-quarks can be measured with sub-percent statistical precision, and the branching ratio for the rare Higgs decay into muons can be measured with 15% accuracy. Masses of SUSY heavy Higgs particles in the 700 to 900 GeV mass range can be determined with $\approx 0.3\%$ statistical accuracy. Similarly, masses of heavy squarks, selectrons, smuons, sneutrinos, charginos and neutralinos can be measured with statistical accuracies better than 1%, while the cross sections for specific decay modes can be measured with a precision better than 3% at a 3 TeV CLIC machine. Top quark production at a 500 GeV CLIC machine can be studied with a high precision using a moderate integrated luminosity of 100 $fb^{-1}$, for example the top mass can be measured with a statistical precision below 100 MeV. The results of the physics measurements at CLIC will depend on the knowledge of the luminosity spectrum, which can be measured in situ using wide-angle Bhabha scattering events.

## Future Detector Research and Development

The CLIC_ILD and CLIC_SiD concepts, described in this conceptual design report, pose technological challenges for almost all detector systems. Many of these challenges are already being addressed in the framework of a broad international linear collider detector R&D effort, other aspects which are more specific to CLIC require a dedicated R&D programme. One important example is the vertex detector, where the combination of small pixel sizes, time-stamping capabilities and ultra-low material budgets is extremely challenging. To achieve the required momentum resolution, the tracking systems require extremely low-mass designs and power dissipation has to be minimised to allow for low-mass cooling approaches. To reduce power consumption it is assumed that many of the detector systems will operate with power pulsing greatly reducing the power consumption in the 20 ms idle time between bunch train crossings. However power pulsing requires extensive R&D, followed by thorough system testing. Similarly, the highly granular calorimeters require advanced R&D towards very compact and cost-effective detection layers with high performance at low power. The high field superconducting magnet systems require R&D on reinforced conductors and on movable service lines and safety systems adapted to the detector push-pull scheme. Several engineering challenges also need to be addressed, such as detailed detector design and integration, detector movements, the integration of the forward region with the accelerator elements, a hybrid beam pipe with thin beryllium and thick steel elements, and also precision alignment techniques. In parallel, a programme of sophisticated detector simulation studies will be pursued to optimise the detector layouts and to reduce further the impact of beam-induced backgrounds. These simulation studies will require continuous improvements to software tools to allow for more detailed studies of specific aspects of the detector design at CLIC.

## Synopsis

This report demonstrates that a future high energy $e^+e^-$ collider, based on CLIC technology, has a very broad physics potential, complementing the LHC, both in discovery potential and in the ability to provide high precision measurements. Detailed simulation studies, based on the CLIC_ILD and CLIC_SiD detector concepts, demonstrate that the CLIC detector goals are achievable even at the highest foreseen CLIC centre-of-mass energy of 3 TeV where background conditions are most challenging. The detector benchmark studies demonstrate that very precise measurements can be made at CLIC for a wide range of physics event topologies and at the highest and also at lower centre-of-mass energies. These studies pave the road to a future TeV-scale $e^+e^-$ collider constructed in a few centre-of-mass energy stages. In a next stage of the CLIC study, the physics potential of CLIC at different centre-of-mass energies will be explored further, closely following the evolution of the physics landscape with the forthcoming LHC results.

# Chapter 1

# CLIC Physics Potential

## 1.1 Introduction

The LHC will allow unprecedented exploration of the mechanism of electroweak symmetry breaking and physical phenomena at the TeV scale. We do not yet know what is lurking at these energy scales, but it is quite likely that the discoveries made at the LHC will alter our present views of the particle world. It is thus not easy to predict with certainty what will be the priorities of particle physics and the main open questions that high-energy physics will have to address after the LHC. And yet, unanswered questions will certainly emerge, since the LHC cannot resolve all the issues related to the TeV region.

In spite of this inherent uncertainty, our present knowledge of the Standard Model (SM) and the many experimental constraints on new theories allow us to envision plausible scenarios for which we can assess the usefulness of CLIC in pushing forward research in particle physics beyond the results that will be reached by the LHC. Thus, the goal of this Chapter is not to give a comprehensive review of the multitude of existing theories beyond the SM as most – if not all – of them will be obsolete by the time CLIC starts operating. Rather the goal is to study a few topical and motivated prototypes of new physics scenarios and draw general lessons on the capabilities of CLIC to address fundamental questions that will likely arise in the post-LHC era.

The Higgs boson is the most plausible anticipated discovery at the LHC, but even if this particle is found, and its mass and couplings are measured, its true nature will still not be known fully. Contrary to the other SM particles, which snugly fit into highly symmetric structures, the quantum numbers and characteristics of the Higgs boson are puzzling when viewed in the context of a unified theory that may supersede the SM. The Higgs discovery will then bring to the forefront questions about the nature of this particle: is it a fundamental particle or a composite? Is it part of a more complicated electroweak sector? Does it universally couple to all matter proportionally to mass? The LHC can only partially answer these questions. As discussed in Section 1.2 (for an elementary Higgs boson) and in Section 1.4 (for a composite Higgs boson), CLIC can explore these issues in much greater depth and unravel these questions by measuring the Higgs couplings to an unprecedented precision.

Supersymmetry is often considered an attractive option to deal with the naturalness problem of the Higgs boson. If supersymmetry indeed lies near the weak scale, the LHC is bound to discover it. However, it is implausible that the LHC can resolve all questions related to supersymmetry. Heavy sleptons, neutralinos and charginos can only be produced copiously at the LHC through decay chains of strongly-interacting supersymmetric particles and, in some cases, these chains do not access all states. On the other hand, CLIC can explore thoroughly the TeV region, looking for any new particles with electroweak charges. The precise mass and coupling measurements that can be performed at CLIC are crucial to address fundamental questions about the mechanism of supersymmetry breaking, about aspects of unification, and about the viability of the lightest supersymmetric particle as a dark matter thermal relic. These issues are discussed in Section 1.3.

It is important to emphasise that the improved level of precision measurements that can be reached at CLIC, with respect to the LHC, is not just a sterile refinement. We have learned from the past that very often in physics a quantitative improvement in a measurement leads to a qualitative jump in the understanding of the underlying physics. For instance, building evidence for grand unification in supersymmetry was made possible only by very precise measurements of the three gauge couplings. As another example, the indirect information about the lightness of the Higgs boson came from refined LEP electroweak data. Similarly, the accuracy that can be reached at CLIC is likely to open new avenues in our understanding of the particle world. For example, discovery of rare top or Higgs decays may give us the necessary hint to crack the flavour puzzle, and precise measurements of new particle properties may be the key to understanding the structure of the new theory and solving outstanding mysteries, such as the nature of dark matter. $Z'$ and Contact Interactions are discussed in Section 1.5. The impact of beam polarisation is discussed in Section 1.6. And, the physics potential of CLIC precision measurements is discussed in Section 1.7.

In the next sections, the physics capabilities of CLIC will be illustrated in several broad categories: Higgs physics, supersymmetry, strongly interacting electroweak theories, and precision physics. In short, the emphasis of this physics study is on a 3 TeV $e^+e^-$ option, with total (top 1% part of spectrum) peak luminosity being $5.9 \cdot 10^{34}\,\mathrm{cm}^{-2}\,\mathrm{s}^{-1}$ ($2.0 \cdot 10^{34}\,\mathrm{cm}^{-2}\,\mathrm{s}^{-1}$). Unlike some other Chapters e.g. in the benchmarking of detectors, these parameters are not always rigorously held to in our discussion, since it is envisioned that on this front the collider will have some flexibility in design and planning. Alternate energy or luminosity needs will be pointed out when it may be helpful for maximal study of new physics phenomena.

## 1.2 Higgs

In the SM, there is only one Higgs particle (for reviews see [1, 2]) with a mass that is expected to lie in the range $M_\mathrm{H} = 114 - 160$ GeV at the 95% confidence level from high precision data and pre-LHC direct searches [3]. Recent data from the LHC at 7 TeV continues to restrict the allowed mass range. In addition, various theoretical arguments such as perturbative unitarity, constrain $M_\mathrm{H}$ to be smaller than approximately 1 TeV. Thus, such a Higgs particle is expected to be observed at the LHC [4, 5]. The added value of an electron–positron collider would be to measure in great detail its fundamental properties [6, 7]: its mass and total decay width, its spin–parity quantum numbers, its couplings to fermions and gauge bosons, and its self couplings that allows one to reconstruct the scalar potential that is responsible of electroweak symmetry breaking. Some of these measurements are very hard, if not impossible, to perform in the complicated environment of a hadron machine [2].

Although an extended Higgs sector may exist in different models of new physics, it is within supersymmetry (SUSY) that it is best justified. Indeed, in supersymmetric extensions of the SM the Higgs sector is enlarged to contain at least five scalar particles: two CP–even h, H bosons, a CP–odd or pseudoscalar A boson and two charged $\mathrm{H}^\pm$ particles. A comprehensive review can be found in [8]. This is the case of the minimal model, the MSSM, where only two parameters are needed to describe the Higgs sector at tree–level: the ratio $\tan\beta$ of vacuum expectation values of the two Higgs fields needed to break the electroweak symmetry and the pseudoscalar mass $M_\mathrm{A}$. In most cases, the lightest neutral CP–even state has almost exactly the properties of the SM Higgs particle but with a mass restricted to values $M_\mathrm{h}^\mathrm{max} \approx 110 - 130$ GeV depending on the radiative corrections that enter the Higgs sector. For a recent analysis, see [9, 10]. The other CP–even state H and the charged Higgs boson are in general degenerate in mass with the pseudoscalar Higgs particle, $M_\mathrm{H} \approx M_{\mathrm{H}^\pm} \approx M_\mathrm{A}$, and for values of $\tan\beta \gg 1$, have the same very strongly enhanced couplings to bottom quarks and $\tau$–leptons ($\propto \tan\beta$) and suppressed couplings to top quarks and gauge bosons. For large enough A masses, $M_\mathrm{A} \gtrsim 0.5 - 1$ TeV, these states could escape detection at the LHC but could be observed and studied at a high–energy $e^+e^-$ collider with a centre-of-mass energy $\sqrt{s} \gtrsim 2 \cdot M_\mathrm{A}$ [6, 7].

Extensions of the MSSM in which some basic assumptions are relaxed, such as the absence of new sources of CP–violation and/or minimal gauge group or particle content, lead to different Higgs phenomenology. Models like the next–to–MSSM where an additional singlet field is present (see e.g. Ref. [11]), or the CP–violating MSSM where new phases alter the Higgs mass pattern, also can be studied at CLIC [12]. Many other supersymmetric and non–supersymmetric extensions of the SM predict a rich Higgs spectrum. Some models predict the existence of new gauge bosons and fermions that can alter the phenomenology of the Higgs particles. This is the case, for instance, of Grand Unified Theories, in which neutral gauge bosons $Z'$ may survive to the TeV scale [13], and extra–dimensional models [14, 15] in which heavy Kaluza-Klein excitations of the gauge bosons are present. Additional gauge bosons arise naturally through the Stueckelberg mechanism of mass generation and can give rise to very narrow resonances at colliders [16]. Within MSSM extensions the Stueckelberg sector mixes with the Higgs sector, and the neutralino sector is extended to include additional mass mixing and kinetic mixings [17]. Extensions of the SM with a Higgs singlet and kinetic mixing lead to narrow resonances [18] and can have significant impact on the Higgs sector [19]. CLIC would have the unprecedented ability to precisely probe the predictions of the models above. In the situation in which these new states have masses below the CLIC centre-of-mass energy, new Higgs production channels such as decays $\rm Z' \to HZ^0$, could occur and would allow the simultaneous study of the Higgs and new gauge bosons.

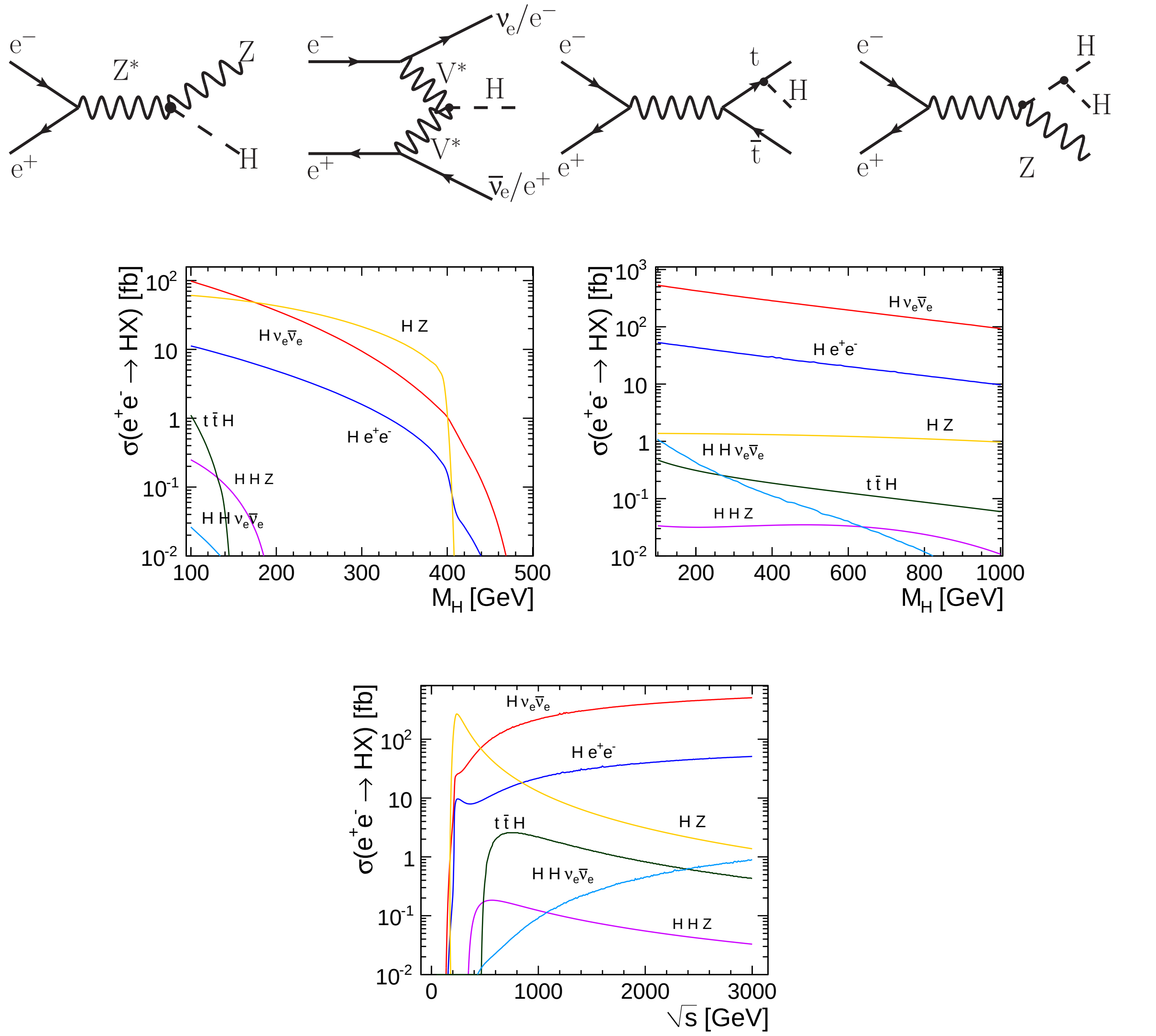

Fig. 1.1: Production mechanisms of the SM Higgs boson at CLIC (top); the total cross sections as a function of $M_{\rm H}$ for $\sqrt{s} = 0.5$ TeV (middle-left), and 3 TeV (middle-right), and cross sections as a function of $\sqrt{s}$ for $M_{\rm H} = 120$ GeV (bottom).

In this Section, we will briefly summarise the potential of CLIC with a centre-of-mass energy up to 3 TeV and with a few ab$^{-1}$ integrated luminosity to study the Higgs sector in the SM and some of its extensions. Some features have been discussed in an earlier CLIC report [20] while for some specific topics, more details will appear in a companion report [21].

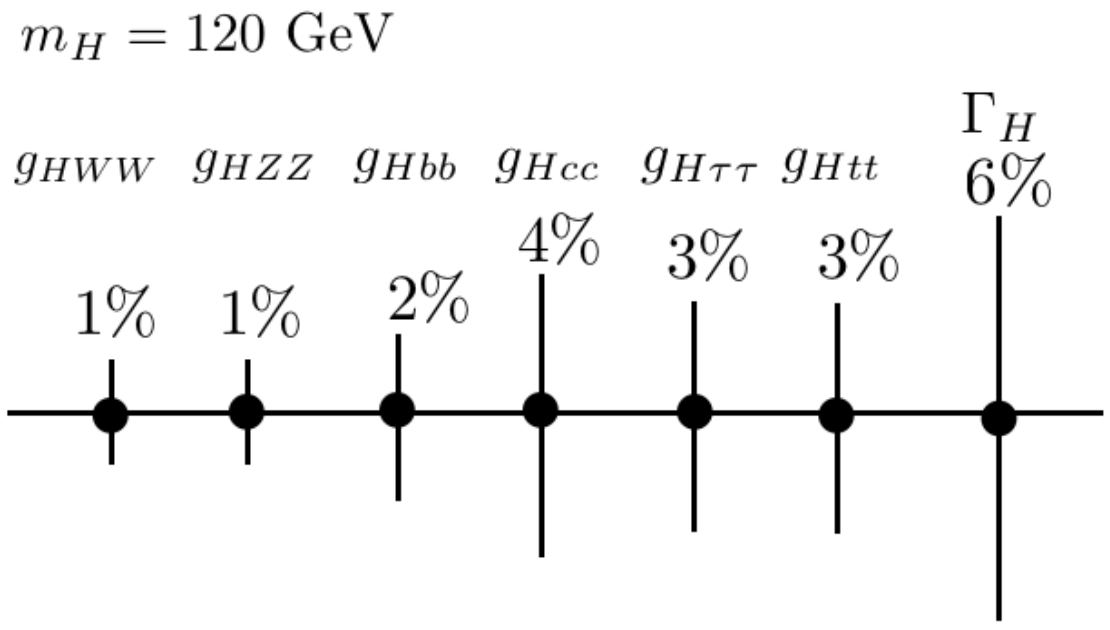

| | Coupling determination(%) | Sensitivity to SM deviation (%) |
|---|---|---|
| Hbb | 2 | 4 |
| Hcc | 3 | 6 |
| H$\mu\mu$ | 15 | 15 |

Fig. 1.2: Relative error in the Higgs boson coupling determination to different particle species. The top diagram is for a Higgs mass of 120 GeV at $\sqrt{s} = 500$ GeV and with 500 $\mathrm{fb}^{-1}$ of integrated luminosity, except $g_{\mathrm{Ht\bar{t}}}$, which is obtained at $\sqrt{s} = 800$ GeV with 1 $\mathrm{ab}^{-1}$. The bottom table gives coupling constant determination and sensitivity to deviations from the SM obtained at CLIC 3 TeV with 2 $\mathrm{ab}^{-1}$ for 120 GeV Higgs boson mass (see text for further explanation).

### 1.2.1 The Higgs Boson in the Standard Model

In $e^+e^-$ collisions, the main production mechanisms for the SM Higgs particle are the Higgsstrahlung and the $W^+W^-$ fusion processes, $e^+e^- \to HZ^0 \to f\bar{f}H$ and $e^+e^- \to \nu_e\bar{\nu}_eH$, see Figure 1.1. Besides the $Z^0Z^0$ fusion mechanism $e^+e^- \to e^+e^-H$, which has an order of magnitude smaller rate than the twin $W^+W^-$ fusion process, sub–leading Higgs production channels are associated production with top quarks $e^+e^- \to t\bar{t}H$, double Higgs production in the Higgsstrahlung $e^+e^- \to HHZ^0$ and fusion $e^+e^- \to \nu\bar{\nu}HH$ processes. Despite the smaller production rates, the latter mechanisms are very useful when it comes to the study of the Higgs properties such as the Higgs Yukawa couplings to $f\bar{f}$ and the Higgs and self–couplings. The production rates for all these processes are shown in Figure 1.1 at centre-of-mass energies of $\sqrt{s} = 0.5$ and 3 TeV as a function of $M_{\mathrm{H}}$. The production cross sections of a 120 GeV Higgs boson are also shown as a function of $\sqrt{s}$.

The cross section for Higgsstrahlung scales as $1/s$ and therefore dominates at low energies, whereas the $W^+W^-$ fusion cross section rises like $\log(s/M_{\mathrm{H}}^2)$ and becomes more important at high energies. At $\sqrt{s} \sim 500$ GeV, the two processes have approximately the same cross sections for the favoured mass range $M_{\mathrm{H}} \approx 115-150$ GeV. In Higgsstrahlung, the recoiling $Z^0$ boson, which can be tagged through its clean $\ell^+\ell^-$ decays, is mono–energetic and the Higgs mass can be derived from the $Z^0$ energy when the initial $e^{\pm}$ beam energies are well-defined. This process allows very accurate determinations of the Higgs properties. It has been shown in detailed simulations [6] that, for $M_{\mathrm{H}} = 115-150$ GeV, running at centre-of-mass energies in the range $350-500$ GeV with a few 100 $\mathrm{fb}^{-1}$ data allows a very precise measurement of the Higgs mass[1], as well as the total decay width, the spin–parity quantum numbers and the Higgs couplings to the (W/Z) gauge bosons, the light (b, c, $\tau$) fermions and to gluons (see Figure 1.2). The $t\bar{t}H$ and the $\lambda_{\mathrm{HHH}}$ Higgs self-coupling are challenging measurements. Studies are underway to determine if the $t\bar{t}H$ and the HHH couplings can be determined at nearly the 3% and 20% levels, as has been previously reported in a different context [20].

[1]See Section 2.2.1, Figure 2.4, for an assessment at a $\sqrt{s} = 3$ TeV CLIC machine.

Some of the measurements above can significantly benefit from an increase of the centre-of-mass energy. At $\sqrt{s} \sim 3$ TeV, the cross section for the $W^+W^-$ fusion process becomes very large and the data sample that can be obtained with luminosities at the $ab^{-1}$ level would allow more precise and complementary measurements. A few examples are given below.

i.) At $\sqrt{s}$ =3 TeV, about 240 $H \to \mu^+\mu^-$ events can be collected for $M_H \approx 120$ GeV with $\mathscr{L} = 2$ ab$^{-1}$ after full reconstruction (Figure 1.3 (left), CLIC_SiD detector model, Section 12.4.2). This will allow the measurement of the Higgs couplings to muons to better than 15%. The di-muon signal can be isolated from the backgrounds with a good statistical significance, see Section 12.4.2. Similarly, the rare loop induced $H \to Z^0\gamma$ decay can also be measured with a reasonable accuracy if a large integrated luminosity is collected [20].

ii.) The $H \to b\bar{b}$ branching ratio becomes very small for $M_H \gtrsim 150$ GeV, and at $\sqrt{s} = 500$ GeV it cannot be determined to better than 10% for $M_H \sim 200$ GeV. At $\sqrt{s} = 3$ TeV, the signal to background ratio is very favourable already at $M_H = 120$ GeV, allowing measurement of $\sigma \times B$ to a precision of 0.22% (see the right plot in Figure 1.3 and detailed analysis in Section 12.4.1). This precision is good enough to not be the primary source of coupling constant extraction uncertainties. Rather, theoretical uncertainties, such as higher order correction computations, and parametric uncertainties, such as quark mass values and $\alpha_s$ determinations, are left as the core barrier to a better Higgs coupling determination. Assuming that no other couplings are allowed to deviate from the Standard Model coupling, the $Hb\bar{b}$ coupling can be determined to a relative uncertainty of $\delta_b = \pm 2 - 4\%$ where 2% refers to only theoretical uncertainties and the 4% refers to theoretical and parametric uncertainties [22]. The latter error has more hope of being reduced in time with more data and a global fit of observables.

iii.) Likewise, if the c quark Higgs process were measured to very high precision the $Hc\bar{c}$ coupling could be determined to relative uncertainty of $\delta_c = \pm 2 - 6\%$ under the analogous assumptions as described for $Hb\bar{b}$. However, in this case, the statistical uncertainty on $\sigma \times B$ is 3.2% (see Section 12.4.1) which leads to an additional uncertainty of $\delta_c = \frac{\delta_{expt}}{2(1-B^c_{SM})} \simeq 1.6\%$ that should be folded into the uncertainty above, and primarily affects the lower bound of the uncertainty. Thus we summarise the charm coupling determination to be $\delta_c = \pm 3 - 6\%$.

iv.) The trilinear Higgs coupling can be measured in the $W^+W^-$ fusion process, $e^+e^- \to \nu\bar{\nu}HH$, for which the cross section is a few fb at $\sqrt{s} \sim 3$ TeV. The sensitivity to $\lambda_{HHH}$ can be enhanced by studying the angle $\theta^*$ of the $H^* \to HH$ system in its rest frame: because of the scalar nature of the Higgs, the $\cos\theta^*$ distribution is flat for $H^* \to HH$ while it is forward–peaked for the other diagrams. From a fit of the distribution one can determine the value of $\lambda_{HHH}$. The quadrilinear Higgs coupling, however, remains elusive even at higher energies.

v.) Finally, anomalous Higgs couplings, some of which grow with $\sqrt{s}$, can be best measured at higher energies by taking advantage of the large cross sections in the $W^+W^-$ fusion process. Particularly interesting would be the probe of the WWH [21, 23], $t\bar{t}H$ [21, 24] and HHH [21, 25] couplings.

The higher energy of the collider can also be very useful in the case where the Higgs boson is very heavy. For $M_H \sim 700$ GeV and beyond, the cross sections for the Higgsstrahlung and $W^+W^-$ fusion processes are small at $\sqrt{s} \lesssim 1$ TeV (see Figure 1.1) and do not allow one to perform detailed studies. At CLIC energies, $\sqrt{s} = 3$ TeV, one has $\sigma(e^+e^- \to H\nu\bar{\nu}) \sim 150$ fb which allows for a reasonable sample of Higgs bosons to be studied. The high energy available at CLIC will also be important to investigate in detail a possible strongly interacting Higgs sector scenario, as will be discussed in Section 1.4.

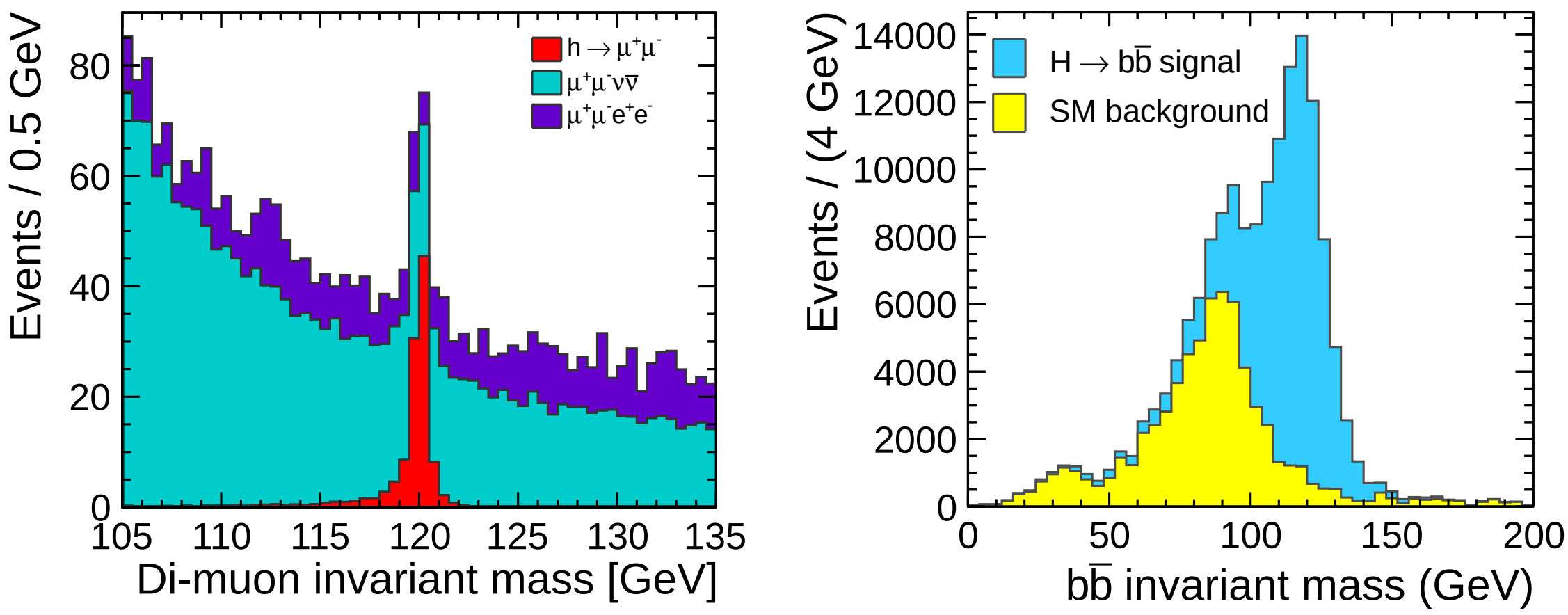

(a) $e^+e^- \to H\nu\bar{\nu}, H \to \mu^+\mu^-$ (b) $e^+e^- \to H\nu\bar{\nu}, H \to b\bar{b}$

Fig. 1.3: Reconstructed sample for two Higgs channels with $M_H = 120$ GeV at CLIC with $\sqrt{s} = 3$ TeV with 2 ab$^{-1}$. The histograms are stacked distributions of signal and background reconstructed using the CLIC_SiD detector (see Chapter 12).

### 1.2.2 The Higgs Bosons of the MSSM

In $e^+e^-$ collisions, besides the usual Higgsstrahlung and fusion processes for h and H production, the neutral MSSM Higgs bosons can also be produced pairwise: $e^+e^- \to A + h/H$. The cross sections for Higgsstrahlung and pair production as well as the cross sections for the production of h and H are mutually complementary. For large values of $M_A$, we are in the decoupling regime where the lightest h boson has the same couplings as the SM Higgs boson and thus large rates are expected in the $W^+W^-$ fusion channel. The A and H states are almost degenerate in mass (see Figure 1.4 left) with vanishing couplings to the SM gauge bosons; the only relevant process is therefore Higgs pair production, $e^+e^- \to HA$ which has a cross section that does not depend on the mixing angle $\alpha$ and $\tan\beta$ and is simply suppressed by the velocity factor $\tilde{\beta}_H^3$ near the kinematical threshold as is typical for two spin–zero particle production. At large $\tan\beta$ values, the A and H states will mostly decay to $b\bar{b}$ and $\tau^+\tau^-$ pairs, with branching fractions of approximately 90% and 10%, respectively. The total Higgs widths are significant (much larger than the $M_H - M_A$ difference) in this case as shown in the central frame of Figure 1.4.

The charged Higgs bosons can also be produced pairwise, $e^+e^- \to H^+H^-$, through $\gamma, Z$ exchange, with a cross section that depends only on the charged Higgs mass; it is large almost up to $M_{H^\pm} \sim \frac{1}{2}\sqrt{s}$. The main decays of heavy $H^\pm$ bosons are into tb and $\tau\nu_\tau$ pairs with again, $\approx 90\%$ and 10% branching fractions. Light $H^\pm$ bosons can also be produced in top decays; in the range $1 < \tan\beta < m_t/m_b$, the $t \to H^+b$ branching ratio and the $t\bar{t}$ production cross sections are large enough to allow for their detection. The cross sections for both HA and $H^+H^-$ production are shown in Figure 1.4 as a function of the masses at $\sqrt{s} = 3$ TeV.

Despite their anticipated tiny cross sections, a few fb at $\sqrt{s} = 3$ TeV as shown in Figure 1.4, the pair production of heavy Higgs bosons, $e^+e^- \to HA$ and $e^+e^- \to H^+H^-$, with their subsequent decays to $b\bar{b}b\bar{b}$ and $t\bar{b}b\bar{t}$, respectively, will give highly characteristic final states [26]. These states can be observed almost up to their kinematical threshold [27].

Since the full energy of the collision is deposited into the detector, the analysis of these processes can benefit from kinematic fits, which generally improve the mass resolution by 30% or more, improving the accuracy on the mass determination and making the Higgs boson total widths accessible down to $\Gamma \sim 10$ GeV. In the case of the $H^+H^-$ process, a further constraint of requiring equal mass for the two decaying bosons can be imposed. This improves both the mass and width determination by another $\simeq 30\%$ for the benchmark scenario studied in Figure 1.5. For the A and H bosons, studied in HA production, the determination of the total width is highly correlated with their mass splitting. However, as shown in Figure 1.4, and in extensive scans of viable points in the MSSM parameter space, the total width of both states is significantly larger than their mass splitting. A detailed discussion of this issue is given in [21].

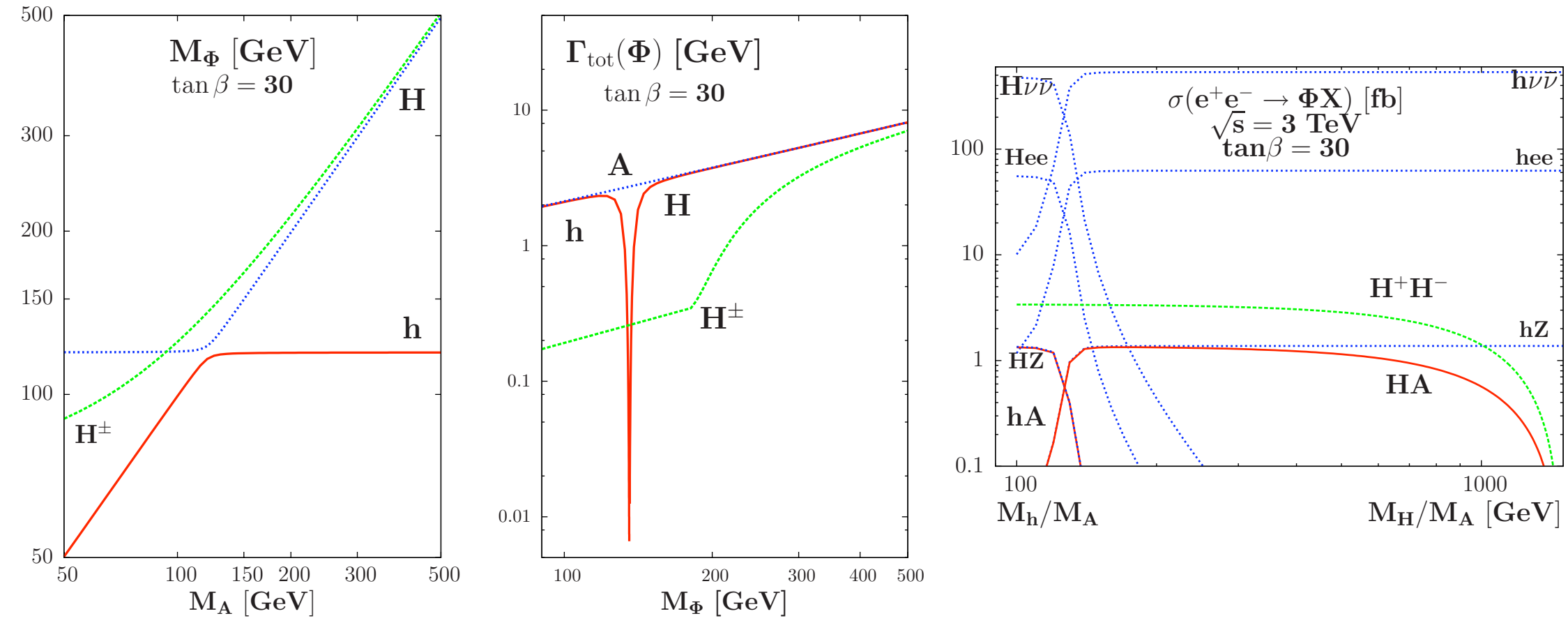

Fig. 1.4: Masses and total decay widths of the MSSM Higgs bosons for $\tan\beta = 30$ and production cross sections in $e^+e^-$ collisions as functions of the masses in $e^+e^-$ collisions at $\sqrt{s} = 3$ TeV; from [8].

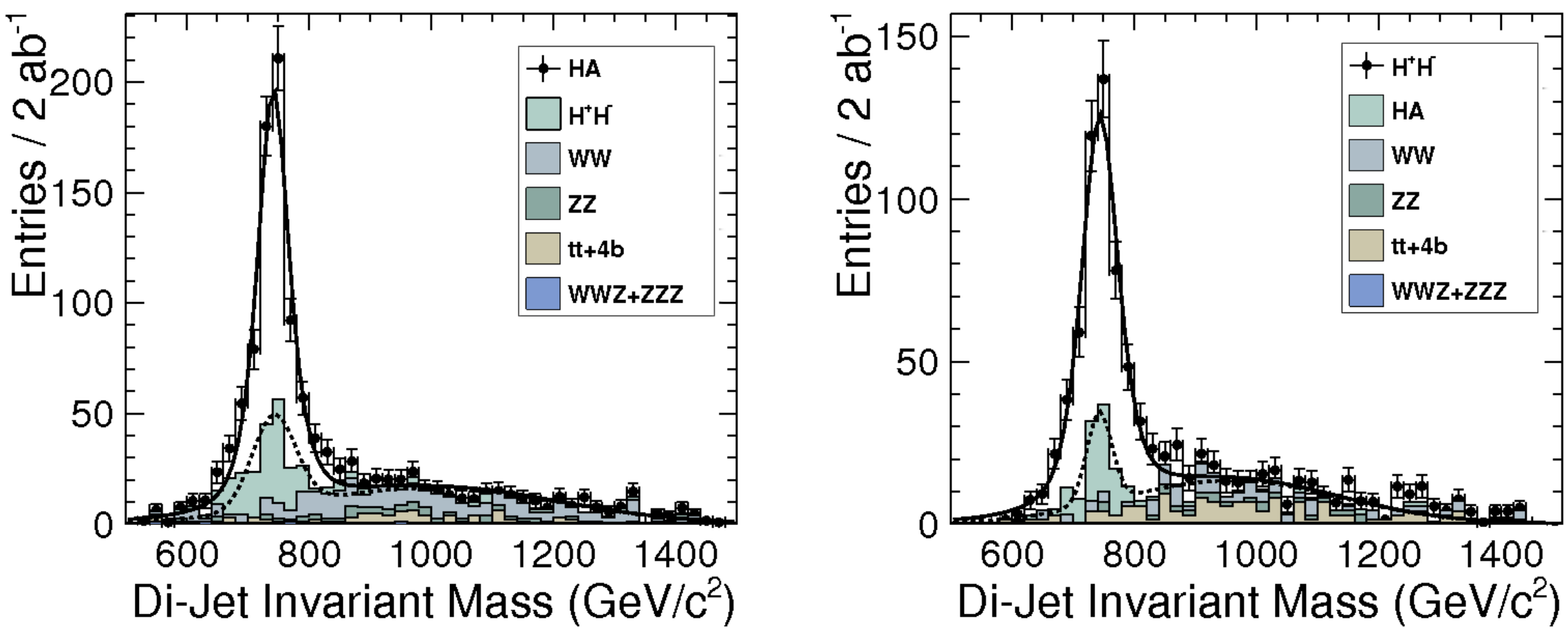

Fig. 1.5: Higgs mass peak reconstruction in the processes $e^+e^- \to$ HA (left), and in $e^+e^- \to H^+H^-$ (right), at a CLIC detector using *model II*, see Section 12.4.3. The corresponding background channels are shown as well. The finite Higgs widths are taken into account.

This justifies the application of equal mass constraints to the HA reconstruction in large $\tan\beta$ scenarios, thus achieving accuracies comparable to those of $H^+H^-$. In summary, the masses and widths of the A/H and $H^\pm$ states can be determined with a relative accuracy of 0.002-0.005 and 0.10-0.15, respectively [28, 29]. In the specific case of *model II* studied as a benchmark reaction for this CDR (see Section 1.3.1), where $\tan\beta$ is large, and $M_\mathrm{A} = 742$ GeV. These constraints on $\Gamma_\mathrm{A}$ and $\Gamma_{\mathrm{H}^\pm}$ also yield a relative determination of $\tan\beta$ to better than 0.06.

The accurate determination of the Higgs masses and widths is essential for outlining the profile of supersymmetry and understanding its connection with cosmology. Indeed, the relation between $M_\mathrm{A}$ and $\Gamma_\mathrm{A}$, in combination with the mass of the lightest neutralino, can determine for this benchmark point the rate of the $\tilde{\chi}_1^0\tilde{\chi}_1^0 \to \mathrm{A}$ annihilation process responsible for the amount of relic dark matter we observe. The accuracies achievable at CLIC on the Higgs masses and widths ensure that the heavy Higgs contribution to the statistical precision in the extraction of the cosmological relic density in the above annihilation channel is of order 0.10.

Finally, we briefly discuss the case where the supersymmetric particle spectrum is light compared to the heavy Higgs sector. First of all, the Higgs bosons could decay into supersymmetric particles and, for instance, the rates of $\mathrm{H} \to \widetilde{\tau}\widetilde{\tau}, \widetilde{\mathrm{t}}_1\widetilde{\mathrm{t}}_1$ decays and those of the $\mathrm{e}^+\mathrm{e}^- \to \mathrm{h}/\mathrm{H}\widetilde{\tau}\widetilde{\tau}$ or $\mathrm{h}/\mathrm{H}\widetilde{\mathrm{t}}_1\widetilde{\mathrm{t}}_1$ associated processes allows for a probe of the trilinear coupling $A_\tau$ or $A_\mathrm{t}$. Decays of the Higgs bosons into the invisible neutralinos $\tilde{\chi}_1^0$ can be detected in the $\mathrm{e}^+\mathrm{e}^- \to \mathrm{h}/\mathrm{HZ}$ process by looking at the recoil Z particle or in mixed $\mathrm{e}^+\mathrm{e}^- \to \mathrm{HA} \to \mathrm{b\bar{b}}\tilde{\chi}_1^0\tilde{\chi}_1^0$ or $\mathrm{t\bar{t}}\tilde{\chi}_1^0\tilde{\chi}_1^0$ decays. Similar channels for the heavier neutralinos and charginos can occur and would allow the simultaneous probe of the Higgs and sparticle sectors. Furthermore, Higgs bosons (at least the lightest h boson) can be produced in cascade decays of sparticles. For instance, one could look for the decays of the second lightest neutralino into the h boson and the lightest sparticle, $\tilde{\chi}_2^0 \to \mathrm{h}\tilde{\chi}_1^0$; this decay could be dominant, as is the case in one of the benchmark scenarios studied in this CDR, and would allow for an additional probe of the neutralino properties. Another possibility would be the decays of the heavier stop into the lighter one and h, $\widetilde{\mathrm{t}}_2 \to \widetilde{\mathrm{t}}_1\mathrm{h}$. If the stops are heavy, this can be probed only at very high energies.

### 1.2.3 Higgs Bosons in other Extensions

As discussed above, a large class of theories that predict new physics beyond the SM suggests additional gauge symmetries that may be broken near the TeV scale. This leads to an extended particle spectrum and, in particular, to additional Higgs fields beyond the minimal set of the MSSM. Especially common are new $U(1)'$ symmetries broken by the vacuum expectation value of a singlet field (as in the NMSSM) which leads to the presence of a Z$'$ boson and one additional CP–even Higgs boson compared to the MSSM; this is the case, for instance, in the exceptional MSSM based on the string inspired $E_6$ symmetry. The secluded $SU(2) \times U(1) \times U(1)'$ model, in turn, includes four additional singlets that are charged under $U(1)'$, leading to six CP–even and four CP–odd neutral Higgs states. Other exotic Higgs sectors in SUSY models are, for instance, Higgs representations that transform as SU(2) triplets or bi–doublets under the $SU(2)_L$ and $SU(2)_R$ groups in left–right symmetric models, that are motivated by the see-saw approach to explain the small neutrino masses and lead, for example, to a doubly charged Higgs boson $\mathrm{H}^{\pm\pm}$. These extensions, which also predict extra matter fields, would lead to very interesting phenomenology and new collider signatures in the Higgs sector.

To give an example of an interesting physics issue that can be discussed at CLIC, we consider a class of “minimal" Z$'$ models, based on the gauge group $SU(2)_L \times U(1)_Y \times U(1)_{B-L}$, with the SM field content enlarged to include three right-handed neutrinos to ensure anomaly cancellation. The additional $U(1)_{B-L}$ symmetry is important to consider as many Grand Unified Theories (GUTs) have this symmetry embedded within, and it could survive down to the TeV scale. The Z$'$ phenomenology is described by 3 parameters [30]: $M_{Z'}$ and two couplings $g_Y$ and $g_{BL}$. The Z$'$ will mix with the SM Z boson, with a small mixing angle given by $\theta_\mathrm{mix} \approx (g_Y/g_Z) \times (M_\mathrm{Z}/M_{\mathrm{Z}'})^2$. If $M_{\mathrm{Z}'}$ is smaller than the centre-of-mass energy, the Z$'$ boson can be produced as a resonance with very large rates. It will decay mainly into SM fermions, but one can also have Z$'$ decays into final states involving a Higgs boson, namely $\mathrm{Z}' \to \mathrm{ZH}$ and $\mathrm{Z}' \to \gamma\mathrm{H}$. The latter process is mediated by a top quark loop. By adapting results for the analogous SM process [31], one finds cross sections at the $\mathcal{O}(0.1)$fb level, see Figure 1.6. In turn, $\mathrm{Z}' \to \mathrm{ZH}$ is a tree-level process generated for $g_Y \neq 0$ by the kinetic term of the SM–like Higgs doublet, with the mixing suppression compensated by a term $\propto (M_{Z'}/M_Z)^2$ in the amplitude squared due to the longitudinal component of the Z boson. Very large $e^+e^- \to ZH$ cross sections, $\mathscr{O}(1)$ pb, i.e. much larger than that of the $W^+W^-$ fusion process, can be obtained, see Figure 1.6. This sample will be more than sufficient to study the Higgs boson and its connection with the symmetry behind the Z′ state. More details on this possibility are given in [21].

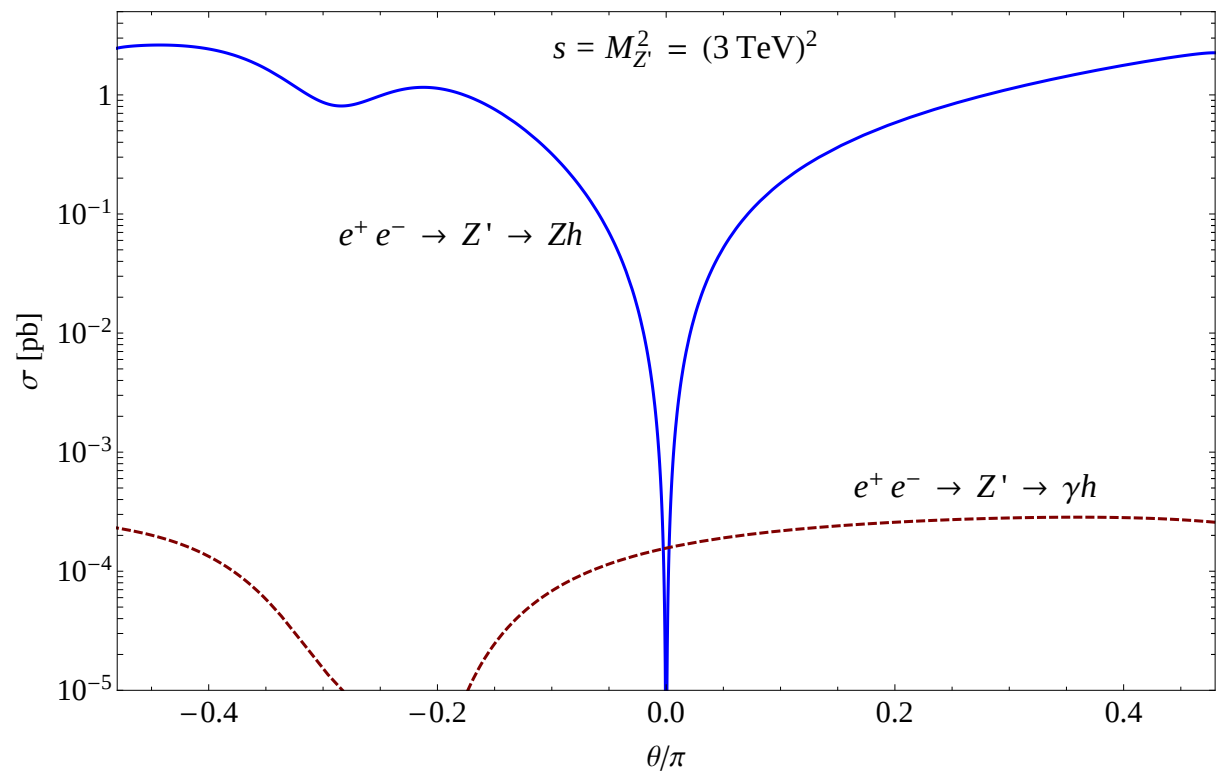

Fig. 1.6: Cross sections for the processes $e^+e^- \to Z' \to ZH$ and $\gamma H$ at CLIC with $\sqrt{s} = M_{Z'} = 3$ TeV as a function of a parameter that scans over minimal Z′ models.

Furthermore, Z′ boson exchange can induce non SM–like Higgs final states, such as $Z'^* \to Z'h, Z'H$ and $Z'^* \to Z'hh$, also via $H \to hh$, as well as $Z' \to \nu_h \nu_\ell h$ via the decay $\nu_h \to \nu_\ell h$, $\nu_{\ell(h)}$ being a light (heavy) neutrino [32]. In the context of the minimal $B-L$ model with no Z–Z′ mixing, the production cross sections are of order 0.1 pb, 0.01 pb and 1 pb, respectively.

## 1.3 Supersymmetry

Supersymmetry (SUSY) is one of the best-motivated theories beyond the SM and certainly the most studied one (for a review see [33]). SUSY predicts a partner particle, a so-called ‘superpartner’ or ‘sparticle’, for every SM state. In its local gauge theory version (supergravity), it also includes spin-2 and spin-3/2 states, the graviton and its superpartner the gravitino, and is hence capable of connecting gravity with the other interactions. From the phenomenological point of view, one expects some of the superpartners to have masses within the TeV energy range. The motivations range from solving the naturalness and hierarchy problems to the unification of gauge couplings, to the existence of dark matter. The search for supersymmetry is therefore one of the primary objectives of experiments at the LHC [5, 34], and the potential synergy with dark matter searches as well as studies of (quark and lepton) flavour violation is enormous.

In the Minimal Supersymmetric extension of the Standard Model (MSSM), in addition to two scalar partners for each quark and lepton, there are four neutralinos and two charginos as superpartners for the gauge and Higgs bosons. Moreover, there are five physical Higgs bosons, h, H, A and $H^\pm$. Remarkably, the MSSM predicts a light Higgs boson with $m_h \lesssim 130$ GeV, as favoured by present precision electroweak data [3]. Indeed $(g-2)_\mu$ prefers light SUSY particles [35], while flavour constraints, e.g. from $B(B \to X_s\gamma)$, limit sparticles from being very light.

A candidate model for new physics is the minimal supergravity model / constrained MSSM (mSUGRA/CMSSM), which is defined by universal boundary conditions for the soft breaking Lagrangian at the unification scale. The parameter space of the model is spanned by a universal soft-breaking scalar mass $m_0$, gaugino mass $m_{1/2}$, trilinear coupling $A_0$, the sign of the Higgsino mass parameter $\mu$, and $\tan\beta$, the ratio of the Higgs vacuum expectation values. ATLAS and CMS searches with

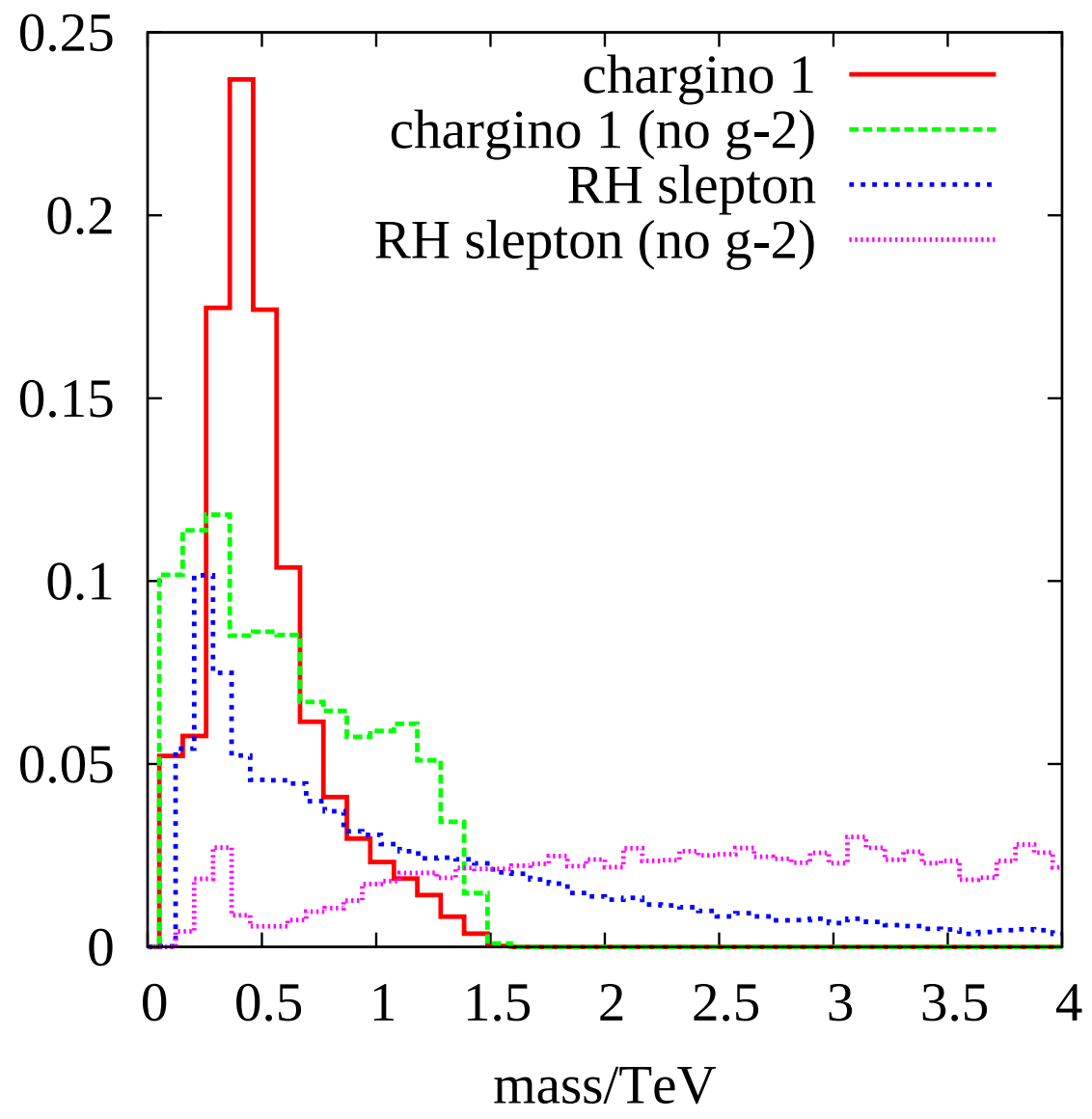

Fig. 1.7: Posterior probabilities from a global fit of the mSUGRA/CMSSM s in [36] (with flat priors in the input parameters), but taking into account the limits from the 2011 SUSY searches based on 1 $\mathrm{fb}^{-1}$ of data.

1 $\mathrm{fb}^{-1}$ of data at $\sqrt{s} = 7$ TeV already put stringent lower limits on squark and gluino masses. If they are of comparable size $m_{\tilde{q}} \simeq m_{\tilde{g}}$ the limits are roughly $m_{\tilde{q},\tilde{g}} \gtrsim 1$ TeV. However, the limits on the gluino mass from the early LHC runs become weaker when squarks are heavier, $m_{\tilde{q}} \gg m_{\tilde{g}}$. Figure 1.7 shows that even with the preliminary LHC constraints, light electroweak superpartners are still allowed by global statistical fits.

If nature is indeed supersymmetric at the weak scale, it is very likely that the LHC will have observed part of the spectrum by the time CLIC comes into operation. However, it is also likely that many of the superpartners will not be detected by the LHC. In particular, Higgsino-like neutralinos and charginos, and even wino-like states and sleptons, depending on their mass hierarchies, can quite easily decouple from the LHC. This can occur even if they are not too heavy, as long as they do not appear with large branching ratios in the squark and gluino decay chains that will provide the dominant LHC signals for supersymmetry. The signals for these particles could be too small to detect or be lost in the squark and gluino backgrounds. Moreover, the LHC may be able to detect, e.g. squarks in a generic sense without resolving the discovery with respect to their flavour and electroweak quantum numbers. In this context, it should be noted that the 3$^{\text{rd}}$ generation squarks (stops and sbottoms) are of particular relevance in understanding the hierarchy problem and the radiative breaking of electroweak symmetry.

To explore the theory fully, we will need to measure accurately the complete sparticle spectrum, just as the measurements of the Higgs boson properties are needed to complete the SM. Indeed, in order to verify SUSY, we will need to determine all the masses, mixing angles, couplings, spins, etc., of the new particles with good precision. This includes

- testing that there is indeed a superpartner for each SM state that differs in spin by 1/2;
- verifying SUSY coupling relations and testing if it is indeed the MSSM or an extended model;
- determining all the soft SUSY-breaking parameters, testing their structures at some high scale, and pinning down the SUSY-breaking mechanism (e.g. whether it is mediated by gravity, gauge interactions or anomalies);
- testing the properties of the lightest SUSY particle (LSP) as a dark matter candidate.

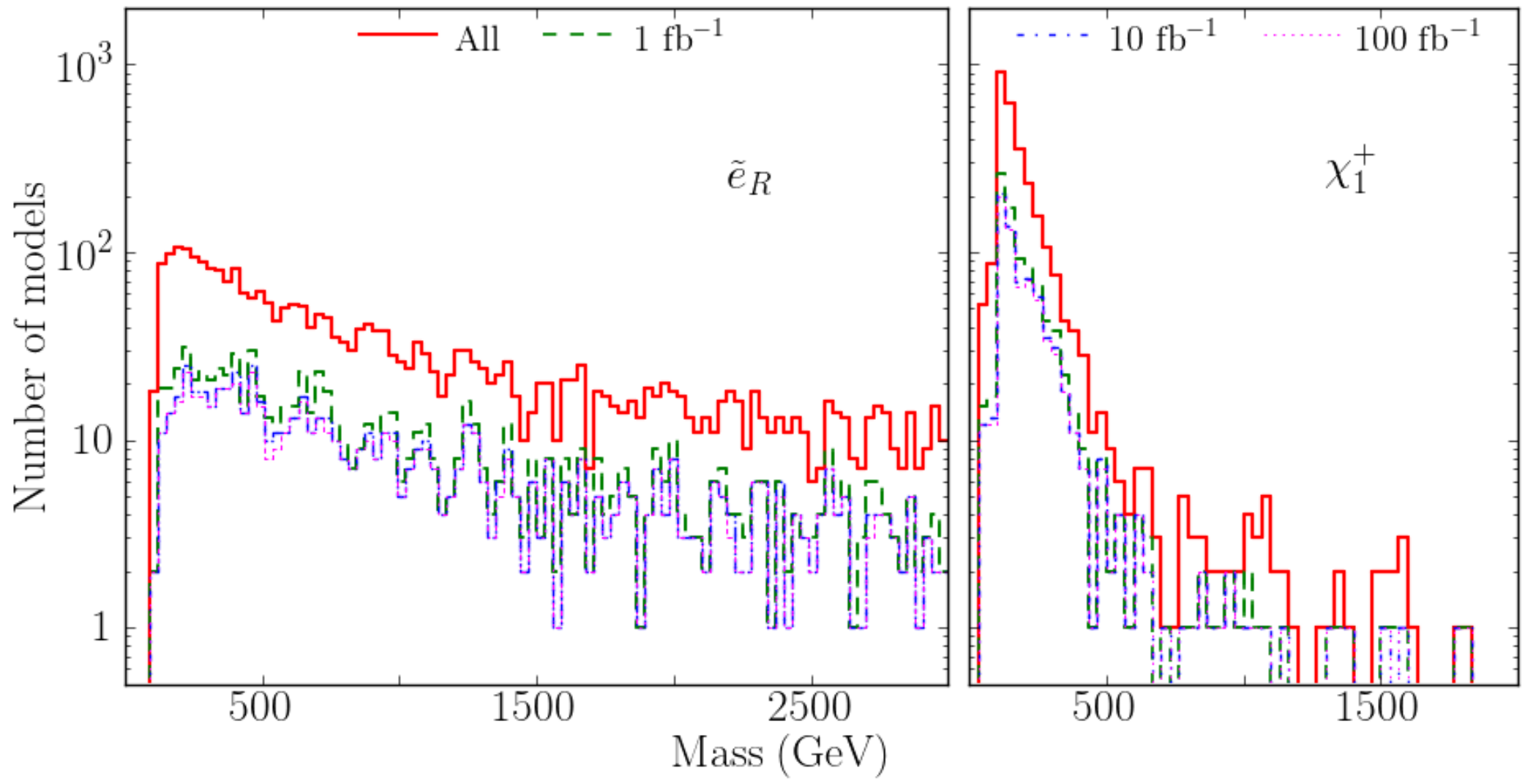

Fig. 1.8: Distribution of $\tilde{\chi}_1^\pm$ (right) and $\tilde{e}_R$ (left) masses of pMSSM points that escape 14 TeV LHC searches with 1 fb$^{-1}$, 10 fb$^{-1}$ and 100 fb$^{-1}$ of integrated luminosity [41]. The top red histograms show the mass distributions in the full model set.

An $e^+e^-$ collider will hence be the perfect tool to complete LHC searches and perform the required high-precision measurements, as has been discussed extensively in the literature, see e.g. [20, 37, 38, 39] and references therein. Of course future findings at the LHC will have an important impact on the physics requirements at a linear collider [40]. At this point it is relevant to ask what are the implications if there is no SUSY signal at the 7 TeV LHC run, or even at 14 TeV. This question was addressed in [41, 42] in the context of the phenomenological MSSM (pMSSM) with 19 free parameters. Figure 1.8 shows the $\tilde{\chi}_1^\pm$ and $\tilde{e}_R$ mass distributions of the pMSSM points from a log-prior scan that escape *all* searches at 14 TeV. Assuming 50% (20%) systematic uncertainty for the SM backgrounds, it was found [42] that 27% (13%) of pMSSM points fail all searches with 10 fb$^{-1}$ at the 14 TeV LHC. This is to a large extent due to difficult kinematical configurations arising in the log-prior scan. Moreover, as can be seen, the coverage of a wide range of masses is essential for complementary searches.

At CLIC, the cleaner experimental conditions of $e^+e^-$ annihilation should enable us to make many important precision measurements, completing the LHC exploration of supersymmetry, and to resolve particularly difficult cases where LHC measurements fail to clearly identify SUSY. Let us stress here that an $e^+e^-$ collider typically covers masses up to the kinematic limit of $m \lesssim \sqrt{s}/2$; it is therefore very likely that the full exploration of SUSY will require a machine in the multi-TeV regime. In the following, we discuss the potential of CLIC for studying heavy neutralinos, charginos and sleptons. To this end, we use for reference a specific high-mass benchmark point in mSUGRA/CMSSM, which is rich in hadronic final states (W, Z, and Higgs bosons). We concentrate in particular on sparticles with masses beyond the reach of the LHC and a TeV-scale linear collider. We also discuss the determination of the underlying SUSY-breaking parameters and their extrapolation to the GUT scale, in order to test unification and to clarify the nature of SUSY breaking. Moreover, we briefly discuss the inference of the neutralino relic density and the interplay with direct dark matter detection experiments.

### 1.3.1 CLIC potential for Heavy SUSY

The CLIC potential for supersymmetry was discussed previously in [20], and ways to contrast it against universal extra dimensions in [43]. Here we discuss the CLIC potential by means of a specific model point with the mSUGRA/CMSSM framework, *model II* ($m_0$ = 966 GeV, $m_{1/2}$ = 800 GeV, $A_0$ = 0, $\tan\beta = 51$, sgn($\mu$) = +, $m_t$ = 173.3 GeV), which is one of two SUSY points chosen for benchmarking the CLIC detector (see Section 2.6 and Chapter 12). The corresponding mass spectrum is given in Table 1.1, and the relevant production cross sections as a function of $\sqrt{s}$ are shown in Figure 1.9.

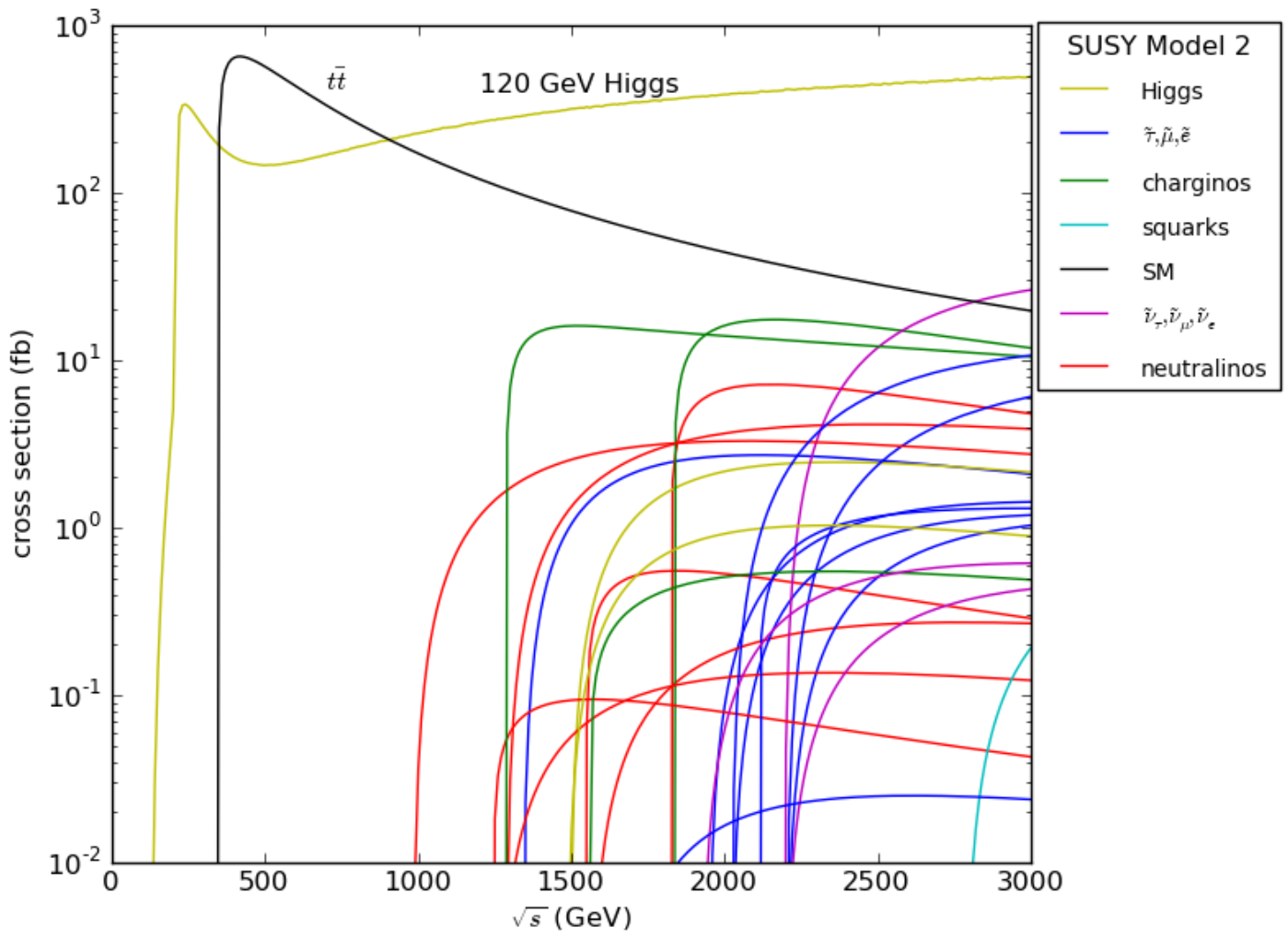

Fig. 1.9: SUSY production cross sections (in fb) of *model II* as a function of $\sqrt{s}$. Every line of a given colour corresponds to the production cross section of one of the particles in the legend, e.g. the three green lines are, per increasing threshold, $e^+e^- \to \widetilde{\chi}_1^- \widetilde{\chi}_1^+$, $e^+e^- \to \widetilde{\chi}_1^\mp \widetilde{\chi}_2^\pm$, and $e^+e^- \to \widetilde{\chi}_2^- \widetilde{\chi}_2^+$ respectively. The first threshold is the $e^+e^- \to$ ZH production.

For this model, all charginos and neutralinos can be observed in $\sqrt{s} = 3$ TeV $e^+e^-$ collisions. Their masses can be determined by measuring the kinematic endpoints of the energy distribution of the W, Z and h bosons produced in two body decays, such as $\tilde{\chi}_2^0 \to h\tilde{\chi}_1^0$. The dominance of these decays is typical for the region of mSUGRA/CMSSM parameter space of the selected benchmark point. To be concrete, for this *model II* we have the following decays [26]: $\tilde{\chi}_2^0 \to h^0\tilde{\chi}_1^0$ (90%), $\tilde{\chi}_1^\pm \to W^\pm\tilde{\chi}_1^0$ (100%), $\tilde{\chi}_{3,4}^0 \to W^\pm\tilde{\chi}_1^\pm$ (50%), $\tilde{\chi}_3^0 \to Z^0\tilde{\chi}_2^0$ (23%), $\tilde{\chi}_4^0 \to h\tilde{\chi}_2^0$ (22%), and $\tilde{\chi}_2^\pm \to W^\pm\tilde{\chi}_1^0$ (12%), $W^\pm\tilde{\chi}_1^0$ (28%), $Z^0\tilde{\chi}_1^\pm$ (26%) and $h\tilde{\chi}_1^\pm$ (24%). The diversity of topologies that emerges is one of the major challenges to resolve[2].

The accuracies for the mass measurements at 3 TeV have been estimated at generator level on an inclusive SUSY data sample corresponding to 2 ab$^{-1}$ of integrated luminosity, taking into account realistic CLIC beam radiation and experimental energy resolution [44]. Results for specific channels have been validated using fully simulated and reconstructed events, using the CLIC_SiD and CLIC_ILD detector geometry described later in Chapter 12. In addition, the accuracy on the measurement of the masses of the heavy Higgs states have been determined using fully simulated and reconstructed events, using the analysis also discussed in Chapter 12. Results are summarised in Table 1.1.

In particular, it is important for this study that all neutralinos can be observed and their masses determined. This allows one to reconstruct the neutralino/chargino sector which depends on four unknown quantities at tree-level: $M_1$, $M_2$, $\mu$ and $\tan\beta$. Since at large $\tan\beta$ the results hardly depend on its precise value, we can fit $M_1$, $M_2$ and $\mu$ keeping $\tan\beta$ fixed. The results are given in Table 1.2. The four-fold sign ambiguity could be resolved using information from polarised cross sections [45]. Using the methods of [46] we obtain nearly the same results. Finally, note that the closure (minimality) of the neutralino/chargino system may be tested through the polarisation dependence of the production cross sections, or by sum rules for the couplings [47, 48], see also [20]. To what extent this can be achieved at CLIC needs further investigation, beyond the scope of this CDR.

[2]This is even more true for studies of stops and sbottoms at higher energies, which lead to additional b and t quarks in the final states from, e.g. $\widetilde{t}_i \to b\tilde{\chi}_j^\pm$, $t\tilde{\chi}_k^0$ ($i, j = 1, 2$; $k = 1 \ldots 4$) and subsequent $\tilde{\chi}$ decays.

Table 1.1: Values of the SUSY particle masses of the chosen benchmark point (*model II*) and estimated experimental statistical accuracies at CLIC, as obtained in the analyses presented in Chapter 12, and also in [20] (indicated with *). All values are in GeV. The last column is either out of kinematic reach or not studied.

| Particle | Mass | Stat. acc. |
|---|---|---|
| $\tilde{\chi}_1^0$ | 340.3 | $\pm 3.3$ |
| $\tilde{\chi}_2^0$ | 643.1 | $\pm 9.9$ |
| $\tilde{\chi}_3^0$ | 905.5 | $\pm 19.0^*$ |
| $\tilde{\chi}_4^0$ | 916.7 | $\pm 20.0^*$ |
| $\tilde{\chi}_1^\pm$ | 643.2 | $\pm 3.7$ |
| $\tilde{\chi}_2^\pm$ | 916.7 | $\pm 7.0^*$ |
| $\tilde{e}_R^\pm$ | 1010.8 | $\pm 2.8$ |
| $\tilde{\mu}_R^\pm$ | 1010.8 | $\pm 5.6$ |
| $\tilde{\nu}_1$ | 1097.2 | $\pm 3.9$ |

| Particle | Mass | Stat. acc. |
|---|---|---|
| h | 118.5 | $\pm 0.1^*$ |
| A | 742.0 | $\pm 1.7$ |
| H | 742.0 | $\pm 1.7$ |
| $H^\pm$ | 747.6 | $\pm 2.1$ |
| **Quantity** | **Value** | **Stat. acc.** |
| $\Gamma(A)$ | 22.2 | $\pm 3.8$ |
| $\Gamma(H^\pm)$ | 21.4 | $\pm 4.9$ |

| Particle | Mass |
|---|---|
| $\tilde{\tau}_1$ | 670 |
| $\tilde{\tau}_2$ | 974 |
| $\tilde{t}_1$ | 1393 |
| $\tilde{t}_2$ | 1598 |
| $\tilde{b}_1$ | 1544 |
| $\tilde{b}_2$ | 1610 |
| $\tilde{u}_R$ | 1818 |
| $\tilde{u}_L$ | 1870 |
| $\tilde{g}$ | 1812 |

Table 1.2: Fitted parameters in GeV from the chargino/neutralino sector. Each column represents a local minimum in the best fit to the data.

| | | | | |
|---|---|---|---|---|
| $M_1$ | $342.1 \pm 3.5$ | $-341.9 \pm 3.5$ | $341.8 \pm 3.5$ | $-342.3 \pm 3.5$ |
| $M_2$ | $655.3 \pm 6.0$ | $655.3 \pm 6.1$ | $654.2 \pm 6.1$ | $654.2 \pm 6.1$ |
| $\mu$ | $924.8 \pm 6.2$ | $924.8 \pm 6.2$ | $-925.5 \pm 6.2$ | $-925.5 \pm 6.2$ |

In the slepton sector things are much simpler as in the cases considered there is a linear dependence on the parameters squared and we obtain $M_{E_1} = 1010 \pm 3$ GeV, $M_{E_2} = 1010 \pm 5.6$ GeV and $M_{L_1} = M_{L_2} = 1098 \pm 4$ GeV. As a technical aside note that all parameters obtained so far are on-shell parameters that need to be converted to the $\overline{DR}$ renormalisation scheme when extracting SUSY parameters at some high scale. However, this implies knowledge of the complete spectrum, plus information on $\tan\beta$ and the $A$-parameters. The corresponding corrections are at most 5% in the electroweak sector.

### 1.3.2 Reconstructing the High-Scale Structure of the Theory

Supersymmetric theories have been shown to be compatible with gauge coupling unification at a high scale. This implies a potential grand unification of forces, which may also enforce the unification of supersymmetric mass parameters at the GUT scale. In order to reconstruct the complete MSSM Lagrangian for *model II* and evolve the parameters to the GUT scale as described in [49], we need to combine CLIC measurements with measurements of squarks and gluinos at the LHC. For this study, we assume that the gluino mass can be determined to 5% precision at the LHC, and then consider different precisions on squarks.

Regarding gaugino masses, at the one-loop level the relative uncertainties do not change when running up to the GUT scale because $M_i/\alpha_i$ is a renormalisation group invariant. At the two-loop level however, the squark masses and trilinear $A$-parameters enter the game, deteriorating accuracies at the GUT scale. Taking into account an uncertainty of 5% (mainly due to the above-mentioned relation between $\overline{DR}$ and on-shell parameters) we find that $M_1$ and $M_2$ unify at the GUT scale to $812 \pm 41$ GeV. Including the gluino with a mass of $1807 \pm 90$ GeV all three gaugino masses unify to $M_{1/2} = 825 \pm 84$ GeV, where we assumed a 50% uncertainty on squark masses from the LHC leading to a 10% uncertainty for $M_3$. This improves to $812 \pm 26$ GeV (3%) and $825 \pm 44$ GeV (5%), respectively, if squark masses are known with 10% precision.

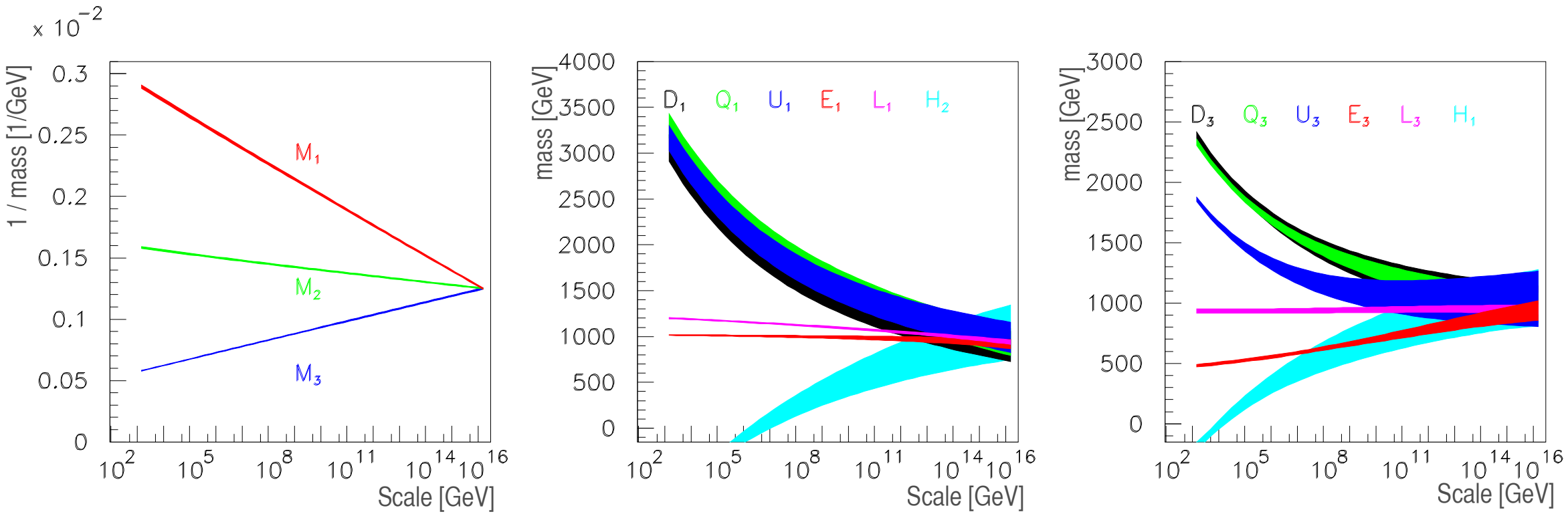

Fig. 1.10: Extrapolation of SUSY-breaking parameters from the electroweak to the GUT scale for *model II*, assuming 3% measurement precision on the physical sfermion masses.

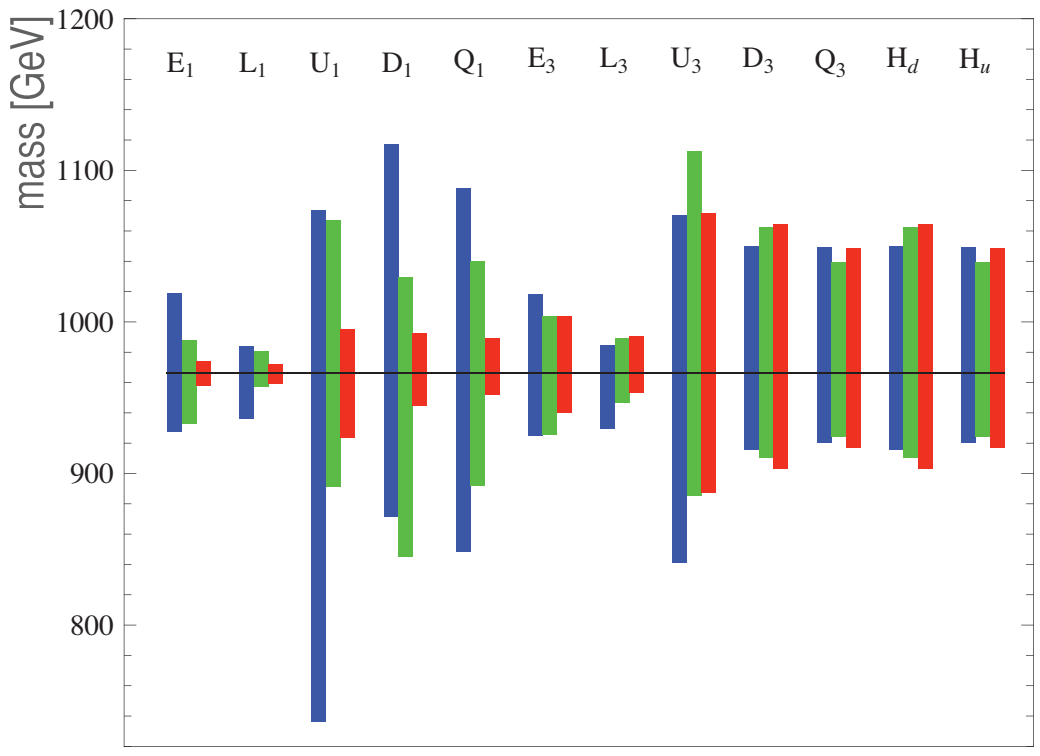

Fig. 1.11: One sigma range of the determined scalar mass parameters at the GUT scale assuming 5% (blue), 3% (green) and 1% (red) measurement precision on the physical sfermion masses. The black line indicates the nominal value $m_0 = 966$ GeV.

The soft-breaking parameters of the Higgs sector and the 3$^{\text{rd}}$ generation sfermions are more difficult because of the coupled nature of their renormalisation group equations. In addition to a precise measurement of $m_{\text{h}}$ (we take $\Delta m_{\text{h}^0} = 50$ MeV) one needs all relevant parameters of the stop/sbottom sector, including the $A$-terms. Figure 1.10 shows the evolutions of the gaugino-mass and the scalar mass-squared parameters, assuming that the masses of all sfermions, including 3$^{\text{rd}}$ generation squarks, can be measured with 3% precision. In addition we have assumed that using cross section measurements one can obtain $A_{\text{t}} = -1250 \pm 125$ GeV, $A_{\text{b}} = -1500 \pm 450$ GeV and $A_\tau = -130 \pm 40$ GeV. As can be seen, there is a clear overlap between all scalar mass parameters, as expected for universal boundary conditions. The power of the test depends, however, very sensitively on independent measurements of all the sfermion masses. The impact of different precisions in the sfermion sector on testing scalar mass unification at the GUT-scale is illustrated in Figure 1.11, while the reconstructed SUSY breaking parameters at the scale $Q = 1.5$ TeV are summarised in Table 1.3.

Table 1.3: Uncertainties on the SUSY breaking parameters at $Q = 1.5$ TeV for different expected precision on sfermion masses.

| parameter | central value | accuracy ($\pm$) | | |
|---|---|---|---|---|
| | | $\Delta m/m$=0.05 | $\Delta m/m$=0.03 | $\Delta m/m$=0.01 |
| $M_1$ | 345.2 | 1.7 | 1.7 | 1.6 |
| $M_2$ | 631.8 | 3.6 | 3.6 | 3.6 |
| $M_3$ | 1723 | 15 | 14 | 8 |
| $M^2_{H_d}$ | $-(435)^2$ | $(109)^2$ | $(109)^2$ | $(109)^2$ |
| $M^2_{H_u}$ | $-(904)^2$ | $(107)^2$ | $(107)^2$ | $(107)^2$ |
| $M^2_{E_1}$ | $(1008)^2$ | $(76)^2$ | $(76)^2$ | $(76)^2$ |
| $M^2_{L_1}$ | $(1096)^2$ | $(85)^2$ | $(85)^2$ | $(82)^2$ |
| $M^2_{D_1}$ | $(1760)^2$ | $(480)^2$ | $(375)^2$ | $(220)^2$ |
| $M^2_{U_1}$ | $(1766)^2$ | $(480)^2$ | $(375)^2$ | $(220)^2$ |
| $M^2_{Q_1}$ | $(1817)^2$ | $(500)^2$ | $(388)^2$ | $(223)^2$ |
| $M^2_{E_3}$ | $(690)^2$ | $(140)^2$ | $(130)^2$ | $(93)^2$ |
| $M^2_{L_3}$ | $(966)^2$ | $(145)^2$ | $(140)^2$ | $(108)^2$ |
| $M^2_{D_3}$ | $(1547)^2$ | $(209)^2$ | $(200)^2$ | $(154)^2$ |
| $M^2_{U_3}$ | $(1361)^2$ | $(138)^2$ | $(136)^2$ | $(128)^2$ |
| $M^2_{Q_3}$ | $(1527)^2$ | $(197)^2$ | $(189)^2$ | $(146)^2$ |
| $\tan\beta$ | 51 | 3 | 3 | 3 |

### 1.3.3 Testing the Neutralino Dark Matter Hypothesis

If supersymmetry is discovered at the LHC, one of the most important questions will be establishing whether the LSP is the dark matter particle which abounds in the universe. A decisive test will be inferring the present thermal relic abundance of supersymmetric particles by computing the annihilation rate of the LSPs in the early universe and then comparing it with the observed value of the dark matter relic density, $\Omega h^2 = 0.110 \pm 0.006$ [50, 51]. This calculation requires precise knowledge of many of the parameters (masses and coupling constants) describing the supersymmetric particles. The number of parameters and the needed accuracy strongly depends on the nature of the dark matter particle, but it is in general very unlikely that the LHC can provide us with sufficient information to make a revealing determination of the relic abundance. Measurements at CLIC then become crucial and the required precisions in the MSSM have been studied in [52]. We show that this is the case in the specific example of our benchmark point, which gives $\Omega h^2 \simeq 0.1$ with the neutralino LSP dominantly annihilating through the pseudoscalar A pole. For this benchmark point, the critical parameters for the determination of the relic abundance are the mass and width of the heavy neutral bosons, which can indeed be measured with high accuracy, see Table 1.1.

The accuracy on the neutralino dark matter relic density, $\Omega_\chi h^2$, from the anticipated CLIC measurements has been calculated with both `micrOMEGAs` [53] and `SuperIso Relic` [54] by means of a scan of the full pMSSM parameter space. The two codes agree on $\Omega_\chi h^2$ to $\lesssim 10\%$ in the relevant region of parameter space. For each valid point in the scan, the values of the sparticle masses are compared to those of the reference benchmark model and the point weighted by its global probability computed using the estimated experimental accuracies on masses and widths. The anticipated CLIC experimental accuracies on the heavy Higgs widths yield the value of $\tan\beta$ to a $\pm 3$ statistical accuracy for this specific benchmark point. The weighted $\Omega_\chi h^2$ distribution obtained with `SuperIso Relic`, representing the probability density function for this quantity, is given in Figure 1.12 (left). It corresponds to a determination of $\Omega_\chi h^2 = 0.098 \pm 0.016$, i.e. a fractional accuracy of $\pm 17\%$, assuming standard cosmology. With `micrOMEGAs` (Figure 1.12 (right)) we get $\Omega_\chi h^2 = 0.111$ with the same fractional accuracy. In the same manner we obtain $\sigma^{\mathrm{SI}}_{\chi p} = (1.307 \pm 0.065) \cdot 10^{-9}$ pb for the spin-independent scattering cross section on protons[3]. The dark matter mass, relic density and direct detection cross section can provide important consistency checks, on the one hand of the neutralino dark matter hypothesis, on the other hand of the thermal hypothesis and astrophysics assumptions [55, 56]. Finally let us note that the case of light cold dark matter in the mass range below 10 GeV might also be probed at CLIC. Extrapolating the results of [57] to $\sqrt{s} = 3$ TeV one finds that in the MSSM with non-universal gaugino masses, for a $100 - 200$ GeV selectron mass, the expected accuracy on a 8 GeV neutralino mass is about $1 - 4$ GeV.

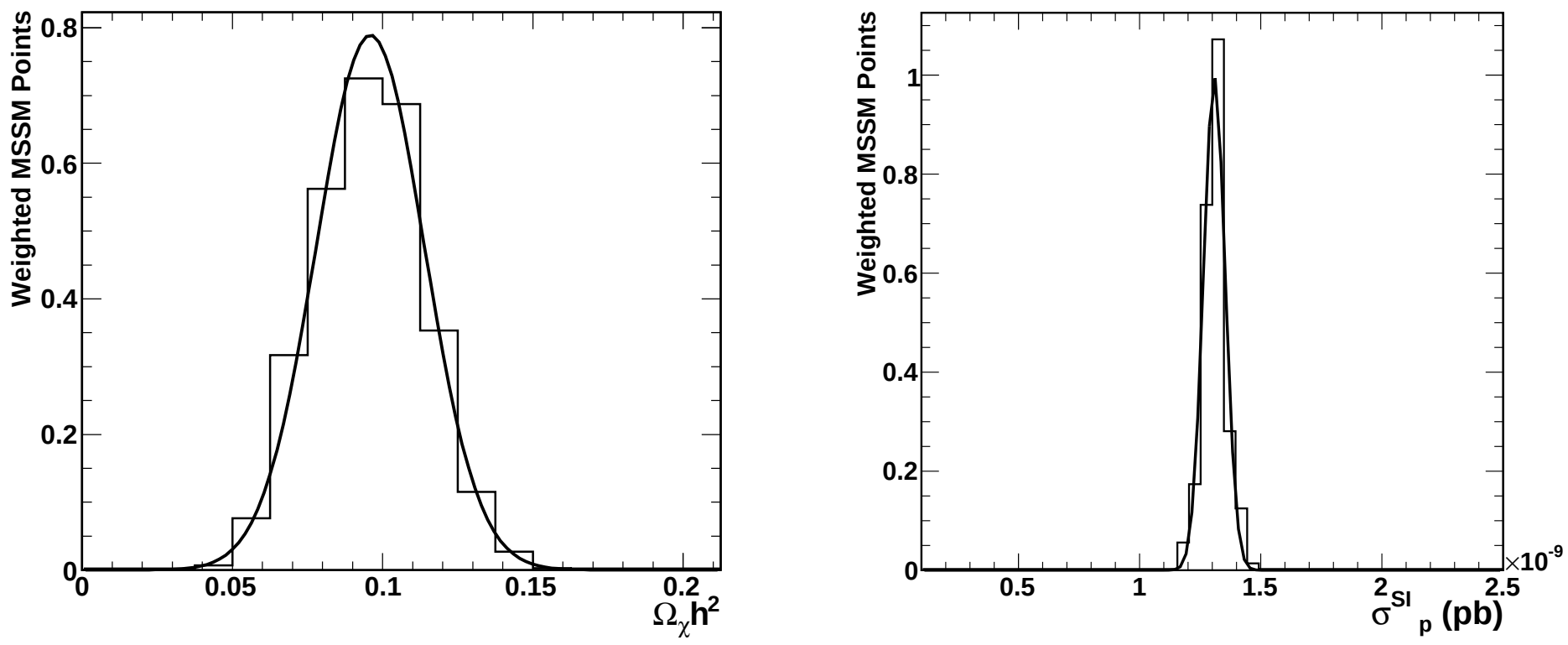

Fig. 1.12: Using probability weighted ensemble of scanned SUSY model points, the reconstructed neutralino relic density (left, using `SuperIso Relic`) and predicted direct dark matter detection cross section (right, using `micrOMEGAs`) for the measurements of *model II*.

## 1.4 Higgs Strong Interactions

If a light Higgs-like scalar h is found at the LHC, it will be crucial to understand its nature and the role it plays in the breaking of electroweak symmetry. A possible scenario, motivated by the hierarchy problem and the LEP electroweak data, is one in which the light Higgs is a composite bound state of new strongly-interacting dynamics at the TeV scale. Such a composite Higgs can be naturally lighter than other bound states of the new dynamics if it arises as the pseudo Nambu–Goldstone (NG) boson of a spontaneously broken global symmetry [58]. Another recently considered possibility is that the role of the Higgs is partially played by a composite dilaton, the pseudo NG boson of spontaneously broken scale invariance [59]. In either case, the light scalar has modified couplings compared to the SM Higgs and becomes strongly interacting at high energy. One obvious process sensitive to the strong interactions is the scattering of longitudinal vector bosons ($V_L V_L \to V_L V_L$, where $V = \mathrm{W}, \mathrm{Z}$), which the scalar exchange fails to fully unitarise. This process is sensitive to the linear couplings of the light scalar to the gauge bosons, but it does not give information on the non-linearities of the Higgs sector. Furthermore, the sensitivity on the measurement of the linear couplings via $V_L V_L$ scattering will be hardly competitive with the information from direct measurements (single-Higgs production processes and Higgs branching ratios determination – see Section 1.2 of this Chapter). It has been recently pointed out that crucial information on the dynamics of the Higgs and EWSB sector also comes from the high energy behaviour of the double Higgs production via vector boson fusion: $V_L V_L \to \mathrm{hh}$ [60, 61]. A signature of composite Higgs models is a scattering amplitude for such a process that grows with the energy (hence a signifi-cant enhancement over the small SM rate), whereas no growth with energy is expected in the case of a dilaton. While the observation of this process is extremely challenging at the LHC due to the large SM background rate [61], a multi-TeV linear collider such as CLIC is able to make a precise measurement of the cross section [62].

[3]The uncertainty quoted is due only to the fit to the supersymmetric model and does not include the dominant source of uncertainty from nuclear matrix elements, which can hopefully be reduced somewhat in the future.

The couplings of a light scalar h to the SM vector bosons and to itself can be characterised in terms of the following Lagrangian (with vacuum expectation value $v \approx 246$ GeV) [61]

$$\mathscr{L} = \frac{1}{2}\left(\partial_\mu h\right)^2 - V(h) + \left(m_W^2 W_\mu^+ W^{\mu\,-} + \frac{m_Z^2}{2} Z_\mu Z^\mu\right)\left[1 + 2a\frac{h}{v} + b\frac{h^2}{v^2} + \dots\right] + \dots, \tag{1.1}$$

where $V(h)$ is the potential for h,

$$V(h) = \frac{1}{2} m_{\mathrm{h}}^2 h^2 + d_3\left(\frac{m_{\mathrm{h}}^2}{2v}\right) h^3 + d_4\left(\frac{m_{\mathrm{h}}^2}{8v^2}\right) h^4 + \dots, \tag{1.2}$$

and $a, b, d_3, d_4$ are arbitrary dimensionless parameters. The dots stand for terms of higher order in $h$. For the SM Higgs boson $a = b = d_3 = d_4 = 1$ and all the higher order terms vanish. The dilaton couplings are characterised by the relation $a = b^2$. The scattering amplitude of $V_L V_L \to \mathrm{hh}$ depends on $a$, $b$ and $d_3$ and can be conveniently written as $\mathscr{A} = a^2\left(\mathscr{A}_{SM} + \mathscr{A}_1 \delta_b + \mathscr{A}_2 \delta_{d_3}\right)$, where $\mathscr{A}_{SM}$ is the value predicted by the SM and

$$\delta_b \equiv 1 - \frac{b}{a^2}, \qquad \delta_{d_3} \equiv 1 - \frac{d_3}{a}. \tag{1.3}$$

At large partonic centre-of-mass energies, $E \gg m_V$, $\mathscr{A}_1$ grows like $E^2$, while $\mathscr{A}_2$ (as well as $\mathscr{A}_{SM}$) is constant. The parameter $\delta_b$ thus controls the energy growth of the amplitude and gives a genuine "strong coupling" signature. On the contrary, $\delta_{d_3}$ determines the value of the cross section at threshold. In an $\mathrm{e}^+\mathrm{e}^-$ collider the scattering $V_L V_L \to \mathrm{hh}$ can be studied via the process $\mathrm{e}^+\mathrm{e}^- \to \nu\bar{\nu}\mathrm{hh}$, whose cross section can be written as

$$\sigma = a^4 \sigma_{SM}\left(1 + A\delta_b + B\delta_{d_3} + C\delta_b\delta_{d_3} + D\delta_b^2 + E\delta_{d_3}^2\right), \tag{1.4}$$

where $\sigma_{SM}$ denotes its SM value. The energy behaviour of the underlying hard scattering is encoded in the coefficients $A, B, C, D, E$. By means of suitable kinematic cuts one can disentangle the high-energy behaviour from the physics at threshold, and extract $\delta_b$, $\delta_{d_3}$ [63].

For simplicity we assume that the two Higgs bosons can be fully reconstructed (for example in the $\mathrm{b\bar{b}b\bar{b}}$ final state where both decay to $\mathrm{b\bar{b}}$), and that the SM background can be reduced to a negligible level. For a given set of kinematic cuts, we have determined the calculable coefficients $A, B, C, D, E$ by performing a Monte-Carlo simulation with the MADGRAPH event generator. We thus consider two sets of cuts:

$$H_T \geq H_T^+ \tag{1.5}$$

$$m_{\mathrm{hh}} \leq m_{\mathrm{hh}}^-, \tag{1.6}$$

where $H_T$ is defined as the scalar sum of the transverse momenta of the two Higgs bosons, $m_{\mathrm{hh}}$ is the invariant mass of the (hh) pair, and $H_T^+$, $m_{\mathrm{hh}}^-$ are numerical values. The first cut, Equation 1.5, selects the high-energy events that undergo a hard, central scattering, and as such it enhances the sensitivity on $\delta_b$; the second cut, Equation 1.6, selects instead the events at threshold, whose rate is controlled by both $\delta_b$ and $\delta_{d_3}$. For example, we obtain the following three sets of coefficients

$$\sigma_{SM} = 0.850\ \mathrm{fb}, \quad \{A,B,C,D,E,F\} = \{4.14, 0.664, 3.10, 14.7, 0.485\}\ [\text{No cuts}] \tag{1.7}$$

$$\sigma_{SM} = 0.080\ \mathrm{fb}, \quad \{A,B,C,D,E,F\} = \{10.9, 0.660, 12.1, 117, 0.665\}\ [H_T \geq 450\,\mathrm{GeV}] \tag{1.8}$$

$$\sigma_{SM} = 0.371\ \mathrm{fb}, \quad \{A,B,C,D,E,F\} = \{3.57, 1.10, 4.36, 6.68, 0.956\}\ [m_{\mathrm{hh}} \leq 550\,\mathrm{GeV}]. \tag{1.9}$$

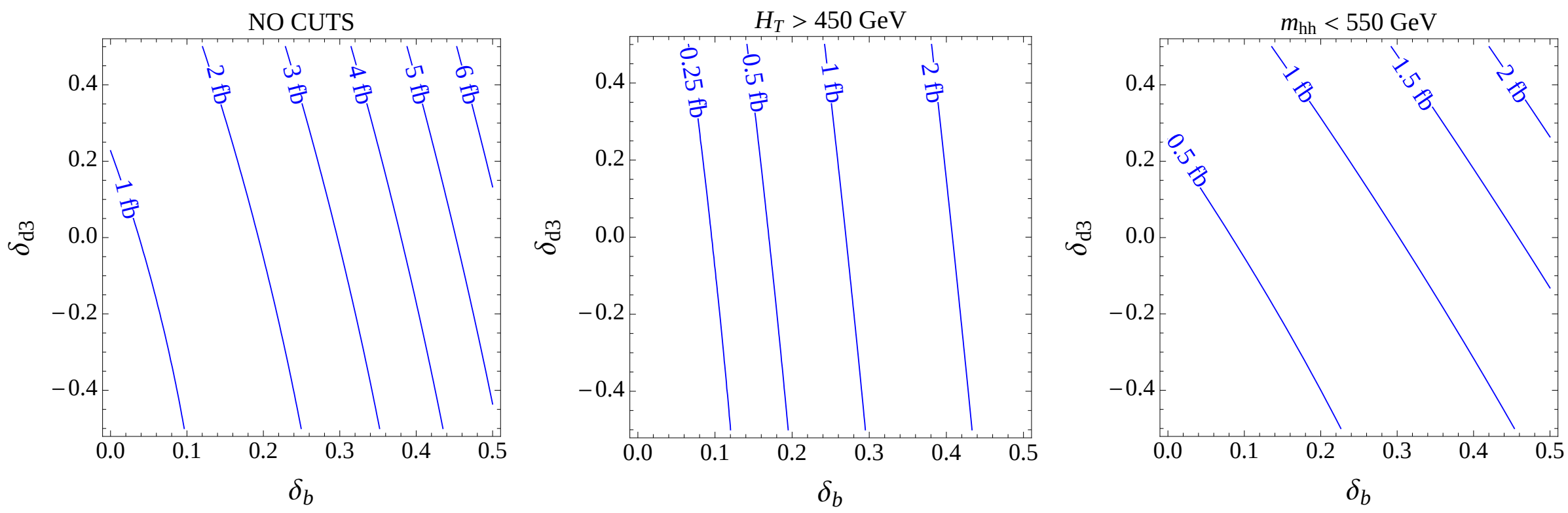

Fig. 1.13: Contours of constant cross section $\sigma(\mathrm{e^+e^-} \to \nu\bar{\nu}\mathrm{hh})$ for $\sqrt{s} = 3$ TeV and $m_\mathrm{h} = 120$ GeV in the plane $(\delta_b, \delta_{d_3})$. The three plots have been obtained by imposing no cuts (left), $H_T \geq 450$ GeV (centre), $m_\mathrm{hh} < 550$ GeV (right) respectively.

Table 1.4: Statistical errors $(\Delta\delta_b, \Delta\delta_{d3})$ on the parameters $\delta_b$ and $\delta_{d_3}$ for $\sqrt{s} = 3$ TeV, $\mathscr{L} = 1\ \mathrm{ab}^{-1}/a^4$ and $m_\mathrm{h} = 120$ GeV. The values of the optimised cuts range in the interval $H_T^+ = 250 - 450$ GeV, $m_\mathrm{hh}^- = 450 - 650$ GeV.

| $\delta_b$ | $\delta_{d_3}$ | | | | | |
|---|---|---|---|---|---|---|
| | -0.5 | -0.3 | -0.1 | 0.1 | 0.3 | 0.5 |
| 0 | (0.02, 0.2) | (0.02, 0.1) | (0.01, 0.08) | (0.01, 0.06) | (0.01, 0.05) | (0.01, 0.05) |
| 0.01 | (0.02, 0.2) | (0.01, 0.1) | (0.01, 0.07) | (0.01, 0.06) | (0.01, 0.05) | (0.009, 0.05) |
| 0.02 | (0.02, 0.2) | (0.01, 0.1) | (0.01, 0.07) | (0.009, 0.06) | (0.009, 0.5) | (0.009, 0.05) |
| 0.03 | (0.01, 0.2) | (0.01, 0.1) | (0.01, 0.07) | (0.009, 0.05) | (0.008, 0.05) | (0.008, 0.05) |
| 0.05 | (0.01, 0.1) | (0.01, 0.09) | (0.008, 0.06) | (0.008, 0.05) | (0.008, 0.05) | (0.008, 0.04) |
| 0.1 | (0.009, 0.1) | (0.008, 0.07) | (0.007, 0.06) | (0.007, 0.05) | (0.007, 0.04) | (0.007, 0.04) |
| 0.2 | (0.008, 0.08) | (0.007, 0.06) | (0.007, 0.05) | (0.007, 0.04) | (0.007, 0.04) | (0.007, 0.04) |
| 0.3 | (0.007, 0.06) | (0.007, 0.05) | (0.007, 0.05) | (0.007, 0.04) | (0.007, 0.04) | (0.007, 0.04) |
| 0.4 | (0.007, 0.05) | (0.007, 0.05) | (0.007, 0.05) | (0.006, 0.04) | (0.006, 0.04) | (0.007, 0.04) |
| 0.5 | (0.007, 0.05) | (0.007, 0.05) | (0.006, 0.04) | (0.006, 0.04) | (0.006, 0.04) | (0.006, 0.04) |

Figure 1.13 shows the corresponding value of the cross section in the plane ($\delta_b$, $\delta_{d_3}$). The parameters $\delta_b$ and $\delta_{d_3}$ are extracted from the values of the cross section $\sigma_+$, $\sigma_-$ after the cuts of Equation 1.5 and Equation 1.6 respectively. For each value of $\delta_b$, $\delta_{d_3}$ the value of the cuts, $H_T^+$ and $m_\mathrm{hh}^-$, is optimised to minimise the statistical error on the parameters. The statistical accuracy on $\delta_b$ and $\delta_{d_3}$ obtained at CLIC with $\sqrt{s} = 3$ TeV and an integrated luminosity $\mathscr{L} = 1\ \mathrm{ab}^{-1}/(a^4)$ is shown in Table 1.4 for $m_\mathrm{h} = 120$ GeV. Details of the $\chi^2$ procedure to determine the entries of Table 1.4 can be found in [63].

While $a$, $b$, $d_3$ and $d_4$ in Equations 1.1, 1.2 can be considered in general as independent, they will be related to each other in specific models. For example, in the $SO(5)/SO(4)$ model of [64] one has

$$a = \sqrt{1-\xi}\,, \qquad b = 1 - 2\xi\,, \qquad d_3 = \sqrt{1-\xi}\,, \tag{1.10}$$

where $\xi = (v/f)^2$ is the dimensionless compositeness scale parameter, and $f$ is the decay constant of the pseudo-NG Higgs boson. Assuming the above relations, one can extract $\xi$ from the value of the $\mathrm{e^+e^-} \to \nu\bar{\nu}\mathrm{hh}$ cross section, which can be written in this case as $\sigma = \sigma_{SM} + \xi\sigma_1 + \xi^2\sigma_2$. The sensitivity on small values of $\xi$ can be enhanced by cutting on $H_T$ and exploiting the $s$-wave growth of $\sigma_{1,2}$ at large energies, as opposed to the behaviour of $\sigma_{SM}$, which is constant at high energies and has a Coulomb enhancement for forward Higgs bosons. Table 1.5 shows the sensitivity on $\xi$ attainable at CLIC with $\sqrt{s} = 3$ TeV and $\mathscr{L} = 1$ ab$^{-1}$ by optimising the cut $H_T \geq H_T^+$ for each value of $\xi$ and $m_h$.

Table 1.5: Statistical relative error $\Delta\xi/\xi$ for $\sqrt{s}$ =3 TeV with $\mathscr{L} = 1$ ab$^{-1}$. The value of the optimised cut $H_T \geq H_T^+$ ranges in the interval $H_T^+ = 300-500$ GeV. The numbers in parentheses report the relative error $\Delta\xi/\xi$ obtained without imposing any cut.

| $m_h$ [GeV] | $\xi$ 0.01 | 0.03 | 0.05 | 0.1 | 0.2 | 0.3 |
|---|---|---|---|---|---|---|
| 120 | 1.1 (1.9) | 0.3 (0.5) | 0.1 (0.3) | 0.06 (0.09) | 0.03 (0.03) | 0.02 (0.02) |
| 150 | 1.1 (2.2) | 0.3 (0.5) | 0.2 (0.3) | 0.06 (0.09) | 0.03 (0.03) | 0.02 (0.02) |
| 180 | 1.2 (2.7) | 0.3 (0.6) | 0.2 (0.3) | 0.07 (0.09) | 0.03 (0.03) | 0.02 (0.02) |

These results can be used to estimate the sensitivity of CLIC on the scale of compositeness of the Higgs boson. Without prior information on the strong sector where the Higgs is emerging from, the compositeness scale can be estimated in terms of the quadratic coupling $b$: $\Lambda \sim 4\pi v/(a\sqrt{\delta_b})$. From Table 1.4 we conclude that CLIC at 3 TeV can reach a sensitivity on $\Lambda$ of the order $\sim (30\,\text{TeV}/a)$ with an accumulated luminosity of 1 ab$^{-1}/a^4$. In the specific case of a pseudo-NG boson Higgs the compositeness scale is given by $\Lambda \sim 4\pi f$. According to Table 1.5, it follows that using double Higgs production alone, CLIC at 3 TeV can probe a 30 TeV compositeness scale with an accumulated luminosity of 1 ab$^{-1}$, which is about half of the scale reached by including single-Higgs production processes [62]. This has to be compared with the sensitivity $\sim 5 \div 7$ TeV expected at the LHC by including both $V_L V_L$ scattering and single Higgs production processes [60, 65] (see also [25] for a recent analysis of double Higgs production via gluon fusion), and the sensitivity $\sim 45$ TeV expected at a 500 GeV linear collider with 1 ab$^{-1}$ by including both $V_L V_L$ scatterings and single Higgs production [62]. Figure 1.14 summarises the LHC and CLIC sensitivities in the determination of the parameter $\xi$ along with the limit of the direct searches for vector resonances at the LHC.

From Table 1.4, one also sees that for $\delta_b = 0$ the precision attainable on $\delta_{d_3}$ is $\sim 0.1$, which means that a measure of the Higgs trilinear coupling in the SM could be possible with a precision of $\sim 10-20\%$. This estimated sensitivity agrees with the one reported in Table 4.3 of [20] and in Section 1.2 of this Chapter.

## 1.5 Z′, Contact Interactions and Extra Dimensions

A multi-TeV linear collider such as CLIC can extend considerably the sensitivity on new heavy particles that can be reached at the LHC or at a 500 GeV linear collider. The effects of new flavour-conserving physics at the scale $\Lambda \gg \sqrt{s}$ on the electroweak observables (total production cross section, forward-backward and left-right asymmetries of $e^+e^- \to f\bar{f}$) can be parametrised in terms of the four-fermion operators

$$\mathscr{L}_{CI} = \sum_{i,j=L,R} \eta_{ij} \frac{g^2}{\Lambda^2} (\bar{e}_i \gamma^\mu e_i)(\bar{f}_j \gamma_\mu f_j) , \tag{1.11}$$

where $g$ is the coupling strength of the new particles. Different models can be defined by choosing $|\eta_{ij}| = 1$ or 0. Figure 1.15 shows the limits on $\Lambda/g$, for the models defined in Table 6.6 of [20]. The sensitivity can extend up to $(\Lambda/g) \sim 80-100$ TeV, depending on the model and the degree of polarisation of the electron and positron beams.

One particular example of new heavy physics that can generate the above contact operators is that of additional spin-1 neutral particles Z′ discussed earlier. Current limits from direct searches at hadron colliders can constrain the Z′ masses of such models above $\sim$ 1 TeV. However in models where the vector boson masses and mixings arise from hidden sector gauge groups, the mass of the Z′ can be well below this scale [16, 18], owing to the narrowness of the predicted resonances. The LHC is expected to explore the region of masses up to several TeV. If a new neutral resonance were to be observed at the LHC it would become interesting to produce it in lepton collisions and accurately determine its properties and nature. A multi-TeV $e^+e^-$ collider, such as CLIC, is very well suited for such a study. By precisely tuning the beam energies to perform a detailed scan, the parameters of the resonance can be extracted with high accuracy [20]. By operating CLIC at its full energy of 3 TeV it is also possible to search for new resonances coupled to the electron current and to perform a first determination of their mass and width using the beamstrahlung and ISR tail through a return scan. An example is given in the left panel of Figure 1.16 where the invariant mass of $\mu^+\mu^-$ pairs in the $e^+e^- \to \mu^+\mu^-$ process is shown for the case of two new neutral gauge bosons arising from an example model with extra dimensions. The fitted masses with 1 ab$^{-1}$ data at 3 TeV are $(1588.7 \pm 3.5)$ GeV and $(2022.6 \pm 1.2)$ GeV.

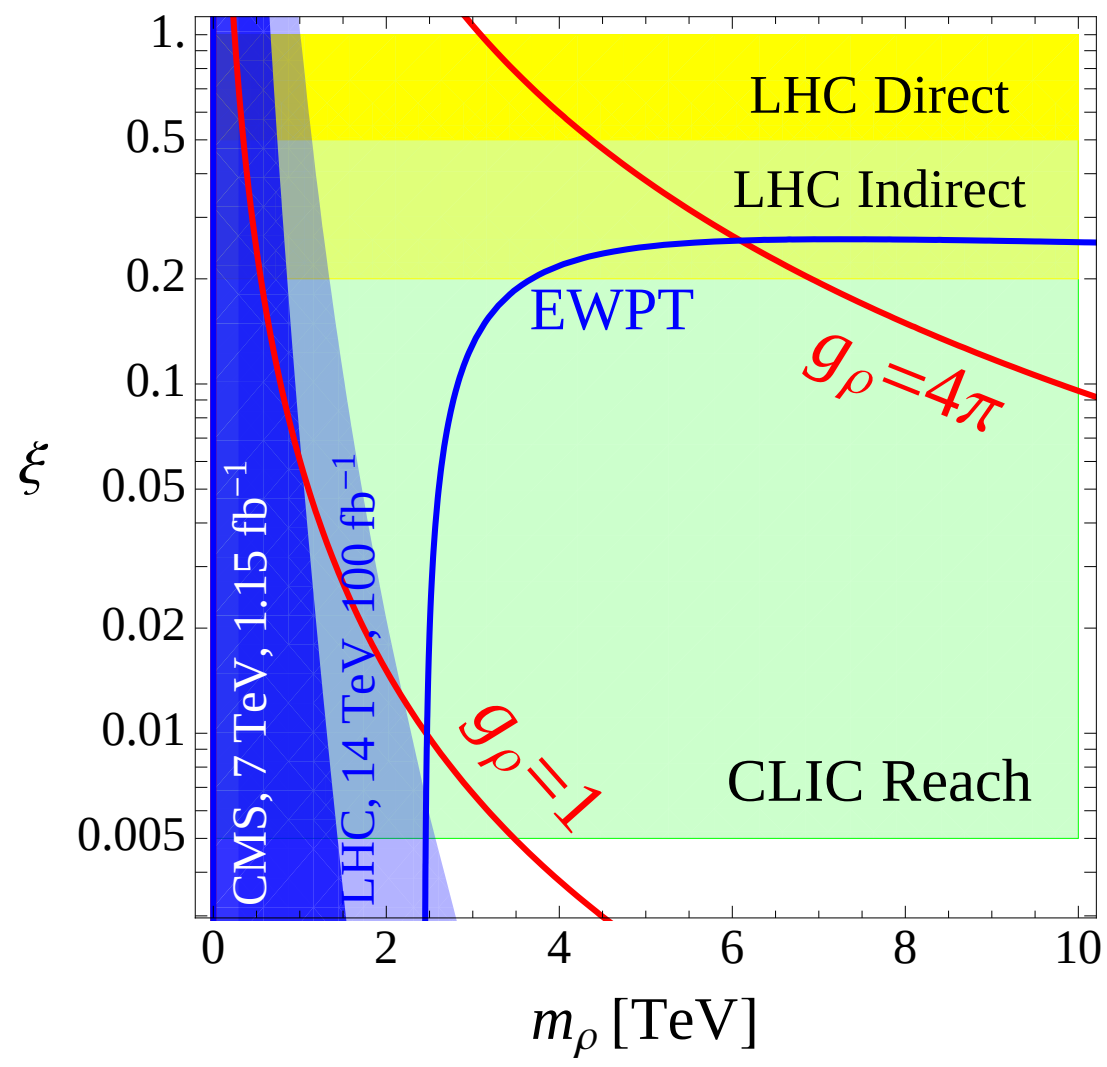

Fig. 1.14: Summary plot of the current constraints and prospects for direct and indirect probes of the strong interactions triggering electroweak symmetry breaking. $m_\rho$ is the mass of the vector resonances and $\xi = (v/f)^2$. The dark yellow band denotes the LHC sensitivity on $\xi$ from $WW$ scattering and strong double Higgs production measurements, while in the light yellow band, $\xi$ can be measured via single Higgs processes. The green band corresponds to the CLIC sensitivity on $\xi$. The dark and light blue regions on the left are the current limit on resonance mass and coupling from the direct search at the LHC in $WZ \to 3\ell$ final state with 1.15 fb$^{-1}$ integrated luminosity at 7 TeV [66] and its proper rescaling for LHC at 14 TeV with 100 fb$^{-1}$. Finally, electroweak precision data favours the region below the blue thick line (the Higgs mass is assumed to 120 GeV and the vector resonance contribution to $\varepsilon_3$ is taken to be $\Delta\varepsilon_3 = 4m_W^2/(3m_\rho^2)$). The domain of validity of our predictions, $1 < g_\rho < 4\pi$, is between the two red lines.

In case no signal is observed, CLIC can still obtain essential information on heavy spin-1 neutral particles from precision study of the electroweak observables. As illustrated above, these are sensitive to the effects of new particles at mass scales well above the collision centre-of-mass energy. For example, the right panel of Figure 1.16 shows how well the “normalised” couplings of a 10 TeV Z′,

$$v_f^N = v'_f \sqrt{\frac{s}{m_{Z'}^2 - s}}, \qquad a_f^N = a'_f \sqrt{\frac{s}{m_{Z'}^2 - s}}, \tag{1.12}$$

defined in terms of its vector and axial couplings $v'_f$, $a'_f$ to a SM fermion f, can be determined at CLIC

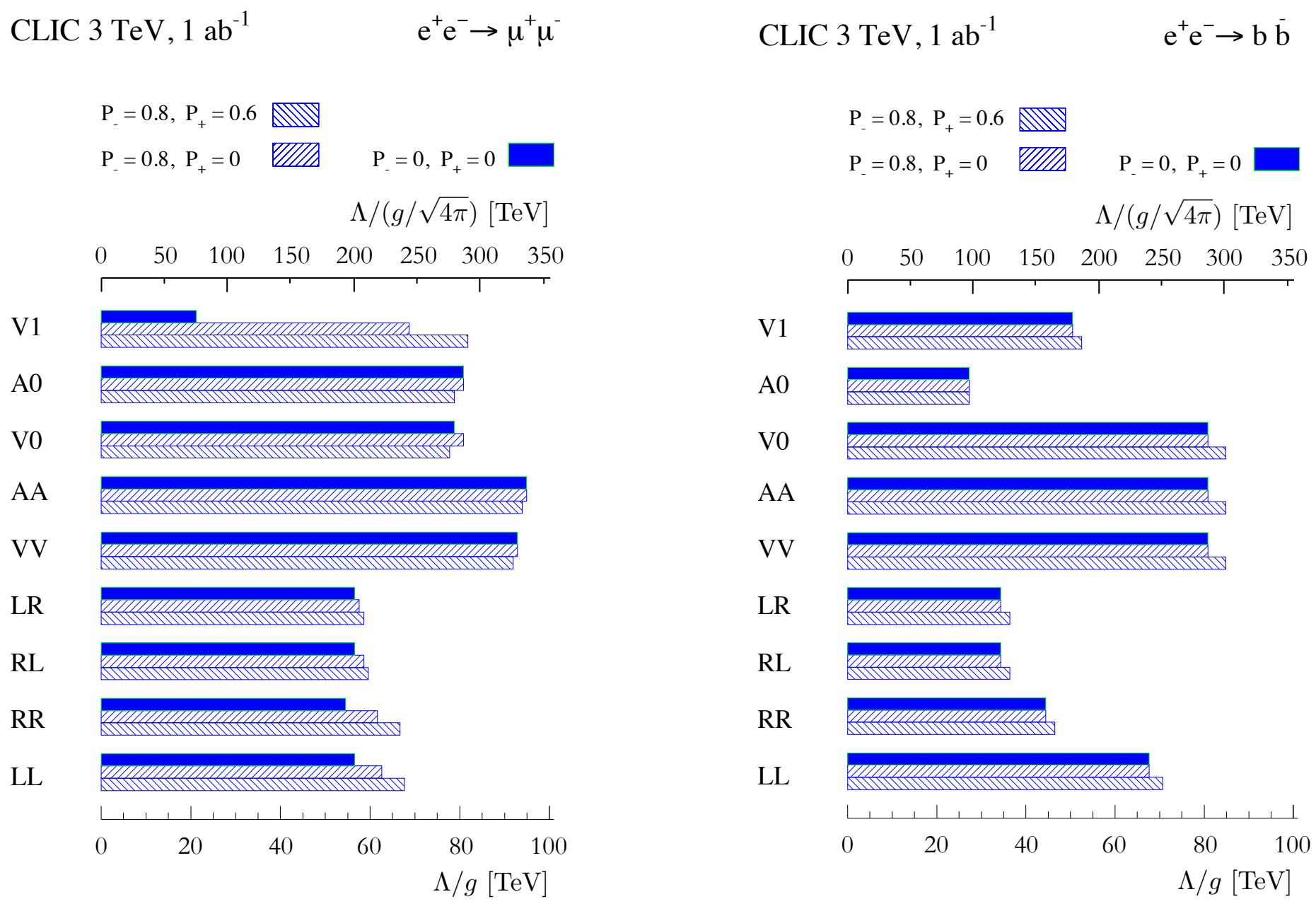

Fig. 1.15: Limits on the scale of contact interactions ($\Lambda/g$) that can be set by CLIC in the $\mu^+\mu^-$ (left) and $b\bar{b}$ (right) channels with $\sqrt{s} = 3$ TeV and $\mathscr{L} = 1$ ab$^{-1}$. A degree of polarisation $P_- = 0, 0.8$ ($P_+ = 0, 0.6$) has been assumed for the electrons (positrons). The various models are defined in Table 6.6 of [20], except the model V1 which is defined as $\{\eta_{LL} = \pm, \eta_{RR} = \mp, \eta_{LR} = 0, \eta_{RL} = 0\}$.

with $\sqrt{s} = 3$ TeV for leptonic final states. In this case the mass of the Z′ is assumed to be unknown, being well beyond the reach of the LHC.

Besides the models considered in the right panel of Figure 1.16, we have studied two other scenarios in detail. The first is a general and model-independent parametrisation of a Z′ boson and its couplings proposed in [68] and generally referred to as *minimal* Z′ model. Its phenomenology at the LHC has been recently studied in [30]. The basic assumption in the model description is the presence of a single Z′ boson originating from an extra $U(1)$ gauge group broken at the TeV scale, and no additional exotic fermions, apart from an arbitrary number of right-handed neutrinos. The requirement of anomaly cancellation and the assumption of flavour universality of the $U(1)$ charges then fix the couplings of the Z′ to the fermions in terms of just two arbitrary parameters, $\tilde{g}_Y$ and $\tilde{g}_{BL}$. Several Z′ models considered earlier in the literature can be incorporated in this framework for specific choices of $\tilde{g}_Y$ and $\tilde{g}_{BL}$.

The second scenario is one in which more than one heavy neutral spin-1 particle exists. This is typical of extra-dimensional extensions of the SM. In particular, we consider the warped/composite two-site model of [69], which represents a qualitatively different scenario where third-generation fermions play a special role. The model can be described as being a “maximally deconstructed” version – i.e. with the extra dimension discretised down to just two sites – of the 5-dimensional Randall–Sundrum custodial model first studied in [70]. In the neutral sector there are three heavy Z′ bosons. Their couplings are controlled by composite-elementary mixing angles, which are generation-dependent. The right-handed top quark, in particular, is fully composite, which implies that the extra spin-1 resonances are strongly coupled to top pairs and are generally broad. The main signatures of the model are large deviations of the top sector observables from their SM expectations. In our analysis we have assumed a universal new vector boson mass $M_*$ and composite coupling $g_*$. Also, we have assumed that the composite fermions have the universal mass scale $m_* = 1.5 M_*$, so that decays of the Z′ particles to the new heavy fermions are forbidden. Our analysis is thus carried out with just two free parameters: $M_*$ and $g_*$. We found only a mild dependence of the final results on the value of the composite Yukawa coupling $Y_{*U33}$ that controls the top mass and the degree of compositeness of $t_L$ and $b_L$.

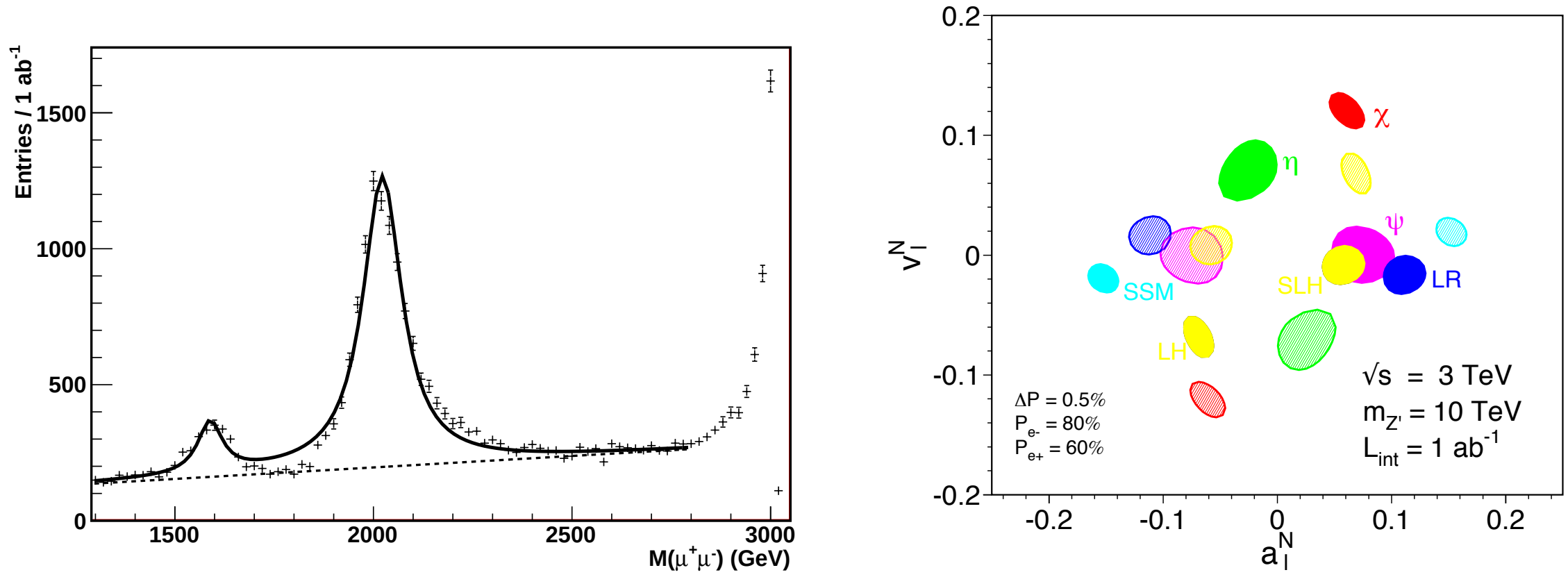

Fig. 1.16: Left: Observation of new gauge boson resonances in the $\mu^+\mu^-$ channel by auto-scan at 3 TeV. The two resonances are the $Z_{1,2}$ predicted by the 4-site Higgsless model of [67]. Right : Expected resolution at CLIC with $\sqrt{s} = 3$ TeV and $\mathscr{L} = 1$ ab$^{-1}$ on the "normalised" leptonic couplings of a 10 TeV Z$'$ in various models, assuming lepton universality. The couplings can be determined up to a twofold ambiguity. The mass of the Z$'$ is assumed to be unknown. $\chi, \eta$ and $\psi$ refer to various linear combinations of $U(1)$ subgroups of $E_6$; the SSM has the same couplings as the SM Z; LR refers to $U(1)$ surviving in Left-Right model; LH is the Littlest Higgs model and SLH, the Simplest Little Higgs model. The two fold ambiguity is due to the inability to distinguish $(a, v)$ from $(-a, -v)$. The degeneracy between the $\psi$ and SLH models might be lifted by including other channels in the analysis ($t\bar{t}$, $b\bar{b}$, ...).

We study the sensitivity of the two models in terms of the discovery regions in their parameter space [71]. The anticipated experimental accuracy on the electroweak observables (total production cross section, $\sigma_{f\bar{f}}$, forward-backward asymmetries and left-right asymmetries, $A_{LR}$) for the process $e^+e^- \rightarrow f\bar{f}$ ($f = \mu$, b, t) is determined from the analysis of fully simulated and reconstructed events, using the same CLIC_ILD detector model and the event reconstruction software adopted for the benchmark analyses discussed in Chapter 12. Beamstrahlung effects are taken into account in the luminosity spectrum, but machine-induced backgrounds are not overlaid on the $e^+e^- \rightarrow f\bar{f}$ events. For polarised observables we assume 80% and 60% polarisation for the $e^-$ and $e^+$ beam respectively. Quark charge is determined using semi-leptonic decays, which are robust against the effect of machine-induced backgrounds. In particular, for $t\bar{t}$ events we tag the top production using the hadronic decay of one top quark and determine the charge using the $W^\pm \rightarrow \ell^\pm\nu$ decay in the opposite hemisphere. The deviations of the nine electroweak observables from their SM predicted values are computed by varying the model parameters in a multi-dimensional grid scan. The sensitivity to a model is defined as the region of parameters for which the $\chi^2$ probability that all the observables are compatible with their SM expected values is below 0.05. Results for the Z$'$ minimal model and the warped/composite model are shown in Figure 1.17 assuming $\sqrt{s} = 3$ TeV. We find that CLIC data are generally sensitive to a mass scale of order 15 TeV with 1 ab$^{-1}$ of accumulated luminosity, which is well beyond the direct accessibility of any current operating collider. In the case of the warped/composite model, the sensitivity is larger for smaller values of $g_*$, since in this limit the couplings of Z' to the leptons and to the bottom quark are larger. For $g_* \gtrsim 1-2$, only the $t\bar{t}$ final state contributes significantly, while the muon and bottom ones are subdominant.

We note that CLIC should also be able to study the signals for spin-2 Kaluza-Klein excitations of extra-dimensional theories through $\gamma + E_T^{miss}$ [72]. Figure 1.18 shows the production cross section as a function of the fundamental gravity scale $M_D$ with the cut $E_{T,\gamma} > 500$ GeV. The SM background rate is 23 fb. With over 1 ab$^{-1}$ of integrated luminosity, the statistical uncertainty in this cross section is of order 0.15 fb. Assuming systematics stay under sufficient control, sensitivity to a signal of greater than 0.5 fb$^{-1}$ is a reasonable assumption, which implies sensitivities to $M_D$ of 14, 9 and 6 TeV for $\delta = 2,4,6$ respectively. A more detailed analysis of the signal versus background for optimised cuts may raise these numbers substantially, perhaps above 20 TeV for the case of $\delta = 2$ [73]. The LHC ($\sqrt{s} = 14$ TeV and 100 fb$^{-1}$ of integrated luminosity) reach for $\delta = 2$ is approximately 9 TeV.

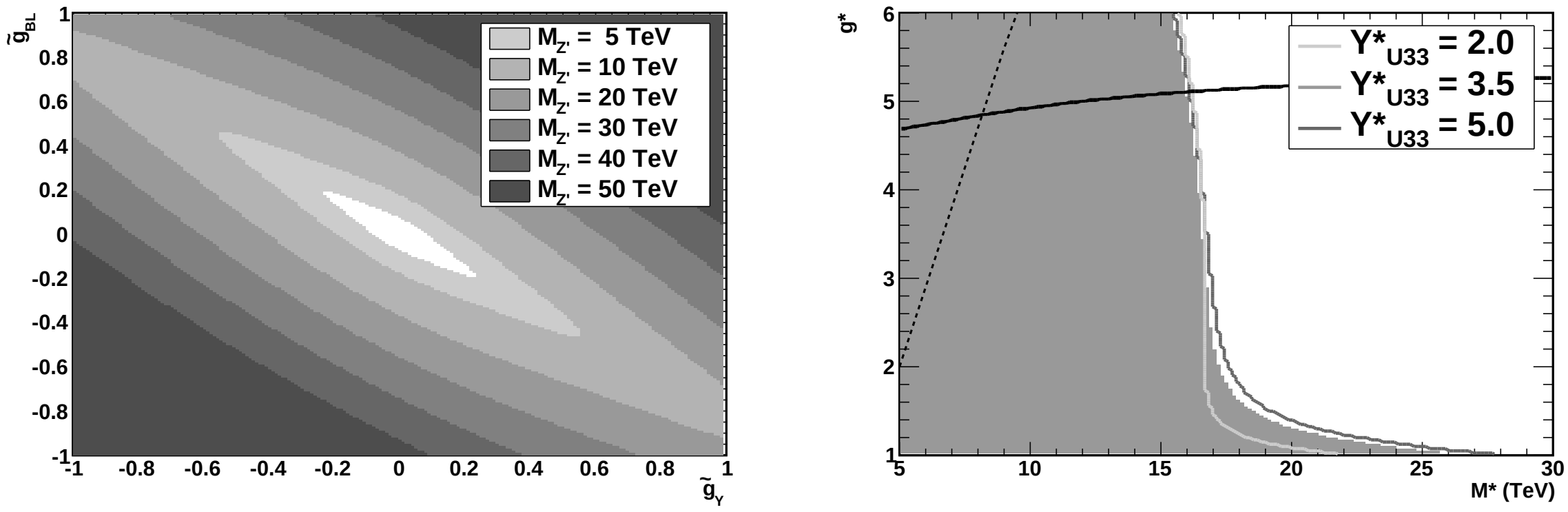

Fig. 1.17: Left: Sensitivity to the Z′ minimal model in the $\tilde{g}_{BL}$ vs. $\tilde{g}_Y$ plane for various values of $M_{Z'}$. Right: Sensitivity to the warped/composite two-site model in the ($M_*$, $g_*$) plane for different values of $Y_{*U33}$. The region above the continuous line has the broader resonance with $\Gamma > 0.5\, M$ and our perturbative calculations cannot be trusted. The region above the dashed line is excluded by present electroweak data for $Y_{*U33} = 2$ but is allowed for larger values of $Y_{*U33}$. In both plots we assume $\sqrt{s} = 3$ TeV with 2 ab$^{-1}$ of luminosity and polarised beams (80% for electrons and 60% for positrons).

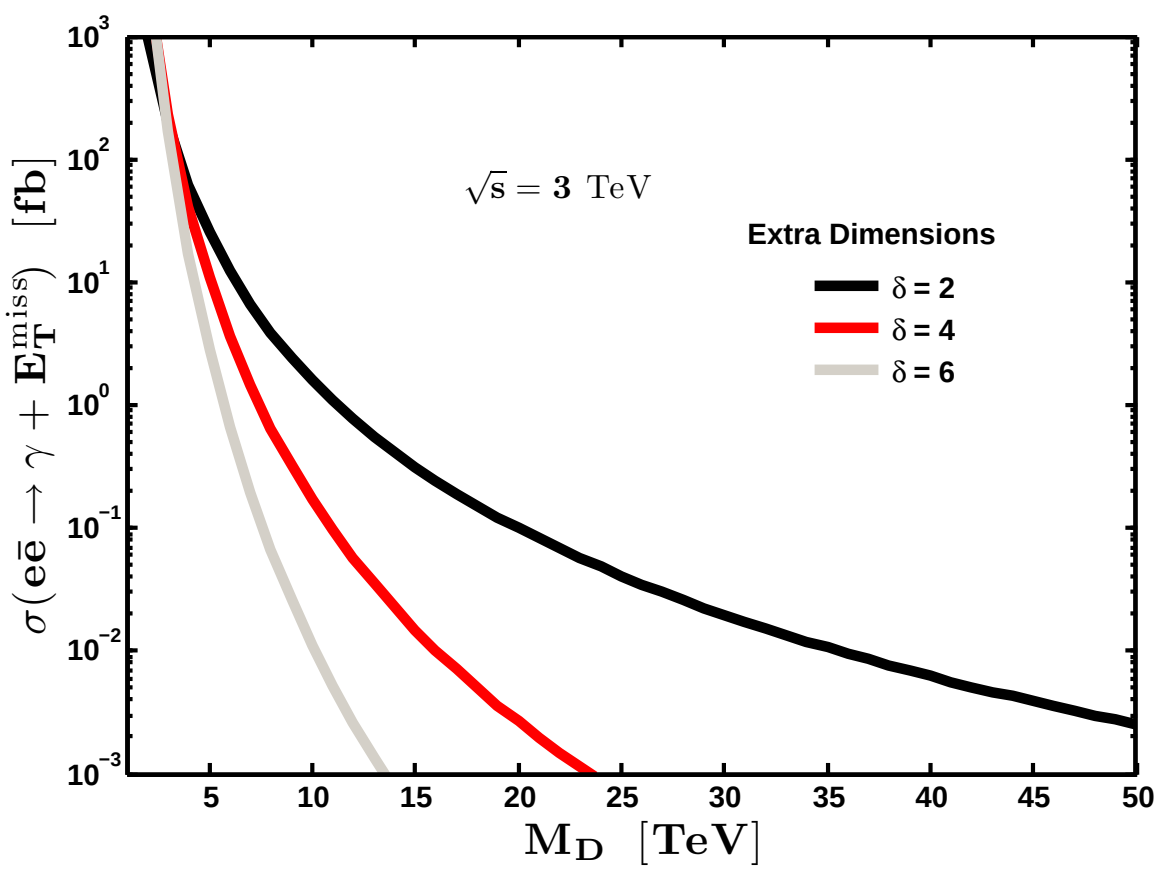

Fig. 1.18: Cross section of photon plus missing energy for ADD flat extra dimensions scenario as a function of the fundamental gravity mass scale $M_D$. Lines are drawn for $\delta = 2,4,6$ extra dimensions. The SM background is approximately 23 fb, which implies sensitivities to $M_D$ of at least 14, 9 and 6 TeV for $\delta = 2,4,6$ respectively with over 1 ab$^{-1}$ accumulated.

## 1.6 Impact of Beam Polarisation

The physics potential of a linear collider is greatly enhanced with polarised beams [45]. On the one hand beam polarisation offers unique access to the chirality of couplings and the structure of interactions. On the other hand, with the appropriate configuration of beam polarisation a more efficient control of background processes can be obtained. This may be also of relevance for disentangling the expected long decay chains, for instance, in Supersymmetry. For example, one of the most dominant SM backgrounds, stemming from $W^+W^-$ pair production, is reduced by a factor of five if the electron beam has a polarisation of 80%. Production cross sections, e.g. for SUSY searches, can be enhanced when using the appropriate polarisation configuration.

The generation and transport of polarised beams, as well as the method used to measure the degree of polarisation, is described in [74] and summarised in Section 2.1.3. The base-line design for CLIC foresees a polarised electron beam at the start of operation, with 80% electron beam polarisation expected. Provisions are made in the design to allow, during a later upgrade phase, for the addition of equipment needed for a polarised positron beam. A recent study at 3 TeV suggests that the measurement of the production of single electroweak bosons in a future CLIC experiment enables one to know the beam polarisation with a precision well below the percent level [75].

Both longitudinal and transverse polarisations are possible for the beams. Effects of transverse polarisation require simultaneously polarised beams and have their usefulness, most especially in the study of CP-violating interactions from physics beyond the SM and the identification of extra-dimension models. However, our focus here is on longitudinal polarisation, which is defined as the ensemble of particles with definite left-handed ($\lambda = -\frac{1}{2}$) or right-handed ($\lambda = +\frac{1}{2}$) helicity [45]:

$$P = \frac{N_R - N_L}{N_R + N_L}.$$

Since the initial leptons can be regarded as being massless, the helicity corresponds to their chirality.

In processes where only (axial-)vector interactions are contributing in $e^+e^-$ annihilation, the dependence on beam polarisation of the cross section can be expressed via the unpolarised cross section $\sigma_0$, the left-right asymmetry $A_{\mathrm{LR}}$, and two polarisation-dependent factors, namely the effective polarisation $P_{\mathrm{eff}}$ and the effective luminosity $\mathscr{L}_{\mathrm{eff}}$. The left-right asymmetry is defined to be:

$$A_{\mathrm{LR}} = \frac{\sigma_{\mathrm{LR}} - \sigma_{\mathrm{RL}}}{\sigma_{\mathrm{LR}} + \sigma_{\mathrm{RL}}},$$

where $\sigma_{LR}$, etc. denotes the cross section for fully left-handed polarised electron and right-handed polarised positron beams.

The effective polarisation is defined to be $P_{\mathrm{eff}} = [P_{\mathrm{e}^-} - P_{\mathrm{e}^+}]/[1 - P_{\mathrm{e}^-}P_{\mathrm{e}^+}]$. And, the effective luminosity is defined to be $\mathscr{L}_{\mathrm{eff}} = \frac{1}{2}[1 - P_{\mathrm{e}^-}P_{\mathrm{e}^+}]\mathscr{L}$, reflecting the number of interacting particles. Utilising these definitions the cross-section can be written as

$$\sigma_{P_{\mathrm{e}^-}P_{\mathrm{e}^+}} = \left(1 - P_{\mathrm{e}^-}P_{\mathrm{e}^+}\right) \frac{\sigma_{\mathrm{RL}} + \sigma_{\mathrm{LR}}}{4} \left[1 - \frac{P_{\mathrm{e}^-} - P_{\mathrm{e}^+}}{1 - P_{\mathrm{e}^+}P_{\mathrm{e}^-}} \frac{\sigma_{\mathrm{LR}} - \sigma_{\mathrm{RL}}}{\sigma_{\mathrm{LR}} + \sigma_{\mathrm{RL}}}\right] = 2\frac{\mathscr{L}_{\mathrm{eff}}}{\mathscr{L}} \sigma_0 \left[1 - P_{\mathrm{eff}} A_{\mathrm{LR}}\right].$$

With the appropriate configuration of beam polarisation a more efficient control of background processes can be obtained. The higher signal-to-background ratio may be crucial for finding manifestations of particles related to new physics and determining their properties. It may also be crucial for disentangling cascade chains from heavier, almost mass degenerate particles.

A prominent example is the suppression of background from W-pairs: with $(P_{\mathrm{e}^-}, P_{\mathrm{e}^+}) = (+80\%, 0)$ $W^+W^-$ production scales by a factor 0.20. Another example is direct graviton production, $e^+e^- \to \gamma G$. The major SM background is determined by $e^+e^- \to \gamma\nu\bar{\nu}$. The contribution from $e^+e^- \to \gamma Z \to \gamma\nu\bar{\nu}$ can easily be eliminated by cutting out the $E_\gamma$ region around the corresponding Z-peak, but there remains a significant, continuous distribution in $E_\gamma$ from $e^+e^- \to \gamma\nu\bar{\nu}$ that has similar behaviour as the signal. Since the neutrino couples only left-handed, the background has nearly maximal polarisation asymmetry and can be effectively suppressed via beam polarisation: with $(P_{e^-}, P_{e^+}) = (+80\%, 0)$ the ratio $S/\sqrt{B}$ is enhanced by about a factor 2.2 and with $(+80\%, -60\%)$ by about a factor 5 [76].

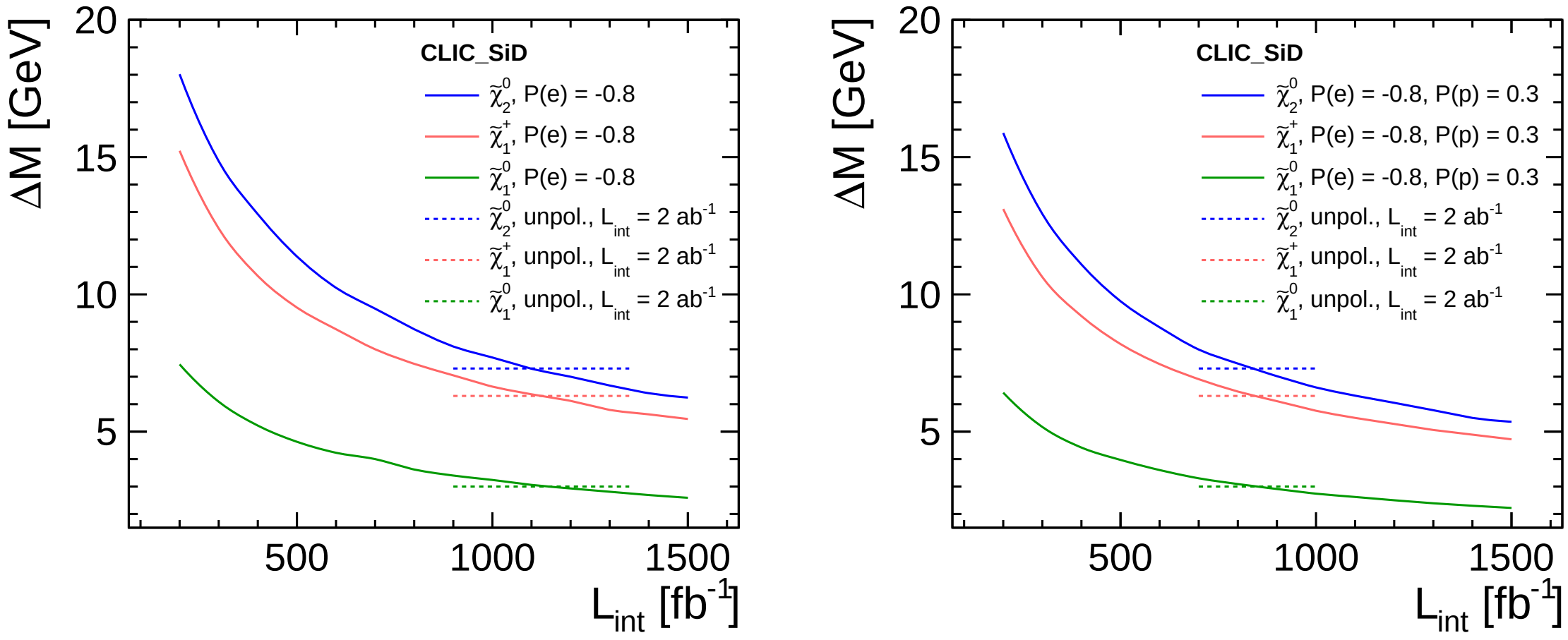

Fig. 1.19: Gaugino mass precision as a function of luminosity for $-80\%$ electron polarisation only (left panel) and for $-80\%$ electron polarisation combined with $+30\%$ positron polarisation (right panel). The horizontal lines represent the achieved mass precision with no polarisation assuming $2\,\mathrm{ab}^{-1}$ of integrated luminosity.

The left-right asymmetry $A_{LR}$ is very powerful in both high precision analyses at lower energies as well as new physics searches at the energy frontier, see [45] and references therein. The relative uncertainty for any left-right asymmetry is given by the expected polarimeter precision for polarised electrons and can be significantly decreased by about a factor 3 if polarised positrons are available.

One of the most promising candidates for physics beyond the Standard Model is Supersymmetry (SUSY). The LHC has a large discovery potential to detect coloured SUSY particles up to 2.5 TeV. The main task of the future linear collider would be to do precision measurements to establish SUSY, and to directly measure the electroweak states (charginos, neutralinos and sleptons) that may be difficult to do at the LHC. The centre-of-mass energy, luminosity, and polarisation needed will depend on the specific choice of SUSY scenario, if indeed nature has chosen the low-scale supersymmetry path.

Although it is well known that the “true” gain in using polarisation occurs for determining chirality, couplings and the structure of the interactions, the first full simulation studies using polarised beams at CLIC concern mass and cross section measurements, which are statistics dominated. These studies have been carried out in the framework of the detector benchmark simulations for our supersymmetry *model II* at 3 TeV (see also Section 2.6 and Chapter 12). The measured masses of supersymmetric particles are determined in part by polarisation. In Figure 1.19 we show the obtained precision on gaugino masses as a function of luminosity for $-80\%$ electron polarisation only (left panel) and for $-80\%$ electron polarisation combined with $+30\%$ positron polarisation (right panel). The horizontal lines represent the achieved mass precision with no polarisation assuming 2 ab$^{-1}$ of integrated luminosity (see Section 12.4.6). We can see that for $P_{e^-} = -80\%$ the same precision can be achieved with only 1.12 ab$^{-1}$. For $P_{e^-} = -80\%$ and $P_{e^+} = 30\%$ the required luminosity decreases further to 0.84 ab$^{-1}$.

In Figure 1.20 we show the precision of gaugino cross section measurements as functions of the polarisation uncertainty. The plots assume luminosities of 1.12 ab$^{-1}$ (left panel) and 0.84 ab$^{-1}$ (right panel), as motivated above. As the polarisation uncertainty is expected to be well below 2%, this leads to excellent cross section determinations.

There are numerous of additional examples where the physics analyses strongly benefit from polarised beams. The ability to control the polarisation of the beams enables additional observables to be measured, giving access to the chirality of the interactions and allowing one to determine the couplings in an accurate and more model-independent manner. Another important example of the utility of polarisation can be found in studies that determine the coupling of exotic $Z'$ bosons coupling to SM fermions. This has been utilised in Section 1.5 to pin down the $a_1^N$ and $v_1^N$ couplings of the $Z'$ boson to muons.

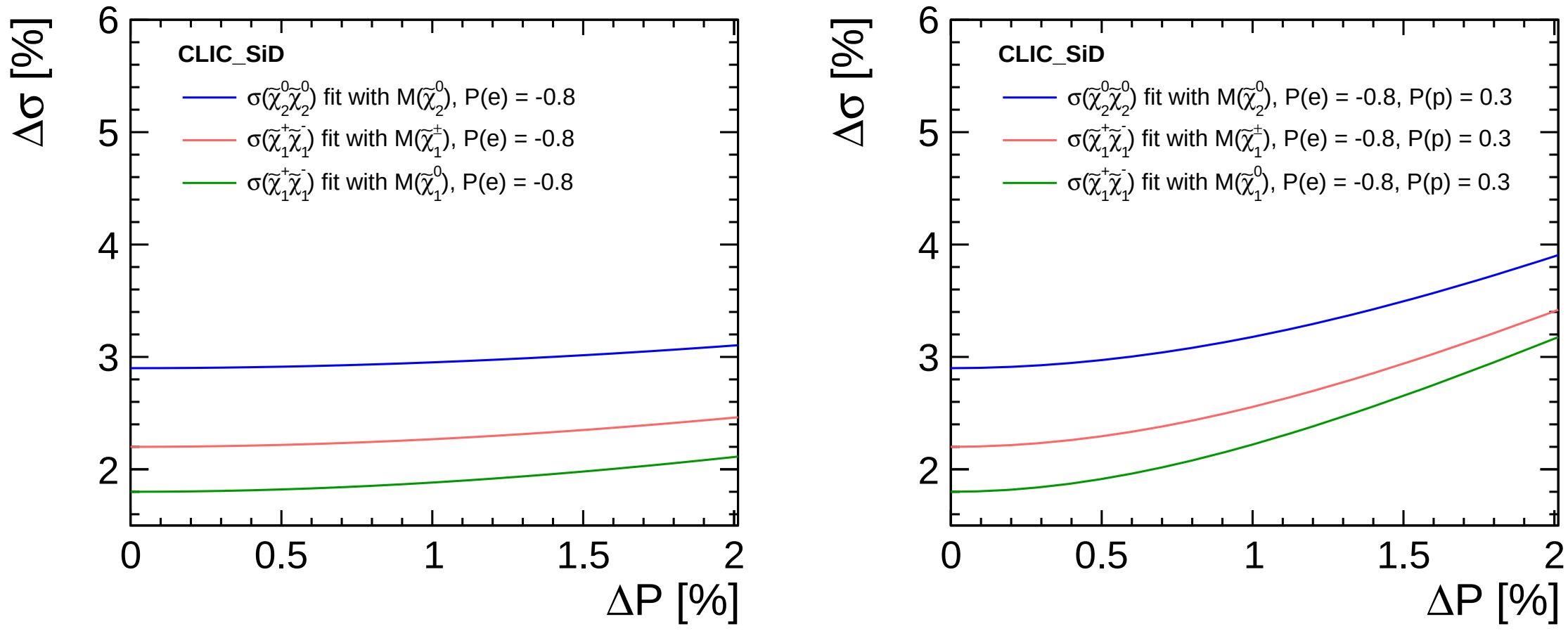

Fig. 1.20: The precision of gaugino cross section measurements as functions of the polarisation uncertainty. The plots assume luminosities of 1.12 $ab^{-1}$ (left panel) and 0.84 $ab^{-1}$ (right panel), as motivated by the discussion in the text.

## 1.7 Precision Measurements Potential

Precision measurements in particle physics are used to determine theory parameters to high accuracy or to test new physics theories indirectly through quantum loop effects or tree-level contributions that are suppressed by heavy-mass terms in propagators or couplings. They are complementary to the direct searches which, however, need to have sufficiently large centre-of-mass energies such that new degrees of freedom can be produced as real particles. The most successful examples for precision measurements in $e^+e^-$ annihilation are LEP-I and SLC running close to the Z-pole. Comparison of data with theory computations predict that the Higgs is light and disfavours some extensions of the SM like the simplest QCD rescaled version of Technicolor [77]. With LEP-II running at higher energies, in some models, neutral gauge bosons were found to be excluded up to around 1 TeV from the direct contributions to fermion pair production [78]. Together with data from rare weak decays and cosmological measurements, precision data from $e^+e^-$ experiments constitute the most important constraints that all new physics models have to pass.

For the ILC at energies between 500 GeV and 1 TeV studies have been done for a variety of processes. Light fermion production is sensitive to an extended gauge sector at higher energies as well as to Kaluza-Klein excitations of SM particles in theories with extra space dimensions. $W^+W^-$ production is sensitive to new physics in theories without a Higgs boson. It is also possible to reach an experimental precision for which there is sensitivity to higher-order loop effects e.g. from Supersymmetry. However, not many theoretical calculations exist for a complete picture. The most sensitive probe of theories without Higgs bosons, where new physics effects regulate vector boson scattering at energies close to the unitarity bound of about 1 TeV, are measurements sensitive to the quartic gauge-boson coupling, namely triple gauge boson production and vector boson scattering. The top quark mass can be measured with very good precision with a line-shape scan close to the top pair threshold while the top quark couplings, which are sensitive to theories where the top quark plays a special role in mass generation, can be measured at higher energies from the determination of branching fractions or production cross sections. In addition, QCD tests are interesting at high energies due to the different interplay of perturbative and non-perturbative contributions compared to lower energy measurements. For short-distance dominated processes the perturbative series converges faster and non-perturbative effects are smaller due to asymptotic freedom. However, accurate measurements become more challenging due to the $1/s$ decrease of the signal and due to the stronger impact of background processes that open up a higher energies.

As discussed in Section 1.5, an important scenario is related to a new neutral gauge boson (Z′) in the energy range directly accessible to CLIC, i.e. with a mass below 3 TeV. In this case CLIC could measure the mass and couplings of the Z′ with an energy scan and several asymmetry measurements in the same way as done for the standard Z boson at LEP-I. The precision is influenced by the knowledge of the beam energy, the absolute luminosity and the beamstrahlung, which affects the average beam energy as well as the effective beam energy spread.

For higher Z′-masses CLIC is sensitive to effects from virtual Z′ exchange. The cross section change relative to the SM one is modified by a term $s/(m^2_{Z'} - s)$, which for masses significantly higher than $\sqrt{s}$ is equal to $s/m^2_{Z'}$. Assuming similar systematic uncertainties as for ILC studies [79] and a luminosity roughly scaling with the centre-of-mass energy, CLIC would be sensitive to Z′ masses above 10 TeV for all studied models. If the mass of the Z′ is known from the LHC, CLIC can do precision measurements of its couplings. Figure 1.16 in Section 1.5 shows the possible precision for which the couplings of a Z′ can be resolved.

Another area of precision tests are triple gauge couplings measured in $W^+W^-$ production. Using the common parametrisation of triple gauge couplings in terms of $g$, $\kappa$, $\lambda$ [80] where the SM expectations are either zero or one, it has been shown that ILC can reach a precision of a few times $10^{-4}$ [81]. Using simple extrapolation laws a precision of better than $10^{-4}$ at CLIC can be expected. Supersymmetry effects in the $10^{-4}$ region are possible, which thus might be visible at CLIC.

If, in models without a Higgs boson, there are vector resonances coupled to longitudinal W-bosons, a Z produced in $e^+e^-$ annihilation could fluctuate into such a resonance in the same way as a photon can fluctuate into a $\rho$ at lower energies. It has been shown that already the ILC is sensitive to resonances with very high masses and also to models without a resonance following the Low Energy Theorem analysing form factors in $e^+e^- \to W^+W^-$ [82]. If vector resonances in the CLIC mass region exist, precision measurements of their masses and couplings should be possible. Unfortunately experimental studies of this process do not yet exist.

Top quark mass measurements at the Tevatron and the LHC are based on comparing experimental distributions such as the reconstructed invariant mass with the Monte-Carlo generator, and thus measure the Monte-Carlo top mass parameter. Since the field theoretic relation of this parameter with Lagrangian mass definitions used for theoretical predictions is unclear, an independence of top mass measurements from Monte-Carlos is desired. Recently, new theoretical developments have made possible first principles factorisation predictions of the top invariant mass distribution in $e^+e^-$ annihilation for centre-of-mass energies much larger than the top mass [83]. These factorisation predictions rely on perturbation theory and on a non-perturbative distribution function that can be determined from massless quark initiated event-shape distributions such as thrust, and allow one to determine the top quark mass directly with full control of the scheme and independent of possible Monte-Carlo artefacts. With dedicated theoretical and experimental effort this method could compete with the threshold scan method that can only be carried out at low-energy runs. An important experimental aspect of such measurements is the separation of top quark initiated events from massless quark initiated events and from background processes. Moreover, good control of the initial state beam effects and of QED radiation are required for a reliable determination of the event shape variables.

The measurement of the Monte Carlo top mass from the reconstruction of the jet invariant mass and using kinematic fits is among the detector benchmark measurements studied in detail in this report (see Sections 2.6.6 and 12.4.7). Compared to the environment at the LHC substantially more accurate determinations of the top quark mass are possible. At 500 GeV centre-of-mass energy uncertainties comparable to similar earlier ILC reconstruction studies were obtained indicating that advanced analysis techniques can overcome the somewhat more challenging CLIC environment. It is conceivable that comparable results can be achieved also at 3 TeV using particle flow event reconstruction, although achieving the same precision is certainly difficult due to the top quark boost and the more challenging background. The results also indicate the potential for top mass measurements based on fits to invariant mass factorisation predictions which provide more control of the theoretical uncertainties.

## 1.8 Discussion and Conclusions

CLIC can play an essential role in second-generation studies of the weak scale that aim at a thorough understanding and a comprehensive completion of the first discoveries made by the LHC. The added value of CLIC will be twofold. On one side, CLIC can extend the *discovery reach* for particles that suffer from low production cross section at the LHC or high background contamination. On the other side, CLIC will offer the opportunity of *precise measurements* of masses and couplings of the new particles discovered at the LHC. Both of these aspects are crucial for interpreting the discoveries, for discriminating among competing models, and for identifying the correct underlying theory. In this Chapter we have given several examples that highlight CLIC's role in the pursuit of solving the mysteries of the particle world at the weak scale, and we summarise here some of the most prominent features.

The Higgs boson is expected to be discovered at the LHC. The foreseen mass precision is $0.1\% - 0.3\%$ for $m_{\mathrm{H}} = 115 - 200$ GeV [5] with 60 $\mathrm{fb}^{-1}$ of integrated luminosity at the 14 TeV collider. If such a particle is discovered, one of the most relevant theoretical questions that we will face is the determination of its nature. In other words, we will want to assess whether the Higgs boson is a fundamental particle or a composite of some new strong force. Part of this information is encoded in the couplings of the Higgs boson, and a resolution of this issue requires highly precise measurements. Roughly speaking, determinations of the Higgs couplings with relative uncertainties $\Delta$ can be translated into probes of the compositeness scale $\Lambda$ up to values $\Lambda \simeq (0.01/\Delta)^{1/2}$ 60 TeV. Since percent precisions are foreseeable, CLIC can test the scale of Higgs compositeness up to about 60 TeV, significantly above the 7 TeV sensitivity of the LHC. This will be sufficient to confirm or rule out any of the models of Higgs compositeness proposed to address the naturalness problem. A crucial test on the nature of the Higgs boson comes from studies of $\mathrm{W}^+\mathrm{W}^-$ scattering and double-Higgs production in vector-boson fusion. These processes probe the couplings of the would-be Goldstone bosons and therefore directly test the compositeness hypothesis. CLIC will greatly extend the LHC sensitivity, as discussed in Section 1.4. In particular, CLIC offers the opportunity to probe the non-linearities of the couplings of the strongly interacting Higgs boson through vector boson fusion into a pair of Higgs bosons, a process which is very difficult to study at the LHC. In this respect CLIC can do for the electroweak breaking sector what LEP did for the gauge sector.

We have emphasised so far the role that CLIC can play in the exploration of the Higgs sector through *precision measurements,* but the CLIC potential *discovery reach* may well play an equally important role. For example, in many theoretically motivated scenarios, the Higgs sector contains more particles than just one neutral state. The two Higgs doublet system of supersymmetric models is the most familiar example. As is well known, preliminary LHC limits on sparticle masses suggest that the minimal supersymmetric Higgs sector approximately splits into a light Higgs boson with nearly SM properties and an almost degenerate Higgs doublet with vanishing vacuum expectation value. If this is the case, discovery of the additional Higgs states beyond the light boson is extremely challenging at the LHC. Heavy Higgs bosons can be discovered only if $\tan\beta$ is fairly large, while for smaller $\tan\beta$ a pseudoscalar Higgs can escape detection at the LHC even if it is as light as 200 GeV. A huge portion of the supersymmetric Higgs parameter space is hidden to LHC searches. The situation is remedied by CLIC which, as shown in Figure 1.21, can cover essentially all the relevant region of supersymmetric models with an impressive gain in the discovery reach.

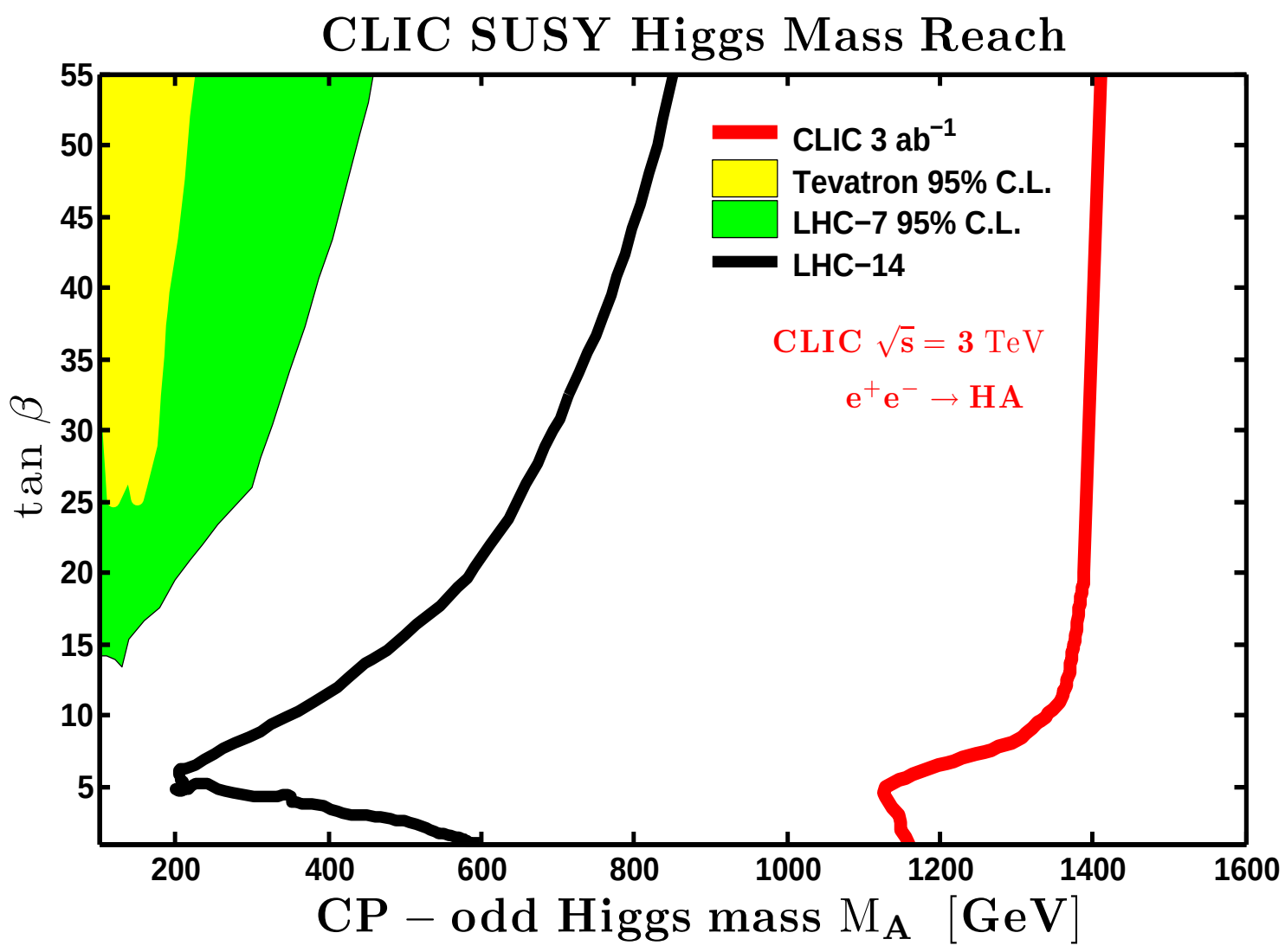

Fig. 1.21: Search reach in the $m_A - \tan\beta$ plane for LHC and CLIC. The left-most coloured regions are current limits from the Tevatron with $\sim 7.5$ fb$^{-1}$ of data at $\sqrt{s} = 1.96$ TeV and from $\sim 1$ fb$^{-1}$ of LHC data at $\sqrt{s} = 7$ TeV. The black line is projection of search reach at LHC with $\sqrt{s} = 14$ TeV and 300 fb$^{-1}$ of luminosity [84]. The right-most red line is search reach of CLIC in the HA mode with $\sqrt{s} = 3$ TeV. This search capacity extends well beyond the LHC. A linear collider at $\sqrt{s} = 500$ GeV can find heavy Higgs mass eigenstates if their masses are below the kinematic threshold of 250 GeV.

We can view supersymmetry as a useful template for discussing theories beyond the SM. If supersymmetry is the solution to the naturalness problem, the LHC will discover it. But even in the most propitious situation, we cannot expect the LHC to unravel all the issues related to the link between supersymmetry and the weak scale. One can easily list some of the questions that are likely to remain unanswered after the LHC has completed its mission. What is the pattern of supersymmetry breaking? (Or, probably more precisely, what is the mechanism for mediating supersymmetry breaking?) Do gaugino masses unify in the same way as gauge coupling constants do? Is the lightest supersymmetric particle the dark matter so abundant in the universe? Do squark and slepton masses become equal at some high-energy scale or do they satisfy special sum rules?

The importance of these questions is as fundamental as the discovery of supersymmetry itself. Actually, the discovery of supersymmetry will remain moot if these questions are not answered. CLIC is the ideal machine to address all these questions. The first aspect is related to the *discovery reach*. While the LHC will efficiently explore any coloured supersymmetric particle with mass below $2.5 - 3$ TeV, the search for supersymmetric particles with only electroweak charges is much more model dependent. Either these particles are produced in decay chains of coloured sparticles or their mass reach at the LHC is limited. Cases of near mass degeneracy can also be very problematic. As a result, it is highly plausible that the LHC will not be able to discover the full set of new particles and many holes will remain in the "supersymmetric periodic table". The complementarity of the CLIC discovery potential is noteworthy. Any supersymmetric particle with electroweak charge and mass smaller than about half the centre-of-mass energy can be efficiently produced and studied at CLIC. Thus, CLIC will plausibly complete the discovery of supersymmetry, finding any missing states.

Also in the case of supersymmetry, the role of *precision measurements* is of paramount importance. If supersymmetry is discovered, the origin and pattern of supersymmetry breaking will become one of the most pressing open questions. At present several different schemes for mediation of supersymmetry breaking are known, and each scheme gives rise to a characteristic mass spectrum. Unfortunately in many cases, the emerging spectra have similar features and their discrimination could be possible only after very precise mass determinations. For example, it is known that there are large regions of parameter space where the superpartner mass spectra of minimal gauge mediated models (mGMSB) overlap with the spectra predicted by minimal supergravity models (mSUGRA). In Figure 1.22 we show one example of this overlap. For the choices of input parameters described in the figure caption, we have overlap of all the masses at a level irresolvable by the LHC. However, the extraordinary capacity of CLIC to measure mass spectra enables us to distinguish between these two models by the careful measurement of the slepton, Higgs and gaugino masses. Shown in the inset is one such example.

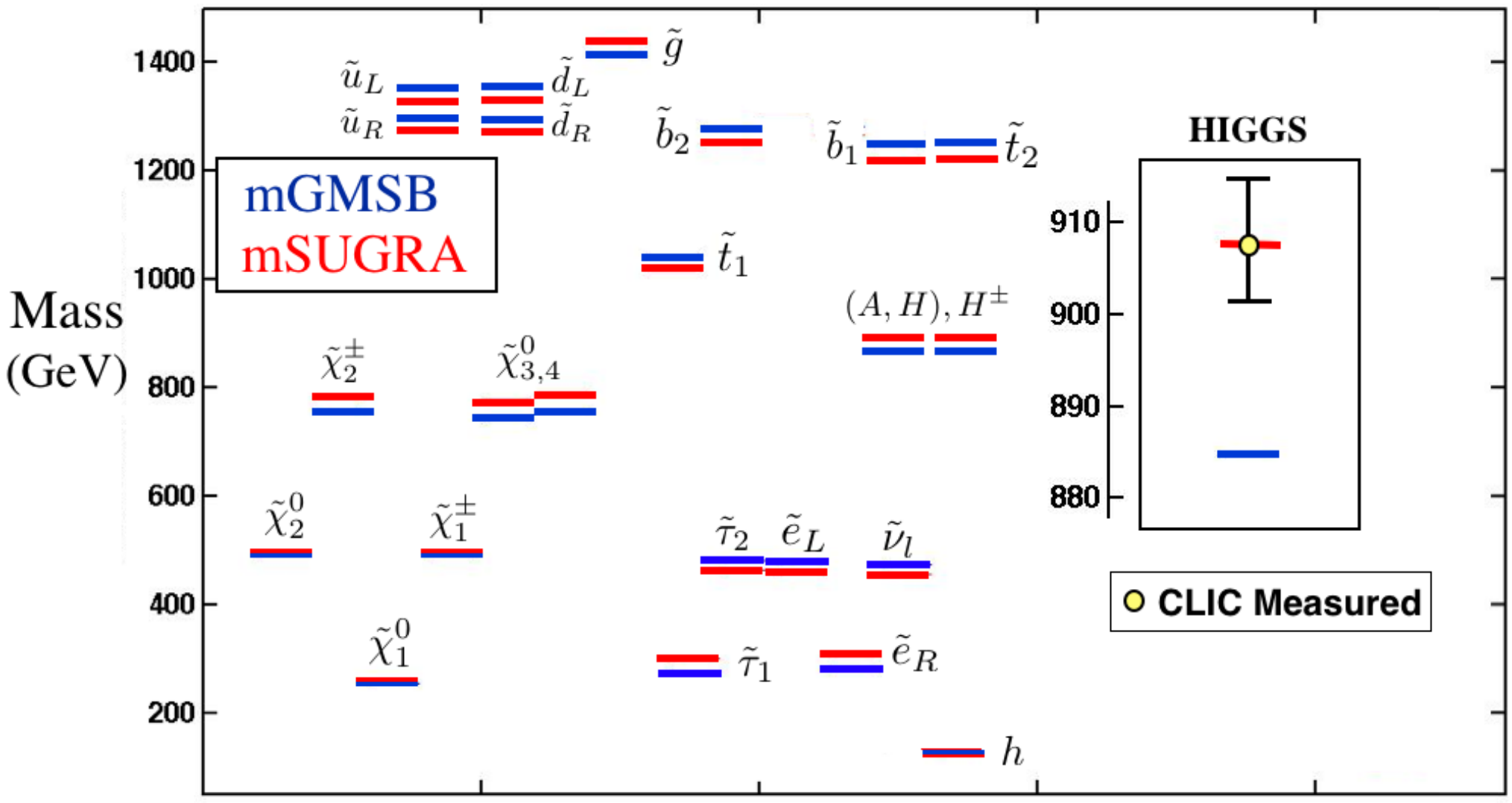

Fig. 1.22: Resolving SUSY breaking models and masses with CLIC: Shown are the nearly degenerate spectra of a mSUGRA model and a mGMSB model. Assuming some of the SUSY particles masses are measured, with a spectrum of the type above predicted by the different models of supersymmetry breaking, CLIC would be able to discern not only some of the slepton masses and the heavier charginos within the two models, but also the SUSY Higgs masses. For mSUGRA the soft masses are $m_0 = 175$ GeV, $m_{1/2} = 645$ GeV, $A_0 = 0$, with $\tan\beta = 10$ and $\mu > 0$. For mGMSB the number of messengers are $n_l = n_q = 5$, and $\Lambda_{\rm SUSY} = 4 \cdot 10^4$ GeV, $M_{\rm Mess} = 10^{12}$ GeV, with $\tan\beta = 10$.

As gauge coupling unification is one of the most attractive aspects of low-energy supersymmetry, the question of gaugino mass unification is particularly important. Studies of gaugino masses can give further evidence in favour of a grand unification of forces and reveal details about the specific unification model. The LHC can provide us with rough indications about unification but, as was the case for the gauge couplings, an increase in the measurement precision of gaugino masses can make a crucial difference. A precise determination of $M_1$ and $M_2$ (the electroweak gaugino masses) at CLIC will allow us to perform two important tests of the idea of unification. First, we will be able to compute the gluino mass from the unification hypothesis and compare the result with the LHC measurement. Second, we will compute the unification scale and compare it with the value obtained from gauge coupling unification. These consistency checks will quantitatively test the idea of grand unification, thereby corroborating our confidence in the far-reaching extrapolation to extremely small distances performed in unification scenarios.

Figure 1.23 shows $M_2$ vs. $M_1$ for various unification scales. The precision with which the gaugino masses can be measured are represented by the full range of the inset box, and demonstrate that the gaugino unification scale can be established to be the same as the gauge coupling unification scale to good accuracy. This plot was made at one loop accuracy for illustration purposes, but two loop effects and superpartner threshold effects, which are at the same order, are calculable and straightforward to include once the full spectrum is known. High-scale threshold corrections are not known, and can have a few percent effect on $M_2$ vs. $M_1$ in unified scenarios, which should be kept in mind when contemplating compatibility with high-scale gaugino unification.

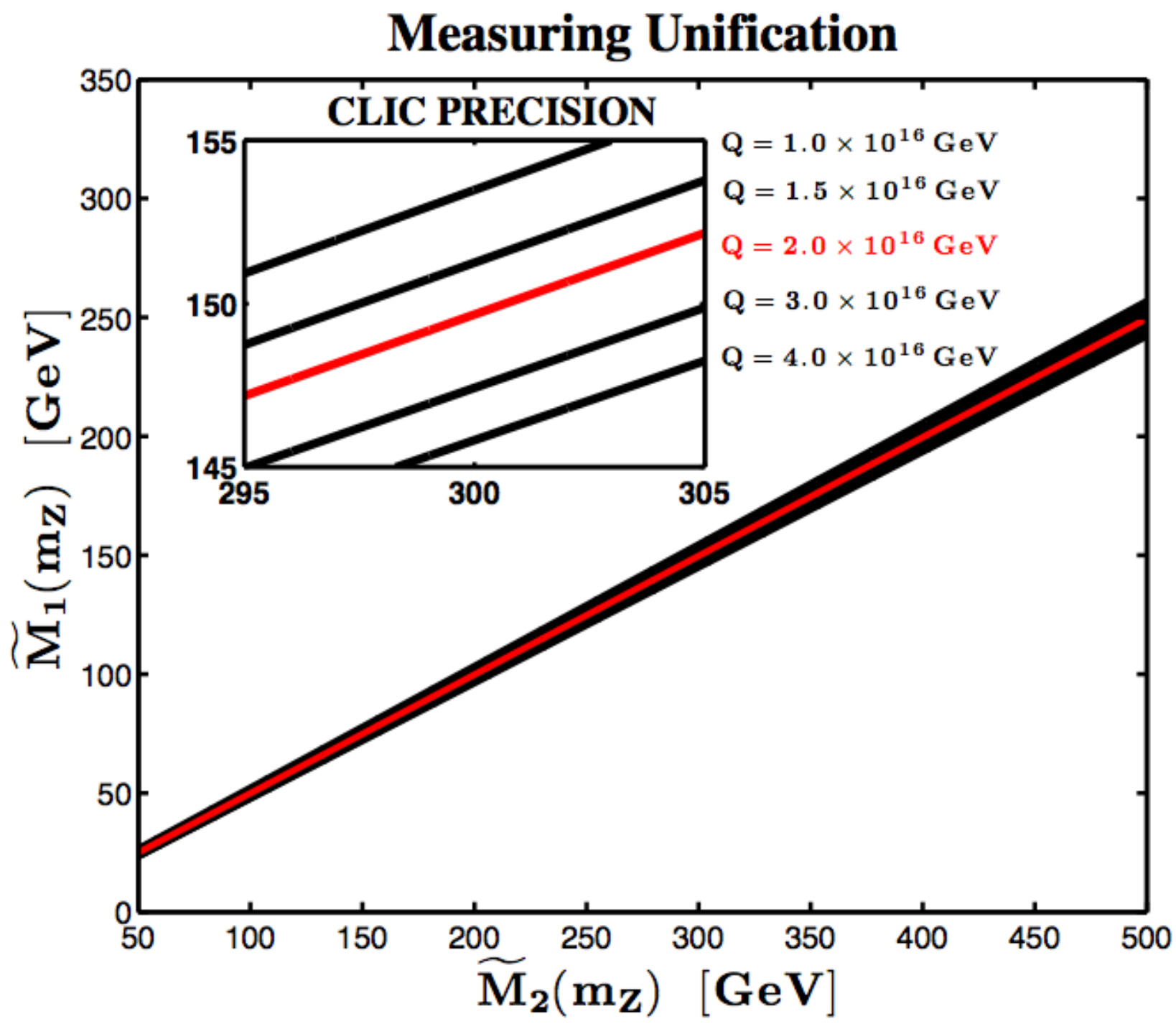

Fig. 1.23: The electroweak gaugino masses $M_2$ versus $M_1$ at the low scale assuming unification of the gauginos at various high-energy scales $Q$. The precision with which the gaugino masses can be measured are represented by the full range of the inset box. This shows that the gauge coupling unification scale of $Q \simeq 2 \cdot 10^{16}$ GeV can be inferred to within a factor of two from gaugino mass measurements alone, which is an excellent independent test of unification.

A fundamental question in particle physics concerns the origin of dark matter. While weak-scale dark matter particles imply anomalous signals with missing transverse energy at the LHC, the converse is not necessarily true. Although an excess of missing energy at the LHC can be interpreted as an indication for artificial manufacturing of dark matter, more solid evidence must come from other sources. One good indicator would be consistency between the properties of the invisible particle observed at the LHC and the properties extracted from signals in direct or indirect dark matter searches. But even in the absence of a positive signal in dark matter experiments, collider searches can provide us with a crucial clue. Through precise measurements of masses and couplings of all new particles, one can reconstruct the thermal relic abundance of the invisible particle and compare it with cosmological observations. It should also be stressed that, even in the presence of direct identification of the dark matter particle, studies of the relic abundance are fundamentally meaningful. In fact, they are the only way in which we can learn whether the dark matter particle is a thermal or non-thermal relic, thus shedding light on the mechanism that is responsible for almost all the mass in the universe.

Although the determination of the dark matter relic abundance is in principle possible at the LHC, in practice it is unlikely to be feasible. The difficulty lies in measuring all the particles participating in the relevant annihilation processes in the early universe and in achieving the required precision in masses and couplings. CLIC is ideally suited to make such measurements and to carry out this program successfully. This claim is substantiated in Figure 1.12 in Section 1.3, where it is shown how, in a special example, one can reconstruct the neutralino relic density from CLIC measurements. In this case, the accurate CLIC measurements will enable us not only to deduce that neutralinos constitute dark matter, but also to identify the process that turned them into fossils of the primordial universe.

Table 1.6: Discovery reach of various theory models for different colliders and various levels of integrated luminosity, $\mathscr{L}$ [73]. LHC14 and the luminosity-upgraded SLHC are both at $\sqrt{s}$ =14 TeV. LC800 is an 800 GeV $e^+e^-$ collider and CLIC3 is $\sqrt{s}$ =3 TeV. TGC is short for Triple Gauge Coupling, and "$\mu$ contact scale" is short for LL $\mu$ contact interaction scale $\Lambda$ with $g = 1$ (see Section 1.4).

| New particle | collider: $\mathscr{L}$: | LHC14 100 $fb^{-1}$ | SLHC 1 $ab^{-1}$ | LC800 500 $fb^{-1}$ | CLIC3 1 $ab^{-1}$ |
|---|---|---|---|---|---|
| squarks [TeV] | | 2.5 | 3 | 0.4 | 1.5 |
| sleptons [TeV] | | 0.3 | - | 0.4 | 1.5 |
| $Z'$ (SM couplings) [TeV] | | 5 | 7 | 8 | 20 |
| 2 extra dims $M_D$ [TeV] | | 9 | 12 | 5-8.5 | 20-30 |
| TGC (95%) ($\lambda_\gamma$ coupling) | | 0.001 | 0.0006 | 0.0004 | 0.0001 |
| $\mu$ contact scale [TeV] | | 15 | - | 20 | 60 |
| Higgs compos. scale [TeV] | | 5-7 | 9-12 | 45 | 60 |

We have focused here on supersymmetry because it provides an explicit and computable setup, but similar considerations can be made for a variety of specimen theories beyond the SM. The extended *discovery reach* and the enhanced *precision measurements* provided by CLIC are likely to be the necessary tools to address many of the fundamental questions about the weak scale left unanswered by the LHC.

The reach of CLIC in comparison with other colliders for a few representative theories is shown in Table 1.6. On the precision side, the ability to measure the Higgs boson couplings at the linear collider is the most well known capability. The excellent sensitivity to new particles and to higher dimensional operators induced by heavy states, is what leads to this unprecedented reach in parameter space at 3 TeV CLIC.

# Chapter 2

# CLIC Experimental Conditions and Detector Requirements

In this chapter the design requirements for a detector at CLIC are described. These requirements are driven both by the physics at CLIC, potentially operating over a range of centre-of-mass energies, and by the machine environment and related backgrounds at the interaction point (IP).

## 2.1 The CLIC Experimental Environment

The main parameters of the CLIC beam of relevance to the physics reach of the machine and the related levels of background at the IP are summarised in Table 2.1 (see also [1]).

The experimental environment at CLIC differs from that at previous $e^+e^-$ colliders such as LEP and also the proposed ILC. In particular, there are three main aspects of the CLIC machine that determine the physics environment and significantly impact the CLIC detector design:

- The high bunch charge density, related to the small beam size at the interaction point, means that the electrons and positrons radiate strongly in the electromagnetic field of the other beam, an effect known as beamstrahlung (similar to synchrotron radiation). Consequently the centre-of-mass energies of the $e^+e^-$ collision have a long tail towards significantly lower values than the notional centre-of-mass energy (discussed in Section 2.1.1).
- There are significant beam related backgrounds. The $e^+e^-$ incoherent pair background has a major impact on the design of the inner region of the detector and the forward region. The pile-up of approximately 3.2 $\gamma\gamma \rightarrow$ hadrons "mini-jet" events per bunch crossing (BX) impacts the timing requirements placed on the individual detector elements and is an important consideration in all physics analyses (discussed in Section 2.1.2)
- The CLIC beam consists of bunch trains of 312 bunches with a train repetition rate of 50 Hz. Within a bunch train, the bunches are separated by 0.5 ns. The short time between bunches means that a detector will inevitably integrate over a number of bunch crossings. This combined with the significant $\gamma\gamma \rightarrow$ hadrons background implies fast readout of all detector elements and excellent time resolution (discussed in Section 2.5).

### 2.1.1 The CLIC Beam

The intrinsic r.m.s. energy spread of CLIC beams is 0.35%. In addition, due to variations of the main linac accelerating RF voltage and phase, the mean beam energy is expected to be subject to a jitter of the order of 0.1%. However, due to beamstrahlung, not all the $e^+e^-$ collisions at CLIC will take place at the nominal centre-of-mass energy. The impact of beamstrahlung on physics measurements is not unlike that of initial state radiation (ISR); the colliding electron and positron may radiate a high energy photon before the collision and, consequently, the centre-of-mass energy of the electron-positron collision $\sqrt{s'}$ is less than the nominal centre-of-mass energy of the machine $\sqrt{s}$. This results in an effective centre-of-mass spectrum (the luminosity spectrum) with a peak at $\sqrt{s}$ corresponding to collisions with no beamstrahlung and a long tail towards lower energies. Figure 2.1 shows the luminosity spectrum for CLIC operating at 500 GeV and 3 TeV. The long tail from beamstrahlung is particularly evident at $\sqrt{s} = 3$ TeV; the effect of beamstrahlung is less evident for the 500 GeV machine.

The impact of the luminosity spectrum on the physics reach of CLIC is not simple to quantify; it depends on the physics process being studied. At the most basic level it reduces the amount of luminosity available at the highest centre-of-mass energies. This is quantified in Table 2.2. For CLIC operating at 3 TeV only 35% of the effective luminosity is within 1% of the nominal centre-of-mass energy.

Table 2.1: The main parameters of the CLIC machine and background rates at the interaction point. The listed variables are: $\theta_c$ the horizontal crossing angle of the beams at the IP; $f_{\mathrm{rep}}$, the repetition frequency; $n_{\mathrm{b}}$, the number of bunches per bunch train; $\Delta t$, the separation between bunches in a train; $N$, the number of particles per bunch; $\sigma_{\mathrm{x}}$, $\sigma_{\mathrm{y}}$, and $\sigma_{\mathrm{z}}$, the bunch dimensions at the IP; $\beta_x$ and $\beta_y$ the beta functions at the IP; $L^*$ the distance from the last quadrupole to the IP; $\mathscr{L}$, the design luminosity; $\mathscr{L}_{0.01}$, the luminosity with $\sqrt{s'} > 0.99\sqrt{s}$; $\Delta E/E$, the average fraction loss of energy from beamstrahlung; $n_\gamma$, the average number of beamstrahlung photons per beam particle; $N_{\mathrm{coh}}$, the number of coherent pair particles per bunch crossing (BX); $E_{\mathrm{coh}}$, the total energy of coherent pair particles per BX; $N_{\mathrm{incoh}}$, the number of incoherent pair particles per BX; $E_{\mathrm{incoh}}$, the total energy of incoherent pair particles per BX; and, $n_{\mathrm{Had}}$, the number of $\gamma\gamma \to$ hadrons events per BX for a threshold of 2 GeV. The backgrounds rates and energy releases are quoted excluding safety factors for the simulation uncertainties.

| Parameter | Units | $\sqrt{s} = 500$ GeV | $\sqrt{s} = 3$ TeV |
|---|---|---|---|
| $\theta_c$ | mrad | 18.6 | 20 |
| $f_{\mathrm{rep}}$ | Hz | 50 | 50 |
| $n_{\mathrm{b}}$ | | 354 | 312 |
| $\Delta t$ | ns | 0.5 | 0.5 |
| $N$ | | $6.8 \cdot 10^9$ | $3.72 \cdot 10^9$ |
| $\sigma_x$ | nm | $\approx 200$ | $\approx 45$ |
| $\sigma_y$ | nm | $\approx 2.3$ | $\approx 1$ |
| $\sigma_z$ | µm | 72 | 44 |
| $\beta_x$ | mm | 8 | 4 |
| $\beta_y$ | mm | 0.1 | 0.07 |
| $L^*$ [a] | m | 3.5 | 3.5 |
| $\mathscr{L}$ | $\mathrm{cm^{-2}s^{-1}}$ | $2.3 \cdot 10^{34}$ | $5.9 \cdot 10^{34}$ |
| $\mathscr{L}_{0.01}$ | $\mathrm{cm^{-2}s^{-1}}$ | $1.4 \cdot 10^{34}$ | $2.0 \cdot 10^{34}$ |
| $n_\gamma$ | | 1.3 | 2.1 |
| $\Delta E/E$ | | 0.07 | 0.28 |
| $N_{\mathrm{coh}}$ | | $2 \cdot 10^2$ | $6.8 \cdot 10^8$ |
| $E_{\mathrm{coh}}$ | TeV | $1.5 \cdot 10^1$ | $2.1 \cdot 10^8$ |
| $N_{\mathrm{incoh}}$ | | $8 \cdot 10^4$ | $3 \cdot 10^5$ |
| $E_{\mathrm{incoh}}$ | TeV | $3.6 \cdot 10^2$ | $2.3 \cdot 10^4$ |
| $n_{\mathrm{Had}}$ ($W_{\gamma\gamma} > 2$ GeV) | | 0.3 | 3.2 |

[a] This value holds for CLIC_SiD, and has been used throughout the accelerator studies for this CDR. For CLIC_ILD, the corresponding value is 4.3 m.

However, this number should not be over-emphasised. Unless a Z′ is discovered, in which the production cross section is likely to be large, physics at CLIC is unlikely to involve operation at the peak of a resonance. Hence the useful luminosity will depend on the threshold of the process being studied; for example, for CLIC operation at 3 TeV the useful luminosity for a process with a threshold of 2 TeV is greater than 75%. In addition to reducing the effective useful luminosity, the effect of beamstrahlung has the potential to distort the reconstructed particle energy spectrum in a number of physics analyses (for example SUSY decays involving the LSP). This aspect is discussed below and later in Chapter 12.

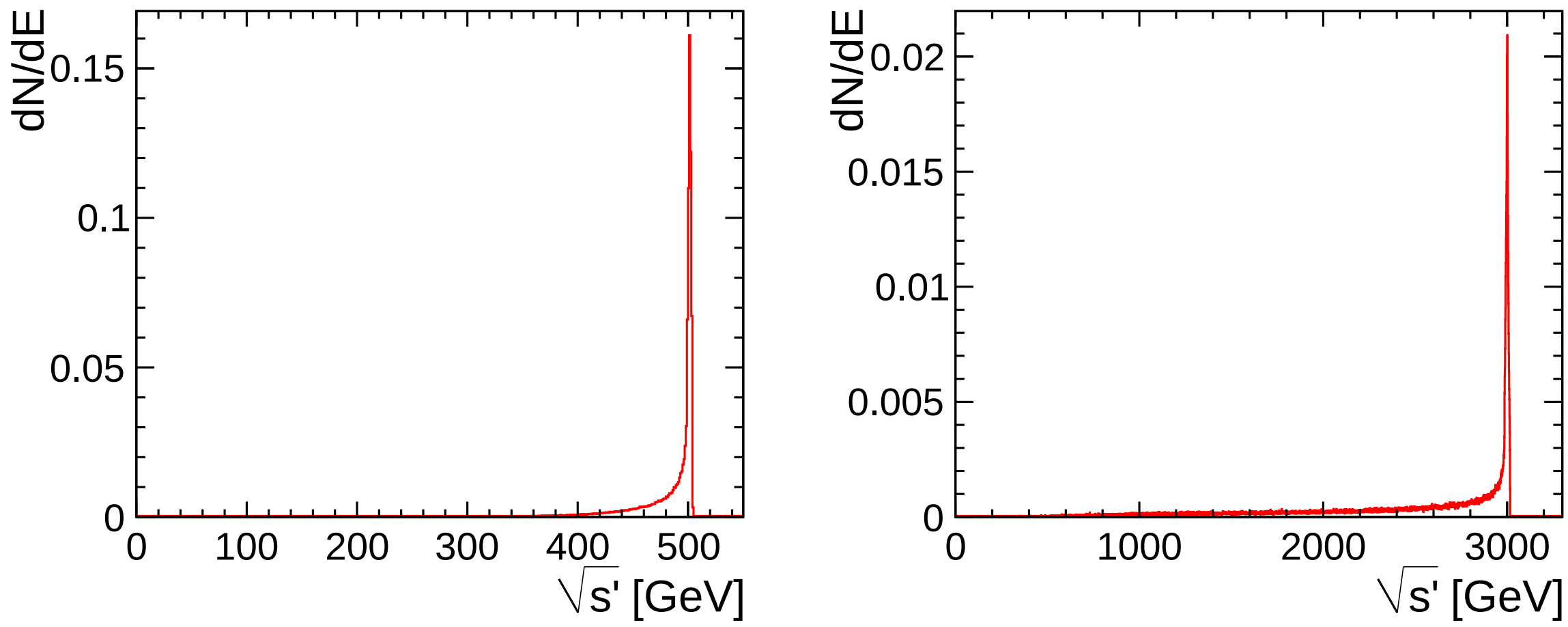

Fig. 2.1: The luminosity spectrum for CLIC operating at (left) $\sqrt{s} = 500$ GeV and (right) $\sqrt{s} = 3$ TeV.

### 2.1.2 Beam-Induced Backgrounds

There are three main sources of beam related backgrounds at CLIC:

- $e^+e^-$ pairs which are predominantly produced with low transverse momenta $p_T$;
- $\gamma\gamma \rightarrow$ hadrons which result in pile-up of low energy particles with $p_T \lesssim 5$ GeV;
- beam halo muons.

Each background has been studied in detail and the impact on the detector design has been carefully evaluated using full GEANT4 [2] simulations of the CLIC detector concepts which are described in Section 2.2 and Chapter 3. The beam-beam backgrounds are estimated from simulation. First, the particles in the CLIC beams are tracked from the beginning of the main linac to the interaction point [3]. Then the resulting distributions are used, without modifications or approximations, as input for the beam-beam simulation code GUINEAPIG [4] which uses the known cross sections for the relevant physical processes [5]. Uncertainties on the simulation of the production rates and of the detector response have been estimated. As a result, safety factors of two for the background rates from $\gamma\gamma \rightarrow$ hadrons and five for the ones from $e^+e^-$ pairs have been estimated. Details are described elsewhere [6, 7]. Throughout this document, results obtained with nominal parameters are presented in most tables and figures, while safety factors are mentioned explicitly in the text.

Table 2.2: Fraction of luminosity above $\sqrt{s'}/\sqrt{s}$.

| Fraction $\sqrt{s'}/\sqrt{s}$ | 500 GeV | 3 TeV |
|---|---|---|
| $> 0.99$ | 62% | 35% |
| $> 0.90$ | 89% | 54% |
| $> 0.80$ | 97% | 68% |
| $> 0.70$ | 99.3% | 76% |
| $> 0.50$ | 99.9% | 88% |

#### *2.1.2.1 Pair Background*

The large flux of beamstrahlung photons will produce $e^+e^-$ pairs in the strong electromagnetic fields of the electron and positron bunches, both by coherent and incoherent pair creation processes [8]. The coherent process consists of the interaction of the real beamstrahlung photons with the collective electromagnetic field of the opposite beam. The coherent production of $e^+e^-$ pairs will increase the total number of colliding electrons and positrons by about 9%. The production of coherent pairs from the virtual photons associated with the beam particles (trident pairs) is roughly an order of magnitude lower than the production of coherent pairs [9]. The incoherent production of pairs arises from the interaction of both real or virtual photons with *individual* particles of the other beam. There are three main physical processes responsible for the production of incoherent pairs: the Breit-Wheeler (BW) process which is the interaction between two real photons from beamstrahlung; the Bethe-Heitler (BH) process of the interaction of a real photon and a virtual photon associated with a beam particle; and the Landau-Lifshitz (LL) process of the interaction between two virtual photons. The GuineaPig calculation for the BH and LL processes uses a Weizsäcker-Williams approach, known as the Equivalent Photon Approximation (EPA). In the EPA, the equivalent spectrum of virtual photons is convolved with the real photon interaction cross sections. The production of incoherent pairs in GuineaPig has been compared to other codes in [5] and [10].

Most pairs are produced with very small angles along the beam axis. In order to avoid significant loss of such particles in the detector, a beam exit line with a half-cone opening angle of 10 mrad is needed, see Figure 3.3. However, depending on the motion of the produced electron and positron with respect to the electron and positron beams they may either be focused or defocused. The effect of this electromagnetic beam deflection gives rise to a component of the pair spectrum with sufficient transverse momentum for it to travel beyond the beam pipe, and thus represent a potential background in the detector volume. The effect of beam deflection on the coherent pairs is relatively small as they are typically very high energy particles which are highly boosted along the beam direction. Consequently, whilst the coherent pair rate is extremely high, $7 \cdot 10^8$ particles per bunch crossing at 3 TeV, almost all of the coherent pairs are collinear with the outgoing beams and thus do not constitute a major detector background.

While the number of incoherent pairs is much smaller than that of the coherent pairs (see Table 2.1), they can be produced at larger angles and potentially provide a significant source of background hits, for example, in the inner layers of the vertex detector. The energy and angular distributions of the pair backgrounds are shown in Figure 2.2. Because of their larger transverse momentum, the incoherent pairs cause more energy deposits in the detector and are a more relevant background source than the coherent pairs, despite the much larger number of coherent pairs.

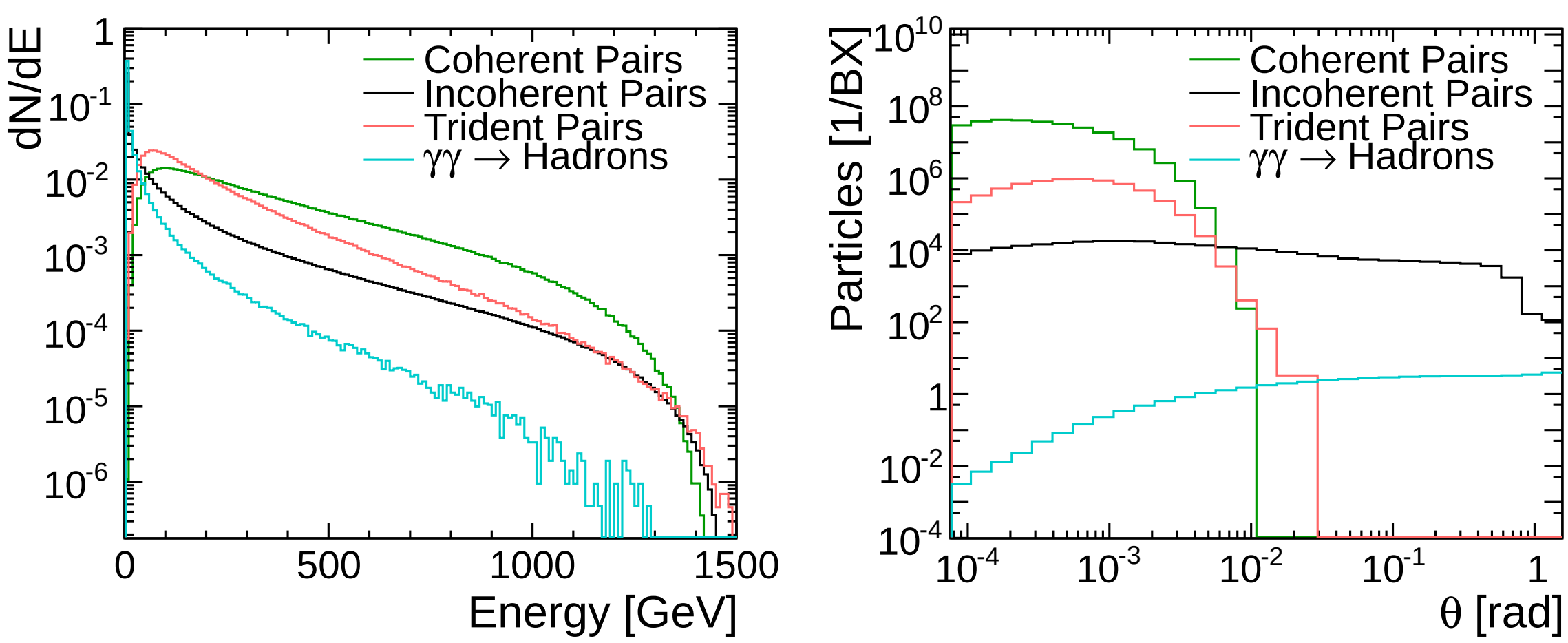

Fig. 2.2: The distributions of the beam related backgrounds: (left) Fraction of energies for the particles of each background source. (right) Angular distribution of the produced background particles. Both plots are for CLIC at $\sqrt{s} = 3$ TeV.

#### *2.1.2.2 Two-photon Background*

The interaction of real and virtual photons from the colliding beams can also lead to multi-peripheral two-photon interactions producing hadronic final states [11, 12]. The energy and angular distributions of $\gamma\gamma \to$ hadrons backgrounds are shown in Figure 2.2. These interactions can produce particles at a large angle to the beam line and constitute the main background for the central tracking volumes and the calorimeters. The simulation of $\gamma\gamma \to$ hadrons uses the photon spectrum from GUINEAPIG and a parametrisation of the total cross section based on [13]:

$$\sigma_{\gamma\gamma}(s_{\gamma\gamma}) = 211\ \text{nb} \left(\frac{s_{\gamma\gamma}}{\text{GeV}^2}\right)^{0.0808} + 215\ \text{nb} \left(\frac{s_{\gamma\gamma}}{\text{GeV}^2}\right)^{-0.4525} \tag{2.1}$$

The predicted number of $\gamma\gamma \to$ hadrons events per bunch crossing within the detector acceptance at $\sqrt{s} = 3$ TeV is 3.2 for a $\gamma\gamma$ centre-of-mass energy greater than 2 GeV. For a $\gamma\gamma$ centre-of-mass energy greater than 5 GeV, 2.8 $\gamma\gamma \to$ hadrons events per bunch crossing are expected [6].

For the purpose of evaluating the impact of the $\gamma\gamma \to$ hadrons background in the detector, the spectrum of colliding photons from GUINEAPIG are used to generate events using the PYTHIA program [14] which simulates the hard interaction and the subsequent hadronisation. The resulting $p_T$ distribution of the produced particles which are within the detector acceptance is shown in Figure 2.3.

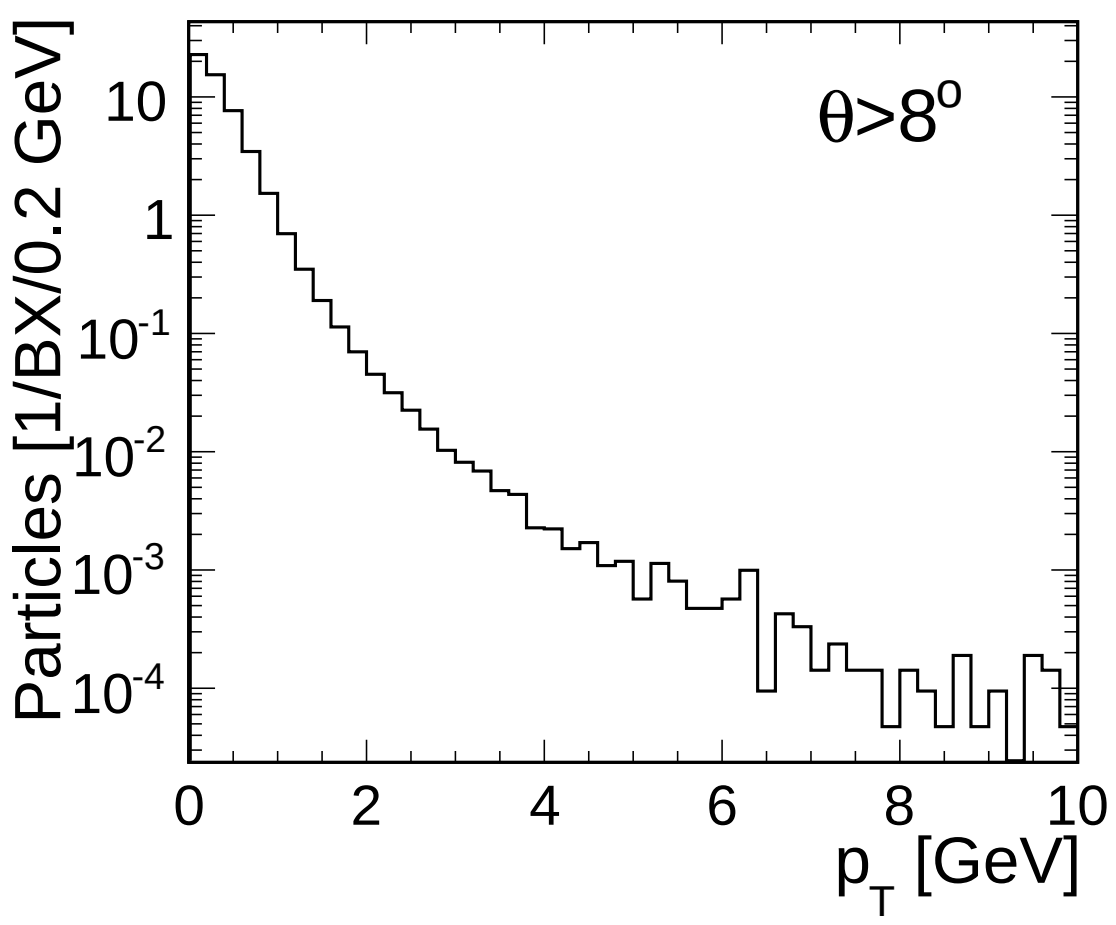

Fig. 2.3: The generated $p_T$ distribution for particles from $\gamma\gamma \to$ hadrons which are within a notional detector angular acceptance of $8^\circ < \theta < 172^\circ$. The rate is normalised to one bunch crossing, excluding safety factors for the simulation uncertainties.

#### *2.1.2.3 Beam Halo Muons*

Machine-induced secondary electrons or positrons, produced, for example, from inelastic collisions with the residual molecules in the beam pipe are a potential source of detector background. The BDS is designed to mitigate the potential effects of this background, with six collimation stations placed at strategic locations to ensure that essentially no beam particles are lost in the last several hundred meters upstream of the IP.

High energy secondary muons are produced in inelastic collisions in the collimation of the beam halo electrons or positrons, and may reach the experimental cavern and the detector. This beam halo muon background can be reduced significantly through adaptations of the collimation scheme and through the placement of passive and/or magnetised iron spoilers in the BDS. Detailed tracking studies of the muons through the BDS have been performed. Preliminary results indicate that it is realistic to aim for an average of 1 muon per 20 bunch crossings (combined from both beams) traversing the detector volume. Further information is given in [1].

The simulated beam halo muon distribution in both energy and position has been used as the input to a number of dedicated detector simulation studies to assess the impact of this background. For example, the occupancy in the TPC due to halo muons was investigated, see Section 5.3, and is found to be negligible compared to the background from $\gamma\gamma \rightarrow$ hadrons.

The beam halo muon background in the calorimeters can locally result in a significant energy deposition due to high energy electromagnetic showers produced by Bremsstrahlung. For an assumed single beam halo muon per bunch crossing – a factor 20 above the expected halo muon rate – the average energy deposition for the entire bunch train would be approximately 3 TeV. Using the "tight" PFO timing cuts (see Chapter 12) this is reduced to approximately 100 GeV in time with a physics event. However, due to the highly granular nature of the proposed CLIC detector concepts, much of the remaining background can be rejected. A preliminary version of a pattern recognition algorithm, designed to remove hits and clusters consistent with coming from beam halo muons, was implemented in the PANDORAPFA reconstruction code. It was demonstrated that the residual calorimeter background can be reduced to the level of approximately 10 GeV. The impact of this background on physics observables was studied by overlaying a full bunch train of halo muons on $W^+W^- \rightarrow q q \mu \nu$ events at 1 TeV (a sample of W–like particles of 500 GeV mass decaying hadronically). For the background assumed for this study, one muon per BX (which implies a safety margin of a factor 20), the impact on the reconstructed W mass distribution was found to be significantly less than that of the $\gamma\gamma \rightarrow$ hadrons background.

In summary, at the expected level of 1 muon per 20 BX, the beam halo muon background does not constitute an important problem for the detector concepts being considered here, i.e., in detectors which have sufficient granularity and timing resolution in the calorimeters.

#### *2.1.2.4 Backscattering from the Post-Collision Line and the Beam Dump*

After collision, the particles are transported through a system of magnets, known as the post-collision line,to the main CLIC beam dump, 315 m downstream of the IP. The beam-beam effect leads to a broad energy spectrum of electrons, positrons and photons, some of which are lost in collimators installed to protect the magnets. Detailed Monte-Carlo simulations have been performed in order to assess the particle flux scattering back from the post-collision line towards the detector at the IP. The model includes the magnets, collimators, vacuum pipe, beam dump and tunnel walls, but not the CLIC detector itself. The average flux of back-scattered particles from each bunch train, crossing a detector plane 3.5 m from the IP, is approximately 20 photons (all with energies below 500 keV) and 4 neutrons (all with energies below 1 MeV) per cm$^2$ [15]. Whilst not explicitly simulated, these low energy particles will be absorbed in the the CLIC detector yoke, and it is concluded that backscattering from the post-collision line and beam dump into the CLIC detector volume is negligible.

### 2.1.3 Beam Polarisation at CLIC

The CLIC accelerator conceptual design includes a source to produce a polarised electron beam, and all elements necessary to transport the beam to the IP without loss of polarisation. An electron beam polarisation of 80% is expected for the CLIC experimental programme. This corresponds to what is already achievable today for lower energy electron accelerators.

Currently, a polarised positron beam is not part of the CLIC baseline, although provisions have been made in the design of the accelerator complex to add this option at a later stage. Conservatively, one may assume 30% polarisation of the positrons after such an upgrade phase.

The degree of polarisation in the beams can be measured in a dedicated section of the Beam Delivery System (BDS) at CLIC, several hundred metres upstream of the IP, using the well-established Compton back-scattering technique. Detailed studies performed for the 500 GeV beams at the ILC have been extrapolated to CLIC, and a statistical uncertainty of better than 0.1% can be expected, even for low intensity beams during initial running of the accelerator complex (cf. BDS section in [1]).

Knowing with high absolute accuracy the degree of polarisation at the time of the interaction is important for a number of physics processes to be studied at CLIC. At 3 TeV centre-of-mass collision energy, this absolute accuracy is hampered by effects of the very strong beam-beam interaction. This produces a large number of beamstrahlung photons and coherent pairs. This creates a spray of background particles in the post-collision line, and makes it impossible to measure the beam polarisation after the IP. Moreover, the beam-beam interaction leads to depolarisation. Simulations indicate that the depolarisation varies throughout the luminosity spectrum [16, 17], starting below 1% around the high energy peak at 3 TeV (i.e. for events with a lower degree of beam-beam losses) and reaching up to 4% at the lower energies (i.e. where the beam-beam effects are strongest). The systematic uncertainty on the absolute degree of beam polarisation is therefore left to future detailed studies.

## 2.2 Detector Requirements for $e^+e^-$ Physics in the TeV-Range

The detector requirements at a 500 GeV $e^+e^-$ collider have been established in the context of the ILC [18, 19]. Assuming a staged approach for CLIC, with the possibility of the initial operation at ILC-like energies, the minimal requirements for a detector at CLIC are that it must meet the ILC detector requirements. However, the detector must also be suitable for physics at centre-of-mass energies up to 3 TeV. In addition, the detector must be able to operate effectively in the CLIC machine environment. Here the most challenging aspect is the 0.5 ns bunch structure combined with the background from $\gamma\gamma \to$ hadrons which results in a deposition of approximately 20 TeV of energy in the calorimeters for the entire train of 312 bunches. This implies not only excellent time resolution for all detector components, but also a highly segmented calorimeter to keep individual cell occupancies to a manageable level.

Chapter 1 presents a wide range of BSM physics scenarios which define the possible goals of a high energy $e^+e^-$ collider such as CLIC. These broad physics goals can be used to define the minimal requirements for the performance of a detector at CLIC at 3 TeV. The possible BSM physics signatures at 3 TeV can be broadly characterised as:

– high multiplicity jet final states, for example, $e^+e^- \to H^+H^- \to t\bar{b}b\bar{t}$;
– multi-jet final states and missing energy, such as $e^+e^- \to \tilde{\chi}_1^+\tilde{\chi}_1^- \to W^+W^-\tilde{\chi}_1^0\tilde{\chi}_1^0$ or signatures for strong EWSB;
– leptons and missing energy, for example, $e^+e^- \to \tilde{\mu}\tilde{\mu} \to \mu^+\mu^-\tilde{\chi}_1^0\tilde{\chi}_1^0$;
– heavy flavour production, for example in $e^+e^- \to h\nu_e\bar{\nu}_e$ where $h \to b\bar{b}$ or $h \to c\bar{c}$;
– exotic final states, for example, non-pointing photons in certain SUSY GMSB models.

The main detector requirements for the reconstruction of physics events at CLIC are discussed below.

### 2.2.1 Track Momentum Resolution

The track momentum goal at the ILC and CLIC is dictated by the Higgs mass determination from the Higgsstrahlung process, $e^+e^- \to Zh$, where the mass can be precisely reconstructed from the mass distribution of the system recoiling against the pair of muons from $Z \to \mu^+\mu^-$ decays. The precision of the measurement is ultimately limited by the beam energy spread. For the ILC operating at 250 GeV and $m_h = 120$ GeV, the momentum resolution needs to be $\sigma_{p_T}/p_T^2 \lesssim 5\cdot 10^{-5}$ GeV$^{-1}$. For higher centre-of-mass energies, where the muons have higher momenta, the requirements are even more stringent. For example, the expected reconstructed recoil mass distribution for $m_h = 120$ GeV with the CLIC beamstrahlung spectrum at $\sqrt{s} = 500$ GeV is shown in Figure 2.4 (left) for different assumed momentum resolutions. Despite the relatively large tail from beamstrahlung in the recoil mass distribution, a clear peak at the Higgs mass can be observed. For the width to be dominated by the beam energy spread requires $\sigma_{p_T}/p_T^2 \sim 2\cdot 10^{-5}$ GeV$^{-1}$.

The track momentum requirement at $\sqrt{s} = 3$ TeV is also driven by the measurement of the Higgs branching ratio to muons. An excellent mass resolution is crucial to distinguish this rare decay from its background channels. Figure 2.4 (right) shows the statistical uncertainty of the cross section times branching ratio measurement of the $\mathrm{h} \to \mu^+\mu^-$ channel depending on the momentum resolution. The numbers are obtained from a fast simulation study similar to the analysis presented in Section 12.4.2, assuming different constant momentum resolutions, independent of the particle momentum or angle. The results corresponding to the nominal detector resolution are consistent with results obtained with full simulation. An average momentum resolution of a few $10^{-5}$ GeV$^{-1}$ is desirable. For even better momentum resolutions the result is limited by the intrinsic statistical uncertainty due to the small number of events.

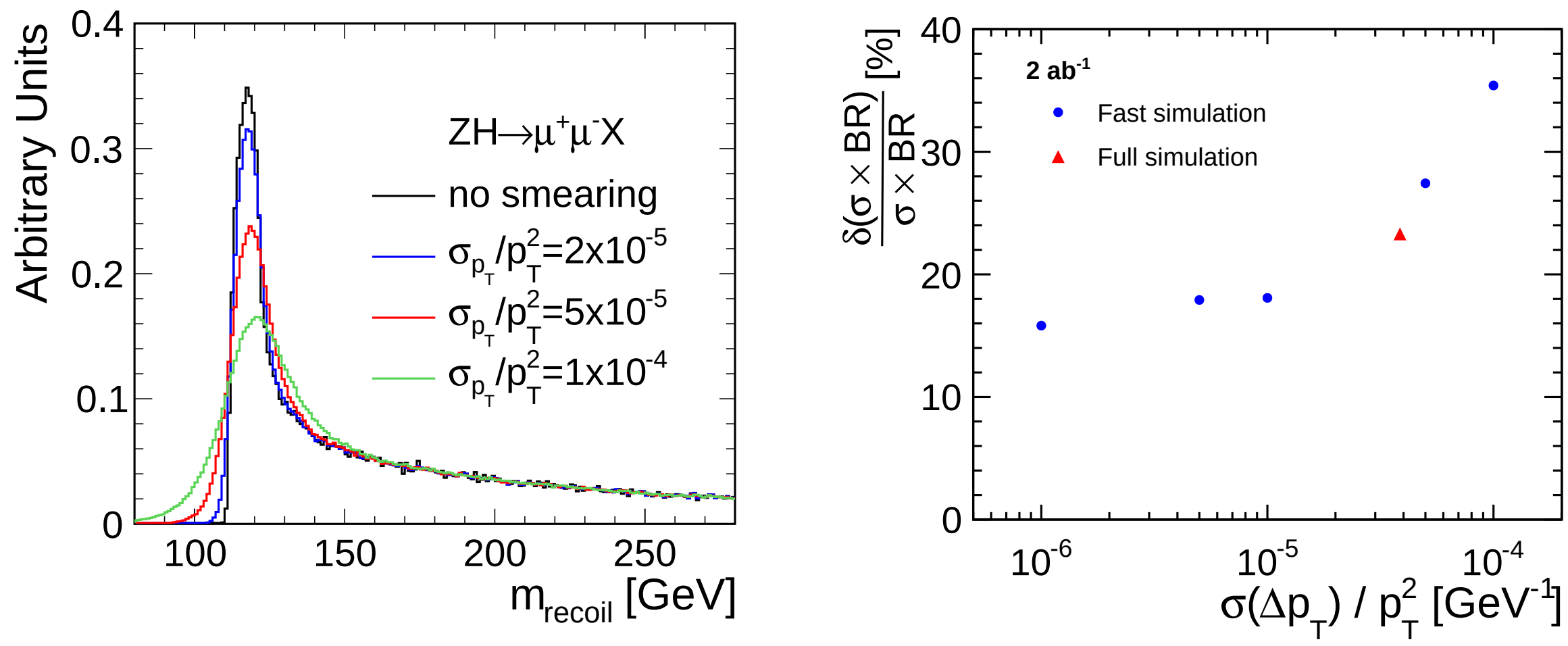

Fig. 2.4: Generator level reconstructed recoil mass distribution in the Higgsstrahlung process $\mathrm{e}^+\mathrm{e}^- \to \mathrm{Zh} \to \mu^+\mu^- X$ from the muon momentum smeared by an assumed track momentum resolution (left). Statistical uncertainty of the $\sigma \times \mathrm{BR}$ measurement of the $\mathrm{h} \to \mu^+\mu^-$ channel depending on the momentum resolution (right). Results obtained from fast simulation are consistent with full simulation results. See Section 12.4.2 for details.

Similar requirements on the momentum resolution follow from the consideration of BSM physics scenarios. One possible example is the determination of the smuon and neutralino masses from the muon momentum distribution in the process $\mathrm{e}^+\mathrm{e}^- \to \tilde{\mu}\tilde{\mu} \to \mu^+\mu^-\tilde{\chi}^0_1\tilde{\chi}^0_1$. Figure 2.5 (left) shows the generator level muon momentum distribution from smuon decays (for the SUSY *model II* described in the Section 2.6) with different values for the assumed momentum resolution. The high momentum part of the spectrum is significantly distorted for a momentum resolution of $\sigma_{p_\mathrm{T}}/p_\mathrm{T}^2 > 4 \cdot 10^{-5}$ GeV$^{-1}$. Figure 2.5 (right) shows the corresponding reconstructed mass resolution for the neutralino and the smuon as a function of momentum resolution.

### 2.2.2 Jet Energy Resolution

Many of the interesting physics processes at CLIC are likely to be characterised by multi-jet final states, often accompanied by charged leptons or missing transverse momentum associated with neutrinos or possibly the lightest super-symmetric particles. The reconstruction of the invariant masses of two or more jets will be important for event reconstruction and event identification. At LEP, kinematic fitting enabled precise invariant mass reconstruction and reduced the dependence on the intrinsic calorimetric performance of the LEP detectors. At CLIC, due to beamstrahlung, kinematic fitting will be, in general, less powerful and the di-jet mass reconstruction will rely more heavily on the intrinsic jet energy resolution of the detector. One goal for jet energy resolution at CLIC is that it is sufficient to provide discrimination between the hadronic decays of W and Z boson. Figure 2.6 (left) shows idealised reconstructed W and Z mass distributions for different assumed mass resolutions. Good separation is obtained for a mass resolution of 2.5%, which corresponds to a jet energy resolution of 3.5%. To obtain a separation corresponding to $2.5\sigma$ implies a jet energy resolution of 3.5% [20] for the entire range of jet energies of interest at CLIC, i.e., 50 GeV – 1 TeV. A jet energy resolution of 5% leads to a $2\sigma$ W/Z separation. The reconstruction of mass-related variables in BSM decays such as $\tilde{\mathrm{q}}_\mathrm{R} \rightarrow \mathrm{q}\tilde{\chi}^0$ will also benefit from good jet energy resolution. Figure 2.6 (right) shows the distribution of the contravariant mass, $M_\mathrm{C}^2 = 2(E_1E_2 + \vec{p}_1 \cdot \vec{p}_2)$, in $\mathrm{e^+e^-} \rightarrow \tilde{\mathrm{q}}_\mathrm{R}\bar{\tilde{\mathrm{q}}}_\mathrm{R} \rightarrow \mathrm{q}\tilde{\chi}^0\bar{\mathrm{q}}\tilde{\chi}^0$, where $E_{1,2}$ and $\vec{p}_{1,2}$ are the reconstructed energies and momenta of the two jets [21]. The location of the high mass edge can be used to determine the mass difference between the $\tilde{\mathrm{q}}_\mathrm{R}$ and the $\tilde{\chi}^0$. The sharpness is determined by both the underlying beamstrahlung spectrum and the jet energy resolution. For $\sigma_E/E$ of $< 5\%$, the measurements are dominated by the effects of beamstrahlung rather than the jet energy resolution.

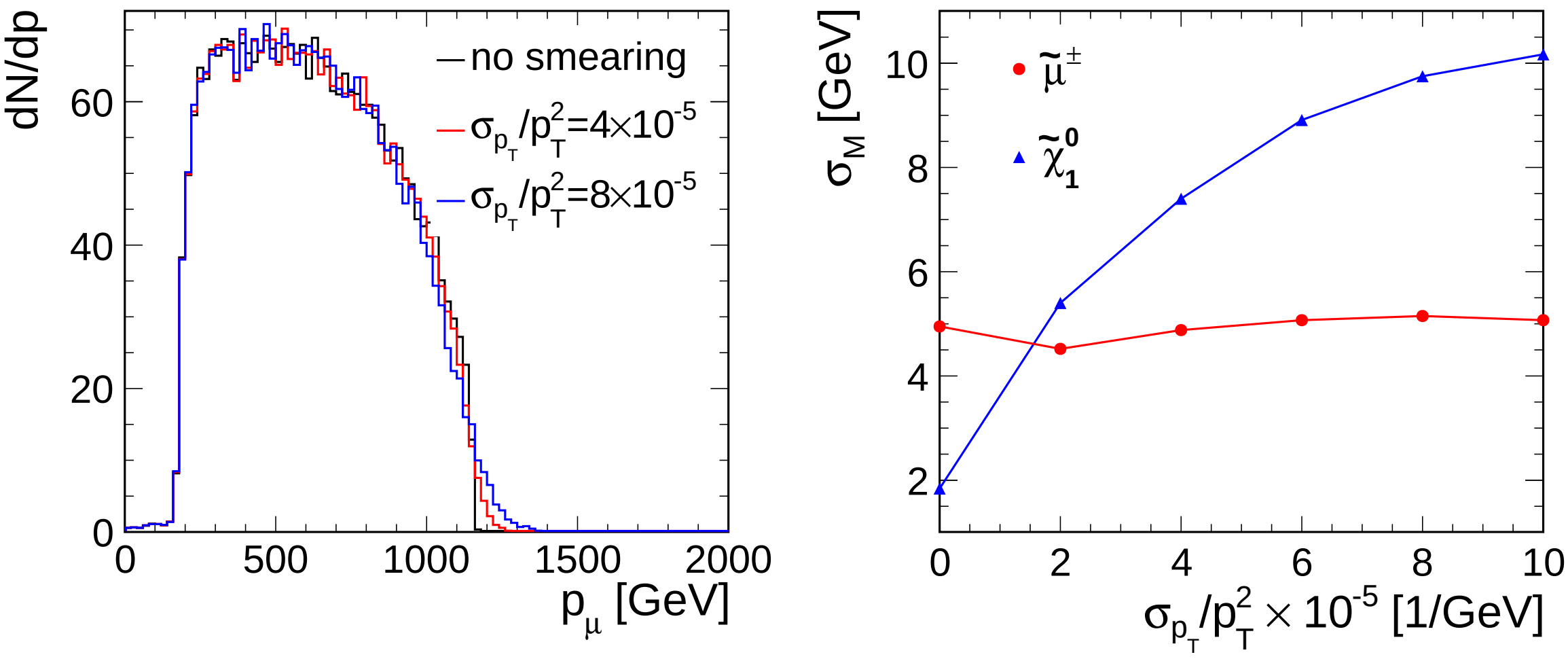

Fig. 2.5: (left) Generator level reconstructed muon momentum distribution for $\mathrm{e^+e^-} \rightarrow \tilde{\mu}\tilde{\mu} \rightarrow \mu^+\mu^-\tilde{\chi}^0_1\tilde{\chi}^0_1$ assuming 2 $\mathrm{ab}^{-1}$ of data at $\sqrt{s} = 3$ TeV in a SUSY model with $m(\tilde{\mu}_\mathrm{L}) = 1100.4$ GeV and $m_{\tilde{\chi}^0_1} = 328.3$ GeV; (right) resolution on the reconstructed smuon and neutralino masses $\sigma_\mathrm{M}$ a function of the assumed momentum resolution $\sigma_{p_\mathrm{T}}/p_\mathrm{T}^2$.

### 2.2.3 Impact Parameter Resolution and Flavour Tagging

Whatever the physics at CLIC, the ability to efficiently tag b-quarks will feature in many physics studies. For CLIC operating in the energy range between $\sqrt{s} = 500$ GeV and $\sqrt{s} = 3$ TeV, it is likely that one of the main physics goals will be the measurement of the couplings of the Higgs. Here the ability to tag both charm and bottom quarks will be important. High performance flavour tagging depends on the ability to identify secondary vertices and tracks which do not originate from the IP. The impact parameter resolution can be expressed in the form

$$\sigma_{d_0}^2 = a^2 + \frac{b^2}{p^2 \sin^3\theta}\,, \tag{2.2}$$

where the constant $a$ depends on the point resolution of the vertex detector and parameter $b$ is related to multiple scattering and thus depends on the amount of material in the inner detector and the geometrical arrangement of the layers. The target values for a detector at CLIC are derived from those for the ILC [19], namely $a \lesssim 5$ µm and $b \lesssim 15$ µm GeV. This represents a factor 2–3 improvement with respect to the SLD vertex detector, both in terms of point resolution and material budget. For CLIC operating at $\sqrt{s} = 3$ TeV, efficient flavour tagging will be essential for final states containing multiple b-jets.

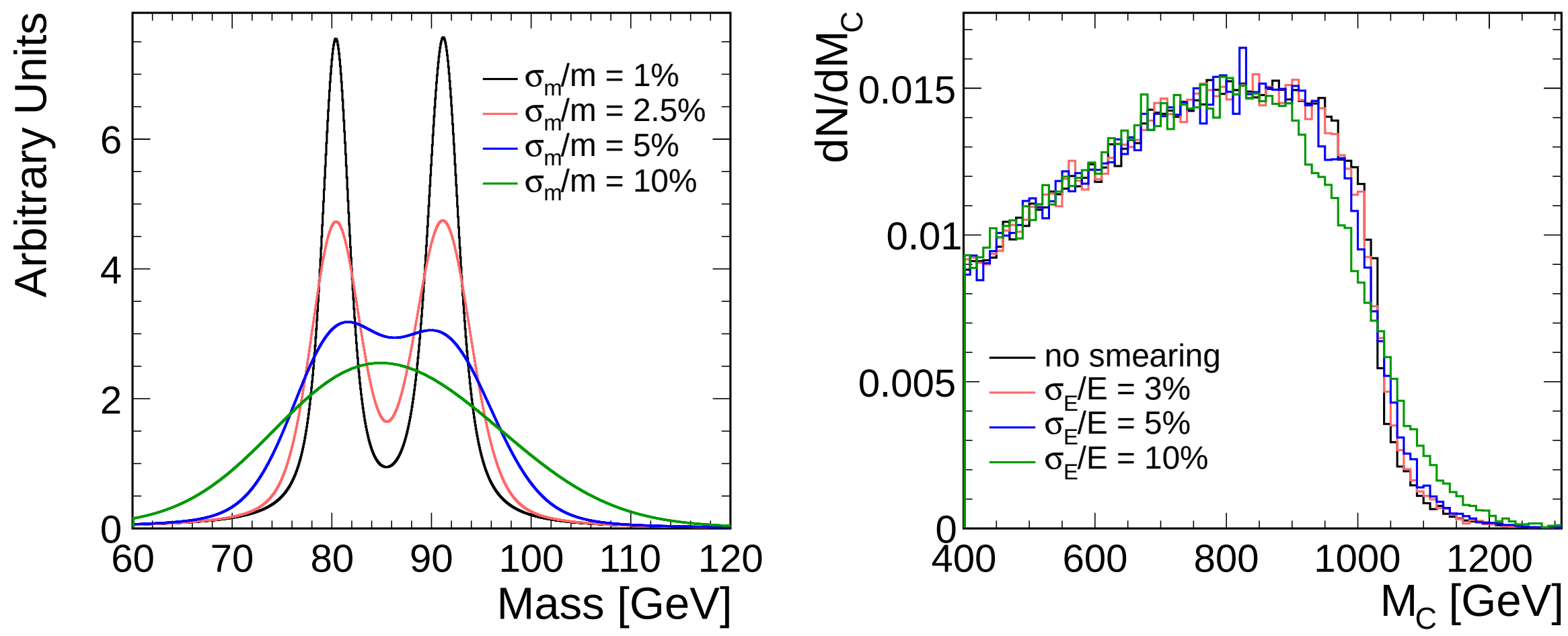

Fig. 2.6: (left) Ideal W/Z separation vs. jet mass resolution obtained using a Gaussian smearing of Breit-Wigner distributions; (right) Reconstructed contravariant mass, $M_{\mathrm{C}}$, for $\mathrm{e^+e^-} \rightarrow \tilde{\mathrm{q}}_{\mathrm{R}}\bar{\tilde{\mathrm{q}}}_{\mathrm{R}} \rightarrow \mathrm{q}\tilde{\chi}^0\bar{\mathrm{q}}\tilde{\chi}^0$ (including the effects of Beamstrahlung) for different jet energy resolutions. The plot was obtained by applying a Gaussian energy smearing to reconstructed jets based on the generator level particles.

### 2.2.4 Forward Coverage

At CLIC many SM processes will result in particles produced at relatively low angles to the beam axis; either due to the boost along the beam axis from beamstrahlung or from $t$-channel processes. To study these processes, on the one hand, or to reduce their impact on BSM physics studies, on the other hand, extending the detector coverage to small polar angles is important [22].

For example, at 3 TeV, approximately 80% of the leptons in the $\mathrm{l^+l^-l^+l^-}$ final state, dominated by gauge boson pairs, are produced at polar angles of $< 30^\circ$ to the beam axis. The forward region is also important for physics signatures with missing energy. It helps to reject background processes like multi-peripheral two photon processes, $\mathrm{e^+e^-} \rightarrow \mathrm{e^+e^-}f\bar{f}$, where the scattered electrons are usually at low polar angles. For example, forward electron tagging is essential to reject the $\mathrm{e^+e^-} \rightarrow \mathrm{e^+e^-}\mu^+\mu^-$ background in the measurement of the Higgs branching ratio into two muons. As shown in Section 12.4.2, it improves the achievable statistical accuracy of this measurement from 23% to 15%, assuming an integrated luminosity of 2 $\mathrm{ab^{-1}}$ and 95% electron tagging efficiency down to $\approx$ 40 mrad polar angle. Another example is the production and decay of stau pairs, $\mathrm{e^+e^-} \rightarrow \tilde{\tau}\tilde{\tau} \rightarrow \tau^+\tau^-\tilde{\chi}^0_1\tilde{\chi}^0_1$, which, in some regions of SUSY parameter space, results in a signal with relatively small missing transverse momentum. In this case, the $\mathrm{e^+e^-} \rightarrow \mathrm{e^+e^-}\tau^+\tau^-$ and $\mathrm{e^+e^-} \rightarrow \mathrm{e^+e^-q\bar{q}}$ background processes need to be rejected by efficient electron tagging at low polar angles. It is therefore important, in general, to provide precision tracking and calorimetry coverage down to small angles, and to extend the forward electron tagging capabilities to very low angles.

### 2.2.5 Lepton ID Requirements

Many of the potential BSM physics signals at CLIC will rely on the ability to efficiently identify high energy electrons and muons, and efficient lepton identification is central to the CLIC detector requirements. For efficient selection of final states with two or more leptons, lepton identification efficiencies of more than 95% over a wide range of momenta are highly desirable. In addition the identification of leptons in jets from semi-leptonic decays of b- and c-quarks will benefit heavy flavour tagging.

### 2.2.6 Summary of Requirements for Physics Reconstruction

From the perspective of the likely physics measurements at CLIC the detector requirements are:

- jet energy resolution of $\sigma_E/E \lesssim 5-3.5\%$ for jet energies in the range 50 GeV – 1 TeV;
- track momentum resolution of $\sigma_{p_\mathrm{T}}/p_\mathrm{T}^2 \lesssim 2\cdot 10^{-5}$ GeV$^{-1}$;
- impact parameter resolution (equation 2.2) with $a \lesssim 5$ µm and $b \lesssim 15$ µm GeV;
- lepton identification efficiency: $> 95\%$ over the full range of energies;
- detector coverage for electrons down to very low angles.

## 2.3 Basic Choice of Detector Concepts for CLIC

The design considerations driving the basic choice of the detector concept(s) for CLIC are clear; excellent track momentum and jet energy resolution, excellent flavour tagging capability and the ability to perform precision physics measurements in the CLIC background environment. This in turn implies excellent time stamping capability for all detector elements. It is the jet energy resolution in the relatively high background environment that has the largest impact on the overall design of a detector concept for CLIC.

Traditionally, jet energies have been obtained from the sum of the energies deposited in the electromagnetic and hadronic calorimeters (ECAL and HCAL) giving a jet energy resolution of the form

$$\frac{\sigma_E}{E} = \frac{\alpha}{\sqrt{E(\mathrm{GeV})}} \oplus \beta .$$

The stochastic term $\alpha$ is usually greater than 60% and the constant term $\beta$, which encompasses a number of effects, is typically a few per cent. For high energy jets there also will be a contribution from the non-containment of the hadronic showers. To achieve the CLIC goal of a jet energy resolution of ∼3.5% or better would require a stochastic term below 30% and a small constant term. This is unlikely to be achievable with a traditional approach to calorimetry. Calorimetry at a future lepton collider has been studied extensively in the context of the ILC; it is widely acknowledged that high granularity particle flow calorimetry is currently the most promising approach to achieving a jet energy resolution of 3.5% [20, 23, 24]. High granularity particle flow calorimetry is also well suited to the relatively high levels of background; it has the potential to separate calorimetric energy deposits from background particles from those of the hard interaction.

Depending on the staging of the machine any detector must meet the physics requirements over a range of centre-of-mass energies, 0.5 TeV – 3.0 TeV. Over the last decade concepts for general purpose detectors which meet the physics requirements for the ILC operating at $\sqrt{s} = 500$ GeV have been developed. In particular two detector concepts, ILD [19] and SiD [18], both based on high granularity particle flow calorimetry, have been studied in detail. Modified versions of these detector concepts (CLIC_ILD and CLIC_SiD) form the basis of the detector concepts for CLIC as discussed in detail in Chapter 3.

### 2.3.1 The Particle Flow Paradigm

On average, after the decay of short-lived particles, roughly 60% of the jet energy is carried by charged particles (mainly hadrons), around 30% by photons, and about 10% by long-lived neutral hadrons (e.g. n, $\bar{\mathrm{n}}$ and $\mathrm{K_L}$). In contrast to a purely calorimetric measurement, particle flow calorimetry requires the reconstruction of the four-vectors of all visible particles in an event. The reconstructed jet energy is the sum of the energies of the individual particles. The momenta of charged particles are measured in the tracking detectors, while the energy measurements for photons and neutral hadrons are obtained from the calorimeters. In this manner, the HCAL is used to measure only about 10% of the energy in the jet. If one were to assume calorimeter resolutions of $\sigma_E/E = 15\%/\sqrt{E(\mathrm{GeV})}$ for photons and $\sigma_E/E = 55\%/\sqrt{E(\mathrm{GeV})}$ for hadrons, a jet energy resolution of $19\%/\sqrt{E(\mathrm{GeV})}$ would be obtained. In practice, this level of performance is not reachable as it is not possible to perfectly associate all energy deposits with the correct particles. This *confusion* rather than calorimetric performance is the limiting factor in particle flow calorimetry. Thus, the crucial aspect of particle flow calorimetry is the ability to assign calorimeter energy deposits to the correct reconstructed particles. This places stringent requirements on the granularity of the ECAL and HCAL. From the point of view of event reconstruction, the sum of calorimeter energies is replaced by a complex pattern recognition problem, namely the Particle Flow reconstruction Algorithm (PFA). Based on detailed simulations of the ILC detector concepts using the PANDORAPFA particle flow reconstruction algorithm it has been demonstrated that jet energy resolutions of approximately 3% can be achieved for jet energies in the range 100 GeV – 1000 GeV [20, 25].

It should be noted that whilst high granularity particle flow calorimetry is a relatively new concept, energy flow and particle flow have been used successfully by a number of collider experiments. OPAL, DELPHI, H1 and DØ obtained improved jet energy resolution using the Energy Flow approach, whereby energy deposits in the calorimeters are removed according to the momentum of associated charged particle tracks. ALEPH and, more recently, CMS used particle flow techniques [26, 27, 28, 29] to attempt to reconstruct the four momenta of the particles in an event.

##### *2.3.1.1 Advantages of High Granularity Calorimetry at CLIC*

The argument for a high granularity detector for the ILC is based almost entirely on the jet energy resolution requirements. This argument still holds at CLIC. However, for a detector at CLIC there is an additional argument for very high granularity calorimetry. At CLIC the hadronic shower development time is longer than the 0.5 ns bunch spacing, the calorimeters necessarily integrate over a number of bunch crossing (discussed in detail in Section 2.5.1). Consequently the calorimeters integrate over a few tens of bunch crossings of background from hadronic two-photon events ($\gamma\gamma \rightarrow$ hadrons). Reduction of this background relies on the ability to temporally and spatially separate energy depositions from the background from those from the physics interaction of interest. The excellent spatial resolution of a high granularity calorimeter designed for particle flow will provide significant additional benefit in the reduction of this background.

### 2.3.2 Detector Design Considerations

The adoption of high granularity particle flow calorimetry significantly impacts the design of the detector at CLIC:

- **ECAL**: the ECAL segmentation has to be sufficient to resolve energy depositions from near by particles in high energy jets. Studies performed in the context of the ILC [19, 20] suggest a calorimeter transverse segmentation of $5 \times 5$ mm$^2$ with approximately 30 longitudinal samplings.
- **HCAL**: the HCAL segmentation has to be sufficient to resolve energy depositions from hadronic showers from different particles. Previous studies [19, 20] suggest an analogue HCAL calorimeter transverse segmentation of at most $3 \times 3$ cm$^2$ with approximately 50 longitudinal samplings. The high degree of longitudinal samplings makes it possible to track particles through the HCAL. The HCAL also needs to be sufficiently thick, about 7.5 $\lambda_I$, to contain the majority of the energy from high energy jets at CLIC.
- **Solenoid**: for the purposes of particle flow reconstruction, the ECAL and HCAL have to be on the inside of the solenoid. A high magnetic field is required to achieve the desired momentum resolution and to separate tracks from nearby particles in high energy jets.
- Overall detector geometry: to increase the separation of particles in the calorimeters a large inner radius for the calorimeters is beneficial although this can, to some extent, be compensated for by a higher magnetic field which tends to increase the mean distance between energy depositions in the calorimeters from neutral and charged particles [20].

In addition to the above considerations, a detector at CLIC must meet the requirements outlined in Section 2.2.

## 2.4 Impact of Backgrounds on the Detector Requirements

The modified versions of the ILC detector concepts certainly meet the main physics performance requirements for CLIC. However the CLIC experimental environment is more challenging than that of the ILC and previous $e^+e^-$ colliders. In particular, the detector must be able to cope with the relatively high levels of background, which in turn dictates the timing and readout requirements for the detector subsystems. Since the background increases with the centre-of-mass energy, the majority of the following discussion focuses on $\sqrt{s} = 3$ TeV where the background is most challenging.

The main backgrounds in the CLIC detector are from incoherent pairs and particles from $\gamma\gamma \rightarrow$ hadrons. Particles from incoherent pairs are the dominant backgrounds in the vertex and in the forward region. The particles from $\gamma\gamma \rightarrow$ hadrons are less forward-peaked and the dominant source of background in the main tracking detectors and in the calorimeters (except at low radii in the endcaps). Whilst the underlying background event rates can be determined from the simulation of the beam alone, the backgrounds in the detector depend on the exact detector design, in particular, on the detailed design of the forward region. The background from incoherent pairs has two components, the particles from the interaction point and backscattered particles from interactions in the very forward region (particularly in the BeamCal) and in the beam pipe itself. The BeamCal is used as an active absorber and provides shielding to nearby elements of the beam delivery system including the final focus quadrupole. Whilst the detector is designed to minimise the backscattered background, it is not possible to eliminate it completely.

To assess the levels of background in the main detector components it is necessary to simulate the entire detector. For this purpose, the BeamCal and beam pipe in the CLIC_ILD and CLIC_SiD detector models were optimised to minimise the backscattered backgrounds in the vertex detector [7]. The background studies described below were obtained for GEANT4 models of the CLIC_ILD detector concept with simulations performed for a full bunch train of incoherent pair and $\gamma\gamma \rightarrow$ hadrons backgrounds. The results for the CLIC_SiD detector are not expected to differ greatly.

### 2.4.1 Impact on the Vertex Detector

In the absence of other constraints the inner layer of the vertex detector would be placed at as small a radius as possible to the beam axis. In practice the minimum radius is limited by the envelop of electrons and positrons from the incoherent pair background. The higher $p_T$ component of the pair background is relatively low in energy and the background particles have helical trajectories in the magnetic field of the detector. The dense core of pair-background particles must not intercept the material of the detector as the resulting interactions would result in a large source of background in the detector volume. This restricts the minimum inner radius of the vertex detector to be approximately 30 mm. The pair background also determines the locations of the forward tracking discs. These constraints and the resulting backgrounds in the vertex detector and forward tracking discs are discussed in detail in Section 4.5.2.

### 2.4.2 Impact on the Central Tracking Detector

The main background in the central tracking detector is due to relatively high $p_T$ tracks from $\gamma\gamma \rightarrow$ hadrons interactions. There are approximately 3.2 $\gamma\gamma \rightarrow$ hadrons events per bunch crossing and each event results in an average of approximately 5 tracks which are reconstructable in the central tracking detector[1]. The mean momentum of these tracks, which are forward peaked, is 1.5 GeV, resulting in a charged particle background with total momentum of 24 GeV per bunch crossing. Integrated over the

[1] Here reconstructable is defined as $\theta > 8^\circ$ from the beam axis, and $p_T > 250$ MeV.

312 bunch crossings in the train there are over 5000 charged particle tracks with a total momentum of 7.3 TeV. The impact of this background on the occupancy in the tracker and the resulting track finding efficiency will depend strongly on the choice of technology used for the central tracking device. This is discussed in detail in Chapter 5.

### 2.4.3 Backgrounds in the ECAL and HCAL

The total energy deposition in the calorimeters from $\gamma\gamma \rightarrow$ hadrons would be expected to be approximately twice that observed in the central tracker. Figure 2.7 shows the energy depositions of the hits in the CLIC_ILD ECAL and HCAL endcaps from an entire bunch train. In the ECAL the contribution from $\gamma\gamma \rightarrow$ hadrons dominates and a clear MIP peak at just below 200 keV can be seen from both background sources. In the HCAL endcaps the background arising from incoherent pairs dominates; this background originates mostly from low energy neutrons which arise from interactions of the large incoherent pair background in the low angle BeamCal.

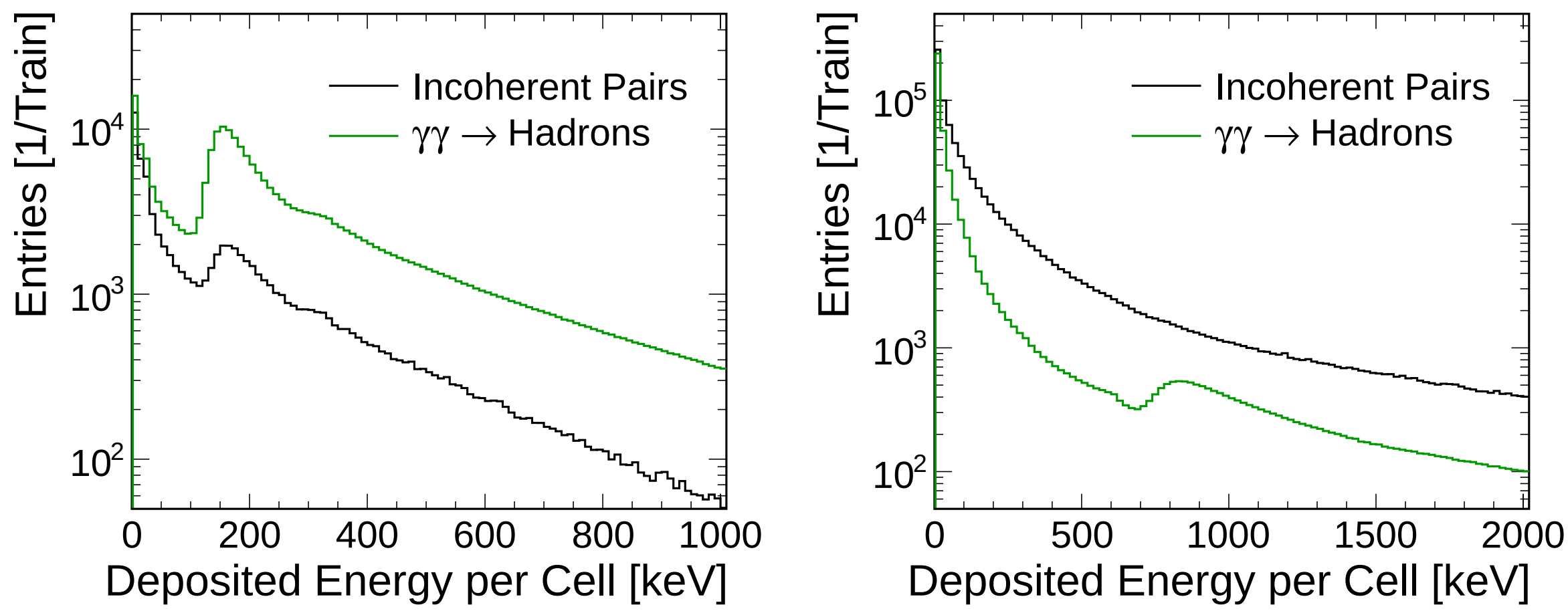

Fig. 2.7: Distributions of the hit energies in the CLIC_ILD ECAL (left) and HCAL (right) endcaps for an entire bunch train of background. The normalisation is applied for nominal background rates, excluding safety factors for the simulation uncertainties.

Figure 2.8 shows the radial distribution of the background in the endcaps. In the ECAL, the background extends out to relatively large radii. In the HCAL the background from $\gamma\gamma \rightarrow$ hadrons also extends out to large radii, but at smaller radii it is swamped by neutrons arising from the incoherent pair background.

Given that the backgrounds for an entire bunch train are high it is clear that the calorimeter read out needs to be able to resolve multiple hits per bunch train. Another important consideration is the level of occupancy per calorimeter cell which, in part, determines the required two-hit time separation. Again this can only be considered in the context of a particular detector design and calorimeter readout cell size. In the simulation of the CLIC_ILD detector, the Silicon sensors in the ECAL are $5 \times 5$ mm$^2$ and the scintillator tiles in the HCAL are $3 \times 3$ cm$^2$. For the occupancy calculation the time window of 300 ns from the start of the bunch train was divided into twelve 25 ns time windows. The mean number of hits above threshold, taken to be 0.3 minimum ionising particle equivalent, is shown for the ECAL and HCAL in Figure 2.9. In the ECAL the occupancies at low radii approach 50% per bunch train and are dominated by the background from $\gamma\gamma \rightarrow$ hadrons. From Poisson statistics this implies that approximately 40% of cells at the inner most radii have at least one background hit per bunch train. In order for this region of the calorimeter to be useful, the ECAL readout must be capable of multiple-hit resolution within the bunch train. In the HCAL the occupancies from $\gamma\gamma \rightarrow$ hadrons are comparable to those in the ECAL reaching a maximum of about one per train at the inner radii of the calorimeter. However, the occupancy in the inner region of the HCAL from neutrons produced by the incoherent pairs in the BeamCal approaches one hit per assumed 25 ns time window.

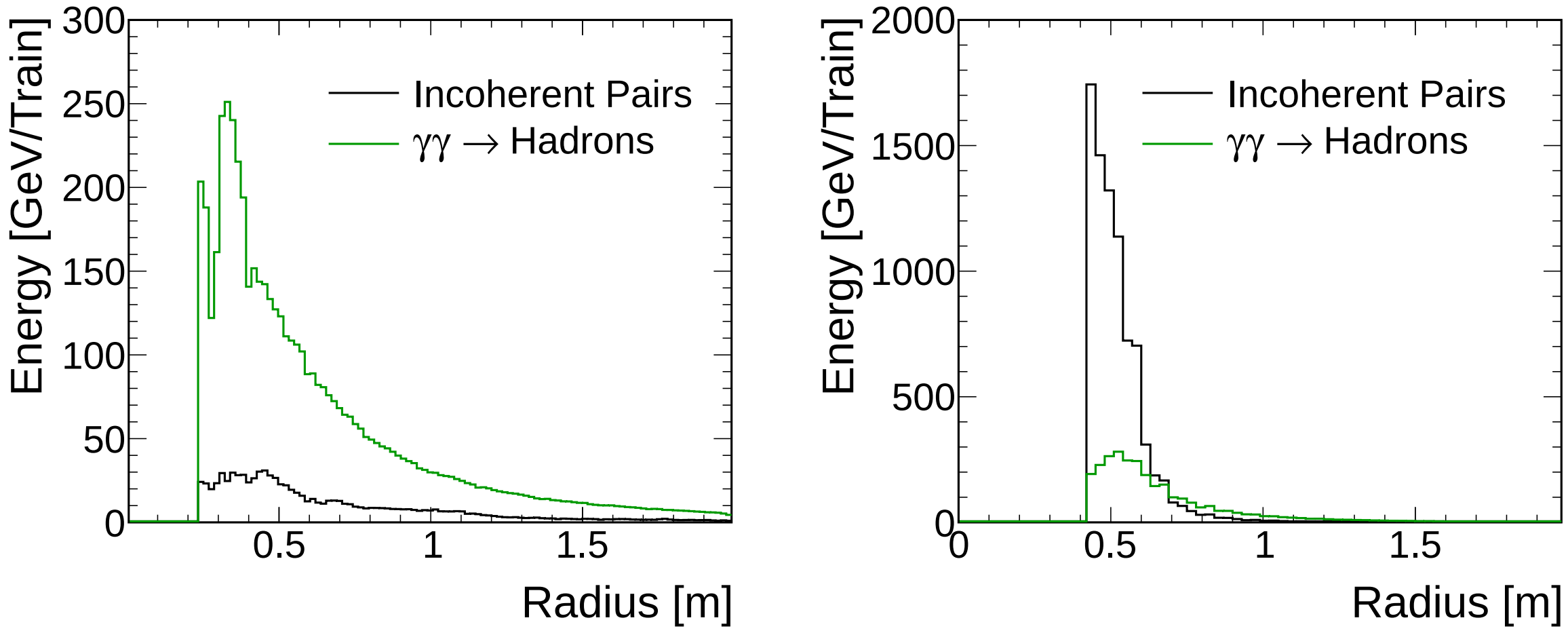

Fig. 2.8: Radial distribution of the calorimetric energy deposition in CLIC_ILD ECAL (left) and HCAL (right) endcaps for an entire bunch train from within 300 ns of the start of the train. The normalisation is applied for nominal background rates, excluding safety factors for the simulation uncertainties. The dip at approximately 0.3 m is due to the relatively thick beam pipe. The structure at 0.40 m in the ECAL corresponds to a small gap between calorimeter components.

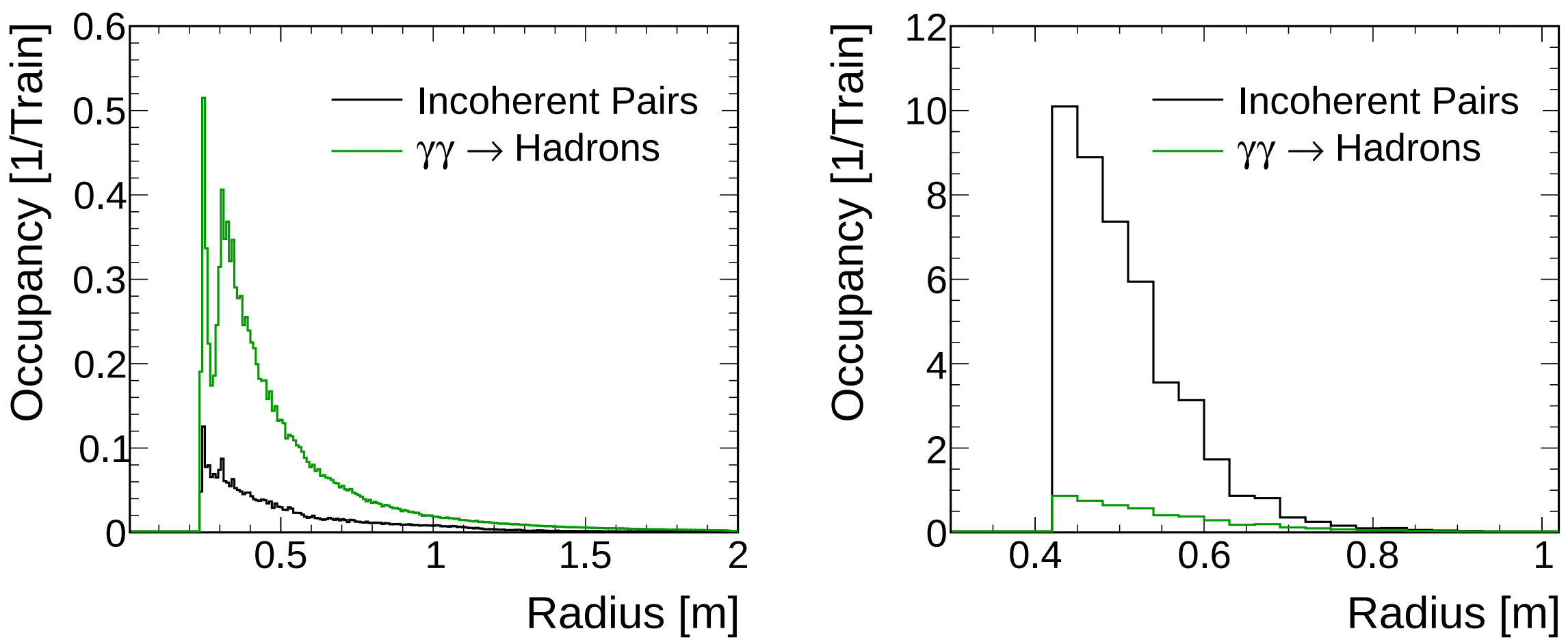

Fig. 2.9: Average cell occupancy in the ECAL (left) and HCAL (right) endcaps of the CLIC_ILD detector. In the case of the ECAL the average is given for layers 5 – 10 which broadly correspond to maximum energy deposit of typical EM showers. In the case of the HCAL the average is quoted for layers 35 – 45 where the maximum activity from the neutron background is observed. The results are obtained for nominal background rates, excluding safety factors for the simulation uncertainties.

The total energy deposition and associated occupancies at the inner radii of the HCAL would pose a serious problem for event reconstruction in these regions. However, the forward region of the GEANT4 models of the CLIC_ILD and CLIC_SiD detector concepts were designed before the impact of the scattered neutron background from the incoherent pairs interacting in the BeamCal was fully understood. The level of background in the low radius region of the HCAL in the current detector models is too high. However, there are a number of possible ways in which to reduce this component of the background. These will be studied in the future and include: increased shielding between the BeamCal and the inner regions of the HCAL; the possibility of using a different active material which, unlike plastic scintillator, is relatively insensitive to neutrons; and increasing the transverse segmentation to resolve the occupancy issue.

For the detector studies presented in Chapter 12 of this CDR, only the backgrounds from $\gamma\gamma \rightarrow$ hadrons are routinely included in the simulation. For all regions of the calorimeters, with the exception of the inner part of the HCAL endcap, this is the dominant source of background both in terms of the number of hits and the total energy deposition.

### 2.4.4 Background Summary

Table 2.3 summarises the simulated background conditions in the CLIC_ILD detector for an entire CLIC bunch train. The total calorimetric energy deposition of 37 TeV is large compared to the centre-of-mass energy and implies strict requirements on the timing resolution of CLIC calorimeters. Even excluding the HCAL contribution from the incoherent pair background, the overall energy deposited in the CLIC detector corresponds to about 20 TeV per bunch train. This is predominantly forward peaked (see Figure 2.8), but nevertheless poses a serious challenge to the design of a detector at CLIC.

Table 2.3: Summary of the background conditions in the CLIC_ILD detector model. The numbers correspond to the background for an entire CLIC bunch train and were obtained for nominal background rates, excluding safety factors for the simulation uncertainties. The reconstructed calorimeter energies are integrated over 300 ns from the start of the bunch train. The backgrounds in the HCAL from incoherent pairs are pessimistic as no attempts to mitigate the effect of neutrons from incoherent pair interactions in the BeamCal have been made.

| Subdetector | Incoherent Pairs [TeV] | $\gamma\gamma \rightarrow$ hadrons [TeV] |
|---|---|---|
| ECAL Endcaps | 2 | 11 |
| ECAL Barrel | – | 1.5 |
| HCAL Endcaps | 16 | 6 |
| HCAL Barrel | – | 0.3 |
| Total Calorimeter | 18 | 19 |
| Central Tracker | – | 7 |

## 2.5 Timing Requirements at CLIC

The backgrounds presented in Table 2.3 are for the full train of 312 bunches separated by 0.5 ns. The most obvious way to reduce the backgrounds associated with a particular physics event is to time stamp the hits from the detector and impose tight timing cuts. The background from $\gamma\gamma \rightarrow$ hadrons is proportional to the number of bunch crossings which are superimposed on the physics interaction. This is determined by the subdetector time integration windows and thus the requirements are driven by the impact of the background on reconstructed physics observables. Whilst the $\gamma\gamma \rightarrow$ hadrons background is high, the majority of the particles have low values of $p_T$ as shown in Figure 2.3 and any tight timing cuts can be restricted to relatively low $p_T$ particles.

The timing requirements at CLIC are driven by the levels to which the background degrades the physics performance of the detector. Provided the occupancies in the elements of the tracking detectors are sufficiently low that efficient track reconstruction is possible, there is unlikely to be a significant impact on the quality of the reconstructed tracks. Hence the main impact of the background will be on the reconstruction of jets. As an example Figure 2.10 shows a generator level study of the W-boson mass resolution for simulated W → qq decays, where the energy of the W-boson is 500 GeV, with different numbers of bunch crossings of $\gamma\gamma \to$ hadrons background superimposed. The jet energy resolution is assumed to be 3.5%. Only particles above a $p_T$ cut are used in the jet finding to suppress the effects of the $\gamma\gamma \to$ hadrons background.

The impact of the background on the reconstructed mass distribution is significant. The reconstructed width increases by approximately 70% when 20 BXs (10 ns) of background are overlaid, equivalent to a factor three reduction in effective luminosity. Figure 2.10 also shows the reconstruction of a high mass di-jet state as a function of the assumed level of background, in this case the reconstructed heavy Higgs mass from the process $e^+e^- \to H^0A^0 \to b\bar{b}b\bar{b}$. From these and other studies it is concluded that the acceptable level of background should correspond to 5–10 BXs, requiring a time resolution of $< 5$ ns. It should be noted that, in reality, simple $p_T$ cuts will be less effective than shown here due to pile-up from multiple background particles faking higher $p_T$ photons and neutral hadrons, thus the plots in Figure 2.10 underestimate the impact of the background.

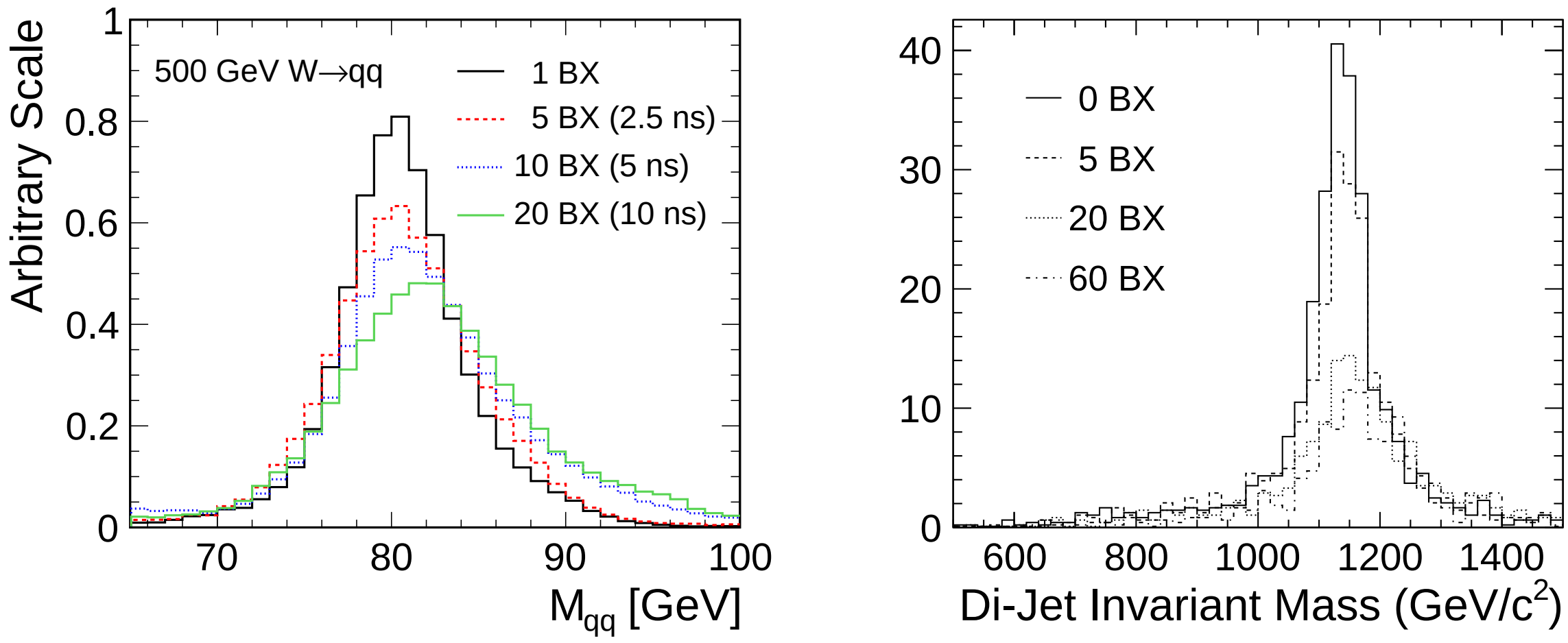

Fig. 2.10: (left) Reconstructed W boson mass distribution as a function of the assumed number of bunch crossing of background superimposed on 500 GeV W → qq decays. Jets are reconstructed as using the $k_t$-algorithm, discussed in Chapter 12, with a cut on $\Delta R = 0.7$. In each case, a $p_T$ cut is applied. The cut value depends on the number of bunch crossings of background overlaid and ranges from 0.5 GeV for 5 BXs to 1.0 GeV for 20 BXs. (right) Reconstructed heavy Higgs mass in the process $e^+e^- \to H^0A^0 \to b\bar{b}b\bar{b}$ for different levels of assumed background. In each case a $p_T$ cut of 0.9 GeV is applied.

From the above arguments it might be concluded that a 5 ns time-stamping capability is required for all subdetectors. However, regardless of the practical considerations, there is a fundamental limitation to the minimum time window from which it is viable to read out the calorimeters; because the propagation speed of the particles in hadronic showers is finite and the energy released in nuclear processes is not instantaneous on the timescale of 1 ns. Consequently a finite time window is required to accumulate the majority of the energy depositions associated with a hadronic shower. This has been studied in a GEANT4 simulation of the CLIC_ILD calorimeters using the high precision `QGSP_BERT_HP` physics list. Calorimeter hit times are corrected for straight-line time-of-flight from the IP.

Figure 2.11 shows the cumulative fraction of the total energy as a function of time for steel and tungsten absorbers. For the HCAL endcap with steel for the absorber material 90% of the energy is recorded within 6 ns (corrected for time-of-flight). For the tungsten absorber assumed for the HCAL barrel, only 82% of the energy is deposited within 25 ns. The calorimetric response of tungsten is slower than that of steel due to the much larger component of the energy in nuclear fragments.

There is a tension between the desire for a short readout window to suppress background and the need to read out the majority of the calorimeter hits associated with the hard interaction. In the studies presented in this document a reconstruction strategy has been adopted that balances these two requirements. It is this strategy, which takes advantage of particle flow reconstruction and the highly granular detectors being considered here, that drives the timing requirements for the CLIC detector.

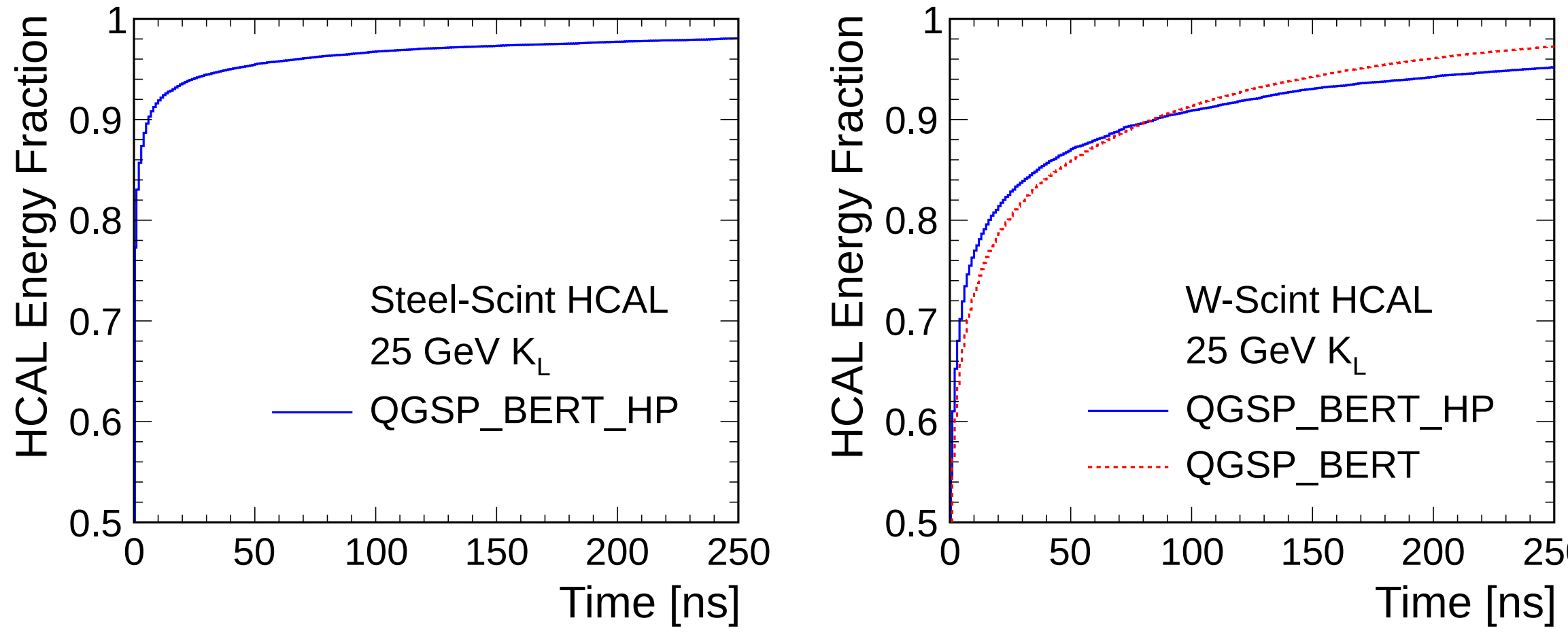

Fig. 2.11: Fraction of total calorimetric energy recorded as a function of time for 25 GeV neutral hadrons as a function in the CLIC_ILD detector (left) for steel and (right) for tungsten HCAL absorbers. The results are based on GEANT4 with the `QGSP_BERT_HP` physics list. Hit times are corrected for the straight-line time-of-flight prior to the cut. In the case of tungsten the plots are also shown for `QGSP_BERT`.

### 2.5.1 Timing in Physics Reconstruction at CLIC

It is assumed that the entire bunch train of data are available for offline reconstruction. Candidates for a hard interaction would be identified within the bunch train and the data in a window around this time would be passed to the event reconstruction. This integration time and the assumed single hit resolutions for the various subdetectors are listed in Table 2.4. The time window for the reconstruction in the calorimeters is driven by the shower development time. The single hit time resolution of 1 ns in the calorimeters allows tighter timing cuts to be applied to reconstructed clusters to further reduce the impact of background. The timing requirements in the Silicon tracking detectors can be looser than the maximum of 10 BXs suggested by studies such as those shown in Figure 2.10, because low momentum background tracks can be rejected using the times of associated calorimeter clusters. An integration time of 20 BXs is chosen, motivated by the need to reduce the combinatorics in the track reconstruction and by the fact that not all tracks will have an associated cluster giving a more precise time stamp. A TPC is assumed to integrate over the entire bunch train. All data in these time windows are passed to the track and PFA reconstruction software giving the reconstructed particle flow objects (charged hadrons, photons, neutral hadrons, electrons and muons).

With the above timing assumptions, it is essential to demonstrate the capability of the CLIC detector concepts to reconstruct physics events in the presence of the background from $\gamma\gamma \rightarrow$ hadrons. For this reason, all the physics studies presented in this report use full GEANT4 simulations of the CLIC detector concepts, including background from $\gamma\gamma \rightarrow$ hadrons overlaid on the physics events. Full track and particle flow reconstruction starting from the digitised hits in the time windows given in Table 2.4 is performed. Monte Carlo information is used at no stage in the reconstruction. Figure 2.12 shows the reconstructed particle flow objects for a simulated $e^+e^- \rightarrow H^+H^- \rightarrow t\bar{b}b\bar{t}$ event at $\sqrt{s} = 3$ TeV. At the reconstruction level, the background from $\gamma\gamma \rightarrow$ hadrons produces an average energy of approximately 1.2 TeV per event, mostly in the form of relatively low $p_T$ particles at relatively low angles to the beam axis. The level of $\gamma\gamma \rightarrow$ hadrons background is roughly 1/15 of that for the entire bunch train (Table 2.3), commensurate with integrating over 10 ns from the total 156 ns. The background can be further reduced by applying tighter timing cuts based on the reconstructed calorimeter cluster time. The cluster time is obtained from a truncated mean of the energy-weighted hit times constituting the cluster. In a fine grained particle flow detector many hits contribute to a single cluster and *cluster* time resolutions of <1 ns are easily achievable. Efficient background rejection is achieved by using tight cuts in the range of 1.0–2.5 ns on the clusters (depending on the type of reconstructed particle and its $p_T$). This procedure is applied to both neutral particle flow objects and to charged objects where the time of the cluster associated to the track, corrected by the helical propagation time, is used. These additional timing cuts are applied to only relatively low $p_T$ particle flow objects. The details of the cuts used are discussed in Section 12.1.4. As a result of the cluster-based timing cuts the average background level can be reduced to approximately 100 GeV with negligible impact on the underlying hard interaction. The use of hadron-collider inspired jet-finding algorithms further reduces the impact of the background of $\gamma\gamma \rightarrow$ hadrons and precision physics measurements are achievable in the CLIC background environment as shown in Chapter 12.

Table 2.4: Assumed time windows used for the event reconstruction and the required single hit time resolutions.

| Subdetector | Reconstruction window | hit resolution |
|---|---|---|
| ECAL | 10 ns | 1 ns |
| HCAL Endcaps | 10 ns | 1 ns |
| HCAL Barrel | 100 ns | 1 ns |
| Silicon Detectors | 10 ns | $10/\sqrt{12}$ ns |
| TPC | entire bunch train | n/a |

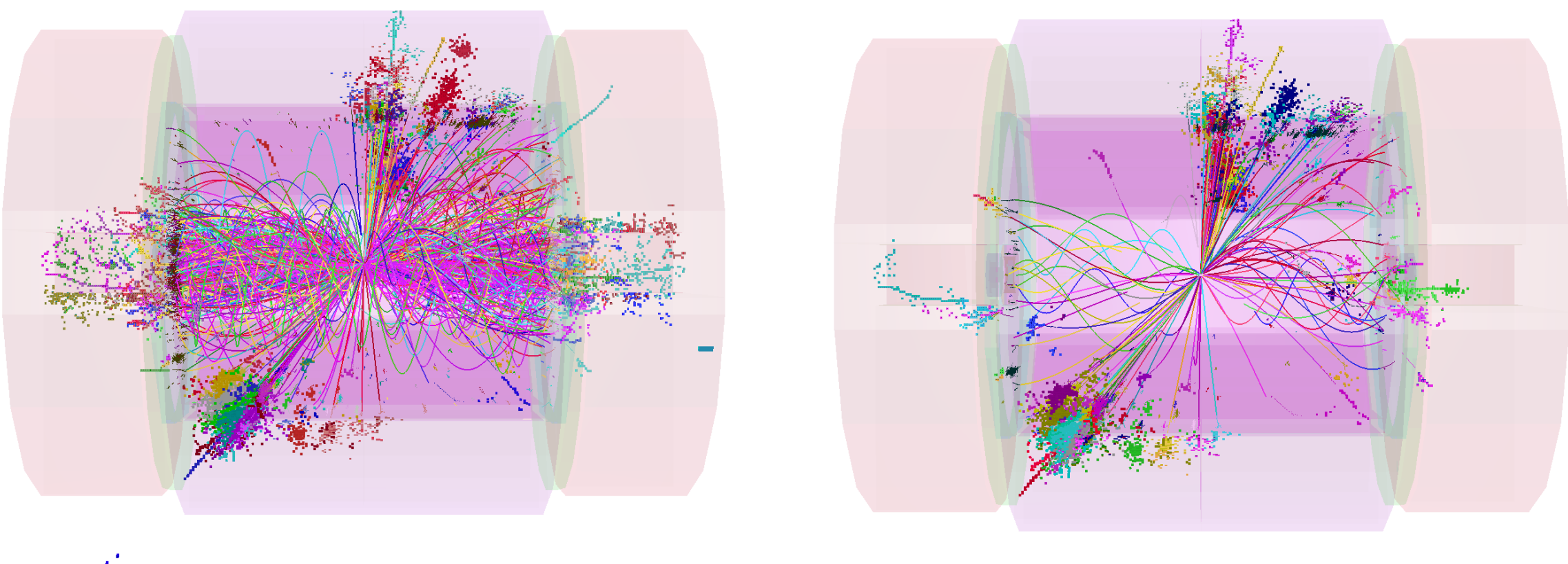

Fig. 2.12: (left) Reconstructed particles in a simulated $e^+e^- \rightarrow H^+H^- \rightarrow t\bar{b}b\bar{t}$ event at 3 TeV in the CLIC_ILD detector concept with background from $\gamma\gamma \rightarrow$ hadrons overlaid. (right) the effect of applying tight timing cuts on the reconstructed cluster times.

## 2.6 Detector Benchmark Processes

One of the main goals of this CDR is to demonstrate that an experiment operating at the CLIC accelerator can deliver the required performance in terms of measuring the physics observables described above. With this aim in mind, a number of benchmark physics processes[2] were chosen, each addressing a particular aspect of the detector performance goals [34]. These benchmark channels were studied using a detailed GEANT4 simulation of realistic detector concepts for CLIC including pile-up from $\gamma\gamma \rightarrow$ hadrons background. The benchmark physics studies were mostly performed for a centre-of-mass energy of 3 TeV. The highest CLIC energy was chosen for most of the channels as it provides the most challenging environment both in terms of the machine background and in terms of the particle multiplicities and jet and lepton energies.

### 2.6.1 Light Higgs Production : $e^+e^- \rightarrow h\nu_e\bar{\nu}_e$

For a light Higgs (with mass $m_h$ = 120 GeV), the dominant production process at 3 TeV is through the $W^+W^-$ fusion process resulting in a large cross section for $e^+e^- \rightarrow h\nu_e\bar{\nu}_e$. This large cross section opens up the possibility to study rare decays. The first CLIC detector benchmark is the measurement of the ratio of $BR(h \rightarrow \mu^+\mu^-)/BR(h \rightarrow b\bar{b})$. The observation of the rare decay $h \rightarrow \mu^+\mu^-$, with a branching ratio of the order of $10^{-4}$, relies on the ability to reconstruct the Higgs mass from the two decay muons with sufficiently good mass resolution to distinguish the Higgs decays from the large and irreducible continuum background from, for example, $e^+e^- \rightarrow \mu^+\mu^-e^+e^-$. Observation of this rare decay thus requires excellent momentum resolution. The reconstruction of $h \rightarrow c\bar{c}$ and $h \rightarrow b\bar{b}$ final states at $\sqrt{s}$ = 3 TeV provides a test of flavour-tagging for relatively low energy jets in the presence of the non-negligible beam related backgrounds. It also requires sufficient jet energy resolution to distinguish Higgs decays from Z decays. Thus this benchmark channel probes the detector performance for:

- muon momentum resolution;
- flavour-tagging for relatively low energy jets;
- jet energy reconstruction for relatively low energy jets.

### 2.6.2 Heavy Higgs Production

Supersymmetric extensions of the SM result in a rich Higgs sector. The study of the heavy Higgs pair production at CLIC requires the reconstruction of high mass, multi-jet final states and thus forms a suitable detector benchmark channel. For the heavy Higgs benchmark study, a SUSY model labelled *SUSY model I* and defined by the following GUT scale parameters was chosen: $M_1 = 780$ GeV, $M_2 = 940$ GeV, $M_3 = 540$ GeV, $A_0 = -750$ GeV, $m_0 = 303$ GeV, $\tan\beta = 24$ and $\mu > 0$. (Other relevant particle masses are given in [34].)

Note that this is not a CMSSM model, since it has non-unified gaugino masses. This allows sleptons to be relatively heavier compared to squarks than the ratio found in mSUGRA within the stau-coannihilation region. These parameters were chosen to be consistent with current data. In particular the contribution to the muon magnetic moment anomaly $a_\mu$ is $\Delta a_\mu = 6 \cdot 10^{-10}$ and they result in $BR(b \rightarrow s\gamma) = 3.0 \cdot 10^{-4}$. It should be noted that for the purpose of the benchmark the detector performance the exact nature of the model parameters are not critical. The essential feature of this model is that it gives rise to a heavy SUSY Higgs sector: $m(h) = 119.13$ GeV, $m(A) = 902.6$ GeV, $m(H^0) = 902.4$ GeV and $m(H^\pm) = 906.3$ GeV. In this model, the heavy Higgs bosons predominantly decay to heavy quark, $BR(H^\pm \rightarrow tb) = 82\%$, $BR(H^0 \rightarrow b\bar{b}) = 82\%$ and $BR(A \rightarrow b\bar{b}) = 82\%$, with the remaining decay modes being dominated by $\tau$-leptons in the final state.

[2] Models are defined and checked using the output of Softsusy 3.1.3 [30] and micrOmegas 2.2 [31]. Branching ratios come from SDECAY 1.1 [32], and cross sections from Madgraph 4 [33].

The reconstruction of the heavy Higgs mass and width in the processes $e^+e^- \to H^+H^- \to t\bar{b}b\bar{t}$ and $e^+e^- \to H^0A^0 \to b\bar{b}b\bar{b}$ is chosen as the second benchmark physics process. This physics benchmark probes the detector performance for:

– flavour-tagging for high energy jets;
– invariant mass reconstruction of high mass states in a high multiplicity environment.

### 2.6.3 Production of Right-Handed Squarks

The production and decay of right-handed squarks in the process $e^+e^- \to \tilde{q}_R\bar{\tilde{q}}_R \to q\tilde{\chi}^0\bar{q}\tilde{\chi}^0$ results in a simple topology of two high energy jets and missing energy and is chosen as the third benchmark process. The same *SUSY model I*, as for the heavy Higgs benchmark channel, is used here. In this model, the squark masses are $m(\tilde{u}_R) = m(\tilde{c}_R) = 1125.7$ GeV and $m(\tilde{d}_R) = m(\tilde{s}_R) = 1116.1$ GeV and the right-handed squarks decay almost 100% to $\tilde{q}_R \to q\tilde{\chi}_1^0$ with $m(\tilde{\chi}_1^0) = 328$ GeV. The reconstruction of the squark mass in the inclusive jets plus missing energy provides a test of:

– jet energy and missing energy reconstruction for high energy jets in a simple topology.

### 2.6.4 Chargino and Neutralino Pair Production

A high energy $e^+e^-$ collider provides a clean environment to study SUSY particles. From the perspective of demonstrating detector performance Chargino and Neutralino pair production provides an interesting benchmark. Since the purpose of the benchmark channels is to demonstrate the detector capabilities for reconstructing typical final states at CLIC rather than providing a full demonstration of the physics reach of the machine, a SUSY model in which the lightest chargino and two lightest neutralinos are dominated by a single decay mode is used. A model labelled *SUSY model II* and defined by the mSUGRA parameters $m_{1/2} = 800$ GeV, $A_0 = 0$, $m_0 = 966$ GeV, $\tan\beta = 51$ and $\mu > 0$ is used. This model has 640 GeV wino-like states and 910 GeV Higgsino-like states. For this benchmark process, the relevant masses are $m(\tilde{\chi}_1^0) = 340.3$ GeV, $m(\tilde{\chi}_2^0) = 643.1$ GeV, $m(\tilde{\chi}_1^\pm) = 643.2$ GeV and $m(\mathrm{h}) = 118.5$ GeV. (Other used masses are given in [34].)

In this model, the $\tilde{\chi}_1^\pm$ decays essentially 100% of the time to $W^\pm\,\tilde{\chi}_1^0$ and the $\tilde{\chi}_2^0$ decays to both the Z and h bosons with branching ratios $\mathrm{BR}(\tilde{\chi}_2^0 \to h\tilde{\chi}_1^0) = 90.6\%$ and $\mathrm{BR}(\tilde{\chi}_2^0 \to Z\tilde{\chi}_1^0) = 9.4\%$. Since the Z, W and h have largest decay fractions to quarks, the signatures for chargino and neutralino production in this model are final states with four jets and missing energy. The reconstruction of the Z, W and h masses from the appropriate di-jet combinations is essential to disentangle the physics signatures. Hence chargino and neutralino production provides a benchmark for the reconstruction of hadronically decaying gauge bosons in a multi-jet environment.

The fourth benchmark process is the reconstruction of the $\tilde{\chi}_1^\pm$, $\tilde{\chi}_1^0$ and $\tilde{\chi}_2^0$ masses in final states with four jets and missing energy from the processes:

$$\begin{aligned}
e^+e^- \to \quad & \tilde{\chi}_1^+\tilde{\chi}_1^- && \to W^+W^-\tilde{\chi}_1^0\tilde{\chi}_1^0 \\
e^+e^- \to \quad & \tilde{\chi}_2^0\tilde{\chi}_2^0 && \to hh\tilde{\chi}_1^0\tilde{\chi}_1^0 \\
e^+e^- \to \quad & \tilde{\chi}_2^0\tilde{\chi}_2^0 && \to hZ\tilde{\chi}_1^0\tilde{\chi}_1^0
\end{aligned}$$

This benchmark process addresses:

– jet energy and missing energy reconstruction in high energy decays;
– di-jet mass reconstruction and separation of Z, W and h hadronic decays.

### 2.6.5 Slepton Production

The production of energetic leptons is a signature for many BSM physics processes and thus the reconstruction of high energy electrons and muons is an essential aspect of a detector at CLIC. Hence the fifth physics benchmark channel, namely slepton production, focuses on lepton reconstruction. For the slepton production, the same *SUSY model II* as for the chargino and neutralino pair production benchmark (see above) is used. In this model, the selectron and smuon masses are: $m(\widetilde{e}_R) = m(\widetilde{\mu}_R) = 1010.8$ GeV and $m(\widetilde{e}_L) = m(\widetilde{\mu}_L) = 1100.4$ GeV, with the right sleptons decaying 100% into electrons and muons, e.g. $\widetilde{e}_R \to e\widetilde{\chi}_1^0$ and $\widetilde{\mu}_R \to \mu\widetilde{\chi}_1^0$, while the left selectrons decay 29% into $\widetilde{e}_L \to e\widetilde{\chi}_2^0$ and 19% into $\widetilde{e}_L \to e\widetilde{\chi}_1^0$. In addition, $m(\tilde{\nu}_L) = 1097$ GeV, and the sneutrino decays in 56% of the cases into $\widetilde{\nu}_{eL} \to e\widetilde{\chi}_1^{\pm}$.

The fifth detector benchmark is the reconstruction of slepton masses from the lepton energy distributions in the processes

$$\begin{aligned} e^+e^- \to \quad & \widetilde{e}_R\widetilde{e}_R \quad \to e^+e^-\widetilde{\chi}_1^0\widetilde{\chi}_1^0 \\ e^+e^- \to \quad & \widetilde{\mu}_R\widetilde{\mu}_R \quad \to \mu^+\mu^-\widetilde{\chi}_1^0\widetilde{\chi}_1^0 \\ e^+e^- \to \quad & \widetilde{e}_L\widetilde{e}_L \quad \to e^+e^-\widetilde{\chi}_2^0\widetilde{\chi}_2^0 \\ e^+e^- \to \quad & \widetilde{\nu}_e\widetilde{\nu}_e \quad \to e^+e^-\widetilde{\chi}_1^+\widetilde{\chi}_1^- \end{aligned}$$

The main aspects of the detector performance which are addressed are:

- reconstruction and identification of high energy leptons;
- energy resolution for high energy electrons and muons, in two leptons, or in two leptons plus four jets topology;
- boson mass resolution.

### 2.6.6 Top Pair Production at 500 GeV

The process of $e^+e^- \to t\bar{t}$ has been extensively studied for the ILC [18, 19, 35]. It is possible that the construction of CLIC will be staged in energy, with the exact construction path depending on the BSM physics uncovered by the LHC. For this reason it is useful to compare the physics reach for CLIC operating at $\sqrt{s} = 500$ GeV to the previous ILC studies. Given the different background conditions it is not *a priori* clear that the same physics sensitivity can be reached. For this reason the measurement of the top mass from direct reconstruction in $e^+e^- \to t\bar{t}$ at $\sqrt{s} = 500$ GeV is chosen as the sixth benchmark process. Both fully-hadronic and semi-leptonic final states are considered, $t\bar{t} \to q\bar{q}b\, q\bar{q}\bar{b}$ (six jets) and $t\bar{t} \to q\bar{q}b\, l\bar{\nu}b$ (four jets + lepton + missing energy). This benchmark channel provides a test of:

- mass reconstruction in a multi-jet final state for low energy jets;
- flavour tagging;
- impact of CLIC beam conditions at 500 GeV compared to those of the ILC.

# Chapter 3

# CLIC Detector Concepts

## 3.1 Rationale

The physics program at CLIC is the exploration and understanding of new physics beyond the Standard Model in the TeV energy range. This programme places stringent requirements on the detector performance. These include precise momentum resolution, vertex reconstruction, particle identification, excellent jet reconstruction and hermetic coverage, as outlined in Chapter 2. To allow for flexibility in the beam energy and possible staging of the accelerator, these requirements have to be met over a range of centre-of-mass energies from 500 GeV extending up to the top CLIC energy of 3 TeV.

For the ILC, with similar detector requirements, albeit at a lower energy, two general purpose detector concepts (ILD [1] and SiD [2]) have been developed into mature designs over the last decade. Both of these concepts were evaluated and validated by the IDAG [3]. They serve as excellent starting points for the CLIC detectors, with modifications motivated by the more challenging experimental conditions at CLIC and by the higher collision energy.

Following the physics requirements, it is not surprising that the main building blocks of the two designs are very similar: a cylindrical detector with tracking and calorimetry inside a solenoid. However, particular choices and the overall approach are quite different. For example, ILD tries to optimise jet reconstruction with calorimetry at large radii to separate the outgoing particles as much as possible at the cost of a lower magnetic field. The consequence is an overall radial size like the CMS detector at the LHC. On the other hand, the SiD design is as compact as possible to provide a cost-optimised detector, resulting in a high magnetic field within a solenoid of minimal radius and with precision all-silicon tracking. Starting from these two designs, two CLIC detector concepts have been developed, referred to as CLIC_ILD and CLIC_SiD, respectively.

## 3.2 High Energy CLIC Environment

The two main differences between CLIC and ILC are the higher centre-of-mass energy and the 0.5 ns between bunch crossings at CLIC. The higher energy leads to more machine related background from incoherent pairs and more hadronic two-photon events mostly in the forward regions. Particle production in the t-channel with particles in the forward region become more important and jets will be more narrow in the detector due to a stronger boost. The ILD and SiD designs are modified to meet the CLIC performance requirements.

## 3.3 Design Principles

The main underlying design principles for the CLIC detector concepts, in line with the ILC ones, are:

- very efficient tracking detectors with excellent momentum reconstruction in a high field solenoid;
- secondary vertex reconstruction with a powerful pixel detector as close as possible to the beam pipe;
- low material budgets in the tracking devices;
- highly segmented electromagnetic and hadronic calorimeters inside the solenoid;
- Particle Flow Algorithms (PFA) for optimal jet reconstruction define the layout of the detector, in particular for the calorimeters;
- hermeticity of the detector, crucial for an excellent determination of missing energy which is an important signature for new physics processes;
- instrumented return yoke for muon identification.

For CLIC, time stamping capabilities of $\mathcal{O}$(1 ns) need to be available for several subsystems. The event readout will integrate over up to 312 bunch crossings. Time stamping could reduce the pile-up from two-photon background events to $\leq 20$ bunch crossings.

## 3.4 Subsystems

We will briefly introduce the CLIC detector concepts, going from small to large radius. Figures 3.1 and 3.2 show longitudinal and transverse cuts of the major components of CLIC_ILD and CLIC_SiD. Table 3.1 compares the key parameters of the CLIC and ILC detector designs. Table 3.2 summarises details of the CLIC_ILD and CLIC_SiD designs.

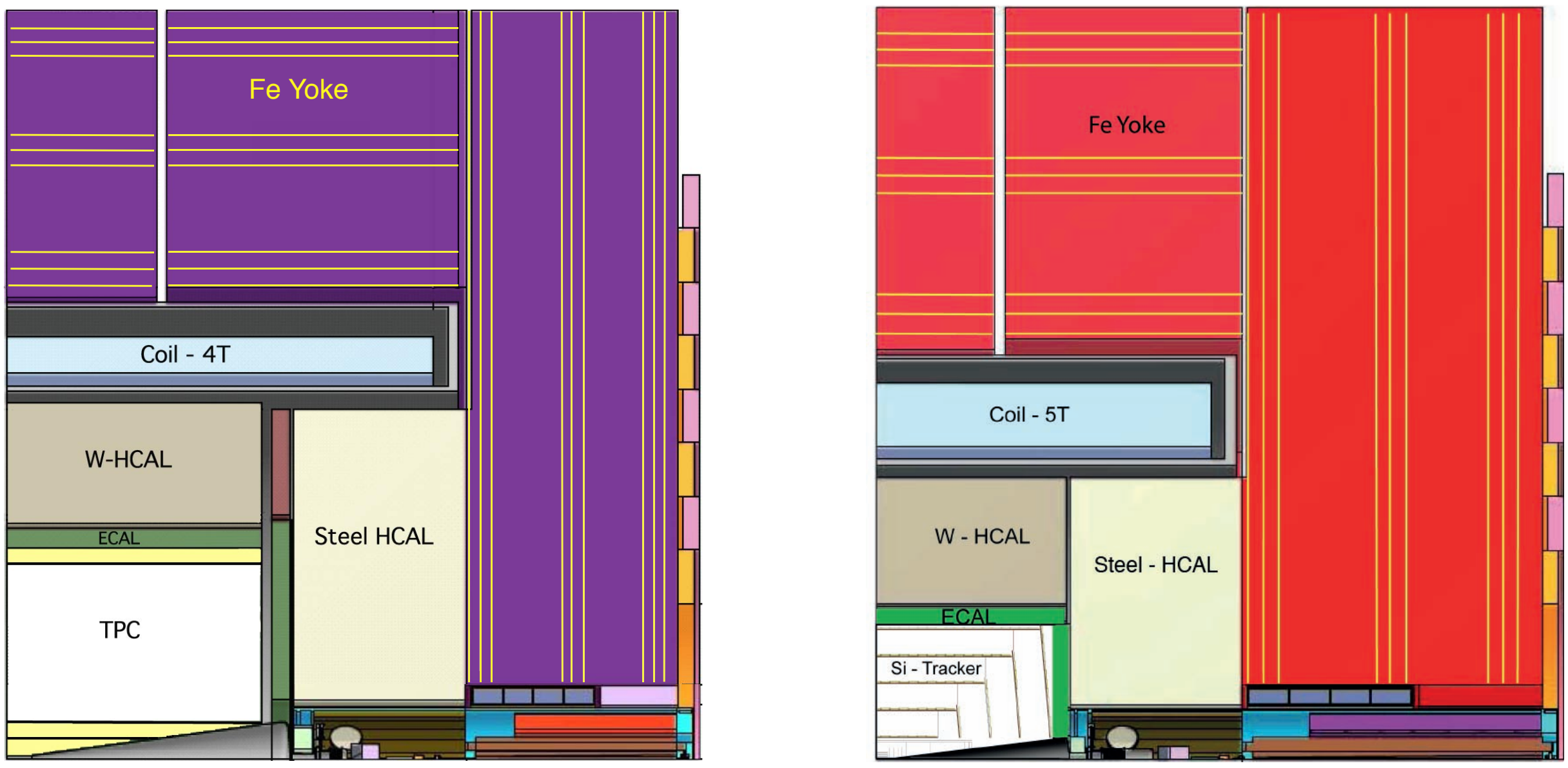

Fig. 3.1: Longitudinal cross section of the top quadrant of CLIC_ILD (left) and CLIC_SiD (right).

Table 3.1: Some key parameters of the ILC and CLIC detector concepts. The inner radius of the electromagnetic calorimeter is given by the smallest distance of the calorimeter to the main detector axis. For the hadronic calorimeter, materials are given both for the barrel (B) and the endcap (E).

| **Concept** | **ILD** | **CLIC_ILD** | **SiD** | **CLIC_SiD** |
|---|---|---|---|---|
| Tracker | TPC/Silicon | TPC/Silicon | Silicon | Silicon |
| Solenoid Field (T) | 3.5 | 4 | 5 | 5 |
| Solenoid Free Bore (m) | 3.3 | 3.4 | 2.6 | 2.7 |
| Solenoid Length (m) | 8.0 | 8.3 | 6.0 | 6.5 |
| VTX Inner Radius (mm) | 16 | 31 | 14 | 27 |
| ECAL $r_{\text{min}}$ (m) | 1.8 | 1.8 | 1.3 | 1.3 |
| ECAL $\Delta r$ (mm) | 172 | 172 | 135 | 135 |
| HCAL Absorber B / E | Fe | W / Fe | Fe | W / Fe |
| HCAL $\lambda_{\text{I}}$ | 5.5 | 7.5 | 4.8 | 7.5 |
| Overall Height (m) | 14.0 | 14.0 | 12.0 | 14.0 |
| Overall Length (m) | 13.2 | 12.8 | 11.2 | 12.8 |

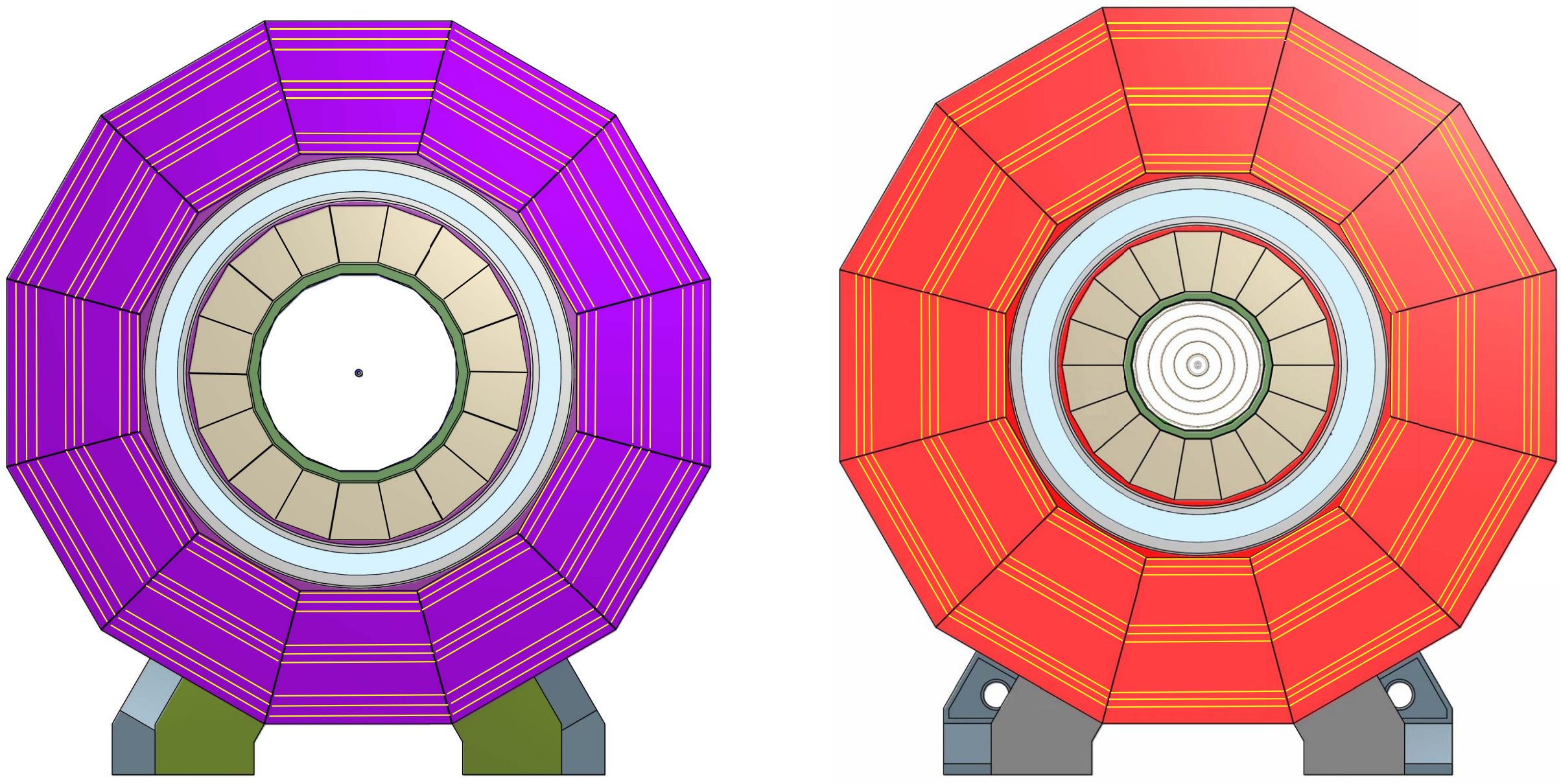

Fig. 3.2: Transverse cross section of CLIC_ILD (left) and CLIC_SiD (right).

The pixel ***Vertex Detector***, VTX, needs to be as close as possible to the beam pipe to obtain optimal secondary vertex reconstruction. The innermost radius is defined by the requirement to stay safely outside the region of high occupancy caused by direct hits from incoherent pair background at low transverse momenta, as discussed in Chapter 4. This requires the first layer to be moved outwards by 15 mm, compared to the ILC detector concepts, for the CLIC detector design. Due to the higher magnetic field, the innermost layer in CLIC_SiD can be at slightly smaller radius than in CLIC_ILD. The geometry is adapted accordingly with the number of layers unchanged, i.e., three double layers for CLIC_ILD and five single layers for CLIC_SiD. In the SiD design the Vertex Detector provides important information for the track finding, hence the choice of five VTX layers for a tracking system with a total number of 10 layers. With a TPC as main tracker in the case of CLIC_ILD, the vertex detector has mainly the role to measure the impact parameter of a track. With double layers one can reduce the material and hence improve the impact-parameter resolution. In the SiD version, the endcap disks are re-arranged to allow for 10 hits/track down to 8 degrees. For CLIC, a different detector than for ILC is required, which provides time stamping on the few nanosecond level. The research and development of suitable detector technologies are discussed in detail in Chapter 4.

For the ***Outer Tracker***, we follow the choices of the two ILC detector concepts: a TPC in the ILD case and a silicon tracker in the SiD case. The most important challenges of a TPC at CLIC are the two-track separation in high energy jets and event identification in the collection of 312 bunch crossings in 156 ns. The supplementary silicon detector layers outside the TPC which are necessary to achieve a high momentum resolution are even more important at high energies.

In the all-silicon tracker version, we follow the SiD choice of five layers of thin silicon strips arranged in a barrel and endcap section. The cooling is assumed to be done by gas flow to keep the material budget as low as possible. The multiple scattering contribution to the momentum resolution can only be ignored for tracks with $p \geq 300$ GeV, higher than the majority of the expected track momenta at $\sqrt{s} = 3$ TeV. Both tracker concepts are presented in detail in Chapter 5.

The ***Electromagnetic Calorimeter*** (ECAL) is taken without modifications with respect to the original ILC concepts. Both the ILD and the SiD ECAL use Silicon-Tungsten sampling calorimeters optimised for particle flow, placing particular emphasis on the separation of close-by electromagnetic showers. This requirement drives the transverse and longitudinal segmentation, which is below the Molière radius $R_M$ and below one radiation length $X_0$ in the front part of the calorimeters, respectively. To achieve maximal compactness, the longitudinal sampling is more coarse in the rear of the calorimeters.

The two concepts differ slightly in their choice of transverse and longitudinal segmentation as well as in overall thickness. While the ECAL of ILD (and of CLIC_ILD) uses active layers with $5.1 \times 5.1$ mm$^2$ pads, SiD (and CLIC_SiD) is using smaller hexagonal pads of 13 mm$^2$. Both concepts use 30 layers in total, with the first 20 layers with a twice finer sampling (below 1 $X_0$) than the last 10 layers. Here, the ILD ECAL uses 2.1 mm and 4.1 mm thick absorber plates for the first and the second detector segment resulting in an overall thickness of 23 $X_0$, while the SiD ECAL uses 2.5 mm and 5 mm thick absorber plates, giving a total thickness of 26 $X_0$. While the ILD design is based on pure tungsten, the SiD ECAL is based on an non-magnetic alloy of tungsten, copper and nickel with a few percent lower density, and thus larger radiation length. In addition to the Silicon-Tungsten option presented here, the use of small scintillator strips with SiPM readout as active medium is being considered, as well as mixed designs using alternating layers of silicon and scintillator.

The ILC ECAL designs satisfy the requirements of a CLIC detector, since they provide granularity significantly below the typical extent of individual showers, needed for the separation of particles within hadronic jets. The increased energy leakage due to higher energy only affects a very small fraction of all particles, and is well controlled by the hadron calorimeter which follows after the ECAL without significant dead material.

The ***Hadron Calorimeter*** (HCAL) represents one of the major changes. The ILC concepts have a total of about 6 - 7 nuclear interaction lengths ($\lambda_I$) for ECAL plus HCAL. This is not enough to absorb the typical hadronic energy of an event at 3 TeV. A thickness of 7.5 $\lambda_I$ for the HCAL is being proposed, resulting in a total thickness of 8.5 $\lambda_I$ for the calorimeter system including the electromagnetic calorimeter. This increased thickness was derived from simulation studies, which show that the particle flow jet energy resolution quickly degrades for a thinner calorimeter system, while only a very moderate improvement can be achieved with an even deeper calorimeter system.

If one does not want to increase the inner bore radius of the ILC detector solenoids, one has to replace the barrel HCAL iron with a more dense absorber. Tungsten has been chosen as absorber for the barrel while for the endcaps steel can be used as absorber since there are no depth restrictions. In this way, one can exploit the faster shower development in steel and limit the usage of expensive tungsten.

For the CDR simulation studies, a highly granular scintillator readout is adopted for the tungsten barrel HCAL in both CLIC detector concepts, using the technology implemented in the ILD detector concept. The active layers have a thickness of 6.5 mm, which could potentially be decreased with additional R&D to fully exploit the increase in average density provided by the use of tungsten. Several alternative readout technologies are being studied, such as gaseous RPC detectors with digital or semi-digital readout as well as Micromegas and GEM detectors.

To mitigate the effect of the high hadronic background from two-photon processes combined with the high bunch crossing frequency, time stamping in the calorimeters is crucial. This applies in particular to the endcap, but to a lesser degree also to the barrel calorimeter. At the cluster (particle) level, time stamping on the nanosecond level is expected to be achievable by combining the information of all contributing cells. The required integration time of the calorimeters, and thus the susceptibility to background pick-up, is driven by the time structure of hadronic showers, which is more extended for a heavier absorber such as tungsten, motivating the choice of different absorber materials for barrel end endcap hadronic calorimeters. More details are given in Chapter 6.

The ***Solenoid***. With the choice of tungsten as HCAL absorber and 7.5 $\lambda_I$ the coil dimensions can be kept similar to those of the ILC concepts. The momentum resolution with the 5 T magnetic field of the compact SiD is sufficient even at high energies, therefore, CLIC_SiD uses 5 T as well. However, for a CLIC_ILD detector at 3 TeV it is advantageous to increase the magnetic field by 0.5 T to 4 T which is possible for the ILD magnet design (see Chapter 7).

The ***Magnet Yoke and Muon Chambers***. The magnetic flux is returned through an iron yoke. The thickness of the iron depends on the mechanical forces due to the magnetic field and on the tolerable fringe field of the two concepts, which are assumed to be the same for ILC and CLIC. The assumption on the maximum allowable magnetic field at 15 m from the detector is 50 Gauss [4]. Due to the different magnetic field strengths of the two concepts, this requirement leads, in a preliminary analysis, to a yoke iron thickness of 230 cm for CLIC_ILD and of 270 cm for CLIC_SiD. To enhance the muon identification capability of the detector the iron is instrumented with track sensitive chambers, either glass RPCs or scintillators. For the sake of simplicity, we chose the same layout of nine muon detector layers for both detector concepts. For reasons of mechanical stability of the yoke and ease of pattern recognition, these layers are arranged in three sets of three layers (see Chapter 8).

The ***Very Forward detectors***. The forward region of a detector at CLIC (Figure 3.3), just as at ILC, has to provide the luminosity measurement and extended coverage with a beam calorimeter. A Luminosity Calorimeter (LumiCal) precisely counts the number of Bhabha events in an angular region between 40 and 100 mrad and allows the measurement of the luminosity spectrum using the acollinearity angle of Bhabha scattering. A Beam Calorimeter (BeamCal) extends the angular coverage of the forward calorimeters down to polar angles of about 10 mrad. Both calorimeters are centred around the axis of the outgoing beams. The forward region also contains masking to prevent particles produced by the beam-beam interaction from backscattering into the main detectors and to protect the equipment downstream of the BeamCal, such as the beam position monitor (BPM) and a kicker of the intra train feedback, and the final focus quadrupole (QD0). The Very Forward detectors are described in detail in Chapter 9.

A few modifications are needed to adapt the detector for CLIC: Most importantly, the solenoid will not be complemented by an anti detector integrated dipole (Anti-DID). An Anti-DID would produce a magnetic field parallel to the outgoing beam-axis to direct low energy background particles out of the detector. The Anti-DID, however, would reduce the luminosity at 3 TeV by about 20% [5]. On the other hand, an Anti-Solenoid, compensating the field on the beam axis in the yoke region, is foreseen around the QD0 as shown in Figure 3.3.

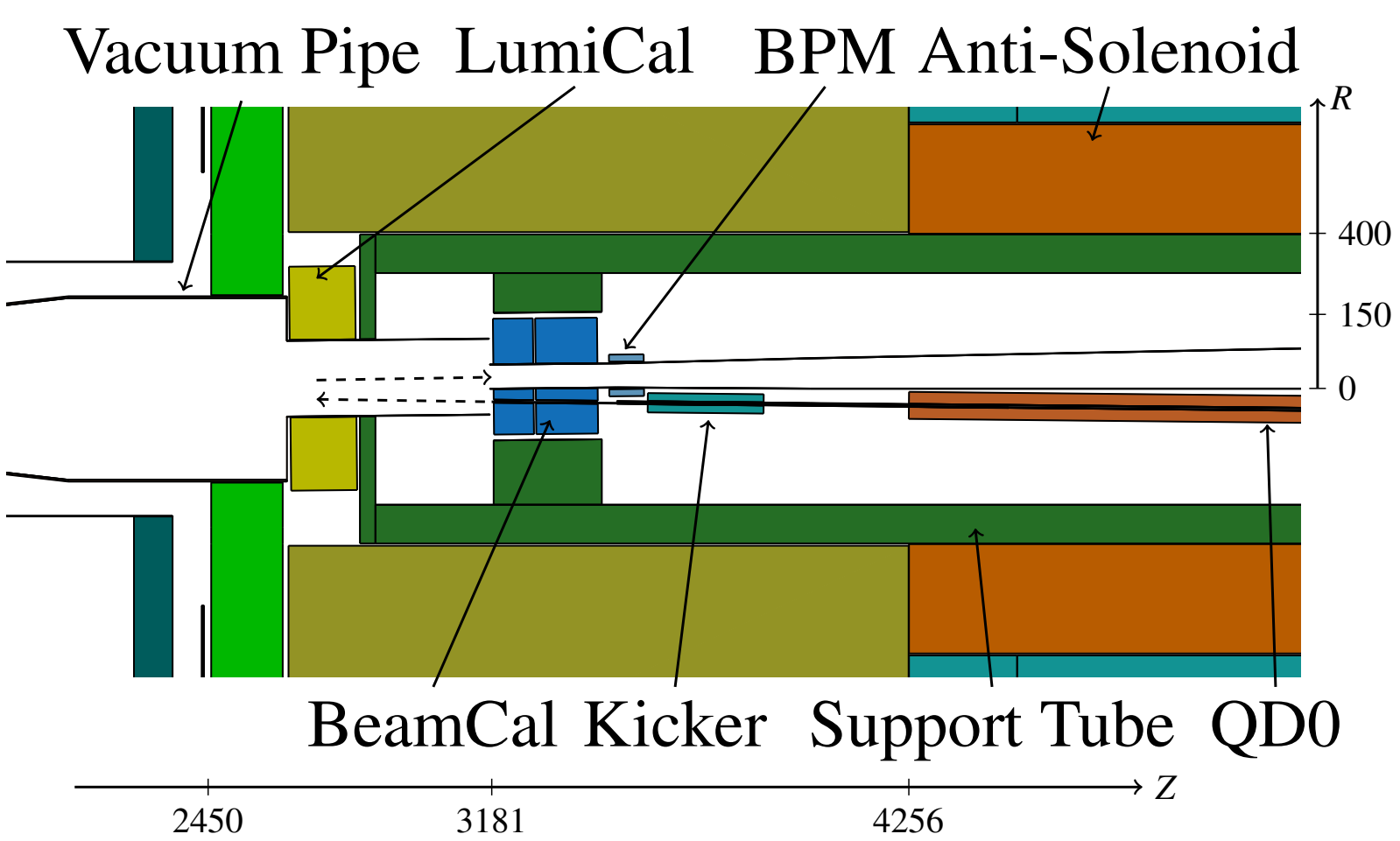

Fig. 3.3: Layout of the forward region, seen from the top. The horizontal axis refers to the interaction point. The dashed arrows indicate the direction of the beams. Dimensions are given in mm.

During the collision the beam develops a large energy spread and a significant amount of beamstrahlung and coherent pairs are created, which have an angular spread of several milliradian due to the beam fields. To safely transport these particles through the detector, the exit line has to have an aperture of about 10 mrad at $\sqrt{s} = 3$ TeV. The crossing angle of 20 mrad needed for the CLIC beams as compared to 14 mrad at ILC leads to changes in the dimensions of the elements in the forward region.

A detailed discussion of the mechanical support of the forward detectors and the integration of the final focus quadrupole can be found in Chapter 11. Due to the different tracker dimensions and different magnetic fields, and consequently different positions for the final focusing elements the two detectors use different focal lengths $L^*$ of 3500 mm at CLIC_SiD and 4340 mm at CLIC_ILD.

## 3.5 Detector Parameters

The main parameters of the two detector concepts for CLIC, CLIC_ILD [6] and CLIC_SiD [7], are summarised in the following table. Some of the parameters are unchanged with respect to the original ILC detector concepts, while others have been adapted as discussed above.

Table 3.2: Some key parameters of the CLIC detector concepts. The inner radii refer to the inscribed circle of the polygon. All dimensions are given in millimetres.

| | CLIC_ILD | CLIC_SiD |
|---|---|---|
| **Overall Dimensions** | | |
| Outer size [W×H×L] | 14000 × 14000 × 12400 | 14000 × 14000 × 12400 |
| Estimated total weight | 12200 tons | 12500 tons |
| Beam-line height | 7900 | 7900 |
| **Vertex Detector** | | |
| Inner radius | 31 | 27 |
| Outer radius | 60 | 77 (barrel), 169 (disks) |
| Max. Z | 125 (barrel), 257 (disks) | 99 (barrel), 830 (disks) |
| Barrel layers | 6 (3 double layers) | 5 |
| Forward disks | 6 (3 double layers) | 7 |
| **Barrel Tracker** | | |
| Technology | TPC (Silicon strips) | Silicon strips |
| Inner radius | 329 (165) | 230 |
| Outer radius | 1808 (1835) | 1239 |
| Max. Z | 2250 | 578 to 1536 |
| Max. samples | 2 (Si), 224 (TPC), 1 (Si) | 5 |
| **Forward Tracker** | | |
| Technology | Silicon strips | Silicon strips |
| Inner radius | 47 to 218 | 207 to 1162 |
| Outer radius | 320 | 1252 |
| Max. Z | 1868 | 1556 |
| Max. samples | 5 | 4 |
| **ECAL: Barrel** | | |
| Absorber | Tungsten | Tungsten |
| Active elements | Silicon pads | Silicon pads |
| Sampling layers | 30 (20 × 2.1, 10 × 4.2) | 30 (20 × 2.5, 10 × 5) |
| Cell size | 5.1 × 5.1 | 3.5 × 3.5 |
| $X_0$ and $\lambda_I$ | 23 and 1 | 26 and 1 |
| Inner radius | 1847 | 1290 |
| Outer radius | 2020 | 1430 |
| Max. Z | 2350 | 1765 |

Table 3.2: continued

| | CLIC_ILD | CLIC_SiD |
|---|---|---|
| **ECAL: Endcap** | | |
| Absorber | Tungsten | Tungsten |
| Active elements | Silicon pads | Silicon pads |
| Sampling layers | 30 (20 × 2.1, 10 × 4.2) | 30 (20 × 2.5, 10 × 5) |
| Cell size | 5.1 × 5.1 | 13 $mm^2$ hexagons |
| $X_0$ and $\lambda_I$ | 23 and 1 | 26 and 1 |
| Inner radius | 270 | 222 |
| Outer radius | 2270 | 1269 |
| Min. Z | 2450 | 1657 |
| Max. Z | 2622 | 1800 |
| **HCAL: Barrel** | | |
| Absorber | Tungsten | Tungsten |
| Sampling layers | 75 × 10 mm | 75 × 10 mm |
| Cell size | 30 × 30 | 30 × 30 |
| $\lambda_I$ | 7.5 | 7.5 |
| Inner radius | 2058 | 1447 |
| Outer radius | 3296 | 2624 |
| Max. Z | 2350 | 1765 |
| **HCAL: Endcap** | | |
| Absorber | Steel | Steel |
| Sampling layers | 60 × 20 mm | 60 × 20 mm |
| Cell size | 30 × 30 | 30 × 30 |
| $\lambda_I$ | 7.5 | 7.5 |
| Inner radius | 400 | 509 |
| Outer radius | 3059 | 2624 |
| Min. Z | 2650 | 1800 |
| Max. Z | 4240 | 3395 |
| **Coil + cryostat** | | |
| Field on central axis | 4 T | 5 T |
| Free bore | 3426 | 2744 |
| Outer radius | 4290 | 3710 |
| Max. Z | 4175 | 3245 |
| **Yoke & Muon System: Barrel** | | |
| Material | Steel | Steel |
| Inner radius | 4404 | 3914 |
| Outer radius | 6990 | 7000 |
| Number of layers | 9 | 9 |
| **Yoke & Muon System: Endcap** | | |
| Material | Steel | Steel |
| Inner radius | 690 | 690 |
| Outer radius | 6990 | 7000 |
| Max. Z | 6200 | 6200 |
| Number of layers | 9 | 9 |

### 3.6 Preparations towards a cost estimate of CLIC Detectors

The estimated cost of future CLIC detectors will be included in a forthcoming document, together with the value estimates for the CLIC accelerator complex. As a first step towards such an estimate, the methodology to be used has been defined and is described in Appendix C. It presents a first look at the relative contributions to the cost from differentsub-detectors and the sensitivity of the cost to the price of a few key materials.

# Chapter 4

# Vertex Detectors

## 4.1 Introduction

The pixel vertex detectors of CLIC_ILD and CLIC_SiD are designed to be integral parts of their respective tracking systems, increasing the precision and efficiency of the track reconstruction in particular for low-momentum tracks. Their primary objective is to deliver efficient tagging of decays involving heavy quarks or tau-leptons by reconstructing their displaced vertices. For the vertex detector mechanical construction, five primary, coupled considerations guide the design. The CLIC vertex detector must have excellent spacial resolution, full geometrical coverage extending to low polar angles $\theta$, extremely low mass, low occupancy facilitated by time-tagging, and sufficient heat removal from sensors and readout. These considerations, together with the physics needs and beam structure of CLIC, push the technological requirements to the limits and imply a very different vertex detector than the ones currently in use elsewhere.

In this chapter, emphasis is put on an improved understanding of the CLIC vertex detector through simulation studies. These studies were carried out to investigate the dependence of the flavour-tagging and tracking precision on the detector layout and on its design options. The results demonstrate that the CLIC vertex detectors require very small pixels when compared with those at hadron colliders, as well as complex on-chip readout and ultra-thin materials. As the vertex detectors are located closest to the interaction point, beam-induced background rates are very high. This leads to high pixel occupancies and the necessity for the vertex detector to contribute to the separation of physics hits from backgrounds hits through precise time-tagging capabilities.

None of the existing technologies are able to fulfil all of the challenging requirements derived from the physics goals and from the constraints given by the running conditions. Therefore at this time it is not possible to chose a specific sensor and readout technology. However, there are several R&D programmes in the silicon-pixel sensor and readout domains as well as for ultra-low mass supports and cooling systems, addressing each of these requirements.

Besides the detector simulations, this chapter addresses integration, assembly and maintenance scenarios for the vertex detector. It also draws possible technology roadmaps that will lead to future implementations of a CLIC vertex detector.

## 4.2 Physics Requirements

To identify heavy-flavour quark states and tau-leptons with high efficiency, a precise measurement of the impact parameter point and of the charge of the tracks originating from the secondary vertex is required. Monte Carlo simulations show that these goals can be met with a constant term in the transverse impact-parameter resolution of $a \approx 5$ µm and a multiple-scattering term of $b \approx 15$ µm, using the canonical parametrisation:

$$\sigma(d_0) = \sqrt{a^2 + b^2 \cdot \mathrm{GeV}^2/(p^2 \sin^3\theta)}. \tag{4.1}$$

These requirements on the measurement precision exceed the results achieved in any of the currently existing full-coverage vertex systems. Simulation results show that a single-point resolution of approximately 3 µm and a material budget of $X < 0.2\%\ X_0$ for the beam pipe and for each of the detection layers are sufficient to reach the required impact-parameter resolution. This material budget corresponds to approximately 200 µm of silicon per layer, including the readout and mechanical support structures. The single-point resolution target of approximately 3 µm can be achieved with pixels of 20 µm $\times$ 20 µm using an analog signal readout.

### 4.3 Simulation Layouts

The *vertex region* is defined here as the volume equipped with pixelated silicon detection layers, whereas the *outer tracking region* refers to the surrounding silicon strip layers and the TPC.

For the current simulation studies we assume that the vertex detectors of the two concepts will be similar. For the geometrical arrangements however, complementary choices have been made where applicable and where dictated by the difference in the magnetic field and the surrounding tracking system.

The vertex detector layers have to fit inside the gap between the beam tube and the surrounding outer tracking detectors. Figures 4.1 and 4.2 show sketches of the CLIC_ILD and CLIC_SiD vertex region simulation layouts, respectively. The main parameters are given in Table 4.1. A detailed description of the simulation layouts can be found elsewhere [1, 2, 3]. The vertex detector models are implemented in the full GEANT4-based simulation [4] of the two concepts (MOKKA [5] for CLIC_ILD and SLIC [6] for CLIC_SiD), as well as in a fast simulation setup based on the LiC Detector Toy simulation and reconstruction framework (LDT) [7].

The inner radius of the central beryllium beam pipe is constrained by the high rates of background from incoherent electron-positron pairs produced in beam-beam interactions coupled with bending in the magnetic field (see Section 4.5.1). For CLIC_ILD ($B = 4$ T), an inner radius of 29.4 mm is chosen, while the larger magnetic field of 5 T in CLIC_SiD allows a reduced radius of 24.5 mm. The sensor layers are surrounding the beam pipes in close proximity, to minimise the effect of multiple scattering and to maximise the acceptance in the forward region.

For CLIC_ILD, a double-layer structure is chosen for both the three barrel vertex layers and the three innermost forward layers. The barrel layers radially extend from 31 to 60 mm. The forward layers extend up to $z = 257$ mm. The two rectangular sensors of each barrel double layer are mounted on either side of a flat central module support structure that extends in $z$ over one half of the barrel vertex region. Full azimuthal coverage is achieved with 13 to 18 overlapping modules per double layer. For the vertex endcap disks, the same double-layer structure as in the barrel is implemented, though without segmentation in $\phi$. The total material budget per double layer is $X/X_0 = 0.18\%$. The double-layer structure minimises the amount of support material and allows for a compensation of deformations perpendicular to the silicon planes. Furthermore, the correlation between close-by hits in the two layers can be used to suppress background not originating from the interaction region. A possible draw back of the double-layer concept is related to highly boosted heavy states decaying after the first double layer. In this case only the original particle may leave hits in the first double layer and therefore the identification of the tracks pointing to the secondary vertex will be less efficient than for a more spread-out distribution of the layers. Further studies are needed to confirm the feasibility and effectiveness of the double-layer layout for CLIC_ILD.

For the CLIC_SiD simulation, a layout with 5 single pixel layers for the barrel and 7 single pixel layers for the endcap disks is chosen. The total material budget per single layer is $X/X_0 = 0.11\%$. The pixel layers in CLIC_SiD cover a larger volume than in CLIC_ILD, in particular for the forward region. The radial extension of the barrel layers is from 27 to 67 mm. The forward layers extend up to $z = 830$ mm. Segmentation in $\phi$ is implemented in the CLIC_SiD simulation framework both for the barrel and the forward region through 18 to 36 rectangular modules and 16 wedge-shaped modules, respectively.

The sensor and readout technology is assumed to be identical for both concepts, with an active thickness of each silicon-sensor layer of 50 µm. The pixel pitch is 20 µm × 20 µm and a single-point resolution of 2.8 µm in both dimensions is assumed, corresponding to an analog readout of the induced signals.

Figure 4.3 shows a comparison between the full GEANT4 and the LDT fast simulation models for the integrated amount of material (including supports and cabling) in the vertex regions of CLIC_ILD and CLIC_SiD. The total amount of material at $\theta = 90\ ^\circ$ is about 0.9% $X_0$ for CLIC_ILD and 1.1% $X_0$ for CLIC_SiD. In the forward region it rises to approximately 3% $X_0$ for CLIC_ILD and 5% $X_0$ for CLIC_SiD. The fast simulation models resemble the main features of the full GEANT4 simulation models of the CLIC_ILD and CLIC_SiD vertex detectors.

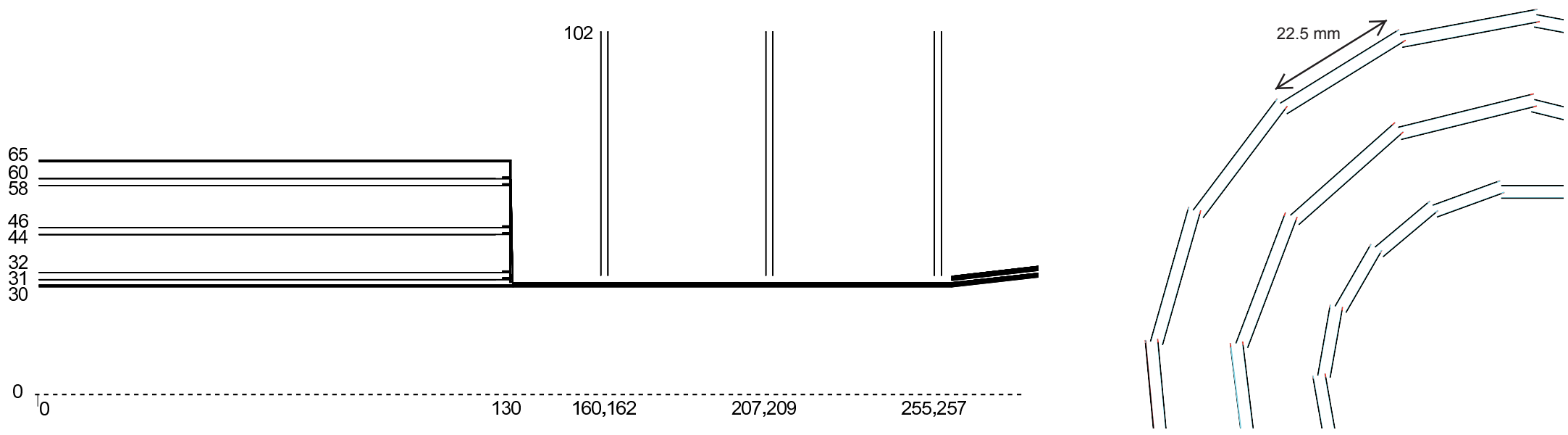

(a) (b)

Fig. 4.1: Sketch of the barrel and forward vertex region of the CLIC_ILD simulation model in the *z*-*r* plane (a) and of the barrel vertex region in the *x*-*y* plane (b). Shown are the double layers of the vertex barrel detectors and of the vertex endcap disks together with the barrel support shell and the central beam pipe. The dashed line corresponds to the detector axis. All values are given in millimetres.

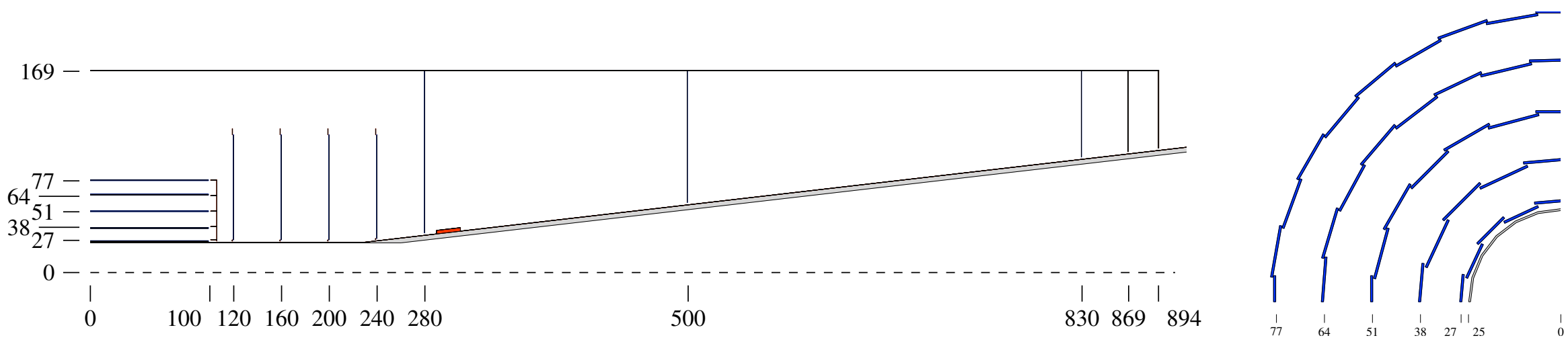

(a) (b)

Fig. 4.2: Sketch of the barrel and forward vertex region of the CLIC_SiD simulation model in the *z*-*r* plane and of the barrel vertex region in the *x*-*y* plane. Shown are the vertex barrel layers, the vertex endcap disks and the forward tracking disks together with the vertex support, cabling and the central beam pipe. The dashed line corresponds to the detector axis. All values are given in millimetres.

## 4.4 Performance Optimisation Studies

The vertex detector performance has been evaluated for the baseline configurations of both concepts in full GEANT4 simulations [8, 9] as well as in fast LDT simulations [10]. In addition, the fast simulation setup was used for geometry optimisation studies and to evaluate the sensitivity of the results on the chosen parameters. The main performance measure was the impact-parameter resolution projected in the transverse plane $\sigma(d_0)$, which is closely linked to the flavour-tagging capability of the detectors, as described in Section 12.3.4. Assessing the impact of the detector geometries and material budgets on the flavour-tagging performance requires dedicated full-simulation studies and will be subject of future R&D.

Table 4.1: Main parameters of the CLIC_ILD and CLIC_SiD vertex region layouts.

| | CLIC_ILD | CLIC_SiD |
|---|---|---|
| Central beam pipe | Beryllium | |
| | $R_i = 29.4$ mm | $R_i = 24.5$ mm |
| | $d = 0.6$ mm | $d = 0.5$ mm |
| Barrel region | 3 double layers | 5 single layers |
| | $\|z\| < 130$ mm | $\|z\| < 98.5$ mm |
| | $R_i = 31,\ 44,\ 58$ mm | $R_i = 27,\ 38,\ 51,\ 64,\ 77$ mm |
| Forward region | 3 double layers | 7 single layers |
| | $z = 160, 207, 255$ mm | $z = 120, 160, 200, 240,$ |
| | | $280, 500, 830$ mm |
| Sensors | 20 µm × 20 µm, $\sigma_{sp} \approx 3$ µm | |
| | $X/X_0 = 0.18\%$ | $X/X_0 = 0.11\%$ |
| | per double layer | per single layer |
| Surface area | 0.736 $m^2$ | 1.103 $m^2$ |
| Number of channels | $1.84 \times 10^9$ | $2.76 \times 10^9$ |

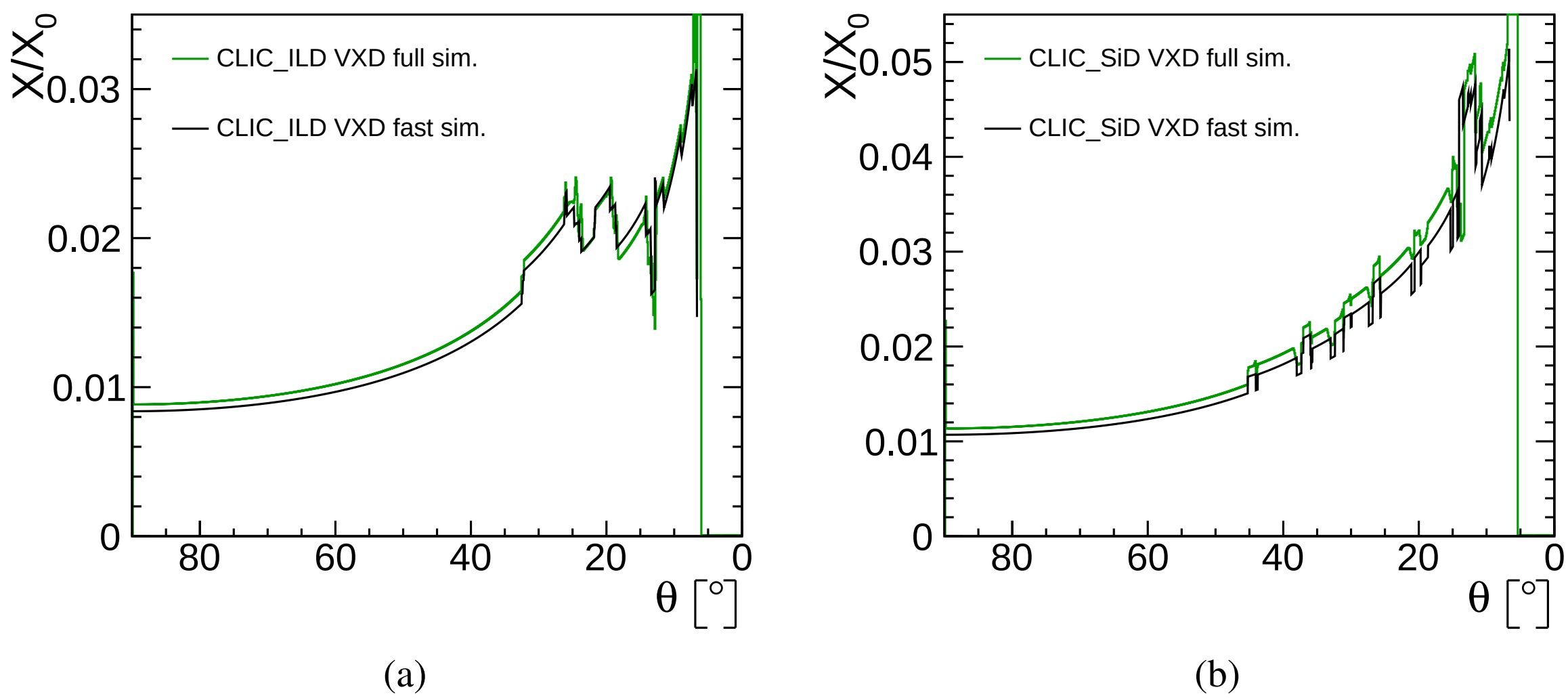

Fig. 4.3: Amount of material within the vertex detector regions for the fast simulation and for a full GEANT4 simulation model of the CLIC_ILD (a) and of the CLIC_SiD (b) detector. Shown is the integrated fraction of a radiation length $X/X_0$ versus the polar angle $\theta$.

### 4.4.1 Performance of the Baseline Configurations

Figure 4.4 shows the transverse impact-parameter resolutions obtained with the baseline configuration for both CLIC_ILD and CLIC_SiD, for isolated muon tracks with momenta of 1, 10 and 100 GeV. The results obtained with the fast LDT simulation setup are compared to the ones from the full GEANT4 simulations. Good agreement is observed for CLIC_ILD, as expected due to the fact that both the full and fast simulation perform a simple Gaussian hit smearing with very similar parameters, to obtain the measurement points in the silicon layers. For CLIC_SiD, a more realistic parametrisation of the charge development and sharing in the silicon sensors is performed during the digitisation phase in the full simulation, resulting in a cluster-size dependent single-point resolution. The fast simulation, on the other hand, uses the same single-point resolution parameters for the pixel layers as for CLIC_ILD. Therefore only qualitative agreement is found between fast and full simulation for CLIC_SiD.

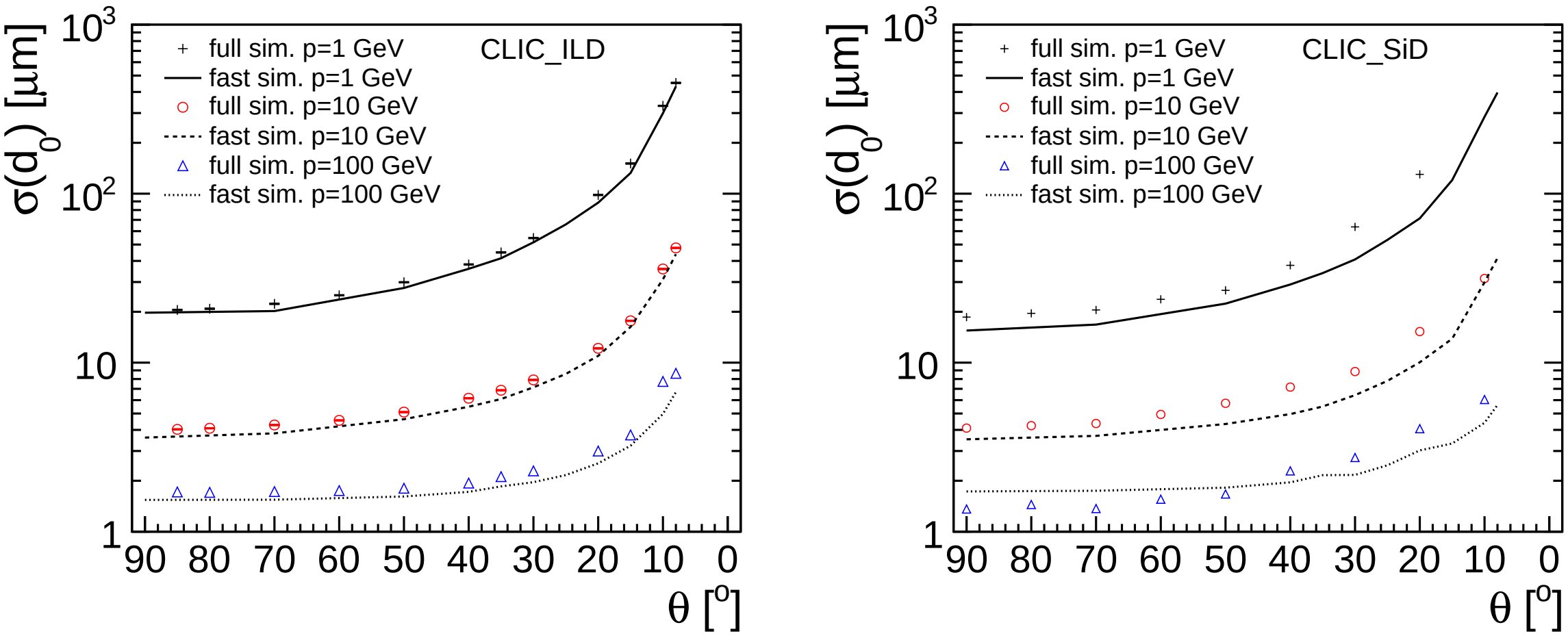

(a) CLIC_ILD (b) CLIC_SiD

Fig. 4.4: Transverse impact-parameter resolutions, obtained with the baseline vertex detector layouts for CLIC_ILD (a) and for CLIC_SiD (b), for tracks with momenta of 1, 10, and 100 GeV. The markers correspond to the full GEANT4 detector simulations, while the lines give the results for the fast LDT simulation. The differences between the simulation results are discussed in the text.

### 4.4.2 Dependence on Single-Point Resolution

The dependence of the transverse impact-parameter resolution on the pixel size was studied by varying the single-point resolution for the simulation of the CLIC_SiD vertex layers by $\pm 1.4$ μm w.r.t. the baseline value of 2.8 μm, corresponding approximately to pixel sizes of 10, 20 and 30 μm. The resulting resolutions for high and low track momenta as function of the polar angle $\theta$ are shown in Figure 4.5. The resolution for track momenta of 100 GeV is found to change by approximately $\pm 40\%$ in the barrel region. Here they exceed the target value for the high-momentum limit of $a \approx 5$ μm for all three pixel sizes, as expected from the corresponding single-point resolutions. For 1 GeV, where multiple-scattering effects dominate, the corresponding variation of the transverse impact-parameter resolution is only $\pm 15\%$. The target value for the multiple-scattering term of $b \approx 15$ μm is approximately reached for all three pixel sizes. It should be noted, however, that the pixel size is also constrained by the background occupancies (see Section 4.5.2) and the ability to separate adjacent tracks in very dense jets in the presence of such backgrounds.

### 4.4.3 Dependence on Arrangement of Layers

The performance of the vertex detector was tested for different geometrical arrangements of the detection layers:

#### *4.4.3.1 Distance to the IP*

The distance of the central beam pipe and barrel vertex layers to the IP was varied by the same amount for all layers of the CLIC_ILD vertex detector. Figure 4.6 shows the resulting transverse impact-parameter resolutions at $\theta = 90^\circ$ as function of the radial distance of the innermost barrel vertex layer to the IP. The dashed line corresponds to the baseline configuration. The sensitivity is about $3.2\%/\text{mm}$ for high momenta (parameter $a$) and $0.8\%/\text{mm}$ for low momenta (parameter $b$).

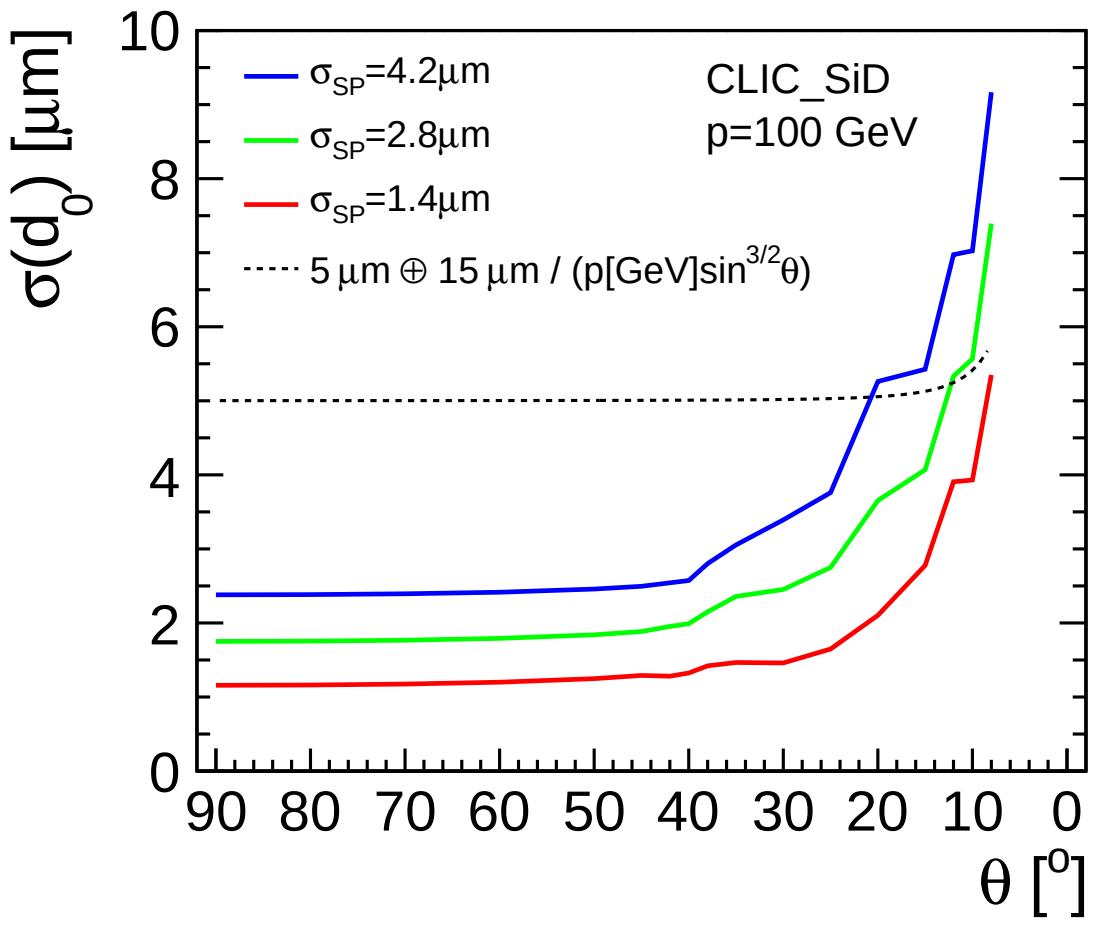

(a)

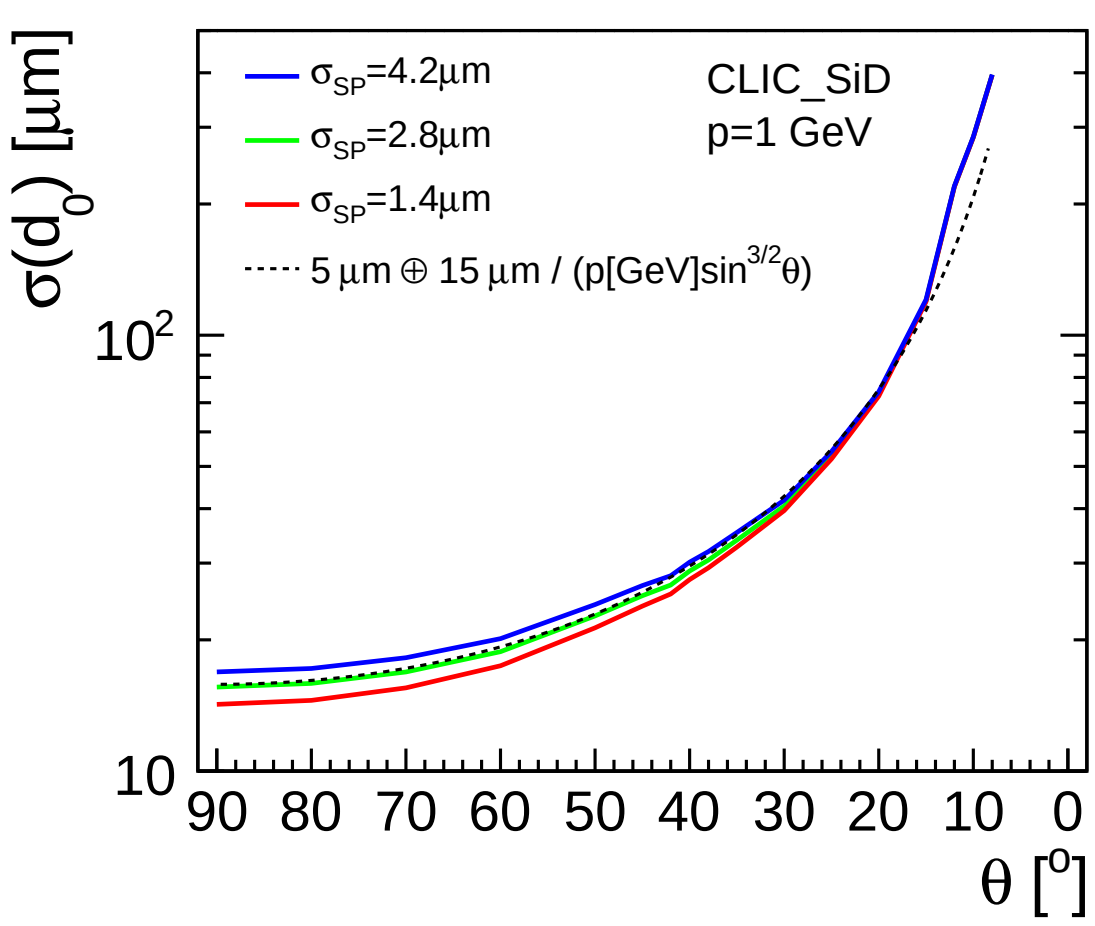

(b)

Fig. 4.5: Transverse impact-parameter resolution as function of the polar angle $\theta$ for three different values of the single-point resolution of the CLIC_SiD pixel layers, as obtained from the fast LDT simulation. Shown are the resolutions for 100 GeV tracks (a) and for 1 GeV tracks (b). Also shown is the parametrisation from Equation 4.1 with the target values $a = 5$ µm and $b = 15$ µm.

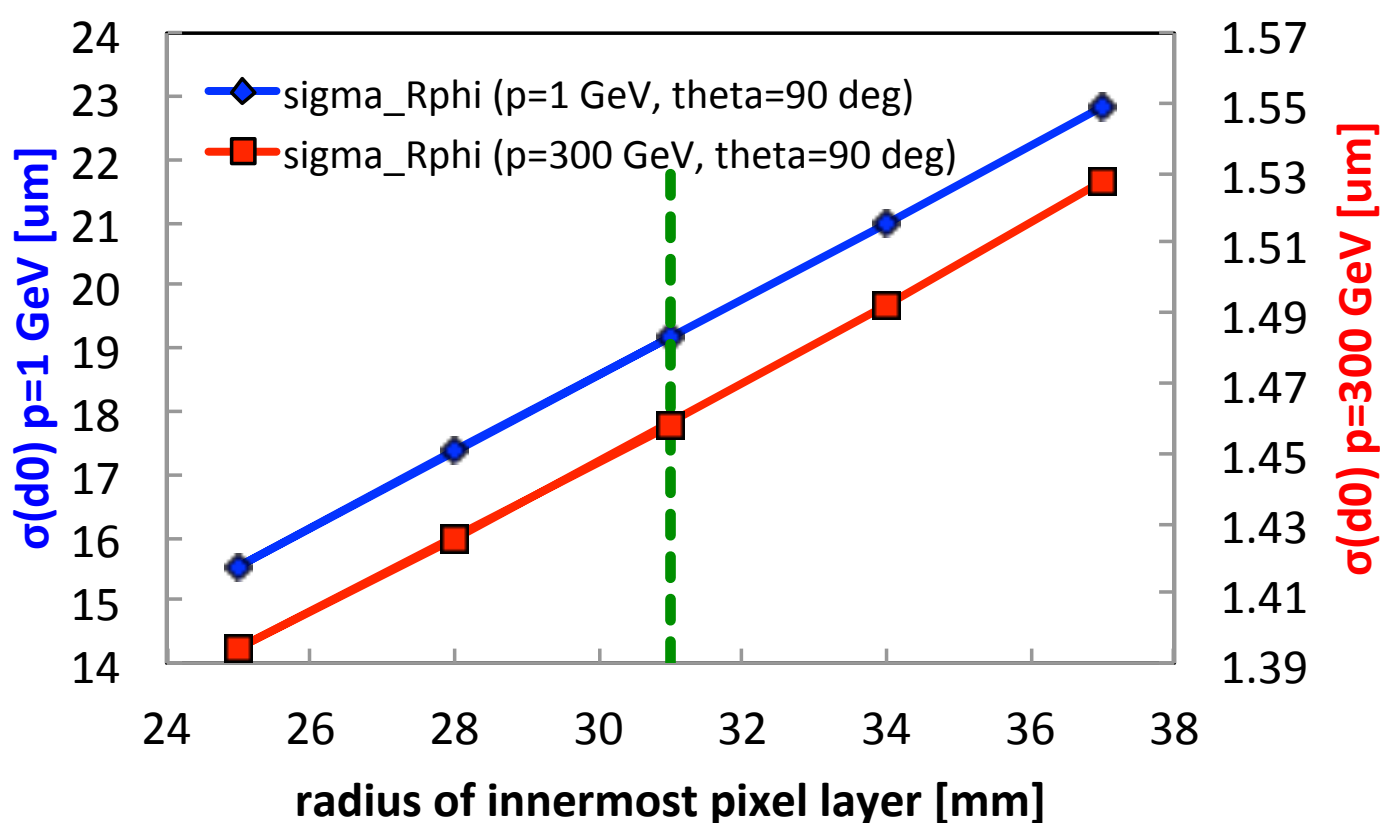

Fig. 4.6: Transverse impact-parameter resolution at $\theta = 90^\circ$ as function of the radius of the innermost pixel layer in the CLIC_ILD model for 1 GeV tracks and for 300 GeV tracks, as obtained from the fast LDT simulation. The radius corresponding to the baseline configuration is indicated with a dashed line.

#### *4.4.3.2 Length of barrel section*

A configuration with an extended length of the barrel layers has been evaluated for the CLIC_SiD vertex detector. For this configuration the first two of the seven endcap pixel disks were removed and the barrel layers were extended accordingly, increasing also the amount of material for cables and support at the barrel endcaps. It was found that a longer barrel section slightly improves the $p_T$ and impact-parameter resolutions in the very forward region ($\theta < 15^\circ$), due to the reduced amount of material before the first detection layers. For the intermediate region ($40^\circ < \theta < 50^\circ$) however, the $p_T$ resolution becomes worse, due to the missing first pixel disk in this region.

#### *4.4.3.3 Cable routing*

In the baseline simulation setup the cable and services routing is foreseen along the beam pipe for both the barrel and endcap disk layers. In this scheme the tracks entering the lower end of the endcap disks have to pass the cables and services for the barrel layers. An alternative routing scheme has been simulated for CLIC_ILD, with the barrel cables and services routed radially up to the first inner-tracking layer, then outwards in $z$ up to the first single forward layer and down in front of this first single layer to the beam pipe. The latter routing improves the transverse impact parameter resolution for the polar-angle region from 12° to 7°. Further studies are needed as soon as more realistic material estimates for the cables and services become available (see also Section 4.6 below).

### 4.4.4 Material Budget

The baseline designs include very small material budgets for the beam pipe as well as for the sensor layers and their support. To assess the sensitivity of the performance on the amount of material, the material budget for the beam pipe and for the detection layers of the CLIC_ILD vertex detector has been varied. The resulting transverse impact-parameter resolutions for low-momentum tracks are shown in Figure 4.7. When increasing the thickness of the beam pipe by a factor of two, the resolution for tracks at $\theta = 90°$ increases by approximately 20%. The same sensitivity is observed for doubling the material in all vertex layers.

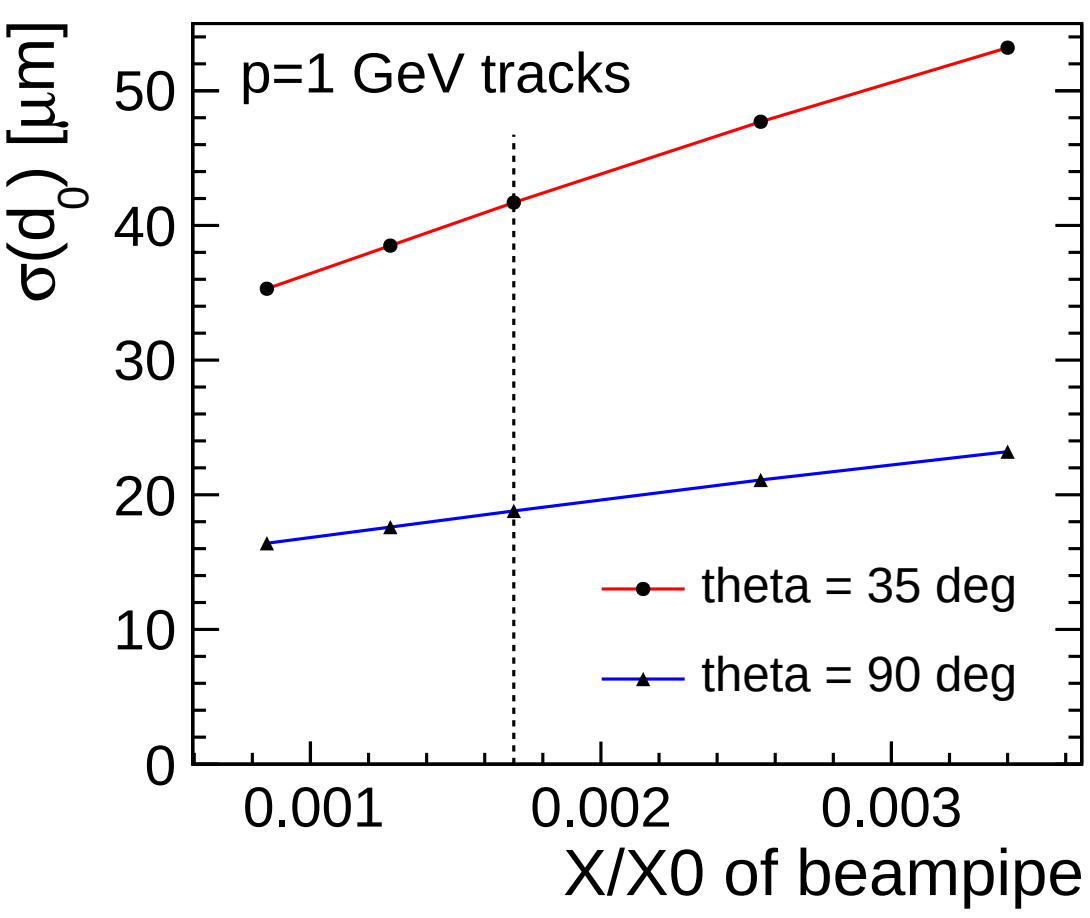

(a) Material in beam pipe

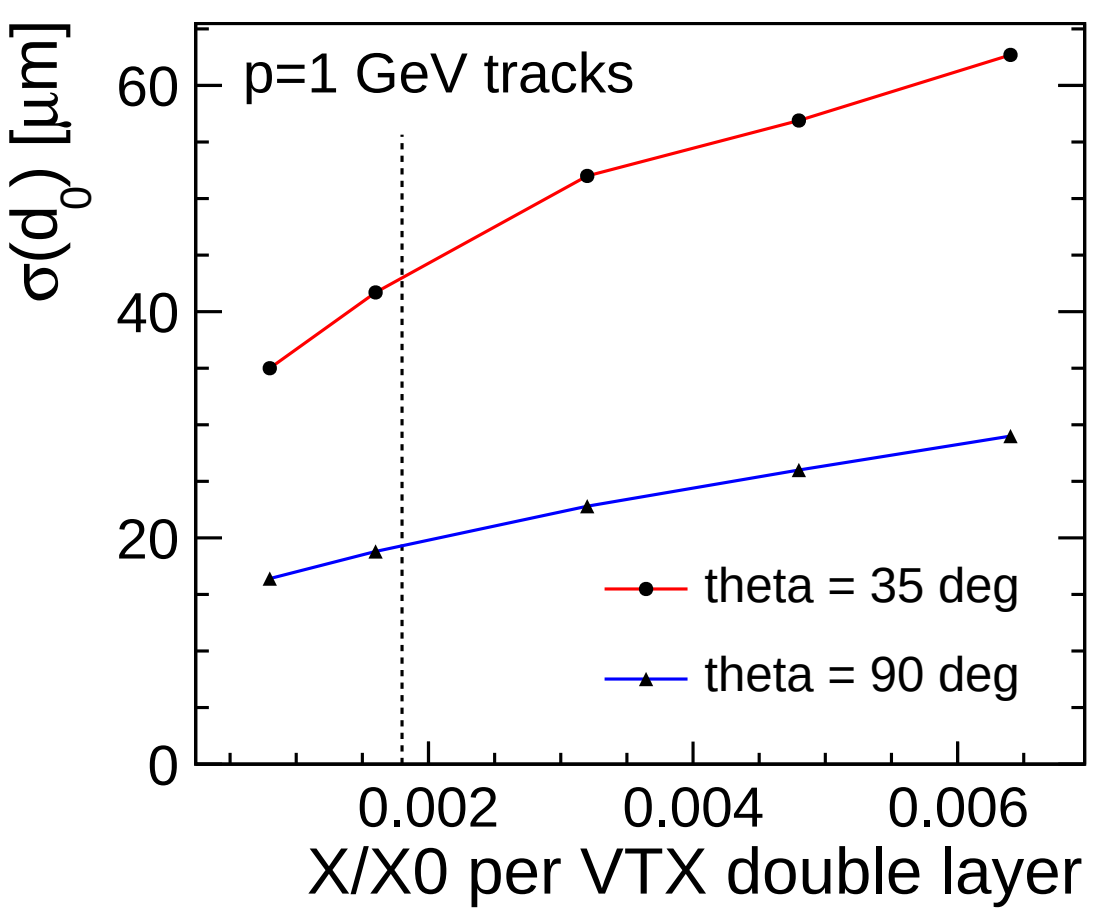

(b) Material in barrel pixel layers

Fig. 4.7: Transverse impact-parameter resolution of the CLIC_ILD vertex detector as function of the amount of material inside the beam pipe (a) and inside the vertex barrel double layers (b), as obtained from the fast LDT simulation. The results are shown for 1 GeV tracks and for polar angles of $\theta = 35$ ° and of $\theta = 90$ °. The material budget corresponding to the baseline configuration are indicated by dashed lines.

## 4.5 Beam-Induced Backgrounds in the Vertex Detector Region

The incoherent creation of electron-positron pairs and the production of hadrons in $\gamma\gamma$ interactions are expected to be the dominating sources of background events originating from the interaction region. These processes have been studied at generator level and with a full simulation of the detector response [11, 12]. In particular the detailed modelling of the very forward detector region allows for a realistic assessment of secondary effects needed to optimise the beam-pipe geometry in view of particles backscattering from the LumiCal and BeamCal into the vertex detector region (see Chapter 9). The studies have focused on the CLIC_ILD detector model and the obtained results have been transferred to the CLIC_SiD detector model and confirmed there.

### 4.5.1 Beam-Pipe Layout and Design

The beam pipe should provide good vacuum at the interaction point, remain outside the background envelope near the interaction region, allow for the placement of silicon elements as close to the beam line as possible, present a low number of radiation lengths for trajectories of interest, and shield against backgrounds originating upstream and downstream of the vertex detector region. It should be noted that the vacuum quality is not critical and that therefore bake-out of the vacuum system is not required inside the interaction region. The design that was developed for the beam pipes is shown in Figures 4.1, 4.2 and 4.8. A straight, beryllium portion near the interaction region minimises the number of radiation lengths before vertex detector elements. Stainless steel conical portions with a wall thickness of 4 mm extend in the forward and backward directions and provide shielding against backscattering upstream and downstream backgrounds. Figure 4.8 shows the density of direct hits in the 4 T field of the CLIC_ILD vertex detector region. A cut-off with a parabolic shape can clearly be seen. With a length of 520 mm and an inner radius of 29.4 mm, the cylindrical section of the beam pipe is located safely outside the region of high hit density, where the production of secondary hits would lead to unacceptably large occupancies in the detectors. Similarly, for a magnetic field of 5 T, the inner radius of the CLIC_SiD central beam pipe was set to 24.5 mm, with a length of 460 mm.

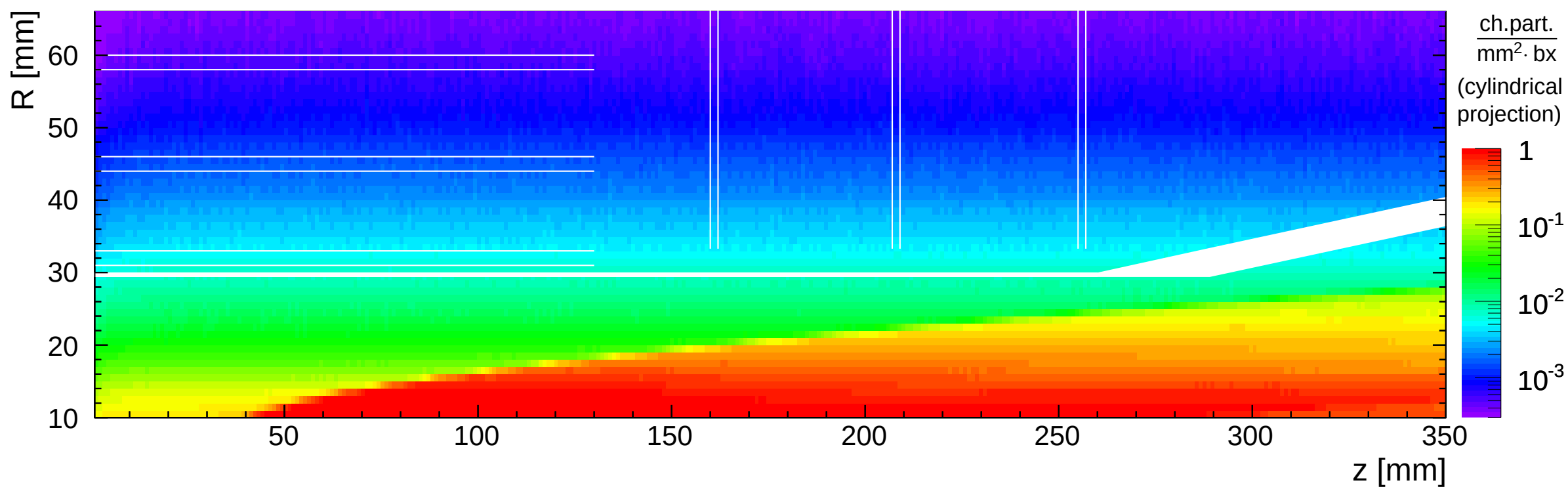

Fig. 4.8: Density of direct hits from incoherent pairs in a cylindrical projection of the vertex detector region of the CLIC_ILD detector. The position of the beam pipe and of the innermost barrel and forward pixel layers are indicated with white lines.

The beryllium wall thickness must be sufficient to address porosity, to resist collapse under vacuum, and to resist forces and moments transmitted from the conical portions. A wall thickness of 0.6 mm was assumed for CLIC_ILD; the corresponding value for CLIC_SiD is 0.5 mm. Those wall thicknesses are conservatively high for vacuum collapse, which depends primarily on elastic modulus, Poisson's ratio, radius, and length. The designs of the Tevatron Run IIb beam pipes demonstrated that a thickness of 0.5 mm is sufficient to address porosity issues [13]. Local stress concentrations will occur at the transition joints to conical pipe portions if abrupt changes in material thickness are allowed. Those concentrations can be minimised by an optimised design of the joint region. R&D in conjunction with potential vendors is expected to be conducted on the transition joints and beam-pipe fabrication methods. A liner of titanium of thickness 25 to 50 µm, may be required for the ILC central beam pipes [14]. However, simulation studies for incoherent-synchrotron radiation originating from the beam-delivery system at CLIC indicate that the radiation envelopes stay within ±5mm for $15\sigma_x$ and $55\sigma_y$ [15]. The impact of radiation emitted in the beam-delivery system needs further studies.

### 4.5.2 Hit Densities in the Vertex Region

Figure 4.9 (a) shows the expected hit densities in the CLIC_ILD barrel vertex detectors for particles originating from incoherent electron-positron pairs and from $\gamma\gamma \to$ hadrons. The results from the full simulation are in good agreement with a fast parametric simulation of the contribution from direct hits. Backscatters from the forward region are largely suppressed by the conical beam pipe sections made of 4 mm thick stainless steel. A description of the forward-region optimisation studies can be found in [12]. The incoherent pairs dominate at small radii and lead to up to $6 \cdot 10^{-3}$ hits/mm$^2$/BX in the innermost barrel layer. The corresponding hit densities in the endcap disk region are shown in Figure 4.9 (b). Also here the hit density from incoherent pairs is, for the inner edge of the disks, an order of magnitude above the one from $\gamma\gamma \to$ hadrons, reaching up to $9 \cdot 10^{-3}$ hits/mm$^2$/BX.

The results do not include safety factors for the uncertainties in the production cross sections, the two-photon luminosity spectrum and the simulation of the detector response. Furthermore they only describe the number of particles traversing the detector, not taking into account the formation of clusters of pixels due to charge spreading and sharing. For the $\gamma\gamma \to$ hadrons background, an overall safety factor of two is sufficient, to take into account the uncertainties on the predicted rate [11]. For the incoherent pairs, backscattering effects in the forward region are of particular importance. Therefore a larger overall safety factor of five should be used [12]. Assuming these safety factors for the simulation uncertainties and an average cluster size of 5 pixels per hit, the resulting maximal occupancy per pixel for the innermost vertex barrel layers during a bunch train of 312 bunch crossings is 1.9%. The corresponding occupancy for the forward vertex pixel layers is 2.9%.

Such high occupancies pose challenging demands on the track- and impact-parameter-finding algorithms. Time stamping of the hits in the vertex detector with a resolution of 5–10 ns will help to reduce possible confusion of the hit assignment to track segments in dense jets and to aid the matching of tracking and calorimeter information especially in the forward region.

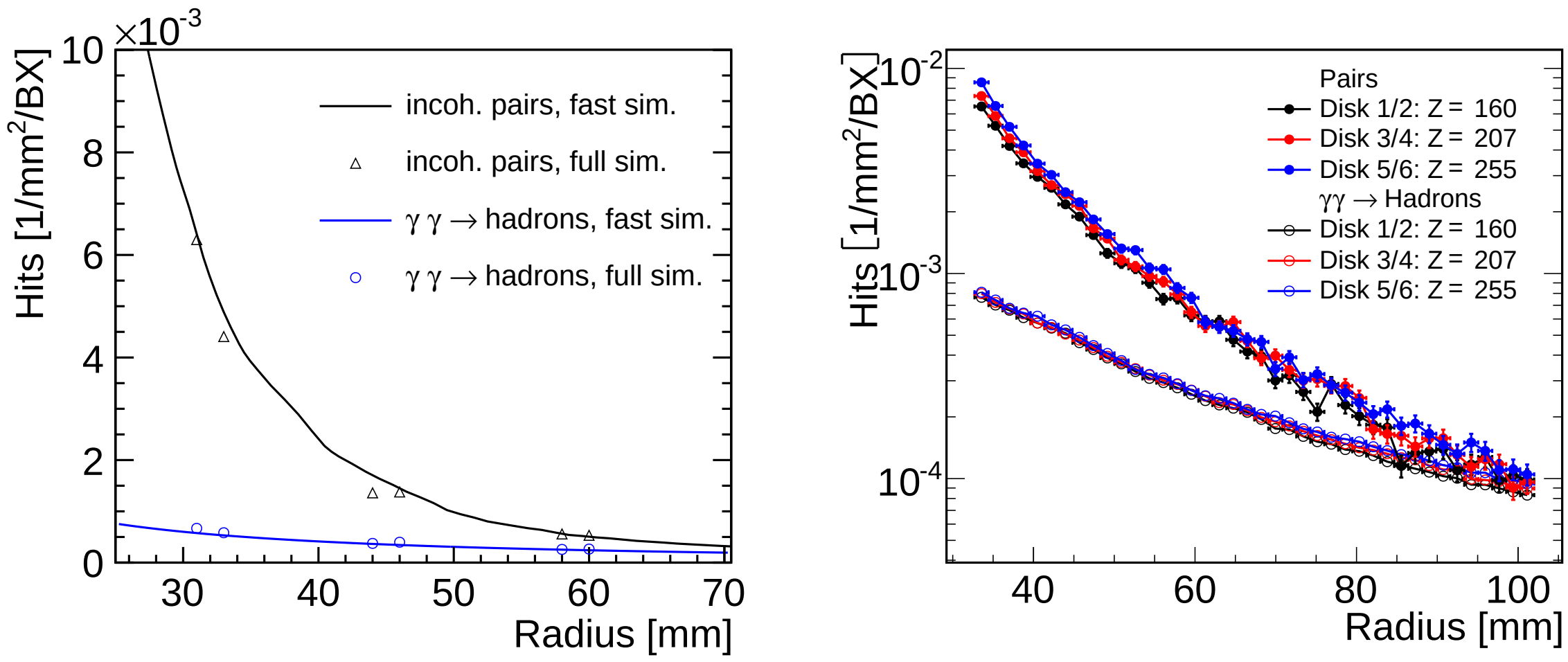

(a) (b)

Fig. 4.9: Average hit densities in the CLIC_ILD barrel (a) and forward (b) vertex detectors for particles originating from incoherent electron-positron pairs and from $\gamma\gamma \to$ hadrons. For the barrel region, the full simulation of the detector response is compared to a fast parametric tracking of the primary particles in the magnetic field. For the forward region, the results are shown for the full simulation only. Safety factors for the simulation uncertainties and cluster formation are not included.

Table 4.2: Expected radiation damage (NIEL and TID) from incoherent pairs and $\gamma\gamma \rightarrow$ hadrons for the barrel pixel sensors (VXB 1–6) and for the lower end of the endcap pixel disks (VXEC 1–6) of the CLIC_ILD detector model. The numbers are quoted without safety factors for simulation uncertainties.

| | Radius [mm] | Pairs NIEL [$10^9 n_{eq}/cm^2/yr$] | Hadr. NIEL [$10^9 n_{eq}/cm^2/yr$] | Pairs TID [Gy/yr] | Hadr. TID [Gy/yr] |
|---|---|---|---|---|---|
| VXB 1 | 31.0 | 3.87 | 11.51 | 39.43 | 4.57 |
| VXB 2 | 33.0 | 2.88 | 8.57 | 27.83 | 4.01 |
| VXB 3/4 | 44.0 | 0.99 | 4.60 | 8.01 | 2.46 |
| VXB 5/6 | 58.0 | 0.45 | 2.92 | 3.30 | 1.66 |
| VXEC 1/2 | 33.6 | 6.17 | 5.64 | 27.99 | 3.10 |
| VXEC 3/4 | 33.6 | 6.72 | 5.79 | 29.25 | 2.96 |
| VXEC 5/6 | 33.6 | 7.83 | 6.14 | 34.12 | 3.13 |

### 4.5.3 Radiation Damage

The high rates of beam-induced backgrounds will be the dominating source of radiation-induced damage in the silicon detectors.

To estimate the expected radiation damage from non-ionising energy loss (NIEL), the hit rates obtained with the full detector simulation have been scaled with a tabulated displacement-damage factor based on the type and energy of the corresponding particle, resulting in the equivalent flux of 1 MeV neutrons leading to the same displacement damage as for the observed spectrum [16]. The total ionising dose (TID) was obtained from the energy deposited in the silicon layers. The TID and NIEL per year of detector operation are obtained assuming an effective runtime of 100 days per year at the nominal luminosity.

Table 4.2 summarises the expected radiation damage per year of detector operation from NIEL and from TID for the pixel layers of the CLIC_ILD detector. The incoherent electron-positron pairs are the dominating source for the total ionising dose. The NIEL damage is dominated by the $\gamma\gamma \rightarrow$ hadrons background for the barrel region. For the forward region, the $\gamma\gamma \rightarrow$ hadrons and incoherent electron-positron pairs contribute equally to the NIEL damage. The quoted numbers do not include safety factors for the simulation uncertainties. Assuming an overall safety factor of two for the $\gamma\gamma \rightarrow$ hadrons background and of five for the incoherent electron-positron pairs, one obtains a maximum flux in the inner vertex layers of about $4 \cdot 10^{10} n_{eq}/cm^2/yr$ and a total ionising dose of up to 200 Gy/yr.

## 4.6 Integration, Assembly and Access Scenarios

### 4.6.1 Assembly and Integration

In the CLIC_SiD concept, the vertex detector is fabricated in two halves so that it can be assembled around the beam pipe. A double-walled outer cylinder made of carbon fibre laminate supports and positions the vertex detector barrels and disks. The outer cylinder itself is supported by the conical portions of the beam pipe and holds the central portion of the beam pipe straight. To do that, load is transferred between the beam pipe and outer cylinder via carbon fibre disks at four longitudinal locations as indicated in Figures 4.10 and 4.11. The outer cylinder also provides longitudinal passages for the distribution of cooling gas.

Each half-layer of the vertex detector barrel is fabricated as a unit. Depending on the sensors chosen and the size in which sensors can be made, sensors will either be glued to one another along their long edges or a carbon fibre framework will be provided to which they can be glued. Carbon fibre end rings complete a half-layer and control its roundness. Completed half layers are held to one another and from the outer cylinder by spoked carbon fibre membranes at each layer end as shown in Figure 4.11.

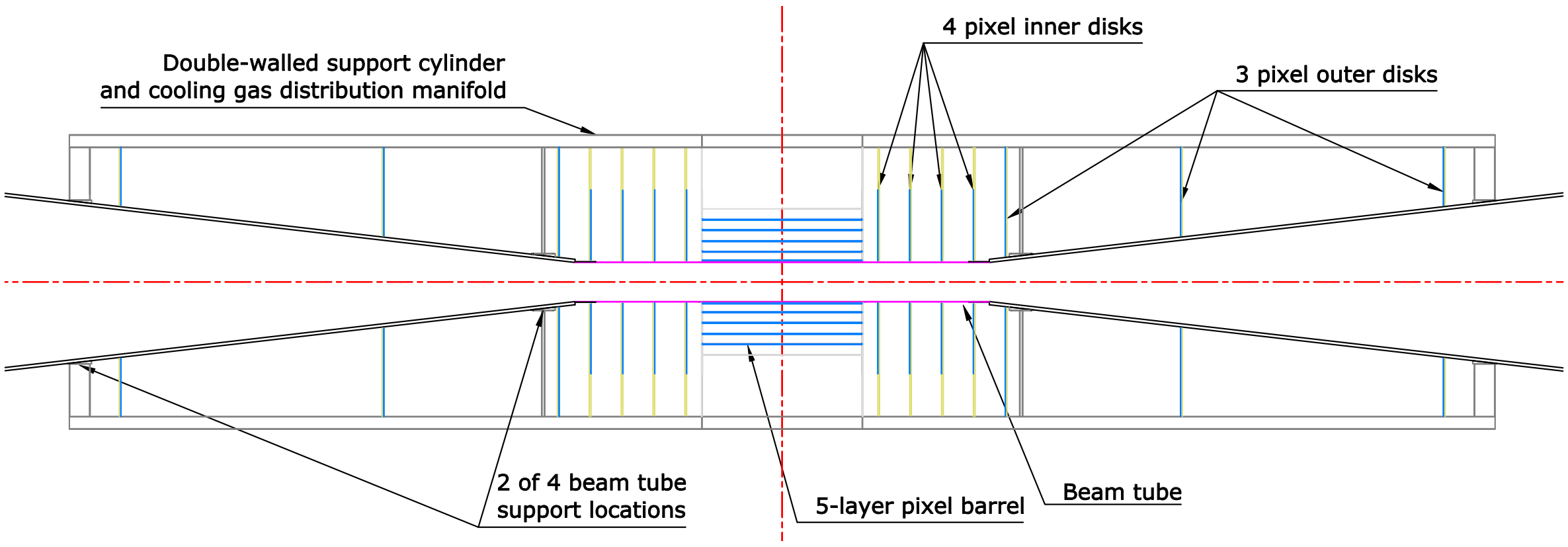

Fig. 4.10: Side elevation of the CLIC_SiD vertex detector region.

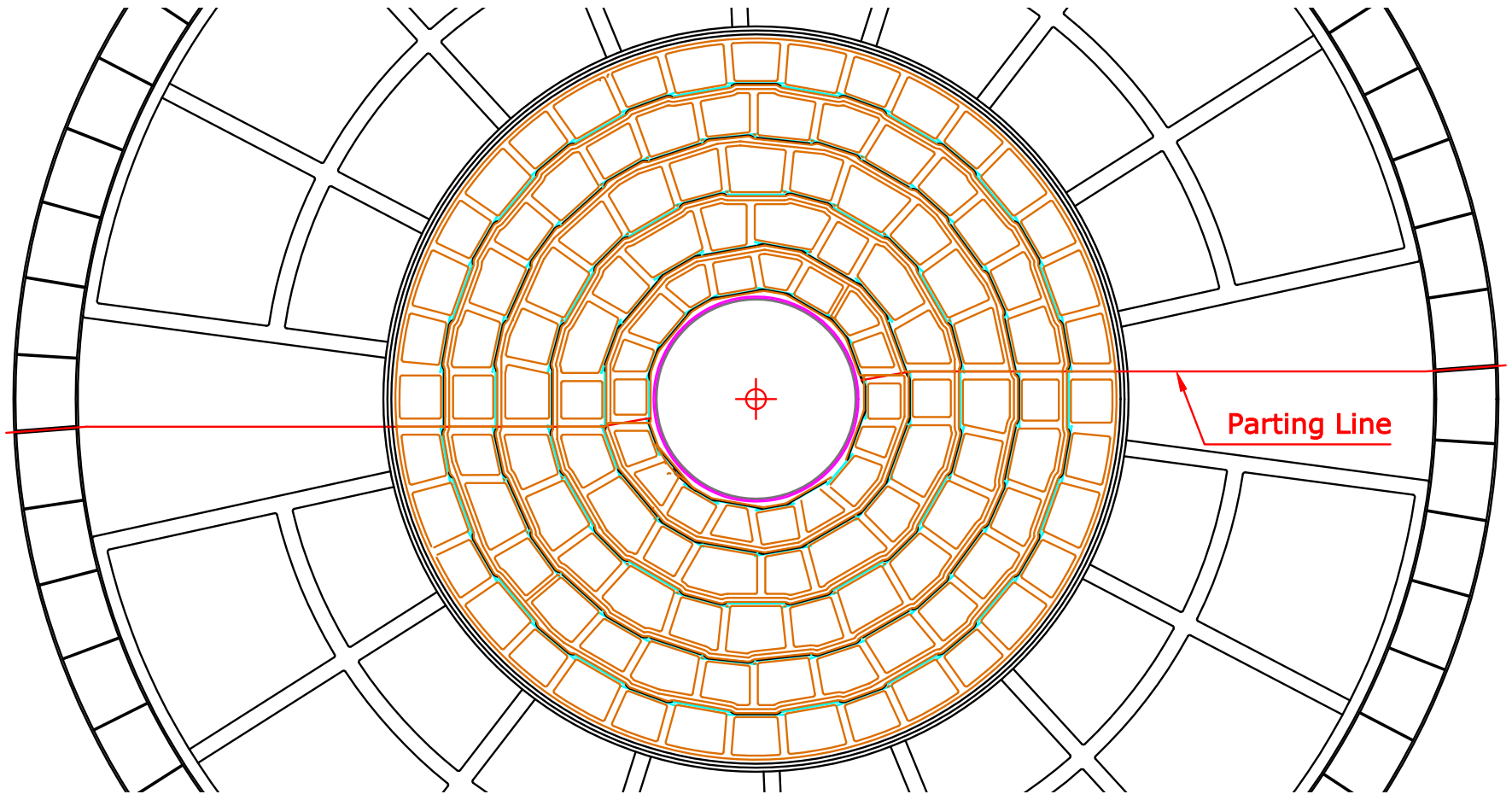

Fig. 4.11: End view of the CLIC_SiD barrel support structure.

Cables would be threaded through these spoked membranes during half-layer fabrication and half-barrel assembly.

Substantial effort is still needed on disk design. Disks are fabricated in two halves and supported via carbon fibre membranes from the outer cylinder. The tiling of disk structures is highly dependent on sensor geometric properties and remains to be determined.

### 4.6.2 Pixel Cooling

The pixel detector design is strongly dependent upon the assumption that sensors and their readout can be adequately cooled by the flow of dry gas (air). The assumption is based upon the low targets for the amount of material in the detector dictated by the low multiple-scattering term given in Section 4.2, and the relatively high mass associated with alternative coolant technologies. Vertex detector studies have mainly concentrated on the CLIC_SiD and CLIC_ILD barrel region. The endcap disk regions pose additional challenges to control air flow, and need further study.

In the barrel region, the flow is envisaged to be primarily longitudinal, from one end of the barrel to the other, but may include a substantial circumferential component. For a given total flow rate, the flow between each pair of barrel layers was adjusted to obtain a common end-to-end pressure difference, and heat transfer from silicon to air is calculated [17]. Both surfaces of a sensor are assumed to participate, and for CLIC_ILD, where foam separates two back-to-back sensors, the thermal impedance of the foam is assumed to be negligible (assuming 6% silicon carbide foam). The heat sources are assumed to be uniformly distributed over the entire surface of each sensor and to contribute a heat load of 0.05 W/cm$^2$ (see Section 10.2.2). Heat source concentrations can lead to higher local temperatures.

To simplify air delivery, an incoming air temperature of 289 K was assumed at the barrel. This minimises the thermal insulation needed for air delivery and corresponds to a dew point that might be expected in a collision hall, although a lower air delivery temperature could be achieved via double-walled cooling lines. In that case, liquid coolant could be used in an outer shell and high-pressure air in an inner line, with the air allowed to expand near the point of use.

The results obtained are shown in Figure 4.12 for the five barrel layers of the CLIC_SiD vertex detector. This figure illustrates the effects of variations in flow division among the various barrel layers. In particular, the temperature of the innermost layer is higher than that of other layers due to the limited cross section available for flow between it and the beam pipe. Supplemental one- or two-phase cooling may be needed for the innermost layer. The variation of temperature with flow rate in all layers illustrates the critical nature of flow division calculations. The transition from laminar to turbulent flow seen in this figure depends on total flow rate, the size of flow passages, the smoothness of surfaces, the extent to which components project into the flow stream, and the way in which flow lines enter and leave the barrel region. Normally, designs assume a relatively low or high flow rate to improve confidence in heat removal predictions and lessen the need for testing of prototypes. Since vibrations are a potential issue, testing is mandatory in any case for CLIC vertex detector structures. Ignoring the region between the beam pipe and layer 1 (L1), flow velocities range from 2.2 m/s at the lowest total flow rate to 12 m/s at the highest total rate. For the range of flow conditions studied, the temperature variation along the length of a module is relatively small: $\pm 0.2$ to $\pm 2.0$ K. Figure 4.12 suggests two options: supplemental cooling for L1 and a total flow rate of about 70 g/s or no supplemental cooling with a total flow rate in the range 150 to 190 g/s. In either case, the possibility of lowering the air delivery temperature to roughly 280 K should be considered.

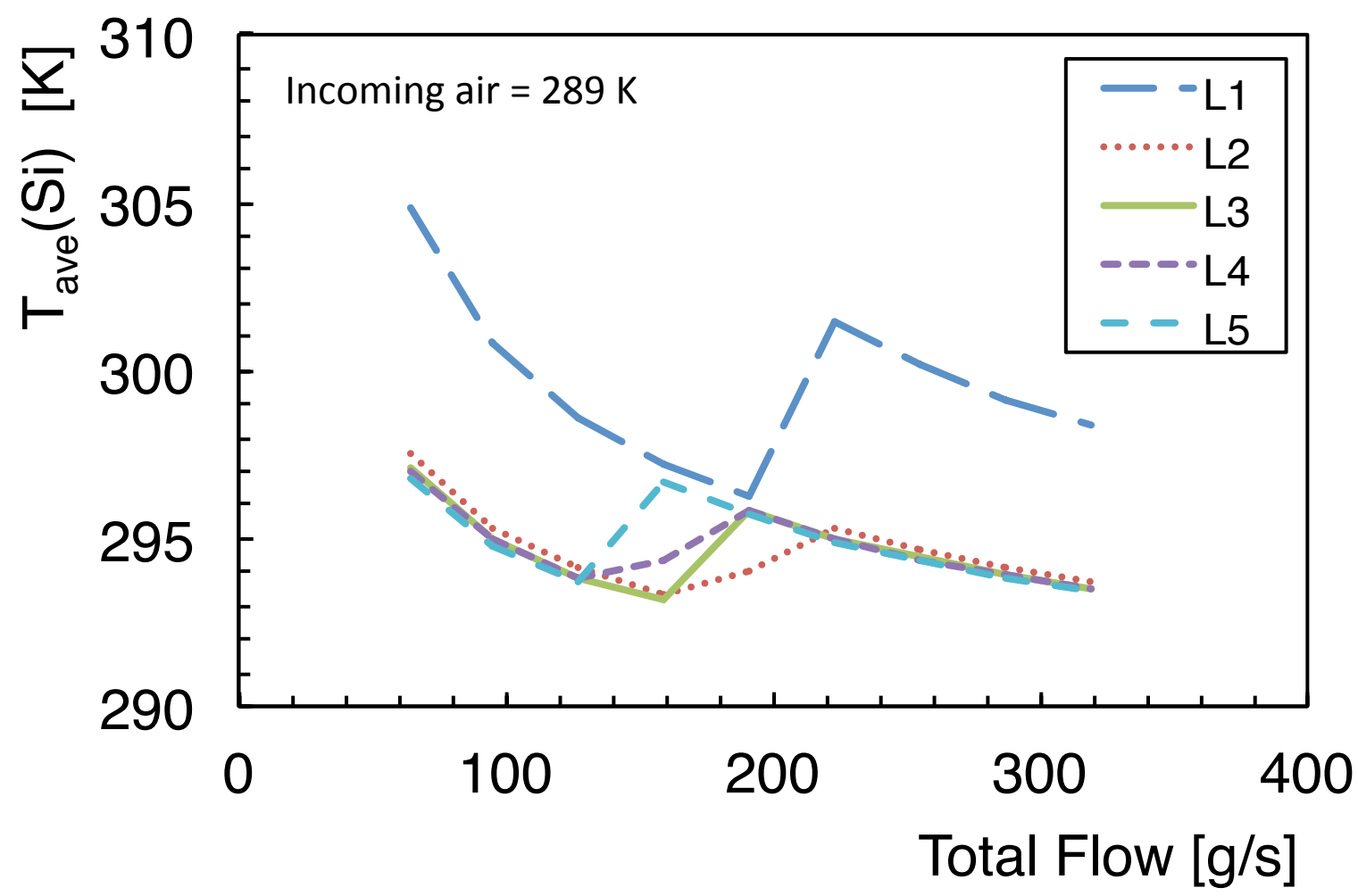

Fig. 4.12: Calculated average temperatures of the five barrel layers of the CLIC_SiD vertex detector as function of the total air-flow rate.

Should a different coolant be required for the disks or supplemental cooling be required for the barrel, one possibility is to use evaporative $CO_2$. Evaporative $CO_2$ cooling generally requires tubing with an inside diameter of 1.5 mm or more for acceptable pressure drop and a wall thickness of 0.15 mm or more for manufacturability and pressure containment; $CO_2$ pressure at room temperature can rise to at least 70 bar; round tubing is normally chosen to avoid shape changes with pressure. Materials considered for tubing include stainless steel, aluminium, and titanium. Stainless steel has the advantages that it resists corrosion well and is straight-forward to form. For stainless steel tubing of those dimensions, the tubing alone, averaged over its width, contributes approximately 2.5% $X_0$. $CO_2$ coolant which averages 75% liquid over the length of the tubing would contribute an additional 0.2% $X_0$. Similar issues occur for coolants such as $C_3F_8$, though the pressure requirement is relaxed to about 10 bar (plus safety margin). Water may be the most promising coolant if air flow alone is insufficient. It can be contained within PEEK tubing in a sub-atmospheric system to limit radiation length contributions and leakage issues. For the same tubing dimensions, the radiation length contribution would be $\approx$0.4% $X_0$ including both tubing and coolant.

Micro-fabricated cooling devices integrated into silicon wafers are currently being developed and investigated as an option for the cooling of the NA62 Gigatracker silicon-pixel detector [18], where they meet the design goal of contributing less than 0.15% $X_0$ per layer to the material budget, with a heat dissipation of 2 W/cm$^2$. Such integrated micro-channel structures help to minimise local concentrations of cooling mass and may be suitable for those regions of the CLIC vertex detectors, where sufficient air flow can not be established.

## 4.7 Sensor and Readout-Technology R&D

### 4.7.1 Requirements of a CLIC Vertex Detector Sensor

The requirements outlined in Section 4.2 present a number of technology challenges for the vertex detector sensors. While several of the main requirements for the detector have already been met individually, the combination of these performances in individual ultra-thin detector modules and subsequently into a full inner tracking detector system will require significant R&D efforts.

For example, in the framework of vertex detector R&D for ILC, several sensor technologies have achieved hit position resolutions of 3–5 µm with pixels sizes in the 20 µm range. Motivated by the desire for ultra-thin detector solutions, technologies have also been developed that combine the sensitive medium and part of the readout functionalities in ultra-thin and integrated semiconductor devices.

One difference with sensor technologies studied for previous applications is in the charge collection. At the ILC, for example, technological solutions could include charge collection by diffusion in a fairly low-resistivity epitaxial silicon layer, due to the relatively long bunch spacing of $>$ 300 ns. The CLIC requirement for precise time resolution of individual hits implies however that charge is collected quickly, through the combination of electron drift in the pixel and with readout capable of marking the arrival time of individual pixel hits. The need for timing information implies a high-resistivity sensitive medium and full depletion, with advanced readout functionality for the individual pixels in the array. As much of this underpinning technology exists, one can think of several developmental roadmaps that can lead to a pixel detector technology for CLIC.

### 4.7.2 Technology Options

One approach is to pursue a so-called “hybrid” solution, composed of a thinned high-resistivity sensor bonded to an ultra compact and thinned readout ASIC in Very Deep Sub-Micron (VDSM) technology. Building on experience with the Timepix ASIC [19] and others [20], many of the CLIC-specific functionalities are currently being implemented in the designs. The Timepix3 chip for example features 55 µm pixel sizes in 130 nm VDSM, and provides a good starting point from which an evolutionary path can be charted [21]. To accomplish the 20 µm pixel sizes required, smaller VDSM feature sizes (90 nm, 65 nm) will need to be investigated, and arrays will need to be designed and prototyped in successive steps to develop a compact implementation of time and pulse height functionalities [22, 23]. The advantage of the hybrid approach is that it profits fully from industrial technology developments for mass-produced consumer products and does not focus on the custom development of integrated solid-state materials and their corresponding custom production technologies. It does however require the integration of a number of technological improvements still in development. These include the power pulsing ideas discussed in Chapter 10, as well as the thinning and interconnect ideas in the next section.

The other approach towards a CLIC vertex detector is to build upon experience with integrated technologies developed for the ILC, by working towards solutions that allow for signal collection through electron drift in a high-resistivity sensitive layer. Several developments have recently started in this domain. In this context, we note the ongoing R&D on CMOS MAPS technologies, where several approaches are being pursued. The CMOS MIMOSA sensors [24], built in 0.35 µm technology and used for the EUDET [25] and STAR [26] pixel detectors, provide a useful starting point for development. Custom high-resistivity epitaxial layers and smaller feature sizes (e.g. 0.18 µm) are being investigated [27]. The aim here is to obtain a fully-depleted resistive epitaxial layer of some 20 µm thickness for fast collection of at least 1000 electrons. A similar approach towards a suitably small pixel with a fully-depleted epitaxial layer is being pursued within the CMOS CHRONOPIXEL project [28]. A somewhat different method towards obtaining a high-resistivity epitaxial layer in MAPS technologies is being pursued via the IN-MAPS [29, 30] approach. In this technology, a deep p-well barrier protects the n-well charge collector interface between the sensitive layer and the readout part. The implementation of this technology has led to more efficient charge collection and opens the possibility to integrate a high-resistivity epitaxial layer and full-featured CMOS MAPS technology. Another approach to obtain faster and more efficient signal collection uses high-voltage CMOS processes [31]. Here the CMOS signal processing electronics is embedded in a deep n-well that is reverse biased and acts as signal-collecting electrode.

Yet another approach will be to develop one of the new, emerging technologies for vertex sensors. These include Silicon On Insulator (SOI) [32, 33] or 3D [34] technologies, where several separately-optimised layers can be included in a single device. The SOI technology makes use of a $\approx$200 nm thin $SiO_2$ isolation layer to separate charge-collecting and readout functionality. Initial challenges of applying bias voltage to the sensitive layer and obtaining efficient charge collection are being met [35], opening another way to fully integrate the sensitive and readout functionality in pixel devices. Recent industrial shift to thinned and wafer-bonded interconnection will also be useful for the development of a fully 3D integrated pixel solution. In these devices, the sensor and readout electronics are separately optimised, thinned, and integrated as "tiers" in a single device. This option can result in, for example, a high-resistivity sensitive tier coupled directly to additional tiers for advanced analog and digital readout functionality [36, 37]. This approach is motivated by industrial interest in mass-produced wafer bonding solutions, and is predicated on the ultimate availability of the process at a relatively low cost.

### 4.7.3 Vertexing Technological Developments

In addition to the developments in CMOS materials, many technological developments must be pursued in parallel to achieve the required readout functionality for individual pixels. These developments are typically needed by more than one of the sensor options outlined above, and while not limited to CLIC, they will be necessary to progress the sensor development paths above. The primary technological challenges to be addressed are in the power delivery, readout, interconnect, and low-mass mechanical construction.

Power pulsing and time-tagging functionalities will need to be incorporated in the designs and developed in parallel. Power-pulsing will be required to maintain low power consumption and to keep cooling needs minimal, so that gas cooling is possible. Power pulsing requirements are discussed in detail in Chapter 10. Similarly, time-tagging of pixel hits will be necessary to reconstruct the details of the events. While the pixel sensor technologies listed previously include the possibility to include timing information, significant R&D will be needed to incorporate this feature in a suitably small pixel. Details of the methods considered (e.g. clock distribution) are discussed in Chapter 10.

Interconnect technologies and pixel connectivity must be studied and optimised. Developments in this area are critical for the success of all pixel options described above, as the cost and complexity of interconnection have become a limiting financial and technical feature of detector construction efforts [38]. The small feature size of CLIC pixels and high-density interconnect requirements of most of the options above imply that a close watch on industrial developments will need to be maintained. In current industrial applications copper pillar bumps and Solid-Liquid Inter-Diffusion (SLID) bumps are routinely applied in “flip-chip” bonding with a 20 µm pitch [39]. For hybrid sensor solutions, smaller readout chips covering $\approx 1$ cm$^2$, are flip-chip connected to larger sensors, covering $\approx 10$ cm$^2$, to form detector modules. In order to achieve seamless tiling without unwanted dead areas, signals to and from the readout chips have to be carefully arranged. Through-Silicon Via (TSV) technology offers flexibility in this regard, and offers the potential to greatly reduce the traditional lateral gaps between the readout chips. Integrated processes combining TSV with SLID bonding of thin sensors to readout chips are currently being developed for large-scale applications [40]. Development of interconnect technologies will be critical also for options such as the 3D devices outlined above, where issues such as wafer planarisation will need to be considered so that the emerging wafer-bonding techniques can be applied with high yield. Finally, lateral inter-connectivity will be an important consideration, to minimise the “dead space” between active sensing arrays. Technologies to be explored include stitched CMOS arrays or “edgeless” hybrid sensors.

Sensor thinning and ultra low-mass material construction will be necessary to stay within the material budget constraints implied by the physics goals outlined in Section 4.2. The low-mass design must extend to the cooling and electrical and readout services as well, since it is not possible to move these completely outside the detector acceptance. Thinning of sensors and ASICs to of order 50 µm will be required for these hybrid detectors. While thinning to such level is routinely achieved in industry for small die, handling, flip chip assembly and mounting of large die such as those used for pixel detectors presents a new level of technical challenge. Wafer-level thinning to much smaller dimensions (10 µm or less) for 3D integration may be needed for interconnect technologies such as wafer bonding, but handling of such thin structures is not anticipated. Low-mass support structures utilising foams or carbon fibre are envisaged, and initial prototyping efforts have been conducted with these materials [41, 42]. R&D will be needed to develop these initial studies into reliable construction techniques.

# Chapter 5

# CLIC Tracking System

## 5.1 Introduction

The tracking systems at CLIC are required to perform momentum measurement of charged particles with a precision that is well beyond the current state of the art. The tracking must cover the full solid angle, except for the incoming and outgoing beams, and must provide excellent tracking efficiency and precision over a wide range of momenta for prompt tracks from the vertex as well as for decay products of long-lived particles. CLIC shares these demanding tracking performance requirements with the ILC case. However, due to the high energies and high beam-induced background levels, tracking is even more challenging at CLIC than at the ILC. Track distances within highly boosted jets are very small, making the separation of tracks more difficult. In addition, cell occupancies are high due to beam-induced background from incoherent pairs and $\gamma\gamma \rightarrow$ hadrons, as well as muon background from the beam delivery system.

Like in the ILC case, the tracking system must be built with minimal material to preserve the momentum resolution, the good lepton identification and the good PFA jet-reconstruction performance. In addition, the readout of all tracking detectors must include precise time-stamping capabilities, complemented with multi-hit readout capabilities for some of the inner strip layers.

In comparison to $e^+e^-$ colliders operating at lower centre-of-mass energies the relevance of the forward tracking region increases significantly. At higher energies two-particle processes tend to appear more often at smaller angles, while multi-particle final states tend to be isotropic and with higher multiplicities. For the latter the probability to have the full event contained in the barrel part of the tracker alone is therefore smaller than for lower centre-of-mass energies. Moreover, many events tend to have a forward boost [1].

The ILC tracking detector concepts represent a good starting point for the CLIC case. Only small geometry adaptations were made towards the CLIC_ILD and CLIC_SiD tracking geometries as described in Chapter 3 and in the following sections. Performance studies carried out so far show that both detector concepts achieve a good track reconstruction efficiency and excellent momentum resolution in the CLIC environment.

## 5.2 Tracker Concepts

The CLIC central tracking concepts are similar to the ones studied for the ILC: either a large Time Projection Chamber (TPC) complemented with a small inner silicon tracker and a silicon envelope surrounding the TPC (CLIC_ILD) or a compact full silicon tracker in a very high magnetic field (CLIC_SiD). Both concepts try to achieve the required performance with different technological approaches.

CLIC_ILD offers a highly redundant, continuous tracking, a high lever arm, good particle identification through $\mathrm{d}E/\mathrm{d}x$ and very little material within the tracking volume itself. It may, however, present more material in front of the calorimeter, in particular in the forward region, and it may be less competitive in track separation within dense jets and in high occupancy situations. In addition, it requires a silicon tracking system to provide independent tracking at low angles, as well as silicon tracking layers surrounding the TPC itself to provide timing and precision points both at the TPC periphery and at the calorimeter entrance.

The all silicon solution of CLIC_SiD offers a well known technology providing very accurate space point resolution and a fast charge collection allowing for time stamping capabilities. The good resolution compensates the smaller lever arm, allowing for a more compact tracker. It has adequate granularity to cope with track separation in dense jets and with hits from the beam-induced background.

The main drawback is the relatively high material density within the tracking system itself (since the services for the barrel detectors will be in front of the endcap tracker), less redundancy, in particular for displaced vertices, and limited $dE/dx$ information.

Despite the pros and cons of both systems, the studies carried out so far indicate that both tracking concepts are able to meet the required specifications. A word of caution is nevertheless appropriate here. Highly performant tracking software is required to fully explore the CLIC experimental environment. The tracking tools, available from ILC and further developed for the CLIC CDR, have allowed to test tracking performances for CLIC physics events with the equivalent of 60 bunch crossings of $\gamma\gamma \to$ hadrons events overlaid. This is the dominant background in most of the tracking region. In a forthcoming phase of the project, these studies will be pushed further to include background from incoherent pairs and integration over a full bunch train. This is of particular importance for the TPC studies. Therefore, as a first step, detailed simulations of signal formation in the TPC were carried out to determine TPC occupancies for the full bunch train and including all background processes (see Section 5.3).

### 5.2.1 The TPC-Based CLIC_ILD Tracking System

The CLIC_ILD concept relies on a mixed gaseous and solid state tracking system. At radial distances from the beam line beyond 33 cm a large Time Projection Chamber (TPC) provides approximately 200 measurement points for tracks with a polar angle between 35° and 145°. The polar angle coverage extends down to $\theta \simeq 12°$ and up to $\theta \simeq 168°$ (10 measurement points). The overall layout of the CLIC_ILD tracking system is depicted in Figure 5.1. A more detailed description of the CLIC_ILD tracking geometry can be found in the corresponding LCD note [2].

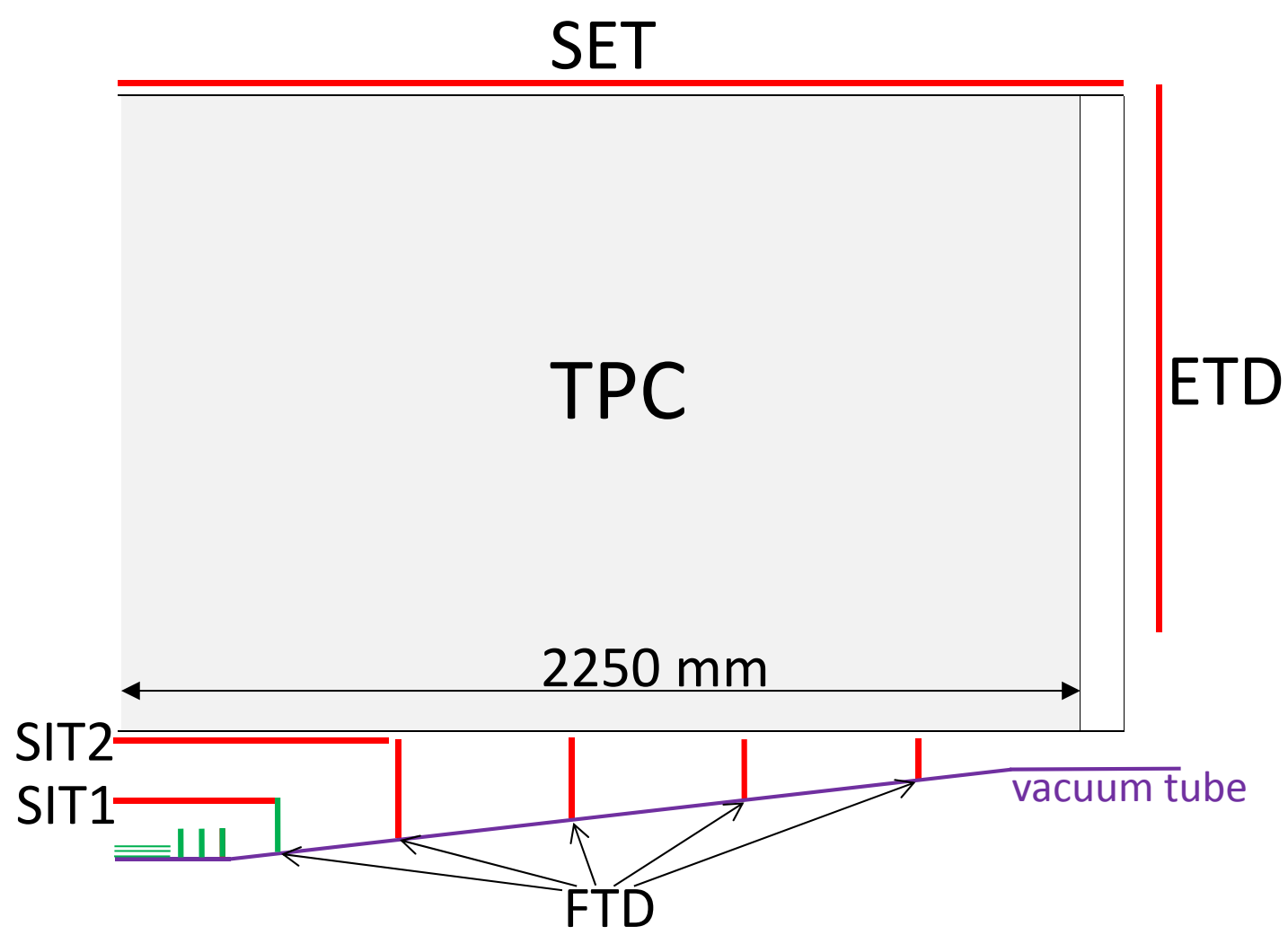

Fig. 5.1: Overview of the CLIC_ILD tracking system, showing the TPC and the silicon trackers in a side view. Detectors in red are implemented in silicon micro-strip technology, while the innermost Forward Tracking Disk (FTD) and the vertex tracker are realised in pixel technology (green).

It has been shown in the ILD Letter of Intent [3] that a TPC as main tracker in a linear collider experiment offers several advantages. Tracks can be measured with a large number of three-dimensional space points. The moderate single-point resolution and double-hit separation are compensated by continuous tracking. The detector will be located in a strong magnetic field of 4 T, parallel to the drift direction of the electrons, which improves significantly the single-point resolution and the double-hit separation due to a large reduction of the transverse diffusion of the drifting electrons when choosing an appropriate gas mixture like $Ar/CF_4/iC_4H_{10}$ (95%/3%/2%) [4]. Continuous tracking allows for easy and precise reconstruction of non-pointing tracks, e.g. from $V^0$s, which is important for studies of long-lived par-ticles and for the particle-flow measurement. The TPC also has the potential to provide good particle identification through the measurement of the specific energy loss $\mathrm{d}E/\mathrm{d}x$, with a resolution of $\sim 5\%$ estimated from past experience at LEP.

#### *5.2.1.1 Design of the TPC*

The choice of a TPC as central tracker in the ILD concept study was based on demonstrated performance in the ALEPH and DELPHI experiments at LEP, for example. The main performance goals for a TPC at CLIC are similar as for ILD at the ILC and are given in Table 5.1. The LCTPC collaboration is actively investigating the design issues related to the performance, like field cage, endcaps, electronics, sensitivity to backgrounds, distortion corrections and alignment. The design is based on a lightweight field cage, with a thin central high-voltage cathode, and the TPC being read out by micro-pattern gas detectors (MPGD) at either end. MPGDs are chosen because they show better performance than the more traditional wire chamber readout. Combined with small (anode) readout pads of $\sim 1$ mm $\times$ 4–6 mm this results in single point resolutions of better than 100 µm in the $r\phi$ plane (perpendicular to the drift direction). A pad height of 6 mm was used in the simulations. Smaller pad heights would reduce the “track-angle effect” and thus improve the resolution for tracks that make an angle with respect to the pad ‘long’ direction. At present a 4 mm pad height is believed to be the lower limit when using standard connector technology on the pad plane. Along the drift direction a single point resolution of $\sim 0.5$ mm should be in reach. MPGDs provide a significant suppression of the flow of positive ions back into the drift volume. Nevertheless, since a TPC will integrate charge over the full CLIC bunch train, an ion gate will be foreseen to eliminate the remaining ion backflow. The momentum resolution of the TPC alone (as given in Table 5.1) assumes a uniform B-field of 4 T. This performance will not significantly degrade due to (small) B-field distortions, provided the main B-field component has been mapped out to a relative precision of $10^{-4}$ and the $B_\phi$ component has been measured to $\pm 0.1$ mT. Further details are given in [5, 6] and Section 7.2.

##### *5.2.1.1.1 Field Cage*

The design of the inner and outer field cages involves the geometry of the potential rings, the resistor chains, the central HV membrane, the gas container and a laser system. They should sustain of the order of 100 kV at the HV membrane, with a minimum of material. The goal for the inner and outer field cage material thickness is 1% $X_0$ and 3% $X_0$, respectively, while the chamber gas adds another 1% $X_0$. The goal of 1% $X_0$ for the inner field cage has almost been reached in the current Large Prototype TPC [7], which has an outer diameter close to the diameter of the inner field cage and material thickness of 1.21% $X_0$. The field cage is made of composite materials, forming a sandwich structure as can be seen in Figure 5.2. The electric drift field is provided by a layer of field strips and a layer of mirror strips at intermediate potential allowing for a drift field homogeneity of $\Delta E/E \sim 10^{-4}$. The outer field cage wall will have a similar structure but with an increased thickness of 60 mm in order to have sufficient mechanical stiffness.

##### *5.2.1.1.2 Endcaps*

The two endcaps will have an area of 10 m$^2$ each. Each endplate is subdivided into many independent MPGD detector modules, which provide near-full coverage of the endplate. The modules will be replaceable without removing the endplate. The endplate should have a low material budget in order not to compromise the jet energy resolution in the forward direction, yet it should be sufficiently rigid to achieve precise and stable positioning of the detector modules to better than 50 µm. Figure 5.3 shows a possible layout of the endplate with 8 rows of detector modules. The material budget of the mechanical structure represents 8% $X_0$. The additional material for the readout planes, front-end electronics and cooling is estimated to be 7% $X_0$, and power cables add an additional 10% $X_0$.

Table 5.1: Goals for performance and design parameters for a TPC with standard electronics.

| | | |
|---|---|---|
| Outer dimensions | | |
| | Radius | 1.8 m |
| | Length | 4.5 m |
| Momentum resolution (4 T) | | |
| | TPC only | $\delta(1/p_T) \sim 8 \cdot 10^{-5}$/GeV |
| | SET+TPC+SIT+VTX | $\delta(1/p_T) \sim 2 \cdot 10^{-5}$/GeV |
| Solid angle coverage | | $12^\circ \lesssim \theta \lesssim 168^\circ$ (10 pad rows) |
| TPC material budget | | |
| | to ECAL $R$ | $\sim 0.05\ X_0$ |
| | Readout endcaps in $z$ | $\sim 0.20$–$0.25\ X_0$ |
| $N_{pads}$ / time buckets | | $\sim 2 \cdot 10^6$ / 1000 per endcap |
| Pad size / $N_{pad\ rows}$ | | $\sim$ 1 mm $\times$ 4–6 mm / $\sim$ 200 (standard readout) |
| $\sigma_{point}$ | | |
| | $r\phi$ | $<$ 100 µm (average over $L_{sensitive}$, modulo track $\phi$ angle) |
| | $rz$ | $\sim$ 0.5 mm (modulo track $\theta$ angle) |
| 2-hit resolution | | |
| | $r\phi$ | $\sim$ 2 mm (modulo track angles) |
| | $rz$ | $\sim$ 6 mm (modulo track angles) |
| $\mathrm{d}E/\mathrm{d}x$ resolution | | $\sim$ 5% |
| Efficiency ($p_T > 1$ GeV) | | |
| | TPC only | $>$ 97% |
| | SET+TPC+SIT+VTX | $>$ 99% |

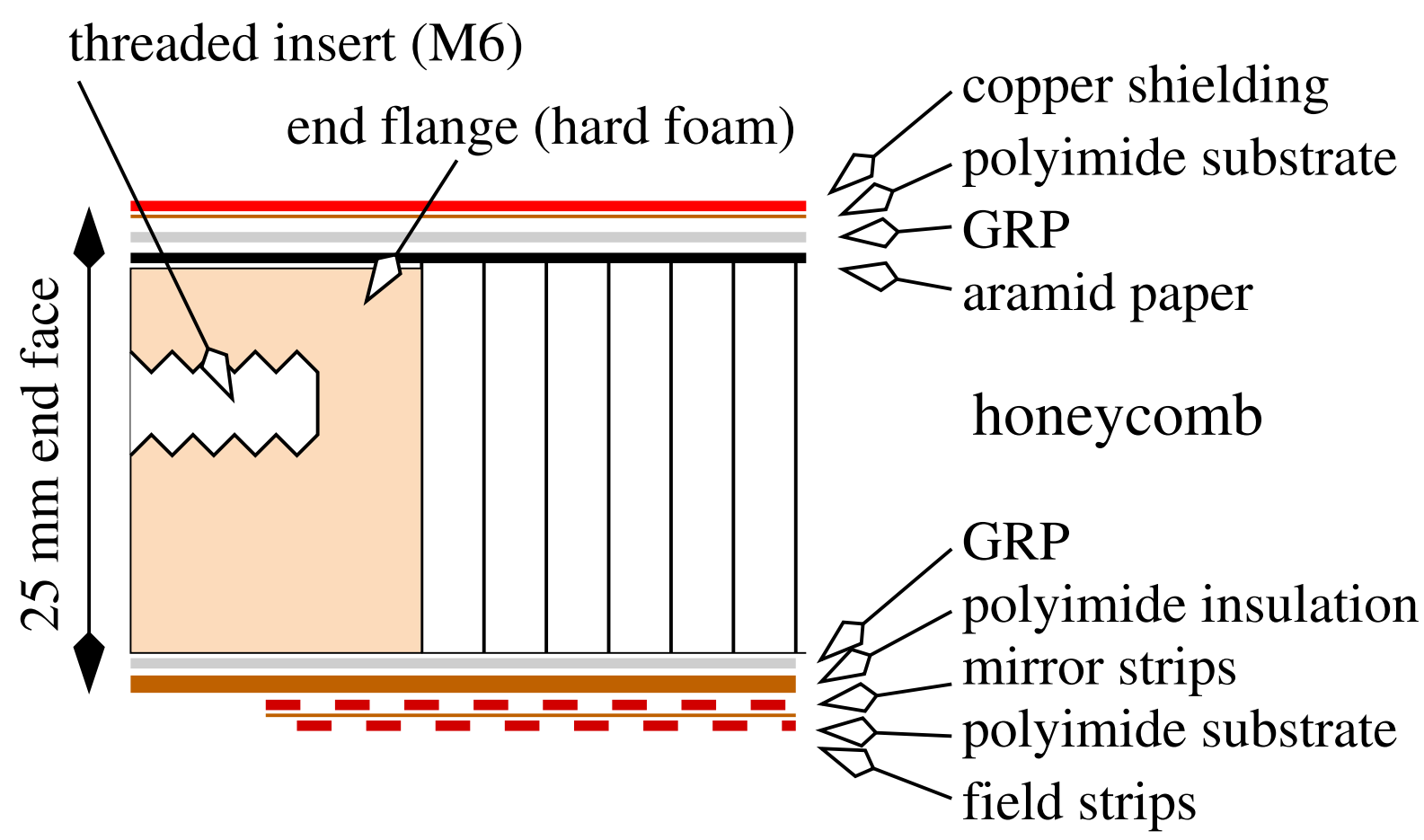

Fig. 5.2: Cross section of the Large Prototype TPC field cage wall. GRP = glass-reinforced plastic. For further details see [7].

##### *5.2.1.1.3 Detector Modules and Electronics*

Each endcap is equipped with some 240 MPGD readout modules with dimensions of $\sim 17 \times 22\ \mathrm{cm}^2$. Both Micromegas and GEM modules of this size have been extensively tested at the Large Prototype TPC [8]. They have been operated with a typical gas gain of $\sim$ 1000. The anode (readout) plane is subdivided into small pads of $\sim 1 \times 6\ \mathrm{mm}^2$. In case of Micromegas the anode will have to be covered by a thin resistive layer of carbon-loaded kapton of a few $\mathrm{M}\Omega/\square$, in order to spread the charge over a few pads. A GEM module will contain a stack of two or three GEM foils, while a single bulk-Micromegas grid [9] gives sufficient gas amplification. A single point resolution (at zero drift distance) of $\sim 60$ µm has been obtained with both technologies. The resolution at larger drift distances increases due to the transverse diffusion and measurements at the Large Prototype TPC in a 1 T magnetic field show that this increase is consistent with the expected diffusion. Earlier measurements with smaller prototypes in a 5 T magnet and with maximum drift distances of up to 60 cm showed that single point resolutions, when extrapolated to a drift distance of 2.25 m, will remain below the goal of 100 µm [8, 10, 11].

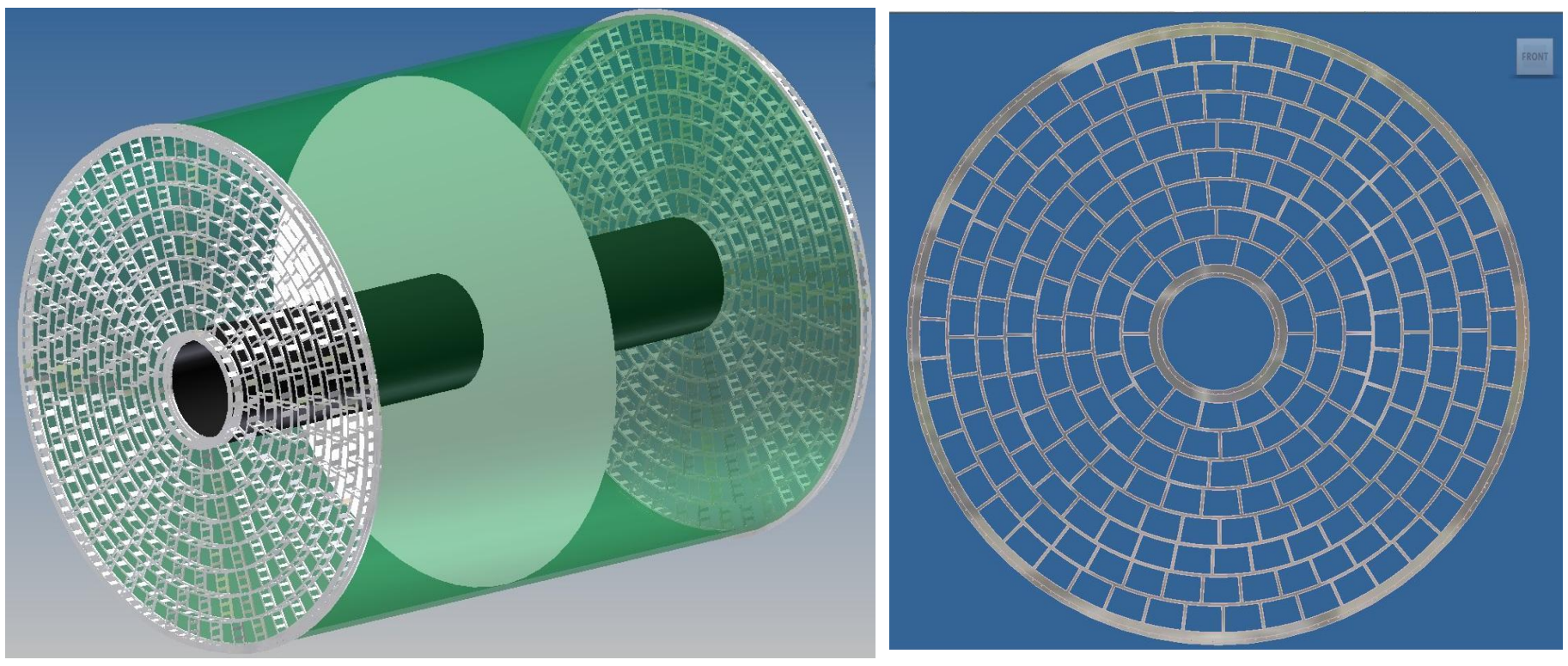

Fig. 5.3: (left): View of the TPC with endcaps made of space frames. The total length is 4500 mm, the inner and outer radius are 329 mm and 1808 mm, respectively. (right): View of the endplate as seen from the inside of the TPC.

Given the large number of readout channels ($\sim 2 \cdot 10^6$ input pads per endcap) deep-submicron electronics integration is necessary. A first full integration on a single ASIC of 16 channels of a Flash ADC-type electronics chain, with a design based on experience with the ALICE TPC, has been developed in 130 nm technology (see Section 10.2.3 and [12]). It is currently under test. Possibilities for a further integration to 64 channels per chip are under investigation. It is foreseen to use power-pulsing of this on-detector electronics; given the 50 Hz bunch train time structure, a power reduction by a factor 25–50 is possible.

A new concept for the combined gas amplification and readout is under development [13, 14]. In this concept the "standard" MPGD is produced in wafer post-processing technology on top of a CMOS pixel readout chip, thus forming a thin integrated device (Ingrid) of an amplifying grid and a very high granularity endcap with all necessary readout electronics incorporated. For a readout chip with $\sim 50$ µm pixel size, this would result in $\sim 3 \cdot 10^9$ pads ($\sim 5 \cdot 10^4$ chips) per endcap. This concept offers the possibility of pad sizes small enough to observe the individual primary electrons formed in the gas and to count the number of ionisation clusters per unit track length, instead of measuring the integrated charge collected. Both an 8-chip (Timepix) + triple-GEM stack module as well as an 8-fold Ingrid module have been successfully tested at the Large Prototype TPC.

Pixelised readout may be mandatory in high-occupancy regions of the TPC. The granularity increases by more than a factor 1000 compared to pad readout, and unlike the case of gas multiplication with GEMs where the electron cloud diffuses over $\sim 100$ pixels, with Ingrid as gas multiplier the electron cloud originating from a single primary electron is collected on a single pixel, thus reducing significantly the occupancy. The current Timepix chip [15] has no pixel multi-hit capability, but future versions of the chip will have this feature.

*5.2.1.1.4 Chamber Gas*

Currently the gas mixture mostly used at the Large Prototype TPC is the one also used by the T2K collaboration: $Ar/CF_4/iC_4H_{10}$ (95%/3%/2%). It has a saturated drift velocity of nearly 8 cm/µs at a drift field of $\sim 300$ V/cm. Due to the large $\omega\tau$ value of this gas, the transverse diffusion coefficient is strongly reduced from $\sim 300$ µm/$\sqrt{\text{cm}}$ at $B = 0$ T to $\sim 30$ µm/$\sqrt{\text{cm}}$ at $B = 4$ T at the drift field of 300 V/cm. At $B = 4$ T the minimum in transverse diffusion is reached for a drift field close to 100 V/cm.

#### ***5.2.1.2 Supplementary Silicon Tracking in CLIC_ILD***

The CLIC_ILD vertex detectors, see Chapter 4, and the TPC are complemented with a silicon tracking system comprising a cylindrical Silicon Internal Tracker (SIT), Forward Tracking Disks (FTD), as well as a Silicon External Tracker (SET) and an Endcap Tracking Disk (ETD). These silicon tracking systems complement the tracking acceptance down to small polar angles of 7° and provide track references at the periphery of the TPC and the ECAL front face. Moreover, the silicon tracking improves the momentum resolution and is foreseen to provide time-stamping of hits within a window of 10 ns. By making use of the TPC drift velocity, the silicon layers surrounding the TPC will also allow to reconstruct the time of TPC tracks with a precision of a few nanoseconds.

An inner tracking system provides a number of high-precision measurements at radii below 33 cm. These tracking layers complement the vertex detector. A view of the inner and forward tracking region is shown in Figure 5.4. For central tracks the cylindrical Silicon Internal Tracker (SIT) layers provide two additional measurements that link the vertex detector to the TPC at polar angles above 26°. The first layer, at a radius of 165 mm, has a half-length of 371 mm, while the second layer has radius of 309 mm and a half-length of 645 mm. In the current simulation studies the SIT detectors are implemented as "false double-sided" [3] silicon micro-strip detectors with a silicon thickness of 275 µm, an $r\phi$ resolution of 7 µm and a resolution in $z$ of 50 µm.

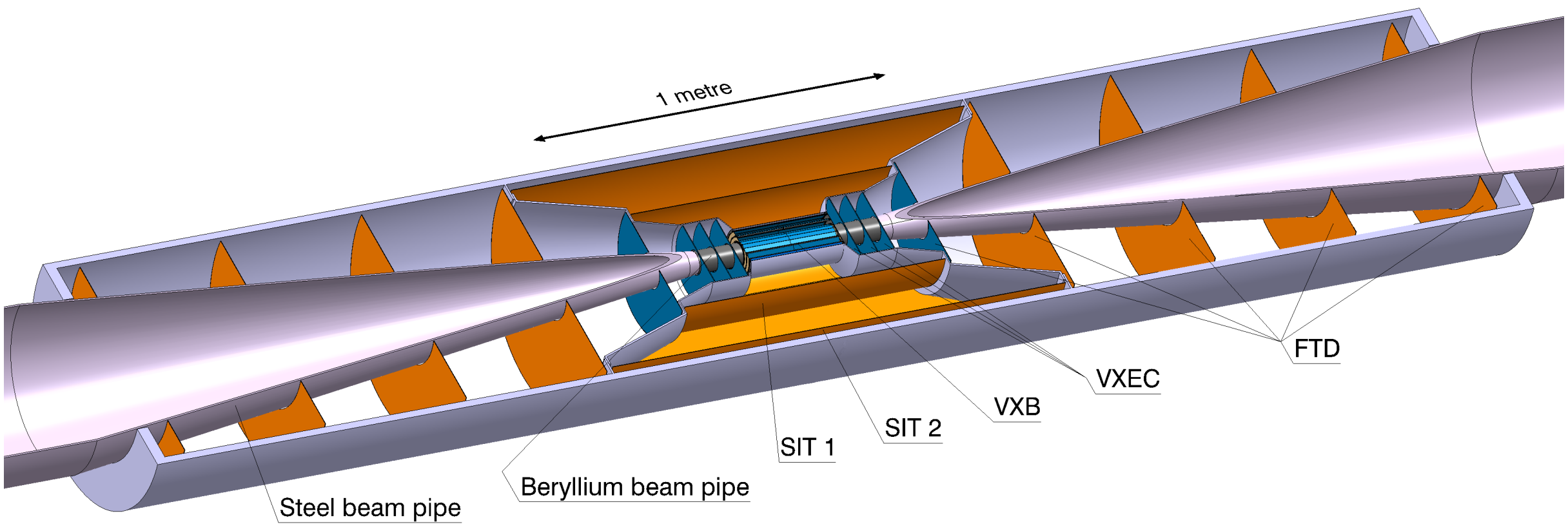

Fig. 5.4: View into the inner and forward tracking region of the CLIC_ILD detector. Shown are the vertex barrel (VXB) and endcap (VXEC) pixel layers, the two inner silicon barrel strip layers (SIT 1/2), the forward tracking disks (FTD), the beam pipe and the support shells for the silicon layers.

At small angles, down to $\theta = 7°$, the silicon tracker comprises five tracking disks, the so-called Forward Tracking Disks (FTD). In the simulation model all FTD disks are implemented as double-layers of silicon micro-strip detectors with an $r\phi$ resolution of 7 µm and a resolution of 50 µm in $r$. However, in view of the high background rates at its inner radius (see Section 5.3), and contrary to the ILC implementation, it is proposed that the FTD disk closest to the IP will be using silicon pixel technology at CLIC. For the other four FTD disks double-sided silicon strip technology is adequate at CLIC, albeit requiring some limited multi-hit capability within the bunch train in addition to the time-stamping capabilities. For these disks, the choice of the optimal stereo-angle represents a compromise between the resolution on the second coordinate, which improves with increasing angle, and the need to keep the hit rate under control.

The Silicon External Tracker (SET) consists of a false double-sided layer of silicon strip detectors, located between the external radius of the TPC and the inner radius of the ECAL. The same detector technology as described above for the SIT is foreseen also for the SET, resulting in an $r\phi$ resolution of 7 µm and a resolution in $z$ of 50 µm. The SET will improve the overall momentum resolution and provide an entry point to the ECAL after the TPC's outer field cage wall.

The Endcap Tracking Disk (ETD) is located between the TPC readout plane and the front face of the ECAL endcap. Three closely spaced single layers of silicon disks with a single-point resolution of $\approx$ 7 µm will be mounted with stereo-angles of 60° with respect to each other, resulting in a symmetric configuration with a single-point resolution in $x$ and $y$ of 5.8 µm.

Micro-strips have demonstrated capabilities in terms of spatial resolution and readout speed. Recent progress is primarily in increasing the radiation hardness of the sensors and front-end (for the LHC programme) and in further integration of the detector elements. The focus of the community involved in future lepton colliders, in particular the SiLC R&D collaboration, is on the latter. Functionalities traditionally performed by separate elements or the readout hybrid can be integrated directly on the sensor. For instance, a detector with pitch adapters integrated in a second metal layer on the sensor has been operated successfully.

Figure 5.5 (left) shows the number of measurement points in the silicon detectors and the fraction of TPC pad rows covered as a function of the polar angle, while Figure 5.5 (right) shows the thickness in units of $X_0$ of the tracking system, including beam pipe, as used in the CLIC_ILD full simulation studies. The sharp peak at low angle corresponds to the conical section of the beam pipe.

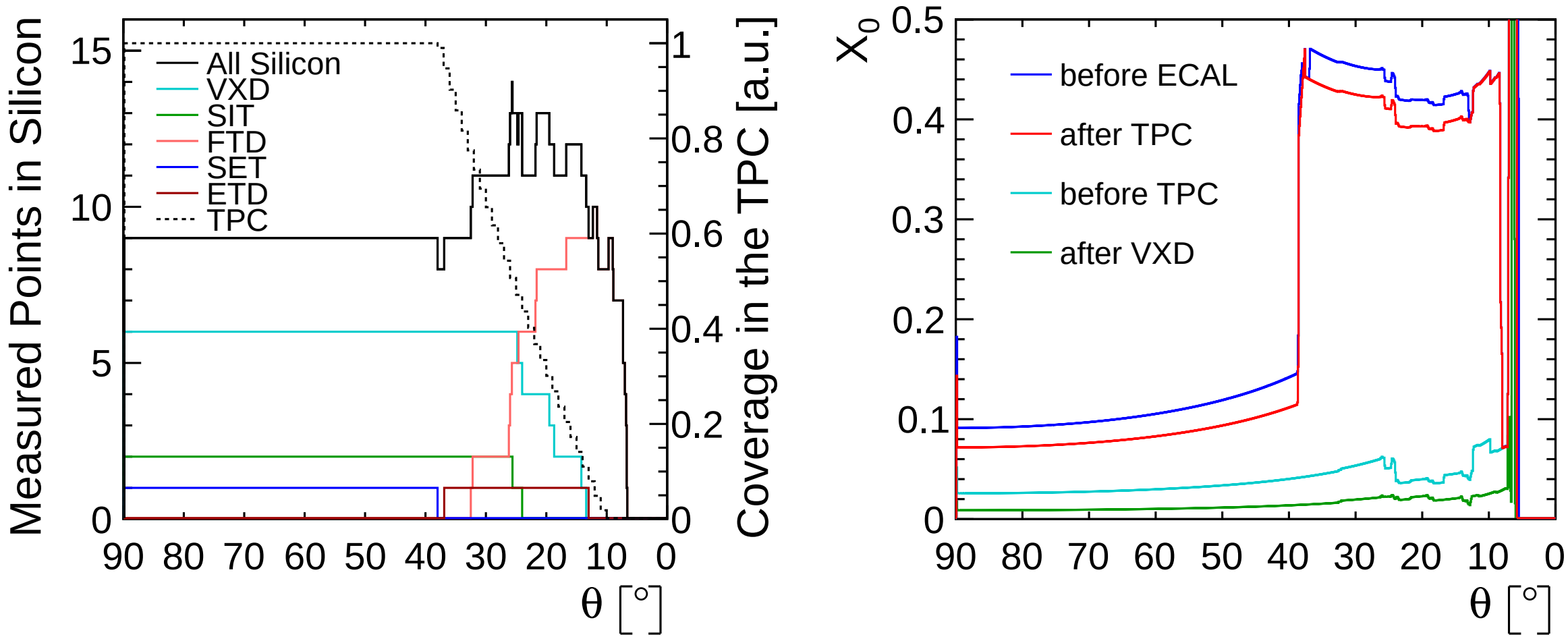

Fig. 5.5: The number of measurement points in the CLIC_ILD silicon detectors and the fraction of TPC pad rows covered as a function of the polar angle (left). The thickness of the tracking system, including beam pipe, as used in the CLIC_ILD full simulation studies (right). The sharp peak at low angle corresponds to the conical section of the beam pipe.

### 5.2.2 The All-Silicon CLIC_SiD Tracking System

In contrast to the ILD concept, the SiD detector relies upon strong integration between the vertex detector and the outer tracker in order to provide pure and efficient tracking for charged particles. The outer tracking system for the SiD concept consists of an all silicon tracker arranged as a set of five nested cylinders in the central region with four disks of silicon stereo-strip detectors on each side, mounted on conical support disks at 5° with respect to the normal (see Figure 5.6). Details of the CLIC_SiD geometry are described in [16].

The integrated tracking system provides at least ten precisely measured points for all tracks down to a polar angle of about 15° and at least six measured points down to a polar angle of about 8°, as shown in Figure 5.7. With five to seven pixel hits from the vertex and forward tracking detectors over the covered polar angle, those detectors provide powerful pattern recognition capabilities on their own. The outer tracker adds important track-finding constraints at larger radii where hits are less dense while providing a very precise momentum measurement with extremely good single-hit resolution in the $r\phi$ plane over a large lever arm.

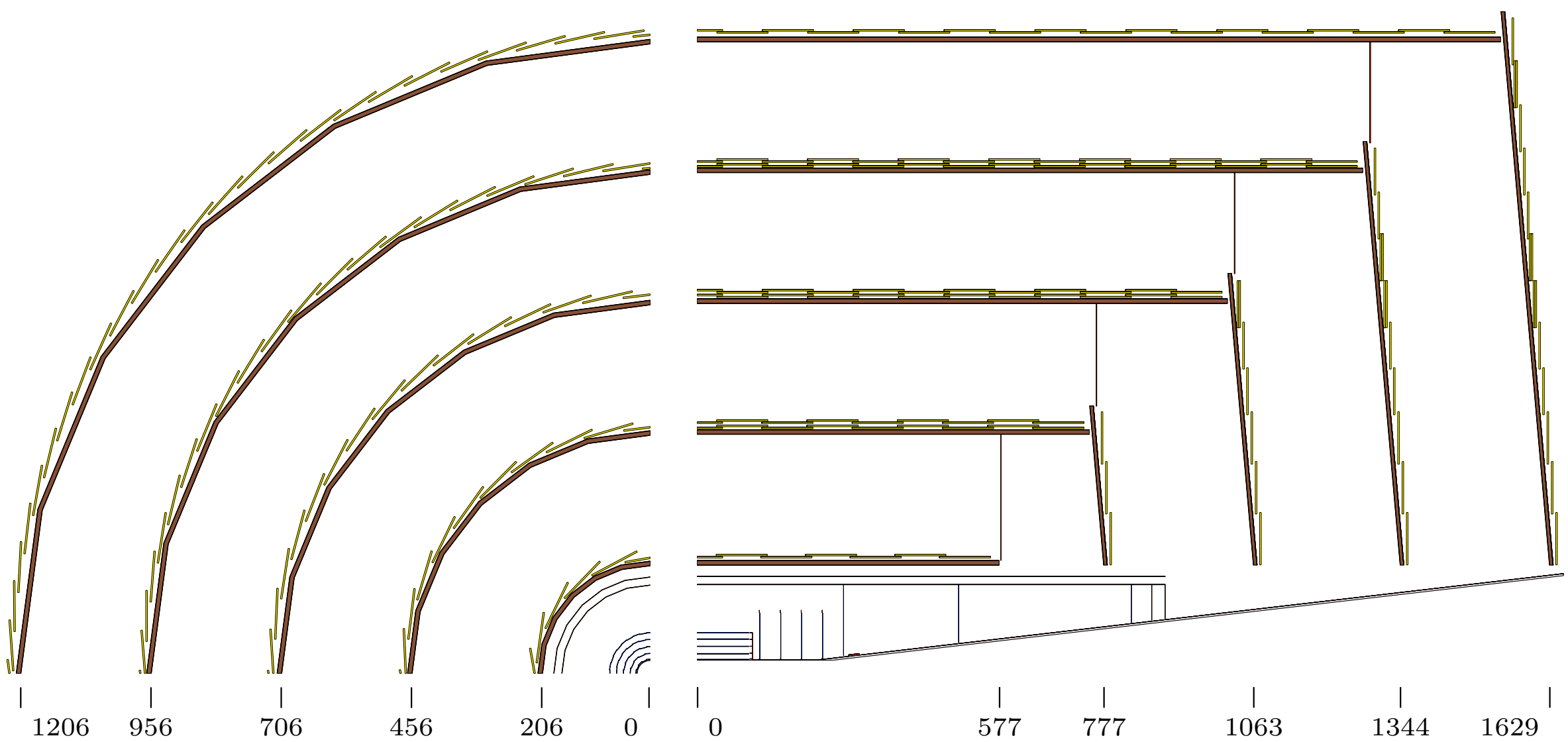

Fig. 5.6: Layout of the CLIC_SiD tracking system in the $r\phi$ plane (left) and in the $zx$ plane (right). The main tracker modules are displayed in yellow and support structures are displayed in brown. Values are given in millimetres.

Each barrel layer is made from several square silicon modules with a size of $97.8 \times 97.8$ mm$^2$. Each module has 0.3 mm of sensitive silicon. The barrel modules are tilted around the direction of the 5 T axial magnetic field so that the Lorentz drift for holes is roughly normal to the surface of the sensor, resulting in a pinwheel layout with overlaps in $\phi$. In the $z$ direction adjacent barrel modules alternate between two slightly different radii to allow for overlap and provide hermetic coverage. The strip pitch of the sensors is 25 µm and every second strip is read out. Therefore the readout pitch is 50 µm.

Each layer of the tracker endcap consists of several rings of trapezoidal modules. Each module has two layers of 0.3 mm of sensitive silicon. The modules are mounted so that the silicon planes are normal to the axial magnetic field. The resulting stepped configuration of the sensors with increasing radius provides overlap in the radial direction while alternating the $z$ of adjacent modules in $\phi$ between inner and outer positions provides overlap in the other dimension. With this configuration, each layer provides hermetic coverage for high-momentum tracks from the origin within the polar acceptance of the detector. The modules used in the inner three rings have a radial extent of 100.1 mm and the modules in the outer rings have a radial extent of 89.8 mm. Like for the tracker barrel, the silicon strips have a width of 25 µm and a readout pitch of 50 µm. The strips in the first sensitive layer are perpendicular to one side of the trapezoid, while the strips in the second sensitive layer are perpendicular to the other side of the trapezoid. The layout of the inner and outer modules is chosen such that in both cases the stereo angle between the two sensitive layers is 12°.

The cylindrical support barrel for each layer consists of sheets of carbon fibre composite sandwiched around a ROHACELL-foam core, fabricated as a single unit. The conical supports for the endcaps are constructed in the same way and mount to the outer faces of flanges which are integrated in the ends of each barrel cylinder, creating a set of closed and rigid units. Spoked, annular rings connect these flanges to the inner surface of the next cylinder outward to create a nested structure which is, as a whole, supported at the ends from the inner radius of the ECAL support structure. Mounted to the outside of the spoked support rings there are readout and power distribution boards which collect data and distribute control signals for the modules in each layer.

The coverage of the CLIC_SiD tracking system as well as the accumulated material budget, including dead material such as readout, cables and support, are shown in Figure 5.7 as a function of the polar angle. At least 6 hits are measured for all tracks with a polar angle down to about 8°. The vertex detector, including the beryllium beam pipe (see Figure 4.2) and the carbon fibre support tube, amount to about 1% $X_0$ for an incident angle of 90°. The total material of all tracking systems corresponds to about 7% $X_0$, which increases to about 17% $X_0$ in the transition region between barrel and disk detectors at polar angles between 30° and 40°.

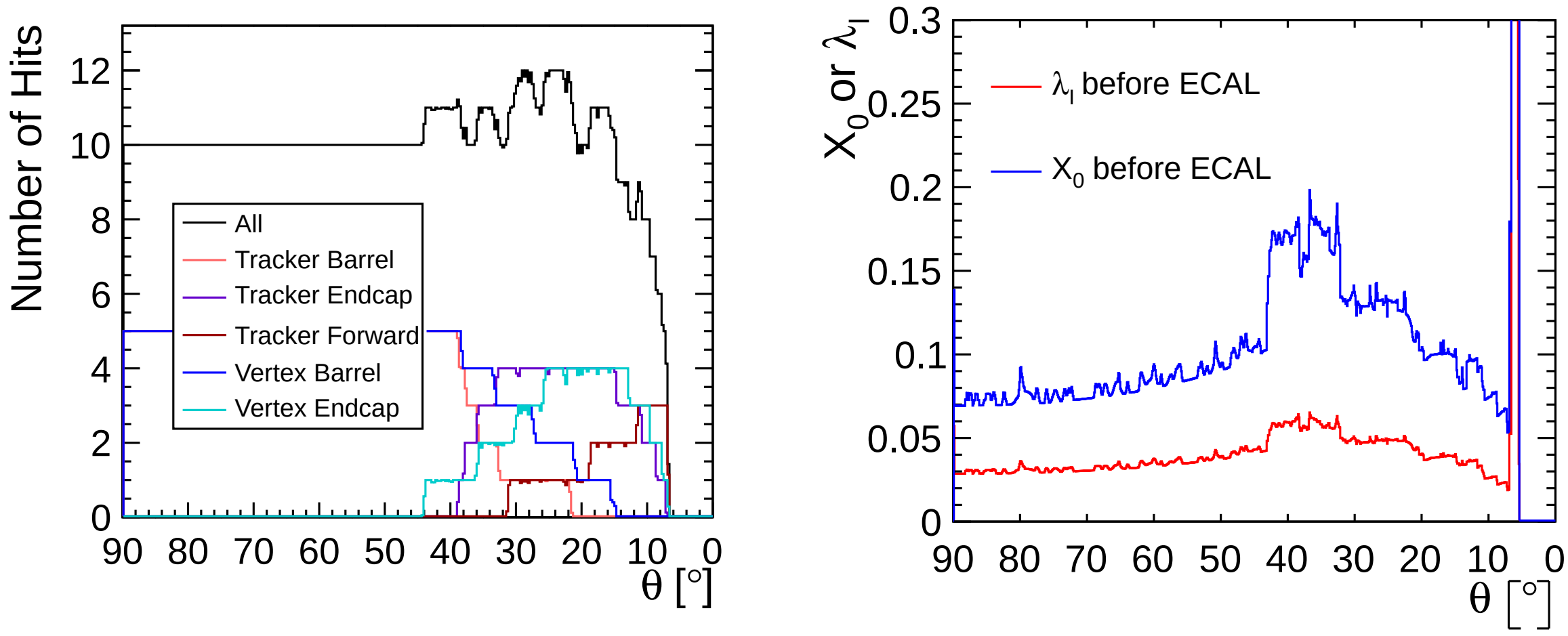

Fig. 5.7: The coverage of the CLIC_SiD tracking system (left) and the amount of material in the tracking region expressed as a fraction of a radiation length $X_0$ (right) as a function of the polar angle $\theta$. The conical beam pipe causes the sharp peak at $\theta \approx 7°$.

A module is the smallest functional unit of the CLIC_SiD tracker. In the barrel, a module consists of a single, square, micro-strip sensor read out by two chips through a single short pigtail cable and glued to a simple low-mass support frame. The endcap modules have the same basic construction, but have identical isosceles-trapezoidal sensors on both sides of the support frame to deliver stereo hit information. The support frame holds the sensors flat and allows for easy handling and stable mounting to the support structures. Each frame is composed of a pair of thin, high-modulus carbon fibre sheets sandwiched around a thin sheet of ROHACELL foam. Large openings are cut in each frame leaving only a pair of cross braces under the sensor to minimise material. Glued to the edges of each module there are three carbon-filled PEEK (poly ether ether ketone) support points which engage a carbon-filled PEEK mounting clip that is glued into the support cylinders. This arrangement allows for easy installation, removal and repair of individual modules.

The barrel sensors, already prototyped by Hamamatsu, are single-sided, AC-coupled, p+ on n-bulk sensors with 25 µm sensor pitch and 50 µm readout pitch. The readout strips are connected by vias to traces on a second metal layer that route the input signals to a pair of bump bonding arrays near the centre of the sensor. A pair of 1024-channel readout ASICs, called KPiX [17], is bonded to those arrays.

The trapezoidal endcap sensors have a similar design, where the readout strips are parallel to one edge of the sensors, at a 6° angle with respect to the plane of symmetry of the trapezoid. This results in a stereo angle between the two sides of 12°. The entire tracker is assembled using only one sensor type in the barrel and two in the endcaps, where the overall dimensions are determined by the largest sensors that can be fabricated on 6" wafers.

The KPiX readout ASIC stores time-stamped hits during each bunch train in analog buffers for digitisation and readout between trains. The version of KPiX that has been developed for the ILC can store up to four hits in each readout channel per train, and provides a time stamp with a resolution of a few hundred nanoseconds, a single bunch at the ILC. For CLIC, it is anticipated that this scheme would be modified to allow hit timing to 5 ns or better while maintaining the same power envelope. With the ASICs bump-bonded directly to the sensor, the double-metal layer on the sensor is used to deliver power and signals for control and readout to the KPiX chips from the short pigtail cable that is bonded to the surface of the sensor between the two chips, see Figure 5.8. Since digitisation and readout occurs only between bunch trains, coupling RF interference from these noise sources to the input signal path is not a concern. The only digital activity during signal integration are a synchronous LVDS clock and comparators firing on the chip when hits are registered. Great care has been taken to minimise the coupling of these pulses to the input path and the susceptibility to them in the KPiX input stages. The power-hungry analog front-end of the KPiX chip is starved for current between bunch trains, reducing power consumption to roughly 20 µW per channel average or 40 (80) mW for a barrel (endcap) module, allowing the tracker to be gas cooled.

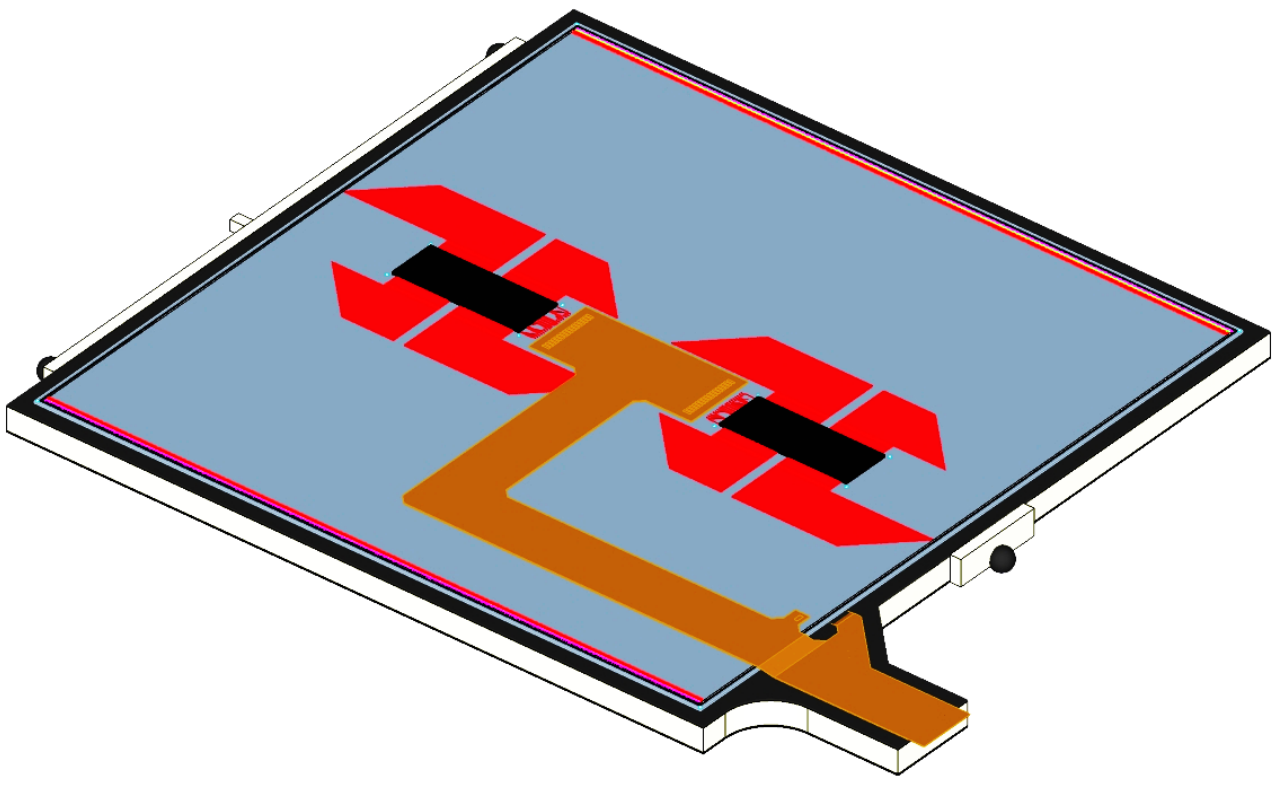

Fig. 5.8: Schematic view of a SiD tracker module placed on a mounting jig. It shows the double-metal readout traces on the sensor (red) that connect the sensor microstrips to a pair of KPiX chips (black). They also route power, readout and control signals from the KPiX chips to a kapton pigtail cable (brown). The kapton pigtail cable also serves to connect the biasing voltage to the sensor back side (bottom-right).

The pigtail cable that reads out each sensor is connected to an extension cable which is routed along the barrel and endcap supports to power and readout distribution boards mounted on the annular support rings that connect successively nested cylinders. The traces on the cables have a layout optimised to minimise Lorentz forces generated by pulsed currents to the modules in a high magnetic field. The distribution boards convert electrical signals to and from the modules to optical signals to the outside DAQ as well as manage DC-DC conversion or serial powering for the connected modules in order to reduce the amount of material and space needed for power and data cables connecting these boards to the power supplies and DAQ located outside of the detector. The radiation length in the CLIC_SiD tracker layout, as used in the CLIC_SiD simulation model, is shown in Figure 5.7.

With unprecedented momentum resolution which relies on very precise single-hit resolution, the stability of detector alignment is critical. In order to monitor the alignment in real time the CLIC_SiD tracker is considering an alignment method based on Frequency Scanned Interferometry (FSI) [18, 19, 20]. This system incorporates multiple interferometers fed by optical fibres from the same laser sources, where the laser frequency is scanned and fringes counted to obtain a set of absolute lengths. Precisions better than 100 nm have been attained using a single tunable laser when environmental conditions are carefully controlled, but precisions under less controlled conditions (e.g. air currents, temperature fluctuations) are an order of magnitude worse.

To overcome this limitation, a dual-laser FSI system is foreseen for the tracker that employs optical choppers to alternate the beams introduced to the interferometer by the optical fibres. Bench tests have achieved a precision of 200 nm under highly unfavourable conditions using the dual-laser scanning technique. Complementary analysis techniques of FSI data can be used either to minimise sensitivity to vibrations in order to determine accurate mean distortions or to maximise sensitivity to vibrations below the Nyquist frequency of data sampling. The latter algorithm could prove especially useful during detector commissioning, since it may help assessing vibration effects, which might arise from pulsed powering in a magnetic field.

## 5.3 Beam-Induced Backgrounds in the Tracking Region

The creation of electron-positron pairs and the production of hadrons in $\gamma\gamma$ interactions are expected to be the dominating sources of background events originating from the interaction region. These processes have been studied with a full GEANT4 simulation of the particle interactions and detector response [21]. The studies have focused on the CLIC_ILD detector model. For the CLIC_ILD TPC, a dedicated study of the occupancies has been performed [22], including also background from beam-halo muons.

### 5.3.1 Occupancies in the Barrel Strip Detectors of CLIC_ILD

Figure 5.9 shows the expected occupancies in the barrel strip detectors, originating from incoherent pairs and $\gamma\gamma \rightarrow$ hadrons. The results shown do not include safety factors for the uncertainties in the production cross sections, the two-photon luminosity spectrum and the simulation of the particle interactions and detector response. Furthermore they only describe the number of particles traversing the detector, not taking into account the formation of clusters of strips due to charge spreading and sharing. For the $\gamma\gamma \rightarrow$ hadrons background, an overall safety factor of two is sufficient to take into account the uncertainties on the predicted rate [23]. For the incoherent pairs, where backscattering effects in the forward region are of particular importance, a larger overall safety factor of five should be used [21]. Assuming these safety factors for the production and simulation uncertainties and an average cluster size of three strips per hit, a readout pitch of 50 µm and a strip-length of 100 mm, the resulting total occupancy for the innermost barrel layer (SIT 1) during a bunch train reaches 2.4 hits/strip. The corresponding occupancy in SIT 2 is about a factor of two smaller. For the two SET layers outside the TPC, the occupancies are almost two orders of magnitude lower than in SIT 1. For comparison, the SIT 1 layer in the CLIC_ILD tracker is located at a radius of 165 mm and extends to $z = 371$ mm, while the inner barrel layer of the SiD tracker is located at a radius of 230 mm and extends to $z = 578$ mm. Due to the higher B-field in CLIC_SiD, the occupancies will be somewhat reduced at a similar radius.

For an efficient hit and track reconstruction, the occupancies have to be reduced to a few percent. Therefore higher segmentation will be required for the two SIT layers. There is scope for implementing a higher segmentation in either space or time. The spatial segmentation can be increased by means of shorter strips. Temporal segmentation may be achieved by implementing multiple readouts per train in combination with time-stamping of hits. The latter option has a better potential to keep the material budget low and will therefore be addressed in an upcoming R&D effort (see Chapter 10).

### 5.3.2 Occupancies in the Forward Strip Detectors of CLIC_ILD

Both the incoherent pairs and $\gamma\gamma \rightarrow$ hadrons backgrounds are strongly forward peaked (see Section 2.1). Therefore the expected background occupancies in the forward region are higher than for the barrel.

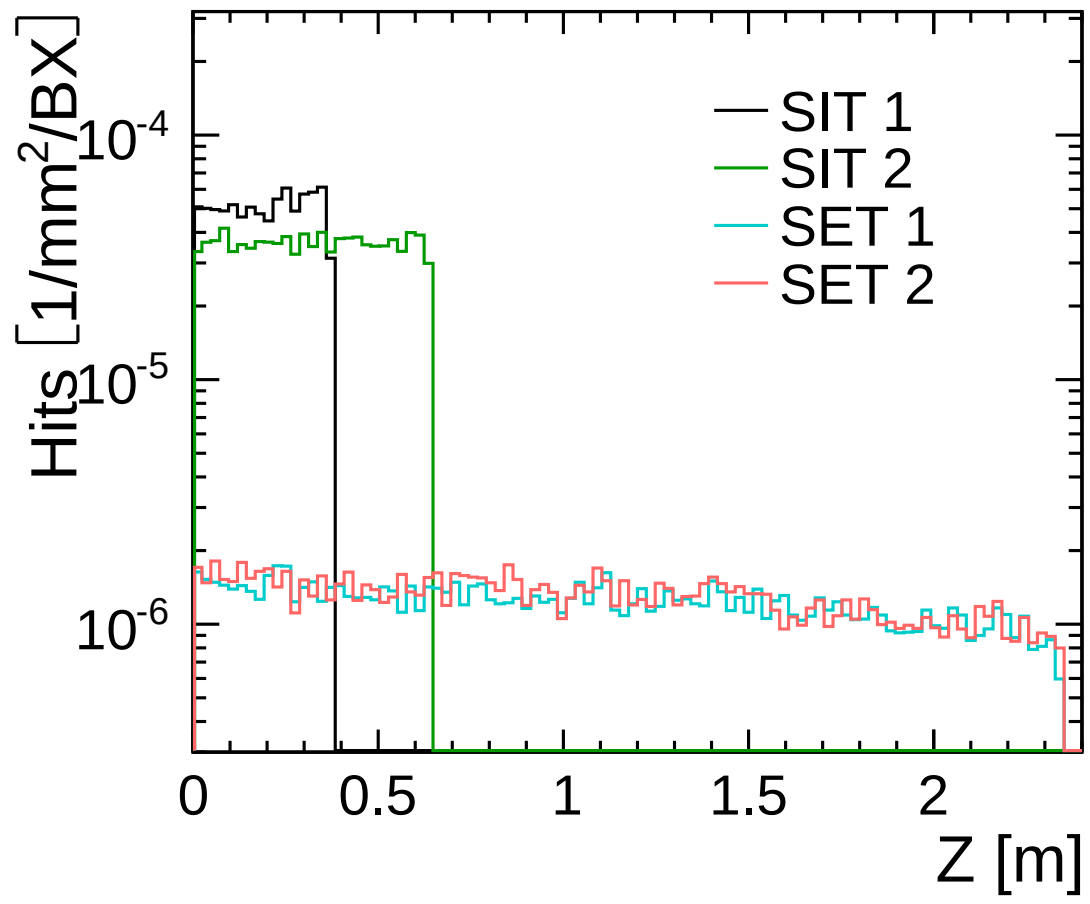

(a) Incoherent pairs

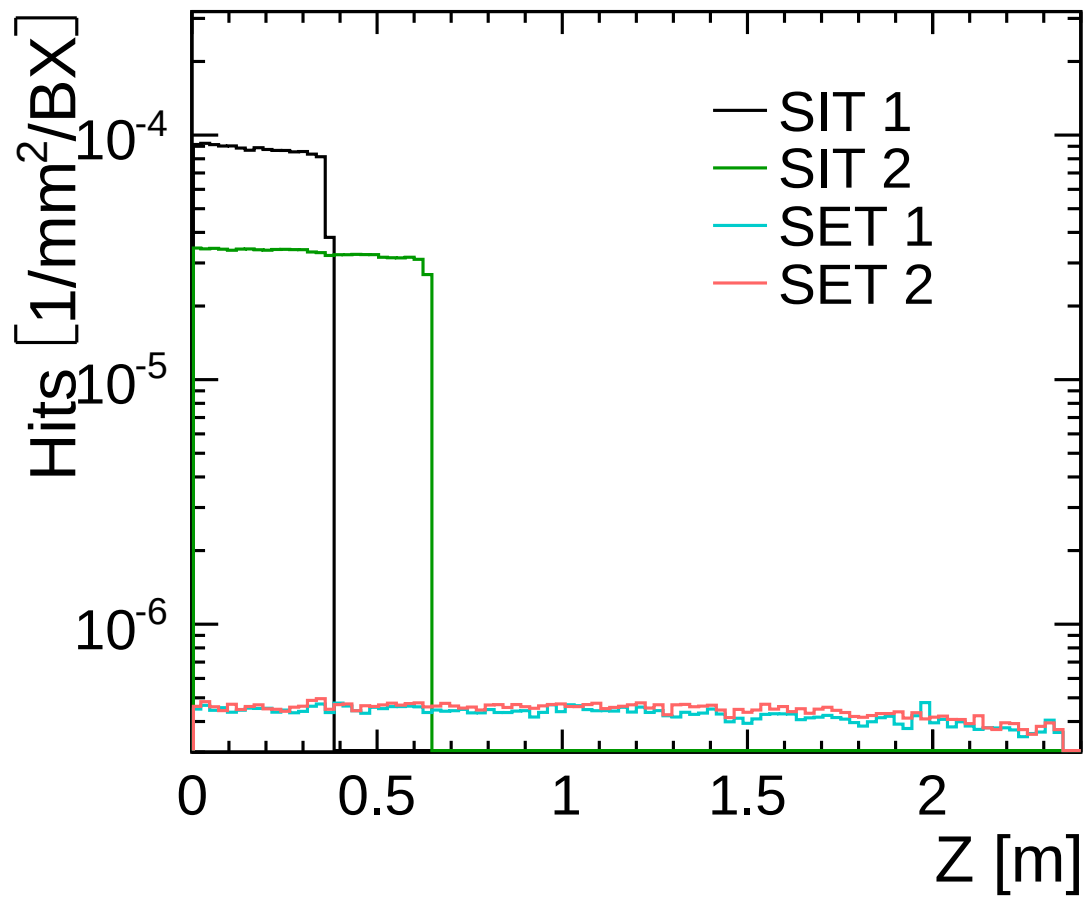

(b) $\gamma\gamma \rightarrow$ hadrons

Fig. 5.9: Hit densities along the ladders in the CLIC_ILD barrel strip detectors for particles originating from incoherent electron-positron pairs (a) and from $\gamma\gamma \rightarrow$ hadrons (b). Safety factors for the simulation uncertainties and cluster formation are not included.

Figure 5.10 shows the radial distributions of the hit densities in the forward tracking disks and in the endcap tracking disks, without safety factors and not taking into account clustering.

Assuming an overall safety factor of five for the incoherent pairs and of two for the $\gamma\gamma \rightarrow$ hadrons, an average cluster size of three strips, a readout pitch of 50 µm and a strip-length of 100 mm, the resulting total occupancy during a bunch train is between 10 hits/strip for the innermost strip disk (FTD 1) and 0.8 hits/strip for the outermost disk (FTD 5). As for the innermost barrel layers, such high occupancies will require additional segmentation also for the FTD layers. For the innermost disk FTD 1, pixel technology is foreseen, reducing the maximal occupancy per train to approximately 1%. For the outer forward disks (FTD 2–5), the occupancies may be reduced to acceptable levels by means of temporal segmentation, as described above for the inner barrel layers. For comparison, in the CLIC_SiD concept similar detector regions are equipped with pixel detectors, as can be seen from Figure 4.2. For the Endcap Tracking Disks (ETD) located behind the TPC endplate the expected background rates are a factor of three below those for the outermost tracking disk due to the larger radial distance from the IP.

### 5.3.3 Occupancies in the TPC

As the readout time of the TPC is much longer than a CLIC bunch train, the TPC integrates the background of the full train. The occupancy is calculated per voxel, where a voxel is a 3D space bucket defined by the readout granularity. The dimensions in the $r\phi$ plane correspond to a pad, in $z$ it is the drift distance corresponding to one ADC time sample. To calculate the number of occupied voxels not only the charge depositions in the fiducial volume have been taken into account, but also effects from charge broadening due to diffusion in the gas, gas amplification and shaping of the electronics. Further details concerning the simulation program used for these dedicated TPC occupancy studies can be found in [24]. Figure 5.11 shows the resulting occupancy for one full bunch train with 312 bunch crossings. For each bunch crossing 3.2 events from $\gamma\gamma \rightarrow$ hadrons, $3 \cdot 10^5$ particles from incoherent pairs (see Table 2.1), and one beam-halo muon are overlaid[1]. $\gamma\gamma \rightarrow$ hadrons are the largest component, followed by the incoherent pairs. The contribution from beam-halo muons is negligible. Using the default pad size of $1 \times 6$ mm$^2$ (Figure 5.11a) the inner pad rows show a high total occupancy of up to 30% of all voxels [22]. With a pad size of $1 \times 1$ mm$^2$ (Figure 5.11b) the occupancy in the inner pad rows can be reduced to 12%. Such a small pad size cannot be implemented with current electronics. Further R&D is required to reach the high readout electronics integration and to assure a sufficient signal to noise ratio, as the signal height per pad is reduced with the pad size. Another option is to chose a pixelised TPC readout, where current R&D shows very promising results (see also Section 5.2.1.1.3).

[1]Note that this number of muons stems from an earlier estimate by the accelerator experts. According to the latest beam halo estimates, a factor 20 fewer muons are expected.

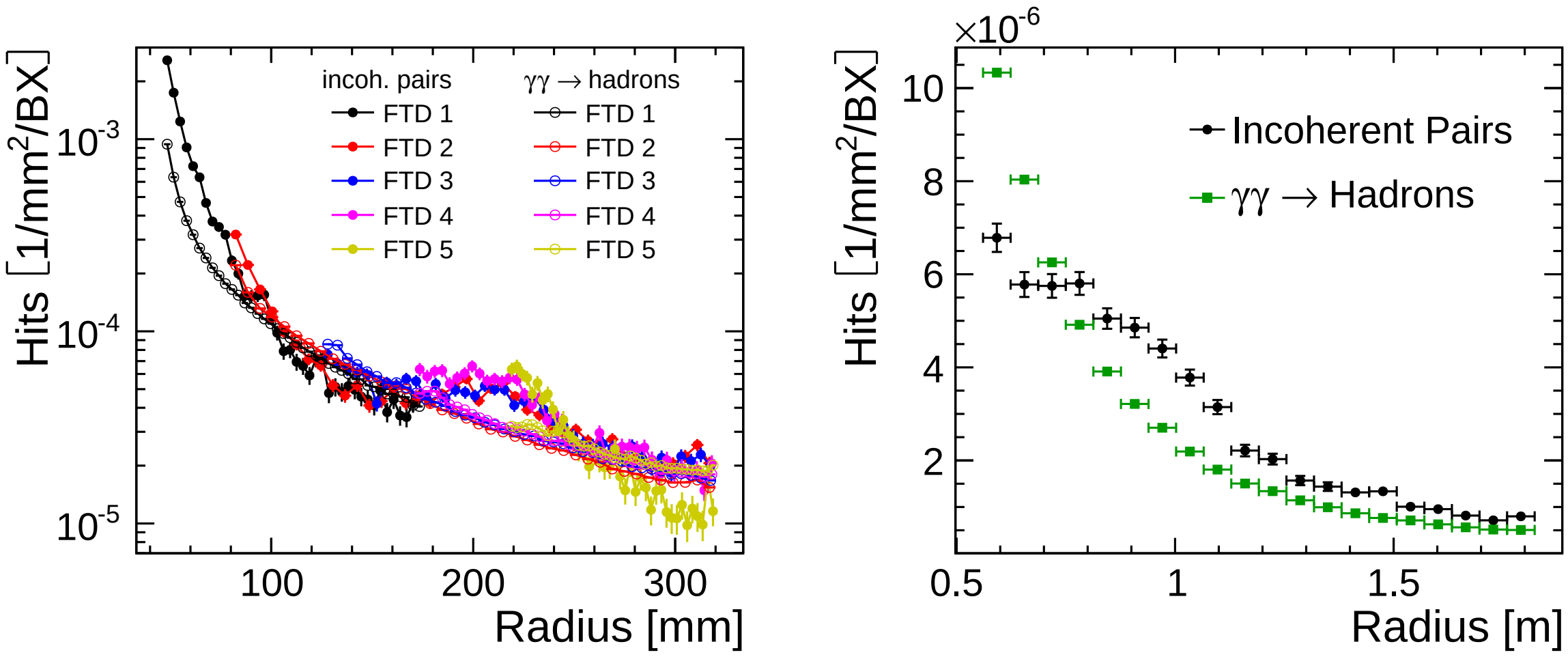

(a) Forward Tracking Disks

(b) Endcap Tracking Disks

Fig. 5.10: Radial distribution of the hit densities per bunch crossing in the CLIC_ILD silicon strip Forward Tracking Disks (a) and in the Endcap Tracking Disks (b) for particles originating from incoherent electron-positron pairs and from $\gamma\gamma \rightarrow$ hadrons. Safety factors are not included.

Further studies will also be needed to assess the occupancy limits for efficient track reconstruction. The current layout of the TPC is not yet optimised with respect to beam-induced backgrounds. A larger inner radius would allow to significantly reduce the background occupancy in the TPC, if needed.

### 5.3.4 Radiation Damage in the Silicon Strip Detectors of CLIC_ILD

The expected radiation damage in the silicon detectors was estimated both for the total ionising dose (TID) and for the non-ionising energy loss (NIEL). The TID was obtained from the energy deposited in the silicon layers. For the NIEL damage, the obtained hit rates have been scaled with a tabulated displacement-damage factor based on the type and energy of the corresponding particle, resulting in the equivalent flux of 1 MeV neutrons leading to the same displacement damage as for the observed spectrum [25]. The TID and NIEL per year of detector operation are quoted assuming an effective runtime of 100 days per year with nominal beam parameters. Table 5.2 summarises the expected maximum flux and doses for the silicon strip detectors in the CLIC_ILD detector model. The values are quoted without safety factors for the simulation uncertainties. For the following discussion, however, we assume an overall safety factor of five for the incoherent pairs and of two for the $\gamma\gamma \rightarrow$ hadrons.

The expected radiation damage from NIEL and TID are moderate in the outer tracking region, where $\gamma\gamma \rightarrow$ hadrons and incoherent pairs contribute similarly. The total expected flux for SIT 1 is $2 \cdot 10^9$ $\mathrm{n_{eq}/cm^2/yr}$. It drops to $5 \cdot 10^8$ $\mathrm{n_{eq}/cm^2/yr}$ for SET 2. The ionising dose amounts to 2 Gy/yr for SIT 1 and 25 mGy/yr for SET 2. For the Forward Tracking Disks (FTD 1–5), the expected flux and ionising dose are higher than in the barrel, reaching between 1 and $2.5 \cdot 10^{10}$ $\mathrm{n_{eq}/cm^2/yr}$ and between 2 and 50 Gy/yr, respectively. The NIEL for the lower part of the ETD is caused mainly by $\gamma\gamma \rightarrow$ hadrons. The expected total flux is $1.5 \cdot 10^{10}$ $\mathrm{n_{eq}/cm^2/yr}$. The maximum ionising dose is 200 mGy/yr.

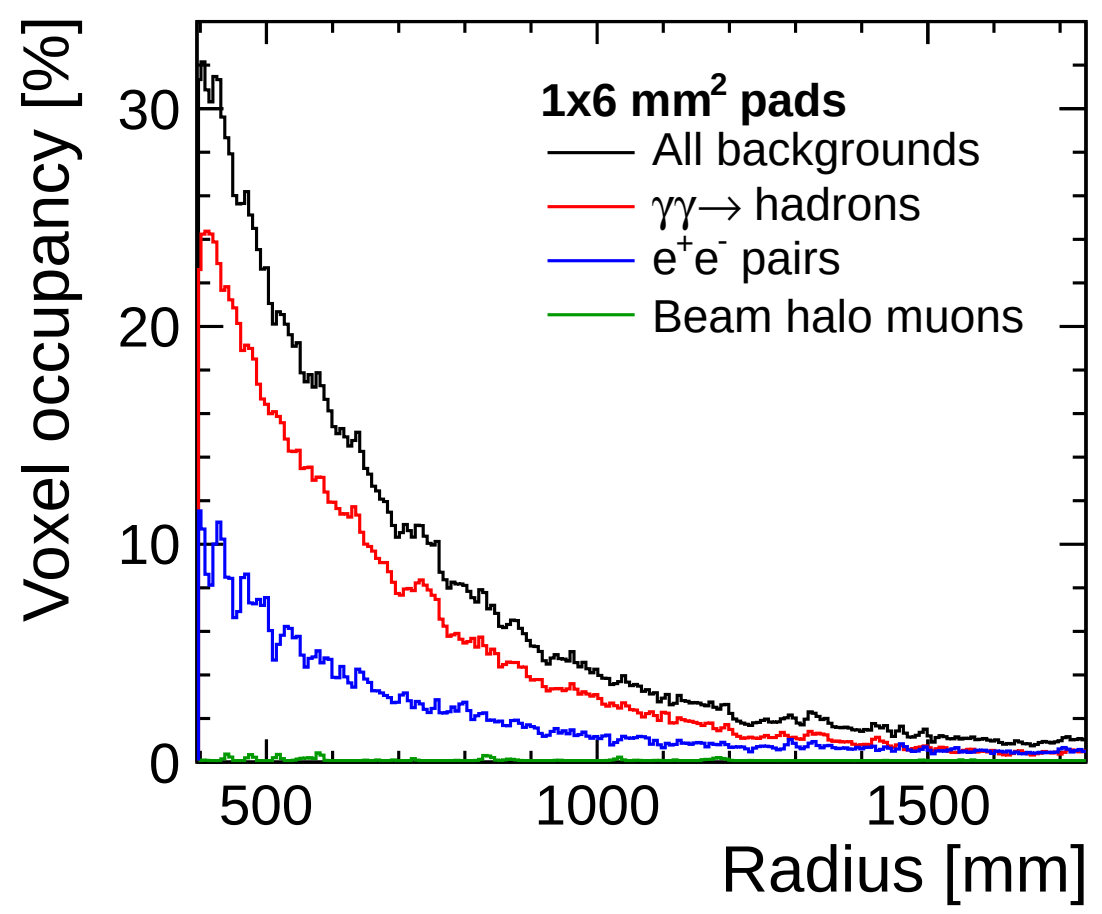

(a) Voxel occupancies for $1 \times 6$ mm$^2$ pads

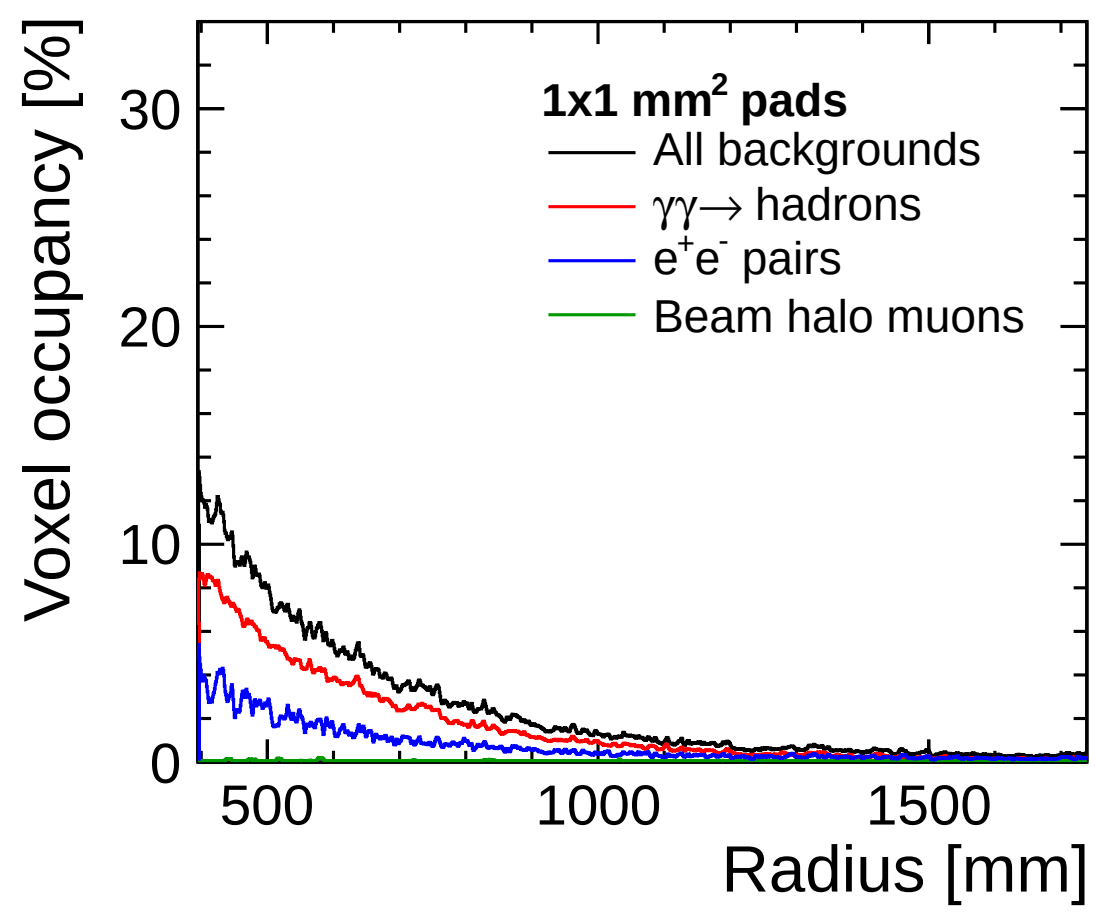

(b) Voxel occupancies for $1 \times 1$ mm$^2$ pads

Fig. 5.11: Voxel occupancies for different pad sizes, averaged per pad row in the TPC for particles originating from $\gamma\gamma \rightarrow$ hadrons, incoherent pairs and beam-halo muons. The data correspond to one complete bunch train and do not include safety factors.

Table 5.2: 1-MeV-neutron equivalent flux and total ionising dose from incoherent pairs and $\gamma\gamma \rightarrow$ hadrons for the silicon strip tracking detectors in CLIC_ILD. The numbers for the barrel-detector elements are averaged over $z$. The values for the forward disks refer to the inner radius of the respective disks. All values are quoted without safety factors for simulation uncertainties.

| | Pairs NIEL [$10^9$ $n_{eq}/cm^2/yr$] | Hadr. NIEL [$10^9$ $n_{eq}/cm^2/yr$] | Pairs TID [Gy/yr] | Hadr. TID [Gy/yr] |
|---|---|---|---|---|
| SIT 1 | 0.154 | 0.504 | 0.226 | 0.429 |
| SIT 2 | 0.211 | 0.249 | 0.154 | 0.165 |
| SET 1 | 0.074 | 0.082 | 0.004 | 0.002 |
| SET 2 | 0.072 | 0.080 | 0.004 | 0.002 |
| FTD 1 | 2.942 | 4.582 | 8.630 | 3.485 |
| FTD 2 | 1.374 | 1.478 | 1.138 | 0.858 |
| FTD 3 | 1.778 | 0.723 | 0.301 | 0.350 |
| FTD 4 | 2.422 | 0.654 | 0.275 | 0.217 |
| FTD 5 | 3.892 | 0.921 | 0.266 | 0.144 |
| ETD | 0.201 | 6.776 | 0.024 | 0.046 |

## 5.4 Performance

### 5.4.1 Tracking Performance of the TPC-based CLIC_ILD Tracking System

With over 200 contiguous readout layers, the TPC as the main tracking device is supposed to allow a very good pattern recognition, even in an environment with a large number of background hits. In addition, the stand-alone tracking capability of the vertex detector enables the reconstruction of low transverse momentum tracks which do not reach the TPC [3].

In the CLIC_ILD detector tracking is done in three steps. In the first two steps, pattern recognition and track fitting is done separately in the TPC and in the silicon sensors. In the third step an algorithm called `FullLDCTracking` [26] is executed, which combines track segments from the two detector components and re-fits the track using a Kalman filter. This also allows to find curling tracks with a transverse momentum of less than 1.2 GeV coming from the interaction point, which can perform many loops, always leaving the TPC through the inner field cage and re-entering. In the TPC stand alone reconstruction this leads to many not-connected helix segments which are combined to one track in the `FullLDCTracking`.

The quality of the reconstructed track depends on the number of hits assigned to the correct Monte Carlo particle. Up to a fraction of 4% of incorrectly assigned hits the resolution does not degrade. Consequently, for all CLIC_ILD tracking studies a track quality cut of at least 96% of correctly assigned hits is applied. For tracks in the forward region with less than 25 hits this cut directly translates into zero incorrectly assigned hits. To avoid a degraded tracking efficiency due to this effect, one hit from a different Monte Carlo particle is always allowed. In order to ensure a good impact parameter resolution, the number of incorrectly assigned hits in the inner silicon detectors (vertex tracker and SIT) is limited to one hit. A more detailed discussion of the track quality cut can be found in [27].

The tracking efficiency has been studied both without any background taken into account and with background from $\gamma\gamma \rightarrow$ hadrons of 3.2 events per BX, where background tracks from 60 bunch crossings have been overlaid for each signal event [27]. For the TPC it would be desirable to simulate a full bunch train of 312 bunch crossings. However, the current reconstruction software cannot cope with the large number of hits.

Figure 5.12 shows the tracking efficiency for tracks in $t\bar{t}$ events as a function of the transverse momentum $p_T$ in the barrel region[2] and as a function of the polar angle $\theta$. The tracking efficiency is $> 99\%$ without background for tracks with $p_T > 2$ GeV, and above 97% down to $p_T > 0.4$ GeV. In the presence of background, mainly consisting of low momentum tracks in the forward direction, the reconstruction efficiency is affected mainly for tracks with $p_T < 1$ GeV, where it drops to 87%. For tracks with $p_T > 2$ GeV the efficiency is not affected by background. The loss of tracking efficiency at higher momenta is correlated with the high rate of badly reconstructed tracks in this momentum range (see below). The dependence of the tracking efficiency on the polar angle $\theta$ is weak and not affected by background (Figure 5.12 (right)). It is between 99% and 98% over the full $\theta$ range. Only for very low angles near the detector acceptance and in the transition region from the FTDs to the SIT between 20° and 25° a slight degradation can be seen.

Figure 5.13 shows the fraction of badly reconstructed tracks in $t\bar{t}$ events for CLIC_ILD with `FullLDCTracking` as a function of the transverse momentum $p_T$ and as a function of the polar angle $\theta$. A badly reconstructed track fails the quality cut described above. For track momenta above 2 GeV the fraction of badly reconstructed tracks is not affected by background. The fraction stays between 1% and 2% up to 20 GeV and is rising for higher momenta up to 12% at 400 GeV. This increase is caused by the silicon tracking (see [27]). Due to the definition of the track quality cut, which only allows one falsely assigned hit in the silicon sensors, confusion in narrow jets causes tracks to fail this stringent requirement. The particle track is still found with a good momentum resolution dominated by the TPC, but track parameters measured by the silicon tracking are not as precise as for well reconstructed tracks. The definition of the track quality cut can be optimised further by including other track observables and energy dependent considerations.

Below 1 GeV, the fraction of badly reconstructed tracks is largely affected by background and is at maximum 10% around 0.4 GeV, which is the most probable transverse momentum of the background tracks. The strong background influence is also visible in the fraction of badly reconstructed tracks as a function of the polar angle $\theta$ for different transverse momentum cuts, $p_T > 0.25$ GeV and $p_T > 2$ GeV. Except for the transition region from the FTDs to the SIT between 20° and 25° the fraction stays below 3% for $p_T > 2$ GeV, whereas the fraction of badly reconstructed tracks is up to 10% for $p_T > 0.25$ GeV.

[2] The $\theta$ angle of the Monte Carlo particle has to be larger than 40°. Particles with $p_T < 1.2$ GeV curl and do not reach the calorimeter barrel.

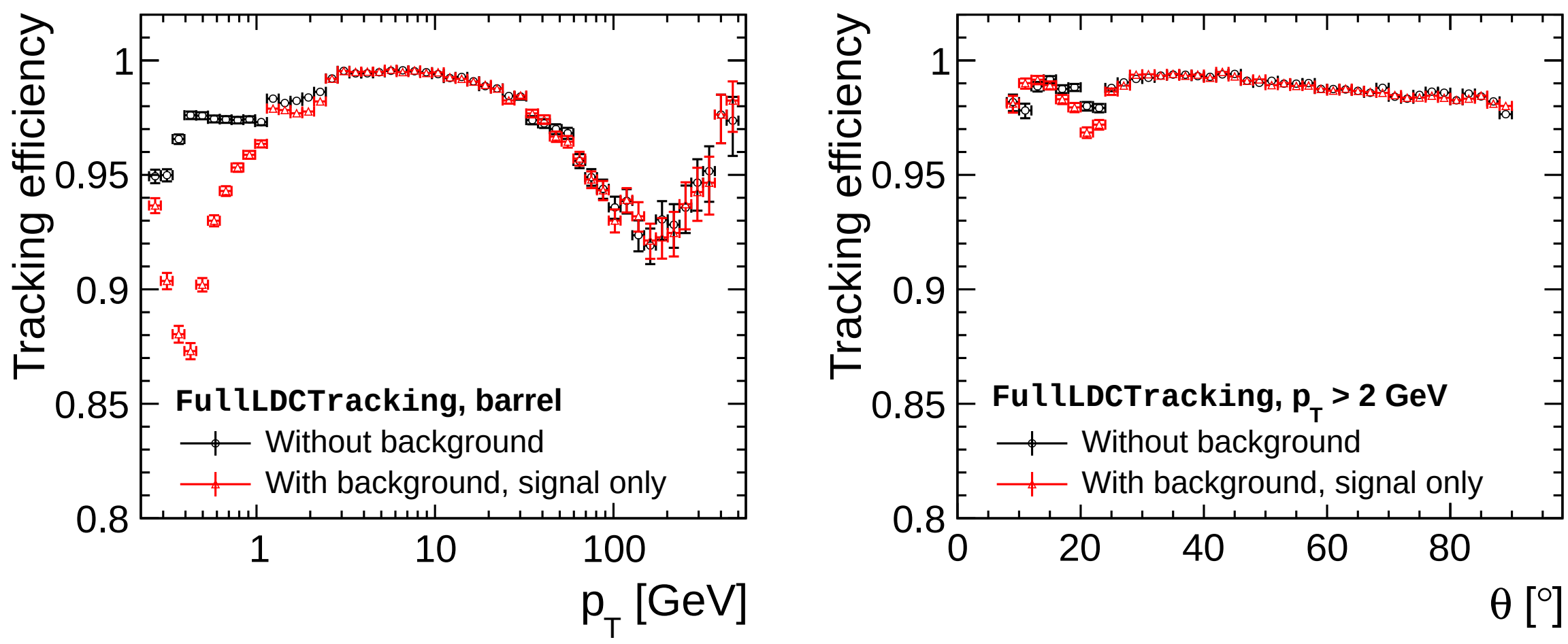

Fig. 5.12: CLIC_ILD `FullLDCTracking` efficiency in $t\bar{t}$ events as a function of the transverse momentum $p_T$ in the detector barrel (left), and as a function of the polar angle $\theta$ (right) for tracks with $p_T > 2$ GeV, with and without background from $\gamma\gamma \rightarrow$ hadrons [27].

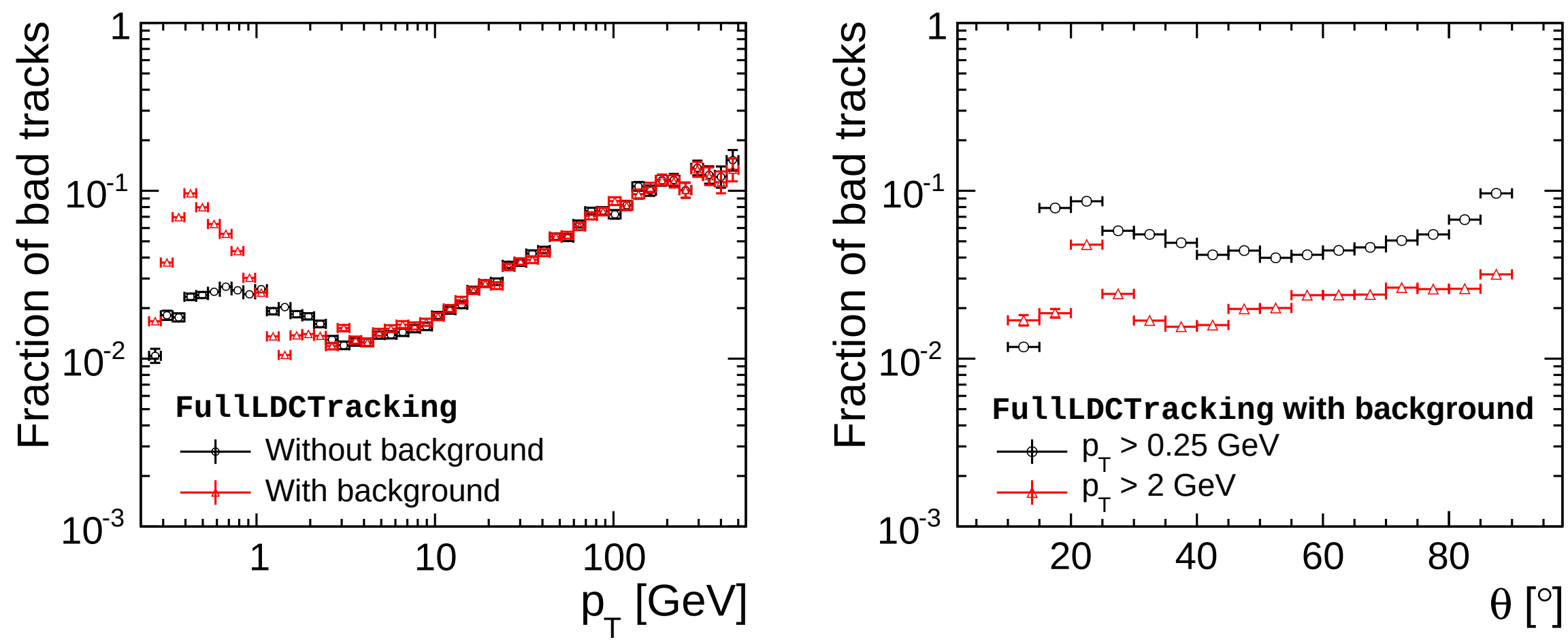

Fig. 5.13: Fraction of badly reconstructed tracks in $t\bar{t}$ events for CLIC_ILD with `FullLDCTracking` as a function of the transverse momentum $p_T$ with and without background from $\gamma\gamma \rightarrow$ hadrons (left), and as a function of the polar angle $\theta$ for $p_T > 0.25$ GeV and $p_T > 2$ GeV with background (right) [27].

The $p_T$ resolution $\sigma(\Delta p_T/p_T^2)$ for single muons is determined from a single Gaussian fit of the distribution $(p_{T,MC} - p_{T,rec})/p_{T,MC}^2$ and is shown in Figure 5.14 as a function of $p_T$ and as a function of the polar angle $\theta$. In the barrel region ($\theta = 40^\circ - 90^\circ$), the $p_T$ resolution is basically constant and meets the design value of $2 \cdot 10^{-5}$ GeV$^{-1}$ for $p_T > 100$ GeV.

The $p_T$ resolution as a function of $p_T$ has been fitted to the parametrisation:

$$\sigma\left(\Delta p_T/p_T^2\right) = a \oplus \frac{b}{p_T} = a \oplus \frac{b}{p \sin\theta} \tag{5.1}$$

where parameter $a$ represents the contribution from the curvature measurement and parameter $b$ represents the multiple-scattering contribution. The fit results for different polar angles are shown in Table 5.3. In the barrel region ($\theta = 80^\circ$), the $p_T$ resolution obtained from the curvature measurement is

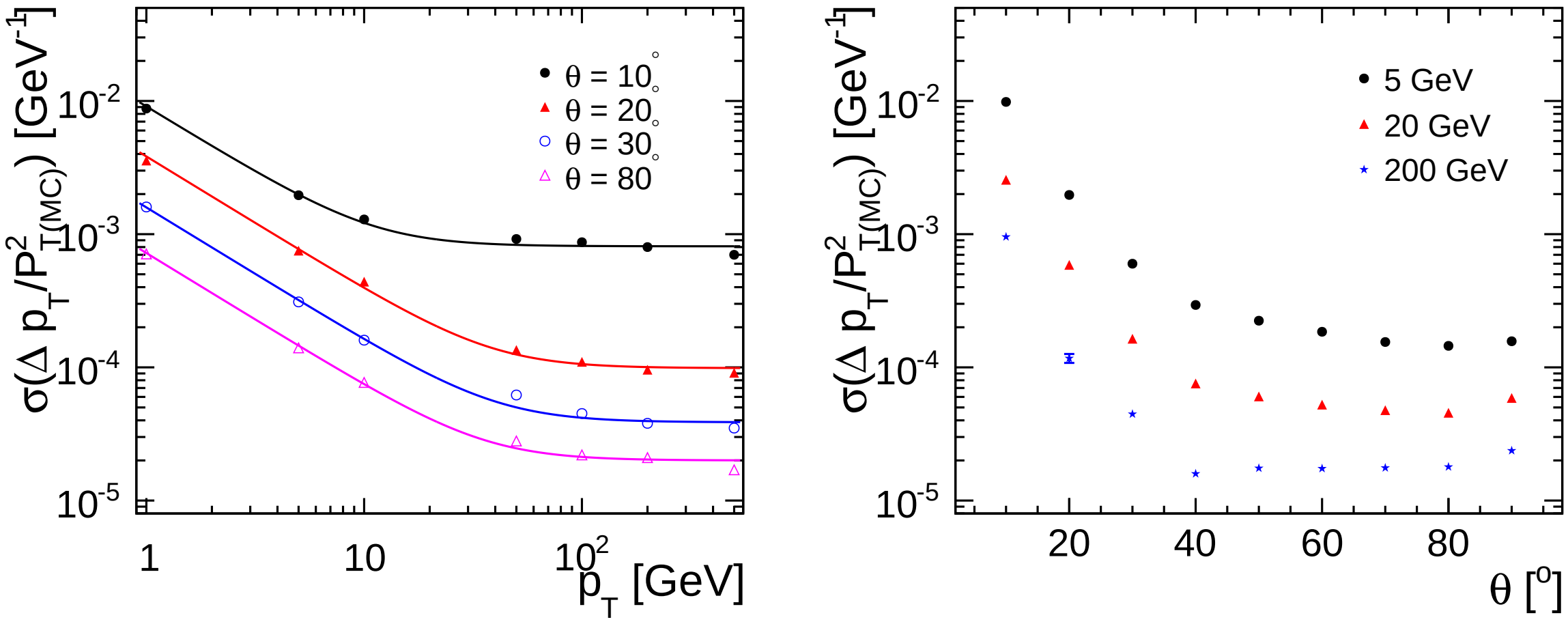

Fig. 5.14: CLIC_ILD $p_T$ resolution $\sigma(\Delta p_T/p_T^2)$ for single muons as a function of $p_T$ at $\theta = 10°$, $20°$, $30°$ and $80°$ (left), and as a function of the polar angle $\theta$ for tracks with $p_T$ = 5, 20 and 200 GeV (right) [28].

$1.97 \cdot 10^{-5}$ GeV$^{-1}$. The $p_T$ resolution has also been studied for tracks in $t\bar{t}$ events and is shown in Figure 5.15. As for single muons, the $p_T$ resolution is fulfilling the design value of $2 \cdot 10^{-5}$ GeV$^{-1}$ for $p_T > 100$ GeV, thus showing no significant deterioration due to the dense environment in $t\bar{t}$ events.

Table 5.3: $p_T$ resolution $\sigma(\Delta p_T/p_T^2)$ in CLIC_ILD for single muons, parametrised by Equation 5.1.

| $\theta$ [°] | $a$ [GeV$^{-1}$] | $b$ |
|---|---|---|
| 10 | $8.19 \cdot 10^{-4}$ | $9.07 \cdot 10^{-3}$ |
| 20 | $9.86 \cdot 10^{-5}$ | $3.83 \cdot 10^{-3}$ |
| 30 | $3.87 \cdot 10^{-5}$ | $1.59 \cdot 10^{-3}$ |
| 80 | $1.97 \cdot 10^{-5}$ | $7.22 \cdot 10^{-4}$ |

#### *5.4.1.1 Time Stamping for Tracks in the CLIC_ILD Detector*

The TPC itself does not directly provide a time stamping possibility for reconstructed hits like the silicon sensors because the arrival time of the charge on the endplate is used to reconstruct the $z$ coordinate, using the drift velocity $v_{\mathrm{drift}}$. The TPC is not able to determine an absolute $z$ position because the arrival time does not only contain the drift time $t_{\mathrm{drift}}$, but also includes the time of the corresponding bunch crossing: $z_{\mathrm{TPC}} = (t_{\mathrm{drift}} + \Delta t_{\mathrm{BX}} \cdot BX)v_{\mathrm{drift}}$. A silicon detector measures the absolute $z$ coordinate, which corresponds to $t_{\mathrm{drift}} \cdot v_{\mathrm{drift}}$ in the TPC. Combining both measurements allows to determine the bunch crossing $BX$ within the bunch train because the time difference between two bunch crossings $\Delta t_{\mathrm{BX}}$ is known [29]. Note that this method does not require any timing information from the silicon sensor, but just a match between the hit in the silicon and the extrapolated TPC track. The Silicon External Tracker has a low occupancy of only $10^{-4}$ hits per mm$^2$ and bunch train (see Figure 5.9). This allows for a mostly unambiguous matching between the silicon hits and the TPC track, which has an uncertainty in $z$ of twice the bunch train length times the drift velocity $\Delta z = 2 \cdot t_{\mathrm{BT}} \cdot v_{\mathrm{drift}} \approx 25$ mm. In case no hit in the outer silicon sensors can be found also a matching with the inner silicon sensors (vertex tracker and SIT) would be possible.

Figure 5.16 shows the difference of the reconstructed bunch crossing and the true Monte Carlo bunch crossing for single muon tracks in the barrel region. The distribution has an r.m.s. of 2.8 bunch crossings, corresponding to 90% of the tracks being reconstructed correctly within ±5 bunch crossings. Details of the impact of the SET resolution on the TPC time stamping can be found in [29].

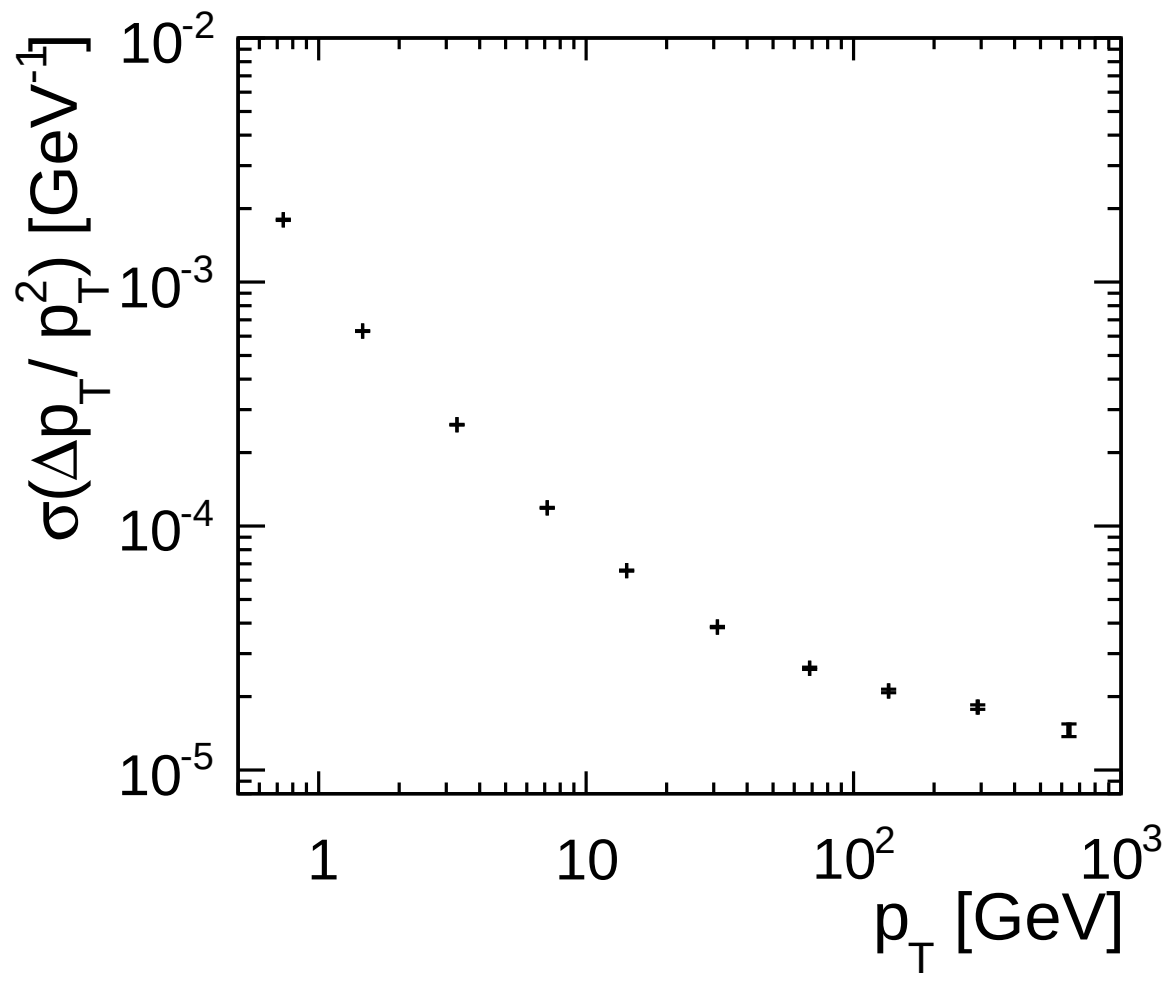

Fig. 5.15: CLIC_ILD $p_T$ resolution $\sigma(\Delta p_T/p_T^2)$ as a function of $p_T$ for tracks in $t\bar{t}$ events using `FullLDCTracking` [27].

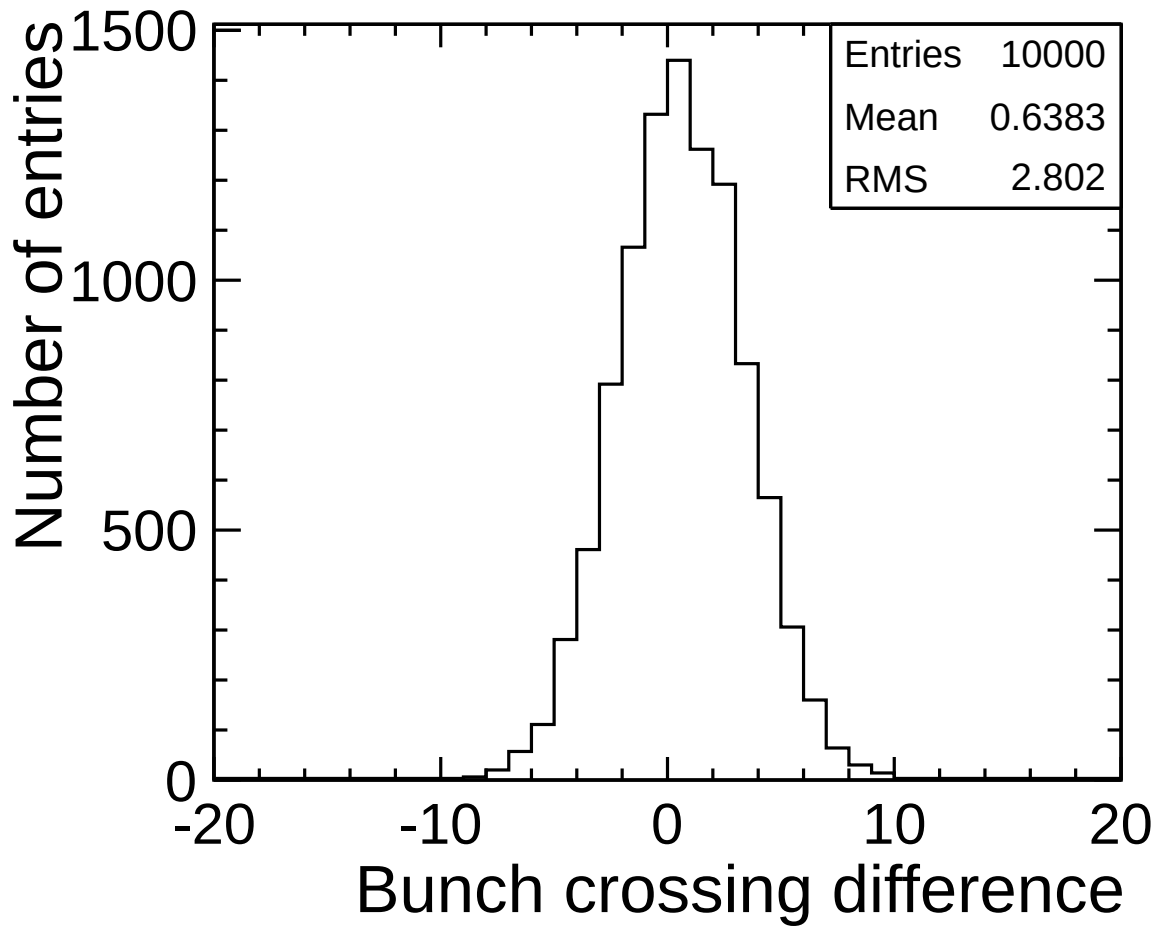

Fig. 5.16: The difference of the reconstructed bunch crossing and the true Monte Carlo bunch crossing for single muons with an energy of 50 GeV at a polar angle of 85° in the CLIC_ILD TPC. The SET $z$ resolution is set to 50 µm [29].

### 5.4.2 Tracking Performance of the All-Silicon CLIC_SiD Tracking System

The tracking performance of the CLIC_SiD detector model [16] has been investigated using the full simulation framework SLIC [30] based on GEANT4 [31]. The detector model assumes a homogenous magnetic field of 5 T parallel to the detector axis throughout the whole tracking region. The reconstruction is done using `org.lcsim` [32] including realistic overlay of $\gamma\gamma \rightarrow$ hadrons [33]. For the background overlay a readout time window of 10 ns is assumed for all silicon detectors, which corresponds to a pile-up of 20 BX of background events. The time window, which starts with the signal event, is placed in the tenth bunch crossing in a small train of 60 BX to assure a realistic background structure also from late hits of earlier bunch crossings. The incoherent pair background is not added in the full simulation, but the two innermost vertex layers are excluded for track seeding since the occupancy in these layers will be a lot higher once this background is added. The tracking performance described below is discussed in more detail in [34].

After the full simulation and the background overlay the simulated tracker hits are passed through a digitisation step. The energy deposition in the silicon is smeared and drifted by a realistic charge simulation. After this, neighbouring hits are clustered together into the final tracker hits. The pattern recognition and track reconstruction is done by the SeedTracker algorithm implemented in `org.lcsim`. The pattern recognition of the algorithm is steered by a set of strategies. Each strategy represents a set of layers in the detector, a minimum number of hits in these layers, and a $\chi^2$-cut. Within a given set of layers, the tracking algorithm tries to find combinations of hits which form a helix. Using strategies, the tracking algorithm is in fact completely decoupled from the actual detector geometry. The algorithm starts by looking for combinations of at least three hits that fulfil a helix fit – once found, this forms a track seed. Subsequently this track seed is extended by successively adding more hits that are consistent with the extrapolation of the seed helix. The number of hits required is user-defined. If fewer than the minimum required number of hits are found, the track candidate is discarded. Tracks may be found by different strategies within the set. If the tracks found in different strategies share more than one hit, only the track with the best fit is kept. In order to reduce the number of possible combinations of hits which could form a track, the algorithm uses a vertex constraint when looking for a track seed. This constraint is usually loose enough to still find all tracks from displaced vertices (see Section 12.3.4). The vertex constraint is only used in the track finding, not in the final track fit.

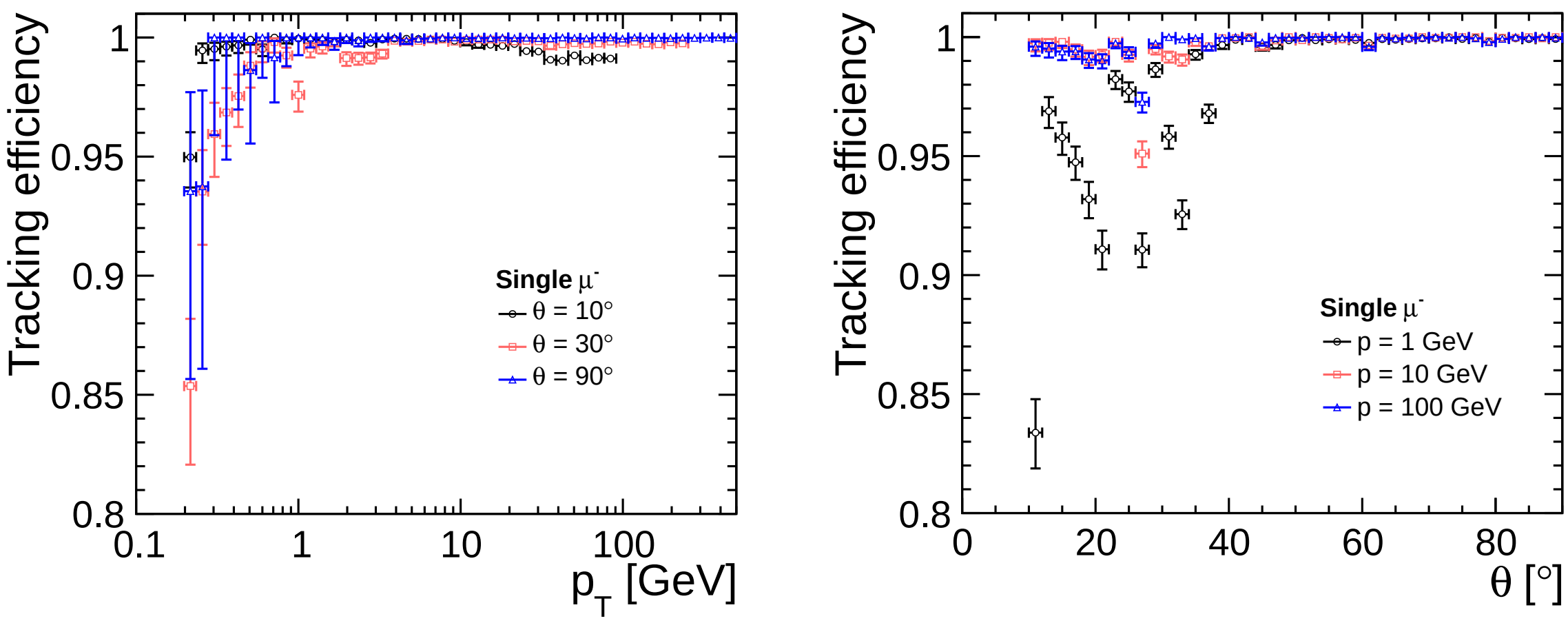

Fig. 5.17: CLIC_SiD tracking efficiency for single muons as a function of $p_T$ at $\theta = 10°$, $30°$ and $90°$ (left), and as a function of the polar angle $\theta$ at $p = 1$, 10 and 100 GeV (right).

A limitation of the current algorithm is that particles originating from long lived decays do not leave enough hits in the tracking system to be found. It has been shown that another algorithm, which uses calorimeter clusters as seeds, can efficiently recover those non-prompt tracks missed by the basic tracking algorithm [35]. Calorimeter assisted tracking has not been part of the reconstruction chain used for the detector benchmark analyses described in this CDR. The set of strategies used for CLIC_SiD requires at least seven hits to form a track, and the $\chi^2$-cut for rejecting track candidates is set to 10. The vertex constraint is fixed at 5 mm in the $r\phi$-plane and at 10 mm in $z$. In order to train the set of strategies, a sample of single muons with a flat distribution in $\phi$, $\theta$ and $p$ was used. A set of strategies which is able to find all particles within the sample that leave at least seven hits in the vertex and tracker was generated. An additional barrel-only strategy, requiring only six hits, was added later to be able to find more particles with lower $p_T$ in the central region.

Similarly to CLIC_ILD, the tracking efficiency is defined as the fraction of findable Monte Carlo particles that can be matched to well reconstructed tracks. In this study, findable Monte Carlo particles are defined as those charged particles whose origin is closer than 50 mm to the interaction point and which lived long enough to travel at least 50 mm along their helical path. A well reconstructed track is defined as a track which has more than 75% of its assigned hits originating from a single particle. The fake rate is defined as the fraction of badly reconstructed tracks. A track is considered badly reconstructed if the number of hits associated with the track which are originating from other particles than the one contributing the most hits exceeds a certain limit. Due to the low number of hits required to form a track (six to seven hits), every single falsely assigned hit can reduce the quality of a reconstructed track significantly. This definition of the fake rate is rather stringent. Defining a fake as a track where every single hit originates from a different particle, such that no particle travelled along the reconstructed helix even partially, would be a less stringent definition.

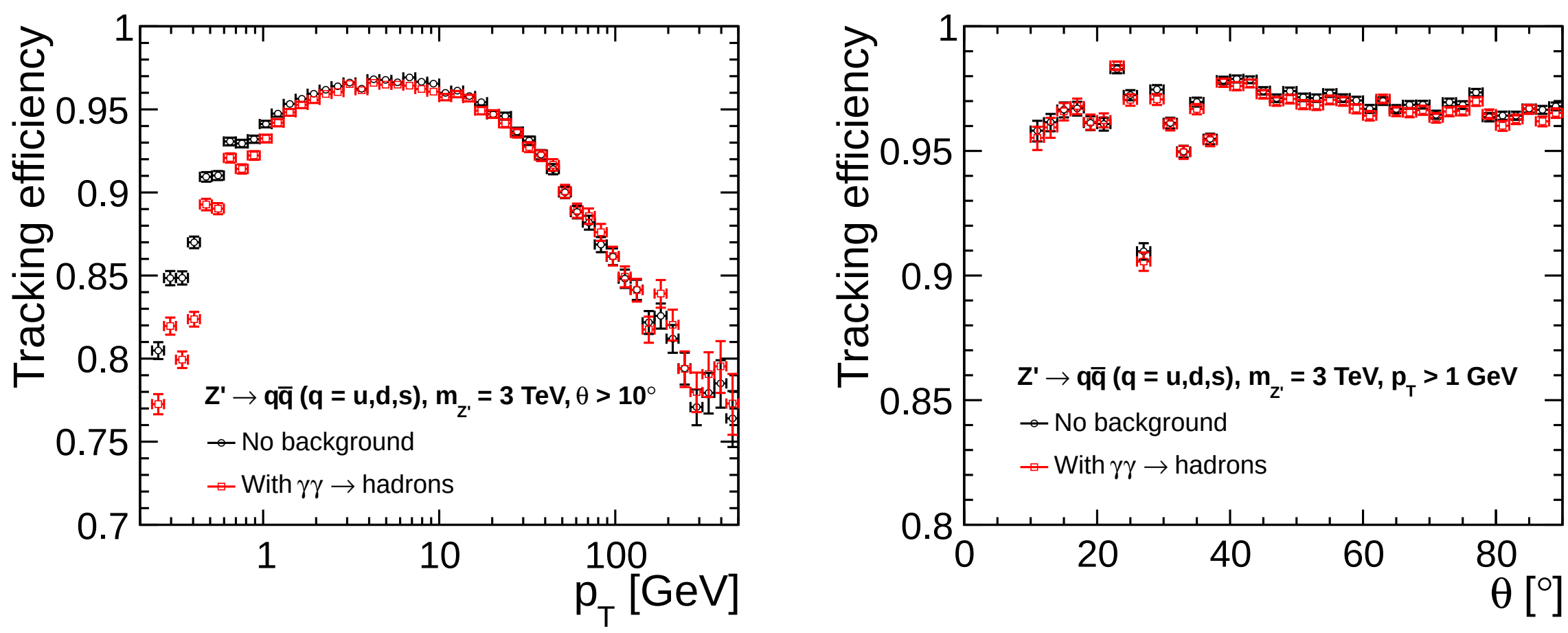

Fig. 5.18: CLIC_SiD tracking efficiency in di-jet decays of a Z-like particle with a mass of 3 TeV with and without background from $\gamma\gamma \rightarrow$ hadrons. Shown is the fraction of findable signal particles that are reconstructed, as a function of $p_T$ (left), and as a function of the polar angle $\theta$ (right).

Figure 5.17 shows the tracking efficiency for single muons in CLIC_SiD. The tracking efficiency is almost 100% for single tracks in the barrel region, even for $p_T$ as low as 0.3 GeV. The efficiency drops slightly for smaller polar angles. In particular, in the transition region between barrel and endcap at a polar angle of around 30° the tracking efficiency drops by 2% even for tracks with very high momenta. For low momentum tracks, the tracking efficiency is lower throughout the whole endcap region because of the higher material budget. The efficiency drops sharply at polar angles of around 8.5° which is the point where less than the required 7 layers are passed (see Figure 5.7).

The tracking efficiency for particles in jets has been studied using decays of a hypothetical heavy gauge boson with a mass of 3 TeV into two light quark jets. In these events the 3 TeV are shared between only two jets, leading to high local track densities. Therefore these events are particularly challenging for the pattern recognition. Figure 5.18 shows the tracking efficiency in this kind of events with and without the addition of $\gamma\gamma \rightarrow$ hadrons background. In case of background overlay the efficiency is referring only to the signal particles. The overall efficiency is about 96% throughout the barrel region for tracks with a moderate $p_T$. Curlers with a $p_T$ of less than 1 GeV in the central region are less likely to pass the required number of different layers to be identified as a track. In combination with the high occupancy within a dense jet this leads to a significant loss of efficiency for this kind of particles. High momentum tracks are usually in the core of the jet which leads to a lower efficiency for these tracks, due to the high local occupancy. Starting at a $p_T$ of about 20 GeV the tracking efficiency is degrading, dropping below 80% for tracks with a $p_T$ of about 400 GeV. This effect can also be seen in Figure 5.19, which shows the tracking efficiency as a function of the closest distance between any of the hits created by the true Monte Carlo particle and any other hit. It can be seen that the efficiency to find a track from a particle is degrading if there is another hit found within 120 µm of any of its hits. This corresponds to six times the pixel size used in the digitisation. Any particle closer than 40 µm will produce a hit in a directly adjacent pixel and the clustering will just create a single tracker hit. Overall the addition of $\gamma\gamma \rightarrow$ hadrons background has little influence on reconstruction efficiency of signal particles.

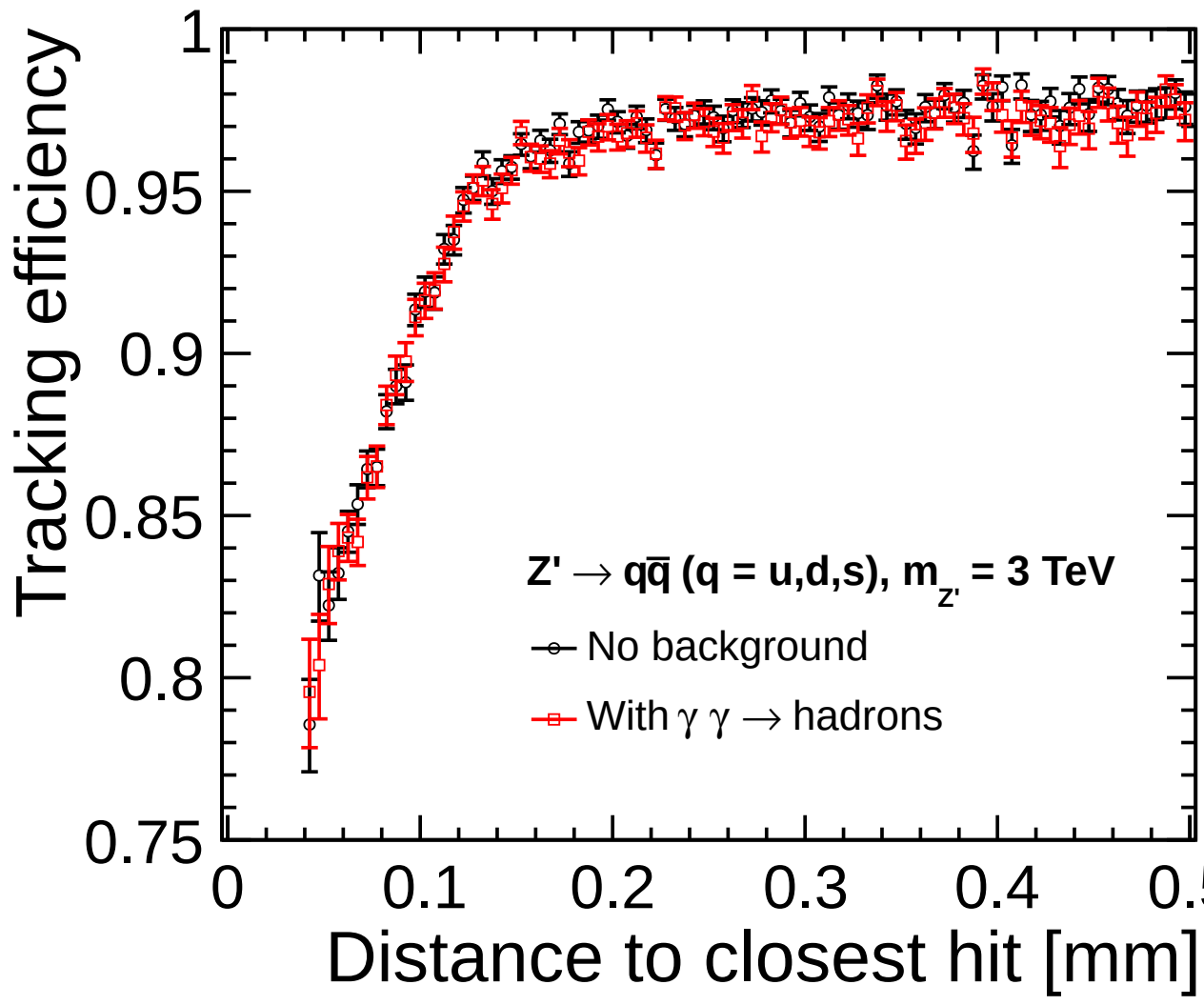

Fig. 5.19: CLIC_SiD tracking efficiency in di-jet decays of a Z-like particle with a mass of 3 TeV. Shown is the fraction of findable particles that are reconstructed, as a function of the closest distance between any of the tracker hits of the true Monte Carlo particle and any other tracker hit.

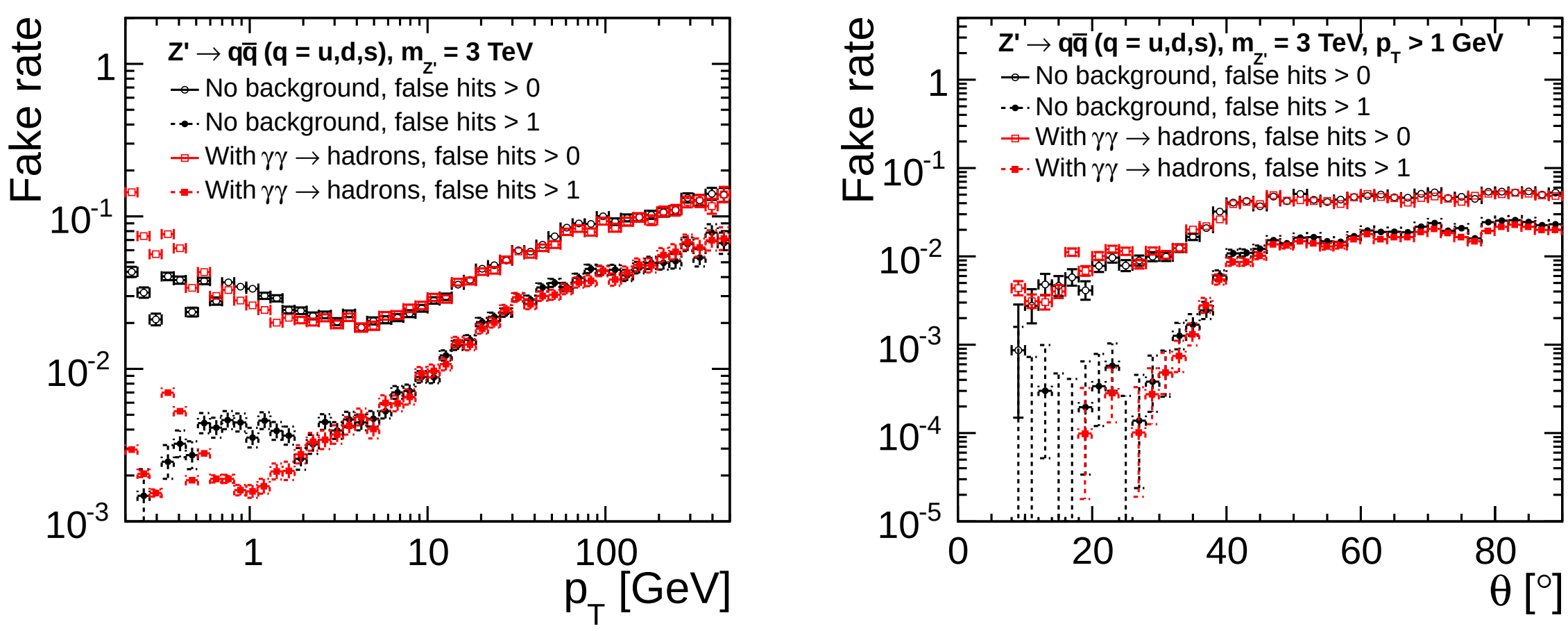

Fig. 5.20: CLIC_SiD fake rate for tracks in di-jet decays of a Z-like particle with a mass of 3 TeV with and without background from $\gamma\gamma \rightarrow$ hadrons. Shown is the fraction of tracks that have more than zero or one hits from other particles assigned to them, as a function of $p_T$ (left), and as a function of the polar angle $\theta$ (right).

Table 5.4: $p_T$ resolution $\sigma(\Delta p_T/p_T^2)$ in CLIC_SiD for single muons, parametrised by Equation 5.1.

| $\theta$ [°] | $a$ [GeV$^{-1}$] | $b$ |
|---|---|---|
| 90 | $7.3 \cdot 10^{-6}$ | $2.0 \cdot 10^{-3}$ |
| 30 | $1.9 \cdot 10^{-5}$ | $3.8 \cdot 10^{-3}$ |
| 10 | $4.0 \cdot 10^{-4}$ | $5.3 \cdot 10^{-3}$ |

The fake rate for tracks in jet events as a function of the polar angle $\theta$ is shown in Figure 5.20 (right). In the barrel detector more than 95% of the reconstructed tracks do not have a single hit from a different particle assigned to them. The fake rate is much lower in the forward region since there are more pixelated layers which largely reduce the local occupancy. Tracks with a higher momentum are more likely to pick up false hits because they are in the centre of the jet, as shown in Figure 5.20 (left). The addition of $\gamma\gamma \rightarrow$ hadrons background does not affect the fake rate in a significant way. Since the background adds a lot of tracks with a $p_T$ of a few GeV, which are evenly distributed throughout the detector, the fake rate actually improves for tracks of these momenta. Only for tracks with a $p_T$ of less than 0.5 GeV the fake rate is worse when adding the background.

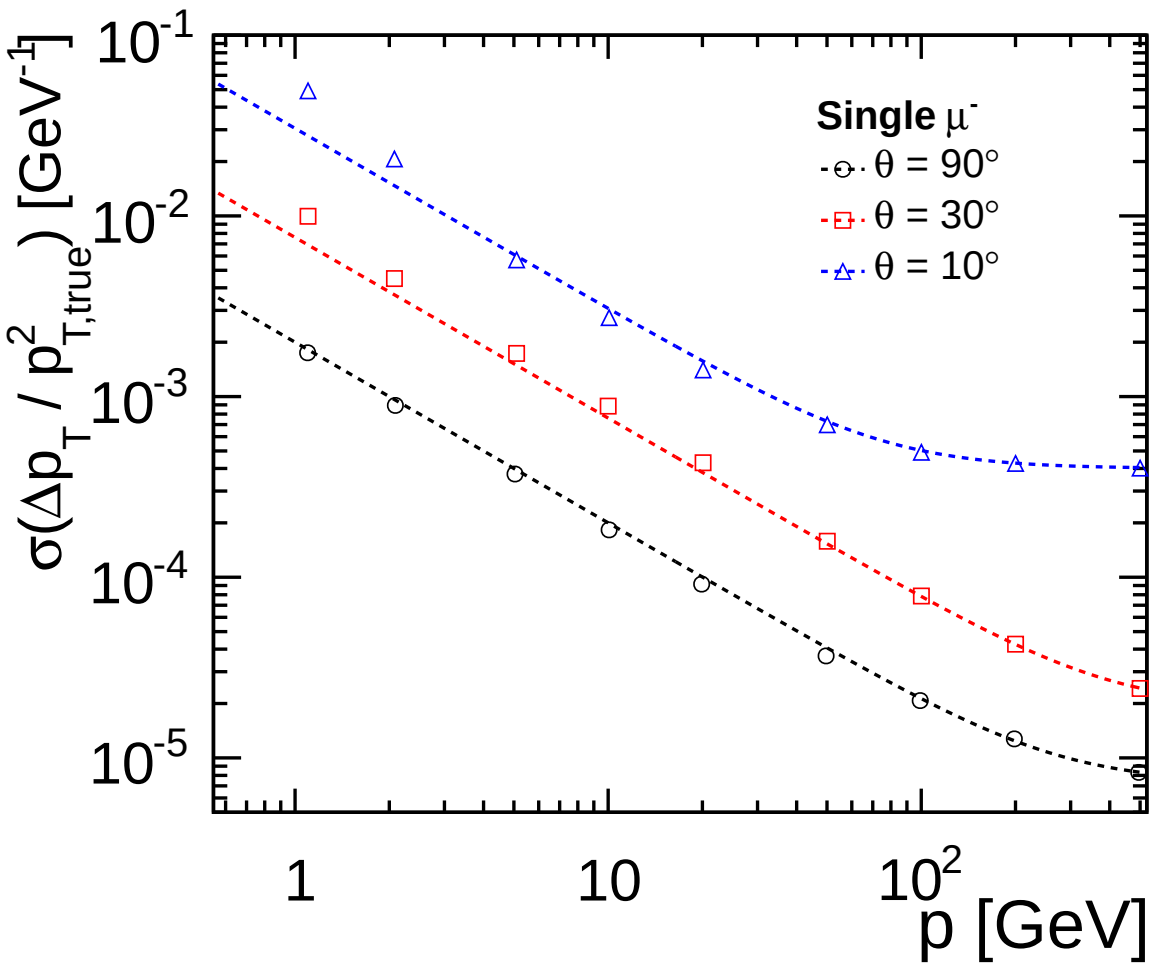

Fig. 5.21: CLIC_SiD $p_T$ resolution as a function of $p$ for single muons with polar angles $\theta$ of 10°, 30° and 90°. The dotted lines show a fit to the parametrisation given in Equation 5.1. The fitted parameters are given in Table 5.4.

Figure 5.21 shows the transverse momentum resolution for single muons of different momenta and polar angles. The values have been obtained by a single Gaussian fit to the $\Delta p_T/p_{T,\mathrm{true}}^2$ distribution. Outliers have been removed before the fit was performed, by applying a cut of five times the r.m.s around the mean of the distribution. The resulting momentum dependence has been fitted to the parametrisation in Equation 5.1. The fit results for different polar angles are shown in Table 5.4. It can be seen that the required $p_T$ resolution of $2 \cdot 10^{-5}$ GeV$^{-1}$ is achieved for tracks with a polar angle down to 30° and even exceeded for more central tracks. For low $p_T$ tracks at small polar angles the values deviate from the parametrisation because the particles pass a lot more material while curling.

# Chapter 6

# Calorimetry

## 6.1 A Particle Flow Calorimeter for TeV Energies

Electron-positron collision events at TeV energies are characterised by multi-fermion final states, which give rise to signatures with six to eight jets. Intermediate production of heavy bosons – W, Z, and presumably Higgs – occurs in many physics channels. These bosons decay predominantly into hadron jets and must be identified by means of the di-jet invariant mass. The particle flow approach allows to satisfy the CLIC requirements on jet energy resolution, described in Section 2.2.2, and will allow in particular to separate W and Z bosons on an event-by-event basis. The basic principles of Particle Flow Analysis PFA are explained in Section 2.3.1. PFA allows to distinguish particles within jets and to optimise the jet energy measurement by making optimal use of the combined information from both the tracking and the calorimetry. An extensive description of the PANDORAPFA particle flow software algorithm and of the considerations for calorimeter optimisation in view of PFA can be found in [1]. When combined with precise timing capabilities, fine-grained calorimeters and PFA software tools also allow to efficiently identify energy clusters originating from beam-induced background events and to separate them from the interesting physics event (see Section 2.5.1 and Section 12.1.4). Consequently, imaging calorimeters and PFA are driving factors in the detector concept designs presented in this CDR.

PFA also forms the starting point of the ILC detector concepts [2, 3] and a broad R&D programme for calorimeter technologies with ultra-fine three-dimensional segmentation [4] is already ongoing in the linear collider context. Since jet and particle energies scale only logarithmically with the centre-of-mass energy, and shower shapes vary only mildly with particle energy, the granularity of the ILC detectors can be considered as close to optimal for CLIC, and are not re-optimised at this stage. Nevertheless, compared to the ILC case, the calorimeter design and technology for CLIC has to address additional challenges which arise from the higher energy and the more intense backgrounds.

### 6.1.1 Tungsten as Absorber for the ECAL and HCAL

The interesting jet energy range at CLIC extends from 50 to 1000 GeV. Leading particles in 3 TeV multi-jet final states will carry sizable fractions of the jet energies, and the single particle spectrum reaches into the few hundred GeV range. The relative importance of the different contributions to the jet energy resolution obtained with the PANDORAPFA particle flow algorithm were studied as a function of the jet energy in [1]. For the range under consideration at CLIC, the intrinsic calorimetric resolution for neutral hadron energies becomes less important, while confusion and leakage increasingly contribute. The most important confusion effect is the wrong assignment of energy deposits between charged and neutral hadrons, followed by charged hadron and photon separation. Concerning the CLIC calorimeters, the imaging power, i.e., the fine segmentation, and a sandwich design which minimises transverse shower extensions, are even more strongly emphasised than in the energy range of the ILC, as one would qualitatively expect from the higher particle density in jets.

Table 6.1: Radiation and nuclear interaction lengths for Fe and W [5].

| Material | $\lambda_I$ [cm] | $X_0$ [cm] | $\lambda_I/X_0$ |
|---|---|---|---|
| Fe | 16.77 | 1.76 | 9.5 |
| W | 9.95 | 0.35 | 28.4 |

To avoid leakage from energetic particles the hadron calorimeter has to provide sufficient depth. Increasing the thickness of the steel structure of the barrel hadronic calorimeters beyond the design values optimised for the ILC at 500 GeV would require an increase of the diameter of the solenoid accordingly. Based on the engineering experience gained with the CMS magnet, this is prohibitive for reasons of cost and operational risks.

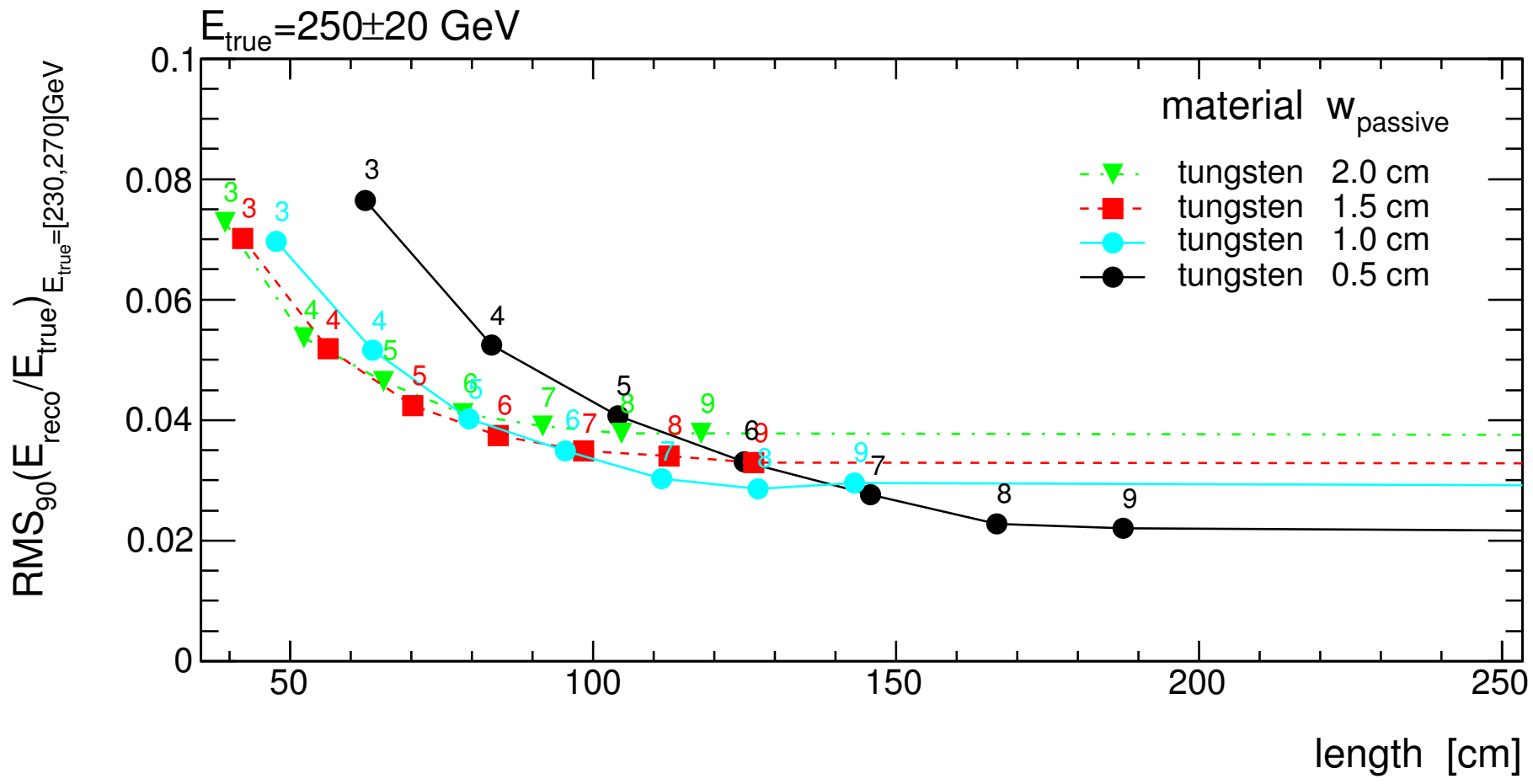

Fig. 6.1: Energy resolution for particles in the tungsten HCAL with an energy from 230 GeV to 270 GeV as a function of the HCAL depth, for different passive layer thicknesses. The numbers next to the data points refer to the total depth in terms of $\lambda_I$. The term $RMS_{90}$ is explained in the text.

Instead, a denser absorber material is required for the barrel HCAL. Uranium, which was successfully used in the ZEUS experiment [6, 7], for example, would be in principle an option, but is disfavoured due to environmental and radiation protection aspects. These considerations lead to the choice of tungsten as absorber material for both the electromagnetic and the hadronic calorimeters at CLIC.

The general design of the electromagnetic calorimeter follows that of the ILC detector concepts very closely, whereas the hadronic calorimeter needs to be re-designed. Tungsten has a nuclear interaction length $\lambda_I$ almost two times smaller than that of steel (see Table 6.1), and a radiation length $X_0$ five times smaller than that of steel. Since the ratio $\lambda_I/X_0$ changes, the sampling structure cannot simply be scaled, but has to be re-optimised. The sampling structure cannot be deduced from existing heavy absorber calorimeters, where there is usually very fine sampling (order of $X_0$) in the longitudinal direction. Such a fine segmentation would minimise sampling fluctuations, but it would not allow for independent readout of every individual layer, as required for particle flow, without increasing the detector volume and the channel count too much.

A GEANT4 study, using the `QGSP_BERT_HP` physics list, was performed in order to compare the performance of tungsten and steel hadronic calorimeters [8]. The absorber layer thickness was varied, while for the active layer a fixed thickness of 5 mm for scintillators plus 2.5 mm for electronics was assumed. To quantify the energy resolution, the measure $RMS_{90}$ has been used, defined in [1] as the root mean square of the smallest interval containing 90% of the distribution. The energy resolution for 250 GeV particles is shown as a function of the total HCAL thickness (labelled "length") in Figure 6.1. The transition from a leakage-dominated range to an asymptotic region governed by the intrinsic resolution can be clearly seen. The stochastic term of the intrinsic resolution improves if the plate thickness is reduced from 2.0 to 0.5 cm, but the total depth needed to reach the region where leakage does not dominate anymore increases from 100 to 170 cm. Figure 6.1 motivates a thickness of 1 cm as optimal choice: Thinner plates give a too large calorimeter, whereas for thicker ones the resolution degrades while gaining only modestly in volume. For steel structures with absorber plate thicknesses around 2 cm, an additional depth of about 50 cm (relative to a tungsten design) would be needed in order to reach the asymptotic region for this particle energy.

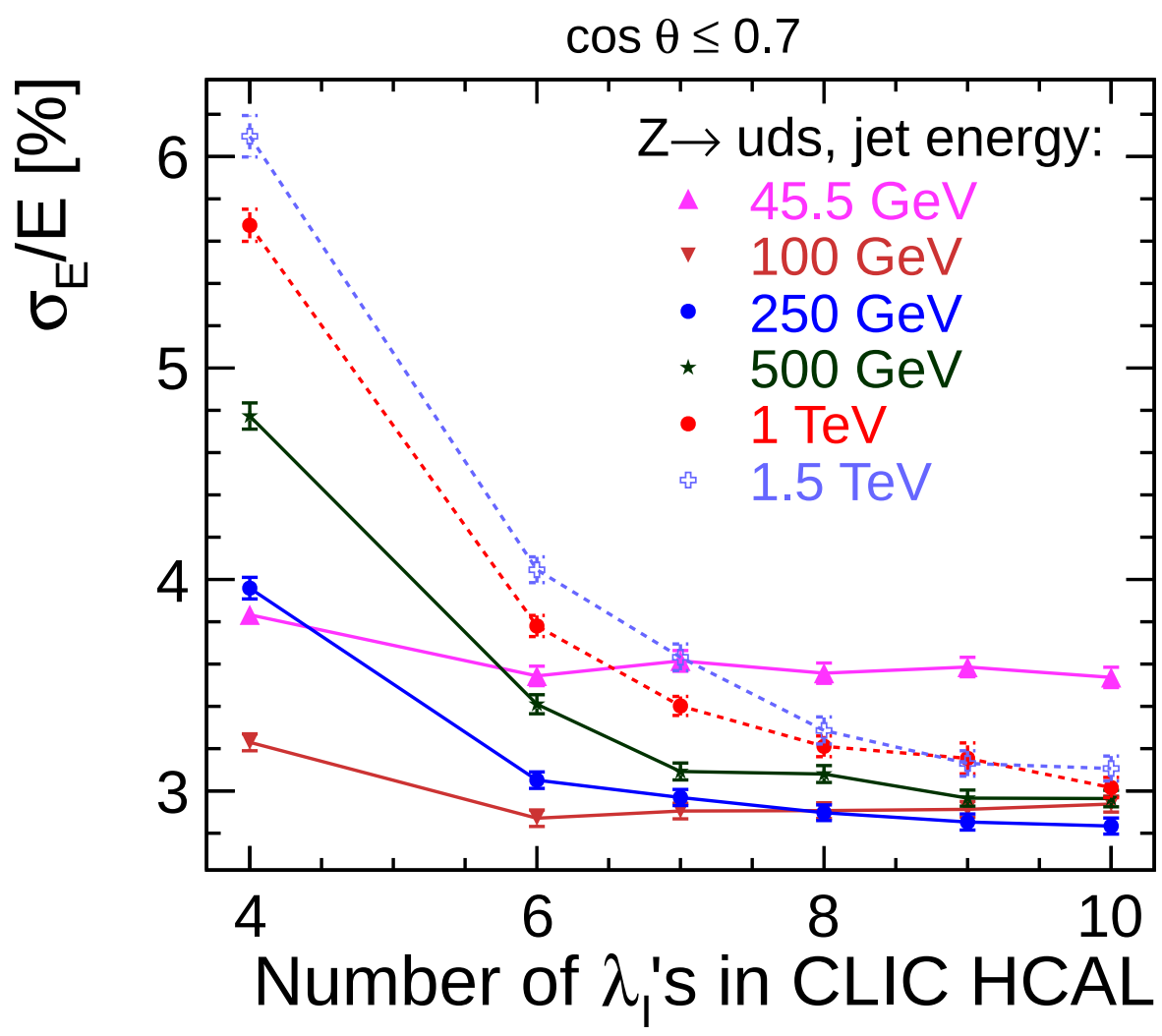

Fig. 6.2: PANDORAPFA jet energy resolution in the barrel HCAL for different jet energies as a function of the tungsten HCAL depth.

To optimise the depth of the calorimeter for the expected particle energy spectrum of CLIC events, the jet energy performance was studied using full simulations of di-jet events and the PANDORAPFA reconstruction algorithm. The result as a function of depth is shown in Figure 6.2, where the HCAL depth is shown as measured in addition to a 1.0 $\lambda_I$ deep ECAL. To guarantee a resolution at or below the required 3.5% for the whole CLIC jet energy range, an HCAL depth of at least 7.5 $\lambda_I$ is required.

### 6.1.2 Time Stamping Considerations

For the calorimeters, $\gamma\gamma$ events and, possibly, beam halo muons are the most critical sources of background. Many of the halo muon induced showers can be recognised and suppressed using their topological signature in the finely segmented calorimeters. However, large so-called "catastrophic" energy depositions by muons may occur and could form an irreducible background if they coincide with a physics event. Hadronic $\gamma\gamma$ final state particles originate from the collision point and produce showers as in $e^+e^-$ events. PYTHIA based simulations [9] predict a rate of 3.2 background events per bunch crossing, which leads to an energy deposition of 6 GeV per bunch crossing in the barrel system, and ten times as much in the endcaps. For a whole bunch train this sums up to about 19 TeV over the entire detector, see 2.5.

Separating this large background from the wanted physics signal is mandatory. This is achieved by topological and timing cuts on fully reconstructed particles (see Section 12.1.4) combined with optimised jet-clustering algorithms. Together, they minimise the impact of the background, while preserving the physics signatures. This puts severe constraints on the readout electronics of the calorimetry systems at CLIC, as a 1 ns time resolution for determining the starting time of the showers is required (see also Section 2.5.1). In addition, multiple hits per cell and per bunch train can be expected in the endcap regions. Forthcoming simulation studies will address the high occupancy in the endcap regions, for example by improved mask design and by adapting the cell sizes in the most forward part of the HCAL. It is therefore expected that final occupancies will not exceed 5 hits per bunch train, including a safety factor of five for incoherent pairs and a factor of two for $\gamma\gamma$ events. As described in Section 10.2.4, the required calorimetry readout performance can be anticipated through continuous fast (40–100 MHz) signal sampling combined with digital filtering techniques.

For hadronic showers, the intrinsic time structure of the shower evolution itself also needs to be considered. Excited nuclei release their energy with de-excitation times extending into the microsecond range, and the signals produced by delayed hard gamma ray emission or evaporation of protons, neutrons and sometimes alpha particles cannot be assigned to a proper bunch crossing in an unambiguous way. Also, low energy and thermal neutrons have a low cross section and can travel for relatively long times until a signal is produced. In this respect, tungsten as an absorber material has draw-backs relative to steel. Iron, in contrast to tungsten, is a magic nucleus with closed shells for protons and neutrons and thus higher excitation energies; tungsten with its higher neutron content releases, on average, a higher rate of neutrons after a hadronic interaction has taken place. Therefore, the time to integrate a given fraction of the energy signal is larger for tungsten than for steel, and so is the fraction of energy with a poor time correlation to the source. This is illustrated in Figure 6.3 for GEANT4 simulations of Fe and W structures, which shows that the hadronic energy resolution is degraded unless the signal collection window is long enough, and that the required time is considerably longer for tungsten than for steel.

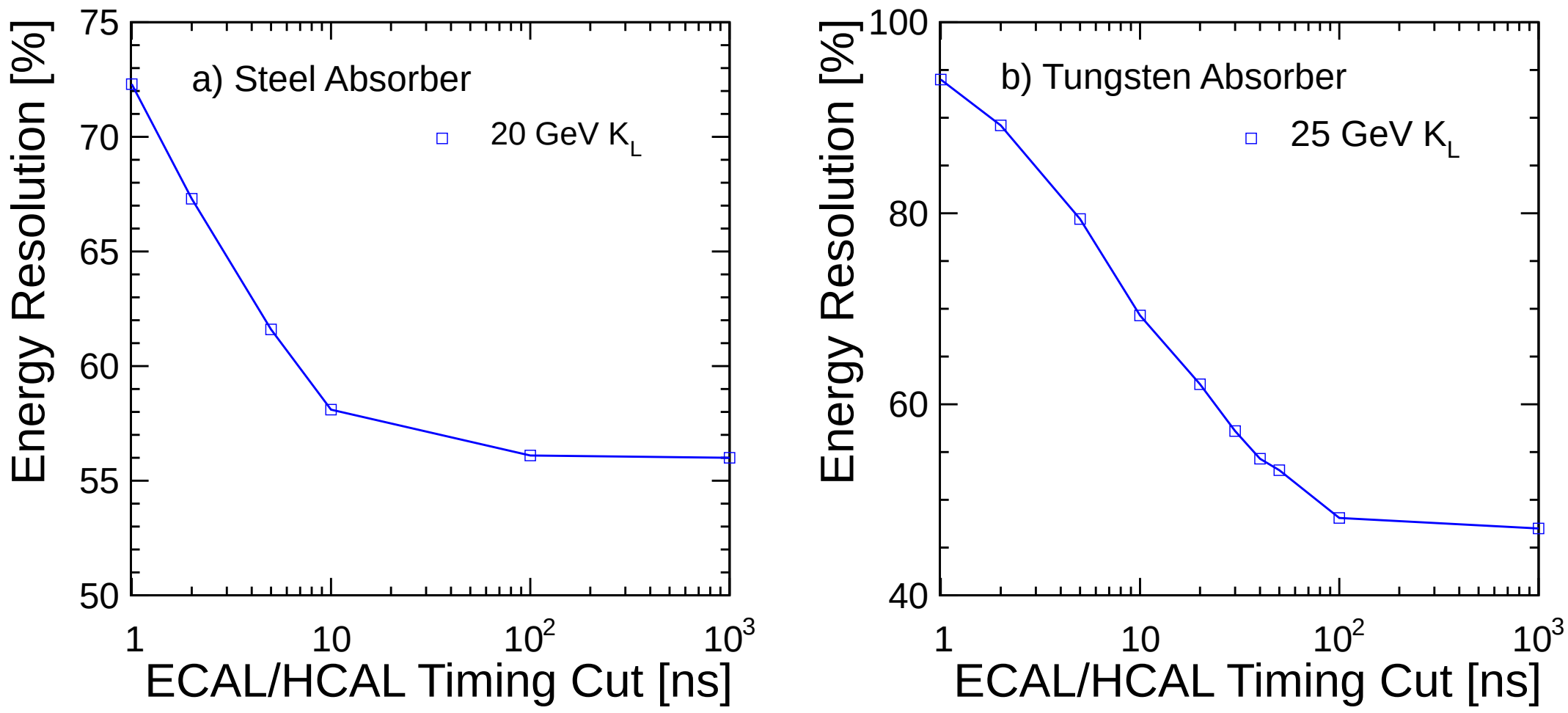

Fig. 6.3: Calorimetric energy resolution for a 20 (25) GeV neutral hadron as a function of the signal collection time (labelled "timing cut") in the CLIC_ILD detector for a) steel and b) tungsten HCAL absorbers. The results are based on the GEANT4 `QGSP_BERT` physics list. Hit times are corrected for the straight line time-of-flight prior to the cut.

In the endcap sections of the HCAL, steel is used as absorber material. On the one hand, this is justified since space is not as tightly constrained as by the solenoid in the barrel region. On the other hand, tighter timing cuts in the analysis, as they are applicable for a steel HCAL, help to reduce the impact of background hits. This is particularly valuable in the forward region of the CLIC detectors, where the background rates are highest.

### 6.1.3 Readout Technologies

The design choices for the active layers of ECAL and HCAL are described in the corresponding sections below. In principle, the same candidate technologies are being considered as for the ILC detectors. Both the high channel count and density and space restrictions demand that the front-end electronics be integrated into the active elements of the calorimeter. The detector signals are processed within the active layer by Application-Specific Integrated Circuits (ASICs) which typically handle between 64 and 1024 individual channels.

For the ECAL, the candidate technologies are silicon pad diodes, or scintillator strips read out by novel Geiger mode silicon photo-diodes (called MPPCs, or SiPMs). The scintillator provides a larger sampling fraction and thus better energy resolution, whereas silicon allows to realise a finer segmentation and a more compact design with smaller effective Molière radius, albeit at higher cost. The compari-son of the physics performance of these technologies is subject of ongoing studies. In particular, the development of particle flow algorithms for the strip geometry is not yet as advanced as for the pad option.

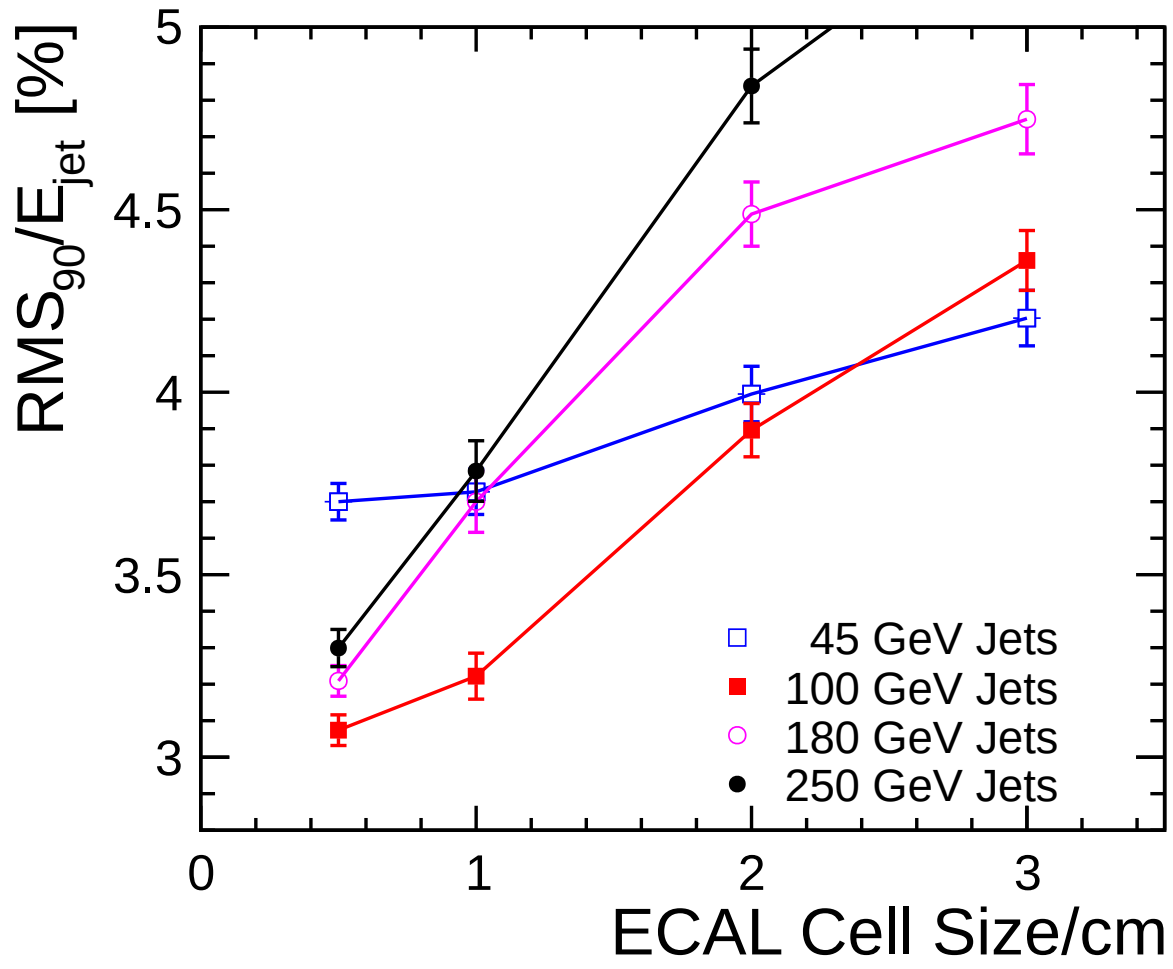

Fig. 6.4: Jet energy resolution obtained with PandoraPFA for different ECAL cell sizes.

For the HCAL, the main choice is between scintillator tiles, again read out by SiPMs, or gaseous devices with different amplification structures, such as resistive plate chambers (RPCs) [10, 11], Micromegas [12] or gas electron multiplier (GEM) [13] foils and pad readout. The scintillator option offers analog electronic readout and thus provides energy deposition information, while the gaseous techniques are read out in digital or semi-digital mode (one or two threshold bits), such that the particle energy is effectively inferred from the number of hit cells. The comparison between the two involves the trade-offs between energy and position resolution of analog and digital readout, possibly higher segmentation of the readout of gaseous media, but also the different response of plastic scintillator and gaseous media to different components of the shower, such as neutrons and photons. The studies require powerful reconstruction algorithms, which are currently not yet fully optimised for the digital option. Validation with beam tests is required separately for steel and tungsten, since the shower composition varies significantly from one to the other. These tests are ongoing and are planned to continue over the next years.

## 6.2 Electromagnetic Calorimeter

In an event reconstruction following the particle flow approach, the role of the electromagnetic calorimeter (ECAL) is primarily to measure the photon energies individually, as well as the early parts of showers initiated by hadrons. However, photons may be located close to each other, and charged hadrons may overlap, too. Therefore the most relevant feature of the ECAL is its fine segmentation. To optimise the separation of the electromagnetic showers from each other and from nearby hadron shower fragments, and thus the jet energy resolution, the lateral segmentation is a crucial parameter.

The effect of the cell size has been studied in simulations using the PandoraPFA algorithm. A result is shown in Figure 6.4 [1], where the jet energy resolution is drawn as a function of the ECAL segmentation cell size. A cell size smaller than 5 mm is favoured. Furthermore, the longitudinal segmentation is also important to achieve the required particle flow performance in the CLIC energy region, since it helps both pattern recognition and electromagnetic energy resolution. Another issue relevant to CLIC is the timing resolution of the ECAL, as discussed above. The requirements for ECAL can therefore be summarised as:

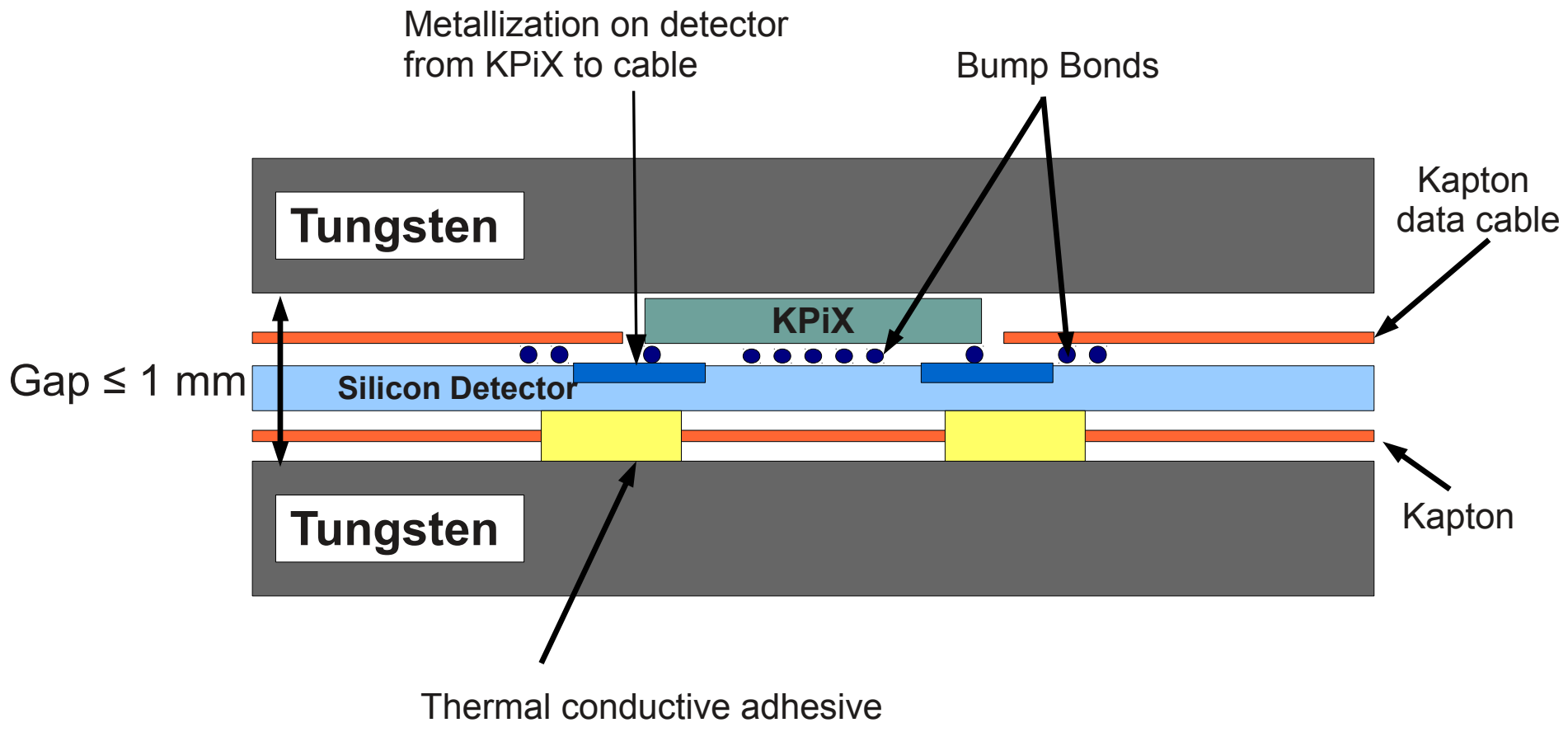

Fig. 6.5: Sandwich layout of the SiD ECAL.

– fine segmentation in the lateral direction of the electromagnetic shower;
– good segmentation in the longitudinal direction;
– good timing resolution for bunch crossing separation.

These requirements can be fulfilled by a sandwich type ECAL with absorber and sensitive material in alternating layers and sufficient overall depth. The choices for CLIC_ILD and CLIC_SiD are similar. The main parameters are given in Chapter 3.

### 6.2.1 ECAL Readout Technologies

The two detector concepts, for both ILC and CLIC, utilise tungsten for the ECAL absorber. SiD has one and ILD has two options for the sensors. One is common for ILD and SiD: the silicon pad sensor, with $5\times5$ mm$^2$ segmentation for ILD, or hexagonal pads for SiD with an area of 13 mm$^2$. The second choice of sensitive material for ILD is scintillator strips coupled to photon sensors for detecting the scintillation light. In addition, though in an early stage of development, a silicon pixel sensor based technology, INMAPS [14], is also being considered for ECAL. The readout concept for all options is to embed the electronics in the sensitive layers, as shown in Figure 6.5 for SiD. The ECAL integration design of ILD is presented in [15].

The signals are processed close to the sensors in custom designed ASICs. There are two activities for developing such ASICs in the ILC context, namely SPIROC [16] and SKIROC [17] in Europe, and KPiX in the USA [3]. In order to assess the dynamic range needed in the ECAL at CLIC, particle energy distributions for different jet energies are compared in Figure 6.6. It is shown that the maximum energies of the particles is indeed higher. It is anticipated that a 12-bit dynamic range of the ECAL readout electronics is required for CLIC. Moreover, the electronics needs to cope with a few (<5) hits within the 156 ns bunch train and provide a hit time resolution of $\approx$ 1 ns. A possible electronics concept is discussed in Section 10.2.

### 6.2.2 ECAL Prototypes

A first group of prototypes for realistic ECALs, referred to as "physics prototypes", has been developed and tested by the CALICE collaboration. The first ECAL prototype [18] utilises silicon sensor pads with a segmentation of $10\times10$ mm$^2$. The second, a scintillator tungsten ECAL [19] prototype has sensors made of plastic scintillator strips of $10\times45$ mm$^2$, which have a small photo-sensor, named MPPC, at its end. The goal for these prototypes was to prove their physics performance rather than the technological and integration issues. Therefore, for simplicity, the readout system was located at the periphery of the detector. A silicon tungsten design for SiD [3] aims to couple the readout electronics with the sensor layer directly. This development is more ambitious and has not yet been exposed to beam tests in a full-system configuration.

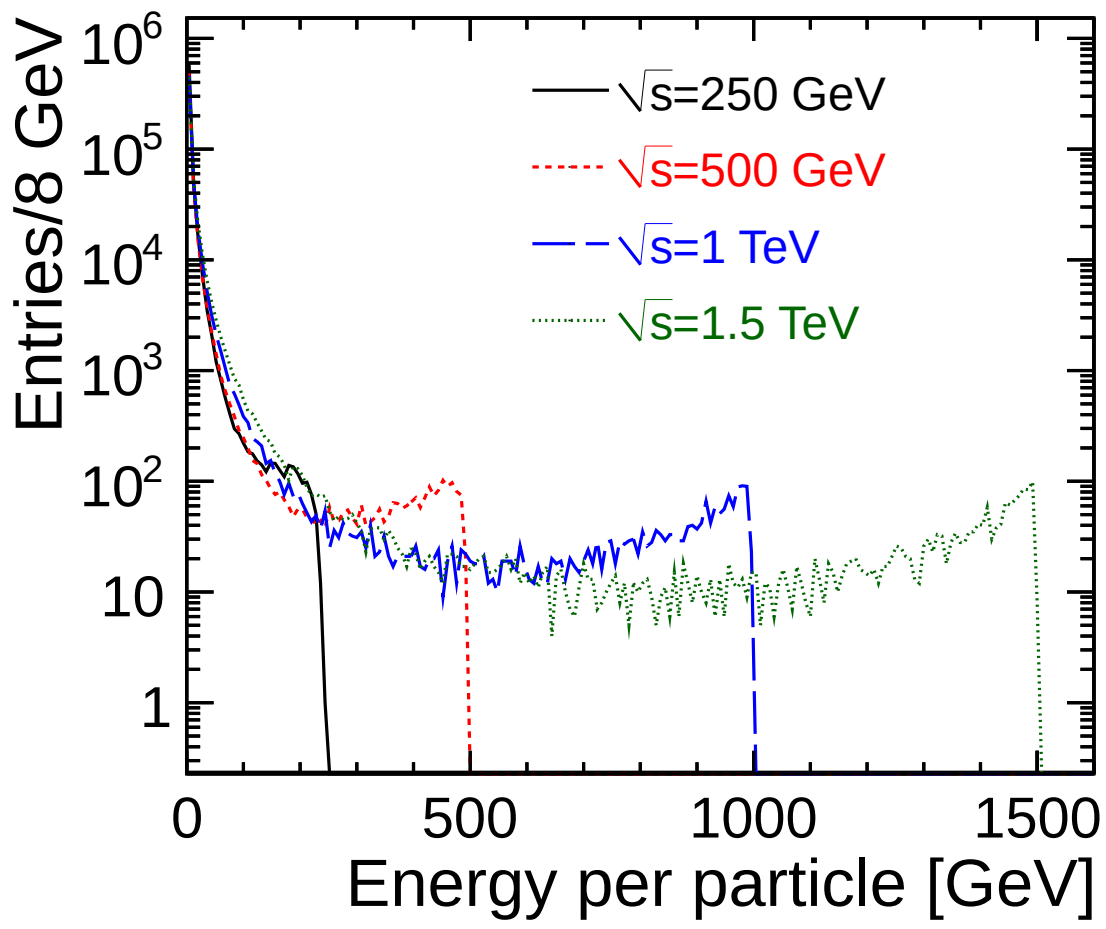

Fig. 6.6: Energy distribution of particles in ECAL for different jet energies.

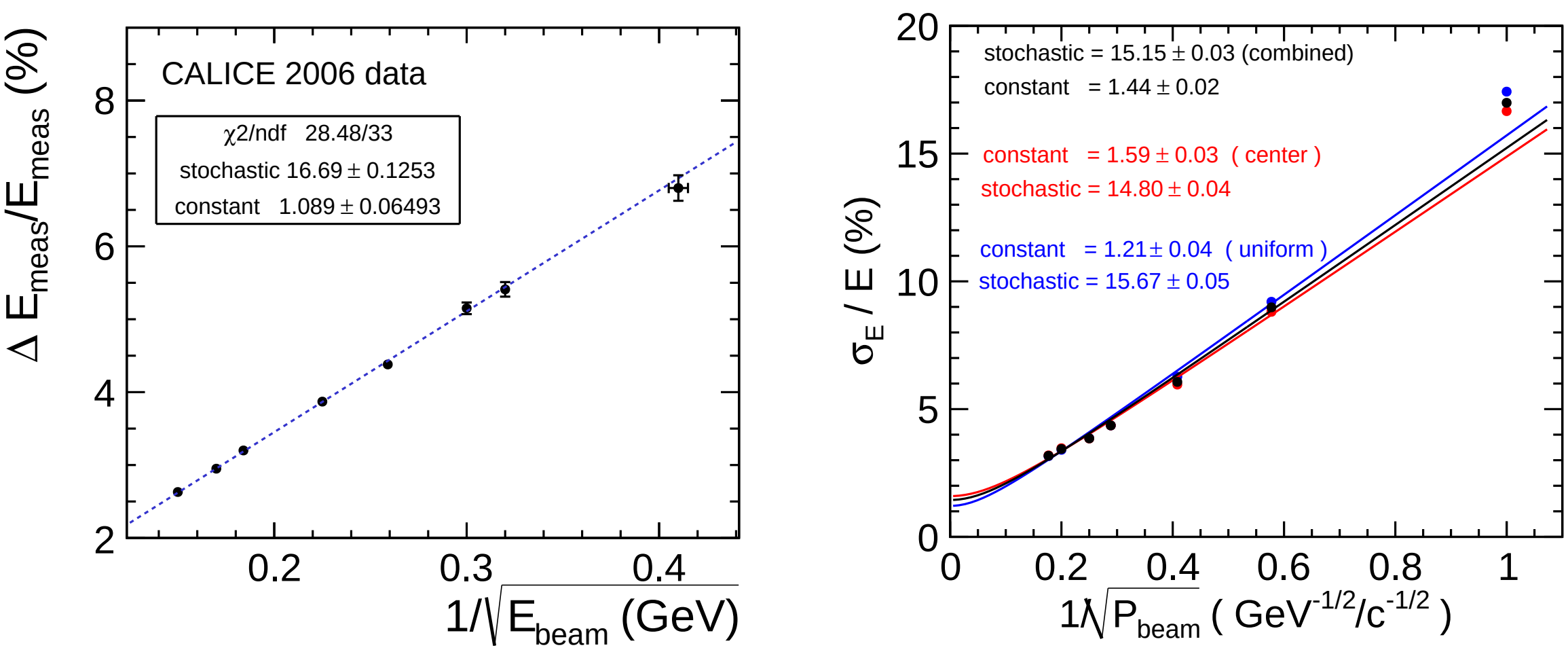

Fig. 6.7: Test beam results obtained with the CALICE silicon (left) and scintillator (right) ECAL prototypes: relative energy resolution as a function of beam energy (left) and momentum (right). The right plot shows the resolution individually for the case where the beam particles were impinging on the centre of a scintillator strip, or uniformly distributed, as well as for the combined data set.

### 6.2.3 ECAL Testbeam Results

Results of electromagnetic energy resolution as obtained in beam tests with the prototypes are shown in Figure 6.7 for the silicon tungsten ECAL [20] and for the scintillator tungsten ECAL [19].

The scintillator tungsten ECAL was also used to study the reconstruction of neutral pions by installing a target in the high energy pion beam line. The two cluster invariant mass distribution is shown in Figure 6.8. The results are consistent with the simulation including the reconstruction efficiency. This indicates that there is a good potential to reconstruct photons. Time resolution is not yet tested in the test beam, because the timing capabilities in the readout electronics are still being commissioned.

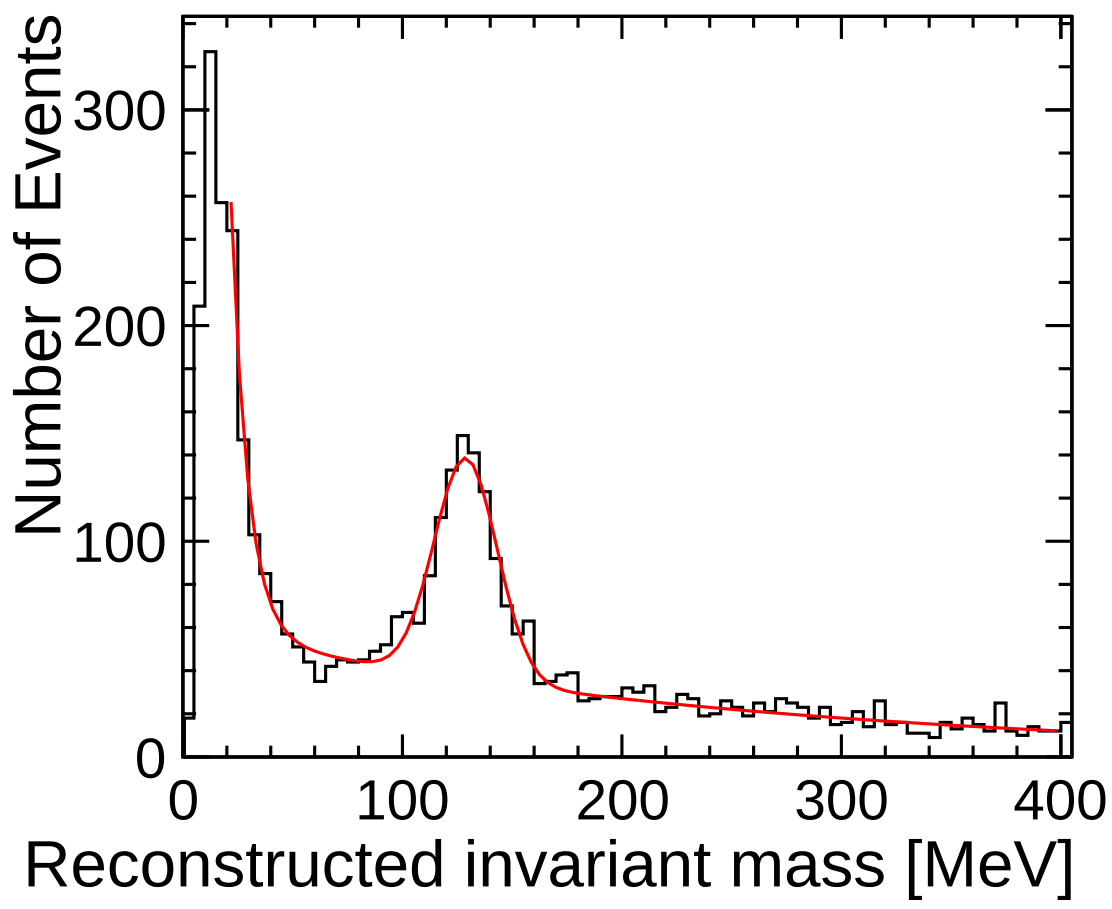

Fig. 6.8: Test beam result obtained with the CALICE scintillator ECAL: Two photon candidate invariant mass distribution.

## 6.3 Hadronic Calorimeter

### 6.3.1 Basic Layout

The CLIC_ILD and CLIC_SiD hadron calorimeters are both imaging devices designed for PFA. They are conceived as sampling calorimeters with tungsten (barrel) or steel (endcaps) as absorber and scintillator tiles (analog HCAL) or gaseous devices (digital HCAL) as active medium. Stainless Steel has been chosen as absorber by the ILD and SiD concepts both for mechanical and calorimetric properties. Due to its rigidity, a self-supporting structure without auxiliary supports (and thus dead regions) can be realised. Iron with its moderate ratio of hadronic interaction length to electromagnetic radiation length (see Table 6.1) allows a fine longitudinal sampling in terms of $X_0$ with a reasonable number of layers in a given total hadronic absorption. As explained in Section 6.1.1, the 2 cm thick steel absorbers proposed for ILC are replaced by 1 cm thick tungsten absorbers in the barrel HCAL. The parameters of the CLIC_ILD and CLIC_SiD hadron calorimeter designs are presented in Chapter 3.

### 6.3.2 HCAL Readout Technologies for Scintillator and Gaseous Options

To allow for the PFA-based approach to calorimetry, the active layers of the hadron calorimeter need to have fine longitudinal and transverse segmentation. This allows to follow charged tracks through the calorimeter with high accuracy, facilitates the separation of hadron showers and their components, and supports the direct measurement of neutral shower energies with adequate resolution. One option for the active layers of the hadron calorimeter are $3 \times 3$ cm$^2$ scintillating tiles, located in a 6.5 mm gap between absorber plates and read out using multi-pixel SiPMs. Another option is RPCs with 1 cm readout pads. The scintillator approach delivers analog output, while the RPCs are either digital or semi-digital (more than one threshold). These readout technologies, and alternatives such as GEM foils and Micromegas, are the subject of intense study and development. The simulation results in this CDR are obtained with the scintillator option, because for this option the simulation and reconstruction software is more advanced.

#### *6.3.2.1 HCAL Technological Prototypes*

In order to understand the issues involved in implementing selected technologies in a linear collider detector, work has started on technological prototypes that incorporate realistic active layers. An example of a technological prototype, the "EUDET prototype" [21], is shown in Figures 6.9 and 6.10. The features to be incorporated include power supply, signal routing, and heat dissipation. Mechanical stability, tol-erances, and integration of controls, supplies and services will also be addressed. For the RPC approach, a prototype for a realistic active layer is being developed as well.

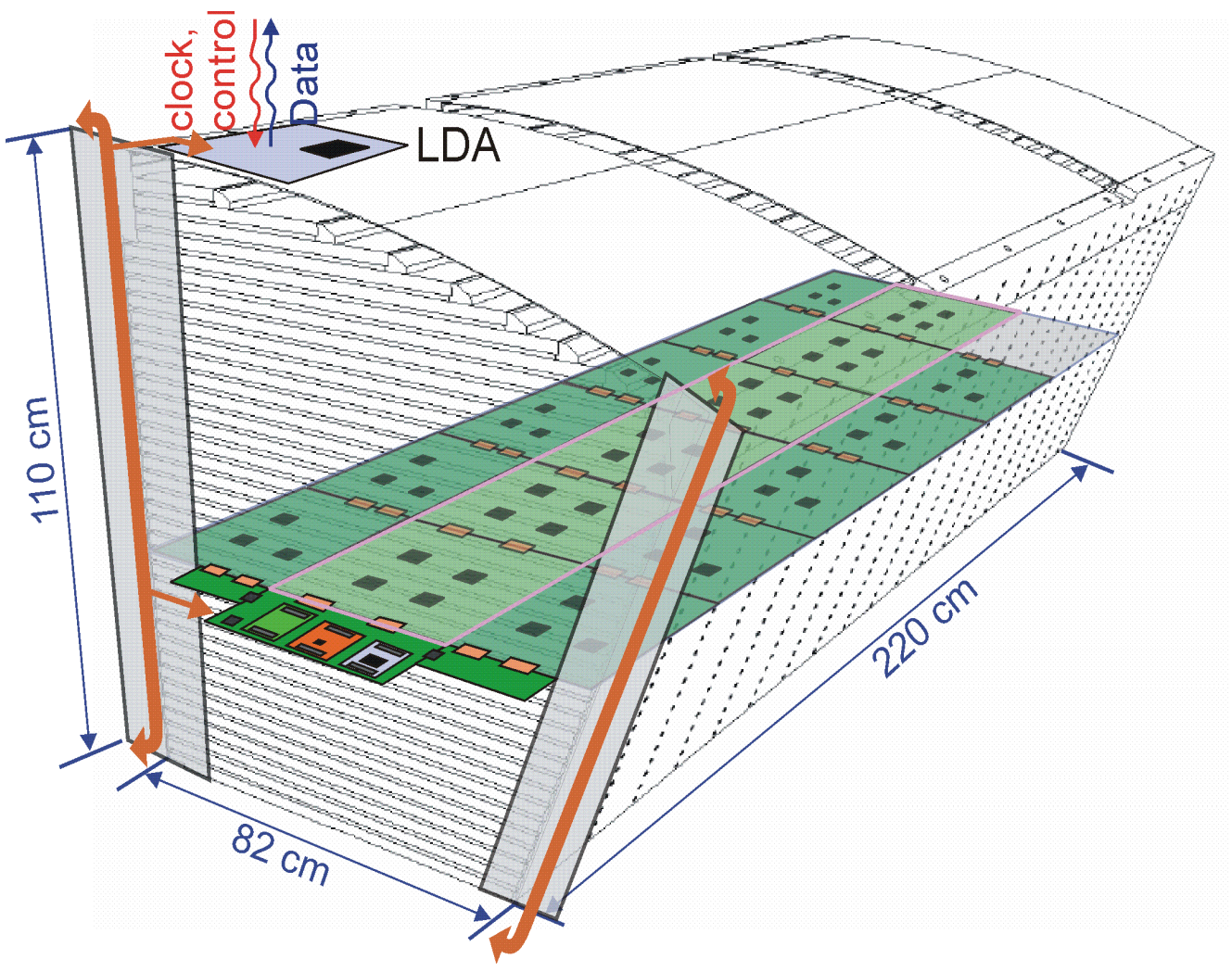

Fig. 6.9: Schematic view of a CALICE AHCAL technological prototype module [21].

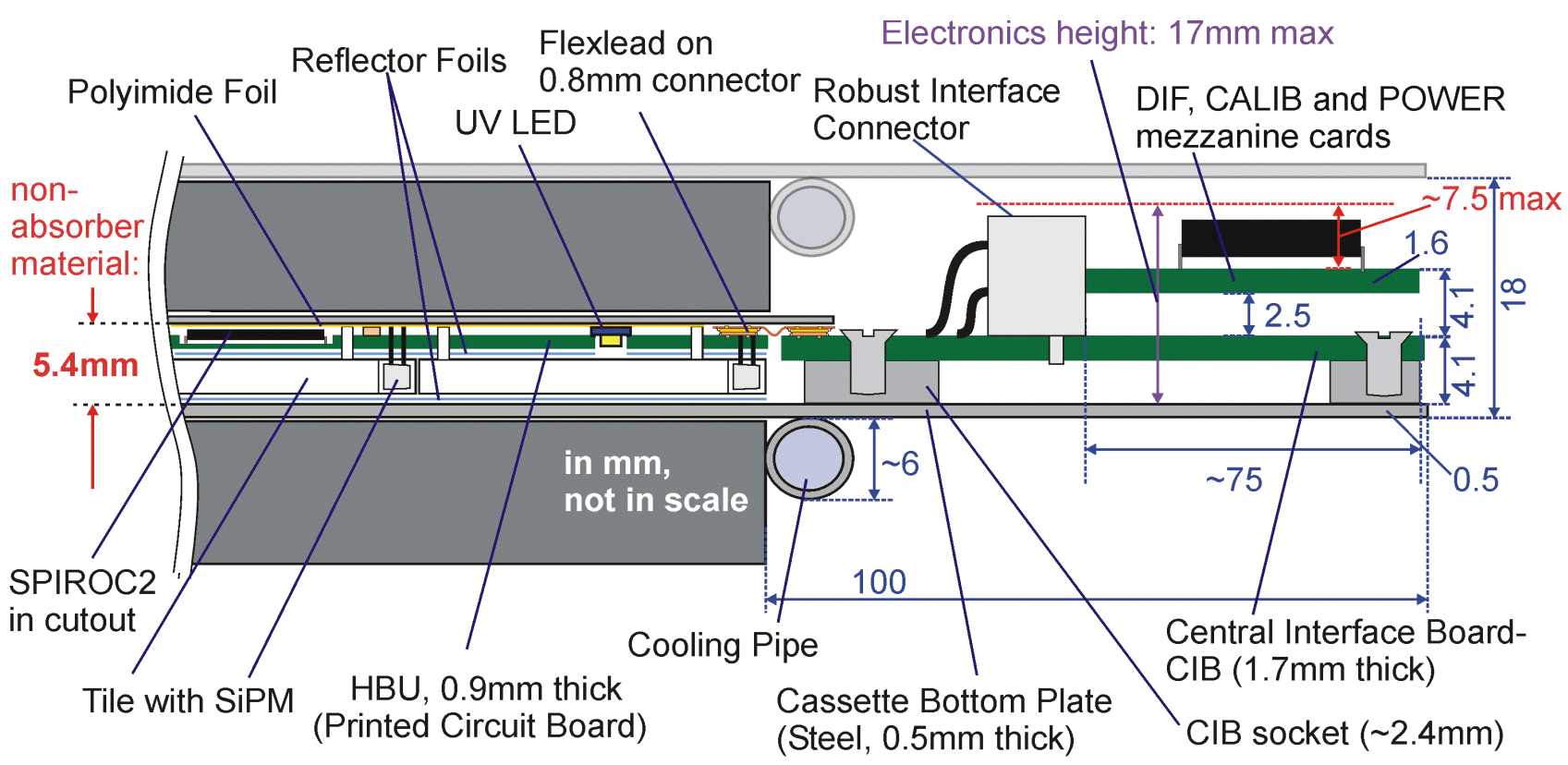

Fig. 6.10: Detailed schematics of the components in the AHCAL technological prototype.

### 6.3.3 HCAL Test Beam Results

#### *6.3.3.1 AHCAL Test Beam Results using Steel Absorbers*

The performance of an analog scintillator/steel calorimeter [22] has been studied by the CALICE collaboration with a 1 m$^3$ stack called AHCAL, exposed to beams of hadrons, electrons [23], and muons, also in conjunction with the ECAL prototypes. This calorimeter is non-compensating and the resolution is affected by fluctuations in the electromagnetic fraction of hadronic showers. Due to the high granularity of the calorimeter, it is possible to apply individual weighting of the shower components, in order to compensate for differences between the hadronic and electromagnetic response as well as for the "invisible" energy depositions. This approach, known as "software compensation", yields a significant improvement in the fitted combined resolution as shown in Figure 6.11 [24].

The measurement of the energy of a neutral particle in the calorimeter can be degraded by the presence of nearby charged particle(s). This issue, often referred to as "confusion", was investigated using test beam data [25]. Figure 6.12 shows the results of a study in which two test beam pion-induced events were superimposed, with one event having its incoming track removed to simulate a neutral particle. The figure shows the probability of PANDORAPFA correctly resolving the situation, within three standard deviations of the true energy, as a function of the distance between the two shower axes. The data are compared with GEANT4 using two different physics lists and are found to be well described by the `QGSP_BERT` list. This corroborates the confidence in the GEANT4 based predictions of the overall detector performance for jet final states, here in the case of an HCAL with steel absorbers.

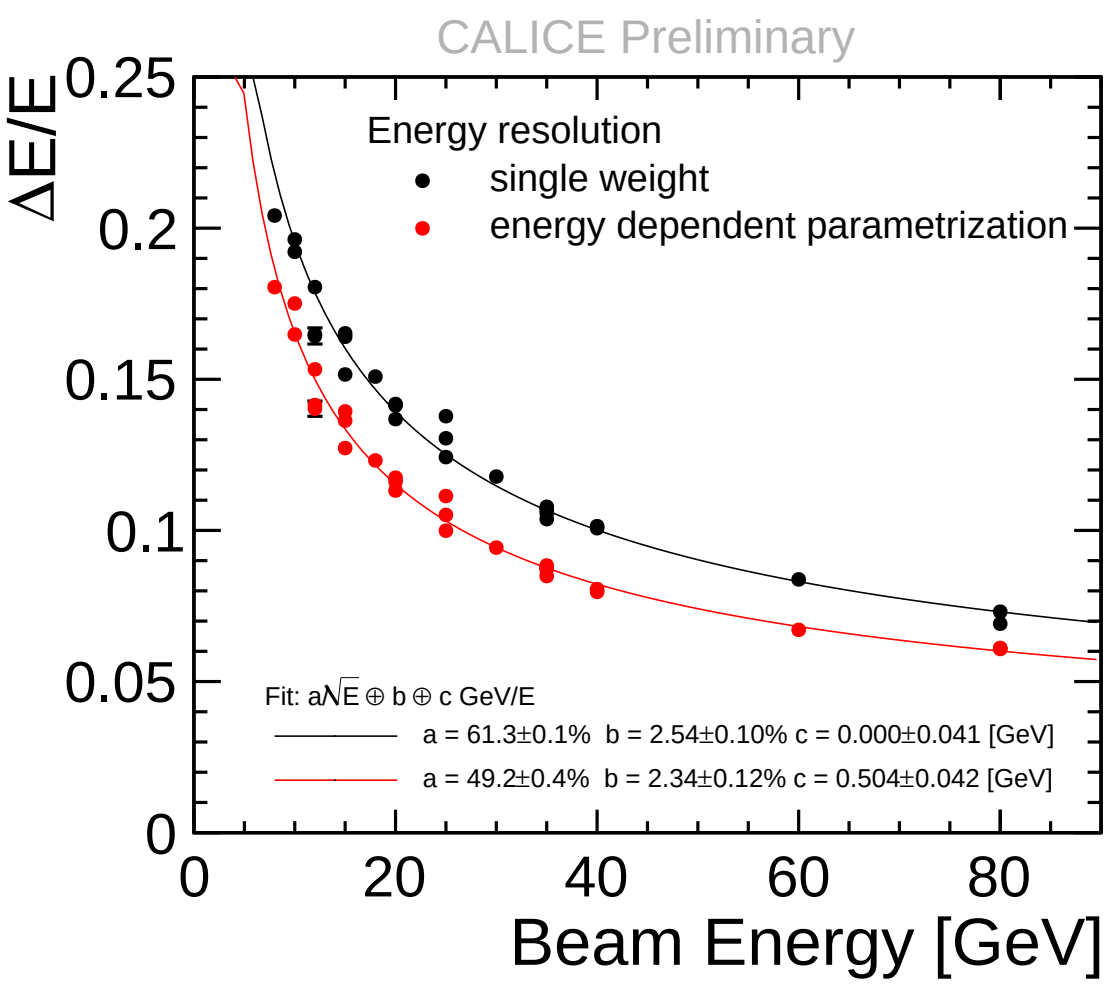

Fig. 6.11: ECAL plus AHCAL combined resolution for pions. The upper curve represents the resolution obtained with a single weight factor for each of the calorimeters, while the lower reflects a simple software compensation approach and uses weights for the hits that depend on the hit amplitude and on the total measured shower energy.

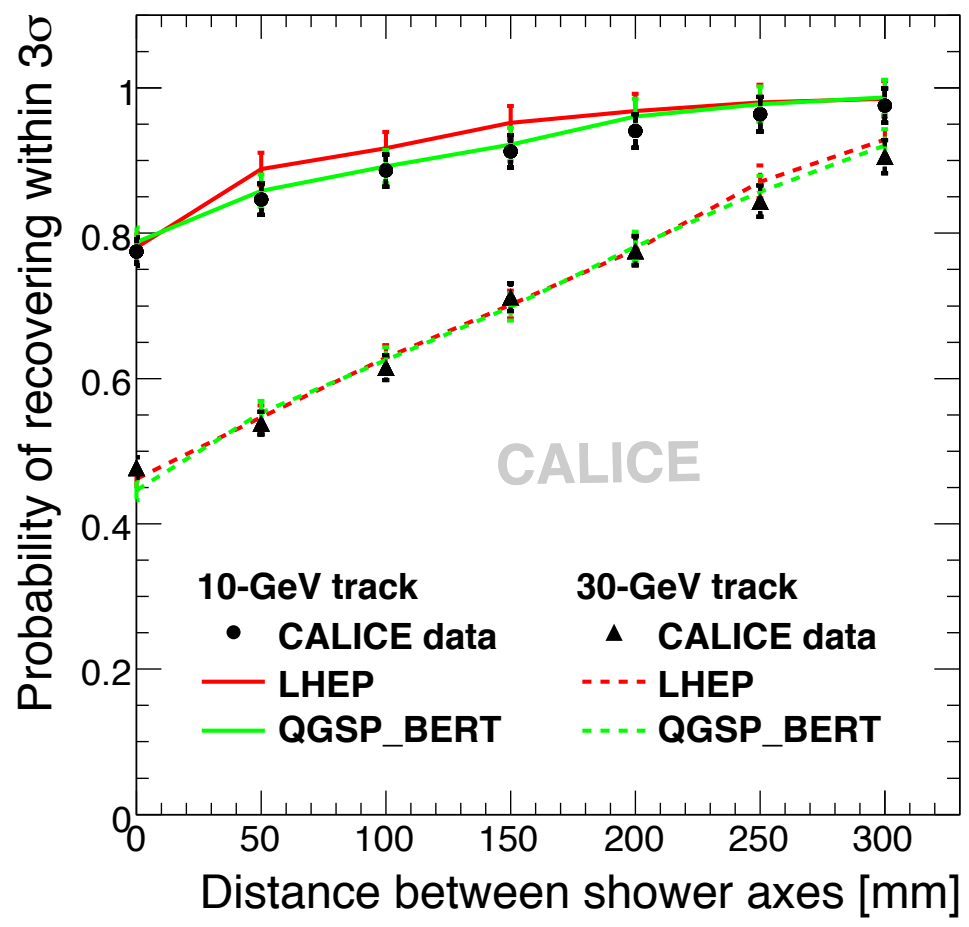

Fig. 6.12: Probability of separating hadron showers: The figure shows the degradation of neutral particle resolution, expressed in terms of the probability to reconstruct the energy within 3 $\sigma$ of its calorimetric resolution, as a function of transverse separation from a second shower induced by a charged hadron.

#### *6.3.3.2 AHCAL Test Beam Results using Tungsten Absorbers*

To test the energy resolution and timing performance of a tungsten-scintillator combination calorimeter, and to validate the corresponding simulation model, a 30-layer (3.9 $\lambda_I$) AHCAL module was constructed and exposed to beam at CERN in 2010. The scintillator tile and readout layers are the same as used by CALICE for a number of earlier tests with steel absorber plates. Figure 6.13 shows the experimental setup and an example of a pion candidate shower in the calorimeter stack.

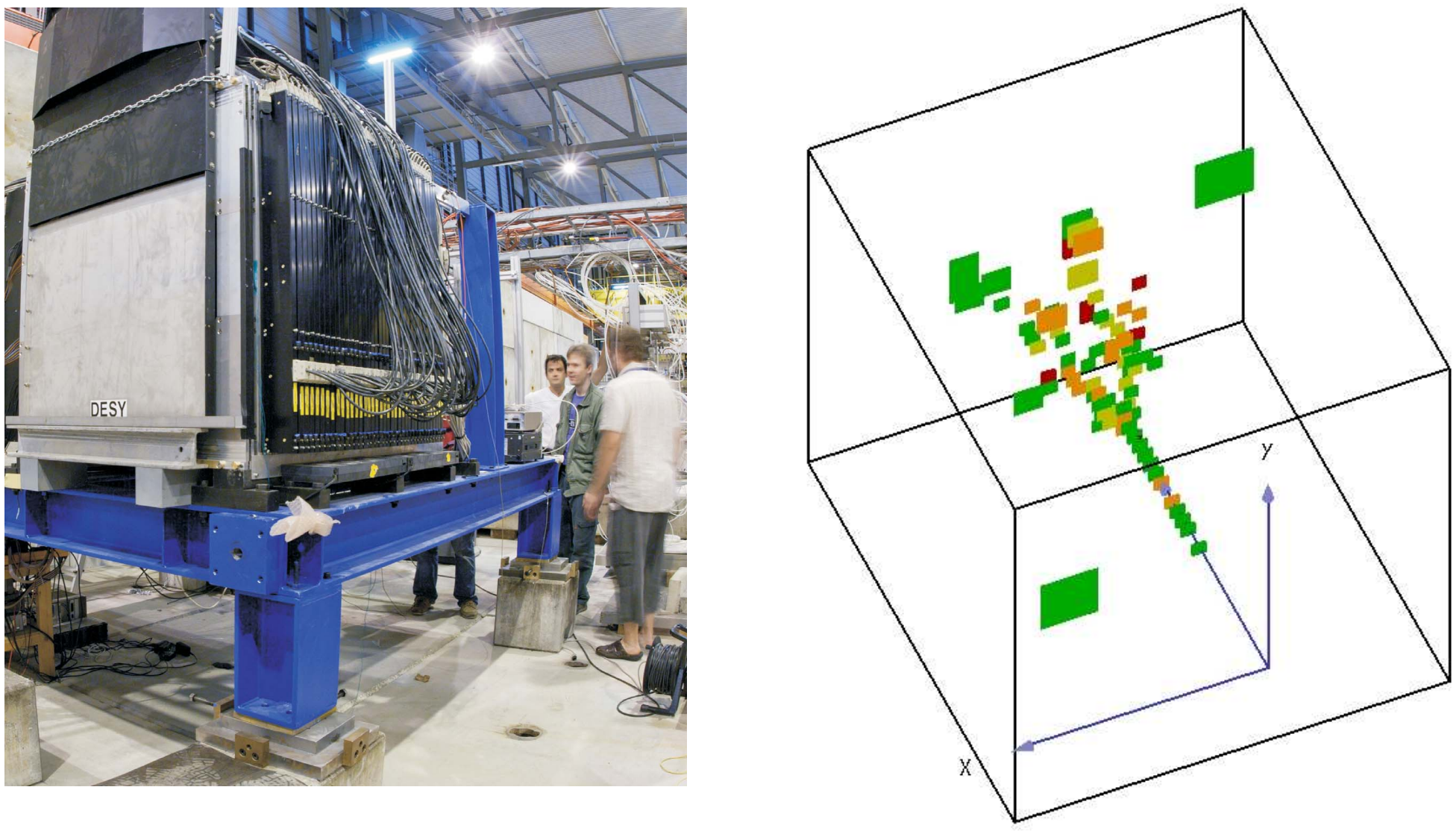

Fig. 6.13: (Left) Tungsten-scintillator module at the test beam. (Right) An example of a pion shower in the 30-layer calorimeter stack.

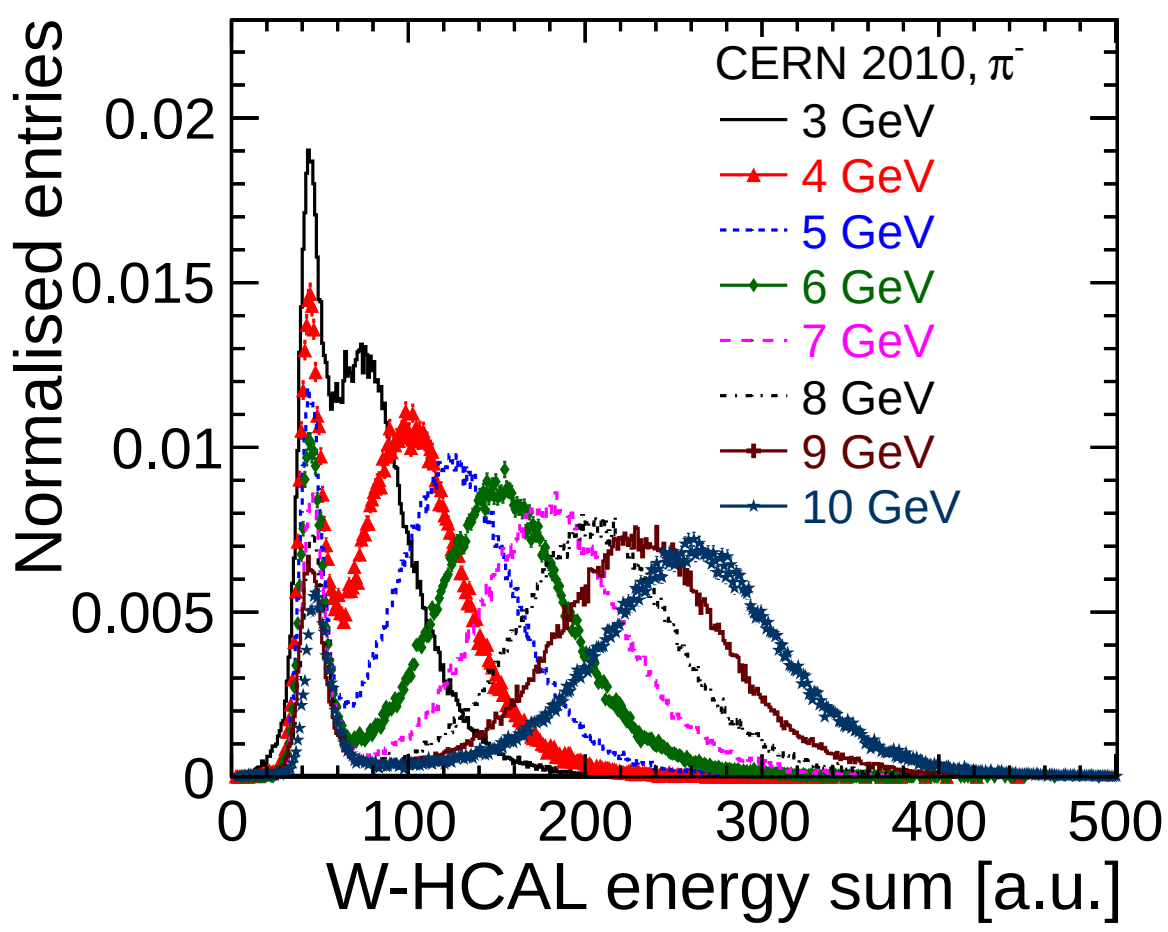

Fig. 6.14: Pion response in the tungsten-scintillator test calorimeter. The peak at the lowest energy is the muon response.

High statistics event samples were recorded for electron, muon, pion, and proton beams with energies from 1 to 10 GeV. Gain calibration was obtained from low intensity LED-pulser runs and the results agree well with previous calibration from runs at Fermilab. MIP calibration was carried out using a muon beam. Examples of calorimeter responses to muons and pions are shown in Figure 6.14. Preliminary results indicate that the electromagnetic resolution is slightly worse than for steel, reflecting the lower sampling ratio relative to $X_0$. Finally, the $e/\pi$ response ratio appears to vary little with the energy. Additional data were taken at pion energies up to 300 GeV in a beam test in 2011, using an extended stack of 38 tungsten-scintillator layers. These data are currently being analysed. Preparations are made to replace the scintillator readout in the tungsten HCAL prototype by RPCs as active layers.

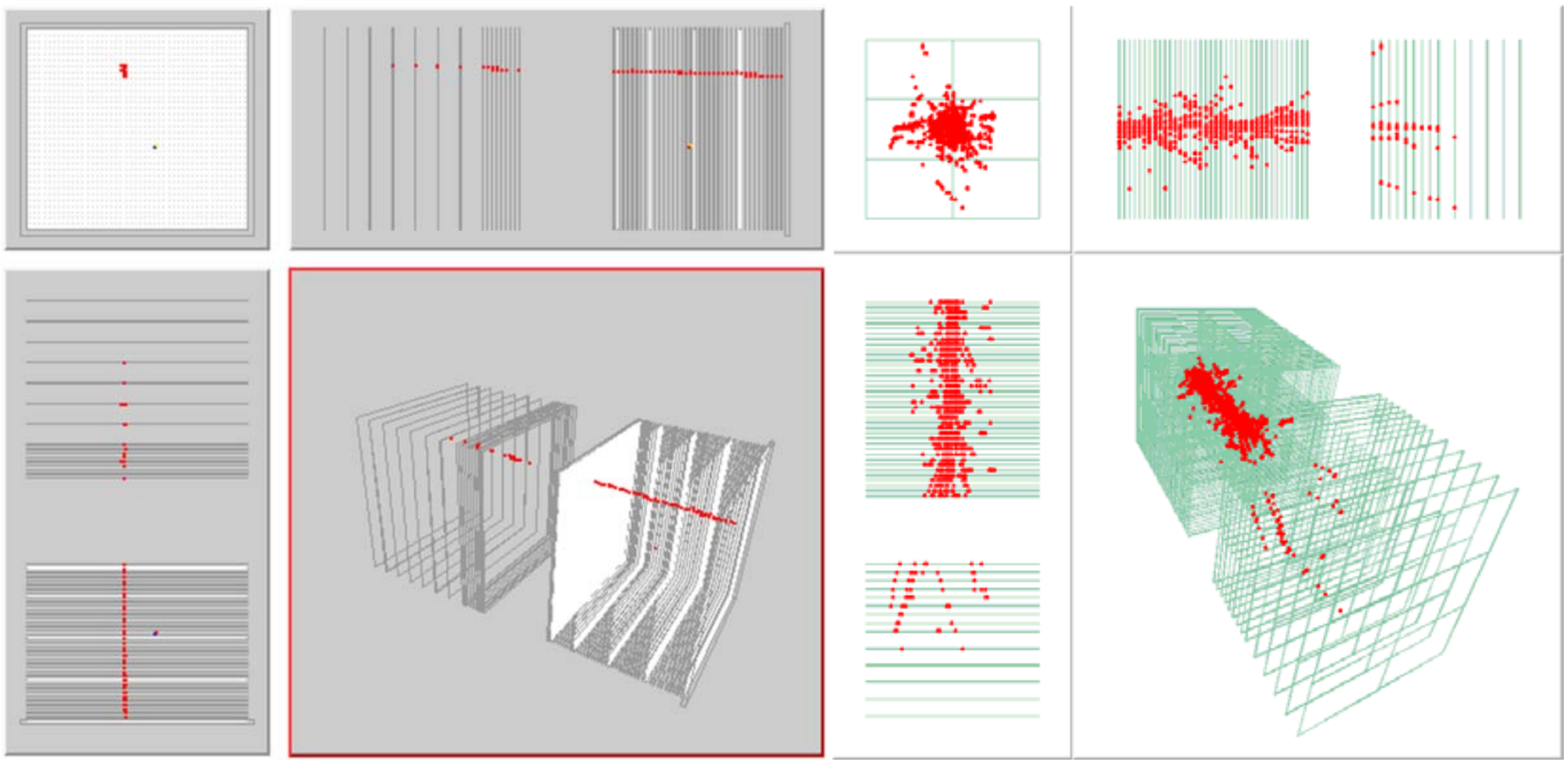

Fig. 6.15: Event display of a muon track (left) and a hadronic shower from a 120 GeV proton (right) in the DHCAL.

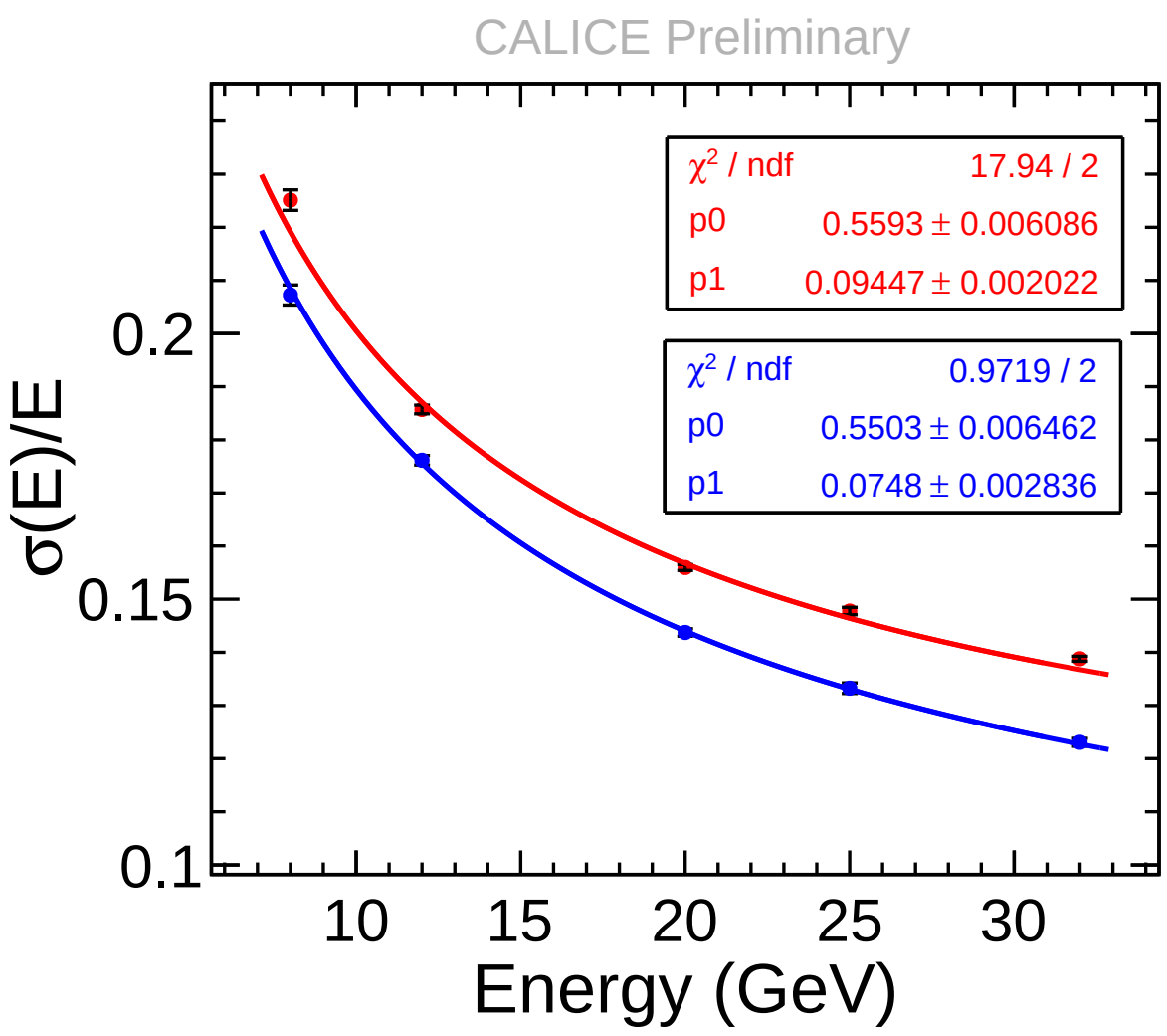

Fig. 6.16: Energy resolution as measured in the DHCAL for pions in the energy range of 8 to 32 GeV. The red (blue) data points are obtained without (with) a cut on any hits in the last two layers of the stack.

#### *6.3.3.3 DHCAL Test Beam Results using Steel Absorbers*

The novel concept of a digital hadron calorimeter with RPCs as active medium is being developed by the CALICE collaboration as well. In order to study the characteristics of such a device, to gain experience with an RPC-based calorimeter and to measure hadronic showers with high spatial resolution, a large prototype, the DHCAL, was assembled with 52 active layers and close to $5 \cdot 10^5$ individual readout channels. Due to the choice of $1 \times 1$ cm$^2$ pads, the calorimeter is compensating in the 6 to 10 GeV energy range, under-compensating at lower energies and over-compensating at higher energies. The prototype was exposed to beams of hadrons, positrons and muons in the Fermilab test beam. To demonstrate the imaging capability of this type of calorimeter, Figure 6.15 shows a muon track and a 120 GeV proton shower in the DHCAL. Note the absence of random noise hits in the muon event. As the energy of a particle is reconstructed to first order as the sum of hits, it is essential to reduce contributions from accidental hits to a negligible level.

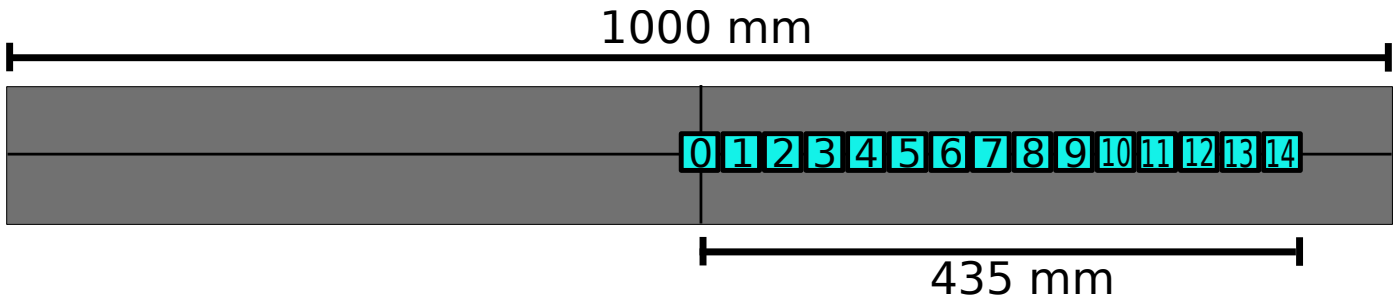

Fig. 6.17: T3B scintillator tile layout.

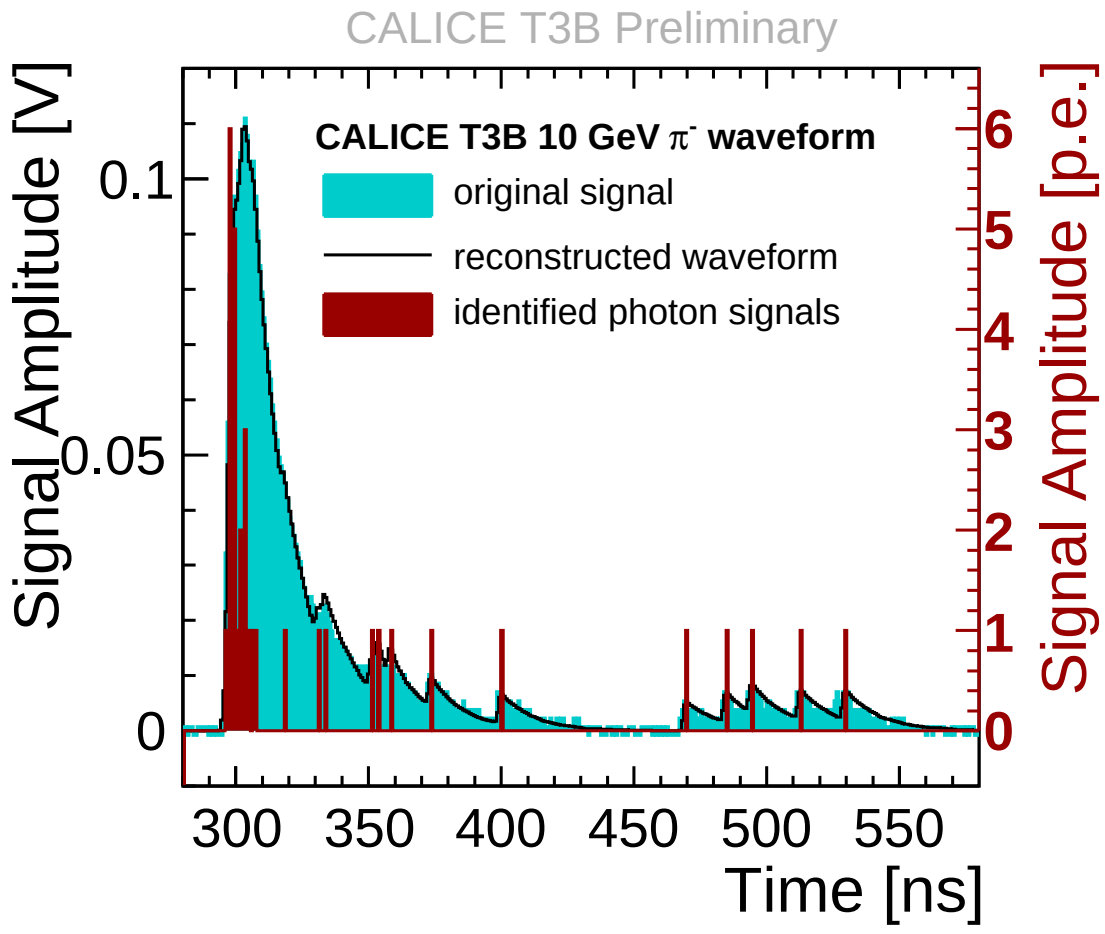

Fig. 6.18: Typical waveform from T3B.

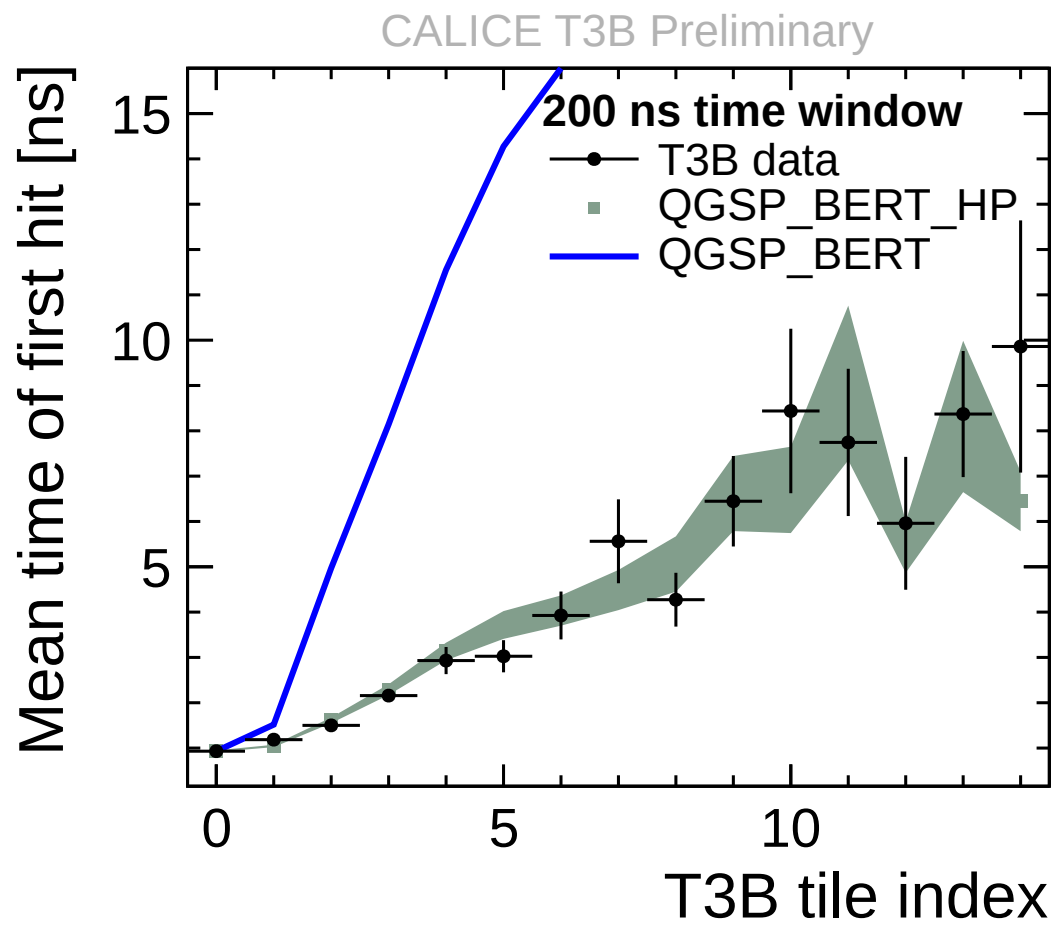

Fig. 6.19: Mean time of the first hit for a 10 GeV $\pi^-$ vs. the radial distance from shower core.

The data analysis is still in its initial stage, as the test beam campaigns were only recently completed. To give a flavour of what to expect in the future, Figure 6.16 shows the measured hadronic resolution for pions in the range of 8 to 32 GeV [26]. In this energy range, the resolution is seen to be comparable to the one obtained with scintillator as active medium. With proper calibration of the response from layer to layer the constant term is expected to decrease. As in the AHCAL case, application of software compensation techniques will further improve the resolution.

#### *6.3.3.4 Test Beam Results on Shower Timing*

Since CLIC has a 2 GHz bunch crossing rate and a large hadron background is expected from $\gamma\gamma$ events, it is important to study the impact of this background on the measurement of hadron showers. The time structure of the shower development in HCAL will be measured with future prototypes featuring time-resolved readout electronics. An initial study of the effects has been made with the Tungsten Timing Test (T3B) system at CERN [27]. As shown in Figure 6.17, a linear array of fifteen $3\times3$ cm$^2$ scintillator tiles was placed at the back of the tungsten-scintillator stack described in Section 6.3.3.2.

The tiles were read out directly (no wavelength shifting fibre) to SiPMs, which were, in turn, read out with 1.25 Gigasamples oscilloscopes with a full time window of 2.4 µs per event to capture the complete time history of showers. A typical waveform, with decomposition into individual photon signals, is shown in Figure 6.18. To understand the possibilities for time stamping of showers, the time of the first hit in a tile for each event has been investigated. The first results are shown in Figure 6.19 and compared with the expectations from two GEANT4 physics lists. As it can be seen, a good agreement with `QGSP_BERT_HP`, which features high precision neutron tracking, is obtained. Further measurements at higher energies have been performed, and data are being analysed. For additional comparison with GEANT4, measurements with steel absorbers are also being prepared.

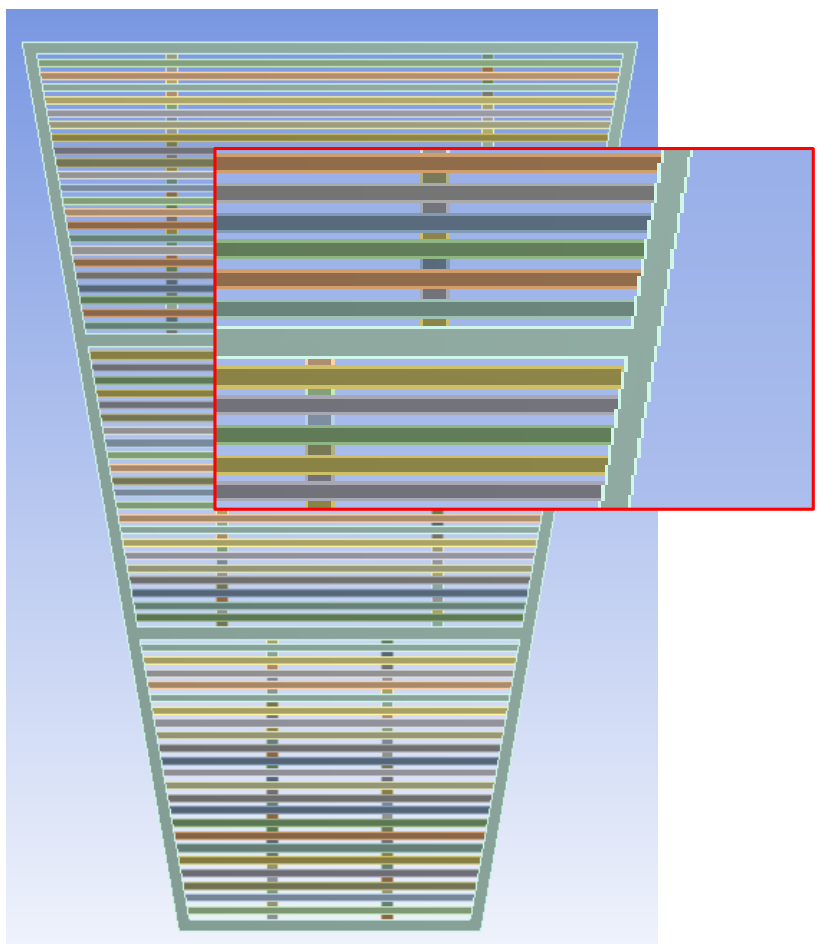

Fig. 6.20: Tungsten HCAL "box design", showing a sector made from 3 trapezoidal boxes of stainless steel. The sector is shown with installed tungsten plates [28].

### 6.3.4 Tungsten Design and Engineering Studies

A conceptual study [28] was carried out to check the mechanical feasibility of a hadron calorimeter based on tungsten absorber plates. Pure tungsten is rather brittle. It is therefore proposed to use a more robust alloy of 93% W, 5.3% Ni and 1.7% Cu with a density of $\rho = 17.5\ g/cm^3$ and a Young's modulus of about 350 GPa. This alloy is non-magnetic and can therefore be installed inside the solenoid. Mechanical finite element analysis in 2D and 3D was carried out on an early version of a CLIC_SiD-type HCAL of approximately 7.5 $\lambda_I$ depth, comprising 60 layers of 12 mm thick tungsten plates, interleaved with 8 mm scintillator plates. In this study, the mechanical detector model, with an outer diameter of 5.8 m, an inner diameter of 2.8 m and a length of 3.5 m, is formed by 18 wedge shaped sectors. The support structure for the absorber and detector planes is made from stainless steel. This way, the required absorber plates have a maximum size of 1 m $\times$ 3.5 m, which can be manufactured with modern production techniques.

Two alternative construction principles for the steel structure have been compared in preliminary analyses and gave similar results. Figure 6.20 shows a "box design", in which the sectors are made by three trapezoidal boxes. The tungsten absorber and detector layers are installed inside the boxes. The tungsten absorber plates participate in the structural behaviour of the sector, since the plates are bolted to the steel support. The total weight of the model detector is some 670 tons, of which the major part (610 tons) is due to the weight of the tungsten absorber plates. The model includes a 75-ton ECAL suspended from the HCAL. Finite element analysis indicates that both designs, under this load, show only relatively small deformations in the range of 1–2 mm. For the calculations the support of the detector barrel has been assumed at the 3 and 9 o'clock positions. The stress levels in the steel lattice remain also below the material limits, respecting the standard safety factors. Therefore the design and construction of a fine-segmented HCAL with a tungsten absorber can be considered conceivable from the mechanical point of view.

## 6.4 Calorimeter Performance under CLIC Conditions

Detailed GEANT4 based simulation studies have been undertaken in order to demonstrate that the conceived calorimeter systems can meet the physics performance requirements at CLIC. This was done by building upon the tools developed in the ILC context. However, dedicated efforts were necessary in order to realistically take the conditions at CLIC into account. First, the extension of the particle flow reconstruction approach to multi-TeV energies was driving the development of the PANDORAPFA algorithm towards optimising its particle separation power in dense and energetic jets. Second, to quantify the effects of background pile-up, a software framework was developed to overlay events [29], taking detector sampling times into account, and investigate the effects on the signal quality.

### 6.4.1 ECAL Performance for High Energy Electrons

Simulation studies to evaluate the performance requirements and the functioning of the existing ECAL designs for electrons in the CLIC energy range are still ongoing. For example, full detector simulation studies carried out in the framework of the heavy slepton benchmark studies (see Section 12.4.5) with the CLIC_ILD detector model indicate that a dynamic range of 12 bits is required for the analog ECAL option. The highest energies are expected for electrons from Bhabha scattering, which are used to measure the luminosity spectrum (see Section 12.2.1). Depending on the required ECAL energy resolution for this measurement an increased dynamic range or a nonlinear amplification curve may be opted for.

While the precision requirements for ECAL at CLIC are still being evaluated, it is clear that the electronics design of ILC detectors needs adaptation in order to accommodate the larger dynamic range. In all technologies, a finer granularity mitigates dynamic range requirements. In the case of scintillator readout, the R&D will be driven towards MPPCs with larger number of pixels. In the case of silicon pads, the dynamic range is presently limited by the front-end electronics, but it is rather straightforward to extend the automatic gain selection schemes implemented in the ROC [16, 17] and KPiX [30] designs. Future studies will define the detailed requirements and lead to optimised solutions for ECAL sensors and readout electronics.

### 6.4.2 Timing Resolution

Background processes leading to pile-up with physics events are a key issue for detectors at CLIC. As described in Section 2.5.1 time stamping is one of the most powerful tools to suppress pile-up and single hit resolutions of ≈ 1 ns are required for both ECAL and HCAL. The time resolution of a subdetector depends on the sensor response characteristics, on the front-end electronics parameters, and, for hadron showers, on the intrinsic time evolution of the delayed processes in the nuclear cascade. The latter also determines the time window over which the signal must be integrated in order to achieve an optimised energy resolution, and over which, as a consequence, pile-up is accumulated. Based on the signal evolution predicted by the GEANT4 simulation shown in Figure 6.3, a window of 100 ns is assumed for the tungsten barrel in the performance studies with background, and 10 ns elsewhere. Gaseous detectors, such as RPCs, have different sensitivities to particles in the shower when compared to scintillation counters. In particular, their sensitivity to neutrons may be much reduced. Therefore, such detectors would most likely measure a different time behaviour of the shower in a tungsten HCAL. The time evolution in hadronic showers will be subject of future experimental investigations.

The sensor itself is fast enough for all the proposed technologies. RPCs are popular for trigger purposes due to their fast response; thin-gap and multi-gap versions are being discussed. For scintillator based detectors, the slowest process is the light conversion in the wave length shifting fibre, which has a time constant of the order of 20 ns. This can be eliminated in the case of a photo-sensor directly coupled to the scintillator as already shown from the T3B test (see Section 6.3.3.4 above). For silicon pads, a somewhat slower shaping time than for SiPMs is required in order to preserve the good signal to noise behaviour. A resolution of 5 ns per hit has already been realised with the existing ASIC ROC scheme, initially not designed with the precise CLIC timing requirements in mind. The KPiX chip, originally developed for the Next Linear Collider (NLC) [31], features fast shaping already. In summary, a single hit time resolution of ≈ 1 ns seems in reach for all technologies. Further details on a possible electronics readout scheme for CLIC calorimetry is given in Section 10.2.

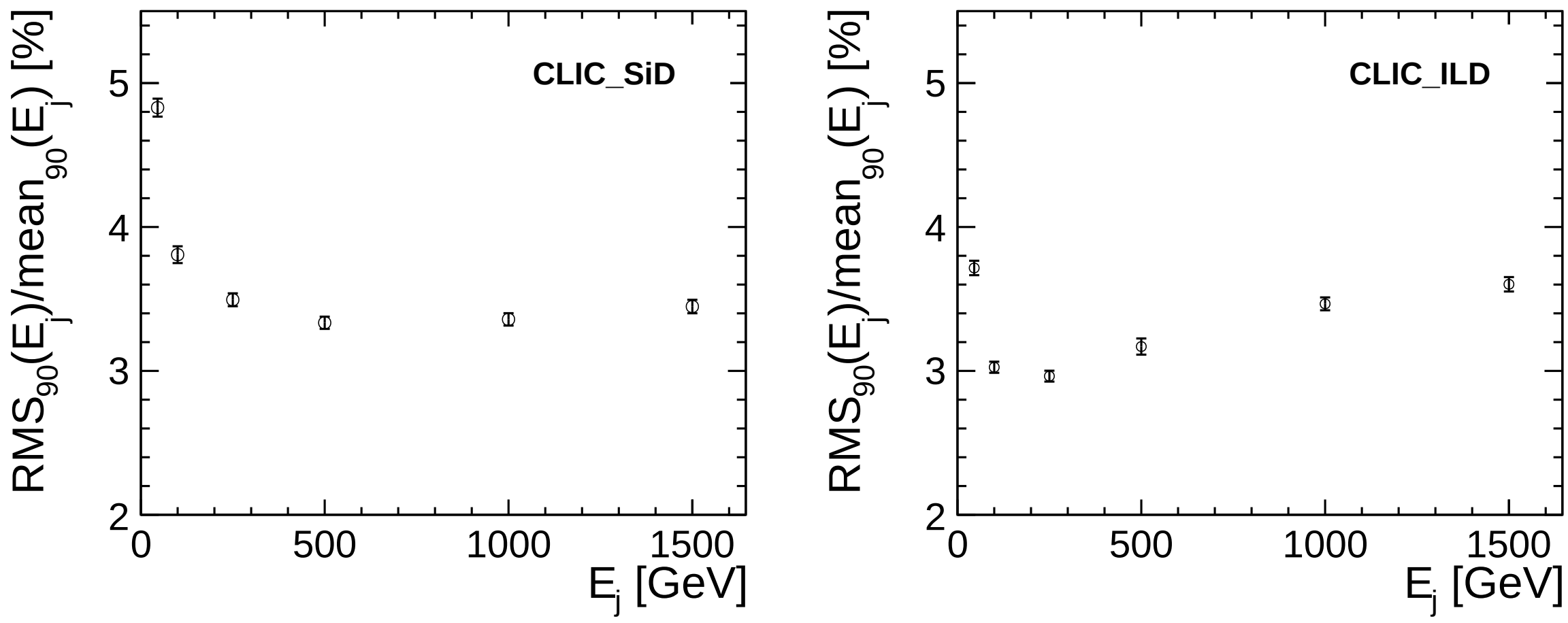

Fig. 6.21: Jet energy resolution as a function of jet energy for CLIC_SiD (left) and CLIC_ILD (right) for the barrel region $|\cos(\theta)| < 0.7$.

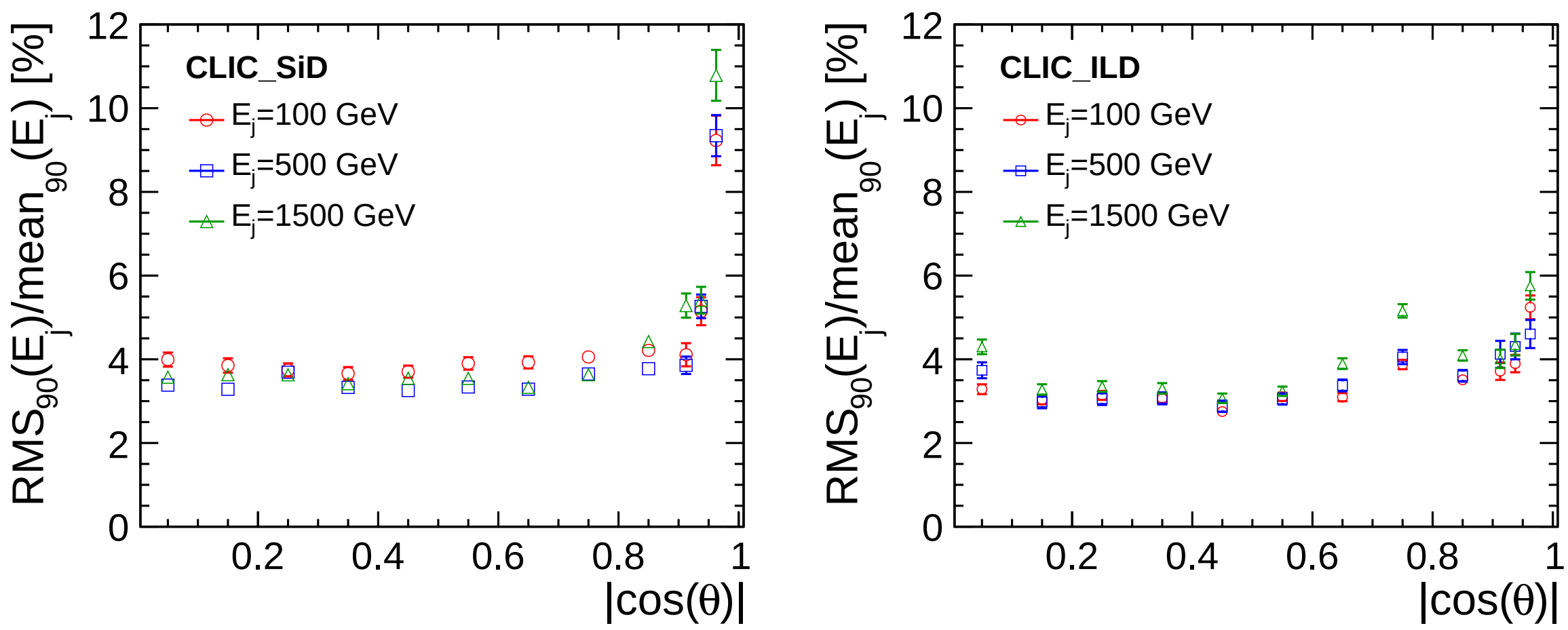

Fig. 6.22: Jet energy resolution dependence on event angle and jet energy for CLIC_SiD (left) and CLIC_ILD (right).

### 6.4.3 Jet Energy Resolution

A study of the particle flow performance for both the CLIC_ILD and CLIC_SiD detector models was carried out. For this study single Z-like bosons of masses ranging from 91 GeV to 3 TeV and decaying at rest into light quarks, hence producing two mono-energetic jets, were generated [32]. No jet reconstruction is carried out at this stage. The full energy deposited in the detector $E_{jj}$ is analysed to avoid a bias from jet reconstruction. The resolution of the jet energy $E_j$ is obtained by calculating the $\mathrm{RMS}_{90}(E_{jj})$ and the $\mathrm{mean}_{90}(E_{jj})$ from the data and then applying a factor of $\sqrt{2}$. Figure 6.21 shows the jet energy resolution as a function of the jet energy. At high energies, particle flow turns into energy flow and confusion becomes dominant. In general both detector concepts show similar performance. For low jet energies, CLIC_ILD shows a slightly better jet energy resolution compared to CLIC_SiD.

In the barrel region, both detectors show only small variation with the angle of the jet as shown in Figure 6.22 for three different jet energies. The performance for CLIC_SiD is somewhat worse in the forward region, especially in the last bin, due to angular coverage. The last bin covers the region $0.95 < |\cos\theta| < 0.975$ which means it goes to angles as small as $12^{\circ}$. Table 6.2 shows that the angular acceptance is smaller for CLIC_SiD, in particular for the HCAL which covers only angles down to 15.5° and therefore misses more energy in the forward region. In the case of CLIC_ILD, a dip in jet energy resolution occurs in the overlap between barrel and forward region. This is due to a gap between the ECAL barrel and ECAL endcap which is bigger in the CLIC_ILD detector than in CLIC_SiD.

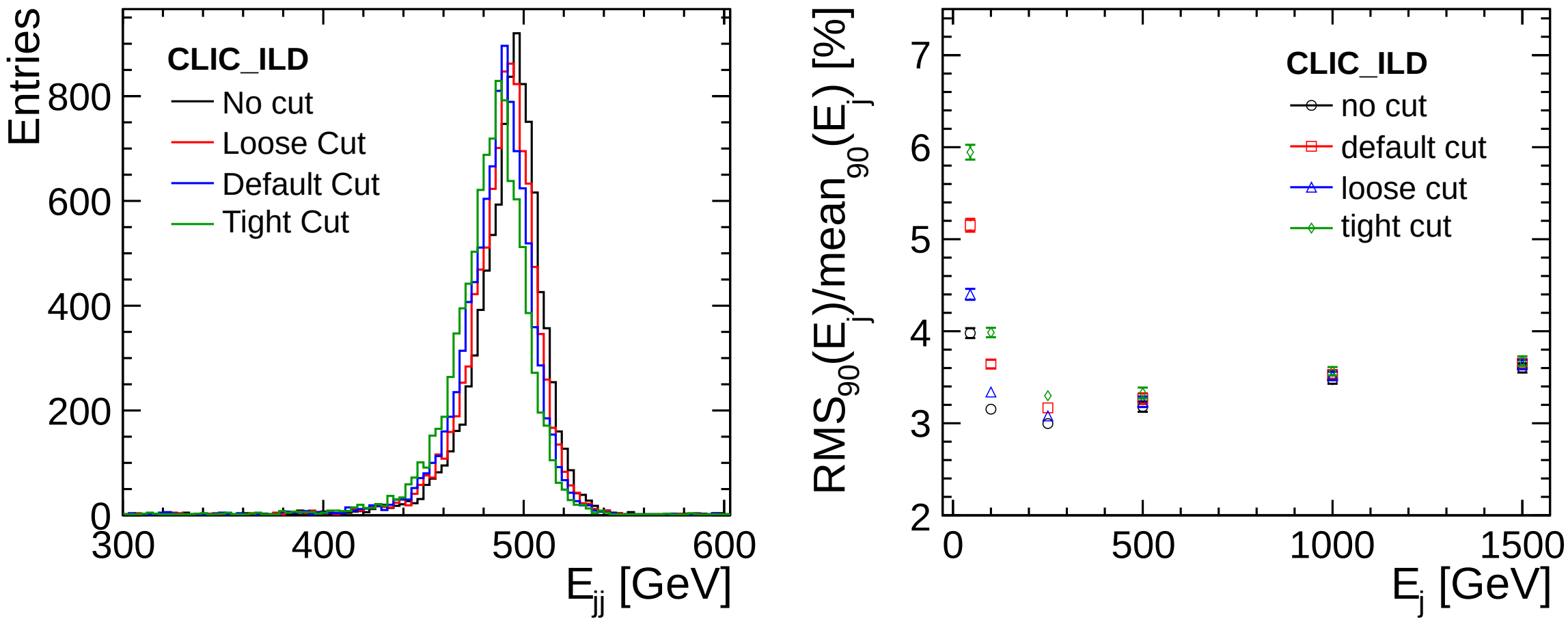

Fig. 6.23: Distribution of the reconstructed energy of a Z-like particle of 500 GeV mass decaying at rest into two light quark jets (left) and jet energy resolution as a function of jet energy (right) both for various timing cuts for CLIC_ILD.

A set of timing cuts has been developed which efficiently suppresses the background, as motivated in Section 2.5.1. First, the impact of the timing cuts on the physics event alone is studied. Figure 6.23 shows the reconstructed energy in the event before applying the timing cuts and for each of the different timing cuts specified in Appendix B. The more stringent the cut the more energy is cut away from the event. The changes get less significant with increasing jet energy. As the combined effect of the cuts and of the residual pile-up depends strongly on the physics channel under study, the effect of background pile-up is presented in more detail in Chapter 12 in the context of the physics performance.

Table 6.2: Acceptance of the calorimeters in the forward region of CLIC_SiD and CLIC_ILD.

| | | $R_i$ [mm] | $Z$ [mm] | $\theta$ [°] |
|---|---|---|---|---|
| CLIC_SiD | ECAL | 210 | 1657 | 7.2 |
| | HCAL | 500 | 1805 | 15.5 |
| CLIC_ILD | ECAL | 242 | 2450 | 5.6 |
| | HCAL | 400 | 2650 | 8.6 |

## 6.5 Future Calorimeter R&D for CLIC

To conclude this chapter, a prioritised R&D programme targeted at a consolidation of the calorimeter designs is outlined. This is not intended to be a complete list of all the work under way or to be performed. A five-year period is anticipated for the next CLIC project phase and a more exhaustive overview of the relevant CLIC-related linear collider R&D topics is given in Chapter 13. In the framework of the overall CLIC detector concept optimisation, there are some outstanding issues with impact on the calorimeter design. As described in Section 2.4.3 the inner regions of the HCAL endcap show high occupancies due to scattered neutrons from incoherent pairs interacting in the BeamCal. Therefore the interaction region and mask design will be re-optimised in the next project phase. This study may also involve adaptations to the calorimeter design in the endcap, such as finer granularity or technology choices favouring active layers which are insensitive to neutrons. Following this study, the requirements and conceptual design of the electronic readout chain can be refined to give better R&D goals. Furthermore, the calibration and reconstruction in the tungsten-barrel to iron-endcap transition region will be studied in more detail.

In addition there are a large number of R&D issues. Among these, the hadronic response of a tungsten absorber calorimeter, using both scintillator and gas based active layers, has a high priority. It is recommended to design readout electronics matched to the CLIC experimental conditions, and to demonstrate its feasibility with a scalable prototype. This electronics needs to have low power budget, by exploiting power pulsing and advanced power distribution schemes. It is proposed to pursue R&D on active devices, in particular on a fast sensor response, for example optimising the SiPM scintillator coupling, or investigating the best suited RPC gap structure. In addition, engineering and integration solutions for the calorimeters are required.

# Chapter 7

# Detector Magnet System

## 7.1 Introduction

The magnet system for the CLIC detector concepts comprises the central solenoid, the ring coils on the endcaps of CLIC_ILD, and the two forward anti-solenoids together with the ancillary systems necessary for their operation. Their design is compatible with the push-pull scenario. While the central solenoids and the anti-solenoids are superconducting magnets, the ring coils are implemented in normal conductor technology. The central solenoids of the two detectors are slightly different in their design parameters, as described in the next paragraph. The most challenging design, the CLIC_SiD, is taken as the reference, as there are no significant technical issues related to using the same approach and assumptions for the CLIC_ILD design. Given its structural rigidity, the cryostat of the superconducting central solenoid is also used as the principal support structure for the barrel calorimeters and tracking detectors.

The conceptual design of the central solenoid and of the two forward anti-solenoids is based on the experience gained in the past fifteen years with the construction and operation of the LHC detector magnets, in particular the ATLAS central solenoid [1] and the CMS solenoid [2, 3, 4]. This experience is complemented with the results of conceptual design studies performed for the ILC detector magnets [5, 6] as well as of R&D campaigns and design studies for high field thin solenoids at KEK [7] and at CERN [8]. The solenoid design, presented in this chapter, is described in more detail in [9].

## 7.2 The magnetic field requirements

The main parameters of the superconducting central solenoids of the two CLIC detectors, CLIC_SiD and CLIC_ILD, are listed in Table 7.1.

Table 7.1: Main parameters of the superconducting solenoids in CLIC_SiD and CLIC_ILD.

| | Nominal magnetic field (T) | Free bore (mm) | Magnetic length (mm) | Cold mass weight (tons) |
|---|---|---|---|---|
| CLIC_SiD | 5.0 | 5480 | 6230 | 170 |
| CLIC_ILD | 4.0 | 6850 | 7890 | 210 |

The nominal magnetic field is the value on the coil axis at the interaction point. The free bore corresponds to the inner diameter of the coil cryostat and the magnetic length to the length of solenoid winding pack. The cold mass weight includes coil windings and the outer support cylinder (see also Section 7.3). Magnetic field calculations using ANSYS [10] have been performed to evaluate the field map (see Figure 7.1) in the detector tracking volume in order to estimate the magnetic field homogeneity. The quality of the field is important, especially in the CLIC_ILD detector, which uses a TPC as the central tracking device. The magnetic field distortion along the z-direction within the size limits $(z, r)$ of the TPC is defined as

$$\Delta l(r) = \int_0^z \frac{B_r(z)}{B_z(r)} dz$$

and is required to be less than 10 mm everywhere. This requirement on the magnetic field quality can be somewhat relaxed assuming that it is possible to precisely measure, at the $10^{-4}$ level, the actual magnetic field map once the magnet is built [11, 12].

The precision of this measurement depends on the resolution of the magnetic field sensors and on the accessibility of the field volume for the field map measurement. At the time of the magnetic field mapping, the experiment shall be in a near-final configuration in order to measure the field map in the correct magnetic environment. Moreover, the magnet power supply has to deliver the nominal current with a precision of a few ppm. The latter was already demonstrated for the LHC detector solenoids.

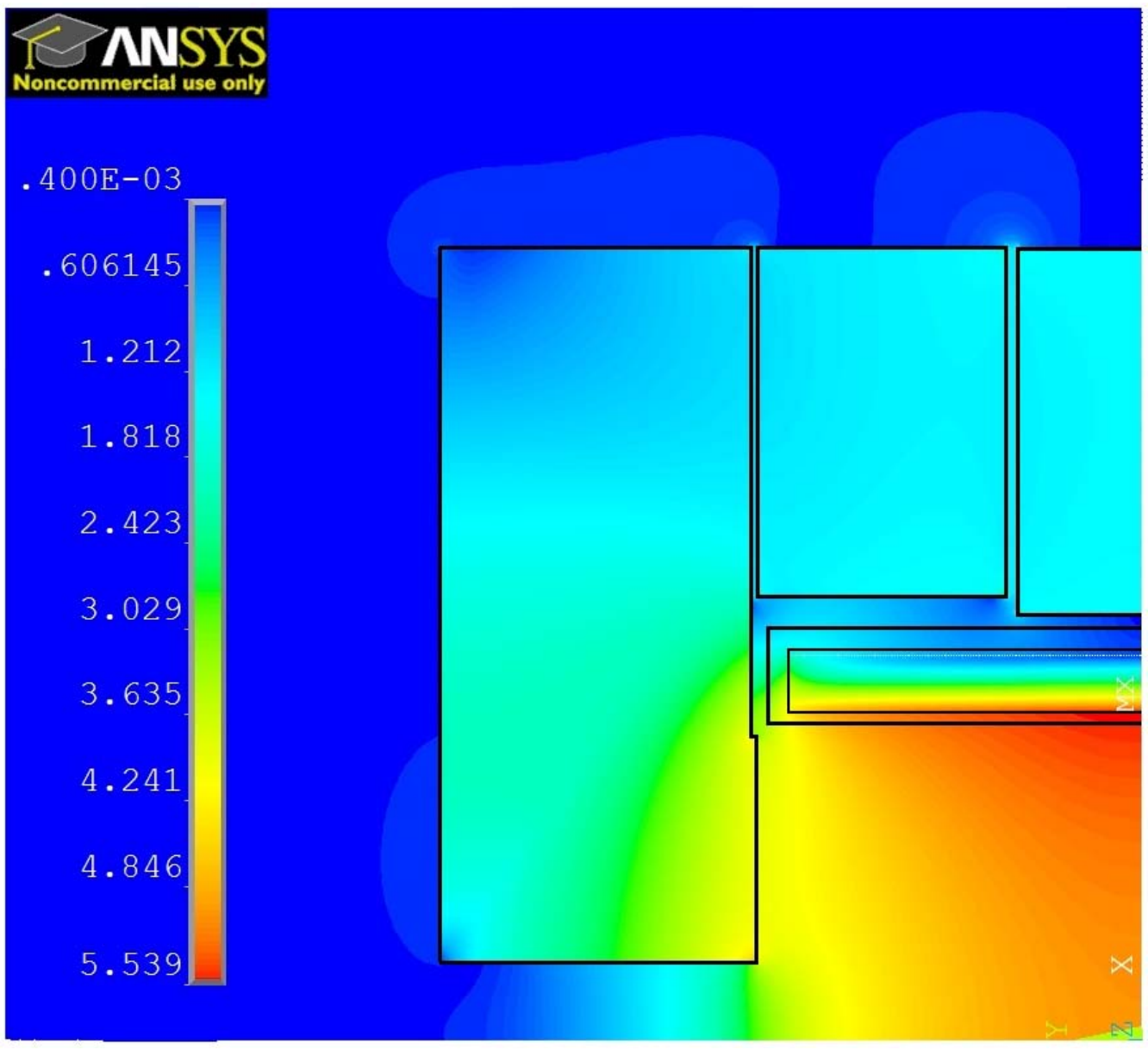

Fig. 7.1: Magnetic field (T) generated by the central solenoid in CLIC_SiD

The magnetic stray field outside the detector is important because the proximity of the iron of one detector can eventually cause perturbations in the magnetic field chart of the other one. The magnetic stray field is reduced by choosing an appropriate thickness for the return yoke, in particular for the detector endcaps. At the same time, the overall weight of the detector shall be limited in order to ease the assembly, facilitate the push-pull operation and keep costs under control. A residual field of 50 Gauss at 15 m distance from the detector is considered acceptable and has been used as guidance to define the yoke thicknesses in the CLIC_ILD and CLIC_SiD detector models. The iron yoke also has a radiation shielding function. This is important because personnel working on the off-beam detector, located in its service cavern at some 28 m distance from the on-beam detector, could be exposed to radiation escaping from the on-beam detector. In addition, provisions for thick shielding doors have been made, between the two detectors, to be implemented in case the measured dose will be higher than expected (see chapter 11). Figure 7.1 shows the solenoid field map, with emphasis on the field quality in the tracking region. In the model shown here, the filling factor of the iron yoke volume is 100%, while in the real detector 10%-15% of the space will be occupied by muon instrumentation. The technical implementation of the solenoid, presented in this chapter, takes this reduced filling factor into account.

The central solenoid is complemented by two smaller coil assemblies, located in the detector forward region and surrounding the forward focusing elements in the yoke endcap region (see Section 11.2, Figure 11.13).

The main function of these coils is to protect QD0 and to locally reduce the magnetic field of the central solenoid in order to limit perturbation on the incoming beam. Because of the beam-crossing angle of 20 mrad, the magnetic field of the central solenoid has a component pointing perpendicular to the incoming beam particles. As a result, the particle trajectories are distorted, causing a reduction in the luminosity. The most severe effects originate from the overlap of the transverse components of the magnetic field of the central solenoid and of the final focusing quadrupole QD0 (see also [13]). The second function of the anti-solenoid coils is to protect the QD0 magnet. QD0 is composed of permanent magnet elements combined with an electromagnet. This allows meeting the requirement of a very high field gradient together with a tuning capability. However, the permanent magnet inserts risk to undergo a partial demagnetisation in the strong field of the central solenoid. The anti-solenoid will mitigate this effect. A side effect of the presence of the anti-solenoid is a distortion of the magnetic field of the central solenoid in the detector tracking region, requiring a careful design of the CLIC detectors in the technical design phase. The anti-solenoid conceptual design is presented in Section 7.5.

The layout of the experimental cavern around the interaction point is optimised in view of the stringent stability requirements for the final focusing quadrupoles QD0. The accelerator tunnel is designed to approach the detector on the IP as much as possible (see chapter 13). This implies that the overall length of CLIC_SiD and CLIC_ILD must be identical. However, given the inner detector (TPC) and calorimeter dimensions, together with the appropriate yoke thickness, the CLIC_ILD detector would be longer than CLIC_SiD. In order to achieve equal length of both detectors, a part of the yoke steel in the endcaps of CLIC_ILD is compensated by the use of ring coils on the endcaps. The conceptual design of this system is described in Section 7.6.

## 7.3 Solenoid Coil Design

The main parameters of the central solenoid of the CLIC_SiD design are listed in Table 7.2. They are similar to those of the CMS solenoid. The design therefore relies on many features successfully tested previously:

- an aluminium stabilised superconductor with a mechanical reinforcement;
- a multi-layer and multi-module coil with an external support cylinder;
- a coil winding technique adapted to a fibre glass wrapping of the conductor;
- vacuum impregnation of each module and a subsequent curing by heat treatment;
- an indirect, conduction based cooling by helium flow in aluminium tubes attached to the surface of the support cylinder;
- operation in a thermo-siphon cooling mode;
- a supporting system with radial and axial tie rods attached to the external mandrel and to the vacuum tank.

Table 7.2: Main parameters of the superconducting central solenoid for CLIC_SiD [9].

| | | |
|---|---|---|
| Nominal magnetic field at the IP | 5.0 | T |
| Peak magnetic field on the conductor | 5.8 | T |
| Free bore diameter | 5.5 | m |
| Magnetic length | 6.2 | m |
| Ampere·turns | 34 | MA·turns |
| Operating current | 18 | kA |
| Stored magnetic energy | 2.3 | GJ |
| Energy/Mass ratio | 14 | kJ/kg |
| Inductance | 14 | H |

The number of layers in the coil windings is kept at a minimum. To facilitate manufacturing, the aspect ratio (height/width) of the conductor is chosen to be below 7, while the thickness of the external support cylinder is limited to 50 mm. As a result a layout with 5 coil layers is chosen in combination with a conductor cross section of 97.4 mm $\times$ 15.6 mm. The total length of conductor is approx. 40 km. The solenoid is split along the $z$-direction into 3 modules. As a result the unit conductor length is some 2.7 km and there are 16 layer-to-layer and module-to-module electrical connections. In order to have maximum temperature margin in these connections they are positioned in a region of low magnetic field on the outer radius of the mandrel, similar to the CMS case. The conductor design fulfils all requirements for good superconducting performance, thermal and electro-magnetic stability, mechanical strength, and quench protection. A cross section of the conductor and a cut through the coil assembly are shown in Figure 7.2.

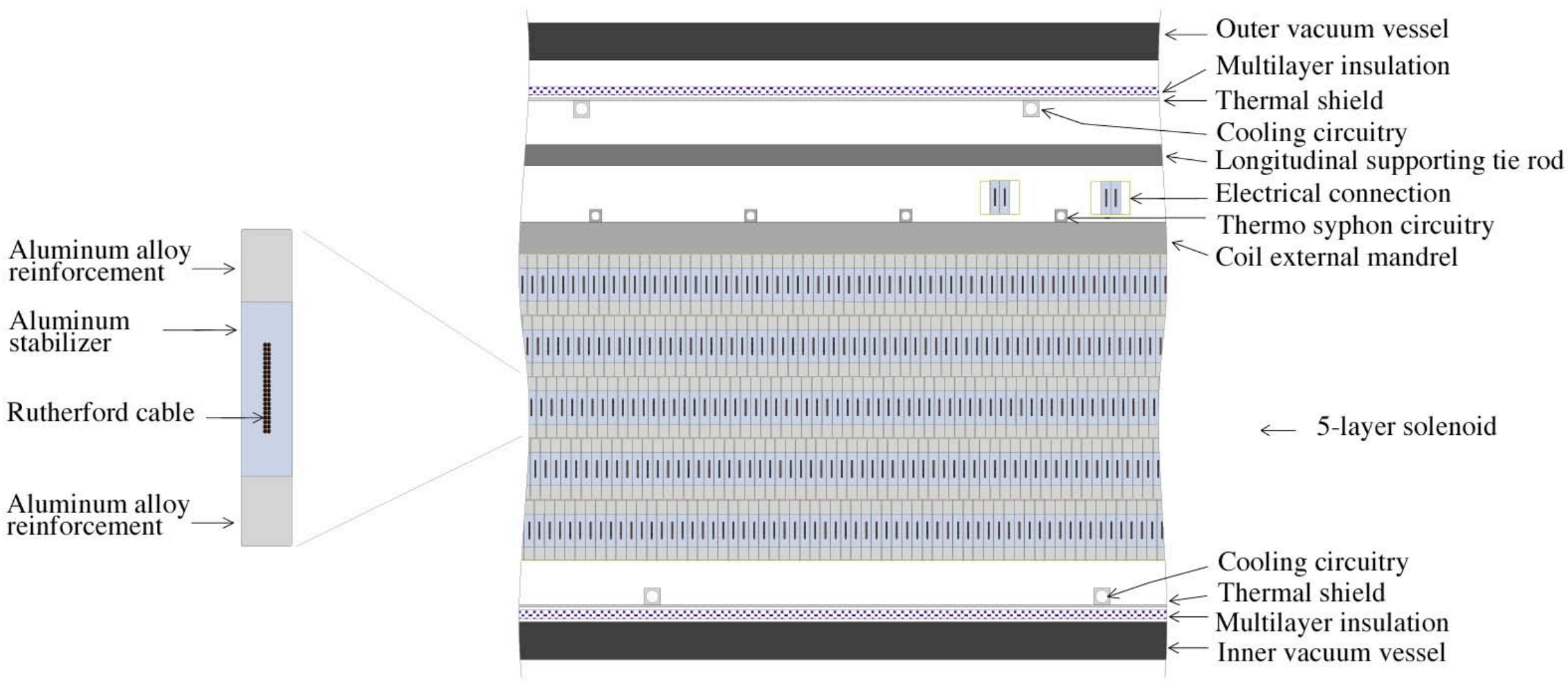

Fig. 7.2: Cross section of the solenoid conductor and cut through the coil assembly.

The core of the conductor is a 40-strand NbTi/Cu Rutherford cable. Based on state-of-the-art NbTi conductors with critical current densities in of 3000 A/mm$^2$ at 4.2 K and 5 T, a temperature margin of 1.5 K is considered realistic. For a current of 18 kA at an operating temperature of 4.5 K and a 5.8 T peak magnetic field in the innermost layer the conductor operates at 32% of its critical current. Under these conditions the enthalpy margin is 2.4 J/m. The 4.5 K operating temperature is compatible with static and transient heat loads. A radiation shield cooled at 80 K by helium gas intercepts the radiation heat load. The thermal shield is covered by multilayer super-insulation. The pressure in the cryostat shall not exceed $10^{-6}$ mbar.

The mechanical reinforcement of the conductor can be realised by applying existing technologies, either by welding one or two structural aluminium alloy bars to the aluminium sheathed Rutherford cable (like the conductor for the CMS solenoid), or by using a micro-alloyed aluminium stabiliser hardened by a final cold working operation (as used for the conductor of the ATLAS central solenoid). A combination of both reinforcement technologies may be the best solution. The dimensions of the conductor and its reinforcement are defined to keep the maximum stress well below the tensile and yield strength of the reinforcing material. Safety margins of typically a third of the tensile strength and two thirds of the yield strength, covering the full area of the coil, are applied. Detailed structural analysis covering the full area of the coil will be carried out to confirm these estimations. In order to guarantee the stability of the superconductor the design ensures that the Residual Resistivity Ratio (RRR)[1] of the stabiliser will amount to at least 300 during the full lifetime of the solenoid.

The coil protection is based on the fast extraction of the stored magnetic energy into an external dump resistor (see also Section 7.6). As soon as the main switch breaker of the quench protection system is opened, the magnet current decreases following the intrinsic L-R time constant of the circuit. Given the mutual coupling between the coil winding and the external cylinder, an electrical current is induced in the cylinder, which then acts as a heater that initiates a quench over the entire neighbouring layer. This so-called quench-back effect limits the temperature gradient in the coil and the associated thermal stresses. The fast discharge voltage will remain below 600 V. The computed temperature gradient between two adjacent layers is less than 7 K, and approximately 60% of the stored energy is extracted. The average temperature of the solenoid cold mass after a fast dump will stay below 70 K. In the ultimate fault case of a normal zone propagation without external energy extraction, the quench-back effect is still efficient and the peak temperature will not exceed 150 K.

[1]RRR is defined as the ratio between the value of the electrical resistivity of the stabiliser at 273 K and the value at 4.2 K.

The inner winding technique, successfully applied to the CMS coil manufacturing [14], is proposed for the coil winding. Vacuum impregnation of the multi-layer coil using fibreglass wrapping and an epoxy-type resin will be applied. The module-to-module coupling is performed by stacking the modules with their axis in vertical orientation while carefully controlling the flatness of the contact surfaces. The coil is then swivelled to the horizontal position and inserted into the outer cylinder of the vacuum tank. This will take place on the experimental area site in a surface building, where the final coil and yoke assembly can be tested.

## 7.4 Conductor Options

To provide the required reinforcement of the aluminium stabilised conductor, the baseline is to adapt the existing strengthening technologies that were applied with success on thin superconducting coils for particle detectors in high energy physics, and to identify and investigate the emerging technologies that may become industrially available at the time the project will be launched. Currently, the options considered are, either exclusively or in combination:

- stabilise the conductor by welding structural bars of aluminium alloy (used in the CMS solenoid);
- use a zinc or niobium-doped, micro-alloyed structural stabiliser (used in the ATLAS solenoid);
- identify and develop a novel doping method for stabiliser reinforcement (e.g. carbon nano-tubes), depending on availability and demonstration of effectiveness.

The conductor cross section discussed in Section 7.3 is shown in Figure 7.3 together with the cross sections of several other aluminium stabilised conductors manufactured for recent large superconducting detector magnets.

Since the feasibility of the conductor technology is crucial for the design, an R&D programme has started to develop conductor options and demonstrate the feasibility of manufacturing the proposed large stabilised conductor. As a first trial, a conductor of 100 m length will be co-extruded and cold-worked using an Al-0.1wt%Ni micro-alloy, identical to the material used in the ATLAS solenoid. The manufacturing of the demonstration sample will allow the measurement of its properties, such as the gain in tensile and yield strength, the RRR of the conductor stabiliser, the uniformity of the properties over the large cross section and the quality of the inter-metallic bonding between the Rutherford cable and the stabiliser. In addition, the superconducting properties will be investigated to check for possible degradation effects in the critical current density, to determine induced ramp losses and to verify the propagation of the normal zone.

## 7.5 Anti-Solenoid Design

The anti-solenoid magnet consists of 6 short coils interleaved with free spaces, surrounding the QD0 support tube over a length corresponding to the detector yoke endcaps. The number of ampere-turns and the current density of each coil have been adjusted to minimise the detector solenoid field along the beam trajectory and in the QD0 region. A maximum current density of up to 80 A/mm$^2$ is reached in the innermost coil, which has 2 MA·turns. The other coils have a much smaller current density. A ferromagnetic disk has been introduced in front of QD0 to lower the radial field in that region. The anti-solenoid coils together with their cryostat and supporting tubes have an inner bore of 1.1 m and an outer diameter of 1.3 m (see Figure 7.4).

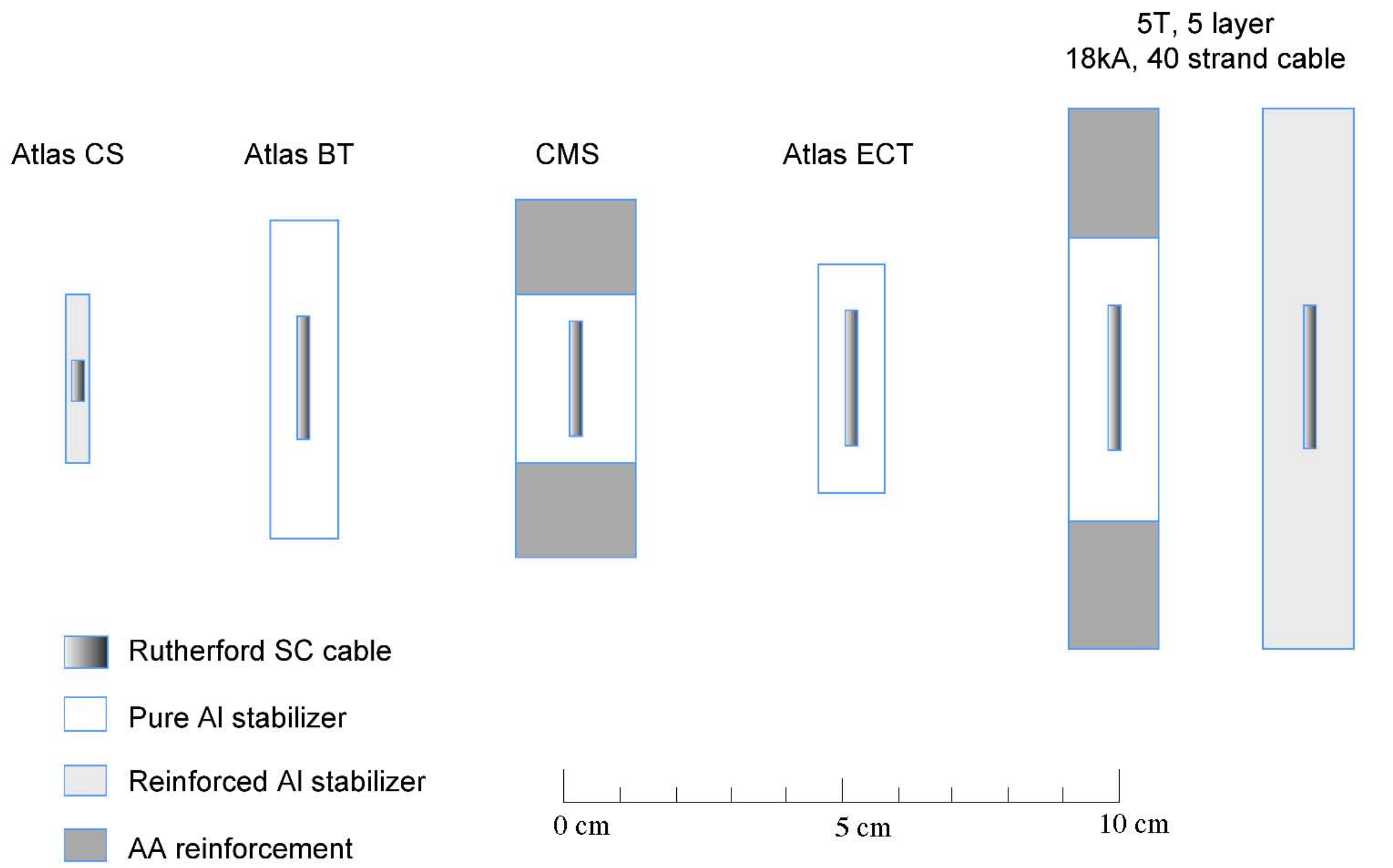

Fig. 7.3: Cross sections of Al stabilised and reinforced conductors used in existing detector systems and the proposed two conductor options for the 5 T solenoid in the CLIC_SiD design.

The magnetic forces resulting from the combination of the magnetic field of the central solenoid and the anti-solenoid have been evaluated using a finite element analysis. Two different load cases have been examined: both magnets are powered ("Load Case 1") or only the anti-solenoid coils are powered ("Load Case 2"). The results are summarised in the Table 7.3. In "Load Case 1", the interaction of the magnetic field of the central solenoid and the anti-solenoid pushes the latter out of the detector along the beam direction with a force of some 680 tons. The anti-solenoid winding is undergoing a high compression stress of up to 110 MPa. Therefore the structural design shall include buckling analysis. In "Load Case 2" the repulsive force is less important and the tension stress on the anti-solenoid winding is limited to 70 MPa.

Table 7.3: Main forces and stresses acting on the anti-solenoid coils

| | Sum of axial force acting on the coils (MN) | Coil maximum hoop stress (MPa) |
|---|---|---|
| Load Case 1 | 6.7 | −110 |
| Load Case 2 | 1.7 | +70 |

The ferromagnetic disk in front of QD0 is subject to an attractive force towards the IP estimated at some 25 tons in "Load Case 1". The numbers show that the mechanical design of the anti-solenoid requires a substantial effort, considering also the limited space for its integration between the endcap yoke and the QD0 support tube. Moreover, a detailed study of the behaviour of the magnetic coupling between the central solenoid and the two anti-solenoids is required to evaluate the best powering, discharge and protection logic to be implemented.

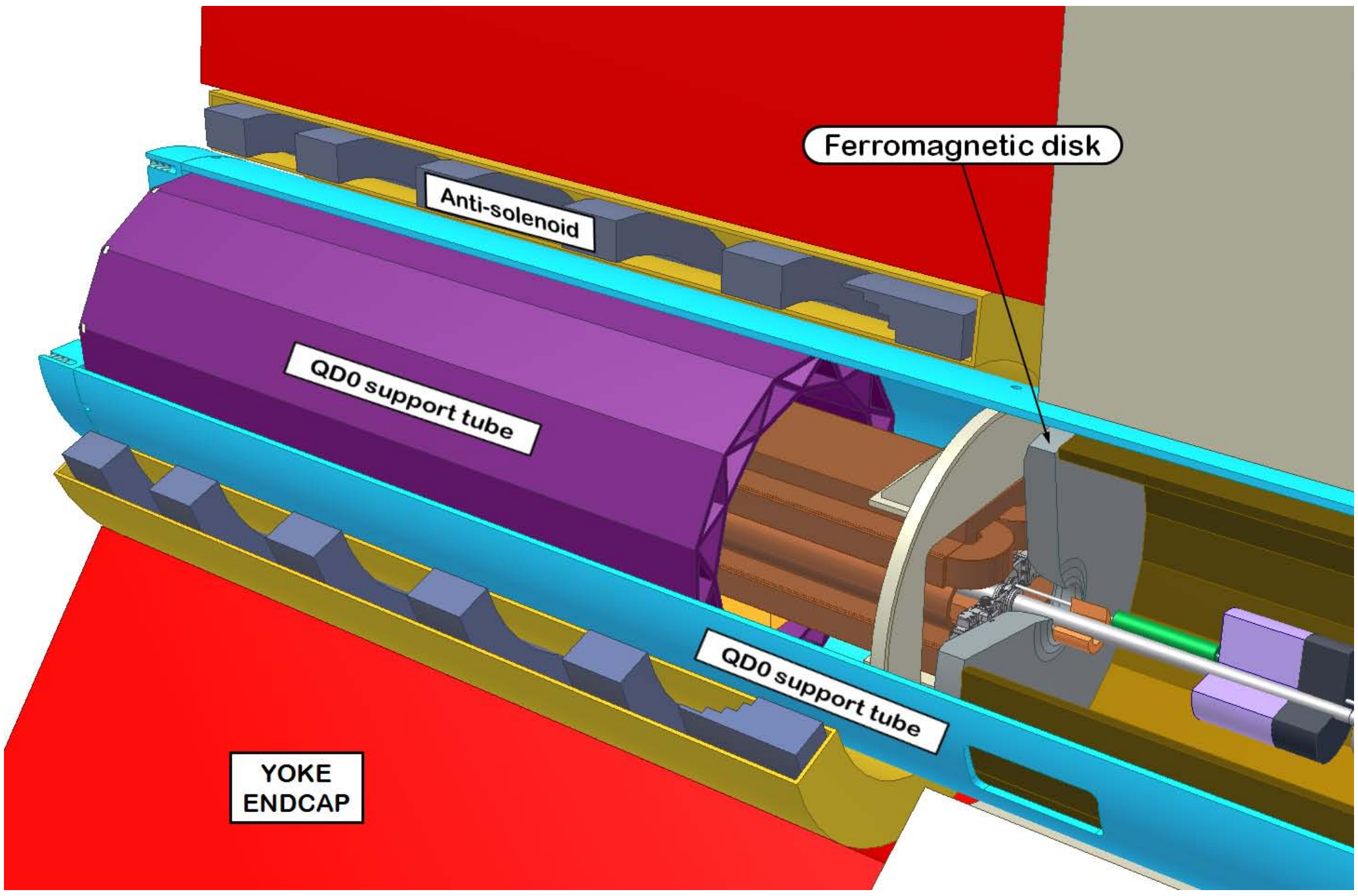

Fig. 7.4: View of the anti-solenoid around QD0 in the forward region of CLIC_SiD

## 7.6 The Ring Coils on the Endcap Yoke of the CLIC_ILD Detector

The two detectors, CLIC_ILD and CLIC_SiD, are assumed to be operating alternately on the IP, using the "push-pull" system, and the two detectors must have the same overall length. Therefore, the yoke endcaps of CLIC_ILD need to be shorter then a classical design would allow. A conceptual study proposes to achieve this by replacing 0.66 metres of iron in each of the endcaps by a set of ring coils. Details are given in [15], the study is summarised below.

The design criterion for the ring coils is to achieve the same field quality as well as the same stray magnetic field as for an all-iron yoke. According to results from POISSON calculations, this can be achieved e.g. by a set of four normal-conducting ring coils per endcap, each with a width in beam direction of 20 cm and radii ranging from 50 to 70 cm. The power consumption of such a system is estimated to be 1.3 MW. The total amount of iron suppressed in the endcaps is 1400 tons. This design proposal will have to be optimised during the technical design phase.

## 7.7 Magnet Services and Push-Pull Scenario

The detector magnets will be switched off during the push-pull operation, however they will be kept at cryogenic temperatures during the displacement. The magnet services (cryogenics and vacuum pipes, powering and protection lines) have to be compatible with constraints imposed by the push-pull scenario. Flexible cryogenics transfer lines and vacuum pipelines are already in use for the ATLAS End Cap Toroids to allow for the opening of the detector and provide access to the inner detectors.

The overall length of the powering line will easily exceed 60 m. To avoid repetitive opening and closing of the circuit via bolted connections during the detector push-pull, with the associated risk of resistance increase and local overheating, a permanent flexible connection is considered. A possible option consists of a high temperature superconducting flexible cable assembly [16] cooled by the helium gas return line at a temperature between 5 K and 20 K. Another option would be a NbTi based flexible cable cooled by the helium inlet at 4.5 K.

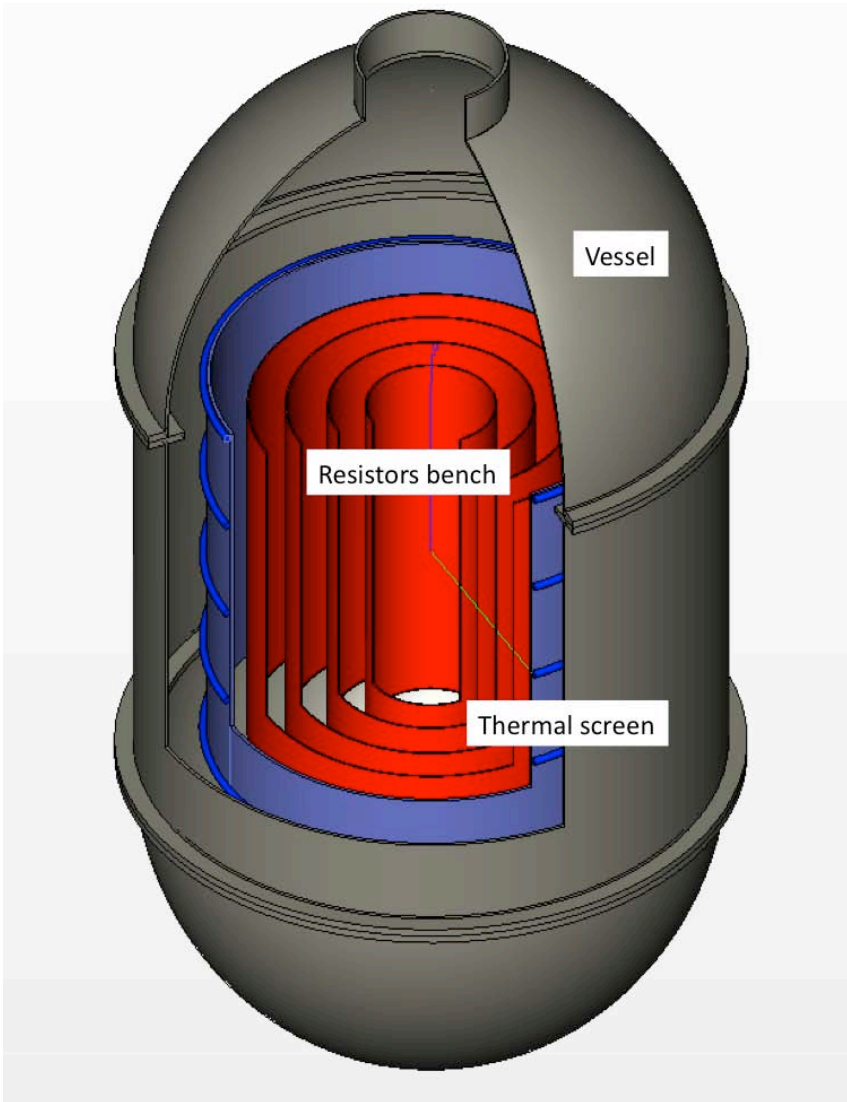

Fig. 7.5: Sketch of the compact dump resistor proposed for the quench protection of the central solenoid.

As any superconducting power line may be subject to an unexpected quench, the detector magnet protection circuit cannot pass through this flexible power line. Therefore a compact dump resistor will be installed close to the coil on the detector platform. The device will protect the superconducting coil in case of a quench by extracting a substantial amount of magnetic energy from the coil itself and converting it into heat in the dump resistor. It is electrically connected in parallel to the coil current leads. Based on CERN experience dated from the nineteen seventies with the BEBC experiment and pursuing the study started at SLAC for SiD in 2007 [17], a water-based compact dump resistor is proposed [18] as shown in Figure 7.5. The main features of this system are highlighted in Table 7.4.

Table 7.4: Main characteristics of the compact dump resistor.

| Water pressure (bar abs) | Water volume ($m^3$) | Enthalpy (15–100°C) (kJ/kg) | Total energy (MJ) | Peak power (MW) |
|---|---|---|---|---|
| 1.0 | 3.6 | 355 | 1.28 | 12 |

The cryogenics and vacuum plant is represented in Figure 7.6. On the surface of the experimental area, the helium gas tank and liquid $N_2$ dewar ensure the storage of the coolants and the helium gas compressor feeds the helium liquefier (cold-box), located in the service area of the experiment cavern. Liquid helium is then sent to a dewar close to the detector magnet and distributed via a valve box. A phase separator, located on top of the magnet, ensures the thermosiphon mode operation, i.e., the cooling helium circulates without any active pump, the driving force being ensured by the difference in density between the liquid and the vapour. The connection between the helium liquefier and the dewar is via a flexible vacuum-insulated cryogenic transfer line that ensures the cooling of the magnet also during the detector movement from the garage to the beam position. The detailed engineering of such a line requires some R&D, namely for what concerns the minimum bending radius and the necessity of having an adequate slope between the equipment on the magnet and the cold-box. An alternative layout, with liquid helium circulators is also envisaged. This solution, although more expensive, has some advantages compared to the thermosiphon cooling mode, in particular in a push-pull scenario, where the pressure drop inside the cryogenic transfer line is critical. The pumping system for the magnet cryostat is also shown in Figure 7.6. The fore-vacuum pumps, which are noisy and a source of vibrations, are located in the service area, far away from the beam position, while the high-vacuum pumps (usually stationary oil-diffusion pumps) are mounted on top of the magnet cryostat. The two systems are connected via a flexible vacuum pumping line that runs parallel to the cryogenic transfer line in a cable-chain above the detectors.

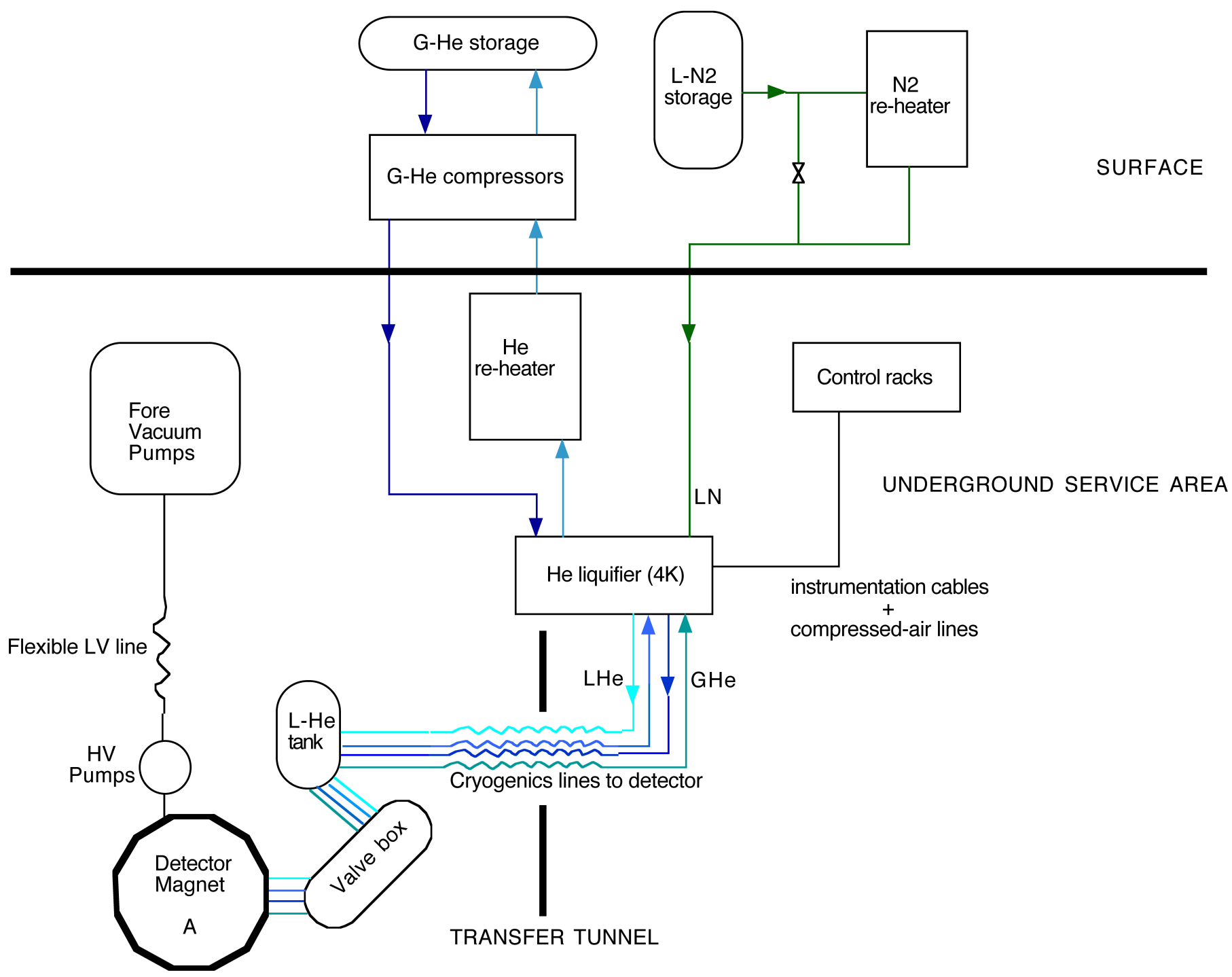

Fig. 7.6: Schematics of the cryogenics and vacuum services.

# Chapter 8

# Muon System at CLIC

## 8.1 Introduction

Many interesting physics processes at CLIC such as $e^+e^- \rightarrow \widetilde{\mu}_R^+\widetilde{\mu}_R^-$ contain muons in the final state. Muon identification with high efficiency and purity is therefore an important requirement for the CLIC detectors. As for many other experiments, including the ILC detectors ILD [1] and SiD [2], the instrumented iron return yoke is used as muon identifier.

In this section the system requirements and the expected background conditions are summarised. Section 8.2 describes the conceptual design of the muon detector, and Section 8.3 summaries the expected performance.

### 8.1.1 Muon System Requirements

Muons from inelastic $e^+e^-$ collisions shall be reconstructed and identified over the largest possible angular range. The physics goals set for the CLIC detectors require that muons are measured with a precision of about $\Delta p_T/p_T^2 = 2 \cdot 10^{-5}$. A momentum resolution of this precision can only be achieved with an excellent tracking system. Due to the large amount of material muons have to traverse before reaching the return yoke, the muon system cannot contribute to improve the momentum resolution. On the other hand, efficient linking of track candidates from the inner detectors with tracks in the muon system is important.

In addition to its muon tagging ability, the first layers of the magnet return yoke will be instrumented to act as a tail catcher for showers developing late in the calorimeters. This slightly improves the energy measurement in the hadron calorimeter. Another aspect of the muon system is the stand-alone identification and reconstruction of beam-halo muons. This requirement has an impact on the muon system granularity and time resolution, which shall be better than 1 ns.

### 8.1.2 Background Conditions

In contrast to the case for other subdetectors, the backgrounds to be taken into account for the muon system are not only the incoherent pairs and $\gamma\gamma \rightarrow$ hadrons, which affect the inner regions of the endcaps, but also the halo muons. These muons are created in the accelerator complex and enter the detector from both sides nearly parallel to the $e^+$ and $e^-$ beam axes, as discussed in Section 2.1.2.3. Their number depends on the amount and type of spoilers used in the beam line. Taking into account a safety factor of 5, for this report we assume that the total number of halo muons entering the detector from both sides is one per 2 ns, hence about 75 in the time interval of one train.

It should be noted that in the simulation of the background hits expected from incoherent pairs and $\gamma\gamma \rightarrow$ hadrons, a considerable fraction originates from multiple interactions in the forward region. Therefore, for both of these channels a safety factor of five is applied in their hit density.

## 8.2 Conceptual Design of the Muon System

Important constraints for the design of the muon system come from the requirements on the iron yoke needed for the magnetic flux return of the detector solenoids. Forces of about 18 ktons have to be absorbed by the yoke. The needed mechanical stability requires that the yoke plate thickness is at least 10 cm. The muon detectors will be interleaved between these plates.

The performance of the muon system has been evaluated by simulating events in the CLIC_ILD concept. To determine the optimal layout, a geometry was created in MOKKA [3] with 19 layers in the barrel and 18 layers in the endcap, at equal distances of 140 mm. After simulating the detector response in MOKKA once, different layouts could be studied by including or excluding layers in the reconstruction phase. For this study the muon layers have been segmented in pads of $30 \times 30$ mm$^2$. Details on the performance study are discussed in [4].

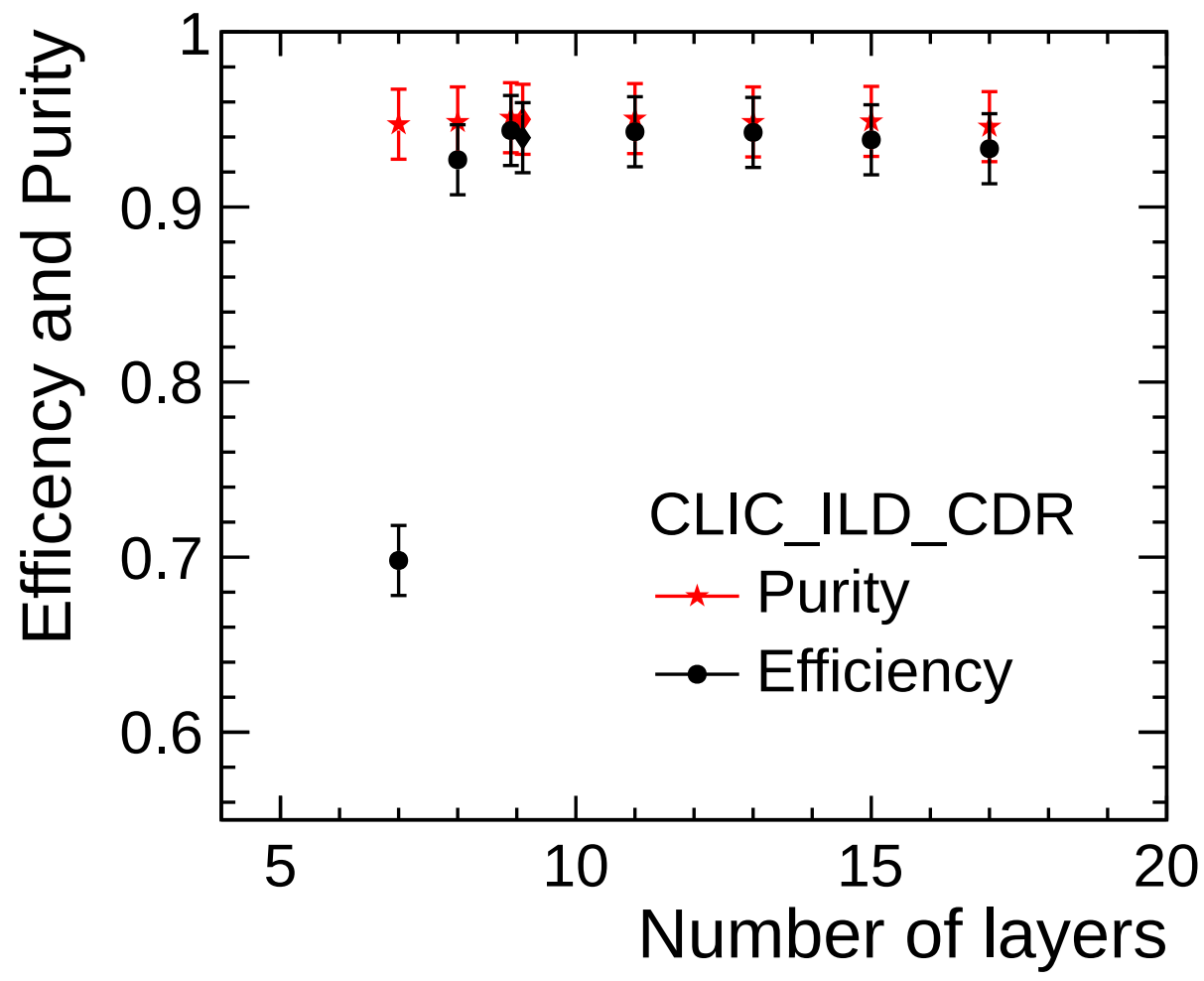

Fig. 8.1: Overall muon identification efficiency and purity of muons in b-jets as function of the number of layers in the tested geometries. Two models with nine layers have been tested: one with evenly spaced layers and one with three groups of three layers. The two layouts perform equally well.

### 8.2.1 Muon System Layers

Several layouts for the muon system have been tested. The first three layers are always included, as they are important to get a muon track segment after the solenoid and provide a minor improvement in the energy resolution of the hadron calorimeter.

The performance of the muon system as a tail catcher depends strongly on the depth of the calorimeters in interaction length and the amount of dead material between the calorimeter and the muon system, which is fixed by the solenoid design. The foreseen hadron calorimeters have a depth of 7.5 $\lambda_I$ and the solenoids in both detector concepts have a thickness of approximately 2 $\lambda_I$. Under these boundary conditions only the first three layers of the muon system slightly improve the energy resolution of the hadron calorimeter [4].

The system performance for the identification of isolated muons hardly depends on the number of layers. The high granularity of the hadron calorimeter in the CLIC detectors allows already to distinguish very well muons from hadrons. The situation is however different for muons in jets. Therefore, a sample of 9000 $e^+e^- \rightarrow Z^* \rightarrow b\bar{b}$ events has been generated with PYTHIA [5]. At a centre-of-mass energy of 1.5 TeV the energy scale of the jets in the di-jet events resembles the energy scale of multi-jet events at 3 TeV. Each event has at least one of the b-quarks decaying semi-leptonically to a muon and a neutrino. With this sample the challenging task of reconstructing muons in dense high-energetic jets at CLIC can be simulated.

In Figure 8.1 the overall efficiency of the muon identification is shown as a function of the number of layers. The purity of the obtained muon sample is also shown. The figure indicates that nine layers are sufficient to reach the best possible performance. The layout with three groups of three layers is selected to have the needed redundancy and coverage in view of mechanical constraints. Moreover, this muon instrumentation layout allows for two yoke plates of 50 cm thickness, which help to absorb the large magnetic forces pulling the endcaps inward.

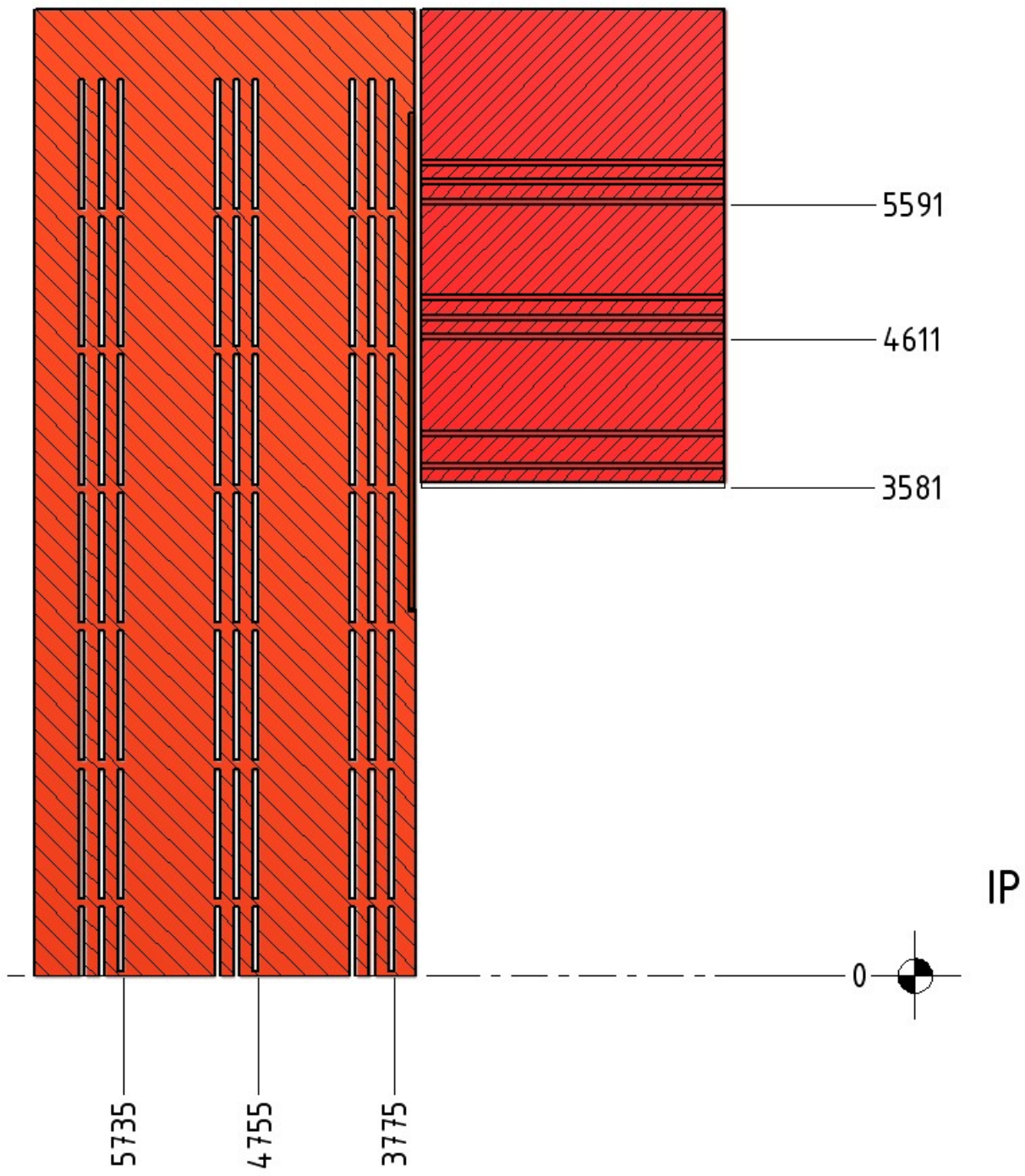

Fig. 8.2: Schematic cross section of the current muon system design of a CLIC_SiD detector quadrant. Instrumentation layer distances from the IP are indicated in millimetres.

In Figure 8.2 the engineering design for one quadrant of the CLIC_SiD yoke and muon system is shown. For simplicity the inner detectors are not shown in the figure. The layout is very similar for the CLIC_ILD concept. Differences between both concepts are in the exact size of the system and in the positions of the muon layers. Due to the larger radius of the CLIC_ILD solenoid, the area to be covered is larger than for CLIC_SiD. The total area for the CLIC_ILD muon system amounts to about 5800 $m^2$, compared to 4600 $m^2$ for CLIC_SiD.

### 8.2.2 Muon Layer Design

For the design of a single muon layer, two aspects have been considered: First, the impact of the granularity on the muon identification efficiency and purity; second, the single channel occupancy, for which the backgrounds of incoherent pairs, $\gamma\gamma \rightarrow$ hadrons, and beam halo muons are considered.

#### *8.2.2.1 Muon Identification Performance for Different Granularities*

The readout granularity implemented in the simulation has a cell size of $30 \times 30$ $mm^2$. This granularity is also used for the analog readout option of the hadronic calorimeter in the ILC detector concepts. However, for a muon system such a granularity is relatively high. Therefore, the effect of a larger cell size on the performance of the muon identification has been investigated using the $e^+e^- \rightarrow Z^* \rightarrow b\overline{b}$ sample mentioned previously.

In Figure 8.3 the efficiency and purity of the muon identification is shown as a function of the cell size. The figure shows that the purity is not significantly affected by going to larger cells. However, the situation is different for the efficiency. While cells of 40 mm instead of 30 mm do not deteriorate the performance, the efficiency starts to drop for larger cells. A careful analysis, including optimisation of the algorithm for each cell size, will be required before adopting cell sizes larger than 40 mm.

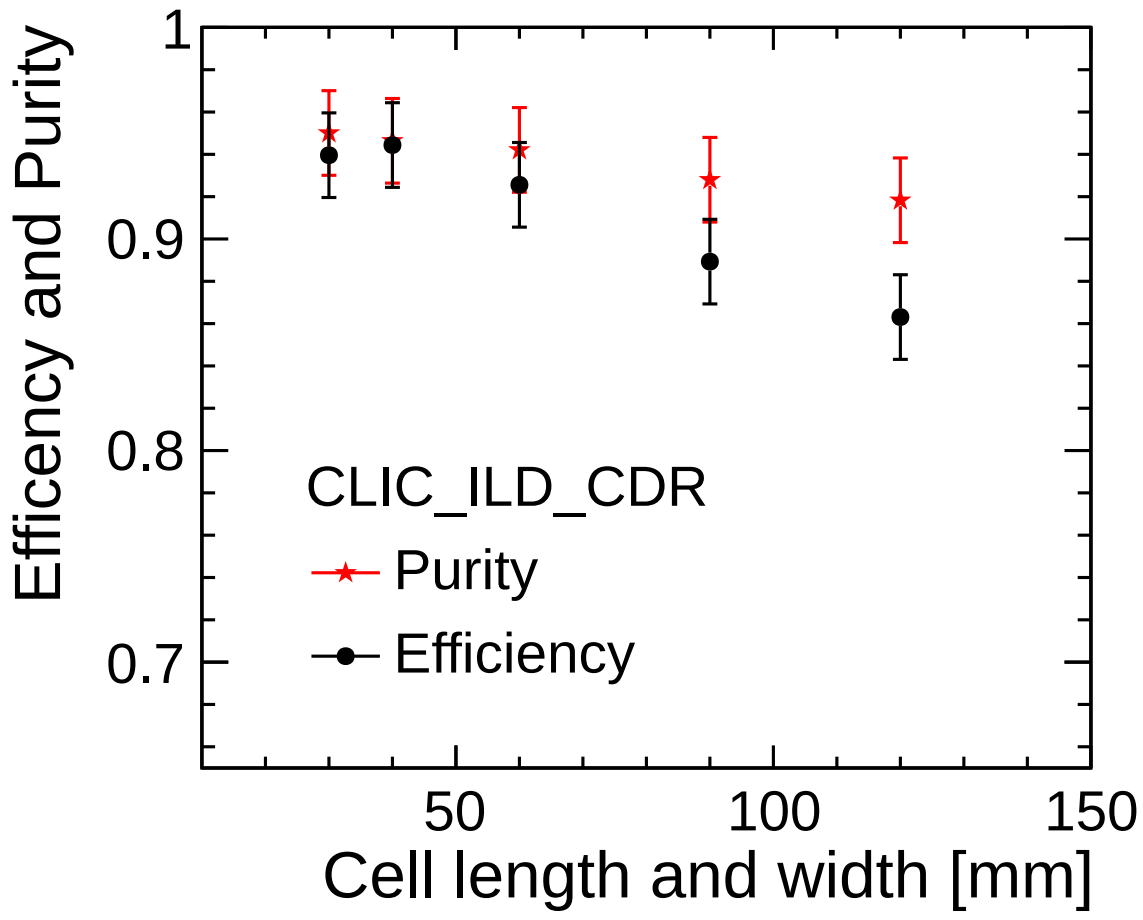

Fig. 8.3: The overall efficiency and purity of the muon identification in the $Z^* \to b\bar{b}$ sample, given for different cell sizes in the muon system.

#### *8.2.2.2 Readout Channel Occupancy and Muon System Technologies*

Several background channels have an impact on the detector occupancy. For the endcap, at a radius of 70 cm, the incoherent pairs and $\gamma\gamma \to$ hadrons form the main background and result in an occupancy of 10% per cell of $30 \times 30$ mm$^2$ per train, including safety factors. At a radius of 1.5 m the background due to beamstrahlung drops by two orders of magnitude and the contribution of 0.002 halo muons per cell becomes dominant. Therefore, pad readout is required in the most inner region, while at a radius larger than 1.5 m crossed readout strips are feasible: strips of 3 cm width and 1 m to 2 m length would result in an occupancy of maximum 10% per strip per train. Besides precise time stamping, multi-hit readout capability within the bunch train will be required.

The situation is different in the barrel region. Contributions from incoherent pairs and $\gamma\gamma \to$ hadrons are negligible, and at a radius of more than 4.5 m the occupancy due to halo muons drops to less than $10^{-4}$ muons per area of $30 \times 30$ mm$^2$ within one train. However, since beam halo muons cross a layer horizontally, all cells in a row of one layer might fire. Given the rather low occupancy in the barrel region, crossed readout strips of 1 m to 2 m length and 3 cm to 4 cm width are feasible.

In case only strip readout is used, the number of electronics channels in the endcaps would be $1.2 \cdot 10^5$ for both detector concepts. For the barrel region the number of readout channels is $6 \cdot 10^4$ for CLIC_SiD and $9 \cdot 10^4$ for CLIC_ILD. These numbers assume strip dimensions as indicated previously. The dimensions have not been optimised from the point of view of detector technology. A careful analysis, taking into account all system constraints, has to be carried out before implementing a crossed strip readout. Considering a full pad readout, as used for the performance studies of the muon system presented in this document, leads to about 10 times more readout channels. However, based on the expected occupancies, pad readout would only be needed in the inner region of the endcaps.

To identify the direction of beam halo muons crossing the detector, especially those crossing the calorimeters, good time stamping in the muon endcap is required. Once beam halo muons are reconstructed, the information can be used for a correction of the energy measurement in the calorimeter. The preferred technologies for the muon system of ILD are Resistive Plate Chambers (RPC) or extruded scintillator strips with SiPM readout [1]. Both technologies offer a very good time resolution. For RPCs, values better than 1 ns have been obtained.

These technologies are also good candidates for the CLIC detectors. To avoid the operational difficulties encountered with RPCs based on Bakelite in past experiments, glass would be the preferred material to be used for the RPC detectors.

## 8.3 Muon Reconstruction Algorithm and System Performance

The muon reconstruction algorithm was developed in the PANDORAPFA framework [6, 7], which provides many functions to investigate cluster topologies. The four different steps of the algorithm are briefly described in the following subsection, followed by a summary of the muon system performance.

### 8.3.1 Reconstruction Algorithm

1. **Identification of yoke track candidates**
   To identify the tracks of muon hits in the muon system, the cone-based clustering algorithm applied also in the hadronic calorimeter is used. However, the algorithm is configured so as to deal with the limited number of layers in the muon system. In particular, the tangent of the angle of the cone used to search for new hits is set to 0.3 and hits are only clustered together if they are in neighbouring layers. With these parameters a cluster reconstructed in the muon system is considered as a muon yoke track if it occupies eight or more layers, yet contains no more than 30 hits. Applying a cut on the maximum number of hits avoids to mis-identify large punch-throughs from the calorimeter into the muon system.

2. **Extrapolation of inner detector (ID) tracks to muon system**
   With the momentum and direction of an ID-track and the magnetic field, the ID-track can be extrapolated to the muon system using a helix. The helix direction is altered at the intersection of the helix with the solenoid middle radius, or when the helix exits the solenoid towards the endcap.

3. **Matching of inner detector tracks and yoke tracks**
   The next step is to find the ID-track matching to the yoke track. For each yoke track all ID-tracks are considered and two variables are calculated: (i) the distance of closest approach of the external helix to the mean position of the hits in the innermost layer of the yoke track; (ii) the angle between the external helix and the direction of the yoke track. For both observables default quality cuts are chosen from the performance of the muon algorithm with different values for the quality cuts: the distance cut of 200 mm is chosen as to keep the purity as high as possible. In order to also obtain a satisfactory efficiency an angular cut of 0.2 rad is chosen.

4. **Identification of calorimeter muon hits**
   If an ID-track is assigned to a yoke track, the muon four-vector can already be constructed. The direction as well as the momentum can be determined from the ID-track. For an optimal jet reconstruction performance it is important to identify also the hits in the calorimeter originating from a muon. All layers between the inner tracking detectors and the muon yoke are considered. This process completes the identification of the calorimeter hits and the track segments in the muon Particle Flow Object.

### 8.3.2 Reconstruction Performance

To illustrate the performance of the muon reconstruction algorithm, Figure 8.4 shows the efficiency and purity obtained with the muons in high multiplicity b-jet events. For comparison the performance for isolated muons is also shown.

For isolated muons with $\theta > 10^\circ$ and energies of more than 7.5 GeV efficiencies and purities of more than 99% are obtained. For muons in high multiplicity events the lower limits are approximately 90%, except for the efficiency at $\theta \approx 60^\circ$: the large distance in the transition region between the endcap and the barrel influences the performance, as the matching of inner detector tracks and yoke tracks is more difficult, in particular for low-energy muons. From the efficiency results we also see that in the transition region from barrel to endcap, at $\theta \approx 40^\circ$, the performance is slightly worse. The described algorithm leads to a system performance matching the requirements.

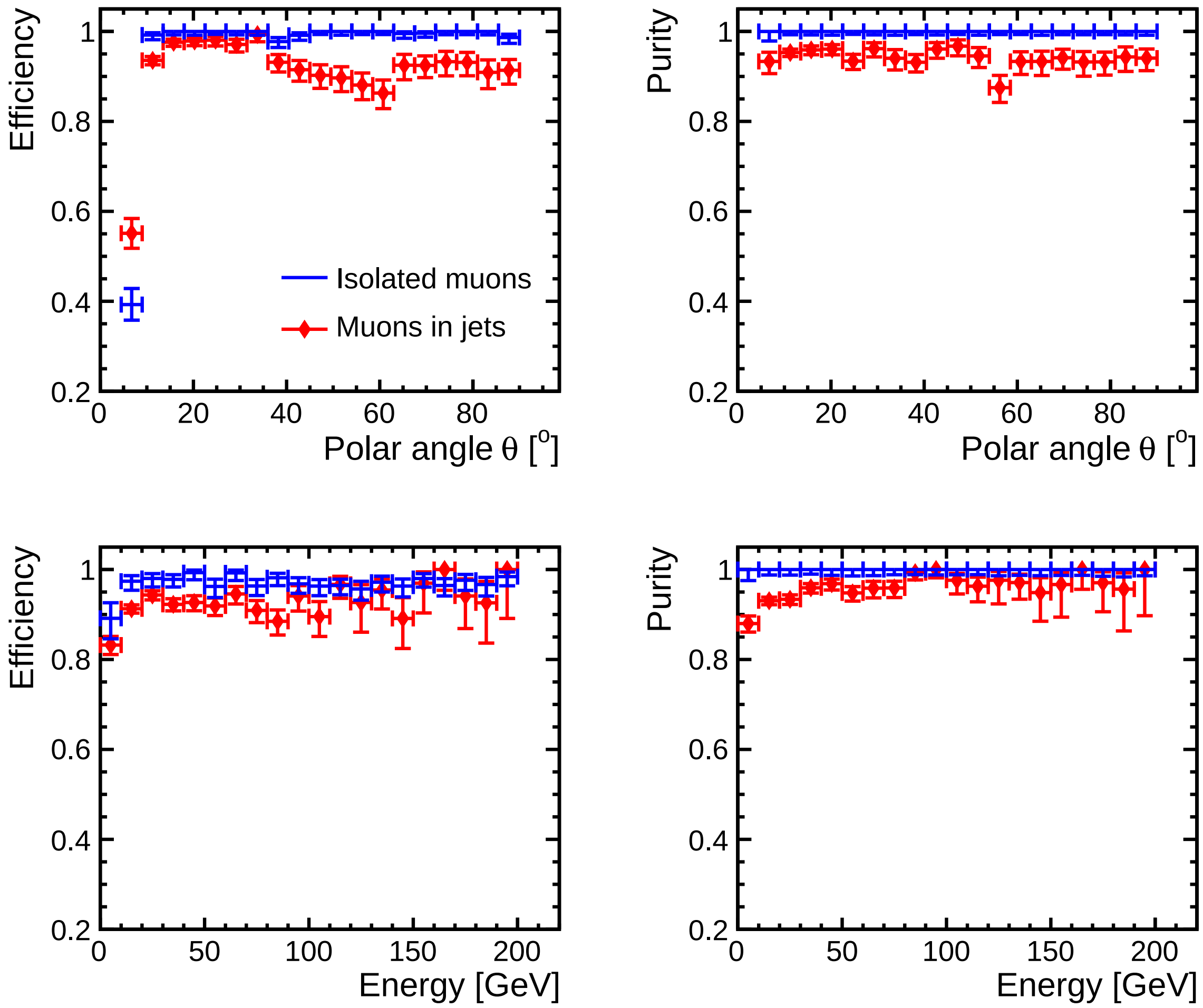

Fig. 8.4: The identification performance as a function of the energy for isolated muons and muons in b-jets. The simulations are performed in the CLIC_ILD detector concept with pads of $30 \times 30\ \text{mm}^2$ in both barrel and endcap muon detectors.

# Chapter 9

# Very Forward Calorimeters

## 9.1 Introduction

Two calorimeters are foreseen in the very forward region of the CLIC detectors: The Luminosity Calorimeter (LumiCal) for the precise measurement of the luminosity and the BeamCal for the fast estimate of the luminosity and tagging of high energy electrons. Both are cylindrical electromagnetic sampling calorimeters, centred on the outgoing beam. The overall layout of the very forward region at CLIC, as well as the conceptual design of the LumiCal and BeamCal, are based on the detailed work performed for ILC and documented in [1]. A comparison of the main geometrical parameters of the two calorimeters in the ILC and CLIC layout is given in Table 9.1.

Table 9.1: Comparison of LumiCal and BeamCal at ILC and CLIC, for the example of the ILD and CLIC_ILD detector concepts.

| | | ILC(ILD) | CLIC_ILD |
|---|---|---|---|
| **LumiCal** | geometrical acceptance [mrad] | 31 – 77 | 38 – 110 |
| | fiducial acceptance [mrad] | 41 – 67 | 44 – 80 |
| | z (start) [mm] | 2450 | 2654 |
| | number of layers (W + Si) | 30 | 40 |
| **BeamCal** | geometrical acceptance [mrad] | 5 – 40 | 10 – 40 |
| | z (start) [mm] | 3600 | 3281 |
| | number of layers (W + sensor) | 30 | 40 |
| | graphite layer thickness [mm] | 100 | 100 |

A conceptual drawing of the very forward region of a CLIC detector is shown in Figure 9.1. The LumiCal is positioned just behind an opening in the forward electromagnetic calorimeter and the BeamCal in front of the final focus quadrupole QD0. The two calorimeters improve the hermeticity of the detector. The amount of particles scattered back into the central detector strongly depends on the design of the very forward region. The current geometry is optimised to keep the flux of backscattered particles small.

To match the expected statistical error for measuring cross sections of most electroweak processes in a typical year (500 $\mathrm{fb}^{-1}$) at CLIC, an accuracy of $10^{-2}$ or better in the absolute luminosity is needed. The luminosity measurement has two components: (1) the luminosity in the high energy peak, using the large cross section for Bhabha events, $e^+e^- \rightarrow e^+e^-(\gamma)$, measured in the LumiCal; (2) the relative shape of the luminosity spectrum (see Section 12.2.1), reconstructed from large-angle Bhabha events measured with the tracker and the ECAL of the CLIC detector. The calibration of the luminosity spectrum is obtained from LumiCal with its good absolute accuracy. Therefore, LumiCal is a precision device with challenging requirements, e.g. on the mechanics and the position control.

The detectors in the very forward region have to tackle high background rates created by the strong electromagnetic fields present at beam collision. The BeamCal in particular will be hit by a large amount of incoherent electron-positron pair particles with most probable energies of a few GeV. These depositions, potentially useful for beam-tuning, will lead to a radiation dose of several MGy per year in the sensors at lower polar angles. Hence radiation hard sensors are needed to instrument the BeamCal.

A small Molière radius is of invaluable importance for both calorimeters. It ensures electron veto capability for the BeamCal even at small polar angles, which is essential to suppress background events in new-particle searches where the signature is large missing energy. In the LumiCal the precise reconstruction of electron and positron showers of Bhabha events is facilitated by a compact design. Each calorimeter consists of 40 tungsten absorber plates, interspersed with sensor layers. Every absorber layer is 3.5 mm thick, which corresponds to one radiation length per layer. To allow installation and removal when the beam pipe is installed, both calorimeters are constructed as two half-cylinders.

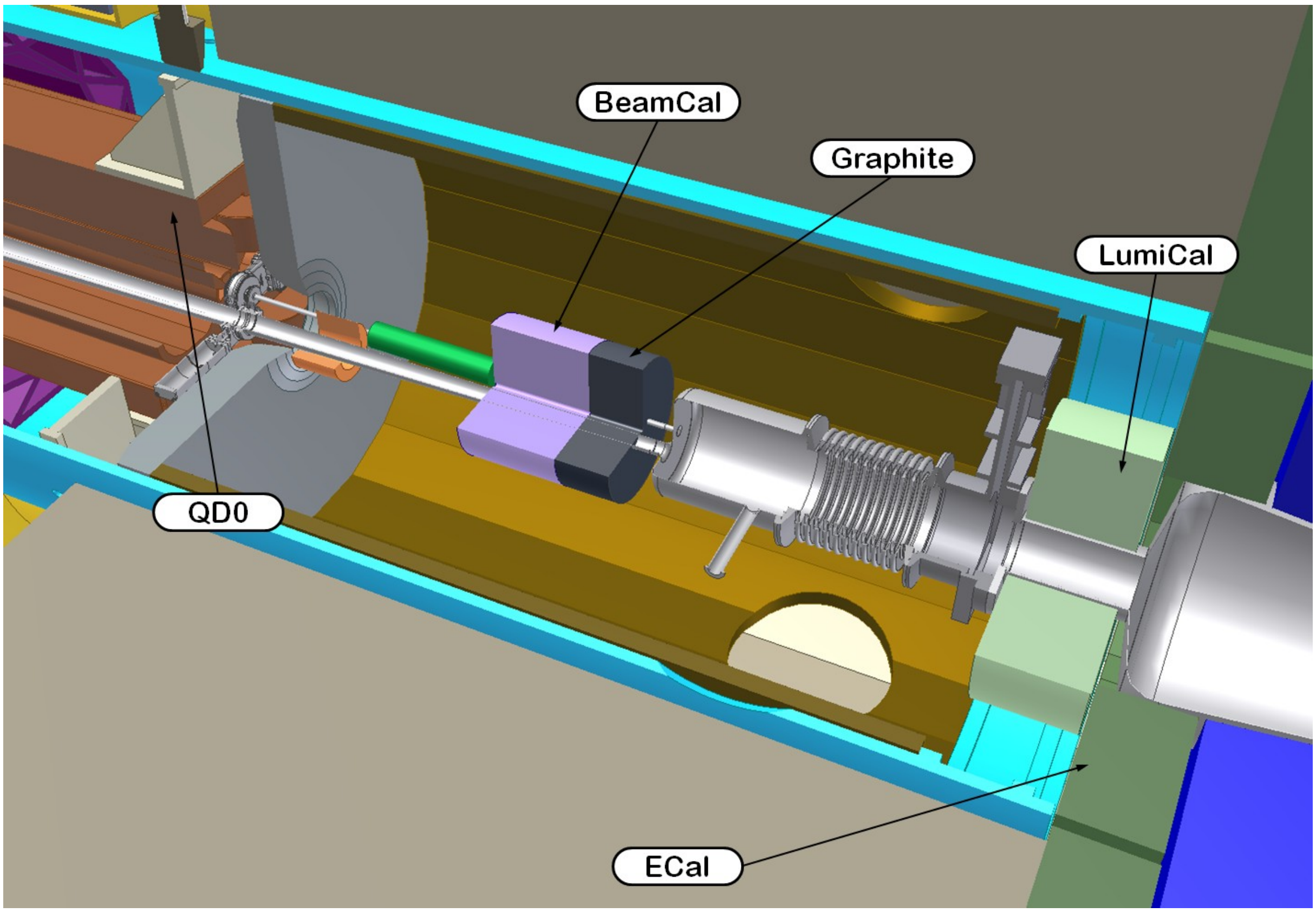

Fig. 9.1: The very forward region of the CLIC detector (engineering layout). The LumiCal, BeamCal and QD0 are carried by a support tube.

Since the time between bunch crossings is 0.5 ns and the duration of the bunch trains is only 156 ns, a trigger-less readout of the BeamCal and LumiCal is envisaged. The occupancy in the BeamCal is very high, which calls for novel solutions in terms of readout electronics. One possibility for the readout is a gated integrator with Correlated Double Sampling [2]. The current pulses integrated continuously by the integrator are sampled in time slices of about 10 ns. The sampling may be performed either by a fast ADC (which would send out converted data immediately) or using an analog memory (which would store analog values and digitise it after the bunch train). Signal amplitudes are obtained subtracting subsequent samples. The subtraction would in addition serve as a noise filtering. Between bunch trains, the integrator is reset. The same readout concept will also be applicable for the LumiCal.

The pad occupancy of the LumiCal estimated from simulations is below 2%. This allows to consider an alternative solution with a readout chain consisting of a fast preamplifier and shaper, followed by a fast sampling digitiser. A deconvolution algorithm applied to the ADC output is being investigated [2]. From the deconvolution result the hit time and amplitude of the pad signals will be obtained. Further details on possible readout schemes are discussed in Chapter 10.

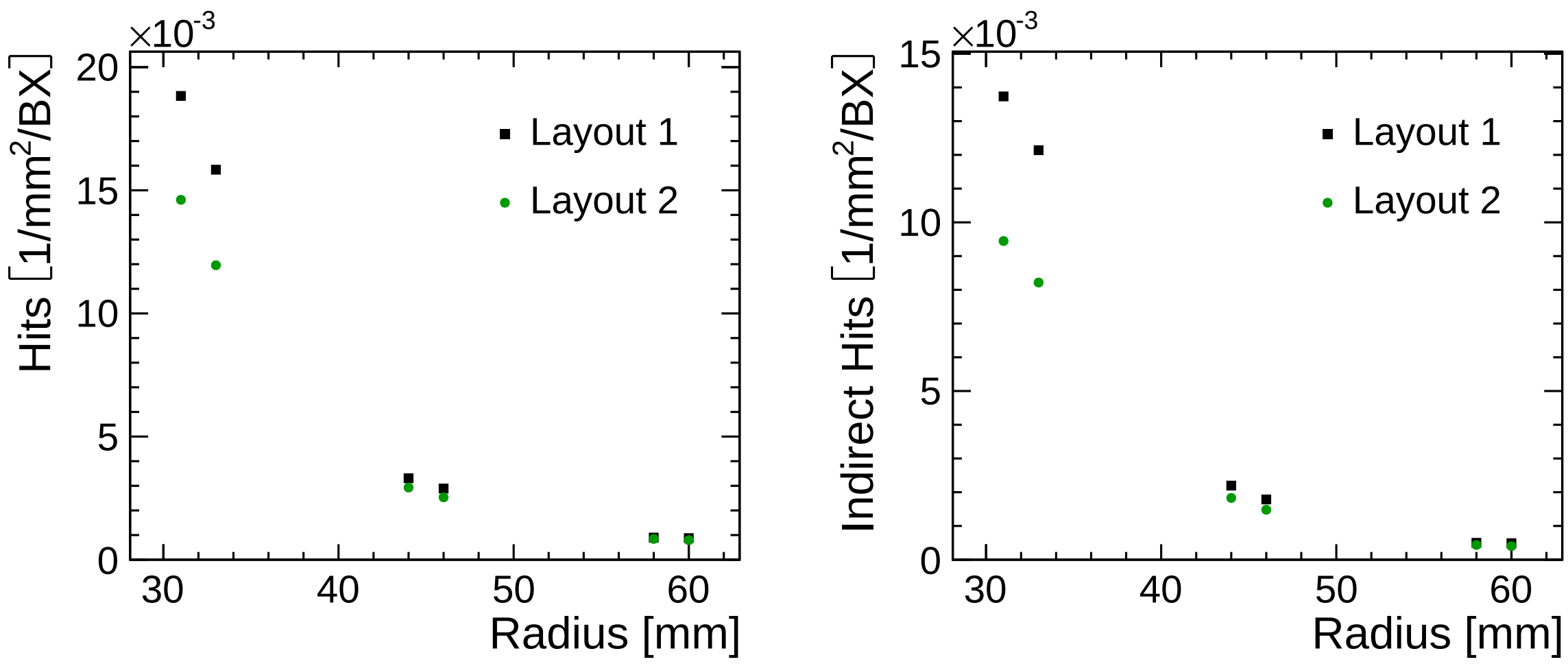

Fig. 9.2: Hit density in the different layers of the vertex detector for two different BeamCal cross sections (layout 1 and layout 2): all hits (left) and backscattered hits only (right). This result was obtained in full simulation of the CLIC_ILD detector, without safety factors.

## 9.2 Optimisation of the Forward Region

At 3 TeV centre-of-mass energy and due to the CLIC bunch parameters, the beam-beam background is severe and demands a careful optimisation of the forward region. Most importantly, several $10^8$ coherent pairs are produced per bunch crossing. Direct hits to the forward calorimeters are avoided by adopting a conical vacuum pipe with half opening angle of 10 mrad for the outgoing beam pipe. This determines the inner diameter of the BeamCal. In addition, some $3 \cdot 10^5$ incoherent pairs per bunch crossing are produced, with a much broader angular distribution. They hit the BeamCal and, to a lesser extent, the LumiCal, and produce backscattering into the inner detectors such as the vertex detector. The geometry of the forward region was optimised to reduce the amount of backscattered particles [3].

In a first step, it is recognised that a low-Z absorber on the IP side of the BeamCal is a powerful means to reduce backscattering. A graphite layer of several centimetres thickness (see Figure 9.1) was introduced in the layout at earlier linear collider studies [4, 5, 6] and at the ILC [7], and a 10 cm thick layer serves its purpose also at CLIC. Further reduction of backscattering hits to the vertex detector was obtained by increasing the diameter of the vacuum pipe wherever possible, e.g. between the LumiCal and the BeamCal, as well as downstream of the BeamCal. An dedicated geometry of the tungsten layers in the BeamCal, leaving as little opening for backscattering from the downstream region as possible, has been introduced in the detector model. The improvement due to this modification is illustrated in Figure 9.2: the hit density in the vertex detector, stemming from backscattered particles, is reduced by about 35%. This corresponds to a reduction of 20% of the total hits, direct hits and backscattered combined.

The number of backscattering particles reaching the inner detector region also depends on the distance between the BeamCal and the IP. As a general rule, it would be preferable to move the BeamCal as far away as possible from the IP. This is however constrained by the position of accelerator components, e.g. the Intra-Train-Feedback system. This system must be as close as possible to the IP to reduce feedback time. On the other hand, the equipment has to be protected from radiation due to incoherent pairs, which is achieved by placing the BeamCal in front of the Intra-Train-Feedback system when seen from the IP. The position of the BeamCal in the present layout of the CLIC detectors is, therefore, optimised considering all constraints.

A further significant reduction of backscatters was obtained by changing the vacuum pipe design on the IP side of the LumiCal [8]. Increasing the thickness of the conical part of the vacuum pipe in this region effectively turns the pipe into a ‘mask’ against backscatters. As a result, the hit density in the inner detectors is now dominated by direct hits from the IP. The resulting hit distributions and rates, as well as the consequences for the vertex detector design are given in Chapter 4.

### 9.3 The Luminosity Calorimeter (LumiCal)

The LumiCal covers polar angles between 38 mrad and 110 mrad around the outgoing beam. In the CLIC_ILD model, for example, it is located at about 2.6 m from the IP, right behind the ECAL, with the edge of the ECAL shadowing the outer perimeter of the LumiCal (see Figure 9.1). The LumiCal will be equipped with silicon sensor planes. In the present conceptual design a gap of 0.8 mm between the absorber plates is foreseen for the sensors. The front-end and ADC ASICs are positioned at the outer radius in the space between the tungsten disks. The mechanical design of the LumiCal is shown in Figure 9.3. A photograph of a prototype sensor is shown in Figure 9.4, showing that the geometry ensures a constant polar angle per sensor ring.

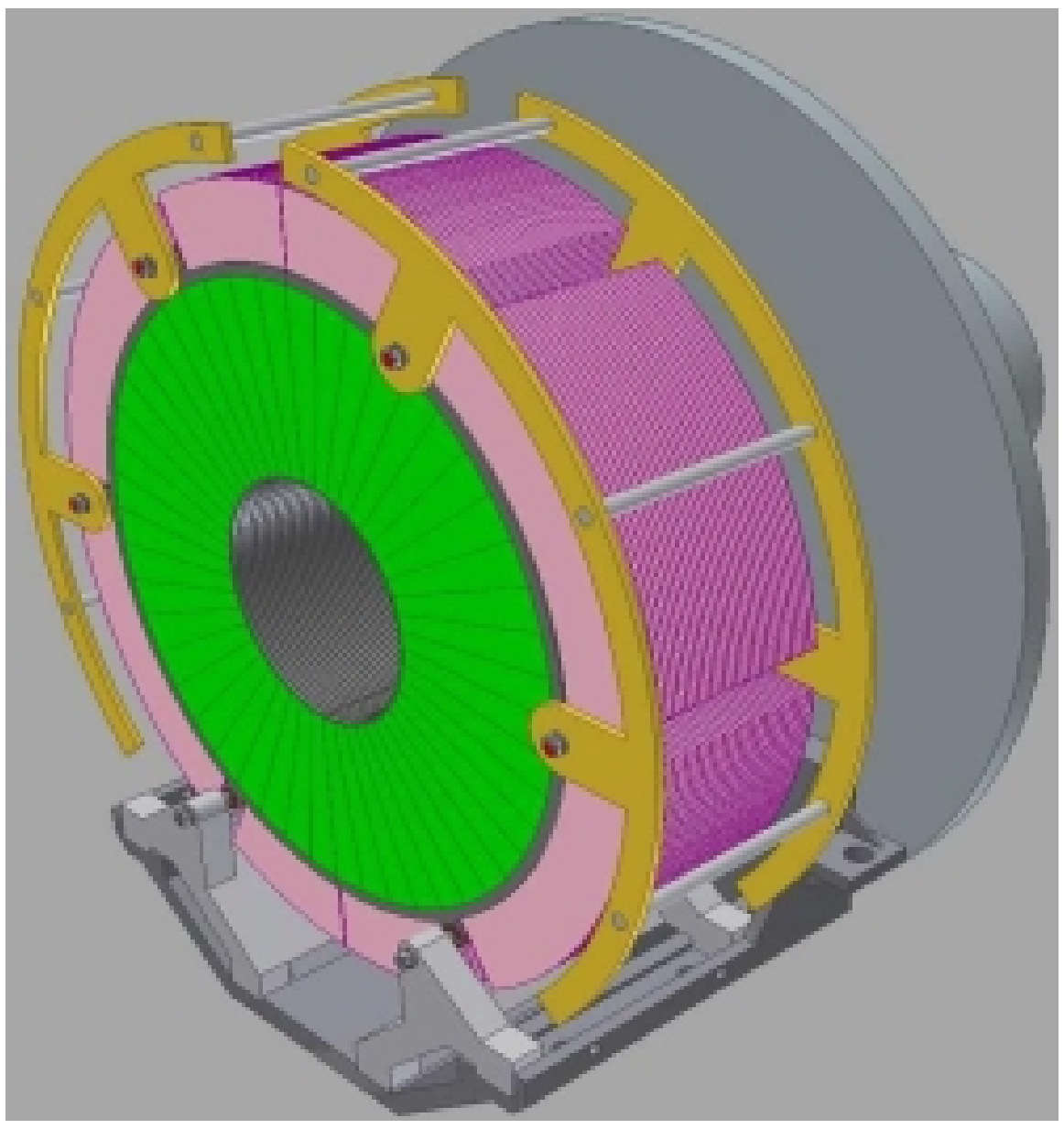

Fig. 9.3: The proposed LumiCal design, with silicon sensor segments shown in green, the tungsten structure in purple and, in yellow, the mechanical frame for the required stability. This picture shows a LumiCal with 30 layers, as proposed for ILC [9], while the CLIC version has 40 layers.

Monte Carlo studies based on the ILC design [1, 10] have shown that a compact silicon-tungsten sandwich-calorimeter is a proper technology for the LumiCal with a typical energy resolution in a stand-alone mode of $\sigma_E/E = 0.21/\sqrt{E(\mathrm{GeV})}$ for contained showers. The requirement of shower containment for the Bhabha scattering final states and the scattering off the ECAL edge limit the fiducial volume of the LumiCal at CLIC_ILD to the range 44 mrad to 80 mrad, which corresponds to a cross section of 62 pb for a centre-of-mass energy of 3 TeV. Assuming an integrated luminosity of 500 fb$^{-1}$ per year, this translates into a relative statistical precision on the Bhabha scattering yearly rate of $1.8 \cdot 10^{-4}$, well below the requirement of $10^{-2}$.

The small Molière radius (1.1 cm) and the finely segmented silicon pad sensors (64 segments radially and 48 azimuthally) ensure an efficient selection of Bhabha events [9] and a precise shower position measurement. The integrated luminosity $\mathcal{L}_{\mathrm{int}}$ is obtained from $\mathcal{L}_{\mathrm{int}} = N_{\mathrm{Bhabha}}/\sigma_{\mathrm{Bhabha}}$, where $N_{\mathrm{Bhabha}}$ is the number of Bhabha events counted in a certain polar angle range and $\sigma_{\mathrm{Bhabha}}$ is the expected Bhabha scattering cross section in the same angular range.

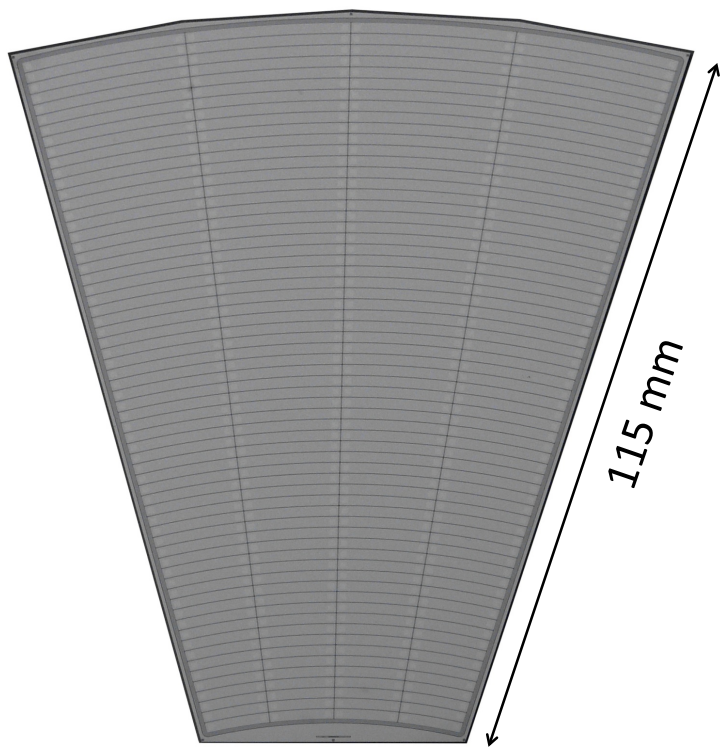

Fig. 9.4: Photograph of a prototype LumiCal sensor – eight such sectors are needed to equip one LumiCal sensor plane.

Since the cross section falls as $1/\theta^3$, the luminosity shift as a function of the polar angle bias can be expressed as

$$\frac{\Delta\mathscr{L}_{\mathrm{int}}}{\mathscr{L}_{\mathrm{int}}} = \frac{2\Delta\theta}{\theta_{\mathrm{min}}}, \tag{9.1}$$

where $\theta_{\mathrm{min}}$ is the minimum polar angle of the fiducial region and $\Delta\theta$ is a bias to the polar angle measurement. To achieve an accuracy of $10^{-2}$, this bias on the angular reconstruction has to be smaller than 0.2 mrad. For the current design and pad layout, this would imply that the inner calorimeter radius of 100 mm must be known with a precision of better than 0.5 mm, and the position of the LumiCal along the beam line has to be known to within 13 mm. With the proposed granularity, the bias in the position reconstruction can therefore be easily controlled to within a fraction of the required precision. The position of the LumiCal will be monitored with a laser system [11] which will provide sensitivity to longitudinal displacements of 100 µm.

The selection of Bhabha scattering events is based on the collinearity of the $e^{\pm}$ candidates provided the total energy deposited in the two arms of the LumiCal system is at least 80% of the nominal centre-of-mass energy. Initial state radiation and beamstrahlung smear the nominal energy and the efficiency of selecting Bhabha scattering events has to be determined from Monte Carlo simulations. The precision with which this correction is known depends on the energy scale uncertainty in the LumiCal. To achieve a precision of $10^{-2}$ in counting the Bhabha events, the energy scale uncertainty has to be below 0.3% [12]. The energy scale will be determined by locating the position of the maximum energy deposition which corresponds to 1.5 TeV. Given the expected detector resolution of 0.5% at 1.5 TeV, it should be possible to achieve the required precision. Depending on the signal collection time, the systematic uncertainty may be dominated by the fluctuations of the beamstrahlung related background.

With a total luminosity of $5.9 \cdot 10^{34}\mathrm{cm}^{-2}\mathrm{s}^{-1}$ about 0.0004 Bhabha scattering events per bunch crossing are expected in the LumiCal, that is about one event per 10 trains. Therefore the main load on the LumiCal readout is due to incoherent pairs from beamstrahlung, and possibly from $\gamma\gamma \rightarrow$ hadrons events. The expected distribution of the background hits from incoherent pairs in the LumiCal is shown in Figure 9.5 as a function of the radial pad number and of the sensor plane number, averaged over 100 bunch crossings. As in the background studies for other CLIC_ILD subdetectors [8], only those hits with a signal above 0.2 MIP are taken into account. The highest occupancy per bunch crossing amounts to 20% and is located in the inner most ring in the back of the calorimeter. For most of the calorimeter volume the occupancy is otherwise below 2%.

The longitudinal profile of electromagnetic showers generated by 1.5 TeV electrons is shown in Figure 9.6. Overlaid is the expected contribution summed over 100 bunch crossings of the incoherent pairs background. The background signals are summed over those pads which lie within the envelope of a 1.5 TeV electron shower.

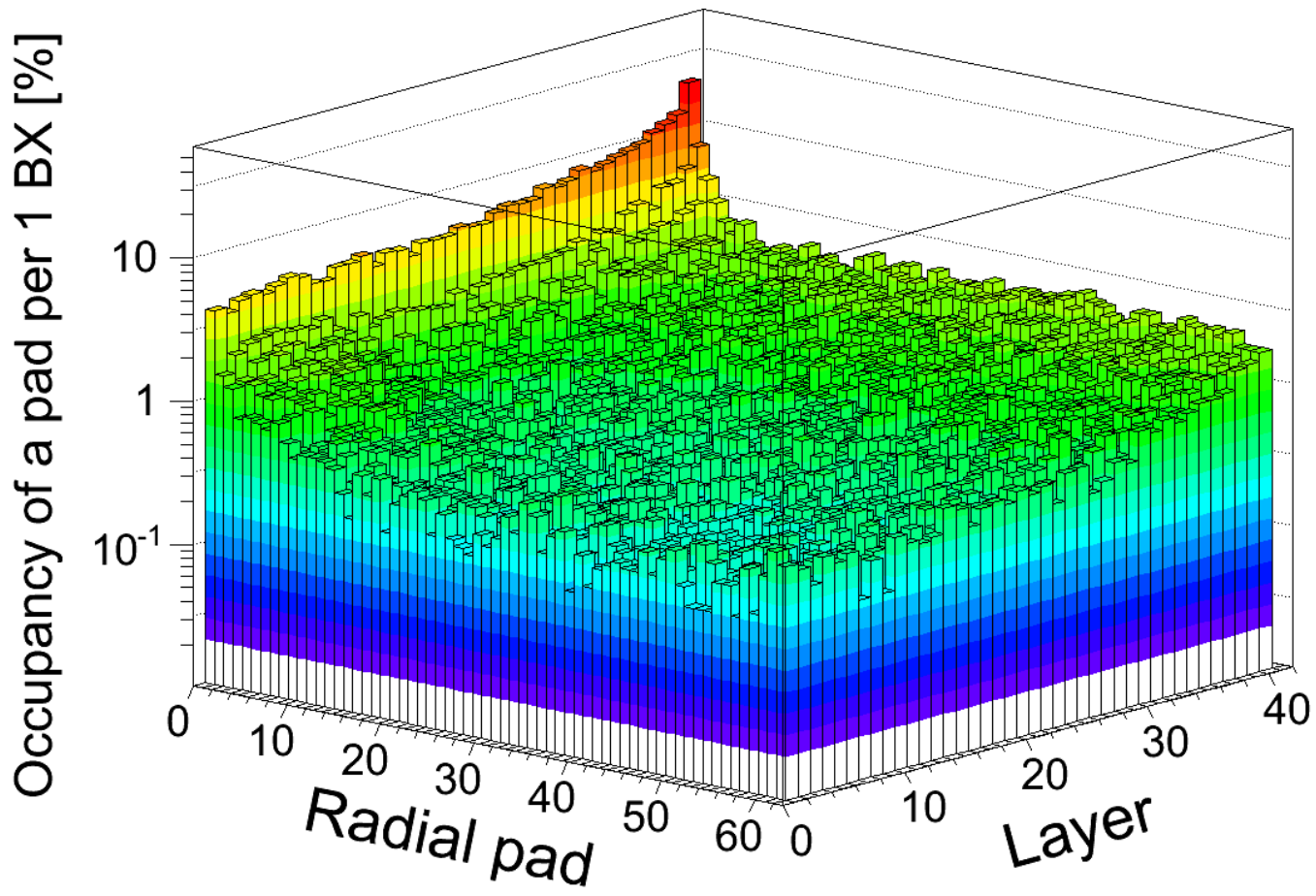

Fig. 9.5: Distribution of the occupancy per bunch crossing in each pad of the LumiCal due to incoherent pairs, averaged over 100 bunch crossings, as a function of the radial pad and sensor plane number. (Note that layer 1 is closest to the IP). This result was obtained in full simulation for CLIC_ILD, without safety factors.

The total signal per bunch crossing due to beam-induced background, summed over all of the active layers of the LumiCal, is about $9 \cdot 10^3$ fC on average, to be compared to a 1.5 TeV electron shower deposit of $780 \cdot 10^3$ fC. The signal deposited by background particles within the envelope of a 1.5 TeV electron shower varies between 90 fC and 450 fC per bunch crossing, including a strong dependence on the scattering angle of the Bhabha electron. When integrated over 10 ns (20 bunch crossings) as anticipated at CLIC, the background contributes 1800 fC to 9000 fC to the electron shower, which is about the same as the expected energy resolution ($\approx$ 4200 fC) for a 1.5 TeV electron. In summary, the impact of the background in the LumiCal may be considerable and needs to be studied further.

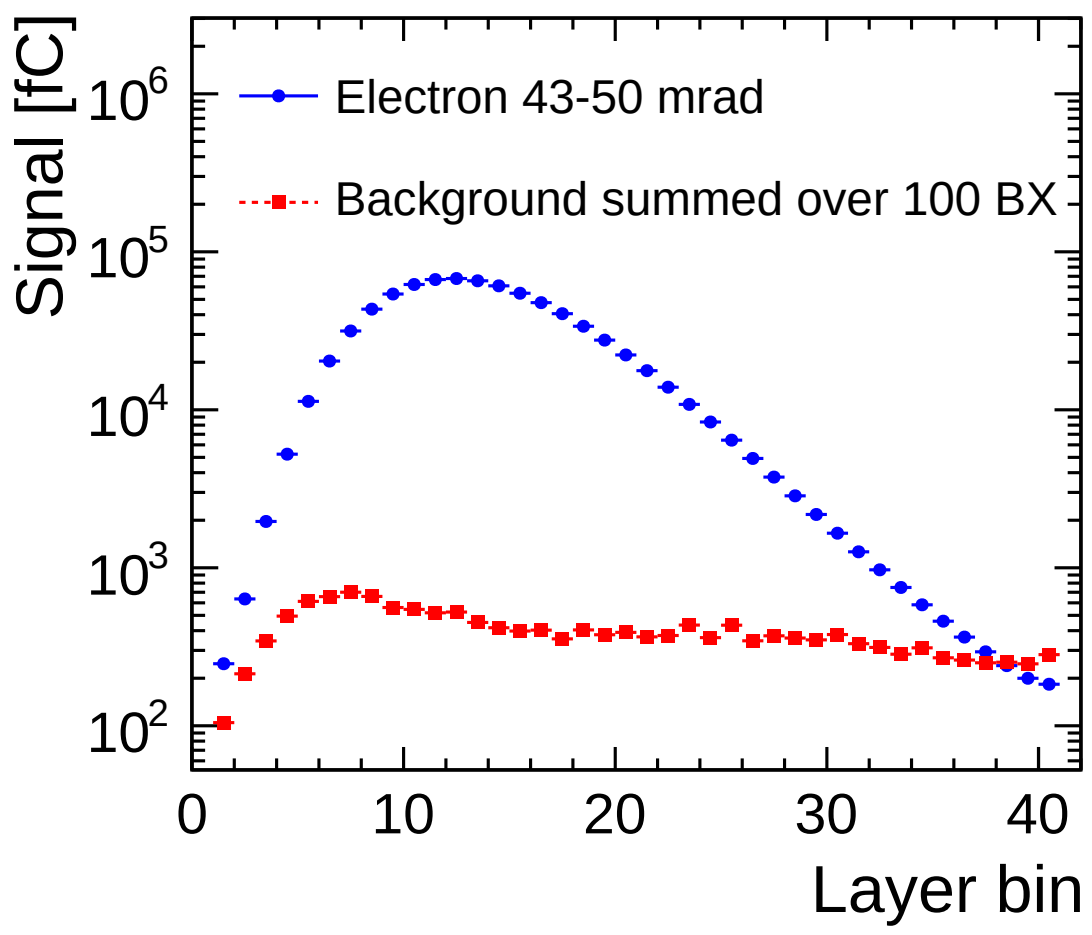

Fig. 9.6: Longitudinal profile of a 1.5 TeV electron shower, for electrons impacting the LumiCal in the angular range of 43 mrad to 50 mrad. Also shown is the expected contribution of the incoherent pairs background integrated over 100 bunch crossings.

### 9.3.1 Remarks on systematic uncertainties to the luminosity measurement

For the LumiCal measurement at the ILC, where an absolute accuracy of $10^{-3}$ is aimed for (compared to $10^{-2}$ for CLIC), a detailed analysis of systematic errors was performed and summarised in Table 1 of [1]. Following that list of systematic uncertainties, preliminary and qualitative remarks for the CLIC case can be made:

- the polar angle resolution at CLIC and ILC should be comparable, for an identical construction of the LumiCal, and should lead to a negligible uncertainty on luminosity (a few $10^{-4}$);
- the polar angle bias, similarly, should be comparable and not give a significant uncertainty at CLIC (a few $10^{-4}$);
- the energy resolution of the LumiCal can be expected to be very similar at CLIC and ILC, and the resulting systematic uncertainty will be of order $10^{-4}$;
- the detailed knowledge of the luminosity spectrum has an impact on the definition of "luminosity in the high energy peak" – since the luminosity spectrum is subject of an on-going detailed investigation, we can not conclude on this contribution to the overall systematic uncertainty here;
- the precise knowledge of the bunch sizes, which directly impacts on the expected uncertainty of the Bhabha suppression effect due to beamstrahlung, is considered one of the major sources of systematic uncertainty at ILC – the corresponding studies at CLIC are in an early stage, therefore the detailed assessment of this source of error has to be postponed to the next phase of the project;
- the physics background to Bhabha events, stemming from four-fermion production, is expected to have a higher cross section at CLIC than at ILC – on the other hand, the electrons from such events are much more forward boosted at CLIC and only a smaller fraction reaches LumiCal, implying that the uncertainty from this source should not exceed $10^{-3}$ [13];
- the energy scale, expected to be known to 400 – 800 MeV at ILC with a resulting systematic uncertainty of $10^{-3}$, could be more uncertain at CLIC due to the impact of incoherent pairs and $\gamma\gamma \rightarrow$ hadrons background, and more studies are needed here;
- the uncertainty on the absolute value of the beam polarisation at CLIC might be a factor of five worse than at ILC, mostly due to large depolarisation effects during the bunch crossings – nevertheless, the resulting uncertainty on the absolute value of the luminosity in the peak should not exceed a few $10^{-3}$.

In summary, the on-going work on the Bhabha suppression due to beam-beam effects, on the experimental determination of the luminosity spectrum at CLIC and on the energy scale error due to backgrounds will provide the necessary input for a more quantitative assessment of the expected systematic uncertainty to the absolute luminosity at CLIC. This detailed work goes beyond the scope of the present CDR and will be documented in the forthcoming phase of the project.

## 9.4 The Beam Calorimeter (BeamCal)

The BeamCal is designed as a sensor-tungsten sandwich calorimeter covering the polar angle range between 10 mrad and 40 mrad around the outgoing beam. This calorimeter has three main functions: (1) it serves as shielding for the accelerator components downstream, e.g. the QD0, and as a "mask" against backscattering into the vertex detector region; (2) it improves the forward hermeticity of the electromagnetic calorimeter system, allowing e.g. to veto low angle high energy electrons from Standard Model processes; (3) it measures a high flux of incoherent pairs, which may be used for luminosity monitoring purposes. A schematic view of the BeamCal, as implemented in the CLIC_ILD simulations, is shown in Figure 9.7. The upstream graphite disk and the tungsten plates are shown with a small hole for the incoming beam and a large opening for the spent beam. The BeamCal sensor layers in between the tungsten absorber plates are made very thin, and they are not visible in the figure. In the simulations, the sensors are represented by $8 \times 8$ mm$^2$ pads arranged in concentric rings. The sensors are equipped with front-end electronics positioned at the outer radius outside the active area of the BeamCal.

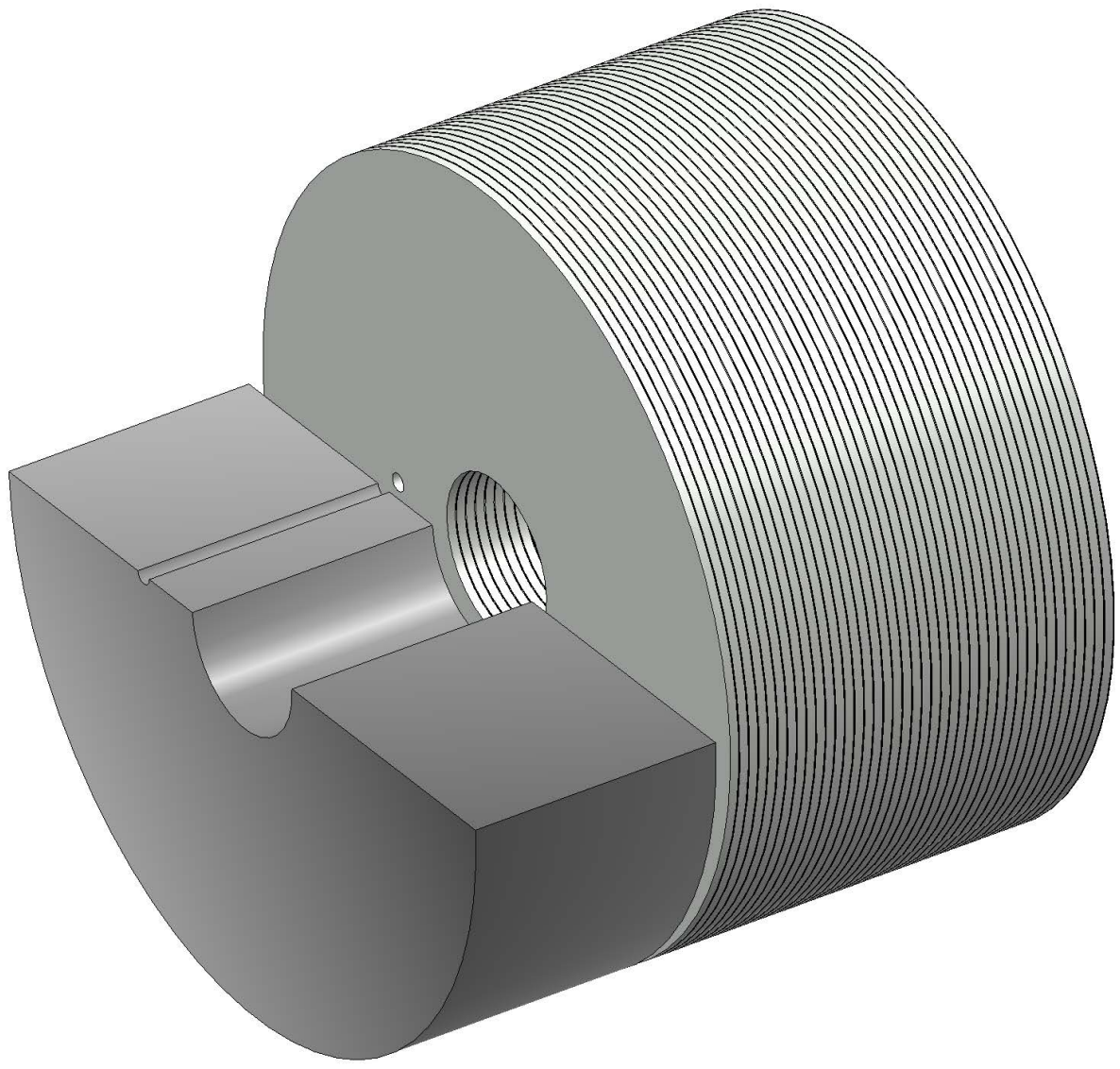

Fig. 9.7: Schematic view of the BeamCal, with the upstream graphite block cut in half for illustration.

As shown in Figure 9.8, the BeamCal will be hit by a large number of incoherent $e^+e^-$ pairs [3]. The size and distribution of these depositions allow a fast luminosity estimate and support beam tuning [14]. The detection of single high energy electrons will be a challenge, but possible. This is illustrated in Figure 9.8 showing the deposition of a single 1.5 TeV electron with an angle of 20 mrad w.r.t. the outgoing beam axis superimposed over ten bunch crossings of incoherent pair background, seen as the red spot on the right side. The high-energy electron results in a significant localised energy deposition on top of the wider spread $e^+e^-$ pair background.

Studies are ongoing to develop an appropriate subtraction of the incoherent $e^+e^-$ pair deposits, and a shower finding algorithm which takes into account the longitudinal shower profile, to detect high single energy electrons with high efficiency.

The challenge for the BeamCal are radiation hard sensors. As can be seen in Figure 9.9, doses up to 1 MGy and neutron fluxes of up to $10^{14}$ per year are expected in the inner rings of the BeamCal. Studies have been performed using polycrystalline Chemical Vapor Deposition (CVD) diamond sensors of 1 cm$^2$ size, and using larger sectors of gallium arsenide (GaAs) pad sensors. Polycrystalline diamond sensors have been irradiated up to 7 MGy and are still operational [15]. GaAs sensors are found to tolerate about 1 MGy [16]. Since large area CVD diamond sensors are very expensive, they may only be used at the innermost part of the BeamCal. At larger radii GaAs sensors seem to be a promising option. These studies will be continued for a better understanding of the damage mechanisms and possible improvements of the sensor materials.

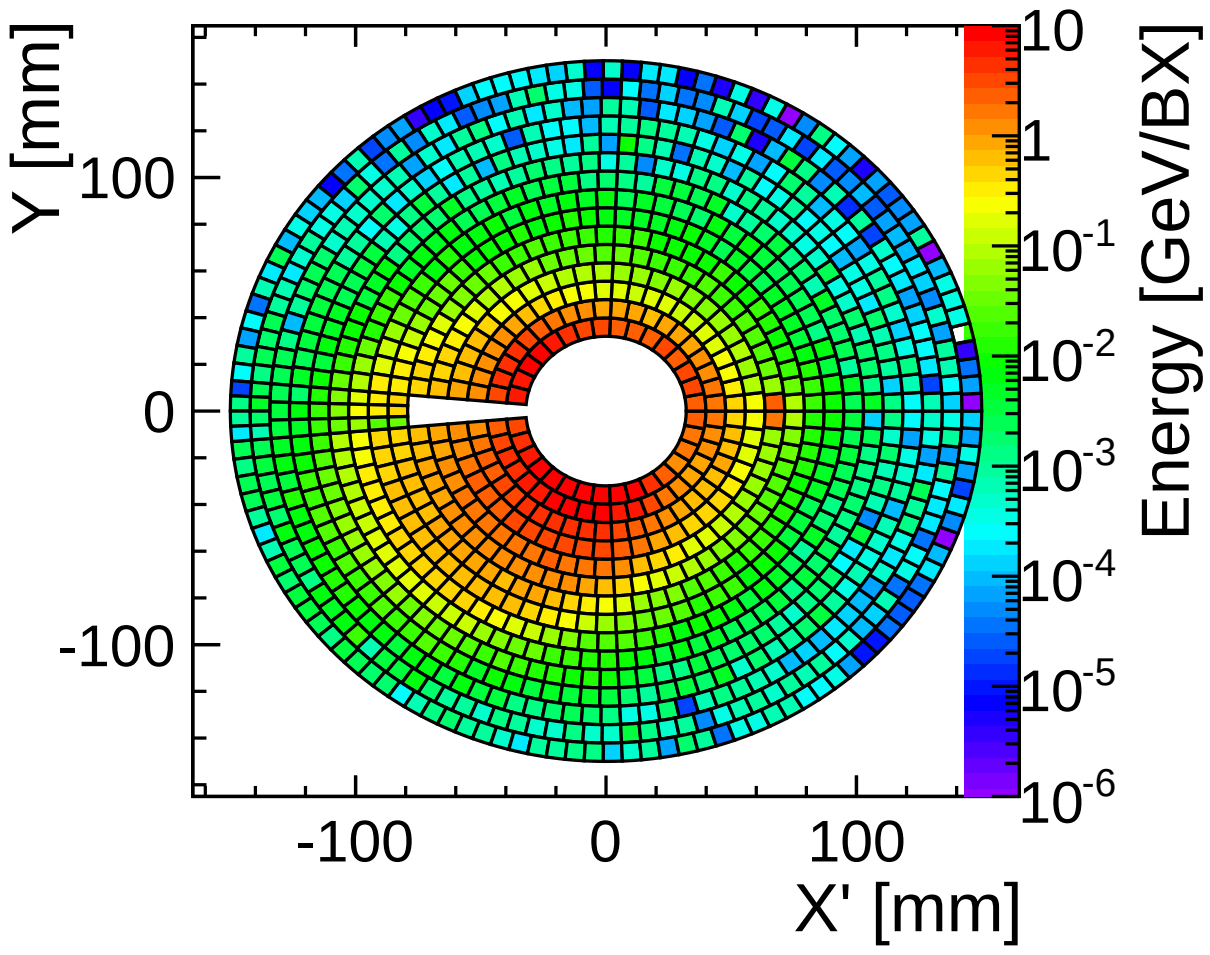

Fig. 9.8: The distribution of energy deposited by incoherent $e^+e^-$ pairs in the tenth layer of the BeamCal, integrated over ten bunch crossings. $X'$ is the horizontal axis, centred on the outgoing beam. Superimposed is the deposition of a single high energy electron at an angle of 20 mrad w.r.t. the outgoing beam axis (red spot at $X' \approx 65$ mm). The white hole and the white sector on the left are the space for the beam pipes.

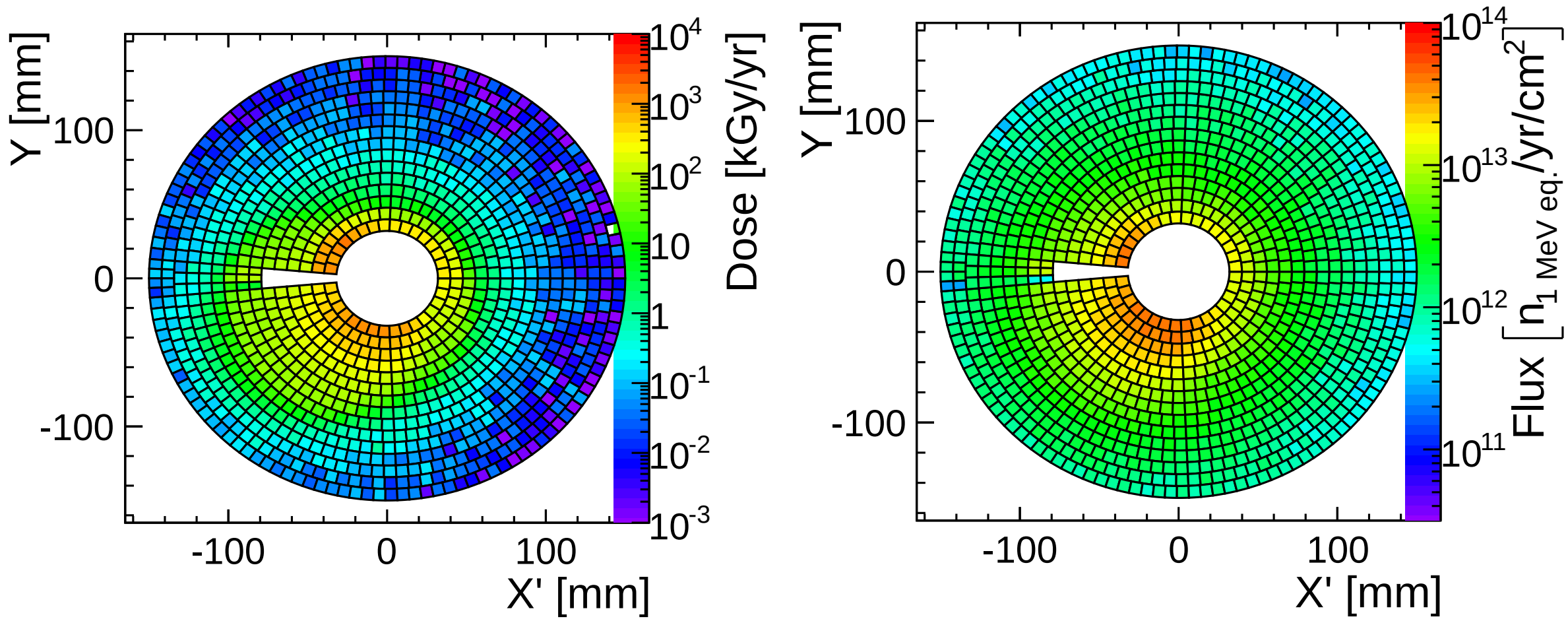

Fig. 9.9: The distribution of the expected total ionising dose (left) and 1 MeV equivalent neutron flux (right) per year in the fifth sensor layer of the BeamCal. These results were obtained for the CLIC_ILD concept, without safety factors .

# Chapter 10

# Readout Electronics and Data Acquisition System

## 10.1 Introduction

This chapter aims to provide an overview of the specific detector readout requirements at CLIC. It indicates how these requirements can be met and provides examples of possible implementation scenarios. It also summarises priority areas where further research and development is needed.

The detector readout requirements at CLIC are, on the one hand, closely linked to the beam structure and to the presence of beam-induced background and, on the other hand, to the needs for high precision measurements. As a result, the CLIC electronics requirements resemble those of LHC detectors for their pile-up and time stamping aspects, albeit with more demanding precision, and those of ILC detectors for the trigger-less readout, the high channel count, the low material budget and the power pulsing aspects. Exposure to radiation is similar to the situation at the ILC and typically a factor $10^4$ lower than at LHC.

The timing requirements for the various subdetectors at CLIC are typically in the range from 1 ns to 10 ns and are principally driven by the beam-induced background, see Section 2.1.2. While interesting high-energy $e^+e^-$ events are rare, at most one event per bunch train, the beam-induced background leads to many additional particles traversing the detectors and to high cell occupancies. In order to measure the physics events with good precision, an excellent separation between physics hits and background hits is required. The 0.5 ns time separation between bunch crossings, and the sequence of 312 bunch crossings in the train makes this a challenging task. The bunch train lasts 156 ns, with the next bunch train arriving 20 ms later. Therefore the readout of data into the off-detector electronics takes place at the end of the bunch train, without use of any trigger selection. The recorded hit data shall therefore contain sufficient timing information to allow for an adequate offline separation between physics and background hits. The amount of pile-up from beam-induced background generally does not require separating physics hits from background hits at the level of a single bunch crossing, but rather at the level of a few bunch crossings. This has been studied more quantitatively in the framework of the detector benchmark simulations, presented in Chapter 12. These physics analyses take typical signal formations in the various subdetectors into account and combine these with precise hit timing functionalities in the electronics. These timing functionalities are optimised for the individual subdetectors with the aim of minimising the impact of beam-induced background on the physics results.

The CLIC detector aims at measuring $e^+e^-$ interactions with very high precision, and thus sets challenging targets for impact parameter, track momentum and jet energy resolution. Therefore, compared to LHC the amount of material in the trackers needs to be reduced significantly. In detector implementations both cooling and power supply connections contribute strongly to the total material budget. It is therefore important to reduce the power consumption in the front-end electronics, as well as the current to be delivered through the power cables traversing the experiment. It is assumed that the progress of micro-electronics technologies and the corresponding reduction in power consumption will be counterbalanced by a higher channel density and by more complex functionalities in the front-end, compared to today's detectors. In order to reduce power consumption one can make efficient use of the low accelerator duty cycle resulting from the 156 ns active time and the 20 ms repetition period. This allows for a power-pulsing scheme of the on-detector electronics. In this concept the electronics is powered only at times when it is needed, such as during the event detection, readout and data transmission. Depending on the effective power-on and power-off times, this scheme allows for large reduction factors in the on-detector power consumption and heat dissipation. In view of the ultra-thin material requirements, most current cooling techniques cannot be applied at CLIC, due to their large mass in piping and liquids. Therefore, technologies such as air cooling or micro channel cooling are considered [1].

An overview of the detector readout channels, together with their basic functionalities, occupancies and readout speeds is given in Section 10.2. Implementation examples for a few subdetectors are presented in this section, in particular concerning the time stamping methods. Powering schemes are described in Section 10.3, while DAQ aspects are presented in Section 10.4.

## 10.2 Overview of Subdetectors and their Implementation Scheme

### 10.2.1 Overview of Subdetectors

Table 10.1 provides an overview of the readout channels for a complete detector [2]. In this table, the CLIC_ILD concept is taken as an example. For each subdetector, the table provides an estimate of the number of channels for a given cell size as well as the expected occupancies [3] during the bunch train per cell. The expected data word length takes the number of bits for time, pulse height and addressing of individual hits into account. Combining the information on the time sampling period, number of channels, average occupancy and number of bits per hit allows to calculate the data volume to be transferred from the front-end electronics for each subdetector per bunch train. From these values the number of optical links for the readout of the detector can then be calculated. The required time-resolution for individual hits and the proposed readout sampling time are given.

In this table the subdetectors have been arranged into five categories, separated by horizontal lines, grouping detectors with similar example readout implementations:

**Silicon pixel detectors:**

Measurement of arrival time and time-over-threshold for one hit readout per train. Zero suppression is applied as described in Section 10.2.2.

**Silicon strip detectors:**

Sampling of pulse height at regular interval. The required time resolution is achieved by signal shape analysis, similar to Section 10.2.4. No zero suppression is assumed at this stage due to the large occupancies.

**TPC:**

Analog pad readout for 1000 time voxels per channel, see Section 10.2.3.

**Calorimeters:**

Sampling of pulse height at regular intervals. The required time resolution is achieved by signal shape analysis, see Section 10.2.4. No zero suppression is applied. Pulse heights are higher than for the strip detectors, therefore the time resolution is better.

**Muon detectors:**

Digital readout with a multi-hit TDC. Zero suppression is applied as well as address decoding.

At the time of writing this CDR, dedicated readout studies for most CLIC subdetectors are just starting. Therefore, the implementation schemes summarised below are based on experience with electronics developments for LHC, ILC or other applications and principally serve as illustrations of feasibility. The following examples were chosen to illustrate three categories in the table: pixel detector readout, TPC pad readout and analog readout for calorimetry.

It is anticipated that, subject to R&D, readout solutions can be found to accommodate up to 5 hits per train in high-occupancy regions of the experiment. Currently, some regions show occupancies exceeding this value. These regions will be subject to further detector optimisation in the next project phase.

### 10.2.2 Implementation Example for a Pixel Detector

The CLIC silicon pixel detector has to meet a challenging combination of performance goals (see Chapter 4). A single point resolution of $\approx$3 μm is required, together with a timing precision of $\approx$5 ns. Meeting these requirements calls for typical pixel sizes of 20 μm × 20 μm combined with limited pulse height information for charge sharing purposes and an adequate Signal-to-Noise ratio, while the material budget has to be kept as low as 0.2% $X_0$ per layer. Power dissipation shall be at the level of 50 mW/cm$^2$ including power pulsing, which will make it possible to foresee low mass air-cooling. The maximum occupancy within the 20 μm × 20 μm pixels will be 3% per bunch train, including a safety factor of five for the incoherent pairs and a safety factor of two for the $\gamma\gamma \rightarrow$ hadrons. The signal will originate from a depleted silicon layer of ≈20 μm thickness, in the case of an integrated CMOS sensor, to ≈50 μm thickness, in the case of a hybrid detector, and will therefore correspond to 1300-3300 electrons.

Table 10.1: Overview of readout details for the various subdetectors of the CLIC_ILD detector concept. Occupancies and data volumes are for a full bunch train and include charge sharing between pixels/strips. Safety factors of five and two are applied to the rates of the incoherent pairs and the $\gamma\gamma \rightarrow$ hadrons, respectively; except for the TPC, for which no safety factors have been applied. Occupancies averaged over entire subdetectors are compared to the maximum values obtained for the regions with the highest backgrounds.

| | time stamping resolution [ns] | time sampling period [ns] | cell size [$mm^2$] | number of channels [$10^6$] | average to maximum occupancy [%] | number of bits per hit [bit] | data volume [Mbyte] |
|---|---|---|---|---|---|---|---|
| **VTX barrel** | $\sim 5$ | 10 | $0.02\times0.02$ | 945 | < 1.5 - 1.9 | 32 | 56 |
| **VTX endcap** | $\sim 5$ | 10 | $0.02\times0.02$ | 895 | < 2.0 - 2.8 | 32 | 72 |
| **FTD pixels** | $\sim 5$ | 10 | $0.02\times0.02$ | 1570 | 0.1 - 1.0 | 32 | 6.3 |
| **FTD strips** | $\sim 5$ | 10 - 25 | $0.05\times100$ | 1.6 | 160 - 290 | 16 | 48 |
| **SIT** | $\sim 5$ | 10 - 25 | $0.05\times90$ | 1.0 | 100 - 174 | 16 | 30 |
| **SET** | $\sim 5$ | 10 - 25 | $0.05\times438$ | 5.0 | 17 - 17 | 16 | 150 |
| **ETD** | $\sim 5$ | 10 - 25 | $0.05\times300$ | 4.0 | 38 - 77 | 16 | 120 |
| **TPC** | –[a] | 25 | $1\times6$ | 3[b] | 5 - 32 | 24 | 500 |
| **ECAL barrel** | 1 | 25 | $5\times5$ | 69.5 | < 3 | 16 | 2090 |
| **ECAL endcap** | 1 | 25 | $5\times5$ | 43.2 | 60 - 150 | 16 | 1300 |
| **HCAL barrel** | 1 | 25 | $30\times30$ | 6.9 | < 5 | 16 | 210 |
| **HCAL endcap** | 1 | 25 | $30\times30$ | 1.8 | 120 - 5200 | 16 | 54 |
| **HCAL rings** | 1 | 25 | $30\times30$ | 0.2 | < 5 | 16 | 6.0 |
| **LumiCal** | 5 | 10 | $5\times5$ | 0.2 | 600 - 6000 | 32 | 28 |
| **BeamCal** | 5 | 10 | $8\times8$ | 0.1 | 15600[c] | 32 | 15 |
| **MUON barrel** | 1 | 25 | $30\times30$ | 1.4 | 0.01 - 0.05 | 24 | < 0.01 |
| **MUON endcap** | 1 | 25 | $30\times30$ | 2.4 | 0.12 - 10 | 24 | < 0.01 |

[a] By combining with different subdetectors in offline reconstruction 2 ns will be achieved.
[b] The 3D TPC reads out 1000 voxels per channel for each bunch train.
[c] All cells measure a signal for each bunch crossing.

While several pixel detector technologies are considered for implementation an example based on a hybrid pixel detector and on the experience with the Timepix1 ASIC [4] is given to evaluate a possible readout concept. Assuming a detector chip organisation as an array of 512×512 pixels with 20 μm × 20 μm dimension, each pixel will include a simple front-end coupled to a discriminator and a digital counter with 4 bits for the arrival time measurement, for 16 time-slices of 10 ns, and 4 bits for Time-Over-Threshold (TOT) measurements. With a 50 μm thick silicon sensor simulations show that setting the overall minimum threshold at 500 electrons (channel noise below 65 electrons) the detection efficiency is larger than 98.5% [5]. Offline, the combination of the arrival time information and the time-over-threshold pulse width allows to derive precise arrival time information with time-walk correction as demonstrated by [6, 7]. The reference clock distributed to all pixels during the bunch train has the same period as the required time binning of 10 ns. The absolute time reference will be given by the time interval from particle arrival to the end of the bunch train. Thus the end of bunch train signal needs to be sent to all pixels with a controlled skew (below 1 ns) in order to ensure the requested time resolution. With such an architecture the reference clock can be distributed as a pixel-by-pixel inverted clock network which minimises coupling and digital power consumption as demonstrated in the Timepix1 chip [8]. Zero-suppressed data readout will be initiated between trains. Each pixel hit will produce 8 bits of timing data and more than 18 bits of address data. In the CLIC_ILD vertex barrel there will be some 945 million pixels and an expected occupancy of 1%, which results in approximately 56 Megabytes of data per bunch train for the full barrel. Given the dense integration of readout functionalities within the very small pixel size, the use of a deep CMOS technology ($\leq$ 65 nm) is anticipated. For this purpose a complete front-end in 65 nm CMOS technology was successfully produced and evaluated recently [9].

The analog circuitry will be the most important component of the power consumption of this detector. With an expected power of 10 µW/pixel it will be consuming approximately 2.5 W/cm$^2$. By using high-Vt digital cells it is expected that the digital leakage consumption will be negligible (<1 mW/cm$^2$). Therefore, the power pulsing feature shall provide a gain factor of $\approx$50 to meet the overall power consumption goal of 50 mW/cm$^2$. The various parts of the circuit will be turned on and off selectively, synchronously following the readout cycle. The analog part is expected to consume power during 40 µs within each bunch train including ramp-on and ramp-off, while the digital part will stay active during some 400 µs. Following the above assumptions the chip would produce some 20 kB of zero suppressed data per train, corresponding to a modest instantaneous data transfer rate of 50 Mbyte/s during the digital power-on time.

In order to minimise the amount of material the silicon detector shall be as thin as possible. The sensor thickness will be determined by the minimum detectable charge, the preamplifier noise and the constraints given by the limited material budget and the detector integration process. Present research targets the application of Through-Silicon Vias (TSV) to reduce the dead area at the periphery of the readout chip for both hybrid pixel detectors and integrated CMOS detectors.

### 10.2.3 Implementation Example for the TPC Pad Readout

The TPC (see Chapter 5) has a drift length of 2.25 m, corresponding to a maximum drift time of $\approx$ 30 µs, depending on the gas mixture used. Gas amplification takes place in a Micro Pattern Gas Detector (MPGD) plane near the endplate. Each end plate comprises 1.5 million readout pads of 1$\times$6 mm$^2$ size. The incoming signals are read out at a pace of 40 MHz during the full 30 µs drift time for each pad and for each bunch train. This yields 10-bit pulse height data for 1000 time-voxels for each individual pad. The TPC hit occupancies vary from 32% of the time-voxels for pads in the inner row to 1% in the outer row, excluding safety factors.

Based on the TPC readout scheme of the ALICE TPC [10, 11] at the LHC, dedicated electronics for pad readout of a linear collider TPC has been developed over the past years and successfully applied in MPGD-based TPC prototypes [12]. The so-called SAltro readout [13] is based on a compact ASIC, comprising a low-noise programmable amplifier, a shaper, a high-speed 10-bit ADC and a set of programmable digital filters for noise reduction and signal interpolation. Following data reduction in SAltro, the total data volume of the TPC will amount to approximately 500 Mbyte per bunch train. The current SAltro-16 compact ASIC with 16 fully integrated readout channels will allow for qualification of TPC prototypes with different gas amplification structures. It will also facilitate first TPC power pulsing studies at the system level. In a next phase, the chip will require an optimised ADC with a reduced power consumption and with a smaller surface, more advanced sequencing of power pulsing as well as the integration of a larger number of channels. This will allow to reduce the current power consumption of 50 mW/channel in continuous mode by a factor of ≈4. An additional factor of 25-50 can be expected from power pulsing. An alternative TPC readout scheme using 55 μm × 55 μm pixels is also under study [8].

As the general TPC concept is based on 3D imaging, making use of the rather slow drift of electrons from ionisation in a large gas volume, time stamping as such is not applicable. However, one can make use of the drift velocity of the electrons along the $z$-direction in the TPC, which amounts up to 75 μm/ns depending on the gas mixture, to measure the offset between each reconstructed TPC track and the precise hit positions for the same track in the outer silicon layers (ETD and SET). As described in Section 5.4 and [14], this allows determining the timing of TPC tracks offline with a precision of better than 2 ns.

### 10.2.4 Implementation Example for the Analog Calorimeter Readout

The readout electronics of the calorimeter has to provide very good energy and time resolution, while occupancies will be high in forward regions. The dynamic range for the pulse height of individual hits corresponds to 12 bits in the ECAL and 14 bits in HCAL, whereas the time of the shower start must be determined with 1 ns precision. The highest occupancies appear in the lower radii of the endcap and originate from $\gamma\gamma \rightarrow$ hadrons events for the ECAL and from backscattered incoherent pairs for the HCAL. These maximum occupancies amount to 50% in the ECAL for $5{\times}5$ mm$^2$ cells and 1100% per bunch train in the $3{\times}3$ cm$^2$ analog HCAL cells, excluding safety factors. Mitigation of these high occupancies will be the subject of further detector optimisation in the next project phase.

Several studies and existing calorimeter readout electronics have shown that one can obtain optimised timing and energy resolutions using a preamplifier-shaper with a peaking time exceeding the required time-stamping resolution and digitising the pulse at a rate slower than the required time resolution. For instance, an optimal digital filtering technique is successfully applied on pulse height samples collected at 40 MHz in the ATLAS liquid argon calorimeter electronics [15, 16, 17]. Thanks to the detailed on-detector calibration of signal shapes for individual channels, an overall hit time resolution below 1 ns was achieved at the system level. A similar approach can be applied at CLIC. The method is schematically depicted in Figure 10.1.

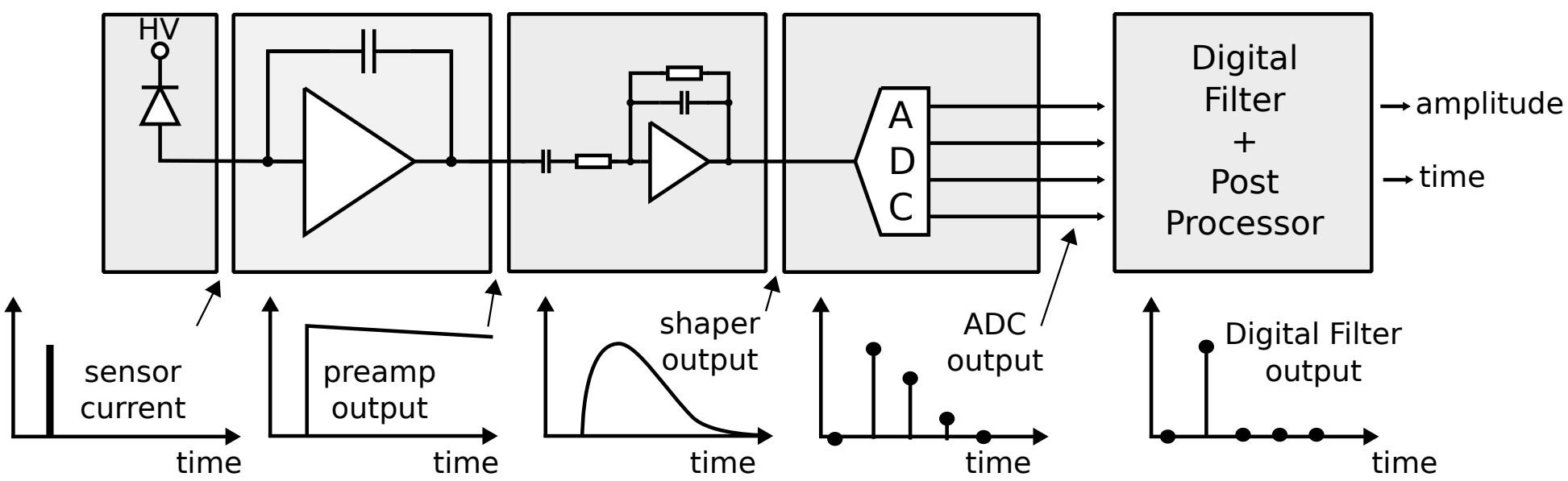

Fig. 10.1: Schematic view of proposed readout scheme for digital filtering. Figure taken from [18].

The basic principle consists of a readout chain with an amplifier-shaper using a RCn-CRp filter delivering a pulse length of about 100 ns duration for a delta-function input signal. This signal is then digitised directly at 40 MHz rate using a fast ADC with 14-16 bits precision [19, 20, 21], covering the required resolution. Alternatively, the signal can be stored at a similar rate using fast analog memory, to be digitised at low speed at the end of the bunch train. Based on present-day techniques, the fast analog memory option can already handle 12-13 bits. This option may therefore call for a multi-stage amplification approach in order to fully cover the required dynamic range. The digital filtering is applied to the digitised data, either in the on-detector or off-detector electronics. Future detailed implementations of the calorimeter front-end electronics for CLIC will be derived from electronics prototypes [22, 23] developed for the ILC experiments in the framework of the CALICE collaboration.

The maximum data rate can be estimated as follows. Assuming no on-detector zero suppression and thus no need for address bits in the data stream, the front-end will deliver 16 bits every 25 ns during the bunch train time of 156 ns extended by 100 ns for the shower development time and 100 ns for the front-end shaping time. Assuming data transmission during 350 ns, 50 times per second for $120 \cdot 10^6$ channels in the electromagnetic and hadronic calorimeter leads to a transmission of 170 Gbyte/s. This can easily be managed with Gigabit links. For comparison, the ATLAS liquid argon calorimeter uses 1600 1.6 Gbit/s links.

The power consumption in the front-end will be dominated by the ADC in case the signal is digitised at 40 MHz. Existing 16-bit 40 MHz ADCs [20], are consuming as little as 70 mW per channel during conversion and require less than 350 µs to recover after a power down; it is expected that future devices will consume even less. The power consumption of the de-randomising buffers and of the Gigabit links will be maintained low by using low power memories and minimising the number of links. In order to limit the power dissipation, a 50 Hz power pulsing scheme will be applied (see Section 10.3). In this way the active time of the ADCs can be limited to $\approx$18 ms every second. As a result, a power consumption of about 150 kW for the ECAL and HCAL combined is anticipated. If an analogue memory storing the input signal is implemented, low speed ADCs can be used. Optimised ADCs targeting the mobile devices market are available and dissipate very low power, typically 150 µW per channel when running at 10 ksamples/s and about 150 nW when no conversion is done [24]. For each train 14 samples per channel need to be digitised, i.e., the ADCs would perform conversions during 70 ms every second. So the power consumption of the ADCs would be about 10 µW per channel, i.e., about 1.2 kW for the ECAL and HCAL combined. This number is to be doubled if a two-gain scheme is used and some small amount of power needs to be added for the analogue memory.

The readout solutions currently foreseen for BeamCal and LumiCal resemble those of ECAL and HCAL. Since the occupancy in BeamCal is very high one possibility for the readout is a gated integrator with Correlated Double Sampling (CDS) [18]. The current pulses integrated continuously by the integrator are sampled in time slices of about 10 ns. The sampling may be performed either by a fast ADC directly or via a fast analog memory. Signal amplitudes are obtained subtracting subsequent samples. The subtraction would in addition serve as a noise filtering. Between bunch trains, the integrator is reset. This concept will also be applicable for LumiCal. For low-occupancy regions in LumiCal an alternative solution, similar to ECAL and HCAL, with a chain consisting of a fast preamplifier and shaper followed by a fast digitiser and a deconvolution algorithm applied to the ADC output is investigated [18].

## 10.3 Power Delivery and Power Pulsing

### 10.3.1 Motivation

In view of the high precision requirements for a CLIC detector, combined with the unprecedented number of channels, efficient power management will be of utmost importance. Large power dissipation in the front-end electronics results in large masses and volumes for cooling systems and cables, which have a negative impact on detector performance. An optimised power management covers the following aspects:

- Power-efficient design of the front-end-electronics;
- Turning power on only when it is needed;
- Efficient power delivery schemes, to reduce thermal losses and the cabling volume from the back-end to the detectors;
- Efficient and low-mass cooling schemes.

A careful design together with the progress of micro-electronics technologies, requiring lower power voltages, will allow for reduced power dissipation in comparison with today's large experiments. However, this advantage will be largely counterbalanced by the huge channel count and by more complex front-end functionalities. As the CLIC beam structure foresees bunch trains of 156 ns duration repeated every 20 ms one can use this low $7.8 \cdot 10^{-6}$ duty cycle to significantly reduce the power consumption by turning power on only when it is needed. Therefore all subdetectors anticipate the use of power pulsing at 50 Hz for the front-end electronics.

In the vertex and silicon tracking systems the power pulsing is driven by the need to reduce the material of the cooling systems. The amount of dead material needs to be kept very small to reduce unwanted multiple coulomb scattering of charged particles. Ideally these detectors will be cooled by air flow. Power pulsing is also foreseen for the CLIC_ILD TPC tracker. Dead material at the location of the TPC endplates is undesirable in order to avoid photon conversions before the ECAL. Nevertheless air cooling is not easily possible here in view of the complex geometry of electronics cards on the large surface of the TPC endplate. Therefore a $CO_2$-based cooling is currently under study. In the electromagnetic and hadronic calorimeters, particle showers have to be kept compact to ensure an optimal separation of clusters for the application of particle flow algorithms. For this purpose, the active layers need to be as thin as possible, while the effective nuclear interaction length of the absorber material has to be as short as possible. To reduce the required service material, power pulsing is therefore also foreseen for the calorimeters. This should allow to limit the cooling infrastructure to cooling lines placed radially along the module sides.

In the domain of efficient power delivery to the front-end, significant research has been carried out in recent years for the upgrade of the LHC experiments. These studies include DC-DC voltage step-down near the front-end, allowing to reduce the volume of cables running from the back-end to the front-end electronics. In the following section it is discussed how power pulsing and efficient power delivery can be combined.

### 10.3.2 Implementation of Powering Schemes for CLIC Detectors

Power switching in the front-end is applied selectively to various parts of the circuit. The analog circuitry can remain off most of the time and turned on only to acquire the physics data during the bunch train. The digital and data transmission electronics operate principally during the bunch train gap. Depending on the data volume of the subdetectors a sufficiently large period is reserved to transmit the data using low power digitisers and communication links. Turning on periodically the front-end electronics results in a periodic demand of peak current, so a voltage regulation circuit is required. A powering branch is therefore composed of a back-end power supply, connected via long cables to a front-end voltage regulator and then to the front-end circuits. For this reason the power gating feature is implemented at the ASIC level. This has the additional advantage of a fast turn-on time of a few tens of µs resulting in the smallest power duty cycle. Because the regulator device continuously provides the power voltage, the configuration data are permanently maintained and do not need to be periodically reloaded. This scheme also allows disabling selectively the most power-demanding ASIC blocks, such as the biasing circuits, the preamplifiers and the ADCs.

When the front-end ASICs are turned on, the current rises from a standby level to several Amperes. The regulating device needs to deliver the peak current during this time while maintaining the voltage regulation. Different topologies making use of linear regulators, DC-DC converters and storage capacitors have been studied. Although the turn-on time and the peak current will vary between systems, typical values of 50 µs turn-on time and a peak current of 10 A are assumed to enable the comparison of two different schemes in a basic simulation [25]. The two regulation schemes are depicted in Figure 10.2. Schemes similar to Figure 10.2 (top) have been applied successfully in linear collider calorimetry applications [26, 27], while the scheme of Figure 10.2 (bottom) makes use of LHC upgrade developments based on DC-DC conversion. For both regulation schemes a peak current is delivered to the front-end electronics during the active time. This peak current is delivered by a storage capacitor close to the front-end system. Provided the storage capacitance is at least 5 mF, the voltage drop in this example resulting from the charge transfer during the active time can be maintained within an acceptable range of 100-200 mV in both regulation schemes simulated. Electrolytic capacitors within this range and up to 400 mF are commercially available [28], however they are large in size and can therefore only be used in places where there is sufficient space.

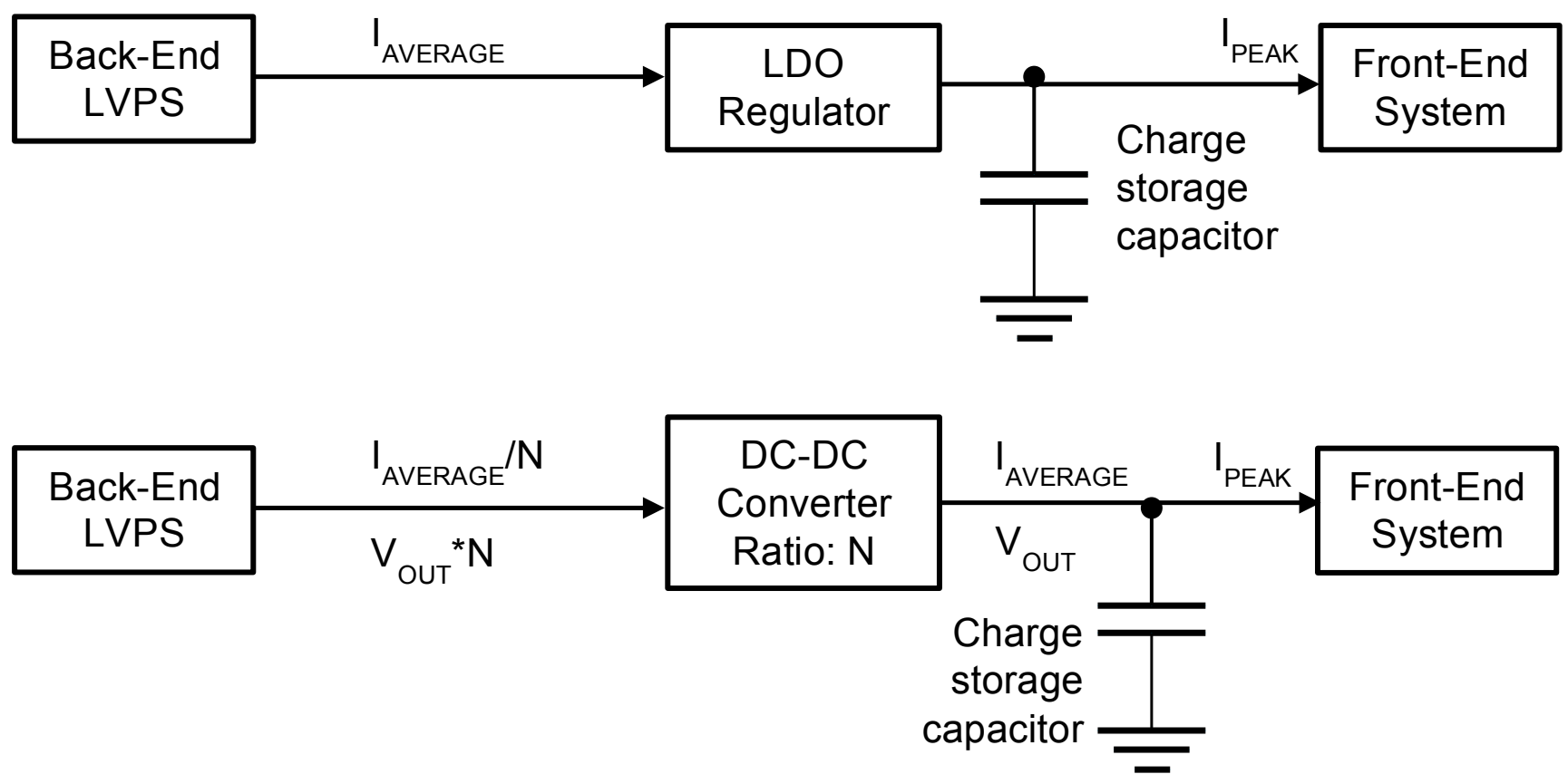

Fig. 10.2: Power pulsing regulation schemes with a selective circuit turning on and off within the ASIC. Voltage and current regulation are based on a low drop-out (LDO) regulator and storage capacitor (top), and on a DC-DC converter and storage capacitor (bottom).

In the case of the low drop-out (LDO) regulator option, the system losses are mainly due to the resistance of the back-end cable and the regulator drop out voltage. Figure 10.3 depicts the behaviour of the load current, storage capacitor current and back-end cable current in the example configuration with an LDO regulator, showing that the load current can be high, for a much reduced current in the back-end cable. The load voltage drop amounts to 100 mV in this case.

Similar results were obtained in the simulation for the DC-DC converter configuration in Figure 10.2 (bottom). It makes use of radiation tolerant buck converter ASICs developed for the LHC experiment upgrades [29, 30]. The buck DC-DC converter is a switching device that delivers a regulated output voltage and output current originating from a higher input voltage with lower input current. The voltage conversion ratio $N$ reduces the input current by $N$ compared to the output current and the losses in the back-end cable are reduced by $N^2$. The conversion efficiency is typically better than 80%, while the losses are dissipated in the DC-DC converter circuit itself. The regulation loop of the converter can be designed specifically to deliver a large output pulse with fast regulation, allowing to reduce the storage capacitor size.

### 10.3.3 Stability and Reliability Issues

There are several reliability issues for pulsed powering schemes that need to be taken into account. First, the proposed regulation topologies based on either an LDO or a DC-DC converter followed by a storage capacitor are still resulting in a non negligible current fluctuation in the backend cable. This current fluctuation not only imposes an oversized cable cross section, but also it exposes the back-end cables to varying induced forces caused by the intense magnetic field of the detector, with the resulting mechanical stress in the cable trays.

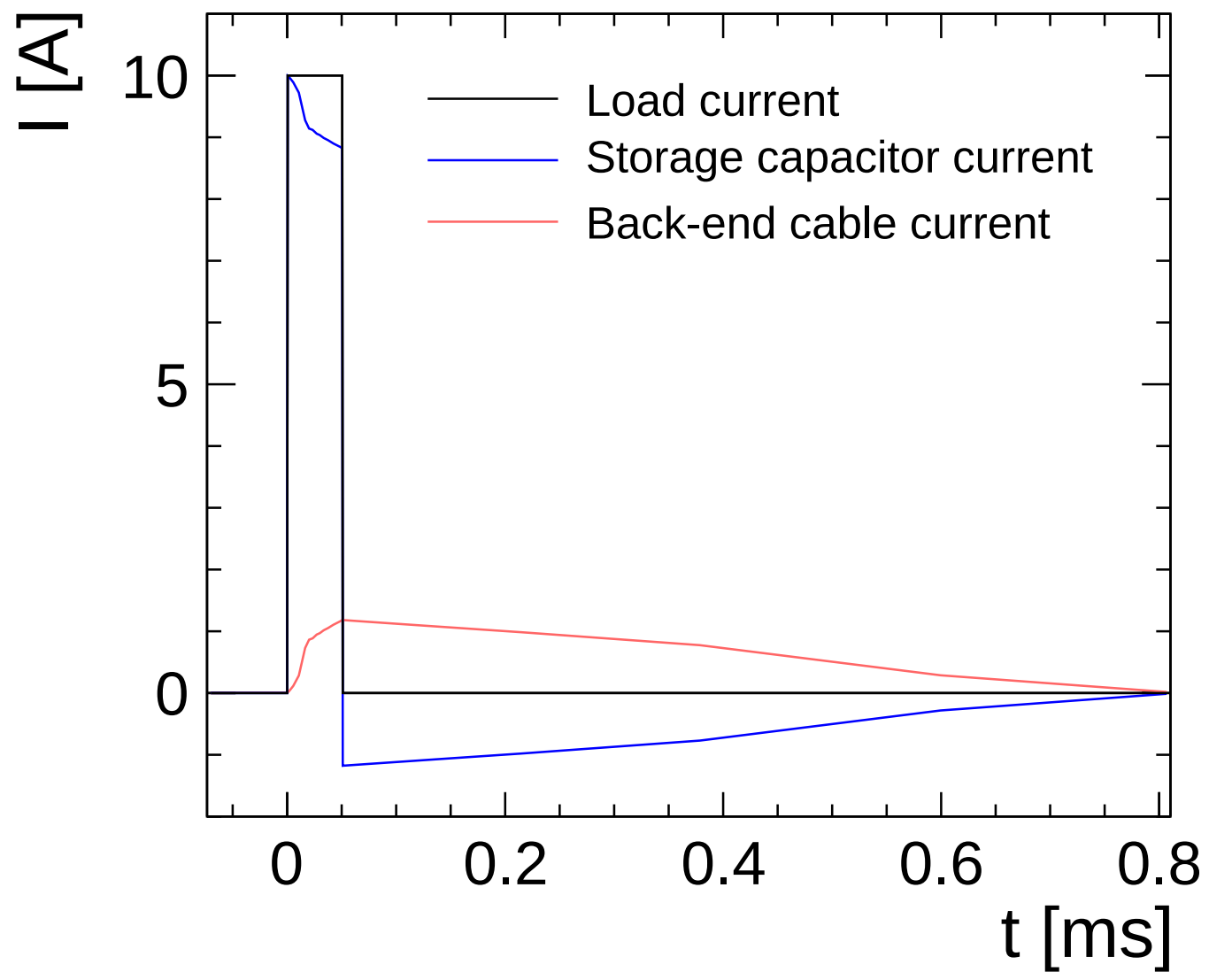

Fig. 10.3: Simulated current response in an example setup of the load current, storage capacitor current and back-end cable current in the LDO-based regulation scheme with 10 A peak current, 50 µs pulse and a voltage drop of 100 mA.

To overcome this problem, the back-end power supply can be configured in constant current mode instead of constant voltage mode. Placing an intermediate storage capacitance before the regulation circuit, a charge is locally accumulated by the constant current source, developing a voltage ramp that is easily regulated by the LDO or the DC-DC converter. This scenario would significantly reduce the current carried by the cables, that can therefore have a reduced cross section, and can virtually eliminate the forces induced by the magnetic field on the back-end cables.

Second, in the gated circuit the instantaneous peak power can be rather large. The front-end system need to be protected against failures that might affect the gating function, particularly in the case of a faulty permanent on state. In this fault situation, the faulty circuit will first sink the power directly available on the output storage capacitance.

The regulation circuit must be sized for delivering a relatively low DC current, sufficient to charge back the storage capacitance during the idle times, but low enough to limit the power delivered in the case of a fault condition. Ultimately the back end power supply should be interlocked on the average power that it is delivering and an interlock condition must be set to prevent fault conditions there too.

Finally, it must be noted that the power gating of front-end circuits will result in temperature cyclic variations that are well know for affecting the system long term reliability. Thermal cycling results in mechanical stress that needs to be taken into account in the assembly of the electronic systems. The reliability of the power pulsed systems will have to be qualified in the frame of a quality assurance plan based specifically on thermal cycling tests.

Initial power pulsing tests, mostly applied to ILC calorimetry applications, have demonstrated the validity of the concept [26, 27]. Extensive R&D is needed, however, to demonstrate the feasibility for application in all subdetectors, and in particular for vertex and tracking applications. The trade-off between additional material in the detector vicinity, due to the implementation of the powering scheme, and the gain in material due to lower power dissipation and cabling needs to be shown. The reliability of the powering scheme and any possible adverse effect on detector performance, stability and alignment need to be demonstrated in full systems tests in magnetic field for all subdetectors.

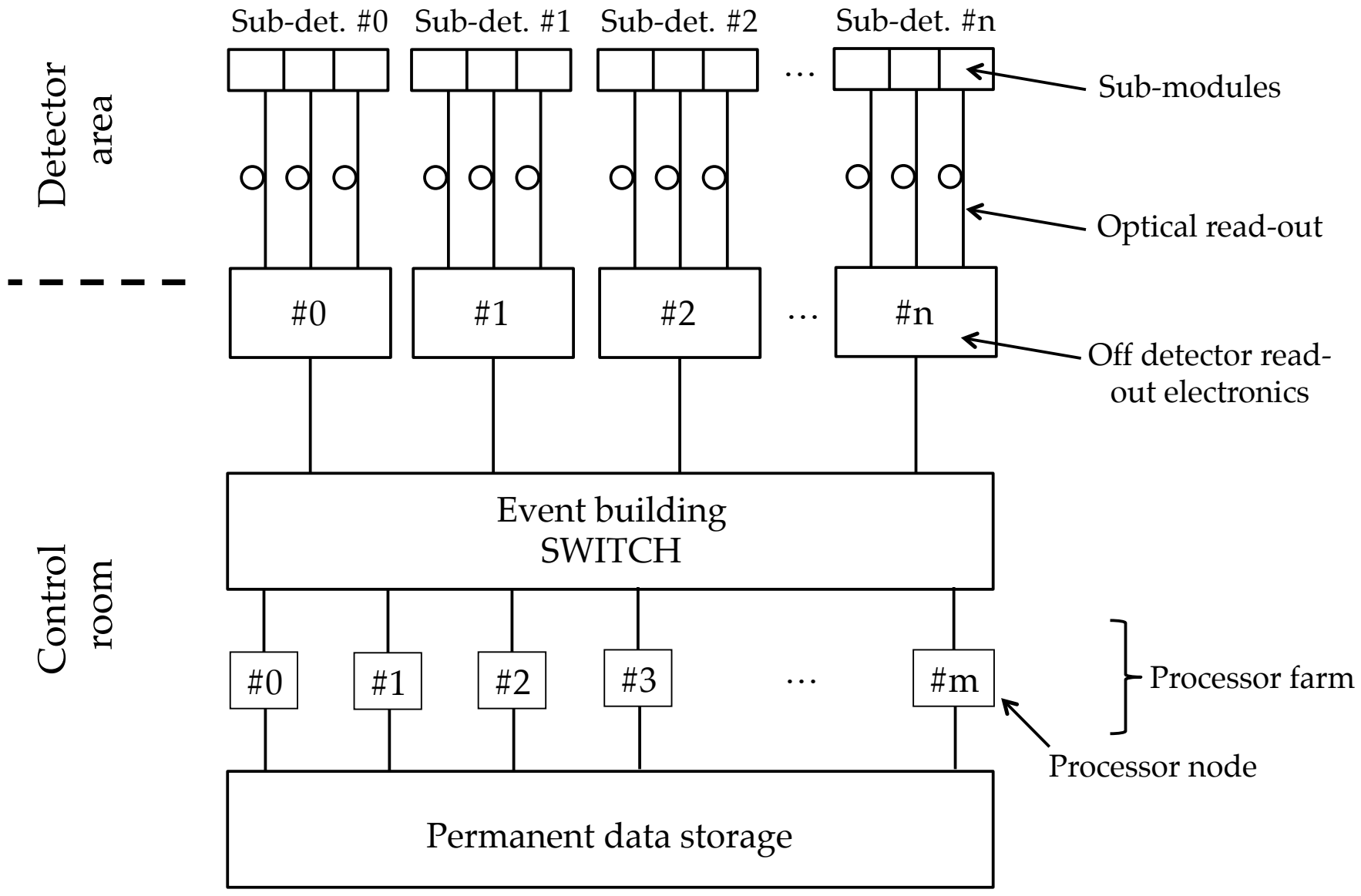

Fig. 10.4: Overview of the DAQ scheme.

## 10.4 DAQ Aspects

This section evaluates the data acquisition system requirements and proposes a possible architecture using currently available technologies in order to make a first estimate of the system dimension, building on experience with LHC upgrades. It is obvious however, that the ongoing improvement of the data communication and processing technologies will offer new opportunities before the DAQ system has to be actually designed and built.

Figure 10.4 shows the readout block diagram where the front-end electronics are directly connected to optical links transferring the data to the off-detector readout systems. The latter may be implemented by custom electronics or commodity computers. There the data are concentrated, re-formatted and possibly zero-suppressed in case this has not already been done in the front-end. The off-detector readout systems forward the data to an event-building switch connected to a processor farm. Apart from zero suppression in the front-end electronics or the readout systems, the processor farm is the first point in the processing chain where the data is filtered by a high-level trigger algorithm.

The total data volume produced by the detector is of the order of 4 Gbyte per bunch train or 200 Gbyte/s considering the bunch train period of 20 ms. Assuming an optical data link bandwidth of 10 Gbit/s, the number of links needed to satisfy the data throughput is 160. However, due to the detector segmentation and limited connectivity between detector sub-modules, the single data link bandwidth may not be used to its full extent. Also, the use of power pulsing in the detector front-end may decrease the effective data rate. In order to minimise the number of data links, powering the readout blocks in the front-end connected to individual links consecutively optimises the data link bandwidth. In the following example a conservative efficiency reduction factor of 10 is assumed, resulting in an effective sustained data rate of only 1 Gbit/s per link. In that case the detector would be connected to a moderate number of 1600 optical links. For comparison, the CMS detector is read out by 80000 links.

It is assumed that one off-detector readout processor feeds the data of 10 optical links to one port on the event-building switch. Therefore some 160 readout processors are needed. With present-day technology, switches are available which come close to this requirement, where 160 10 Gbit/s input ports can be routed to some 300 output ports [31]. Depending on the number of required processors in the processor farm at the time of implementation, each of these ports is connected either directly to a processor node or via a distributor node consisting of a processor and a sub-switch.

Based on the current reconstruction algorithms and the required CPU time to process a bunch train the number of required processing nodes has been estimated. A conservative average of time to fully reconstruct one bunch train is 3000 seconds on one Intel® Xeon® E5520 CPU with 2.27 GHz and less than 2 Gbyte RAM. This corresponds to 150000 seconds processing time in a second of beam operation. With today's commodity technology of four cores per CPU and two CPUs per node some 19000 processing nodes would be required. Due to the expected technology advancement until the time of implementation, an event farm with much less than 7500 processor nodes, which corresponds to the CMS event farm size in 2011, will be needed. Also, in an online farm, full reconstruction will not be needed, and developments of faster online algorithms and improvements in the available technology are expected to reduce the required amount of nodes by far.

The study on the data suppression factor has shown that contributions from $\gamma\gamma \to$ hadrons events to the data amount can easily be reduced by a factor of 10 by dismissing out-of-time hits. Identifying hits from incoherent pairs are expected to suppress the data volume further by a factor 5 to 10 for an overall reduction of 15 to 20. This leads to a sustained data rate of some 10 Gbyte/s to be written to storage. Even today this data rate to a permanent storage system is feasible. For example, in 2011 ALICE wrote 5 Gbyte/s to disk.

## 10.5 Summary

The present state of the detector definition allows to derive preliminary specifications for detector front-end and data acquisition systems (see Table 10.1). The front-end electronics specifications are driven by the short bunch crossing interval and the particle arrival time stamping as well as the reduction of material budget. The silicon tracker system needs to minimise material budget, while the calorimeter electronics need to be compact in physical space.

For the silicon vertex detectors technologies and approaches are available to pursue research on systems combining the position resolution, time-stamping capabilities and low material budget constraints. In particular, research on the sensor and front-end electronics level as well as in the field of low mass integration, interconnect, opto-electronics, power pulsing and cooling needs to be conducted to define a full system concept.

For the silicon tracker systems further detector optimisation together with research on silicon tracker techologies are required to optimise cell sizes and cell occupancy values in order to define low-power multiple-hit time-stamping tracking systems.

Concerning the TPC, both the pad readout and the pixel readout are already in an advanced development stage. In a next phase the pad readout will profit from technology trends to achieve reduced power consumption, while the next TPC pixel readout will also have more advanced functionalities on the chip.

In contrast to the LHC calorimeters the CLIC detector also requires the reduction of material in the active calorimeter systems which translates to low power implementations. Further studies and research are required to optimise detector layout, occupancy, front-end signal and shaping time to achieve the required time resolution with low power systems especially in high occupancy regions.

For all detectors the implementation of power pulsing is planned. A combined research effort is required in the domain of front-end ASIC design and on the power pulsing system level. Detector specific studies need to be undertaken to define the implementation feasibility and the system requirements. A coherent power pulsing control and communication system design including advanced safety features needs to be implemented. Especially in the inner tracking systems the integration issues are highlighted, including the positioning of the power pulsing components.

The present studies show that due to the expected low physics interaction rate the demands on the off-detector readout schemes and data acquisition system are manageable and no further research is required at present.

# Chapter 11

# CLIC Interaction Region and Detector Integration

## 11.1 Introduction

The Beam Delivery System at CLIC will have a single interaction point where the two CLIC detectors, CLIC_ILD and CLIC_SiD will share the beam time in a so called "push-pull" mode, alternating between data taking at the IP and being in caverns off the IP. Both detector layouts, CLIC_SiD and CLIC_ILD, are described with their overall parameters like dimensions, magnetic field, self-shielding and integration of services in Section 11.2. This is followed by a description of how to push-pull 14,000 ton detectors with their platforms, and the requirements and constrains resulting from it in Section 11.3. The overall underground experimental area is described and illustrated in Section 11.4. A number of accelerator-related issues near the IP are described in [1].

For proper machine operation and, in particular, for luminosity optimisation, the final focusing quadrupoles (QD0) of the CLIC accelerator need to be close to the IP. This distance, called $L^*$ , has been chosen to be less than 5 m, resulting in a position inside the detector volume. The very small vertical beam spot of only 1 nm r.m.s. requires pre-alignment of QD0's to better than 10 µm and a stabilisation of the QD0's position to well below 1 nm. These requirements have had a significant impact on the design efforts at different levels and dictated choices that had to be made for QD0 technology, support tube considerations etc. This will be described in Section 11.5.

The procedures developed for Opening and Maintenance of the detectors are finally described in Section 11.6.

## 11.2 Detector Layout

The principal structural elements of the two CLIC experiments, CLIC_ILD and CLIC_SiD, are: a superconductive solenoid and an iron return yoke consisting of two endcaps and a barrel section split longitudinally in 3 rings. These will be referred to as "barrel-rings" in the following. This layout allows for assembly at the surface, with subsequent lowering of barrel-rings, endcaps and solenoid into the underground experimental area. It also allows pre-commissioning of the solenoid on the surface. This procedure was successfully employed for the CMS detector [2] and it will be adopted for the CLIC detectors.

The central barrel-ring will support the solenoid. The calorimeters and the tracking detectors are situated within the free bore volume of the solenoid. The main differences between CLIC_ILD and CLIC_SiD are in the value of the magnetic field, the inner diameter of the solenoid, the choice of the inner detector technology and a different $L^*$. Figures 11.1 and 11.2 summarise the main dimensions of both detectors.

The thickness of the return iron is defined by the requirements of the magnetic field quality and limitations on the allowable magnetic fringe field, as well as self-shielding for radiation due to any accidental beam loss. The thickness of iron in the endcaps is, in addition, constrained by the requirements of compactness along the beam line to accommodate both detectors on the IP and to provide vibration immunity of the QD0s by keeping their support tubes as short as possible. All of this is described in more detail in Section 11.2.2.

### 11.2.1 Overall Dimensions and Weights

The two detectors have an approximate weight of 11000 and 13000 tons, of which about 90% is in the iron yoke of the magnet. The outer shape of the detectors is a dodecagon with a diameter of 14 m in the flat portions of the outer surface and a total length of 12.8 m. The two detectors have different approaches for the tracking region: CLIC_ILD has a TPC tracker in a 4 Tesla solenoidal field whereas CLIC_SiD has a Silicon tracker in a 5 Tesla field. The large volume of the TPC explains the larger diameter and longer length of the CLIC_ILD coil. Table 11.1 summarises the main parameters of the two CLIC detector concepts.

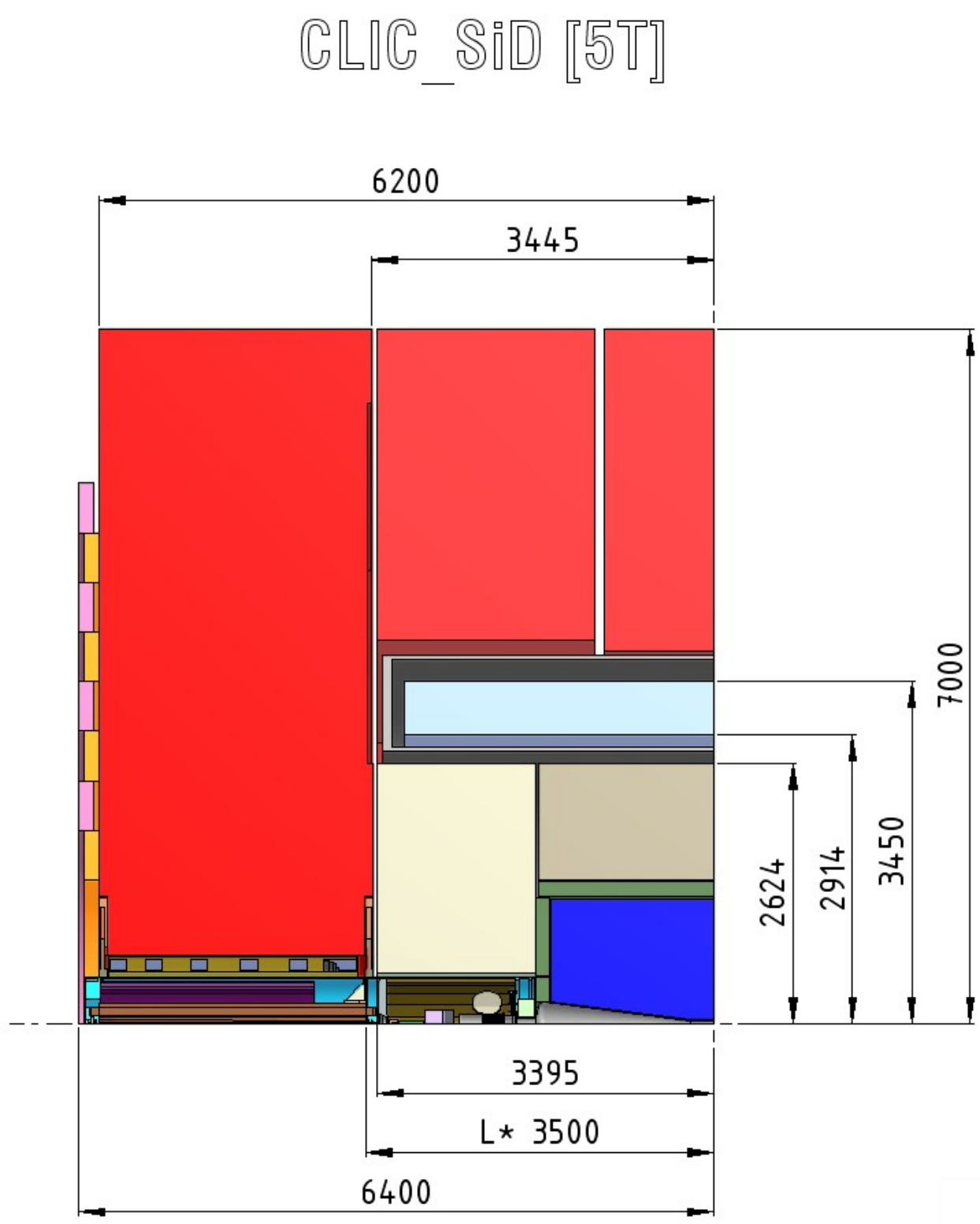

Fig. 11.1: Quarter View of CLIC_SiD

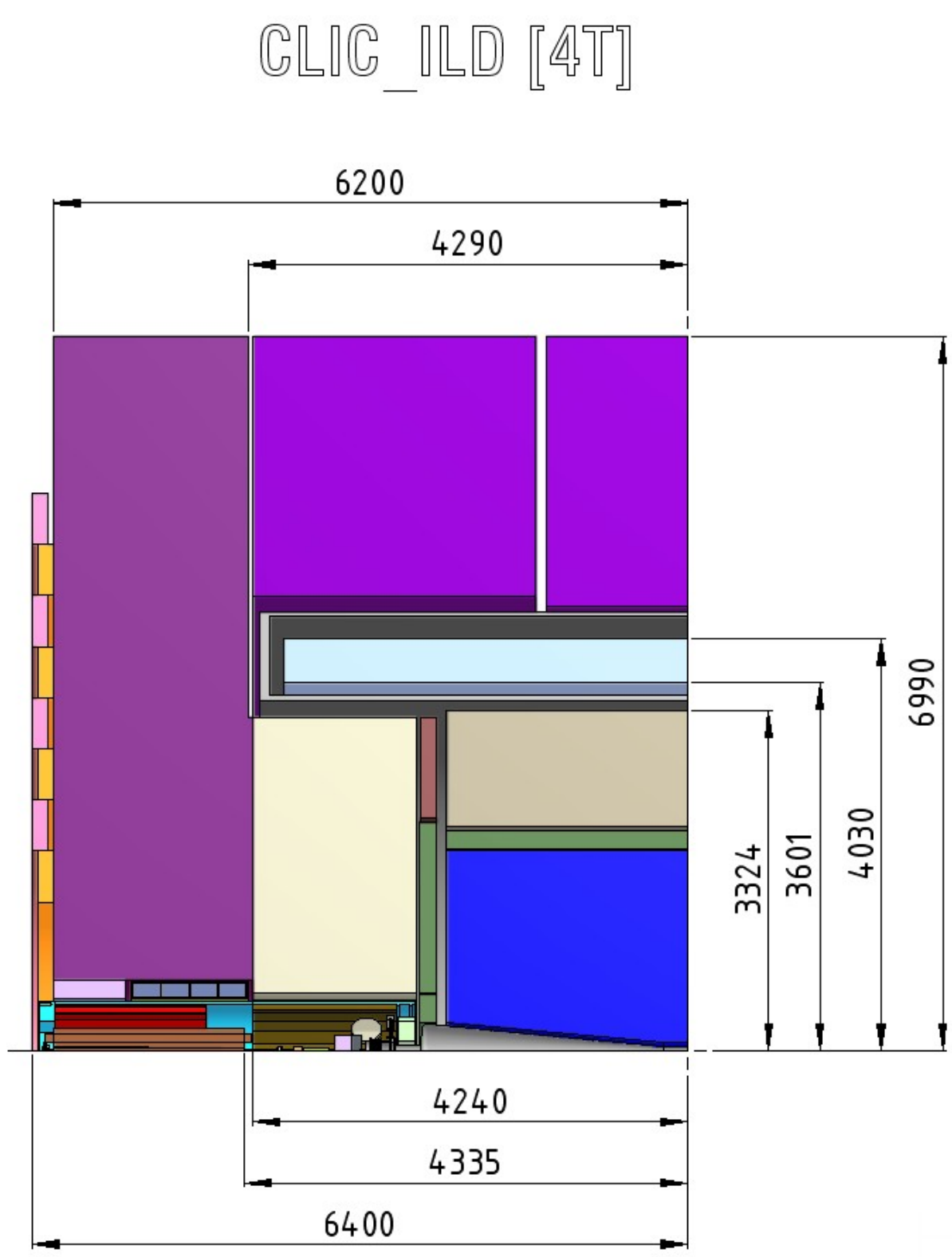

Fig. 11.2: Quarter View of CLIC_ILD

Table 11.1: Main dimensions and weights of both detectors

| Parameter | CLIC_SiD | CLIC_ILD with end-coils |
|---|---|---|
| Magnetic Yoke length | 12400 mm | 12400 mm |
| Detector overall length | 12800 mm | 12800 mm |
| Detector diameter | 14000 mm | 14000 mm |
| Free bore inside vacuum tank | 5488 mm | 6852 mm |
| Coil inner diameter | 5828 mm | 7202 mm |
| Coil outer diameter | 7008 mm | 7888 mm |
| Coil length | 6230 mm | 7890 mm |
| Coil weight | 201 tons | 173 tons |
| Vacuum Tank weight | 128 tons | 173 tons |
| Radial height vacuum tank | 1020 mm | 828 mm |
| Vacuum Tank length | 6690 mm | 8350 mm |
| **L$^*$** | **3500 mm** | **4340 mm** |
| Free bore in Endcap for support tube and anti-solenoid | 1380 mm | 1380 mm |
| Single Endcap weight | 2900 tons | 2100 tons |
| Barrel weight | 5000 tons | 4700 tons |
| Complete return yoke | 10800 tons | 8900 tons |
| **Total weight of detector** | **12500 tons** | **10800 tons** |

### 11.2.2 Magnets, Shielding and the Return Yoke

As already mentioned above CLIC_SiD has a 5 T central field and CLIC_ILD has a 4 T field. These numbers refer to the field value on the beam-axis at the IP. There are two additional requirements related to the magnetic field that impact the size of the iron return:

1. The field homogeneity within the tracking volume requires that the integral $\int(B_r/B_z)dz$ over the tracking volume, is less than 10 mm (see also Section 7.2).

2. The magnetic stray field outside the yoke at 15 m should be below 50 Gauss. This is discussed in more detail below.

Measurements of the CMS stray field and tests in the experimental cavern have shown that interventions e.g. for maintenance are increasingly difficult in fields exceeding 50 Gauss [3]. Therefore the return yoke must be designed to be magnetically self-shielding to assure that this value is not exceeded at x = 15 m from the beam axis. The distance between the axes of the two detectors along the push-pull direction is 28 m. The distance from the beam axis to the beginning of the cavern wall is 15 m. In push-pull operation, while the first detector is taking data on the beam, the second will be in the garage position, thus imposing shielding constraints for the protection of the working personnel against exposure to the magnetic fringe field and the radiation dose induced by accidental beam losses. The issue of magnetic self-shielding is also important when the off-beam detector performs magnetic tests in its cavern. This should not distort the field quality of the on-beam detector by more than 0.01% inside its tracking volume (ILC criteria).

The iron return yoke is important not only for field quality but also for resisting magnetic forces and for radiation protection purposes. It also serves as the mechanical backbone for the detector. It is a 12-sided structure (dodecagon) with the barrel yoke subdivided along the beam axis in three barrel rings, each 2.2 m long for CLIC_SiD and 2.7 m for CLIC_ILD.

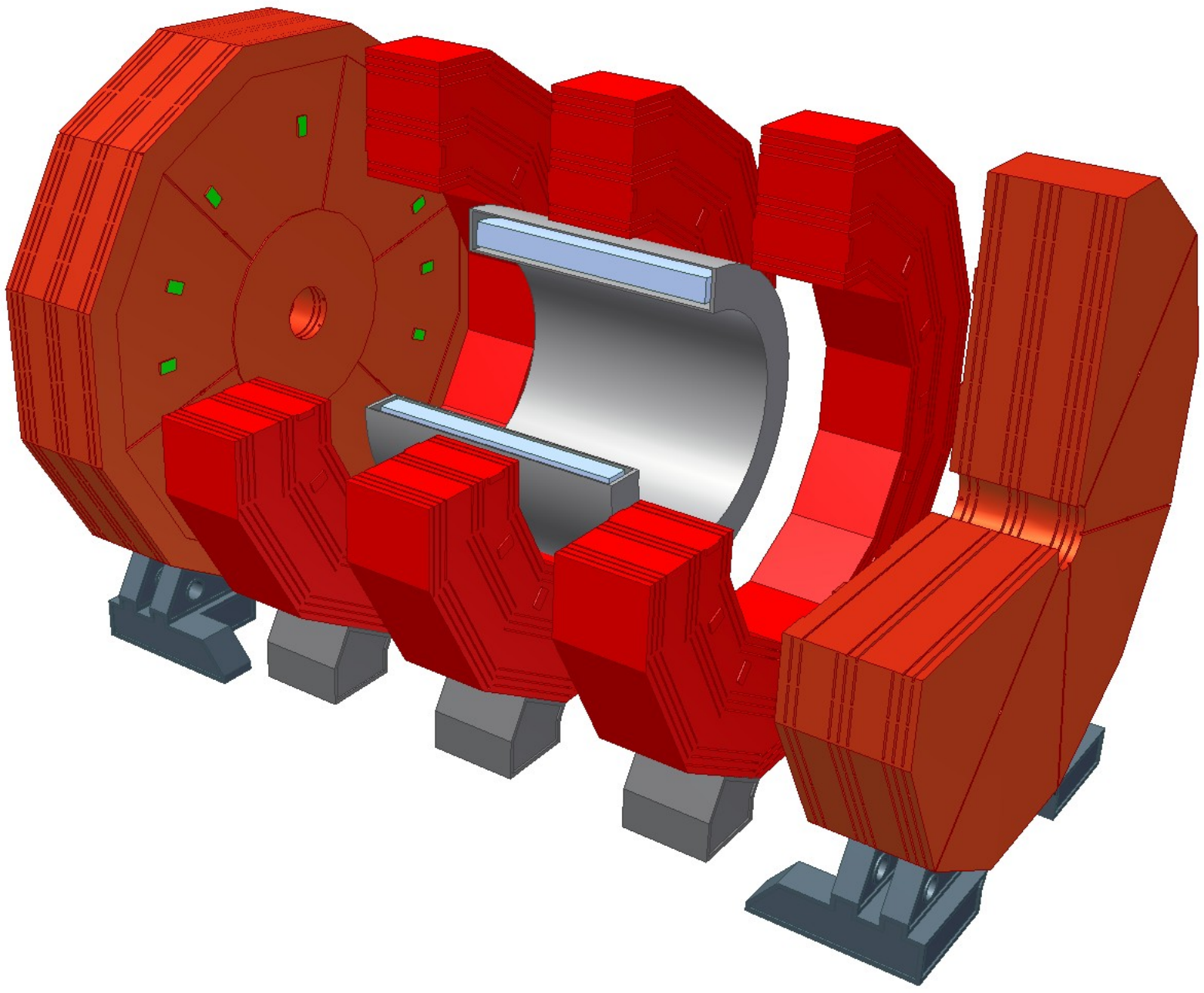

Fig. 11.3: Iron return yoke, solenoid CLIC_SiD

The central barrel ring supports the solenoid and the inner detectors. The two outer barrel rings can slide along the beam direction, to allow the insertion and maintenance of the muon detectors. A gap of 100 mm between the rings is used to route the services out of the detector. The gap between the barrel and endcap will be 50 mm as there are fewer cables in this region. Figure 11.3 shows the return yoke with the central barrel supporting the solenoid, outer barrel and endcaps. The endcap has the forward electromagnetic and hadronic calorimeters attached to its front (not shown in Figure 11.3). Integrated in the actual design are nine slots for the muon detectors.

The endcap needs to resist about 18000 tons of attractive forces on its poles [4], when the solenoid is on. These reaction forces are transmitted by non-magnetic plates situated in the gaps between the barrel rings and between barrel and endcap. The deformation under 18000 tons of force was calculated by FEA and is of the order of 3.5 mm. Figure 11.4 shows the ANSYS [5] result. The requirements for many ancillary support systems, e.g. the moving system, hydraulic forces, tooling, assembly, push-pull motors and the platform are determined by the yoke's dimensions and mass.

The concern of maximum exposure to radiation allowed for the personnel working in the cavern during the beam operations comes from potential beam losses. The iron yoke will provide enough shielding for beam losses inside the detector but the most likely location where losses may happen is in the region of the final focusing magnets, at the interface between the endcap and the cavern wall. Sufficient shielding is provided by concentrically arranged shielding rings on the backside of the endcap iron. The shielding consists of rings located in fixed positions on the endcaps and rings which are movable by pneumatic or hydraulic jacks, and which are pressed against the wall. Those rings will move about 10 cm from their position thus creating a chicane that closes the gap between the endcap and the tunnel wall. Figure 11.5 illustrates this detail with the movable shielding in its "on" and "off" position.

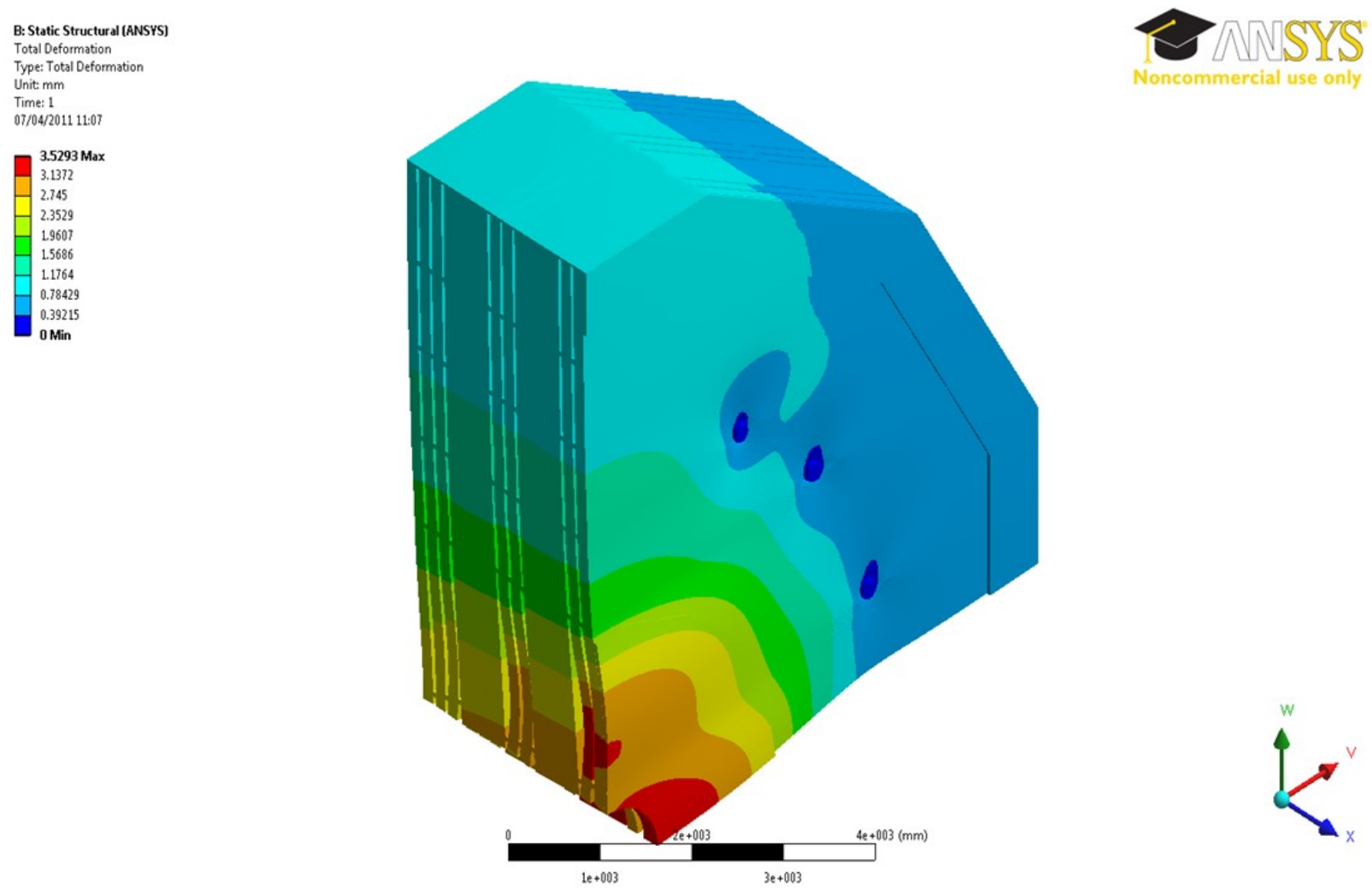

Fig. 11.4: Deformation of the endcap due to magnetic forces at 5 Tesla

To be fully magnetically self-shielding would require for CLIC_ILD an extra length of 660 mm for the iron in each endcap; this is avoided by adding end-coils which allow to keep the two detectors at the same length. To address this CLIC_ILD will be equipped with end-coils [6, 7] such that it will provide the same level of fringe field matching as CLIC_SiD and still fit at the IP. The end-coils are placed on the rear face of the endcap of CLIC_ILD. The conceptual design of these end-coils including field calculations can be found in [6]. Only this approach allows short lever-arms between QD0 supports in the tunnel and the experiments. First results of calculations done with POISSON with full iron and then the reduced version applying endcap ring coils show this is a feasible solution [6]. For CLIC_ILD the end-coils are, in fact, the fixed shielding rings (depicted in Figure 11.6).

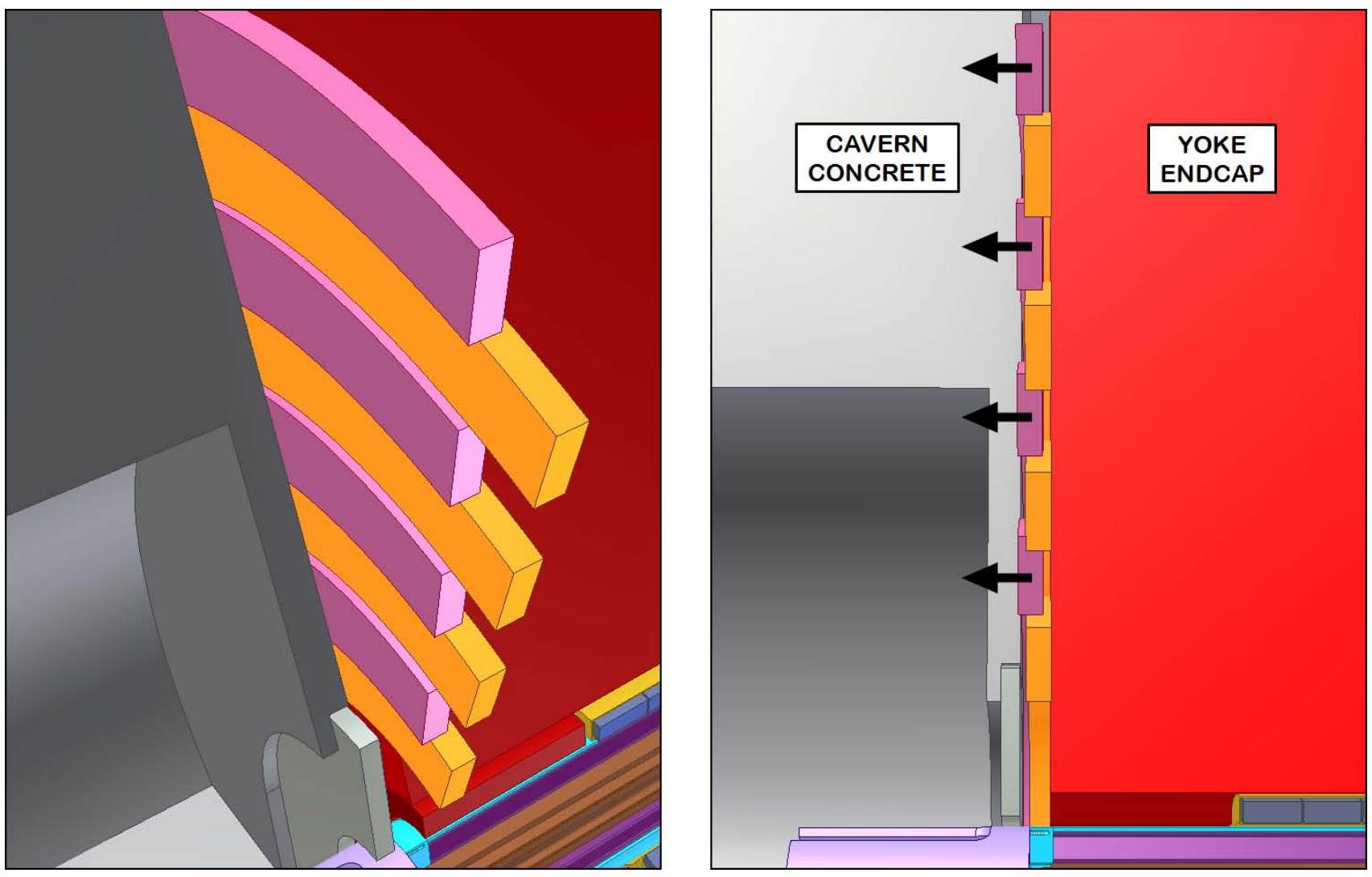

Fig. 11.5: Radiation chicane made of concentric ring modules

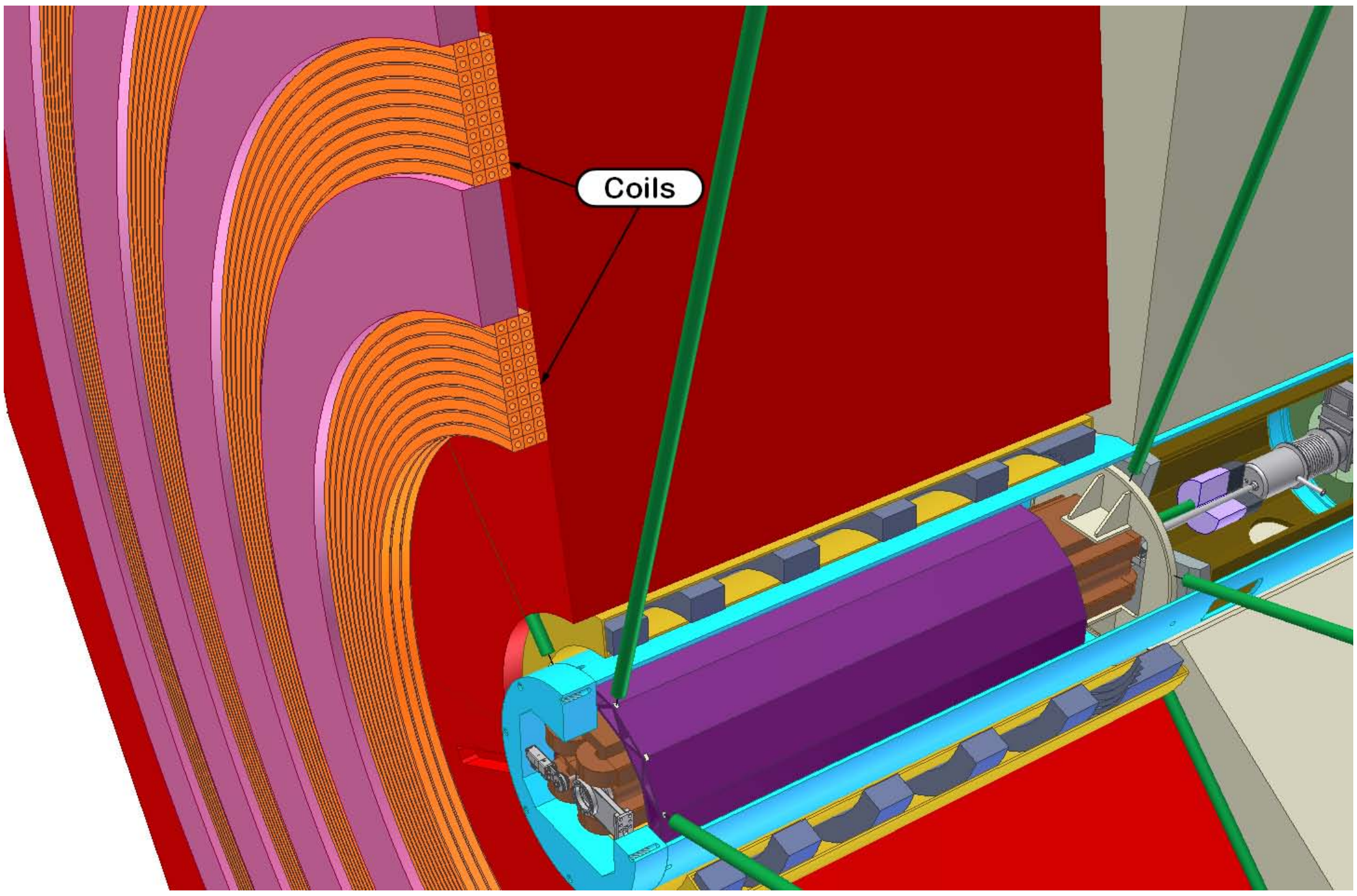

Fig. 11.6: Radiation chicane in the case of CLIC_ILD

The technology chosen for the QD0 magnets is a combination of electro-magnetic coils, "permendur" pole tips and permanent magnets. Permanent magnets are chosen to avoid cooling fluid with turbulent flows which could jeopardise the stabilisation efforts. The electro-magnetic coils will work with low current density thus not needing water cooling. A draw-back of this solution is the need of an anti-solenoid to shield the permanent magnets from the main solenoidal field. At the same time the anti-solenoid is crucial to keep luminosity losses due to the beam crossing angle of 10 mrad within reasonable limits. A general layout of the forward region, showing these details, is in Figure 11.13.

### 11.2.3 Services Integration

All the services (power, signal, cooling and gas) will be routed to/from the subdetectors, through the gap between the barrel rings and the barrel-endcap , to racks and manifolds placed on balconies at the external periphery of the iron yoke, where they will be regrouped inside flexible cable chains. Short cable runs on the top of the detector, between the barrel rings, will allow a fast and reliable opening of rings for maintenance without disconnecting the services. At the bottom of the detector three large cable chains, one for the barrel and one for each endcaps, will connect the detector to the remote counting rooms. The cable chains will be run underneath the platforms to allow for easier access of personnel around the detector. This concept has been already validated in the CMS experiment.

Magnet services, the He liquefier, the power supply and the vacuum pumps are placed at dedicated locations at the extremity of the cavern in order not to be affected by any stray-field and not to introduce vibrations to the detector. The solenoid will be serviced by two chimneys, one for liquid Helium and one for current leads, located respectively at the two opposite sides of the central barrel ring. The external barrel rings will have notches to accommodate the passage of the chimneys. A large buffer dewar will be placed on a scaffolding tower moving with the detector and a flexible cryo-transfer line will run to the cold box in the cavern. The solenoid will remain permanently connected to the power supply in the cavern via a HTS line as developed presently for the LHC upgrade (see Section 7.6).

## 11.3 Push-Pull Operation

Both detectors are installed on independent platforms made of reinforced concrete, with a size of approximately 13 m × 16.5 m × 2.5 m, corresponding to the footprint of the detectors. The design will be similar to the plug of the PX56 shaft at CMS and weigh about 1500 tonnes. Such a plug has been successfully operated, supporting statically up to 2500 tonnes and having a span of 20 m between the rails. For the present application the gross weight of the detector plus platform will be ≈14000 tonnes and the free span between the supports will be much smaller, in the order of 4 m. Figures 11.7 and 11.8 show the dense steel reinforcement of the CMS plug and the completed plug after on site during civil construction. At rest, the platforms will be in contact with the floor through a set of anti-seismic supports that will redistribute the total load to ground. First FEA calculations confirm the thickness of about 2.5 m and that the local stress is well below the admissible values.

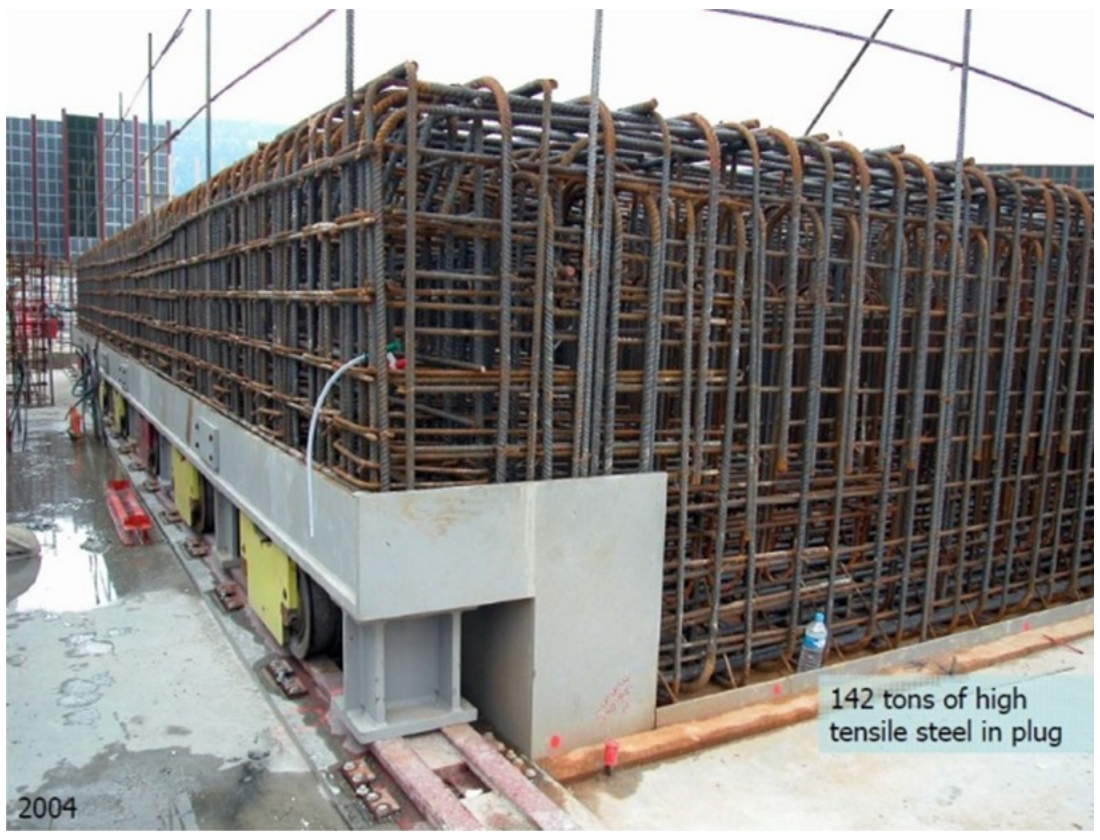

Fig. 11.7: Steelwork of the CMS plug

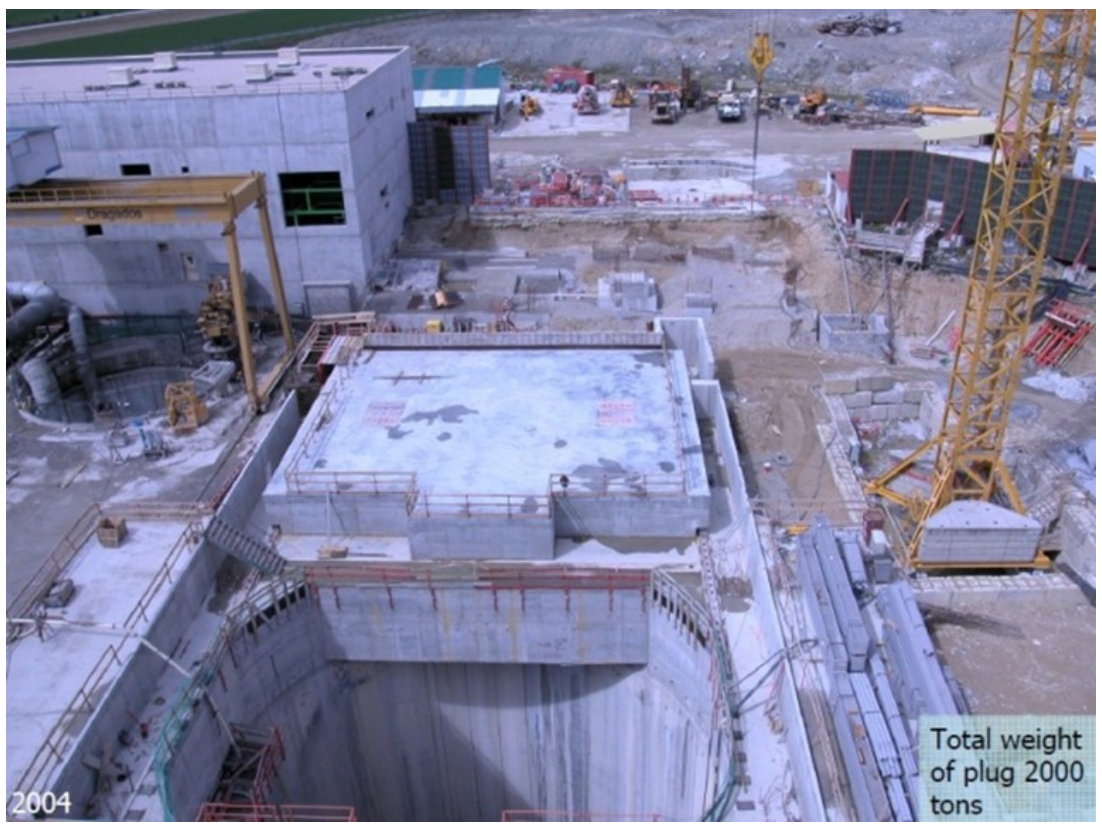

Fig. 11.8: The CMS plug after concreting

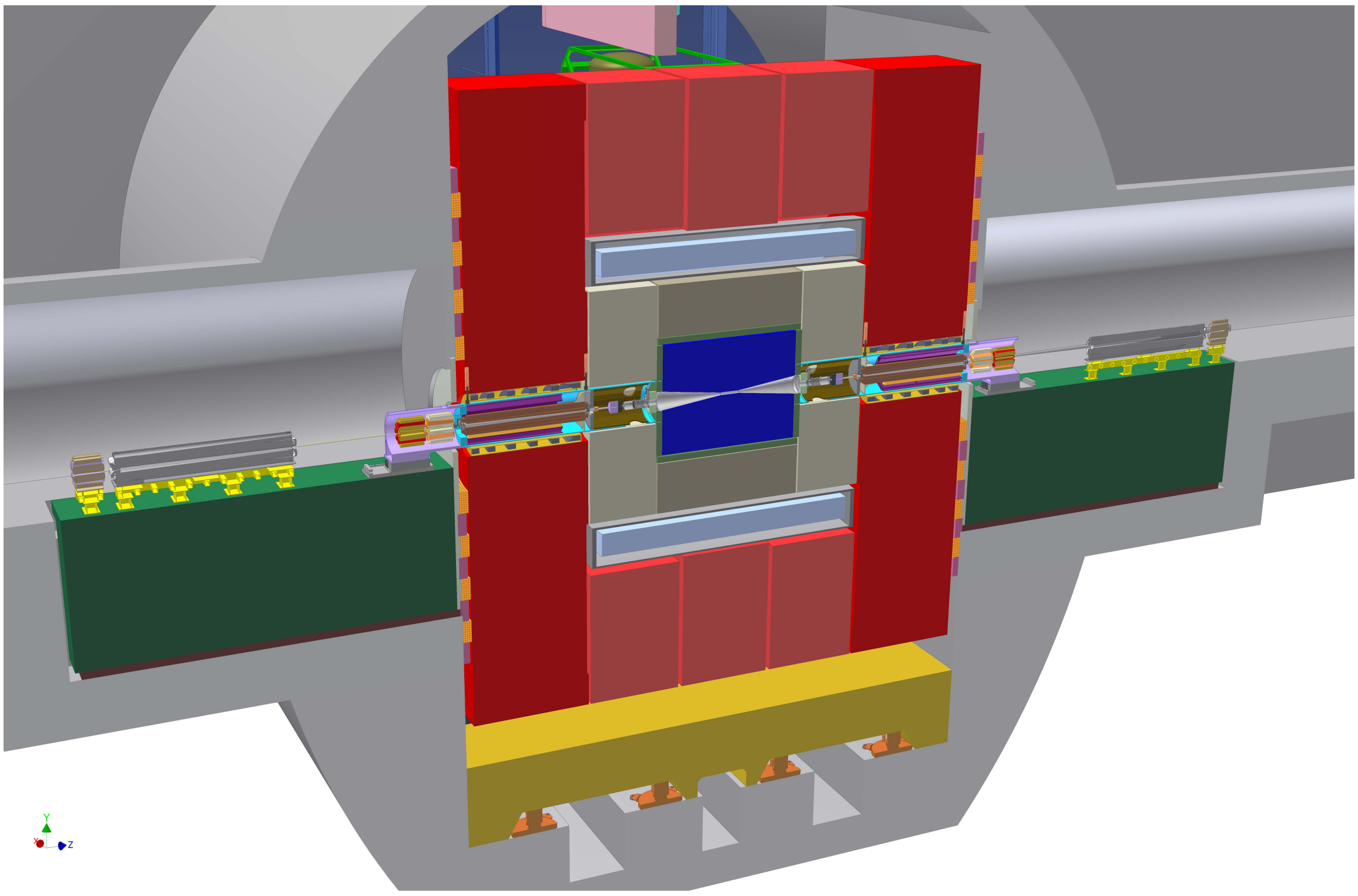

Fig. 11.9: Vertical cut through the experiment

The moving system will be designed to move a total mass of 14000 tonnes and use either air-pads or heavy-duty rollers. The friction factor including stick-slip will be 1.5% and 5% respectively. Most likely, there will be two to four hydraulic jacks situated on the left and right side underneath the platform, with a capacity of a hundred tonnes.These devices are commercially available and can be integrated in the overall design of the cavern without difficulties. A guiding rail system with positive indexing capability at the interaction point will be employed to achieve the required alignment precision on the beam of $\pm 1$ mm, between consecutive push-pulls. Given the typical performances of industrial moving systems, the estimated time, to travel $\approx$30 m from garage position to the interaction point, will be less than one hour. This excludes preparation time for last minute un-cabling of some racks, pumping the air pads, putting the platform into equilibrium, safety checks, etc. The floor along the trajectory of the platforms will contain deep trenches to host the cable chains and provide access for the maintenance of the air pads or the heavy-duty rollers. Figure 11.9 shows the experiment on a platform, the cable trays underneath and the connections to the accelerator.

## 11.4 Underground Experimental Area

The overall conceptual layout of the underground cavern is driven by the needs to accommodate two detectors and operate them in "push-pull" mode. An overall three dimensional view of the underground experimental area is shown in Figures 11.10.

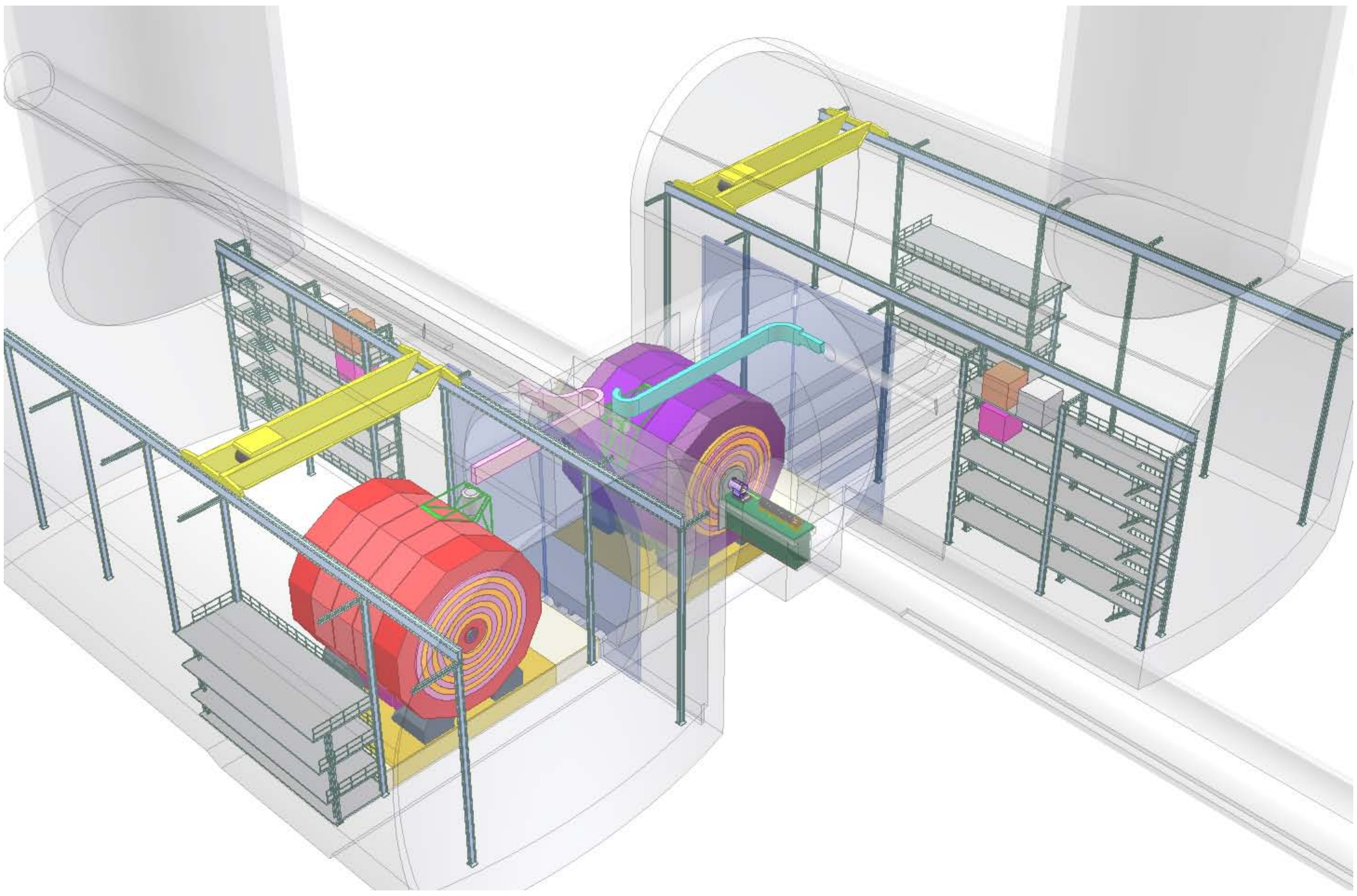

Fig. 11.10: General View of the interaction region at CLIC

The underground interaction region has to satisfy many requirements like minimising the excavating volume, the costs, services integration, personnel accesses, ventilation, survey galleries and general safety. At the stage of the CDR we consider an assembly of the detector at the surface with lowering of large units [2] into the underground area. This technique has been successfully used by the CMS experiment at LHC. Given this, only cranes with a capacity of the order of 40 tonnes are foreseen in the underground area. Each experimental cavern has its own access shaft. For the time being, this access shaft is situated at the extremity of the cavern outside the region covered by the opened experiment. The largest detector pieces require a diameter of 14 m for the shaft. Adding approximately 1.5 m on each side for additional structures, an 18 m diameter shaft seems reasonable. Because the elements to be lowered are much larger in one dimension than in the other one, the lift, ventilation ducts and the emergency staircase can be put inside the same shaft. Figure 11.11 and 11.12 depict the main dimensions.

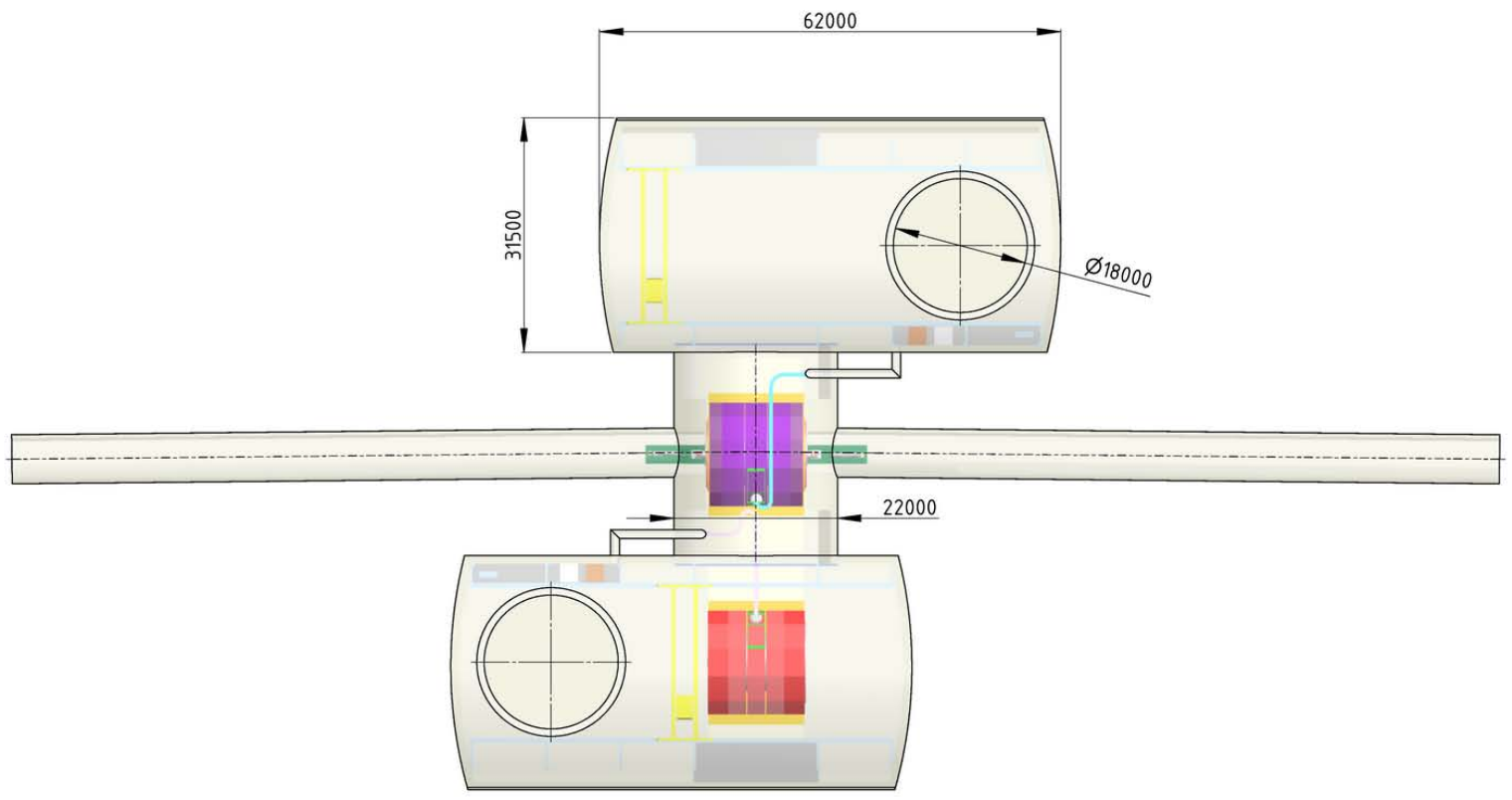

Fig. 11.11: Top view with dimensions.

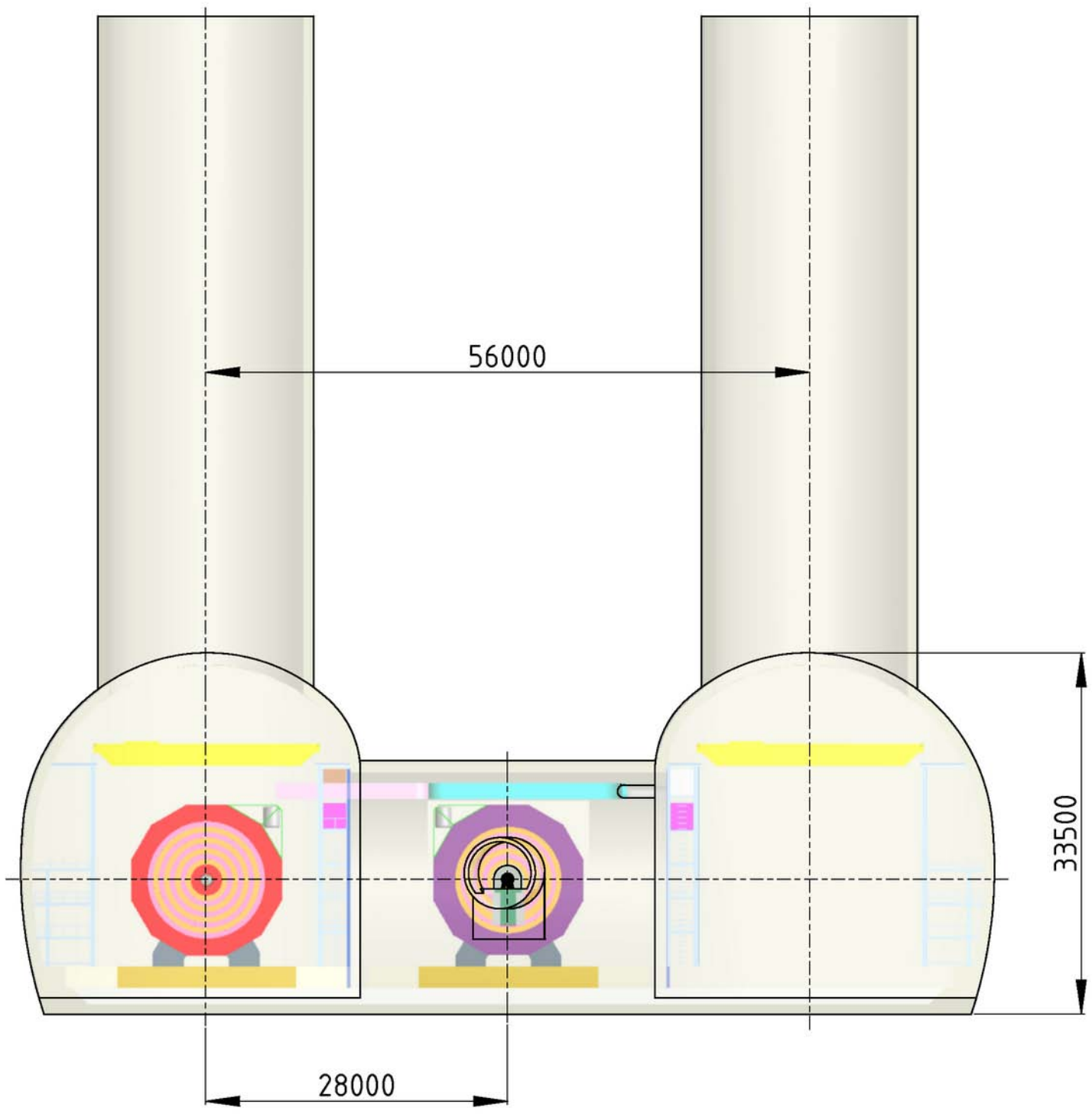

Fig. 11.12: Side view with dimensions.

The experimental caverns are shaped such that enough rack space is available near the experiment and the transition region to the transfer tunnel has vertical walls [8] to accommodate sliding doors. These doors could be just thin doors separating the assembly and maintenance area from the beam area for ventilation and safety purposes. If however, at the time of operation, safety requirements demand more than just a “self-shielding” detector, these doors may also be re-enforced (up to 2 m) and serve as an additional shielding wall. This may be designed for both radiation and magnetic purposes. In the beam direction the transfer tunnel, connecting the two detector caverns, is kept to the strict minimum to keep volume, distances and lever arms as small as possible.

## 11.5 Forward Region

For proper accelerator operation and, in particular, for luminosity optimisation, the location of the final focusing quadrupole QD0 of the CLIC accelerator needs to be close to the IP. This so-called $L^*$ value has been chosen to be 3.5 m for CLIC_SiD and 4.3 m for CLIC_ILD, implying a position inside the detector volume. In addition, the very small vertical beam spot with an r.m.s of 1 nm requires a stabilisation of the QD0 position with an r.m.s of 0.15 nm above 4 Hz.

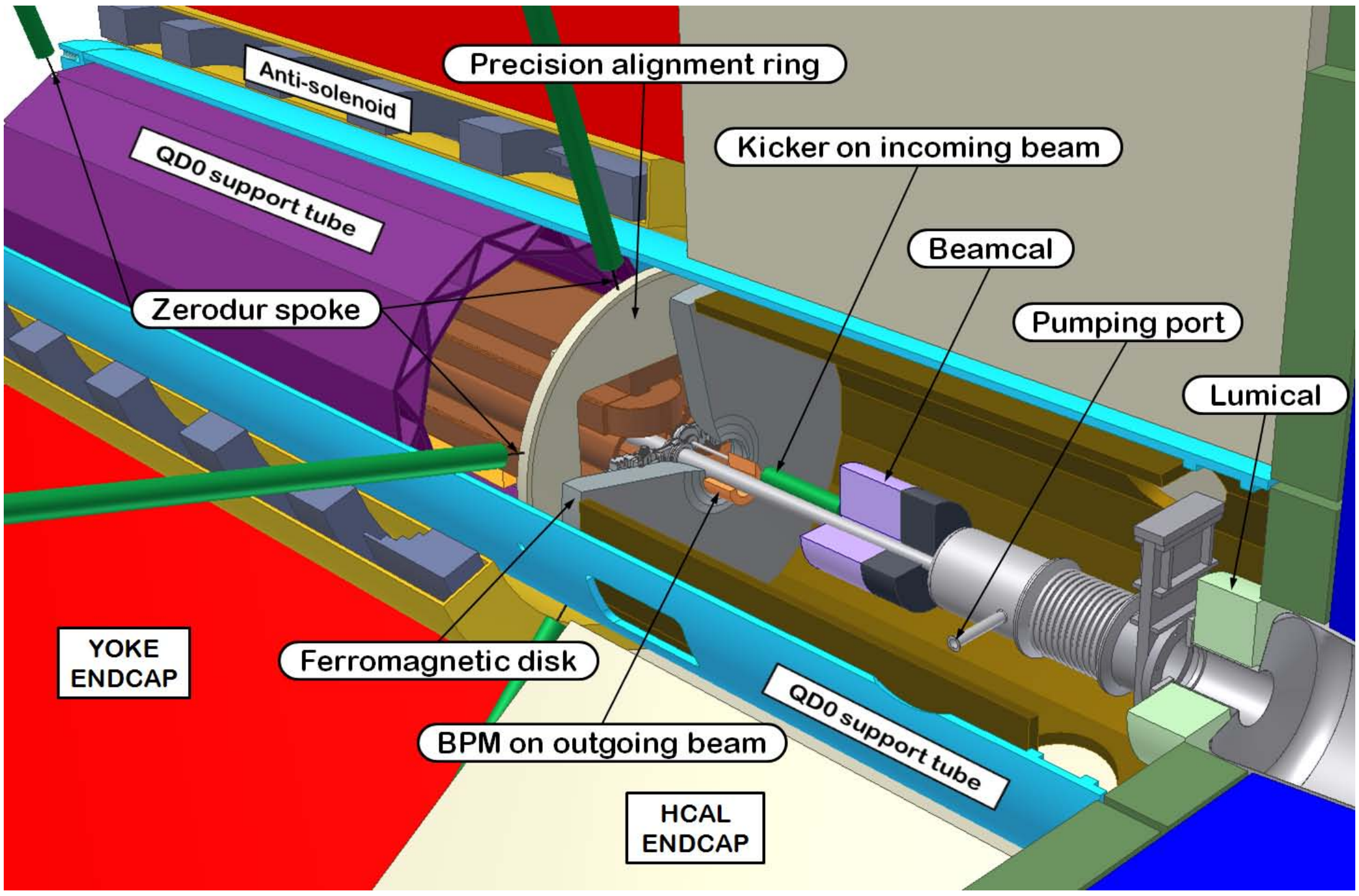

Fig. 11.13: Isometric view of inner and outer support tube and the equipment installed in this region.

To meet these stringent stability requirements the following guiding principles for the layout have been adopted [9]:

(a) A strategy of non-opening on the IP allows a compact detector and transfer tunnel design without bulky shielding between the experiment and the tunnel;

(b) Implementing a moving ring-chicane shielding at the interface between the experiment and the accelerator tunnel-wall (see Section 11.2.2);

(c) Supporting the QD0 from a pre-isolator in the tunnel;

(d) Choosing a hybrid QD0 technology with permanent magnets needing little or no cooling;

(e) Designing a two-in-one support tube for QD0 and forward calorimetry with fine-tuned eigenfrequencies, allowing to support the coils independently of the yoke;

(f) Using an active stabilisation system underneath the QD0 in combination with a robust active pre-alignment system crossing the detector.

This section describes the forward region of the detector and how the above requirements and constraints have been implemented in the design. This region of the detectors is essentially the extension of the accelerator into the detector area. The forward region, illustrated in Figure 11.13, includes several important components with quite different functionalities. Two independent support tubes with distinct functions, stiffness and eigenfrequencies, do provide the mechanical support for the forward region. Also shown are elements belonging to the alignment system and the anti-solenoid (ferromagnetic disk). All this is described in more detail below.

### 11.5.1 Forward Region Layout

The forward region is the heart of the machine-detector-interface. The design drivers are QD0, BPM and kicker as accelerator equipment and LumiCal, BeamCal and alignment structures from the detector side. In addition there are vacuum valves and pumping ports as well as tungsten shielding against backscattering from the IP and the ferromagnetic disk, to reduce the external field close to zero in the QD0 region.

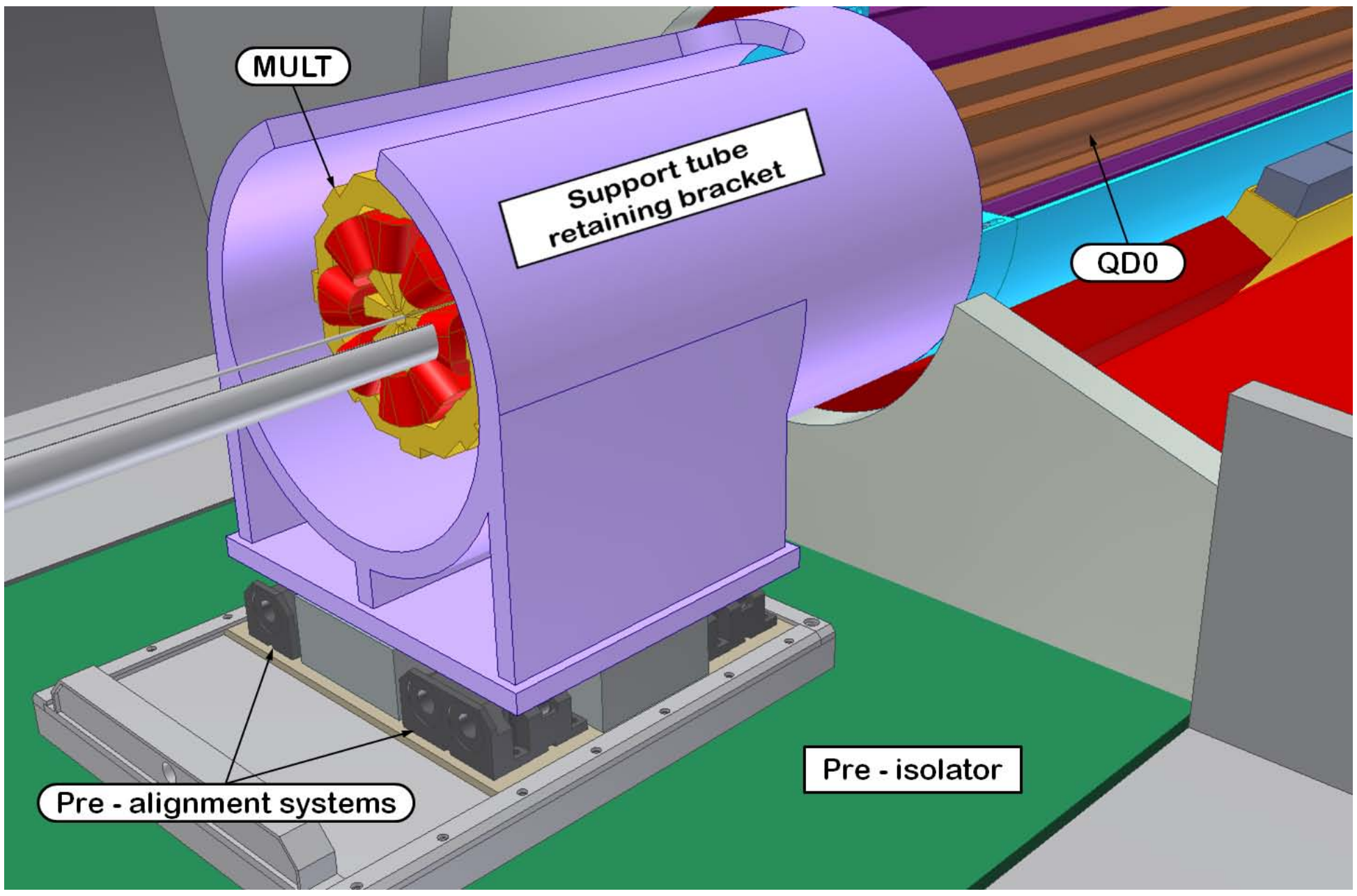

Fig. 11.14: Retaining bracket and pre-alignment underneath.

During beam operation this entire assembly is cantilevered from a massive retaining bracket mounted on the pre-isolator in the tunnel [6, 9, 10]. This element has a stiff flange that allows a bolted connection from the QD0 support tube to the support tube retaining bracket, connected to the pre-alignment system. This whole system sits on a pre-isolator and is illustrated in Figure 11.14.

Figure 11.15 depicts more of the details in the front part of the QD0 support tube and shows the access holes for bellows and valves. Additional integration problems arise due to the 20 mrad crossing angle, but QD0 is aligned with respect to the incoming beam.

The push-pull procedure requires breaking the vacuum for each operation. Therefore valves, creating separate vacuum sectors, are installed on the beam pipe, between QD0 and BeamCal as well as between BeamCal and LumiCal, for quick, safe and reliable vacuum operations.

### 11.5.2 Alignment

A pre-alignment to a precision of 10 µm of QD0 to the beam is required after each reconnection of the detector following a push-pull operation.

To monitor the relative position of the two QD0s the following approach is adopted: A very precise reference ring is mounted at the extremities of the quadrupoles along the beam axis and their relative position is measured on a 3-D measuring machine before installation. By means of 6 mm thick radial spokes made of ZERODUR, housed in 60 mm channels in the endcap and passing through the outer support tube, the position is transferred to a 6-folded Rasnik alignment system [11] traversing the whole experiment, including endcaps (see Figure 11.13). The information obtained by this system is then used to actively intervene with the pre-alignment mechanics installed under the support tube retaining bracket (see Figure 11.14) in the tunnel. This function is provided by a 5 degrees-of-freedom mechanical system based on excentrical cams and stepper motors, with 8 µm precision in all directions.

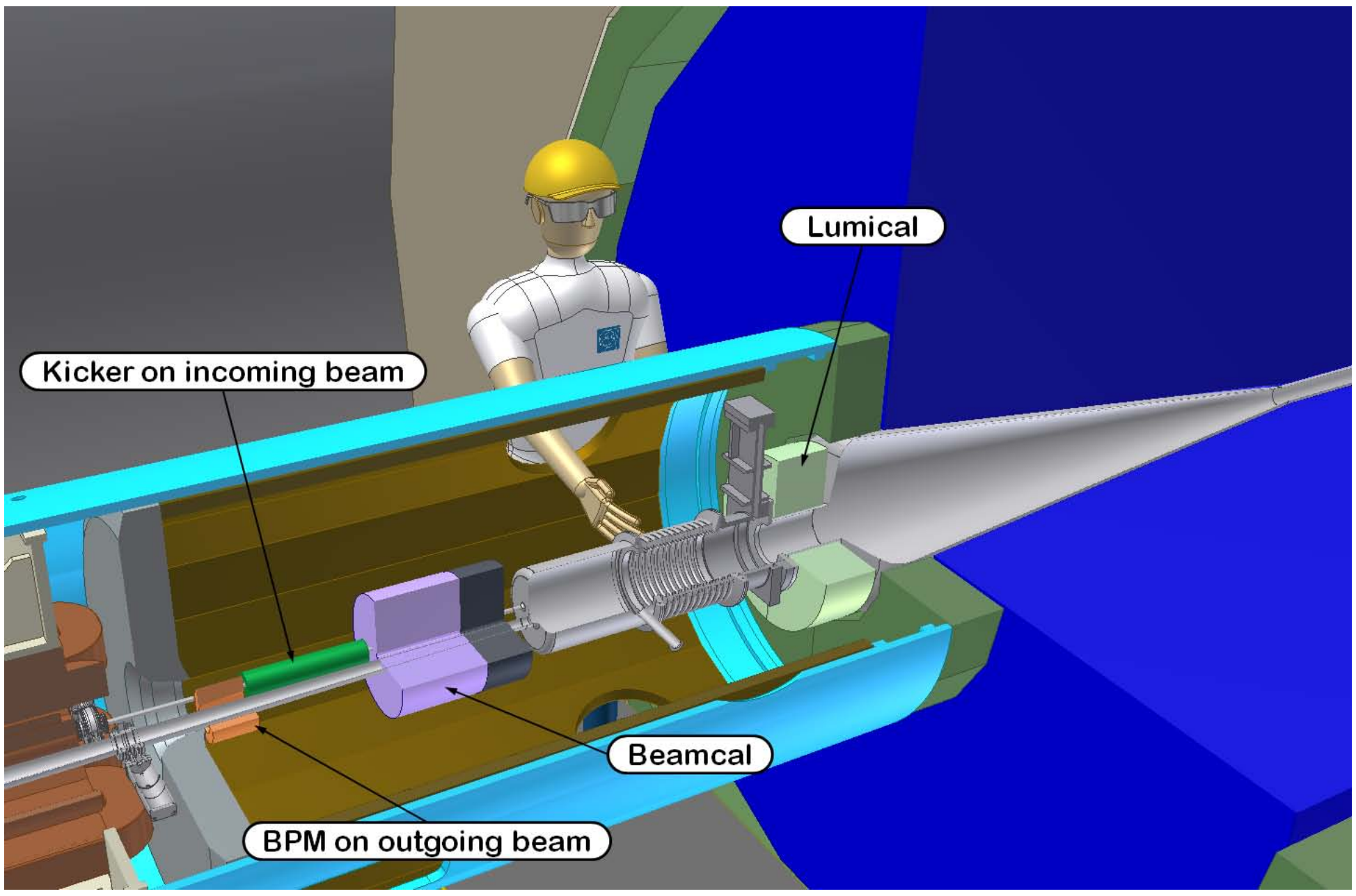

Fig. 11.15: Support tube with QD0, BPM and kicker, vacuum valve and BeamCal.

### 11.5.3 QD0 Stabilisation Requirements

The stabilisation requirements for the QD0s at CLIC are very challenging. To avoid luminosity losses the vertical position of the quadrupole must be stabilised to 0.15 nm r.m.s. for frequencies above 4 Hz. The approach is based on warm QD0s (i.e., not superconducting) supported by the innermost of two concentric tubes of different stiffness and eigen-frequencies. The eigenfrequency of the inner support tube for QD0 is tuned to be in phase with the bunch train frequency of 50 Hz. Located in the supports underneath QD0 is an active stabilisation system based on piezo-actuators and capacitive gauges combined with V-shaped concentric elastomeric strips for guidance. This is illustrated in Figure 11.16. To suppress the high frequency part of the perturbation from the ground motion beyond 4 Hz, a pre-isolation system with high mass and low stiffness springs is proposed as the baseline for the supports of QD0. It is a completely passive device and will act as a low-pass filter. This approach is widely used in nanotechnology labs as a first defence against vibration. This system is under development and its parameters are being optimised, but first calculations show that a cut-off eigenfrequency of 1 Hz can be achieved with a mass of 50-80 tonnes.

## 11.6 Detector Opening and Maintenance

The key to success for a fast and reliable push-pull is a sequence of operations with minimum interruption of sensitive systems of the forward region like vacuum and shielding. A preliminary procedure is described here, assuming that the detector is initially located on the IP. The procedure can be inverted when moving from the garage position to the IP position.

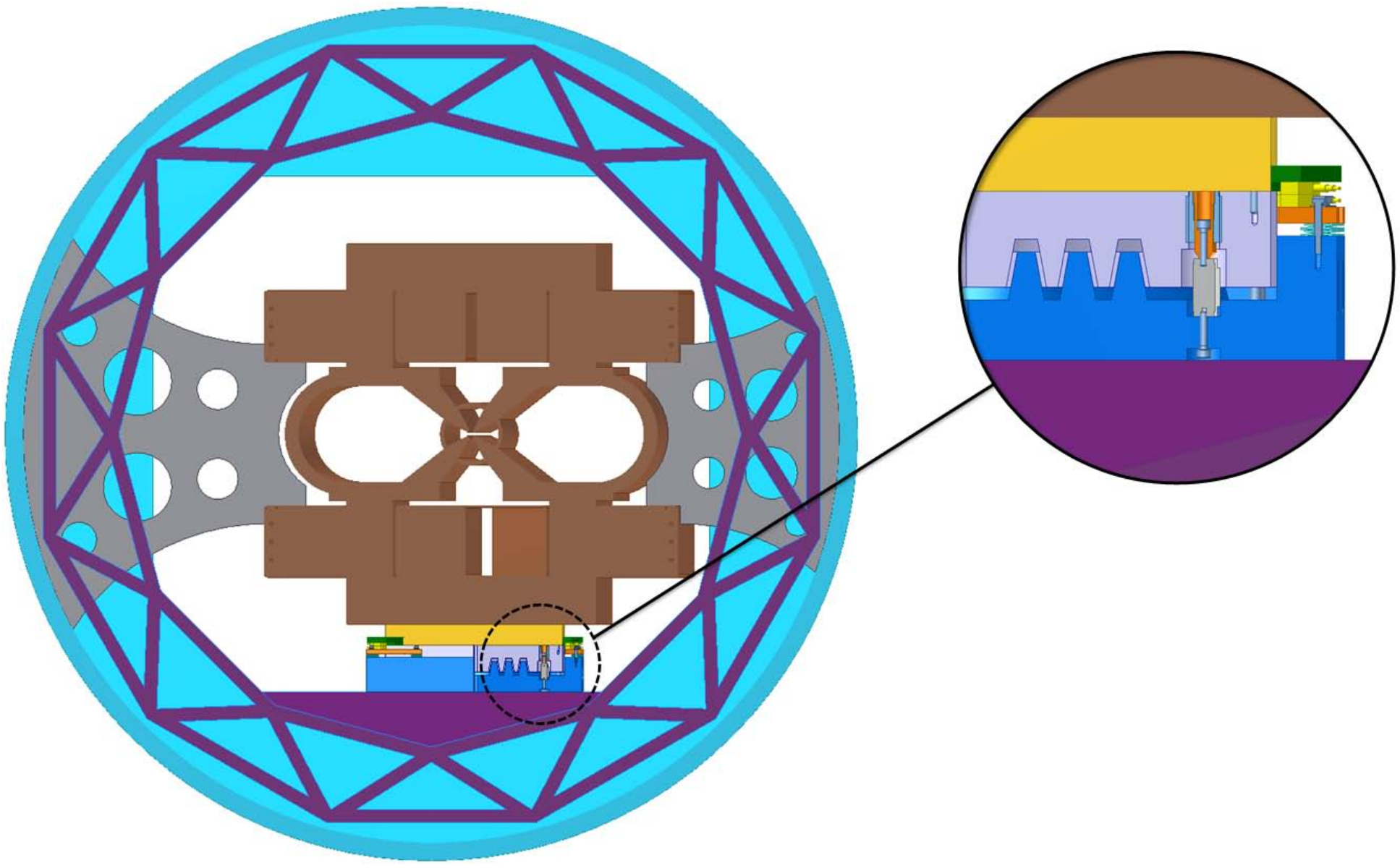

Fig. 11.16: QD0 support detail, copper coils are supported by outer tube.The active stabilisation system is shown in the insert.

1. The shielding rings on the back-face of the endcap retract, creating space ($\approx$ 5-10 cm) between the endcap and the transfer tunnel wall.
2. Close all vacuum valves. This creates three sectors inside the detector: section 1 corresponds to QD0, section 3 belongs to inner detectors and section 2 is in between (see Figures 11.17 and 11.18). It is worth noting that the vacuum quality in this region is not critical and that bake-out of the vacuum system is, therefore, not required.
3. The load of the QD0 support tube is transferred to the endcap bore surface, by the activation of small jacks inside the endcap bore.
4. With the jacks now holding the weight of the support tube, bolts connecting the retaining bracket in the tunnel to the support tube can be unscrewed.
5. The retaining bracket itself is then slid backwards on grease pads (Figure 11.19 giving access for the removal of the bellows. Now the experiment is disconnected from the accelerator and ready to move to the cavern.
6. With the detector in the garage position the QD0 installation/extraction tool is installed behind the endcap. This tool allows cantilevering the support tube from behind, releasing the jacks in the bore and opening the endcap (see Figure 11.20).
7. The endcap is slid back giving access to the other main vacuum connection between support tube and the inner detector.
8. LumiCal and ECAL plug are opened sidewards and the whole QD0 support tube can be retracted some 20 cm by the extraction tool as a first step and then be removed as a whole by a crane. Figures 11.20, 11.21, 11.22, 11.23 illustrate this scenario.

It must be emphasised that this sequence of operations is intended to demonstrate the feasibility of push-pull operation between CLIC_SiD and CLIC_ILD. A number of engineering issues still remains to be investigated, partly in the interplay between accelerator and detector. For example, in the technical design phase of CLIC, a robust pre-alignment scheme of the complete BDS, and in particular the region of the QD0, SD0 and MULT magnetic elements, must be studied.

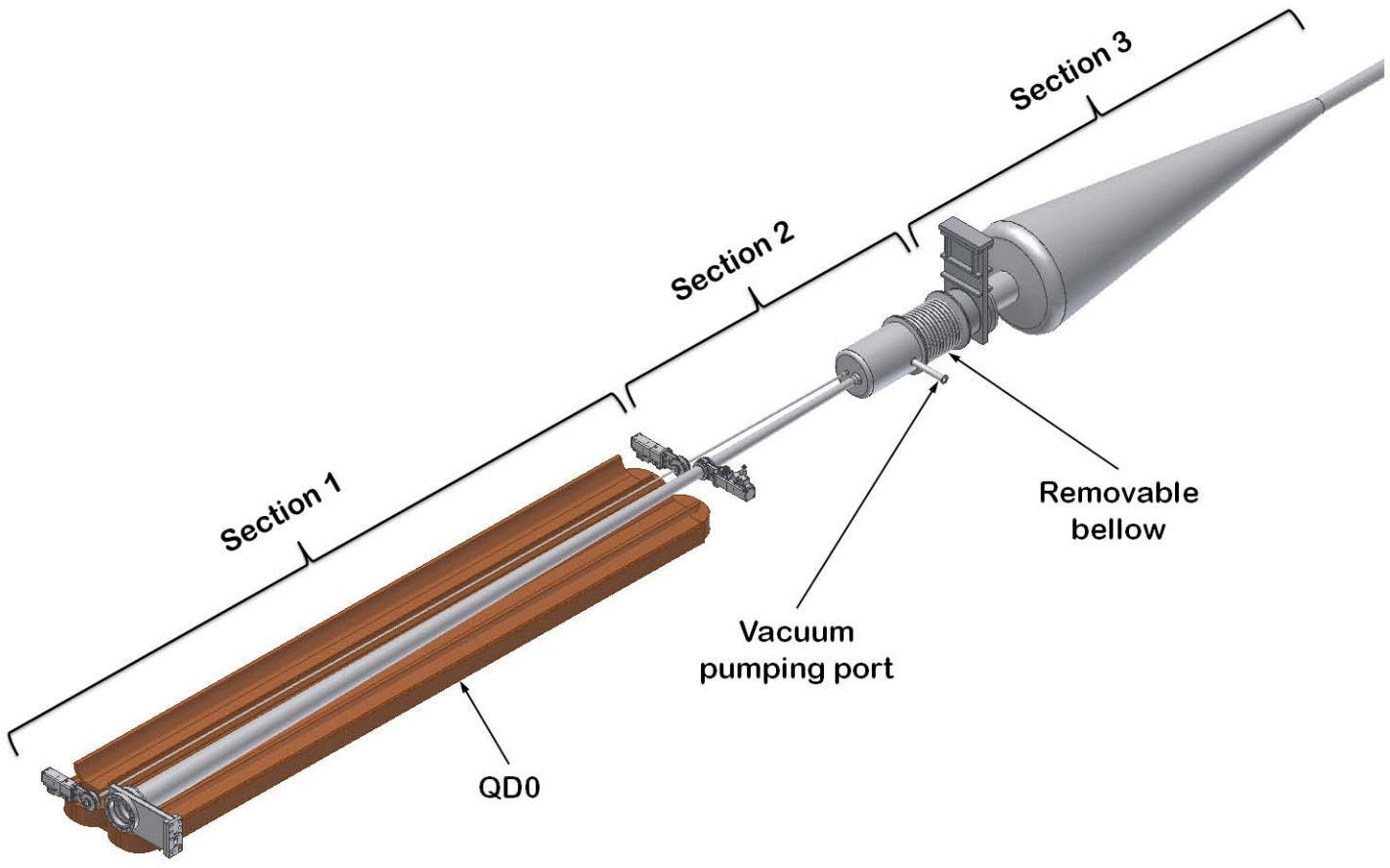

Fig. 11.17: The sectors of the vacuum system near the IP.

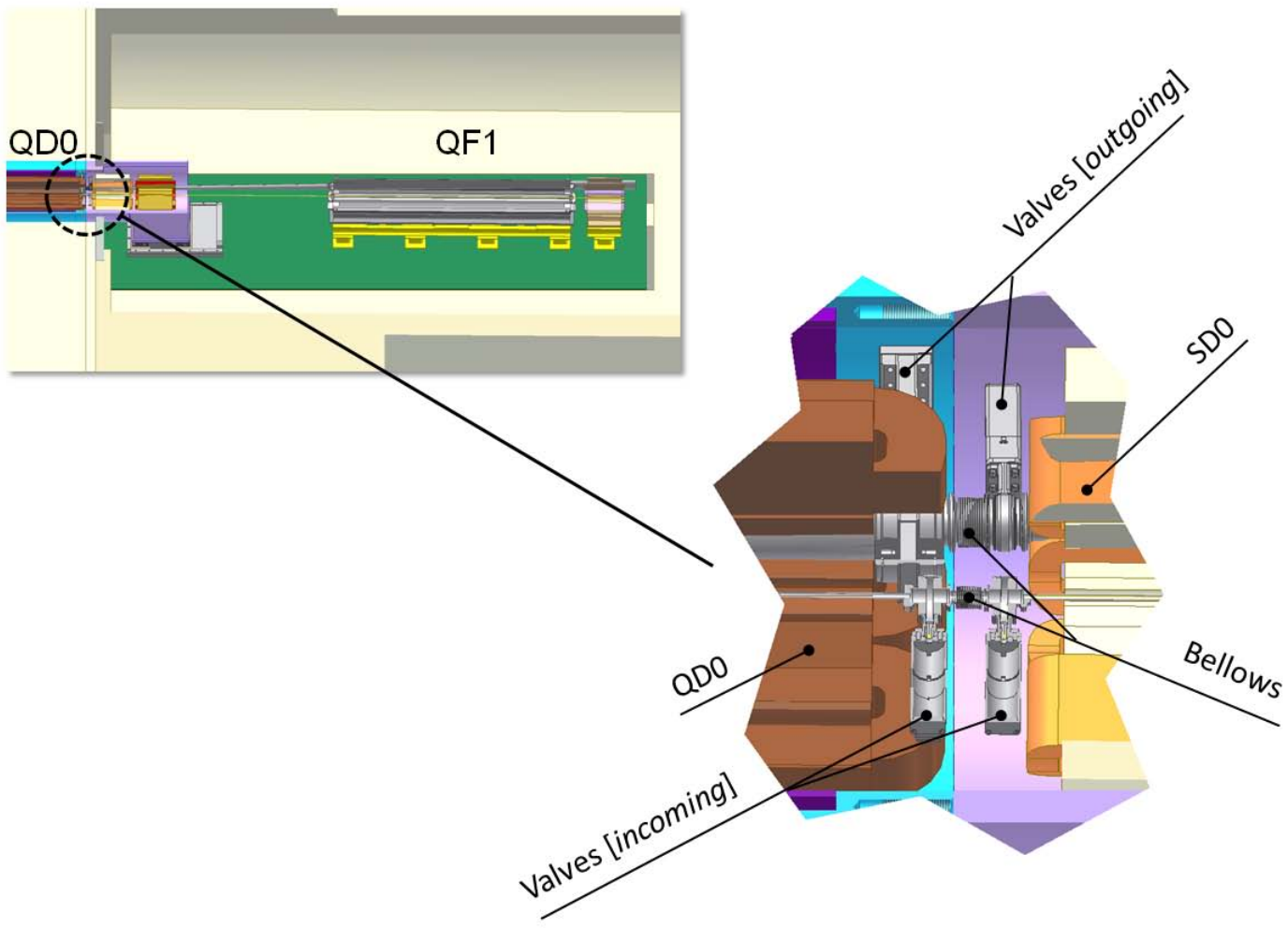

Fig. 11.18: Conditions of vacuum valves and disconnection.

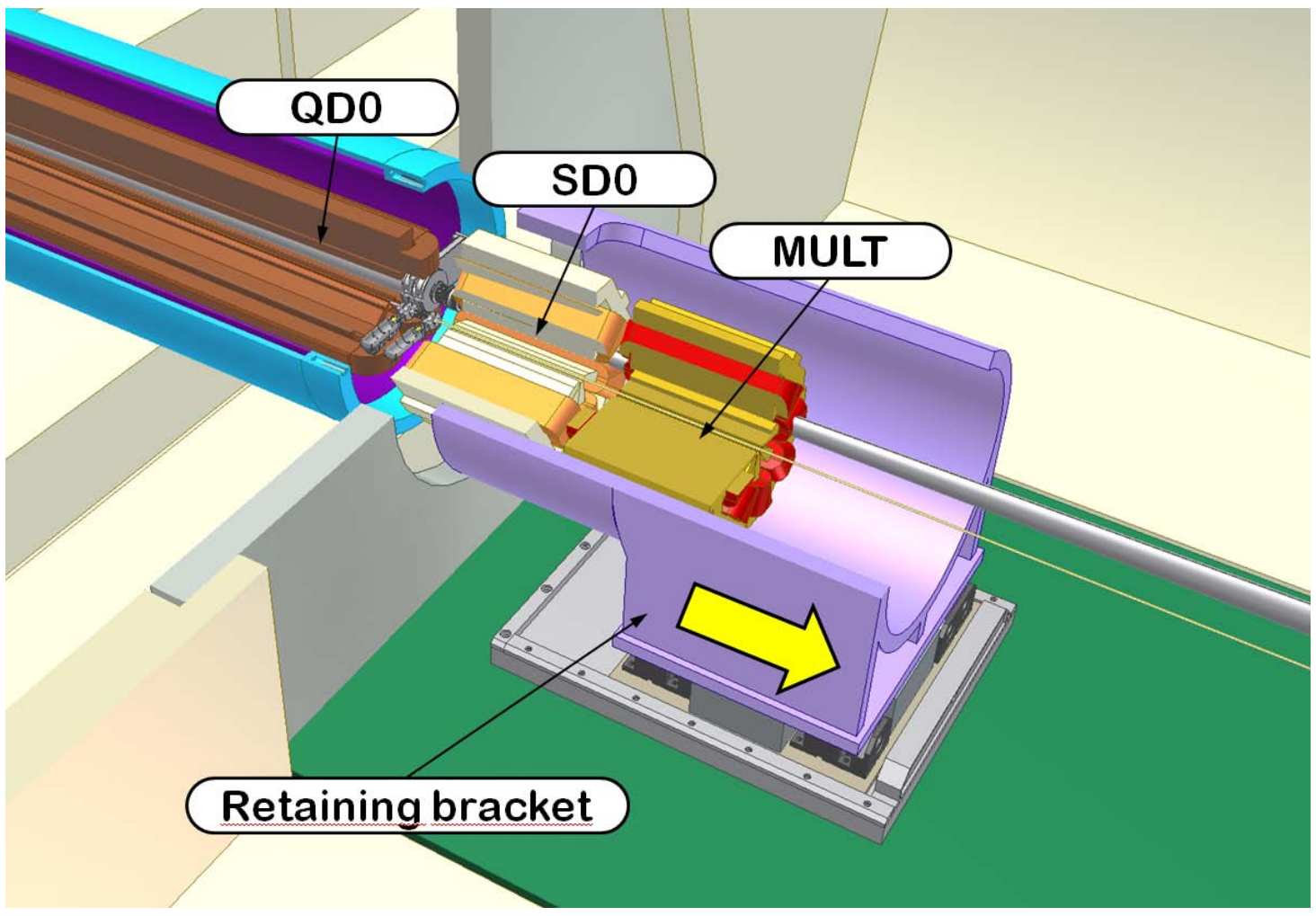

Fig. 11.19: Retracted retaining bracket.

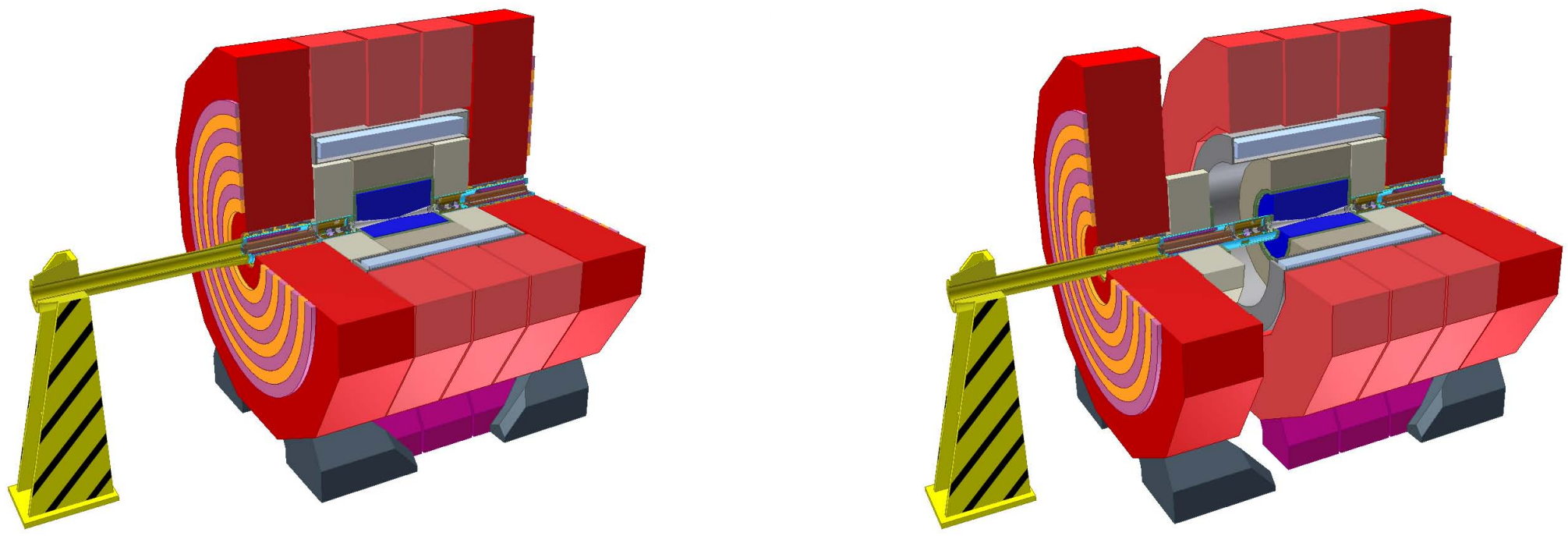

Fig. 11.20: Extraction tool for QD0 support tube (left) and backward move of the detector endcap (right).

This may, possibly, best be performed by combining the three elements into one common enclosure. However, such a detailed technical study goes beyond the scope of the present CDR.

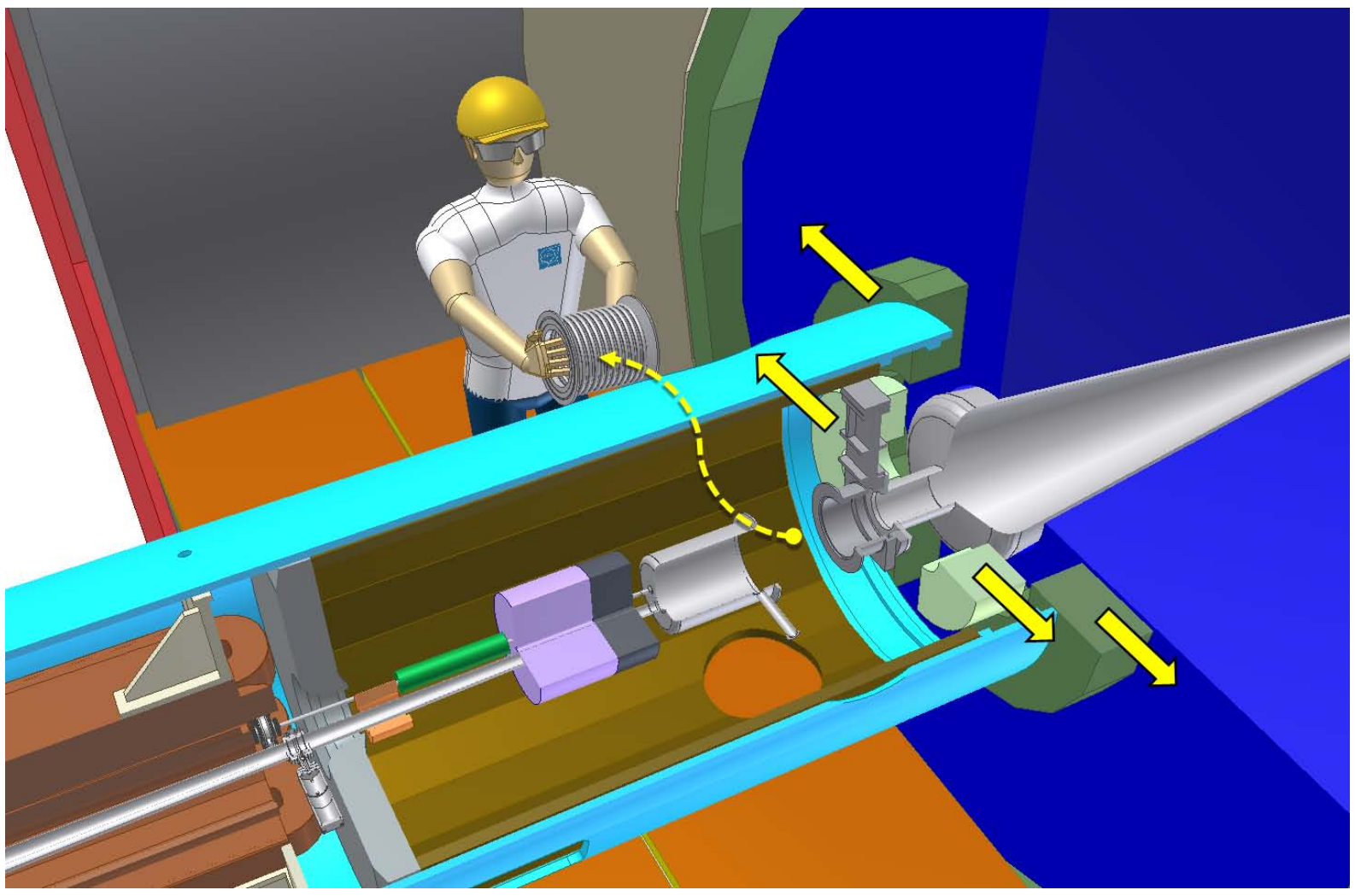

Fig. 11.21: Opening LumiCal and ECAL plug for the passage of the valve.

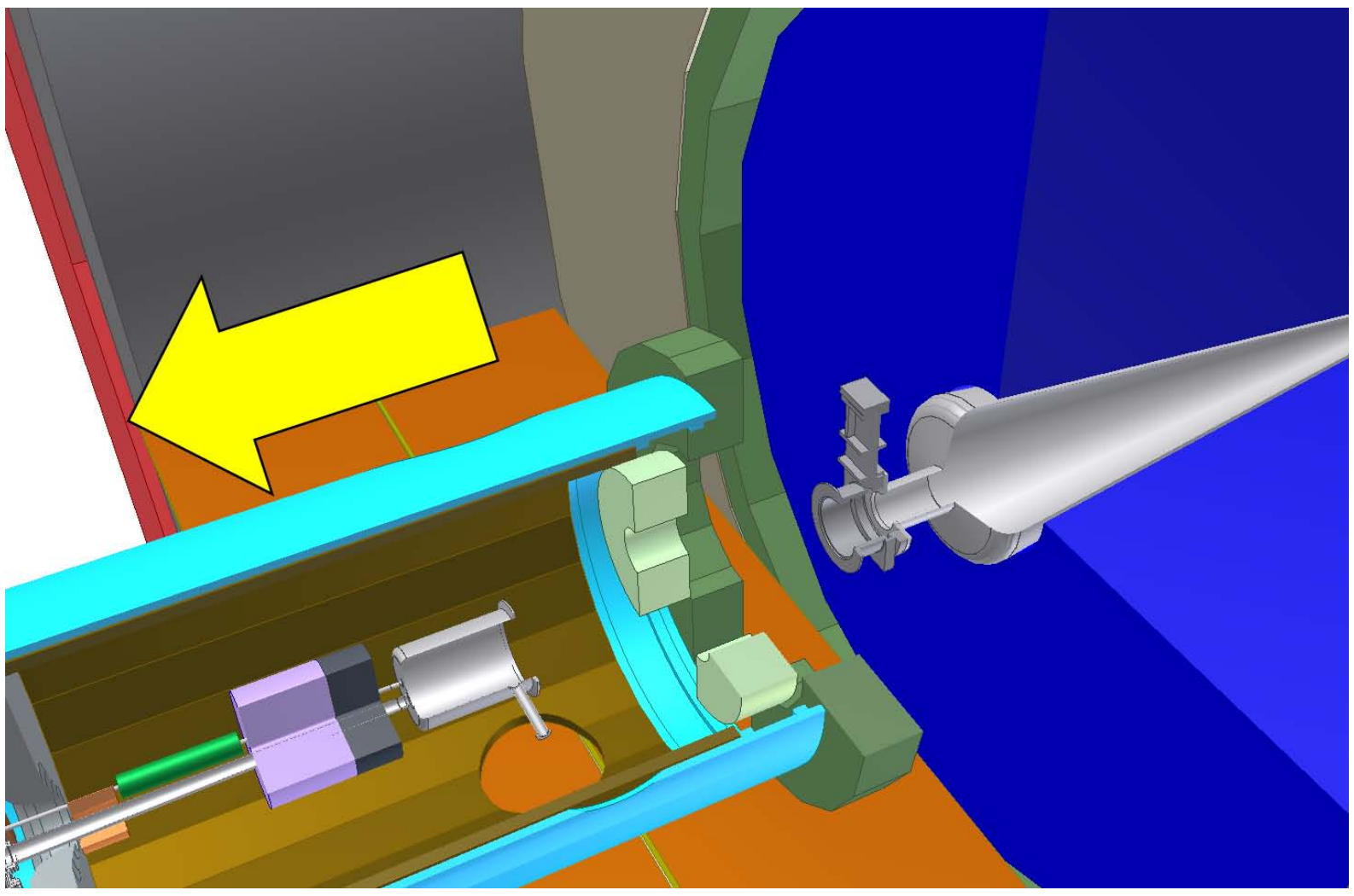

Fig. 11.22: Retraction of the support tube.

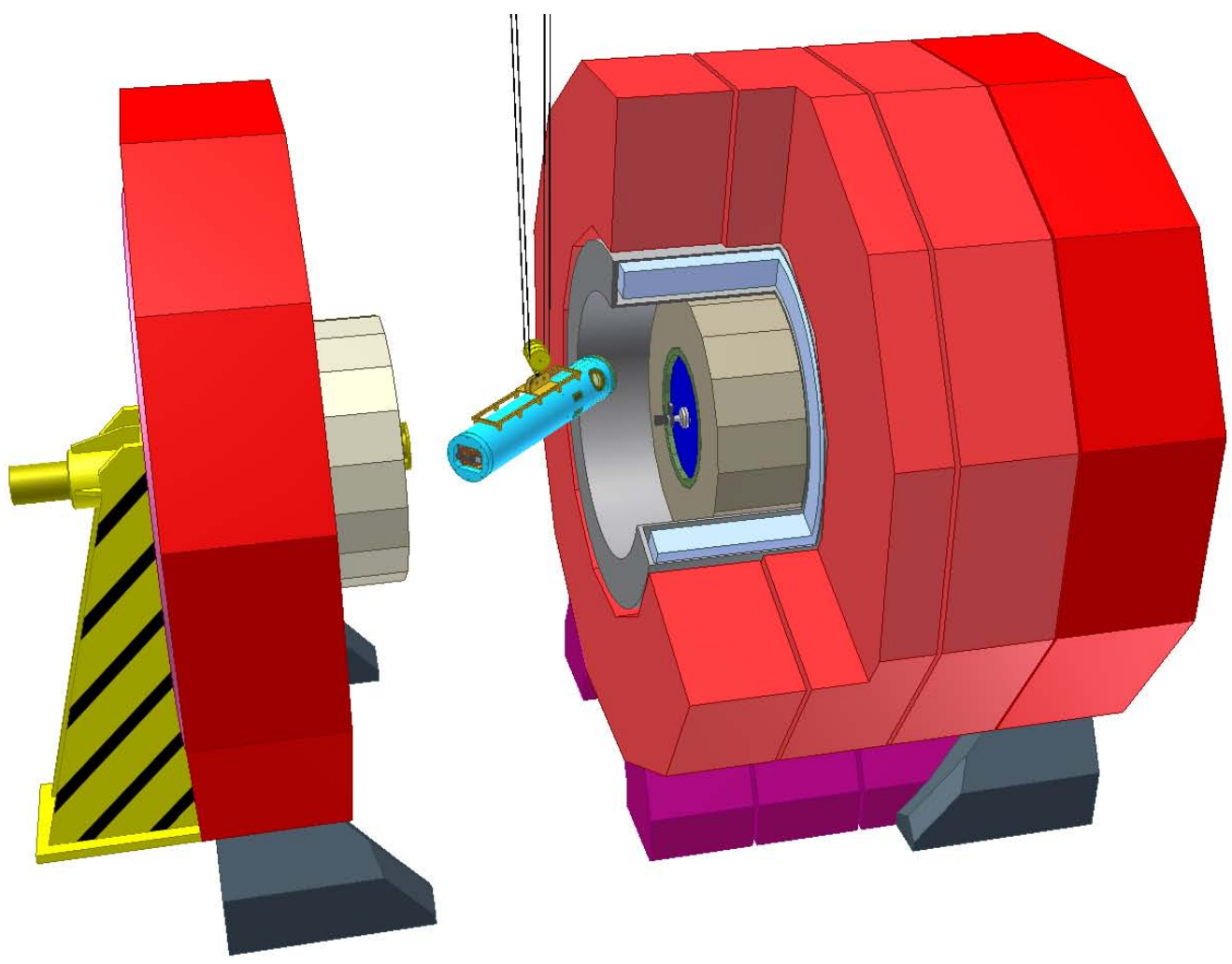

Fig. 11.23: Final removal of the support tube by crane.

# Chapter 12

# Physics Performance

This chapter presents the results of detailed GEANT4 [1, 2] simulation studies using the CLIC_ILD and CLIC_SiD detector concepts. These studies have been carried out under realistic CLIC conditions including, e.g., the luminosity spectrum and the overlay of $\gamma\gamma \rightarrow$ hadrons background events taking into account the time-structure of the bunch train. Results are based on full event reconstruction including tracking, particle flow analysis and flavour-tagging. First, the simulation and reconstruction tools used for the study are presented. In particular the methodology applied to the reconstruction of physics signals in the presence of background is described. Subsequently, the detector performance for basic physics observables is presented. Contrary to most observables presented in the preceding subdetector chapters, the physics observables discussed here are reconstructed by combining information from several subdetectors. Most observables discussed in this chapter are linked to the detector benchmark processes, described in Section 2.6, and the quality of their reconstruction in the presence of background is studied. Thereafter the simulation and analysis of the physics benchmark processes themselves are studied in detail, with the aim of assessing the performance of high precision physics measurements at CLIC.

## 12.1 Simulation and Reconstruction

The detector simulation and reconstruction programs used for the results presented here are based on those developed by the linear collider community over the past decade. They were already used in the preparation of the letters of intent for the ILD and SiD detector concepts.

### 12.1.1 Event Generation

The Monte Carlo event samples for the benchmark physics studies, introduced in Section 2.6, were generated mostly using the WHIZARD program [3, 4], assuming zero polarisation of the electron and positron beams. Parton showering, hadronisation and fragmentation is performed using the PYTHIA [5] program, with the fragmentation parameters tuned to the OPAL data taken at LEP, listed in Table B.5 of Appendix B. The decays of $\tau$ leptons are handled using TAUOLA [6] and the default PYTHIA treatment of Final State Radiation (FSR) is used. The luminosity spectrum, described in Chapter 2, generated using GUINEAPIG [7], is interfaced to WHIZARD. The effects of Initial State Radiation (ISR) are included in WHIZARD, with the ISR photons always being collinear with the incoming beam directions. For the generation of processes involving supersymmetric particles, the SUSY parameters are put into WHIZARD using the Les Houches format [8]. The parameters for the two SUSY models [9] that are studied here are used by PYTHIA to obtain the appropriate branching fractions for decays involving SUSY particles. For several analyses, the generation of Standard Model background is restricted to the signal-like region by using generator level cuts.

If unstable particles (e.g., W, Z, h, or t quarks) are required to be in the final state, WHIZARD assumes their natural width to be zero. Hence, processes like $e^+e^- \rightarrow W^+W^-$, $e^+e^- \rightarrow Z^0Z^0$, and $e^+e^- \rightarrow t\bar{t}$, are generated using a standalone version of PYTHIA. In this case the luminosity spectrum is included using CALYPSO [7] and the ISR photons then have a non-zero $p_T$.

It is essential to include the background from $\gamma\gamma \rightarrow$ hadrons in order to study the detector performance under realistic conditions. The $\gamma\gamma \rightarrow$ hadrons events are hadronised using PYTHIA, as described in Section 2.1.2.2.

### 12.1.2 Detector Simulation

The GEANT4 package was used to provide a detailed simulation of the response of the two CLIC detector concepts. The simulations are based on the detector parameters presented in Section 3.5. With a few relatively minor exceptions, the simulated detectors correspond closely to the global parameters of the engineering design [10]. The GEANT4 simulation of the CLIC_ILD [11] and CLIC_SiD [12] detector concepts use the MOKKA [13] and SLIC [14] programs, respectively. In both cases the `QGSP_BERT` physics list is used to simulate the detailed development of hadronic showers in the detector. Since head-on collisions are generated, the crossing angle of 20 mrad is introduced in the simulation by applying a corresponding Lorentz boost to all particles. Both the MOKKA and SLIC programs output a list of generated particles and detector hits which are stored as SIO files [15] using the LCIO [16] event data model. The use of a common data format enables the use of common particle flow and flavour tagging programs.

### 12.1.3 Event Reconstruction

The MARLIN framework is used for the digitisation, reconstruction, and analysis of events simulated with the CLIC_ILD detector. The main steps in the reconstruction are: digitisation of the simulated hits, TPC pattern recognition based on algorithms developed at LEP, track finding in the silicon detectors and merging of the silicon tracks with the TPC tracks. The reconstruction of simulated data in the CLIC_SiD concept uses the `org.lcsim` [17] framework to digitise the detector hits and to perform track pattern recognition. Detailed descriptions of the reconstruction software can be found in [18, 19, 20] and references therein. A number of significant developments were made for the studies presented here. In particular, the inclusion of $\gamma\gamma \to$ hadrons background for all generated events necessitated modifications to the tracking software for the CLIC_ILD and CLIC_SiD concepts. The track reconstruction is described in more detail in Section 5.4. For both concepts, particle flow reconstruction is performed using the PANDORAPFA [21] event reconstruction package, producing a list of reconstructed Particle Flow Objects (PFOs). Improved particle identification in PANDORAPFA was developed for this report [22]. Hits generated in the LumiCal and in the BeamCal were not included in the reconstruction.

### 12.1.4 Treatment of Background

Background from $\gamma\gamma \to$ hadrons was routinely included for the studies presented in this report. The hits from simulated $\gamma\gamma \to$ hadrons events were added to those from the underlying simulated $e^+e^-$ collision prior to digitisation, track finding and particle flow reconstruction [23, 24]. The inclusion of background was restricted to 60 bunch crossings (BX) in a time window of $-5$ ns to $+25$ ns around the generated physics event, with a time of 0.5 ns in between two BX, mimicking the CLIC train structure. Assuming subdetector readout capabilities as detailed in Section 2.5.1, 60 BX are sufficient to account for most of the impact of the background. For each BX, the number of $\gamma\gamma \to$ hadrons background events included was drawn from a Poisson distribution assuming a mean of 3.2 events per bunch crossing. This rate corresponds to the nominal simulation results and excludes safety factors for the simulation uncertainties. The robustness of the jet reconstruction has been confirmed for one example with a safety factor of two applied for the background rate, as discussed in Section 12.3.3.

The detector readout was accounted for by assuming readout windows of 10 ns for all detectors, apart from the TPC and the barrel HCAL (see Section 2.5.1, Table 2.4). In the barrel HCAL a window of 100 ns is assumed to account for the shower development in tungsten (see Figure 2.11). In the TPC all hits are kept. The readout windows are corrected for the straight line time-of-flight to the centre of each readout cell. In case of multiple hits in the same readout cell, the time stamp of the first hit inside the readout window is used for the whole cell. Hits outside of the readout windows are rejected.

The surviving hits (primarily from the physics interaction and approximately 20 BX of $\gamma\gamma \to$ hadrons background) are passed to the digitisation and then to the reconstruction software. Track recon-struction and the subsequent particle flow reconstruction are performed using these hits as input.

The output of PANDORAPFA is a list of PFOs. Of the 19 TeV of energy deposited in the calorimeters in a full bunch train there is, on average, 1.2 TeV of reconstructed energy from $\gamma\gamma \rightarrow$ hadrons that are in the same readout window as the physics event (see Section 2.5). This energy presents itself mostly in the form of relatively low $p_T$ PFOs. The impact of this background is reduced by applying $p_T$ and additional timing cuts to the fully reconstructed PFOs. It is important to note that $p_T$ cuts alone are not sufficient, and that the use of time information in this reconstruction step is crucial for the reduction of beam-induced background. Three levels of timing cuts were applied, `default`, `loose` and `tight`. In each case a separate list of PFOs was written for use in the subsequent analyses. The additional timing cuts are based on the truncated mean time of the calorimeter hits forming the cluster of the respective PFO which is calculated by first determining the median time of all hits in the cluster. The outlying 10% of hits in the cluster are then neglected and the remaining hits are used to calculate an energy-weighted mean value. If sufficient hits are available, timing information from the ECAL region may be used preferentially to that from the HCAL region. If the PFO also contains a track, a helix fit to the track is used to calculate the arrival time of the track at the surface of the ECAL, otherwise a straight line time-of-flight is assumed. This value is subtracted from the measured cluster time. The timing cuts, listed in detail in Tables B.1, B.2 and B.3 of Appendix B, depend on the particle type, $p_T$ and $\cos\theta$ of the PFO. Table 12.1 shows the impact of the timing cuts on the reconstructed $\gamma\gamma \rightarrow$ hadrons background and on a physics process, in this case a W of 500 GeV energy, where $W \rightarrow u\bar{d}$. The impact of a simple $p_T$ cut is shown for comparison. Separate PFO-based timing cuts were developed for CLIC operation at $\sqrt{s} = 500$ GeV to account for the lower energy of the signal particles. They are described in detail in Table B.4.

Table 12.1: Comparison of the impact of the different PFO selections on the reconstructed energy of $\gamma\gamma \rightarrow$ hadrons, and of a hadronically decaying W boson of 500 GeV energy, with the impact of a simple $p_T$ cut.

| Cut | $\gamma\gamma \rightarrow$ hadrons Energy (GeV) | 500 GeV di-jet Energy (GeV) | 500 GeV di-jet energy loss |
|---|---|---|---|
| `No cut` | 1210 | 500.2 | 0% |
| `Loose` | 235 | 498.8 | 0.3% |
| `Default` | 175 | 498.0 | 0.5% |
| `Tight` | 85 | 496.1 | 0.8% |
| $p_T > 3.0$ GeV | 160 | 454.2 | 9.2% |

## 12.2 Luminosity Spectrum

There are several effects which may change the centre-of-mass energy away from the nominal, e.g. 3 TeV, before an interaction (see Section 2.1.1): The beam-energy spread from the main linac, the beamstrahlung due to beam-beam interactions, and the initial state radiation (ISR). ISR can be calculated with high accuracy and is generally accounted for in the event generators such as WHIZARD. The energy spread and beam-beam effects lead to the luminosity spectrum (also referred to as differential luminosity). For precision physics results at CLIC, the accurate determination of the luminosity spectrum is a key ingredient.

### 12.2.1 Luminosity Spectrum Measurement using Bhabha Events

The beam energy spread is measured in an energy spectrometer upstream of the IP with sufficient accuracy. The energy loss from beamstrahlung can be calculated from strong-field QED, but it strongly depends on parameters such as the beam position offsets and angular rotations, and the details of the particle distributions in the bunches. All of these parameters are changing continuously during operation of CLIC, and many of them cannot be measured directly. The luminosity spectrum must, therefore, be determined through the measurement of a physics channel.

The luminosity spectrum is the same for all physics events. Therefore, a well-known physics channel such as wide-angle Bhabha scattering can be used to deduce the luminosity spectrum from measurements [25]. Bhabha events are measured using the high angular precision of the tracking detectors and the energy resolution of the calorimeters. The accuracy of reconstructing the luminosity spectrum by this method is assessed in generator-level Monte Carlo simulations. A systematic study of this reconstruction has recently been performed for the ILC at 500 GeV [26]. The method is being adapted for CLIC at 3 TeV. The status of this work is summarised here.

Similarly to the ILC study, a parametrisation of the energy distribution of the colliding particles at CLIC has been developed. This parametrisation takes the correlation between the two beams into account. The Bhabha events are simulated using BHWide [27], and a fitting technique is used to extract the parameters based on measurable observables: the energy of the electrons, positrons and FSR photons measured in the ECAL and the angles of the electrons and positrons measured in the tracker.

Figure 12.1 shows the resulting luminosity spectrum from the re-weighting fit using Bhabha events generated by BHWide. The data generated for this analysis correspond to an integrated luminosity of 21 $\text{fb}^{-1}$. These results are compared to the luminosity spectrum as obtained directly from GUINEAPIG. Deviations of up to 15% are observed in the peak, and the region below 800 GeV shows significant differences. Most importantly, detector resolution effects which might impact on the quality of the fit are not yet included.

The results presented here are to be considered a first iteration. In future work, the validity of the parametrisation needs to be assessed by introducing smearing in the four-vectors of particles produced by BHWide. In a further step, the Bhabha events will need to be simulated and reconstructed using a full detector simulation. This will allow to assess the effects of realistic energy resolution and tracking on the accuracy of the measured luminosity spectrum. In addition, the effect of overlaying beam-beam background events needs to be investigated. Finally, the accuracy of the measured luminosity for longer runs (better statistical precision) needs to be assessed.

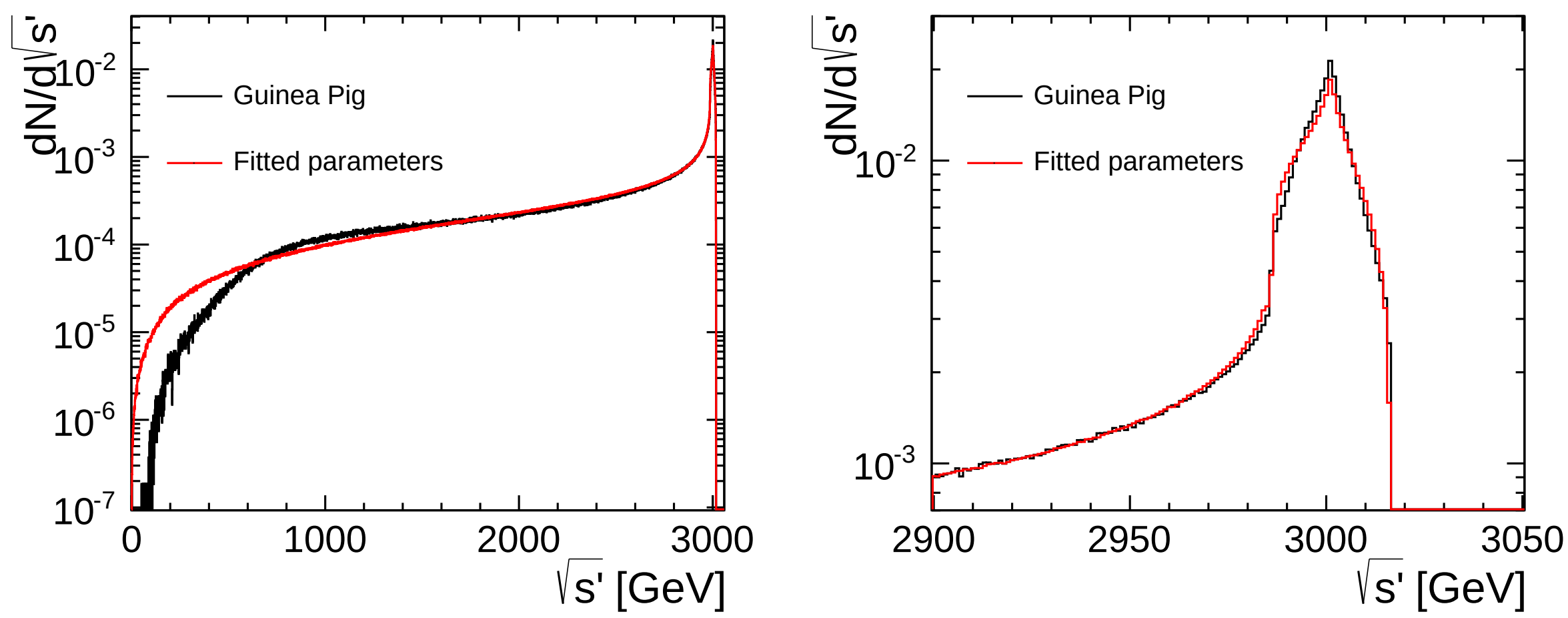

Fig. 12.1: Comparison of the luminosity spectrum deduced from Bhabha events with the distribution generated by GUINEAPIG, over the entire energy range (left) and in the region of the high energy peak (right). Details of the analysis are described in [28].

### 12.2.2 Systematic Effects due to Uncertainty of the Luminosity Spectrum

Detailed studies of the accuracy with which the luminosity spectrum can be reconstructed from Bhabha events are still on-going. Nevertheless, a preliminary assessment of the sensitivity of the detector benchmark studies, presented in this CDR, to uncertainties in the knowledge of the luminosity spectrum has been made.

Two *ad hoc* variations of the luminosity spectrum were simulated, corresponding to a change of $\pm 5\%$ in the normalisation of the number of events in the high energy peak of the energy distribution for each beam relative to the numbers of events in the tails. This corresponds to a changing the average $\sqrt{s}$ by $\pm 1\%$. In both cases, the integral of the luminosity spectrum was unchanged. These *ad hoc* variations, which are much larger than the expected precision achievable from measurements of the acollinearity distribution of wide angle Bhabha scattering events, provide a mechanism for determining the sensitivity of a physics analysis to large variations in luminosity spectrum. The corresponding changes in the physics observables such as masses and cross sections, extracted using the distorted luminosity spectra, are compared to the values obtained for the nominal luminosity spectrum. The results are given in Section 12.4.4 for the study of right-handed squarks, in Section 12.4.5 for the slepton searches and in Section 12.4.6 for the chargino and neutralino pair production measurement.

## 12.3 Performance for Lower Level Physics Observables

The ability to accurately reconstruct leptons and jets determines the experimental sensitivity in the majority of physics measurements. The CLIC detector concepts are designed for excellent momentum and jet energy resolution, high efficiency flavour tagging and excellent lepton identification capability. The results presented in this section, which all use the full GEANT4 simulation of the CLIC_ILD and CLIC_SiD detector models and full event reconstruction, summarise the performance obtained for these lower level physics observables.

### 12.3.1 Particle Identification Performance

Particle identification (particle ID), and in particular lepton ID, will be central to many physics studies at CLIC. The performance of the particle ID of the PANDORAPFA package was studied using samples of single particles and isolated leptons in simulated physics events with and without $\gamma\gamma \rightarrow$ hadrons background. The particle ID efficiency is obtained by matching reconstructed particles (PFOs) to the generated particles:

$$\text{particle ID Efficiency} = \frac{\text{matched particles}}{\text{findable particles}},$$

where matched particles have a reconstructed PFO of the same particle type and charge, within a cone of $2^\circ$ around a generated particle. Findable particles are defined to be generated particles of a particular type with energy $> 7.5$ GeV and a polar angle $8^\circ < \theta < 172^\circ$ [29].

The particle ID efficiency for single isolated electrons, photons, pions and muons has been studied for the CLIC_ILD detector concept using samples with a single particle generated isotropically with a uniform energy distribution in the range 0–400 GeV. As an example, Figure 12.2 shows the particle ID efficiency (which includes both the efficiency for reconstructing a PFO and correctly identifying the particle type) as a function of energy for electrons and photons. The average particle identification efficiencies for single particles are approximately 93% for photons, 97% for electrons, and 99% for muons. Similar efficiencies are obtained with the CLIC_SiD detector [29]. Charged particles reconstructed by PANDORAPFA are assumed to be pions unless they pass the lepton ID requirements or are associated with a reconstructed displaced vertex ($V^0$) or particle decay in the tracking volume. Figure 12.3 shows the efficiencies for single pions versus energy and versus the polar angle.

The effect of the background from $\gamma\gamma \rightarrow$ hadrons events on the muon identification efficiency was studied using $e^+e^- \rightarrow \tilde{\chi}_1^+\tilde{\chi}_1^- \rightarrow W^+W^-\tilde{\chi}_1^0\tilde{\chi}_1^0$ events generated at $\sqrt{s}$ = 3 TeV using the CLIC_SiD detector model. Here, the muons arise from semi-leptonic and hadronic decays of the W boson, and are, in general, not isolated. The muon identification efficiency, shown in Figure 12.4, is $> 90\%$ over almost the whole energy range, even in the presence of background. Similar results are obtained for the CLIC_ILD detector model [30].

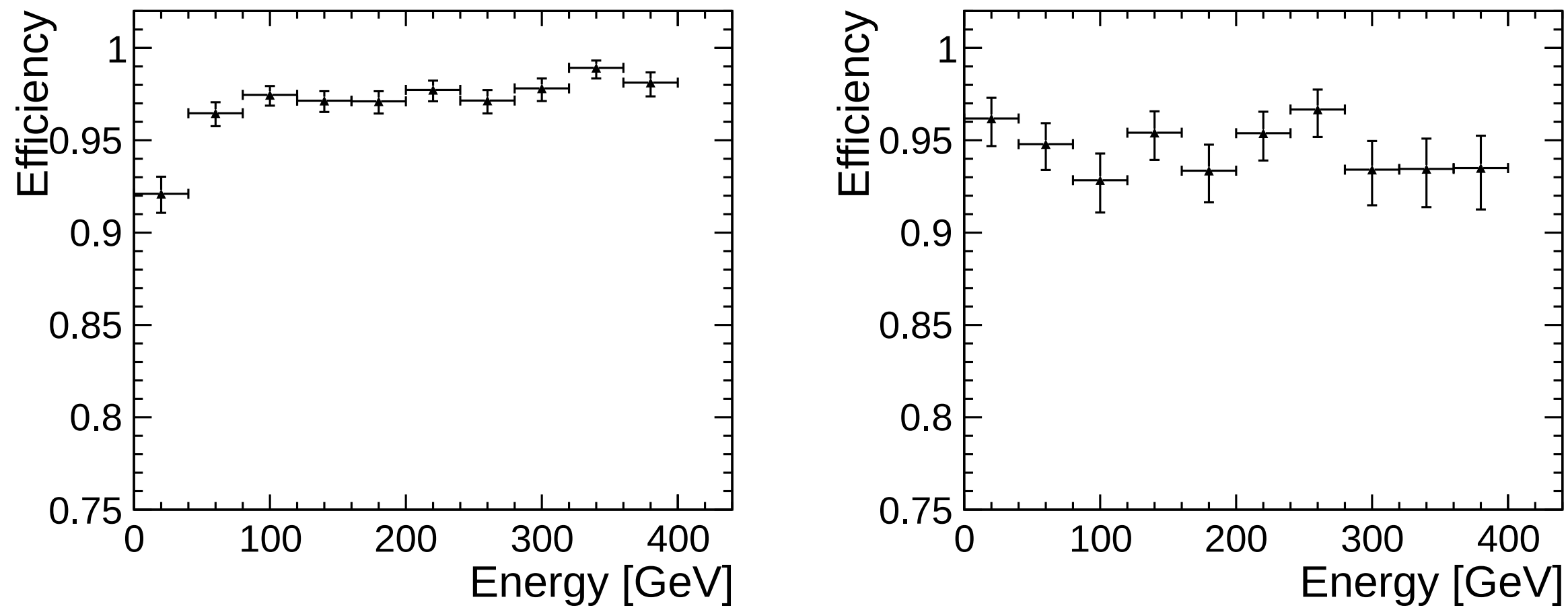

Fig. 12.2: Particle ID efficiency for single electrons (left) and photons (right) in the CLIC_ILD detector as a function of the generated particle energy.

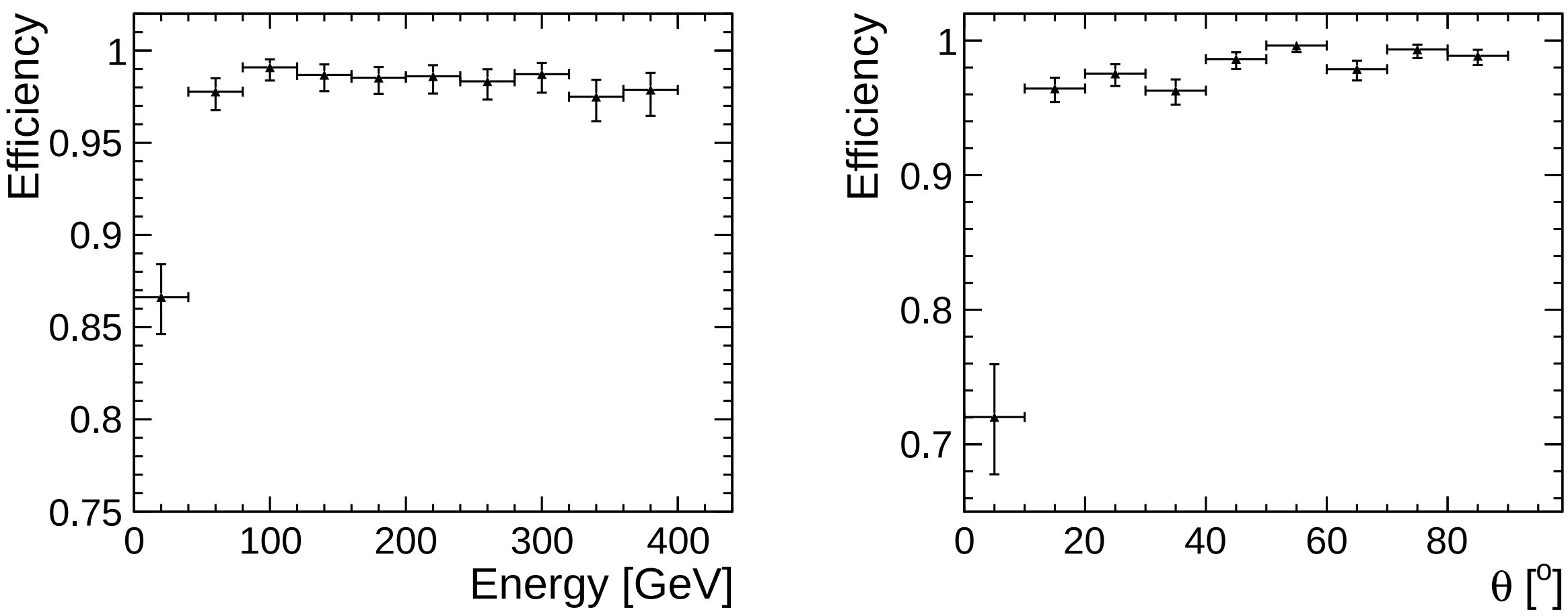

Fig. 12.3: Particle ID efficiency for single pions in the CLIC_ILD detector as a function of the generated particle energy and polar angle.

### 12.3.2 Muon and Electron Energy Resolution

The tracking systems of the CLIC detectors are designed to provide excellent momentum measurement for charged particle tracks. The reconstruction of momentum for high energy leptons is studied in SUSY processes with two high energy leptons in the final state. The performance for *isolated* leptons is studied in the process $e^+e^- \to \tilde{\ell}_R^+ \tilde{\ell}_R^-$, where each slepton decays to a high-energy lepton of the same family and the lightest neutralino. The events were simulated using the CLIC_ILD detector model.

The energy of the lepton is reconstructed from the momentum of the charged particle track corrected for final state radiation and bremsstrahlung; the energy of photons and $e^+e^-$ pairs from conver-sions within a cone of 20° around the reconstructed lepton direction is added to the energy from the track. The lepton energy resolution is characterised using $\Delta E/E^2_{\mathrm{True}}$, where $\Delta E = E_{\mathrm{True}} - E_{\mathrm{Reco}}$ is the difference between the lepton energy at generator level (before final state radiation or the effects of bremsstrahlung), $E_{\mathrm{True}}$, and the reconstructed lepton energy including any associated photons or conversion pairs, $E_{\mathrm{Reco}}$. Figure 12.5 shows the energy resolution obtained for high energy muons in $e^+e^- \to \tilde{\mu}_R^+\tilde{\mu}_R^-$ events and for high energy electrons in $e^+e^- \to \tilde{e}_R^+\tilde{e}_R^-$ events. The resolution is parametrised using the sum of two Gaussian functions. The muon energy resolution is described by a Gaussian with a width of $\Delta E/E^2_{\mathrm{True}} = 1.5 \cdot 10^{-5}$ GeV$^{-1}$. The central region of the distribution is defined in between $\Delta E/E^2_{\mathrm{True}} = \pm 0.5 \cdot 10^{-3}$ GeV$^{-1}$. Only 4.1% of the events are outside of the central region; these are well described by a Gaussian of width $\Delta E/E^2_{\mathrm{True}} = 4.9 \cdot 10^{-5}$ GeV$^{-1}$. The electron energy resolution is described by a Gaussian peak with the same width as that for muons, $\Delta E/E^2_{\mathrm{True}} = 1.4 \cdot 10^{-5}$ GeV$^{-1}$. However, even with bremsstrahlung recovery, about 30% of the events are outside the central region. These are due to cases where final state radiation and bremsstrahlung are not sufficiently well accounted for by the present method; they are reasonably well described by a Gaussian of width $\Delta E/E^2_{\mathrm{True}} = 7.7 \cdot 10^{-5}$ GeV$^{-1}$.

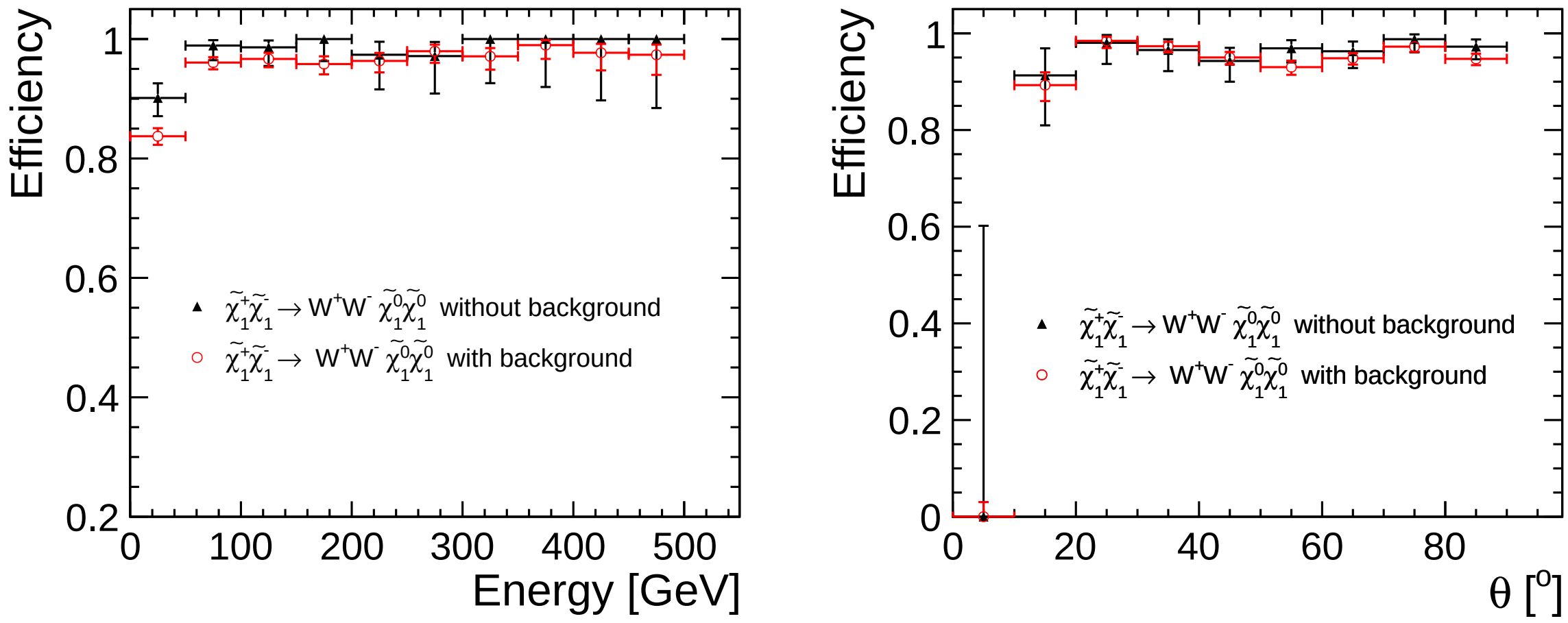

Fig. 12.4: Muon identification efficiency in the CLIC_SiD detector as a function of the generated particle energy (left) and of the polar angle (right). The $e^+e^- \to \tilde{\chi}_1^+\tilde{\chi}_1^- \to W^+W^-\tilde{\chi}_1^0\tilde{\chi}_1^0$ events were simulated at $\sqrt{s}$ = 3 TeV with (red points) and without (black points) $\gamma\gamma \to$ hadrons background.

The correction of the reconstructed energies for possible final state radiation and bremsstrahlung introduces a potential bias for more complicated final states. This possibility is studied using the processes $e^+e^- \to \tilde{\nu}_e\tilde{\nu}_e$ and $e^+e^- \to \tilde{e}_L^+\tilde{e}_L^-$, which are characterised by high energy electrons, missing energy and jets from chargino or neutralino decays. The reconstructed particles (PFOs) from events simulated in the CLIC_ILD detector are clustered into jets using the inclusive anti-$k_t$ algorithm from the FASTJET [31] package. Jets are required to have a minimum energy of 20 GeV. The electron energy resolution is studied for events where six "jets" are reconstructed and where two of the jets are identified as isolated leptons. Figure 12.6 shows the lepton energy resolutions for the two lepton and four-jet final state. Despite the presence of the four jets, the electron energy resolution in these events is consistent with the energy resolution obtained for the isolated leptons (see Figure 12.5b).

To investigate the effect of background from $\gamma\gamma \to$ hadrons, the energy resolution obtained in the samples without background is compared to that from equivalent samples including background from $\gamma\gamma \to$ hadrons. For samples with only two leptons and missing energy in the final state, essentially all the background can be removed by rejecting reconstructed particles with $p_T < 4$ GeV. Thus, there is no impact on the lepton energy resolution, although the event selection efficiency is reduced by 1.0% and 4.6% for $e^+e^- \to \tilde{\mu}_R^+\tilde{\mu}_R^-$ and $e^+e^- \to \tilde{e}_R^+\tilde{e}_R^-$, respectively (see Table 12.2).

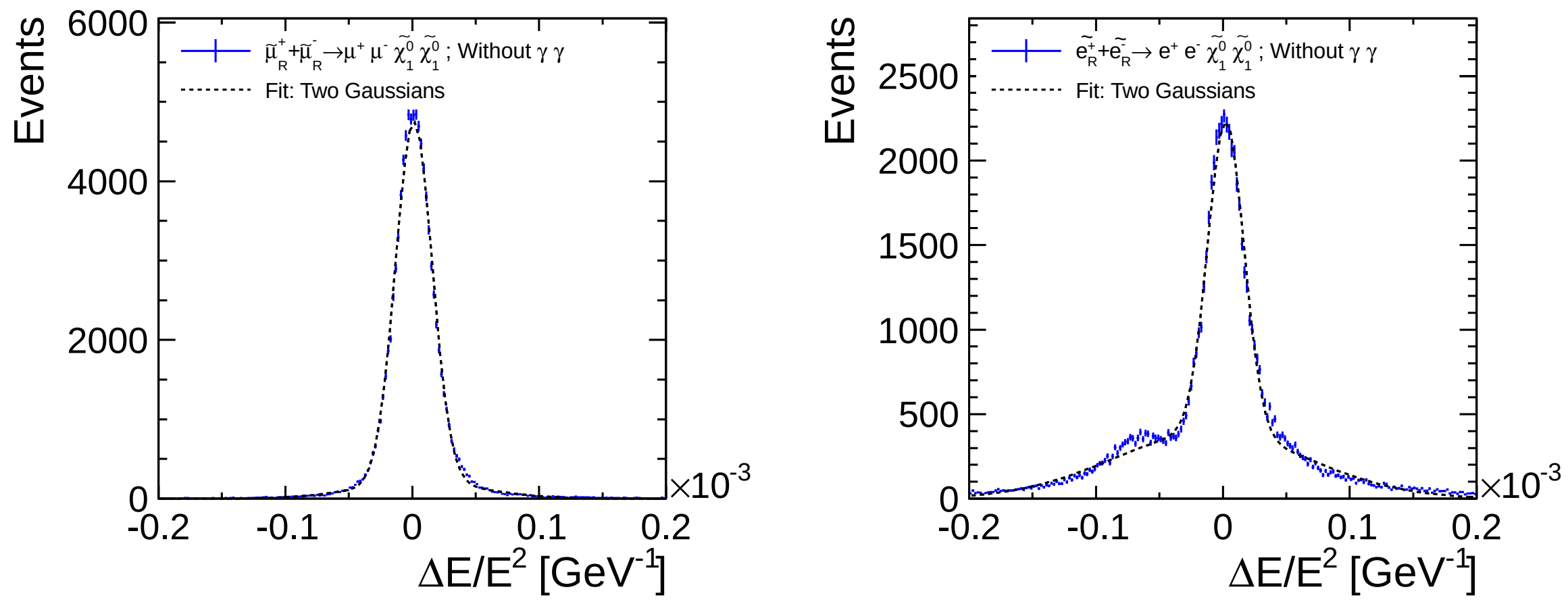

(a) $e^+e^- \to \tilde{\mu}_R^+\tilde{\mu}_R^-$

(b) $e^+e^- \to \tilde{e}_R^+\tilde{e}_R^-$

Fig. 12.5: Lepton energy resolution for processes with missing energy and two isolated leptons. The left plot shows the muon energy resolution obtained from $e^+e^- \to \tilde{\mu}_R^+\tilde{\mu}_R^-$ events while the right plot shows the electron energy resolution observed for $e^+e^- \to \tilde{e}_R^+\tilde{e}_R^-$ events. Both samples were simulated without $\gamma\gamma$ background.

Table 12.2: Event reconstruction efficiency, $\varepsilon_R$, without and with inclusion of $\gamma\gamma$ background in the simulation for different SUSY slepton signal processes. The statistical error on the efficiencies is typically about 1%.

| Process | Decay mode | $\varepsilon_R$ | |
|---|---|---|---|
| | | No background | $\gamma\gamma$ background |
| $e^+e^- \to \tilde{\mu}_R^+\tilde{\mu}_R^-$ | $\mu^+\mu^-\tilde{\chi}_1^0\tilde{\chi}^0$ | 0.98 | 0.97 |
| $e^+e^- \to \tilde{e}_R^+\tilde{e}_R^-$ | $e^+e^-\tilde{\chi}_1^0\tilde{\chi}_1^0$ | 0.95 | 0.90 |
| $e^+e^- \to \tilde{e}_L^+\tilde{e}_L^-$ | $\tilde{\chi}_1^0\tilde{\chi}_1^0 e^+e^-(h/Zh/Z)$ | 0.67 | 0.63 |
| $e^+e^- \to \tilde{\nu}_e\tilde{\nu}_e$ | $\tilde{\chi}_1^0\tilde{\chi}_1^0 e^+e^-\ W^+W^-$ | 0.49 | 0.46 |

In final states with four jets and two leptons, the background from $\gamma\gamma \to$ hadrons cannot be removed using a similar $p_T$ cut, as this would significantly degrade the jet energy reconstruction. Figure 12.7a shows the bias in the reconstructed electron energy when the $\gamma\gamma \to$ hadrons background is included. This bias is due to additional background particles being associated with the electron in the attempt to account for FSR and bremsstrahlung.

Figure 12.7b shows the equivalent distribution after applying `tight` timing cuts at the PFO level. The bias is essentially removed and the original energy resolution is recovered. However, due to the reduced lepton-ID efficiency, the presence of background from $\gamma\gamma \to$ hadrons reduces the overall selection efficiency for the processes with a final state of two electrons, four jets and missing energy by 6% (see Table 12.2).

### 12.3.3 Jet Reconstruction

As discussed in Section 2.2.2, the goal for the jet energy resolution at CLIC is to distinguish hadronic decays of W and Z bosons. This is challenging even without the presence of beam-induced background. The underlying physics interactions of interest are accompanied by significant additional energy from the $\gamma\gamma \to$ hadrons background. For this reason it is not possible to use the jet-clustering algorithms

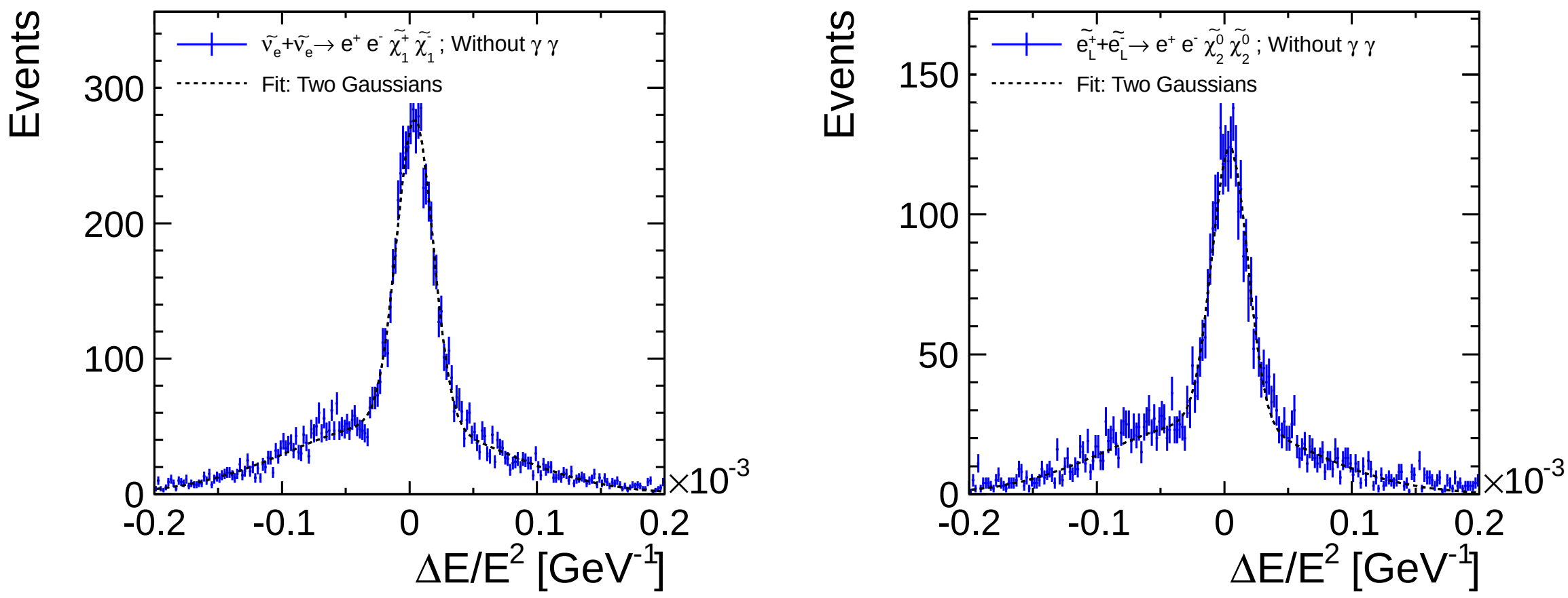

(a) $e^+e^- \to \tilde{\nu}_e\tilde{\nu}_e$ (b) $e^+e^- \to \tilde{e}^+_L\tilde{e}^-_L$

Fig. 12.6: Lepton energy resolution for processes with two leptons, four jets and missing energy. The left plot shows the lepton energy resolution obtained from $e^+e^- \to \tilde{\nu}_e\tilde{\nu}_e$ events while the right plot shows the lepton energy resolution observed for $e^+e^- \to \tilde{e}^+_L\tilde{e}^-_L$ events. Both samples were simulated without $\gamma\gamma$ background.

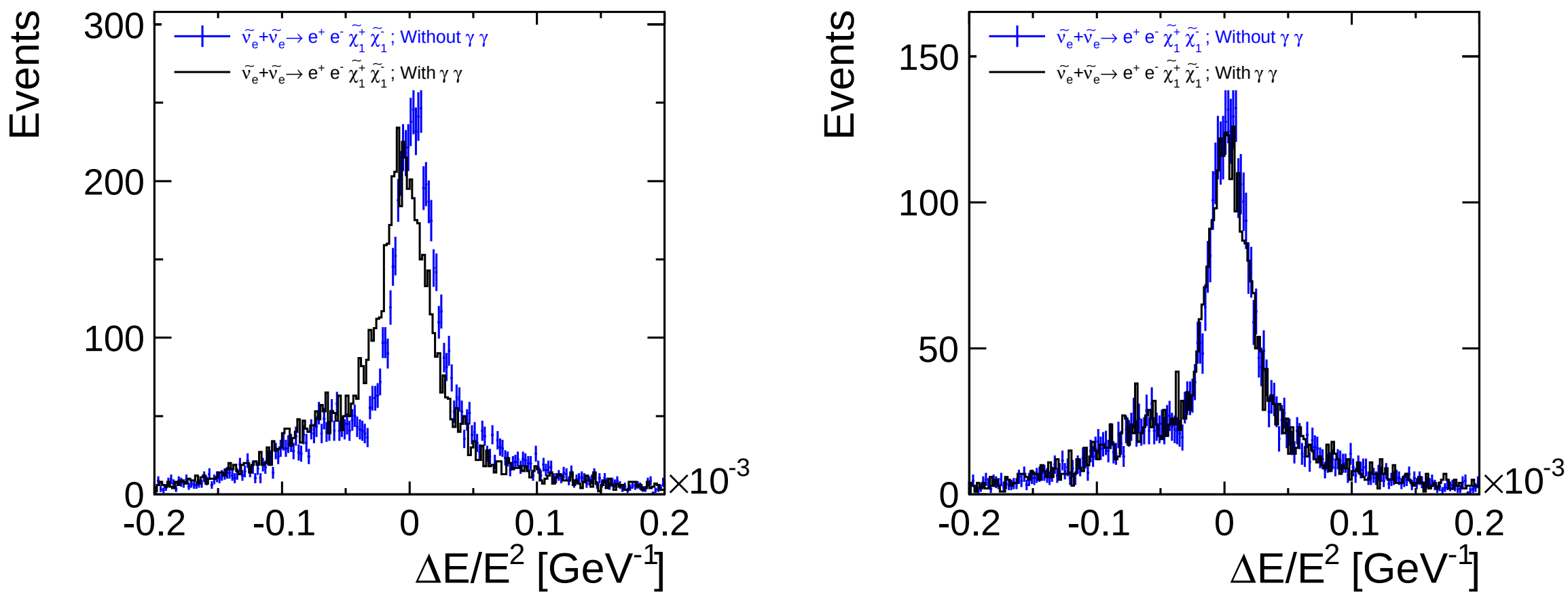

(a) $e^+e^- \to \tilde{\nu}_e\tilde{\nu}_e$, `No cut` selection (b) $e^+e^- \to \tilde{\nu}_e\tilde{\nu}_e$, `tight` selection

Fig. 12.7: Comparison of the electron energy resolution with (black histograms) and without (blue points) $\gamma\gamma$ background in events with two leptons, four jets and missing energy. The left plots shows the distributions obtained without a PFO selection while the right plot shows the distributions for `tight` cuts.

developed for LEP which combine all particles into jets [32]. The jet finding at CLIC was studied using the FASTJET [31] package. Studies carried out in the context of the squark and heavy Higgs benchmark analyses found that the $k_t$ and anti-$k_t$ algorithms [33] developed for hadron collisions are more suitable [34]. Using these algorithms, with a distance parameter $R$ based on $\Delta\eta$ and $\Delta\phi$, leads to a better performance since this increases distances in the forward region and thus reduces the inclusion of background particles into the jets from the $e^+e^-$ interaction. Figure 12.8 compares the reconstructed visible energy observed with the `ee_kt` (Durham) algorithm (left) to that from the $k_t$ algorithm with $R =$ 0.7 (right). For the Durham algorithm, commonly used at LEP, 1.2 TeV of energy from the background (integrated over approximately 10 ns) is added to the reconstructed jets. Whilst this can be reduced using the timing cuts at the reconstructed particle level, the impact is non-negligible. For the $k_t$ algorithm, the impact of the $\gamma\gamma \to$ hadrons background is substantially reduced even without the additional timing cuts.

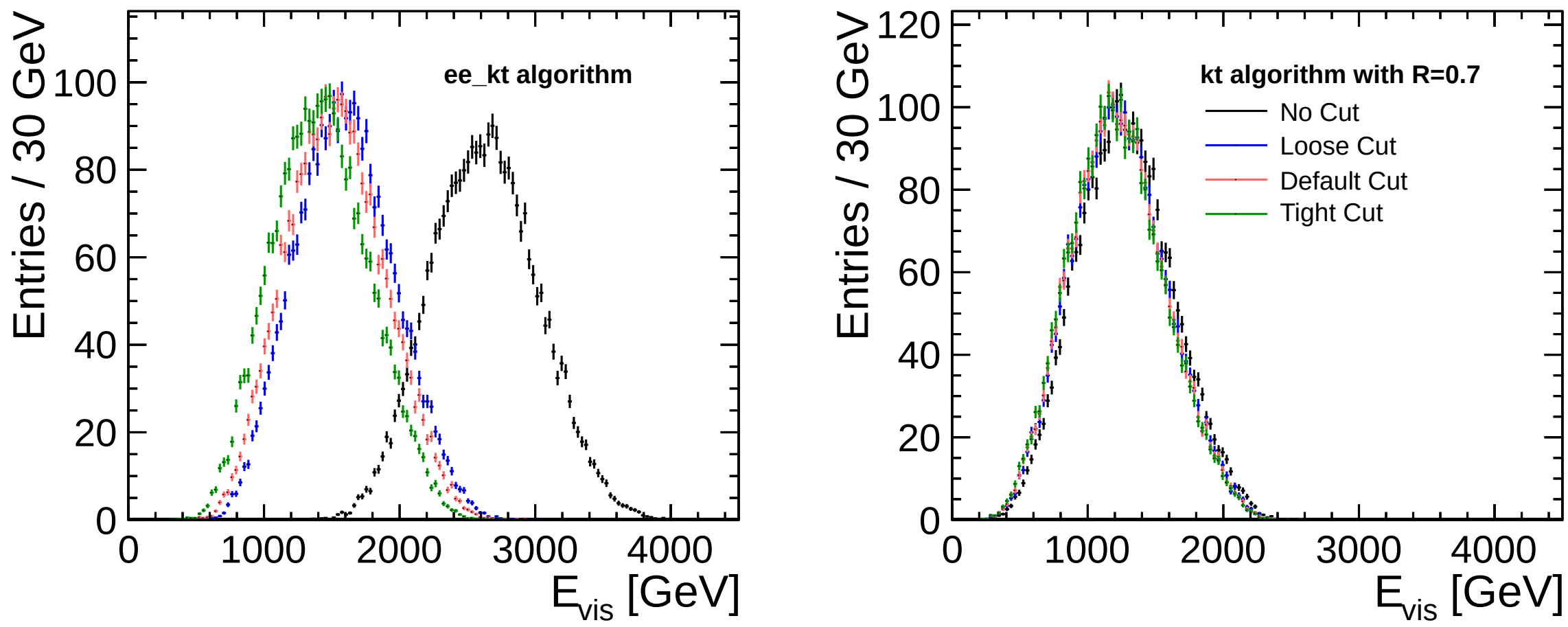

Fig. 12.8: The visible energy, $E_{\text{vis}}$, in $e^+e^- \to \tilde{q}_R\tilde{q}_R \to q\bar{q}\tilde{\chi}_1^0\tilde{\chi}_1^0$ events for different timing cuts (`tight`, `default` and `loose`) and for PFOs without timing cuts. The electron-positron variant of the $k_t$-algorithm (left) picks up significantly more of the $\gamma\gamma \to$ hadrons pile up, while the hadron variant of the $k_t$-algorithm (right) is almost insensitive to this background.

By using the appropriate jet-finding algorithm and applying timing cuts at the reconstructed particle level, most of the background from $\gamma\gamma \to$ hadrons can be removed. The jet energy resolution is a critical component in most of the analyses presented here, either through the measurement of an invariant mass or through the reconstruction of kinematic edges. The timing cuts applied to reduce the impact of the background have the potential to remove particles from the physics interaction of interest and thus degrade the jet energy resolution. The combined effect of jet-finding algorithm and timing cuts is studied on events with a W boson decaying to two jets [35].

Figure 12.9 (left) shows the energy distribution of a reconstructed W with an energy of 500 GeV. The distribution is shown with nominal background (60 BX) when no timing cuts are applied and for the tight timing cuts. For comparison the corresponding distributions in the case of no background and for twice the nominal amount of background are displayed as well. Without applying any timing cuts too many background particles remain in the event and are reconstructed as part of the jet, shifting the energy distribution to higher values.

Figure 12.9 (right) shows the energy resolution of the reconstructed W as a function of the W energy without background (0 BX) and for nominal (60 BX) as well as for twice the nominal ($2\times60$ BX) amount of $\gamma\gamma \to$ hadrons background. When background is included, the tight timing cuts are used. The degradation of the energy resolution in the presence of background at lower energies is significant. With increasing jet energy the effect of background becomes less dominant. The effect of a factor two more $\gamma\gamma \to$ hadrons background is only visible at lower energies and the resolution is only slightly worse.

The ability to separate W, Z and h bosons in the CLIC background environment is demonstrated qualitatively for the gaugino benchmark analysis where the final state corresponds to four jets and missing energy. Figure 12.10 shows the reconstructed di-jet invariant masses of W, Z and h candidates in simulated $\tilde{\chi}_1^+\tilde{\chi}_1^- \to W^+W^-\tilde{\chi}_1^0\tilde{\chi}_1^0$, $\tilde{\chi}_2^0\tilde{\chi}_2^0 \to hh\tilde{\chi}_1^0\tilde{\chi}_1^0$ and $\tilde{\chi}_2^0\tilde{\chi}_2^0 \to hZ\tilde{\chi}_1^0\tilde{\chi}_1^0$ signal events including $\gamma\gamma \to$ hadrons background and using the $k_t$ jet finder. Separate peaks corresponding to the hadronic decays of the $W^+W^-$, h h and h Z final states can be clearly identified. The horizontal band for $M_{jj,2} \approx M_h$ and $M_{jj,1} < M_{\mathrm{h}}$ is caused by $\tilde{\chi}_2^0\tilde{\chi}_2^0 \rightarrow \mathrm{hh}\tilde{\chi}_1^0\tilde{\chi}_1^0$ events, where one of the h bosons is only partially reconstructed. No corresponding vertical band is visible due to the way the jets are ordered in the analysis.

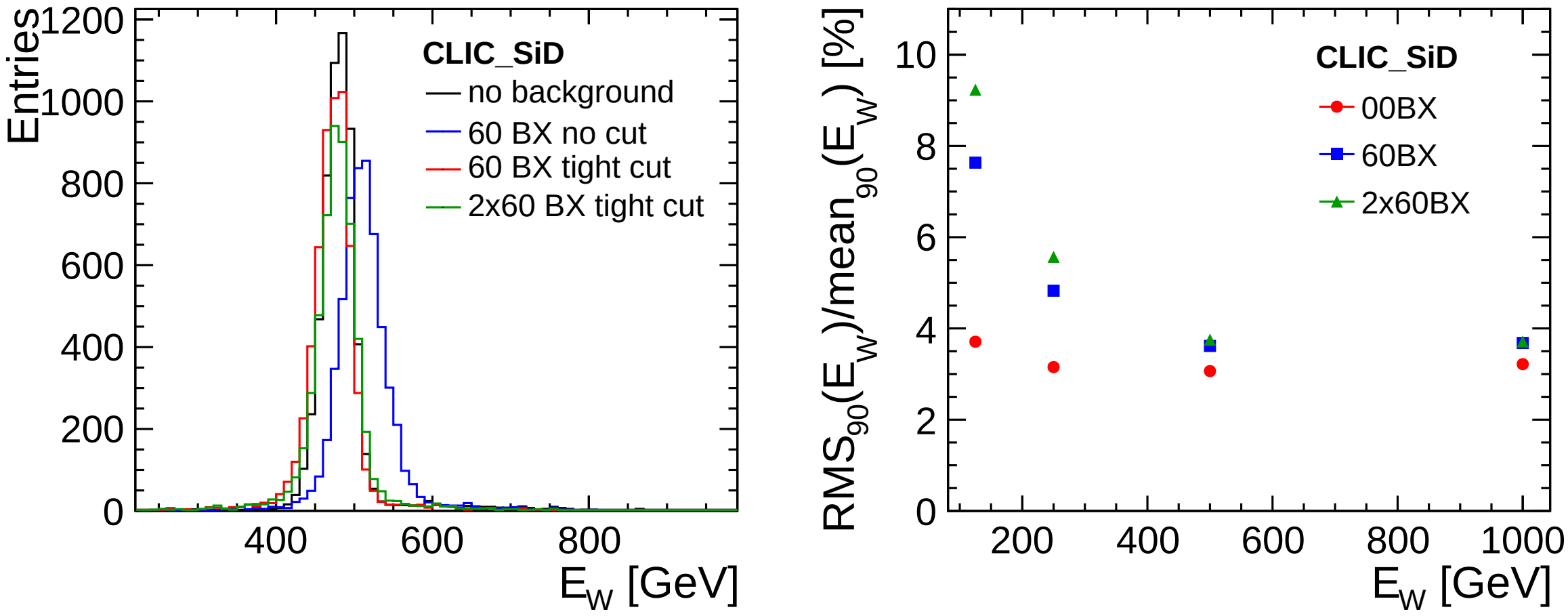

Fig. 12.9: Left: Energy distribution of the reconstructed W with an energy of 500 GeV for various amounts of $\gamma\gamma \rightarrow$ hadrons background overlaid (no background, 60 BX and 2×60 BX) and for different timing cuts (no cut and tight timing cuts). Right: Energy resolution of the reconstructed W as a function of the W energy for various amounts of $\gamma\gamma \rightarrow$ hadrons background overlaid. In case of background the tight timing cuts are used. The results in both figures are obtained for the CLIC_SiD detector model.

The mass resolution of W bosons is studied in the framework of the slepton production analysis in events with two high-energy leptons, two hadronic W decays and missing energy. This is shown with and without background in Figure 12.11a. With the `tight` PFO selection timing cuts, the impact of the $\gamma\gamma \rightarrow$ hadrons background is greatly reduced. To quantify the degradation, the distributions were fitted with a Breit–Wigner function convolved with a Gaussian. The fitted width of the Gaussian increases slightly from 4.1 GeV without $\gamma\gamma \rightarrow$ hadrons background to 4.7 GeV with $\gamma\gamma \rightarrow$ hadrons background [36].

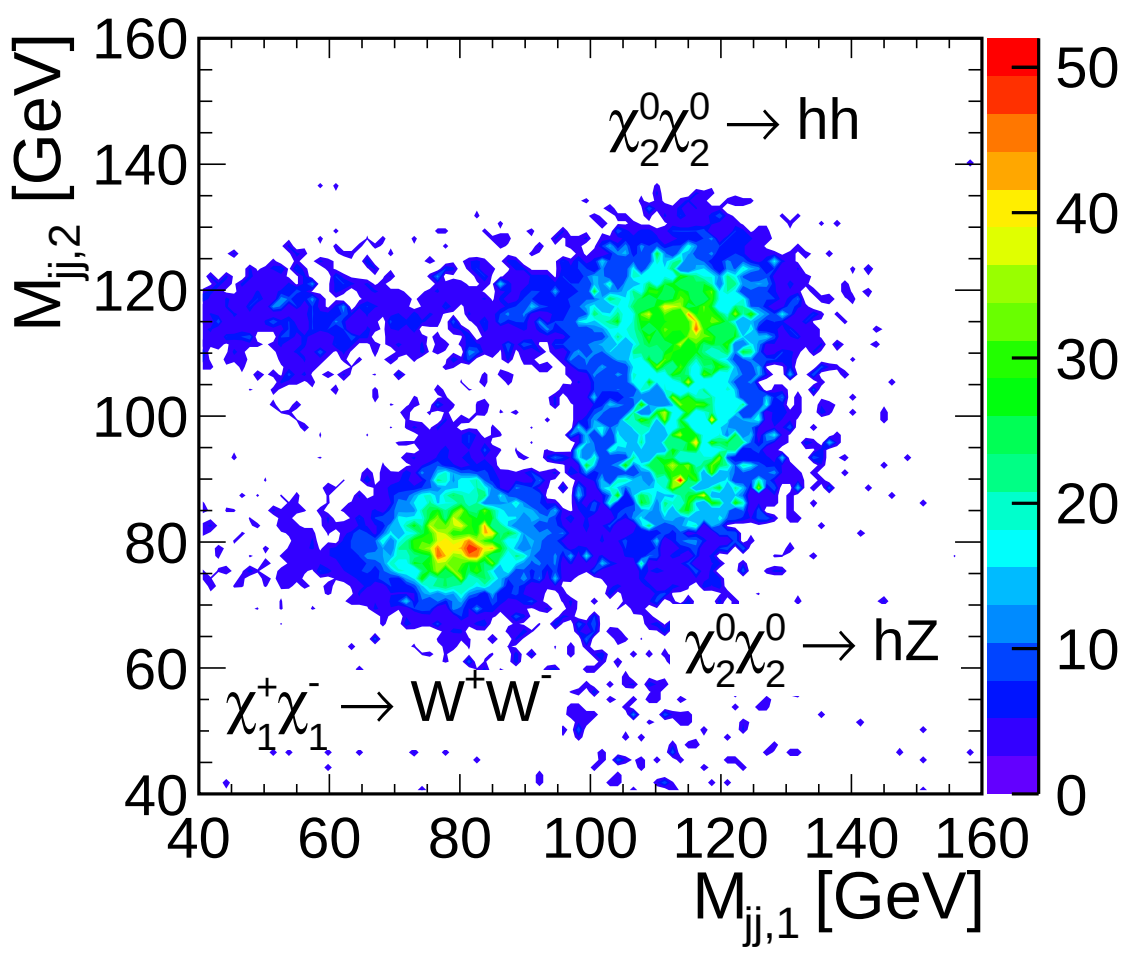

Fig. 12.10: Scatter plot showing the reconstructed di-jet invariant masses of W, Z and h candidates in simulated $e^+e^- \rightarrow \tilde{\chi}_1^+\tilde{\chi}_1^-$ and $e^+e^- \rightarrow \tilde{\chi}_2^0\tilde{\chi}_2^0$ signal events including $\gamma\gamma \rightarrow$ hadrons background. The peaks corresponding to the individual chargino and neutralino decays are indicated. The event samples were scaled to have a similar number of events for each channel.

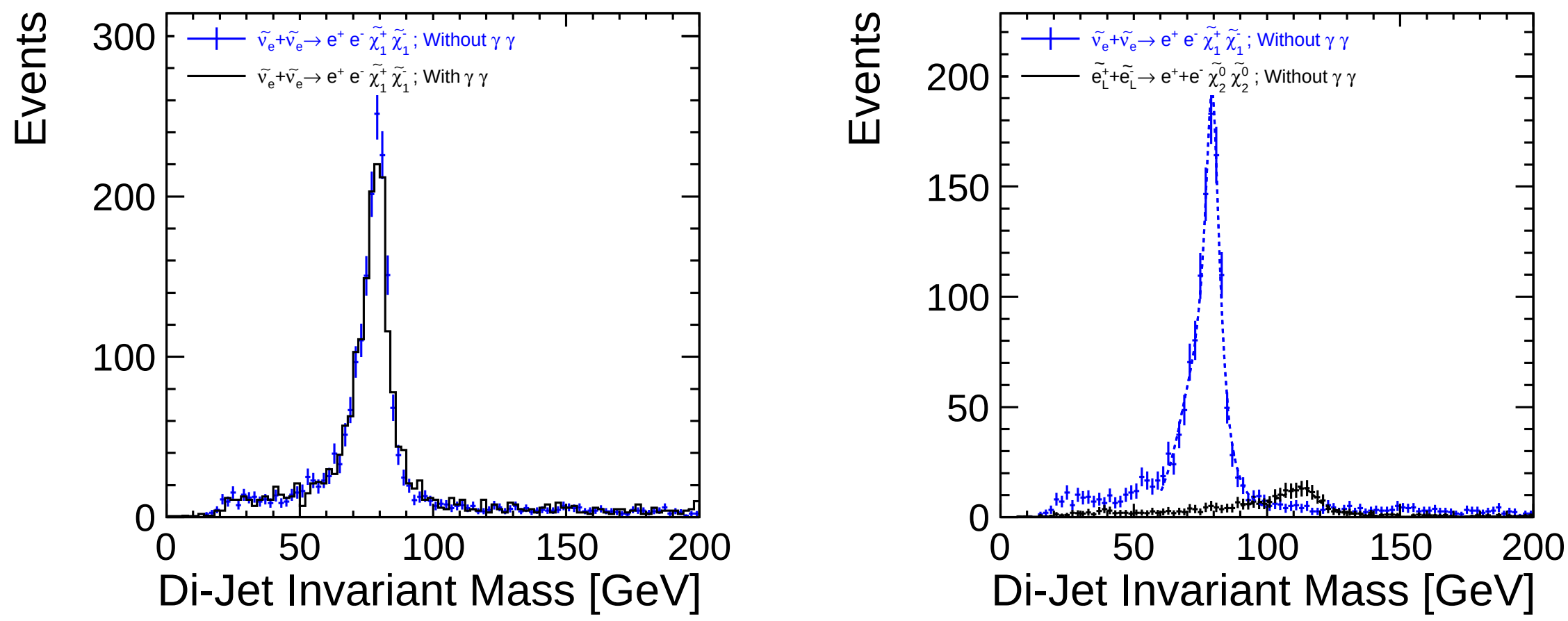

(a) $e^+e^- \to \tilde{\nu}_e\tilde{\nu}_e$ (b) $e^+e^- \to \tilde{\nu}_e\tilde{\nu}_e$ and $e^+e^- \to \tilde{e}^+_L\tilde{e}^-_L$

Fig. 12.11: Reconstructed di-jet mass distributions for processes with two isolated leptons, four jets and missing energy. The W candidate distributions obtained for $e^+e^- \to \tilde{\nu}_e\tilde{\nu}_e$ events without (blue points) and including (black histogram) the simulation from $\gamma\gamma \to$ hadrons interactions are compared in the left plot. The `tight` cuts were applied to the PFOs with background. In the right plot, luminosity-scaled distributions of the boson masses for decays to W bosons (blue points) and h and Z bosons (black points) in $e^+e^- \to \tilde{\nu}_e\tilde{\nu}_e$ and $e^+e^- \to \tilde{e}^+_L\tilde{e}^-_L$ events are compared.

Finally, Figure 12.11b shows the reconstructed boson mass distribution for the processes $e^+e^- \to \tilde{\nu}_e\tilde{\nu}_e$ and $e^+e^- \to \tilde{e}^+_L\tilde{e}^-_L$ corresponding to an integrated luminosity of 2 ab$^{-1}$. The distributions are fitted with two Breit–Wigner functions. The mass distribution of the Higgs boson is broader than that of the W boson, due to a 10% background component from Z boson decays and due to semi-leptonic heavy flavour decays in the $h \to b\bar{b}$ process. In a realistic analysis, the flavour tagging will help to separate W and h final states.

### 12.3.4 Flavour Tagging

An important criterion in the design of the inner tracking detectors is the resolution of secondary interactions from the primary vertex, and the identification of bottom and charm decays. The flavour identification package developed by the LCFI [37] collaboration consists of a topological vertex finder ZVTOP, which reconstructs secondary interactions, and a multivariate classifier which combines several jet-related variables to tag bottom, charm, and light quark jets.

The vertex finder identifies regions in space where two or more tracks overlap. These vertex candidates are then fit with kinematic constraints. Figure 12.12a shows the resolution in $xy$ (green line) and $z$ (blue line) of the primary vertex position versus the number of tracks in the vertex in $e^+e^- \to q\bar{q}\nu\bar{\nu}$ events with a mean jet energy of 130 GeV. The events were simulated in the CLIC_SiD detector and $\gamma\gamma \to$ hadrons backgrounds were overlaid. The lines show functions of the form $1/N_{\mathrm{Tracks}}$ which provide an empirical parametrisation of the resolution. The reconstructed primary vertex position in the $xy$ plane for vertices with more than 20 tracks is shown in Figure 12.12b.

Displaced vertices are the most significant characteristic of b quark decays, thus several vertex-related variables are combined in the tagging classifier. The rate with which jets of one flavour are tagged as a different flavour is used to assess the performance of the package. Figure 12.13a shows the mis-tag rate for c-jets (blue line) and light jets (green line) as b-jets versus the b-tag efficiency, while Figure 12.13b shows the mis-tag rate for b-jets (red line) and light jets (green line) as c-jets versus the c-tag efficiency. The presence of $\gamma\gamma$ backgrounds is found to reduce the flavour tagging performance, although the effect is limited.

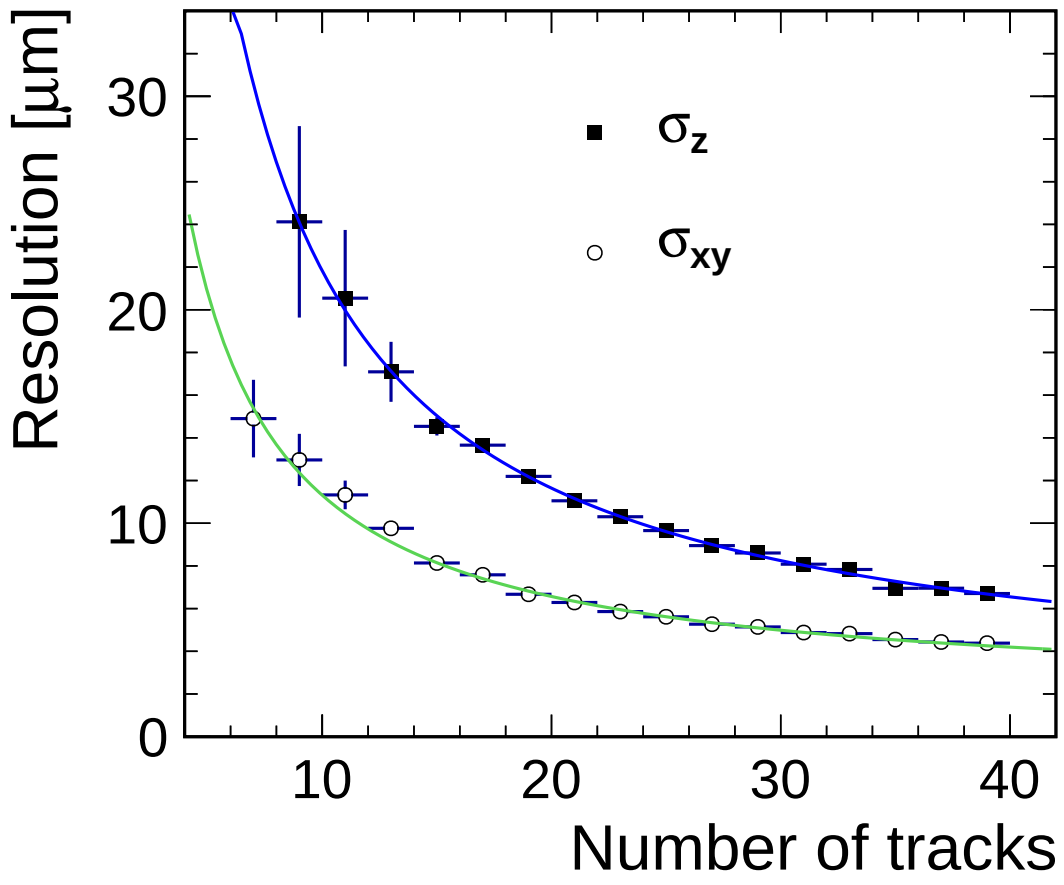

(a) $x-y$ and $z$ resolution

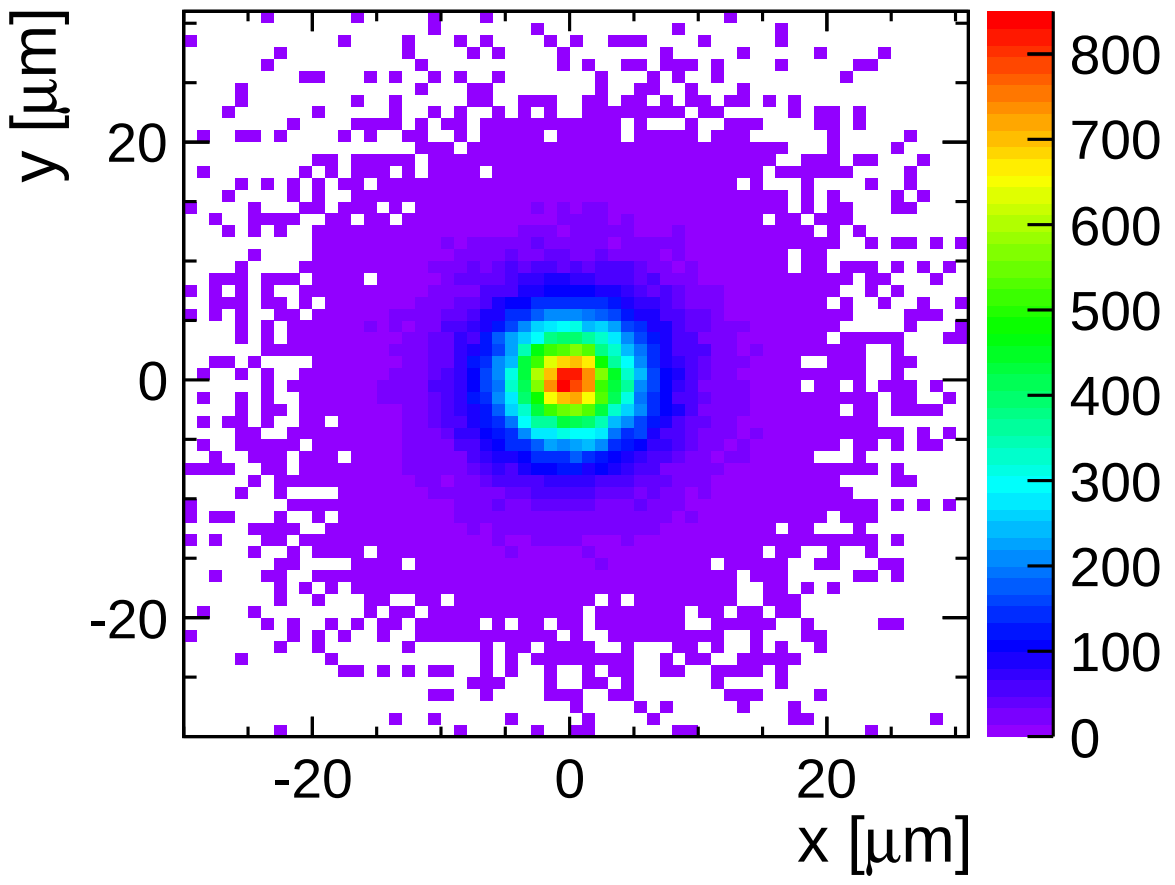

(b) $x-y$ view

Fig. 12.12: Position resolution of the primary vertex in the $x-y$ (green line) and $z$ (blue line) direction as a function of the number of tracks in the vertex (left). The $x-y$ position of primary vertices with more than 20 tracks (right).

Because of the large boost of b jets at typical CLIC energies, approximately 30% of the secondary tracks originate from a decay beyond the innermost layer of the vertex detector. Jets with no secondary reconstructed vertex account for 20–30% of energetic b jets and over 90% of light quark jets. Those from b jets are due to decays with less than two charged decay products, decays beyond the first vertex layer or failure of secondary vertex reconstruction. The heavy Higgs analysis therefore complements the use of variables from the secondary vertex search with a track-based reconstruction of the secondary system and with tagged lepton observables to recover cases in which no detached vertices are reconstructed. The performance is illustrated in Figure 12.14 showing the mis-tag rate for light jets as b as a function of the b-tag efficiency.

## 12.4 Detector Benchmark Processes

The six detector benchmark processes, described in detail in [9] and in Section 2.6, have been simulated and reconstructed using the CLIC_ILD or CLIC_SiD detector concepts. As explained in Section 12.1.1, signal and physics background events are generated at centre-of-mass energies of 3 TeV for five processes and at 500 GeV for one process. The analyses at 3 TeV assume an integrated luminosity of 2 ab$^{-1}$ corresponding to four years of operation of a fully commissioned machine running 200 days per year with an effective up-time of 50%. The study at 500 GeV assumes an integrated luminosity of 100 fb$^{-1}$. The luminosity spectrum as well as initial and final state radiation are taken into account and background from $\gamma\gamma \rightarrow$ hadrons events is overlaid before the digitisation stage. After full detector simulation and event reconstruction, event selections are applied and the physics signals are extracted. The following subsections describe the event selection and analysis results.

### 12.4.1 Light Higgs Decay to $\mathrm{b\bar{b}}$ and $\mathrm{c\bar{c}}$

A fundamental test of the Standard Model Higgs mechanism is the predicted scaling of the Higgs couplings to fermions in proportion to their masses. In $e^+e^-$ annihilations, at 3 TeV the dominant Standard

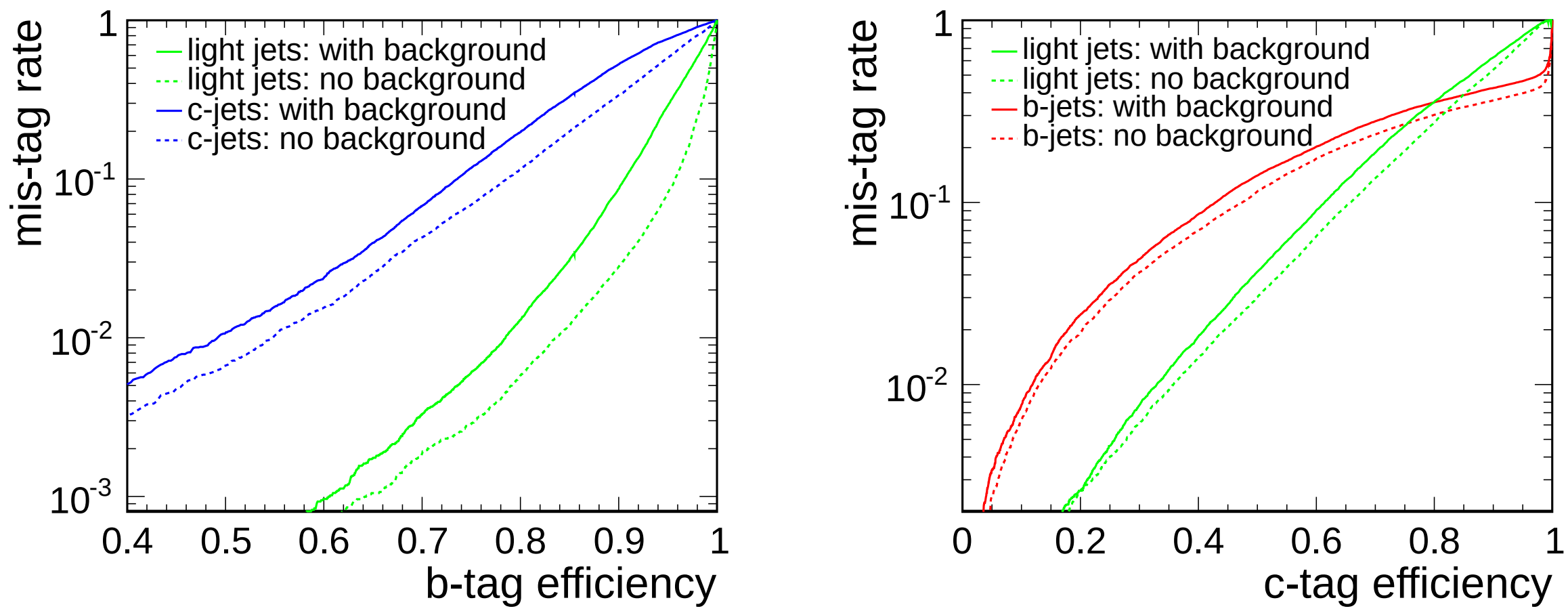

(a) b-tag

(b) c-tag

Fig. 12.13: In the left plot, the mis-tag rate in the CLIC_SiD detector for charm (blue) and light (green) jets as a function of the b-tag efficiency is shown. The right plot shows the mis-tag rate for bottom (red) and light (green) jets as a function of the c-tag efficiency. The mean $p_T$ of the jets is 70 GeV while the mean energy is $\sim$ 130 GeV.

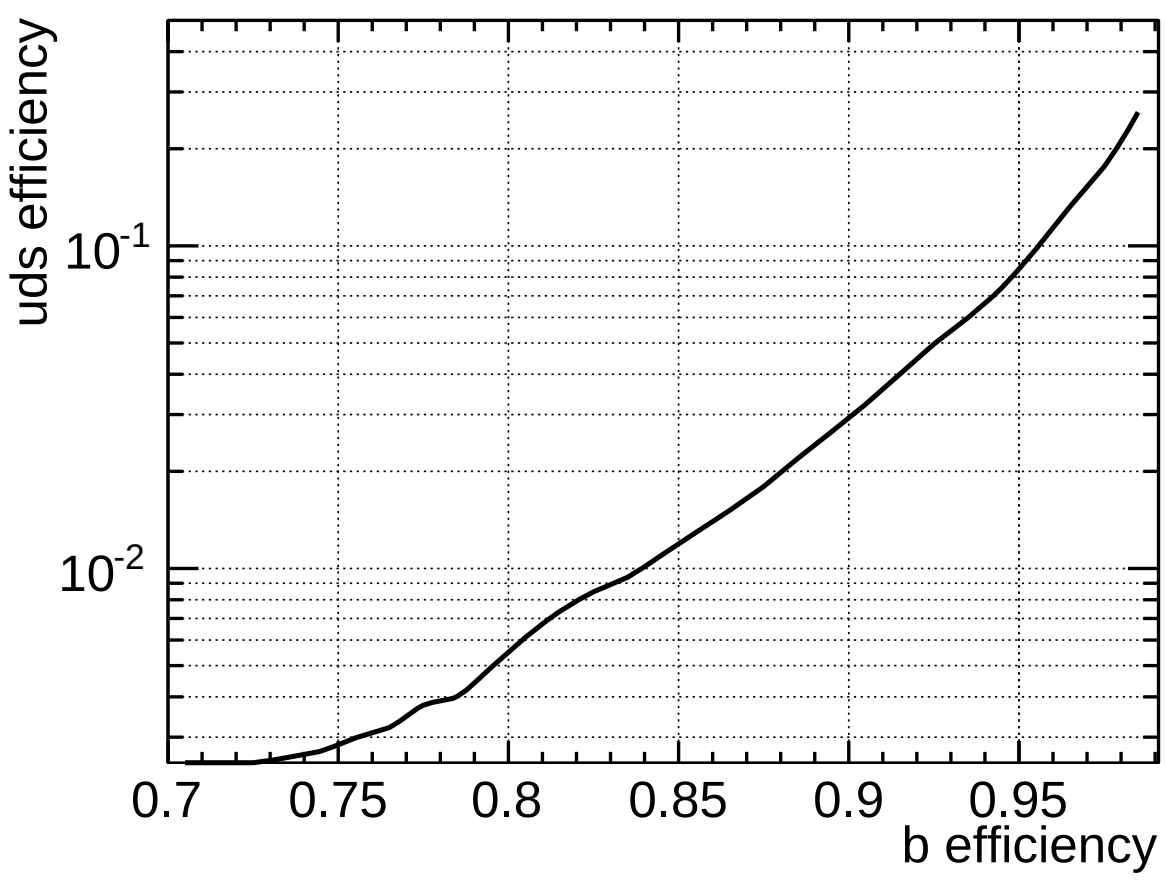

Fig. 12.14: Efficiency of b tagging for b jets in HA $\longrightarrow b\bar{b}b\bar{b}$ events as a function of the misidentification probability for light flavour (u, d and s) jets of the same kinematics.

Model Higgs production mechanism is via $W^+W^-$ and $Z^0Z^0$ fusion (see Section 1.2). For the light Higgs studies the Higgs mass is assumed to be 120 GeV and the corresponding production cross sections are 420 fb, for the $W^+W^-$ fusion, and 42.6 fb, for the $Z^0Z^0$ fusion. The production through $W^+W^-$ fusion was investigated for this analysis. With a total integrated luminosity of 2 ab$^{-1}$, one expects to produce about $8.4 \cdot 10^5$ $h\nu_e\bar{\nu}_e$ events, allowing an accurate measurement of the branching ratios to heavy quarks as well as the measurement of the branching ratio of rare decays. The statistical accuracy on the cross section times branching ratio measurement is investigated in the decay channels $h \longrightarrow b\bar{b}$ and $h \longrightarrow c\bar{c}$. These have branching ratios of $6.8 \cdot 10^{-1}$ and $3.6 \cdot 10^{-2}$ respectively.

The analysis is performed in the framework of the CLIC_SiD concept. It is presented in detail in [38]. The data samples considered for this analysis are listed in Table 12.3.

Table 12.3: Production cross sections for the signal processes and for the backgrounds considered for the analysis of $h \to b\bar{b}$ and $h \to c\bar{c}$ decays.

| Type | Final state | Cross section $\sigma$ (fb) |
|---|---|---|
| Signal | $h\nu_e\bar{\nu}_e$, $h \to b\bar{b}$ | 285 |
| Signal | $h\nu_e\bar{\nu}_e$, $h \to c\bar{c}$ | 15 |
| Background | $q\bar{q}\nu\bar{\nu}$ | 1305 |
| Background | $q\bar{q}e\bar{\nu}_e$ | 5255 |
| Background | $q\bar{q}e^+e^-$ | 3341 |
| Background | $q\bar{q}$ | 3076 |

#### *12.4.1.1 Event Selection*

After reconstruction, the events are forced into two jets using the exclusive $k_t$ algorithm of the FASTJET package [31], where the parameter $R$ is set to 0.7. The LCFI vertexing package [37] is used to identify jets according to their quark content: b, c and light quarks. It computes the corresponding jet flavour tag values.

For each event, the following variables are computed and used as input to the neural network, which performs the event classification:

- the invariant mass of the di-jet system;
- the sum of the LCFI jet flavour tag values;
- the maximum of the absolute values of jet pseudorapidities;
- $R_{\eta\phi}$, the distance of jets in the $\eta - \phi$ plane;
- the sum of jet energies;
- the total number of leptons in an event;
- the total number of photons in an event;
- acoplanarity of jets.

Figure 12.15 shows the distributions of the two most discriminating variables, the invariant mass of the di-jet system and the sum of the LCFI jet flavour tag values.

For the neural network training the data samples are scaled to the same integrated luminosity. The ability of the neural network to separate the Higgs signal from the background is illustrated in Figure 12.16, where the region below a classifier value of 0.15 is not shown due to the very large number of background entries per bin. The arrow indicates the chosen event selection cut.

The signal cross section uncertainty and the signal purity depend on the choice of this selection cut. Figures 12.17a and 12.17b show the signal cross section measurement uncertainty and the signal purity as a function of the signal selection efficiency, respectively. The uncertainty distribution has a wide flat bottom, thus allowing for the selection of an optimal working point over a wide range of selection efficiencies. The cut value is chosen such as to obtain the lowest statistical uncertainty on the cross section measurement. This uncertainty is proportional to $\sqrt{S+B}/S$, while the signal purity is given by $S/(S+B)$, where $S$ and $B$ are the total number of selected signal and background events, respectively. The cross section uncertainty does not include systematic uncertainties on the purity and on the signal efficiency.

#### *12.4.1.2 Results*

The results are summarised in Table 12.4. The statistical uncertainty on the cross section times branching ratio of the decay $H \to b\bar{b}$ is 0.22% with a corresponding signal selection efficiency of 54.6%. Using the c-flavour tag value instead of the b-flavour tag value to identify the signal, the same analysis procedure is used to derive the statistical uncertainty on the cross section times branching ratio for the decay $h \to c\bar{c}$. It was found to be 3.24%, with a signal selection efficiency of 15.2%.

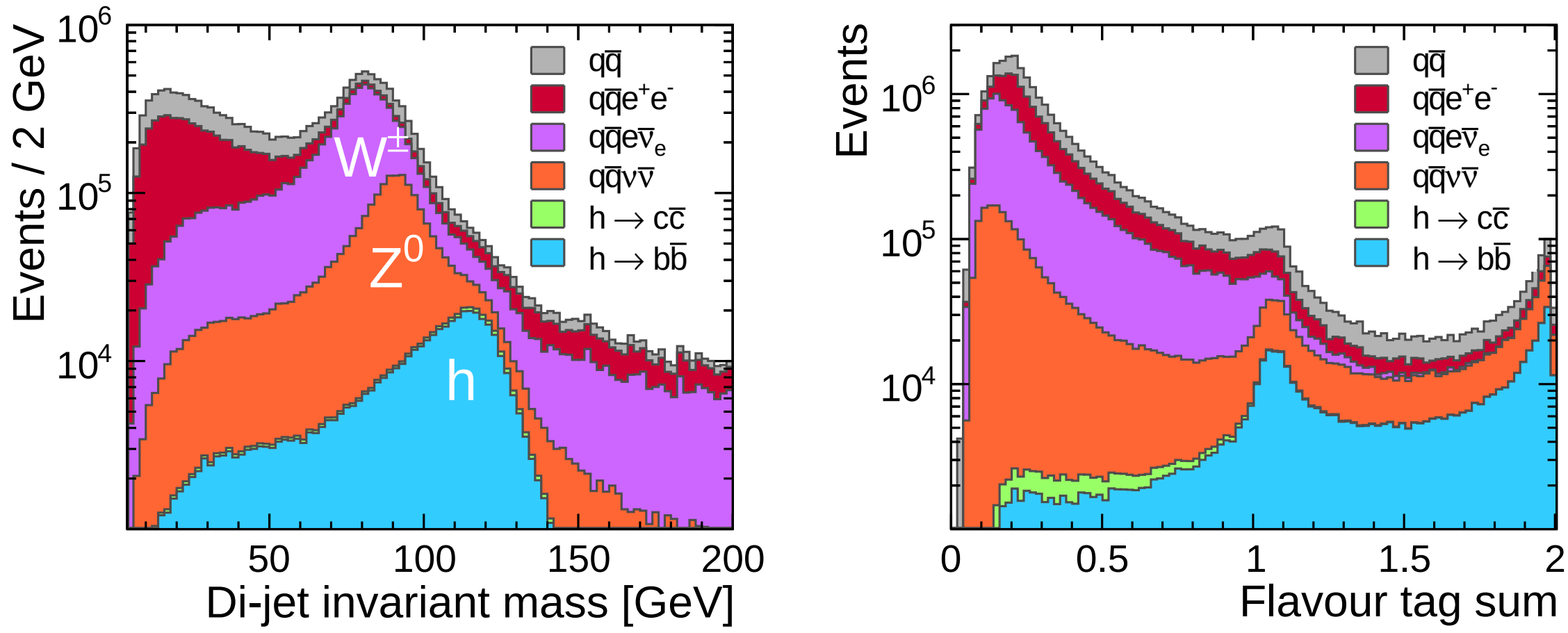

Fig. 12.15: Distributions of the di-jet invariant mass (left) and sum of the LCFI jet flavour tag values (right) for the $h \to b\bar{b}$ and $h \to c\bar{c}$ signals and for the individual backgrounds. All contributions are stacked and scaled to an integrated luminosity of 2 $ab^{-1}$.

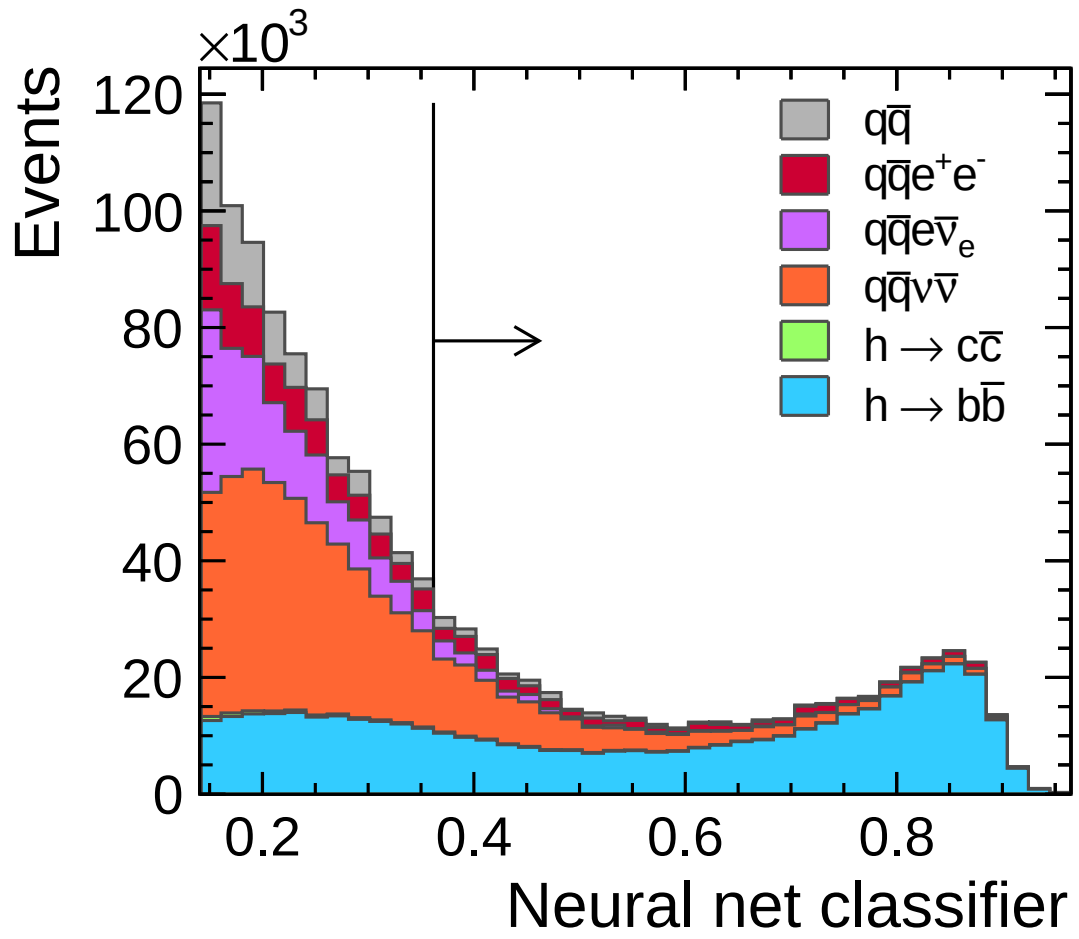

Fig. 12.16: Distribution of the event selection neural network classifier for the $h \to b\bar{b}$ and $h \to c\bar{c}$ signals and for the individual backgrounds. The arrow shows the event selection cut applied in the analysis.

### 12.4.2 Light Higgs Decay to Muons

The measurement of the small $h \to \mu^+\mu^-$ branching ratio provides an important test of the expected linear relation between the mass of a fermion and its coupling to the Higgs boson. As introduced in Section 12.4.1, the Higgs production at CLIC is largely dominated by gauge boson fusion processes. For a Standard Model Higgs boson with a mass of 120 GeV, produced through $W^+W^-$ fusion, the cross section times branching ratio into muons is 0.12 fb, making its measurement very challenging. The two main background channels are the irreducible background $e^+e^- \to \mu^+\mu^-\nu\bar{\nu}$ with a cross section of 132 fb, and the process $e^+e^- \to \mu^+\mu^-e^+e^-$ with a cross section of 5.4 pb, where the electrons are mostly very forward. Other channels that share the signal topology are $e^+e^- \to \mu^+\mu^-$ (350 fb), $e^+e^- \to \tau^+\tau^-$

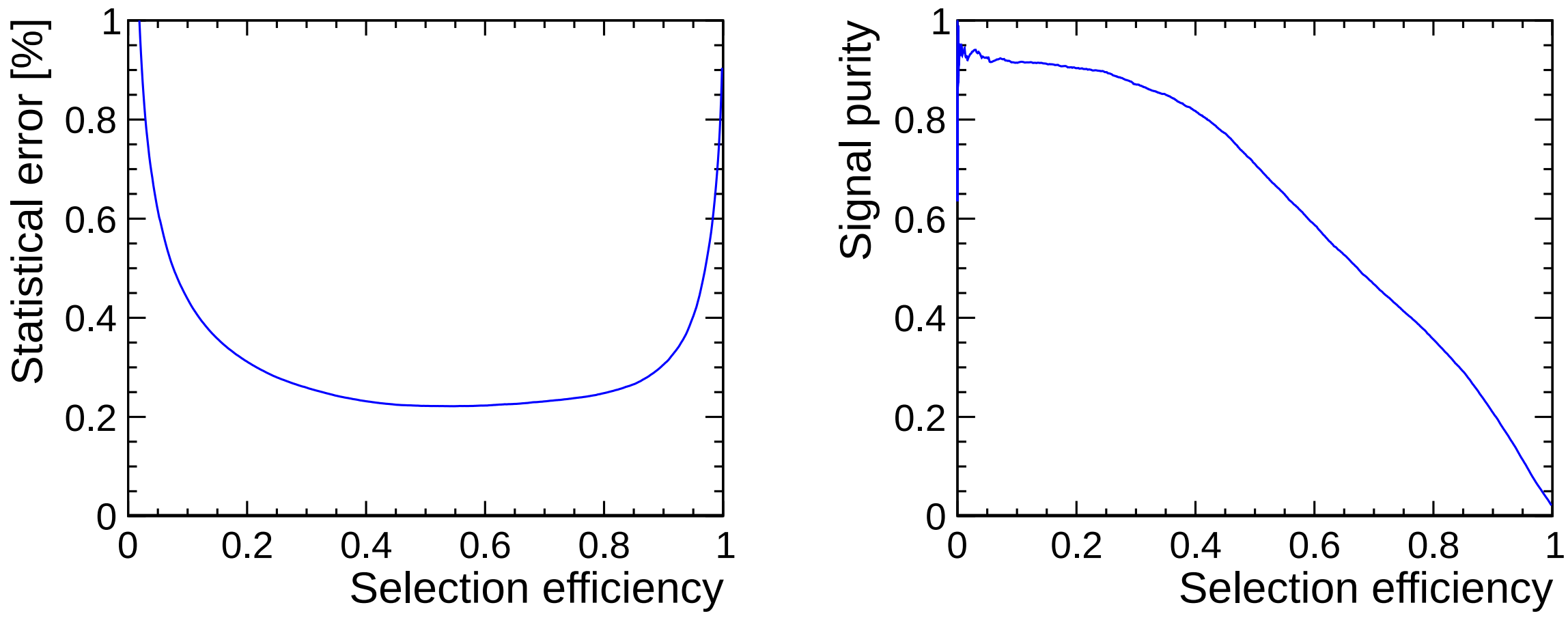

(a) (b)

Fig. 12.17: Statistical uncertainty of the extracted cross section for $h \to b\bar{b}$ (left) and signal purity in the selected sample (right) as a function of the signal selection efficiency.

Table 12.4: Signal purities and efficiencies as well as measured cross sections with statistical uncertainties for $h \to b\bar{b}$ and $h \to c\bar{c}$ decays. All numbers are obtained assuming an integrated luminosity of 2 $ab^{-1}$.

| | $h \to b\bar{b}$ | $h \to c\bar{c}$ |
|---|---|---|
| Signal purity | 65.4% | 24.1% |
| Signal efficiency | 54.6% | 15.2% |
| cross section statistical uncertainty | 0.22% | 3.24% |

(250 fb), $e^+e^- \to \tau^+\tau^-\nu\bar{\nu}$ (125 fb) and the beam-induced incoherent muon pair background (20 pb)[1].

The analysis presented here is performed in the CLIC_SiD framework and is documented in full detail in [39].

#### *12.4.2.1 Event Selection*

For the initial selection two reconstructed muons are required. The average reconstruction efficiency for muons with a polar angle of 10° or larger is 99.6% but drops quickly for muons at lower angles. Due to this limit in the acceptance, the reconstruction efficiency of the signal sample is 78.7%. An additional cut on the invariant mass of the di-muon system, 105 GeV $< M(\mu\mu) <$ 135 GeV reduces the signal efficiency to 74.1%.

When adding $\gamma\gamma \to$ hadrons background, the muon reconstruction efficiency is reduced from 99.6% to 98.4%. Only those reconstructed particles with a transverse momentum larger than 5 GeV are kept to guarantee a clean event topology also in the presence of beam-induced backgrounds. This cut does not remove any of the muons in the signal events, but removes most of the particles coming from $\gamma\gamma \to$ hadrons events. The $\gamma\gamma \to$ hadrons background does not affect the momentum resolution and reconstructed kinematic quantities in a significant way. Therefore the large physics background samples produced for this analysis are not simulated with $\gamma\gamma \to$ hadrons events.

[1] The cross section of the muon pair background includes a generator-level cut of 100 GeV $< M(\mu\mu) <$ 140 GeV.

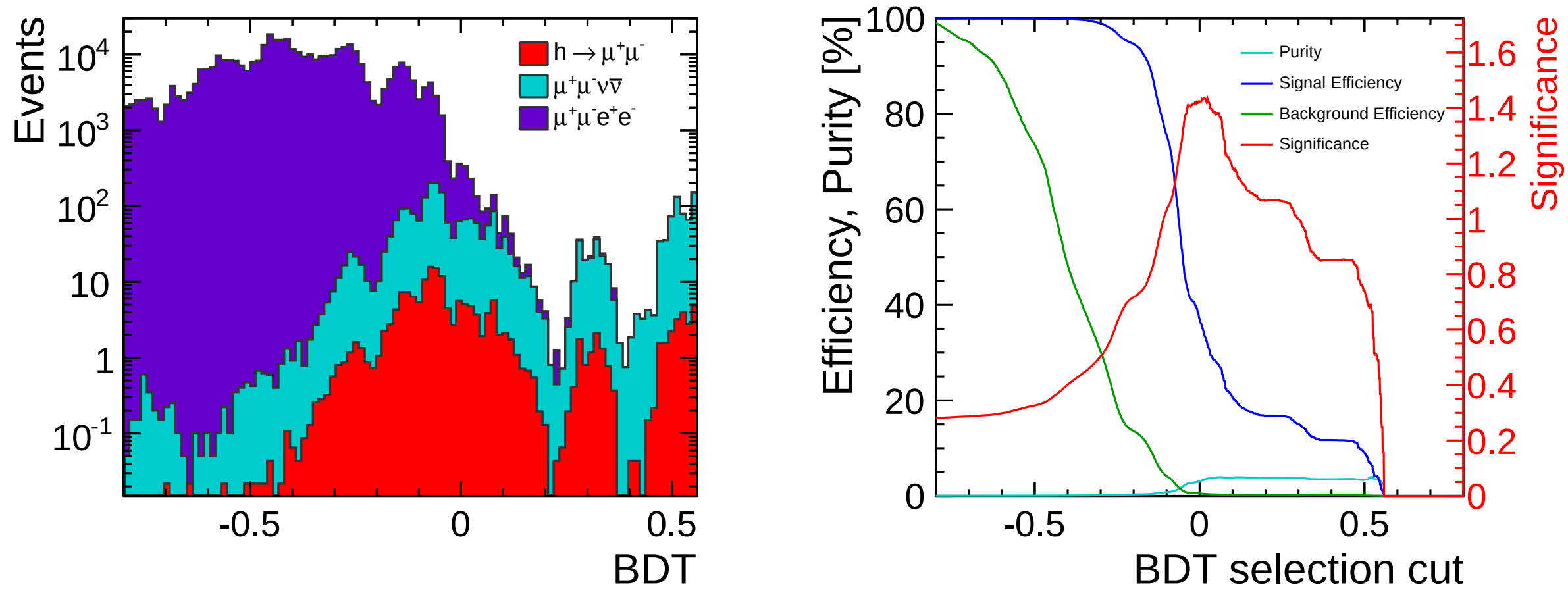

Fig. 12.18: BDT classifier distribution for $h \to \mu^+\mu^-$ events and for the two most prominent backgrounds, stacked and scaled to an integrated luminosity of 2 $ab^{-1}$ (left). Resulting significance, purity, signal efficiency and background efficiency depending on the selection value (right).

Aside from the signal, the two remaining backgrounds after these selection cuts are $e^+e^- \to \mu^+\mu^-\nu\bar{\nu}$ and $e^+e^- \to \mu^+\mu^-e^+e^-$. They still add up to approximately 200 $fb^{-1}$, but they are further reduced using the boosted decision tree classifier (BDT) implemented in TMVA [40]. The following kinematic variables are used as an input to the BDT:

- $E_{vis}$ - the sum of the energy of the reconstructed particles after PFO-selection without the two muons;
- $p_T(\mu_1) + p_T(\mu_2)$ - the scalar sum of the transverse momenta of the two muons;
- $p_T(\mu\mu)$ - the transverse momentum of the di-muon system;
- $\theta(\mu\mu)$ - the polar angle of the di-muon system;
- $\beta(\mu\mu)$ - the relativistic velocity of the di-muon system;
- $\cos\theta^*$ - the angle between a muon in the di-muon rest frame and the di-muon system in the lab frame. The muon chosen is the one with the highest energy in the lab frame.

The training of the BDT is performed using large statistically independent samples of signal events, as well as background events from the $\mu^+\mu^-e^+e^-$ channel. The $\mu^+\mu^-\nu\bar{\nu}$ background is not used in the training, since it is indistinguishable from the signal in the relevant invariant mass region. The response of the BDT is shown in Figure 12.18 (left). The cut on the BDT response is chosen to be such that the highest signal significance $S/\sqrt{S+B}$ is obtained, where $S$ and $B$ are the number of selected signal and background events, respectively. The maximum significance is obtained at a BDT selection cut value of 0.0135, see Figure 12.18 (right). Together with the reconstruction efficiency and detector acceptance this results in a total signal selection efficiency of 25.2%.

The impact of the $\gamma\gamma \to$ hadrons background on the measurement has been studied by removing the visible energy from the BDT, replacing it with a cut $E_{vis} < 150$ GeV and applying the selection procedure to the signal sample with $\gamma\gamma \to$ hadrons events and to the physics backgrounds without. This step accounts for the fact that the overlaid $\gamma\gamma \to$ hadrons events alter the distribution of the visible energy. As removing a variable reduces the discriminating power of the BDT, the effect is likely overestimated. This method results in a slightly worse signal and background separation and the BDT selection cut with the highest significance yields a total signal selection efficiency of 21.7%.

The impact of electron tagging in the forward calorimeters to reject $\mu^+\mu^-e^+e^-$ events is studied by assuming an ad-hoc electron tagging within the fiducial volume of the LumiCal, between 44-80 mrad (see Table 9.1). Incoherent electron pairs produced from beamstrahlung lead to an occupancy in the LumiCal of between 2% and 20% per readout cell and bunch crossing, depending on the polar angle (see Section 9.3). The LumiCal is not used in the full simulation. Nevertheless, the low energy deposited by the incoherent electron pairs compared to a typical electron energy of several hundred GeV in $\mu^+\mu^-e^+e^-$ events makes a single electron tagging efficiency of 95% plausible. To simulate electron tagging, events of the $\mu^+\mu^-e^+e^-$ sample are randomly rejected according to the tagging efficiency, if the true polar angle of one of its electrons lies within the fiducial volume of the LumiCal. Afterwards, two independent BDTs are trained on the reduced background sample and the signal sample with and without $\gamma\gamma \rightarrow$ hadrons background, as outlined above. The signal and background separation by the BDTs is largely improved by the a priori removal of the background events. The BDT trained on signal events without $\gamma\gamma \rightarrow$ hadrons background yields a signal selection efficiency of 49.3% at the maximum signal significance, while the BDT trained on signal events with $\gamma\gamma \rightarrow$ hadrons events yields a signal selection efficiency of 50.0% and a slightly lower maximum signal significance.

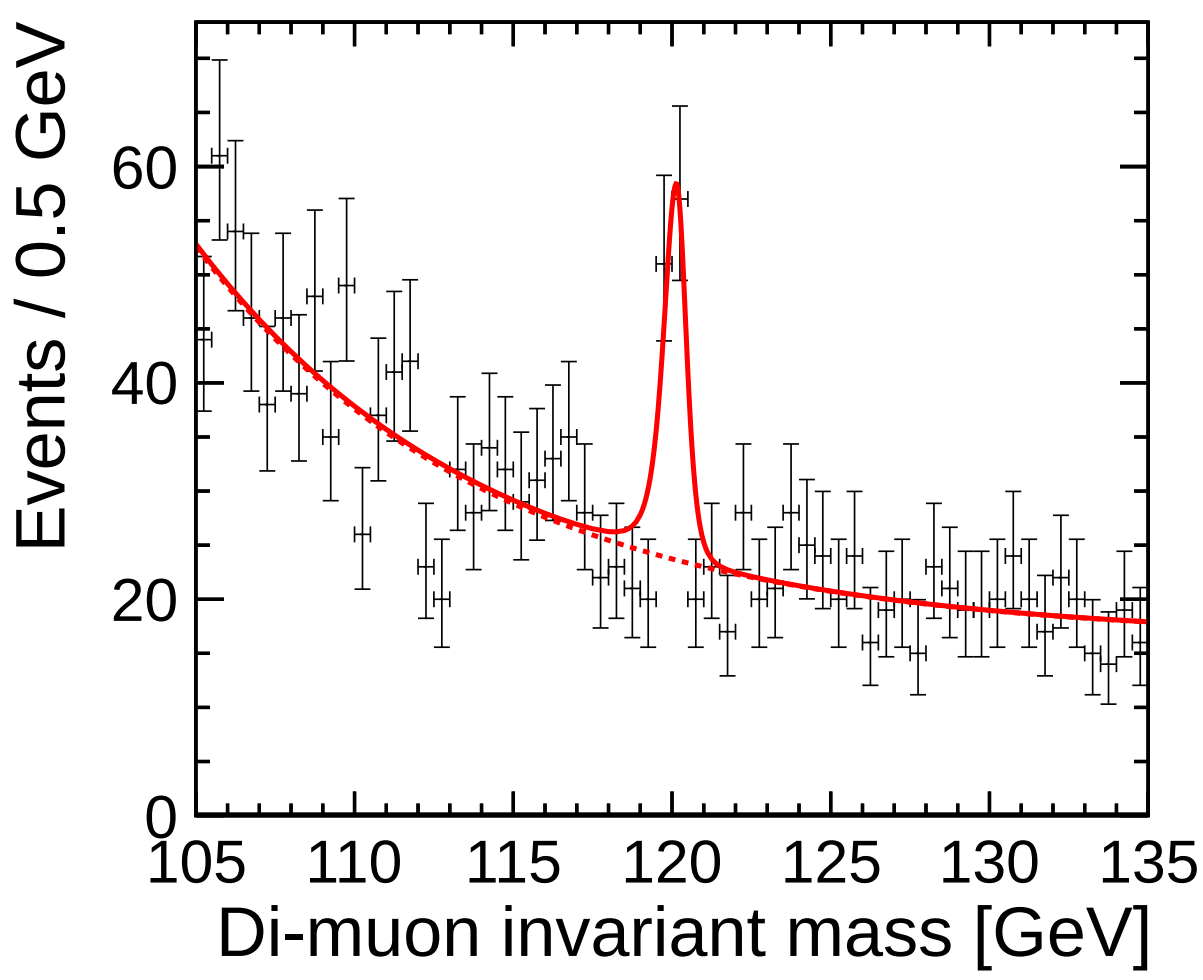

Fig. 12.19: Toy Monte Carlo sample for an integrated luminosity of 2 $ab^{-1}$ with a fit to the signal and background hypothesis shown in red. The red dashed line shows the background contribution resulting from the fit.

#### *12.4.2.2 Higgs Mass Fit*

After the application of the cut on the BDT response, the resulting invariant mass distributions for the different channels are fitted individually using the RooFit framework [41] to obtain their PDFs.

Each of the two background channels is modelled by an exponential function plus a flat contribution, which leads to two free parameters each, one for the exponential function and another for the ratio of the exponential and the flat components. After the event selection the $\mu^+\mu^-e^+e^-$ background is mostly flat, while the $\mu^+\mu^-\nu\bar{\nu}$ background is dropping exponentially with increasing mass. The signal distribution is Gaussian with two asymmetric tails. It is modelled by two half Gaussian distributions with two independent tail parameters, which, together with the mean position, leads to a total of five free parameters. Assuming that this measurement will be a test for the SM, all parameters that define the shapes are fixed afterwards. The three models are added and only three normalisation parameters are left as free parameters.

The expected measurement accuracy for an integrated luminosity of 2 $ab^{-1}$ is determined as the average from 100 toy Monte Carlo fits. In order to simulate a measurement, random events are picked for the three contributions according to the desired luminosity and their respective cross section. The background events are generated randomly from their PDFs, while the signal events are picked randomly from the fully simulated event sample.

The randomly generated data sets are then fitted with the combined PDF of signal and background using an un-binned likelihood fit. The fitted scale factor together with the fixed shape of the signal PDF yields the number of signal events $N_s$, which translates to the measured value of $\sigma \times \mathrm{BR} = N_s/(\mathscr{L}\varepsilon_s)$dividing by the luminosity $\mathscr{L}$ and the selection efficiency $\varepsilon_s$. One example measurement is shown in Figure 12.19.

#### *12.4.2.3 Results*

The selection efficiencies and statistical uncertainties for the different measurements of the $\mathrm{h} \to \mu^+\mu^-$ cross section times branching ratio are summarised in Table 12.5. The obtained value of $\sigma_{\mathrm{h}\nu_e\bar{\nu}_e} \times \mathrm{BR}_{\mathrm{h}\to\mu^+\mu^-}$ is in all cases consistent with the generated value. Overlaying the $\gamma\gamma \to$ hadrons background degrades the achievable relative statistical uncertainty from 23% to about 26%.

The result can be improved by using the forward calorimeters to tag electrons and reject more $\mu^+\mu^-\mathrm{e}^+\mathrm{e}^-$ events. By randomly rejecting $\mathrm{e}^+\mathrm{e}^-\mu^+\mu^-$ events, corresponding to an electron tagging efficiency of 95% in the LumiCal, a relative statistical uncertainty of 15.0% without and 15.7% with $\gamma\gamma \to$ hadrons background can be achieved. Dedicated full simulation studies are needed to verify the feasibility of the assumed electron tagging efficiency.

All given uncertainties are statistical only. Based on the LEP measurements of $\mathrm{Z} \to \mu^+\mu^-$ [42] we estimate that the systematic uncertainty due to detector effects is negligible compared to the statistical uncertainty.

Table 12.5: Selection efficiencies and statistical uncertainties of the $\mathrm{h} \to \mu^+\mu^-$ cross section times branching ratio measurement. All numbers are obtained assuming an integrated luminosity of 2 ab$^{-1}$.

| | No forward electron tagging | | 95% tagging in LumiCal | |
|---|---|---|---|---|
| | No background | With $\gamma\gamma \to$ hadrons | No background | With $\gamma\gamma \to$ hadrons |
| Signal events | $62 \pm 14$ | $53 \pm 14$ | $120 \pm 17$ | $122 \pm 19$ |
| Signal efficiency | 25.2% | 21.7% | 49.3% | 50.0% |
| Stat. uncertainty | 23.3% | 26.3% | 15.0% | 15.7% |

### 12.4.3 Heavy Higgs Production

The precise determination of the masses and widths of the neutral CP-odd and CP-even and charged heavy Higgs bosons is an important part of the study of an extended Higgs sector in new physics models. A non-minimal Higgs sector is one of the simplest extension of the Standard Model and it acquires a special relevance in SUSY. In this case the detailed study of the heavy Higgs sector is crucial to assess the relation between new physics and cosmology through dark matter. A high-energy lepton collider is particularly well suited for such a study even in those regions of the parameter space where the sensitivity of the LHC becomes marginal. In particular, at CLIC the pair production processes, $\mathrm{e}^+\mathrm{e}^- \to \mathrm{HA}$ and $\mathrm{e}^+\mathrm{e}^- \to \mathrm{H}^+\mathrm{H}^-$ give access to all four heavy Higgs states almost up to the kinematic limit [34, 43].

In this benchmark study we consider two SUSY models *SUSY model I* with $M_A$ = 902 GeV and *SUSY model II* with $M_A$ = 742 GeV. For the chosen sets of parameters the cross sections for $\mathrm{e}^+\mathrm{e}^- \to \mathrm{HA}$ and $\mathrm{e}^+\mathrm{e}^- \to \mathrm{H}^+\mathrm{H}^-$ pair production are given in Table 12.6. The dominant decay modes are $\mathrm{H} \to \mathrm{b\bar{b}}$, $\mathrm{A} \to \mathrm{b\bar{b}}$ and $\mathrm{H}^\pm \to \mathrm{tb}$ leading to $\mathrm{b\bar{b}b\bar{b}}$ and $\mathrm{t\bar{b}b\bar{t}}$ final states.

Table 12.6: Production cross sections for the $e^+e^- \to HA$ and $e^+e^- \to H^+H^-$ signal processes and for the dominant backgrounds.

| Process | $\sigma$ (model I) (fb) | $\sigma$ (model II) (fb) | Generator |
|---|---|---|---|
| HA | 0.5 | 0.7 | ISASUGRA 7.69 PYTHIA 6.215 |
| $H^+H^-$ | 1.1 | 1.6 | ISASUGRA 7.69 PYTHIA 6.215 |
| Inclusive SUSY | 77.1 | 84.9 | ISASUGRA 7.69 PYTHIA 6.215 |
| $W^+W^-$ | 728.2 | 728.2 | PYTHIA 6.215 |
| $Z^0Z^0$ | 54.8 | 54.8 | PYTHIA 6.215 |
| $t\bar{t}$ | 30.2 | 30.2 | PYTHIA 6.215 |
| $b\bar{b}b\bar{b}$ | 5.8 | 5.8 | WHIZARD |
| WWZ | 32.8 | 32.8 | CompHEP PYTHIA 6.215 |
| ZZZ | 0.5 | 0.5 | CompHEP PYTHIA 6.215 |

#### *12.4.3.1 Event Reconstruction and Selection*

The event simulation and reconstruction is done in the CLIC_ILD framework. The event selection is based on the identification of four heavy parton final states in spherical, large visible energy events with equal di-parton invariant masses. Most of the analysis criteria are common to both the $b\bar{b}b\bar{b}$ and the $t\bar{b}b\bar{t}$ channels. The analysis starts with a cut-based event pre-selection. Jet clustering is applied to pre-selected events, followed by b- and t-tagging. A kinematic fit is performed to improve the di-jet invariant mass resolution, mitigate the impact of machine-induced backgrounds on the parton energy resolution and reject the remaining physics backgrounds.

In order to reject particles which are poorly reconstructed or which are likely to originate from $\gamma\gamma \to$ hadrons events, a set of minimal quality cuts is applied. Only particles with $p_T > 1$ GeV are considered. Charged particles are also required to have at least 12 hits in the tracking detectors and the relative momentum error $\sigma_p/p < 1$. The event selection proceeds as follows: First multi-jet hadronic events with little or no observed missing energy are selected. Events are required to have at least 50 charged particles, a total reconstructed energy exceeding 2.3 TeV, an event thrust between 0.62 and 0.91, a sphericity between 0.04 and 0.75, the transverse reconstructed energy exceeding 1.3 TeV and $3 \leq N_{\text{jets}} \leq 5$, where $N_{\text{jets}}$ is the number of jets reconstructed using the Durham clustering algorithm [32] with $y_{cut} = 0.0025$. These cuts remove all the SUSY events with missing energy and the $e^+e^- \to f\bar{f}$ events. For events fulfilling these criteria, the final jet reconstruction using the anti-$k_t$ clustering algorithm in cylindrical coordinates [33], implemented in the FASTJET package [31], is performed. The choice of cylindrical coordinates is optimal, since the $\gamma\gamma \to$ hadrons events are forward boosted, similarly to the underlying events in pp collisions at the LHC, for which the anti-$k_t$ clustering has been conceived and optimised. For each event, the minimum $R$ value at which the event has exactly four jets with energies in excess of 150 GeV is used for the clustering. The di-jet invariant mass is computed from pairing these jets. Since there are three possible permutations for pairing the four energetic jets and the pair-produced bosons are expected to be (almost) degenerate in mass, the combination minimising the difference $\Delta M$ of the two di-jet invariant masses is chosen, requiring $|\Delta M| < 160$ GeV and $|\Delta M| < 150$ GeV for the HA and $H^+H^-$, respectively. Since the signal events are predominantly produced in the central region while the $\gamma\gamma \to$ hadrons and most of the SM background processes are forward peaked, only events for which the jet with the smallest polar angle, $\theta$, fulfils $|\cos\theta| < 0.92$ are accepted.

The b-tagging represents the single most effective event selection cut to separate signal events with four b hadrons from the SM backgrounds. The irreducible SM $b\bar{b}b\bar{b}$ has a cross section of only 0.5 fb and is effectively reduced by the equal di-jet mass constraint and by kinematic fitting. The b-tagging is based on the response of the vertexing variables of the ZVTOP algorithm [37]. These characterise the kinematics and topology of the secondary system in the jet. They are supplemented by the corresponding kinematic observables for the secondary system built on the basis of particle impact parameters and not vertexing, when the ZVTOP algorithm does not find any secondary vertex. This procedure allows to increase the efficiency for b jets at the higher end of the kinematic spectrum in signal events. Tagging observables are combined into a discriminating variable using the BDT classifier implemented in the TMVA package [40].

Table 12.7: Summary of the mass and width fit results for model I and II. The numbers extracted without and with background from $\gamma\gamma \rightarrow$ hadrons interactions are compared. All numbers are obtained assuming an integrated luminosity of 2 ab$^{-1}$. The given uncertainties are statistical only.

| | | SUSY *model I* | | SUSY *model II* | |
|---|---|---|---|---|---|
| | State | Mass (GeV) | Width (GeV) | Mass (GeV) | Width GeV |
| Without $\gamma\gamma$ | A/H | $902.1 \pm 1.9$ | $21.4 \pm 5.0$ | $742.7 \pm 1.4$ | $21.7 \pm 3.3$ |
| Without $\gamma\gamma$ | $H^{\pm}$ | $901.4 \pm 1.9$ | $18.9 \pm 4.4$ | $744.3 \pm 2.0$ | $17.0 \pm 4.7$ |
| With $\gamma\gamma$ | A/H | $904.5 \pm 2.8$ | $20.6 \pm 6.3$ | $743.7 \pm 1.7$ | $22.2 \pm 3.8$ |
| With $\gamma\gamma$ | $H^{\pm}$ | $902.6 \pm 2.4$ | $20.2 \pm 5.4$ | $746.9 \pm 2.1$ | $21.4 \pm 4.9$ |

In the case of charged Higgs bosons, top tagging is performed. First the event is reconstructed as a four jet event and jets are tested for their compatibility with the top mass. Then a de-clustering procedure is applied to the jets to study possible jet substructure arising from the $t \rightarrow Wb \rightarrow q\bar{q}'b$ decay. This follows the procedure originally developed for identifying highly boosted top quarks at the LHC [44, 45].

#### *12.4.3.2 Heavy Higgs Mass Fit*

In order to improve the di-jet mass resolution, a constrained kinematic fit is applied, using the port of the PUFITC kinematic fit algorithm [46] to the MARLIN framework. PUFITC was originally developed for $W^+W^-$ reconstruction in DELPHI at LEP and it has been successfully used for the reconstruction of simulated linear collider events at lower energies [47]. The kinematic fit adjusts the momenta of the four jets as $p_F = ap_M + bp_B + cp_C$, where $p_M$ is the jet momentum from particle flow, $p_B$ and $p_C$ are unit vectors orthogonal to $p_M$ and to each other and $a$, $b$ and $c$ are free parameters. For these analyses, the constraints $p_x = p_y = 0$, $E \pm |p_z| = \sqrt{s}$ and $M_{\mathrm{jj1}} = M_{\mathrm{jj2}}$ are imposed, where the second condition accounts for beamstrahlung photons radiated along the beam axis. Only events with a kinematic fit $\chi^2 < 5$ are accepted. After the kinematic fit, the relative jet energy resolution $\mathrm{RMS}_{90}/E_{\mathrm{jet}}$ for b-jets improves to $0.094 \pm 0.002$ without background and to $0.103 \pm 0.002$ with $\gamma\gamma \rightarrow$ hadrons background overlaid. The di-jet invariant mass resolution improves by more than a factor of two to $\sigma_M$ = $27.7 \pm 4.8$ GeV using the semi-inclusive anti-$k_t$ method. The use of a kinematic fit also mitigates the effect of the overlaid $\gamma\gamma \rightarrow$ hadrons events on the di-jet mass resolution. Since the nominal centre-of-mass energy is imposed, allowing for beamstrahlung, jet energies are rescaled in the fit to be consistent with $\sqrt{s}$ = 3 TeV. The di-jet invariant mass resolution is $130 \pm 15$ GeV for the raw particle flow objects with $\gamma\gamma \rightarrow$ hadrons background overlaid. It improves to $39.2 \pm 6.2$ GeV using the kinematic fit and to $36.6 \pm 4.3$ GeV using the kinematic fit and the `default` timing cut selection. Imposing the equal mass constrain reduces it further down to $19.3 \pm 3.0$ GeV.

#### *12.4.3.3 Results*

The di-jet invariant mass distributions for the $b\bar{b}b\bar{b}$ and $t\bar{b}b\bar{t}$ final states of model I and II are shown in Figure 12.20 and Figure 12.21 respectively. The masses and widths of the heavy bosons are extracted fitting these distributions with the sum of two Breit–Wigner functions, describing the signals, folded with a Gaussian resolution term. Results are summarised in Table 12.7, showing that the heavy Higgs mass can be measured to a statistical accuracy at the 0.3% level. This accuracy is achieved even in presence of $\gamma\gamma \to$ hadrons background by applying the anti-$k_t$ jet clustering and kinematic fitting.

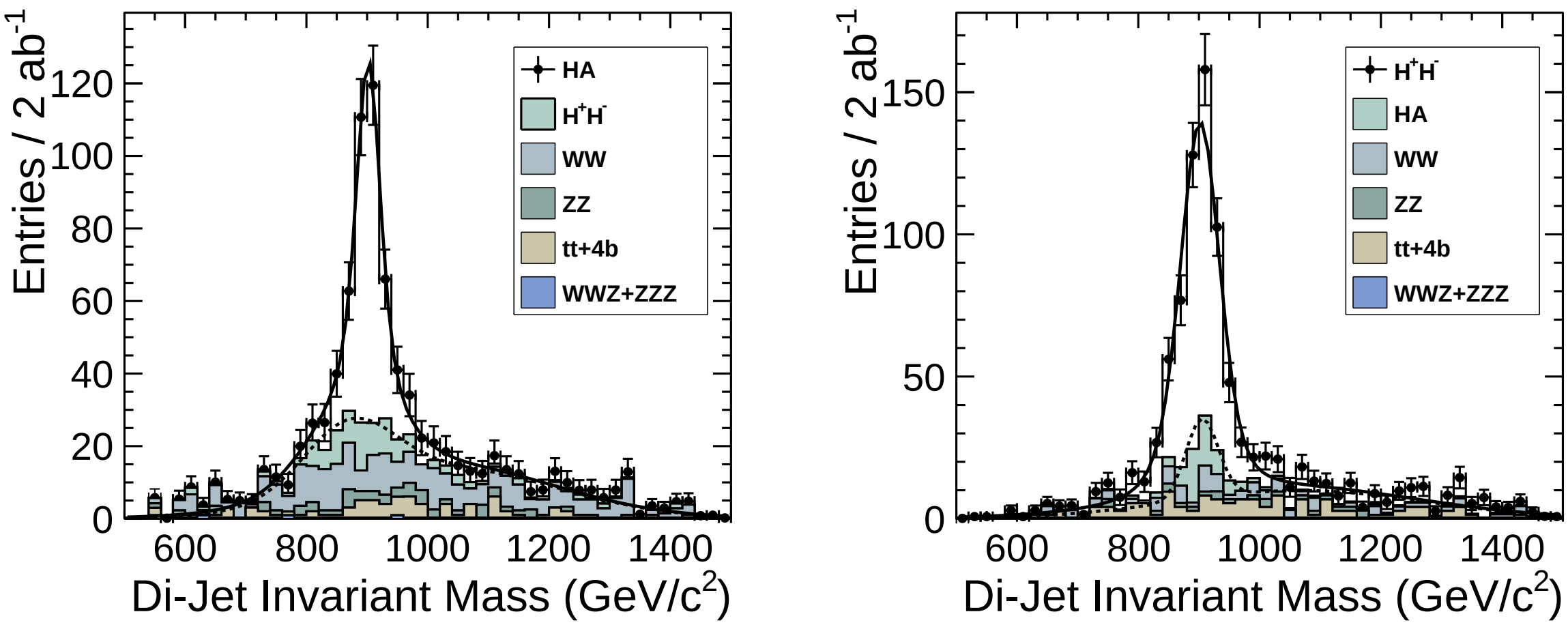

(a) $e^+e^- \to b\bar{b}b\bar{b}$ (b) $e^+e^- \to t\bar{b}b\bar{t}$

Fig. 12.20: Di-jet invariant mass distributions for the $b\bar{b}b\bar{b}$ (left) and $t\bar{b}b\bar{t}$ (right) final states for model I. The distributions for the $e^+e^- \to HA$ and $e^+e^- \to H^+H^-$ processes and for the individual backgrounds are shown separately.

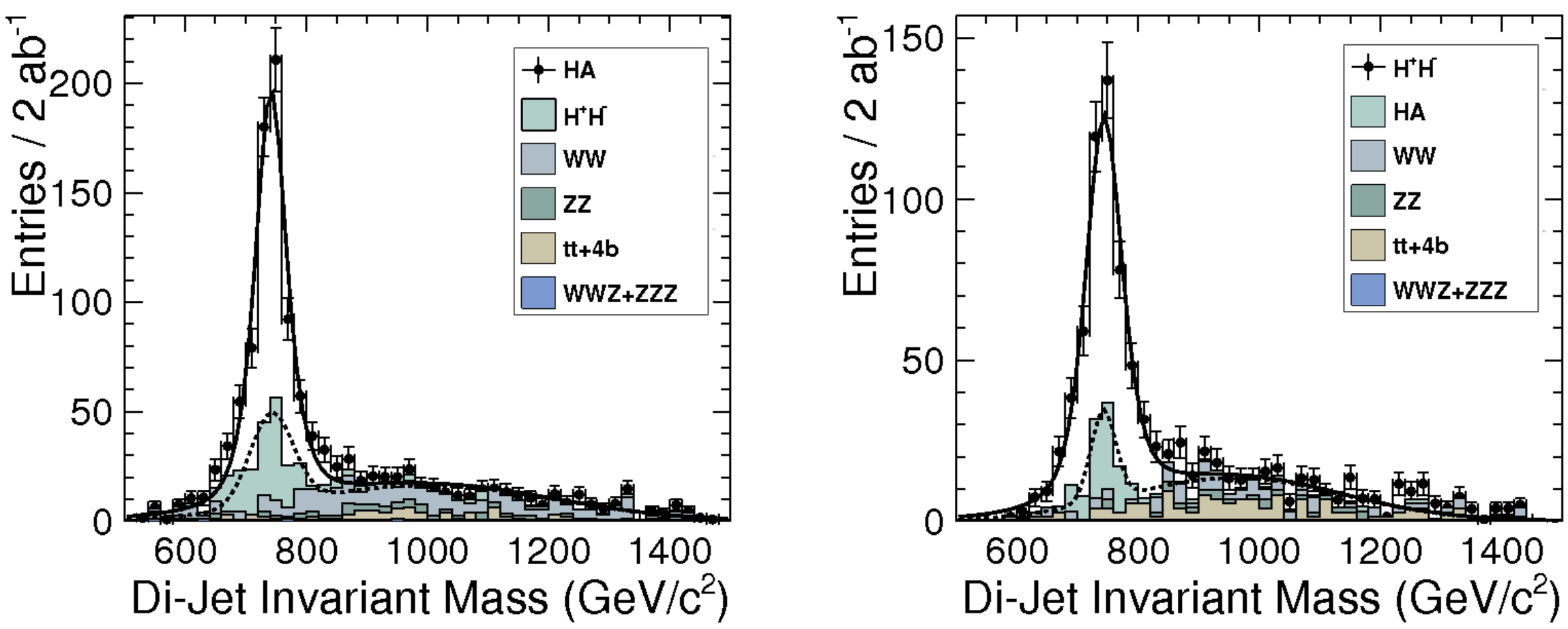

(a) $e^+e^- \to b\bar{b}b\bar{b}$ (b) $e^+e^- \to t\bar{b}b\bar{t}$

Fig. 12.21: Di-jet invariant mass distributions for the $b\bar{b}b\bar{b}$ (left) and $t\bar{b}b\bar{t}$ (right) final states for model II. The distributions for the $e^+e^- \to HA$ and $e^+e^- \to H^+H^-$ processes and for the individual backgrounds are shown separately.

### 12.4.4 Production of Right-Handed Squarks

This benchmark process, introduced in Section 2.6, provides a test of the jet energy and missing energy reconstruction for highly energetic jets in a simple topology, and for the capability to select signal events in an environment with large cross-section Standard Model backgrounds. The analysis is documented in detail in [48].

Table 12.8: Cross sections of the signal and main SM background processes used in the squark analysis.

| | Process | Cross Section |
|---|---|---|
| Signal | $e^+e^- \to \tilde{q}_R\tilde{q}_R \to q\bar{q}\tilde{\chi}_1^0\tilde{\chi}_1^0$ | 1.47 fb |
| SM background | $e^+e^- \to q\bar{q}\nu\bar{\nu}$ | $\sim$ 1500 fb |
| | $e^+e^- \to q\bar{q}e^\pm\nu$ | $\sim$ 5300 fb |
| | $e^+e^- \to \tau^+\tau^-\nu\bar{\nu}$ | $\sim$ 130 fb |

Light-flavoured squarks are typically among the heaviest particles in supersymmetric models. Right-handed squarks of the first two generations decay almost exclusively into their Standard Model partner quark and the lightest neutralino, resulting in a very general event signature of two energetic jets and missing energy. For the parameters of *SUSY model I*, $m_{\tilde{u}_R} = m_{\tilde{c}_R} = 1125.7$ GeV, $m_{\tilde{d}_R} = m_{\tilde{s}_R} = 1116.1$ GeV. The production ratio is close to 4:1, essentially given by the square of the charge ratio of up- and down-type squarks. The combined cross section for the process considered here, $e^+e^- \to \tilde{q}_R\tilde{q}_R \to q\bar{q}\tilde{\chi}_1^0\tilde{\chi}_1^0$ ($q = u, d, s, c$), is 1.47 fb at 3 TeV, taking into account the CLIC luminosity spectrum. The squark production study is performed at 3 TeV using the CLIC_ILD detector framework; it reports the expected accuracy of the measurement of the squark production cross section and of the squark mass.

Table 12.8 summarises the cross sections of the signal and the relevant background processes. Detailed studies show that SM processes without missing energy can be rejected very efficiently with a missing energy cut, leaving the SM four fermion processes with hadronic final states and neutrinos as main background. To optimise the computing resources, a cut on the missing transverse momentum $p_T^{miss} > 530$ GeV is applied on generator level during event generation.

#### *12.4.4.1 Event Selection*

After reconstruction, the events are forced into two jets using the exclusive $k_t$ algorithm of the FASTJET package [31], where the parameter $R$ is set to 0.7.

To reject a substantial fraction of the background, a $p_T^{miss} > 600$ GeV requirement is imposed, compatible with the generator-level cut during event generation. Requiring large missing transverse momentum alone is insufficient to reduce the Standard Model background to a manageable level. Thus, a BDT from the TMVA toolkit [40] is introduced as an additional event selection stage. The BDT uses a total of 18 variables: jet energies, jet masses, event geometry and leading particle energies and types to classify events either as signal or as background. Following the BDT, a significance of $S/\sqrt{S+B} = 25.8$ is obtained. The performance of the background rejection is illustrated in Figure 12.22, which shows the $M_C$ distribution (introduced below) of signal and background processes, after the missing $p_T$ cut (left) and after the additional cut on the BDT (right). The overall signal efficiency of the event selection is 36%.

#### *12.4.4.2 Squark Mass Determination*

The CLIC energy spectrum and the resulting uncertainty of the centre-of-mass energy as well as the low cross section and high SM background make the use of the distribution of individual jet energies for the reconstruction of the squark mass impractical. Instead, a modified invariant mass,

$$M_C = \sqrt{2\,(E_1E_2 + \vec{p}_1 \cdot \vec{p}_2)}, \tag{12.1}$$

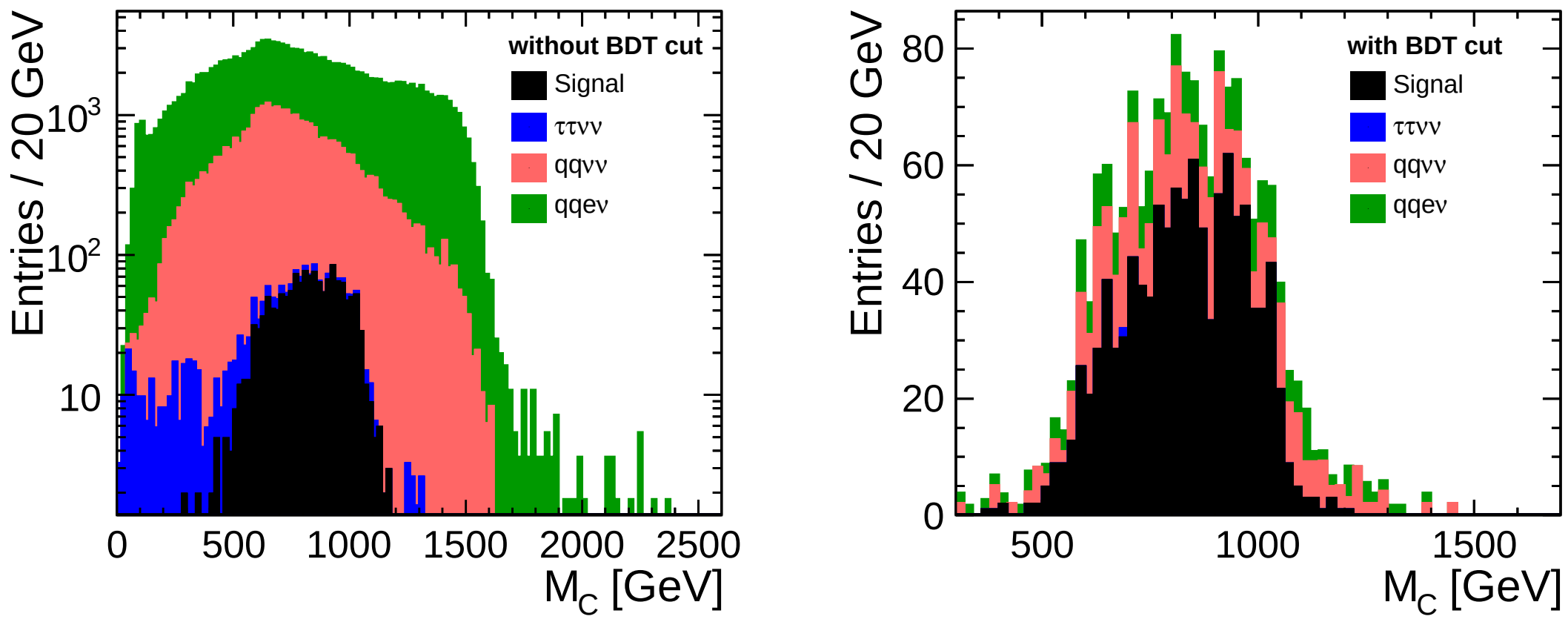

(a) $p_T > 600$ GeV (b) $p_T > 600$ GeV and BDT Cut

Fig. 12.22: The $M_C$ distributions for the signal and the individual backgrounds. The histograms are stacked and normalised to an integrated luminosity of 2 ab$^{-1}$. In the left plot only a simple $p_T > 600$ GeV cut is applied while the right plot shows the distributions after an additional cut on the output of the BDT.

which is invariant under contra-linear boosts of equal magnitude of the two heavy parent particles, and thus to first order independent of the centre-of-mass energy [49, 50], is used. The distribution of $M_C$ reaches its maximum at its high-mass endpoint, given by $M_C^{max} = (m_{\tilde{q}}^2 - m_{\tilde{\chi}}^2)/m_{\tilde{q}}$. It is assumed here that the neutralino mass is known from the slepton measurements presented in Section 12.4.5, thus the distribution of $M_C$ provides direct sensitivity to the squark mass.

The right-handed squark mass is extracted from the $M_C$ distribution by means of a template fit. For the fit, templates with varying squark masses in 3 GeV steps from 1050 GeV to 1248 GeV are generated, assuming a mass splitting of 10 GeV between up- and down-type squarks. Each of the templates contains 50000 generator-level events. After jet finding, the energy of the jets is convolved with a Gaussian distribution with a width of 4.5% to account for the detector resolution. This factor was determined by comparing the $M_C$ distribution in a fully simulated, high-statistics signal sample with the corresponding generator-level distribution after convolution with various resolution factors. The value that gave the best Kolmogorov–Smirnov score between the generator-level information with Gaussian smearing and the fully simulated sample is chosen. The $p_T^{miss}$ and BDT cuts are then applied to the templates, resulting in realistic $M_C$ distributions. Since the templates do not include effects from $\gamma\gamma \rightarrow$ hadrons background, which results in a slight upward shift of the edge of the $M_C$ distribution due to the additionally picked up energy, the mass scale of the templates is calibrated with a high-statistics fully simulated signal sample.

Parametrised background contributions, determined with a fit to a statistically independent background sample, are subtracted from the final $M_C$ distribution before the template fit. The template fit itself is performed by calculating the $\chi^2$ for all templates compared to the background-subtracted final $M_C$ distribution. The squark mass, given by the weighted average of up- and down-type squarks, is determined from the minimum of the resulting $\chi^2$ distribution, shown in Figure 12.23a. The template with the lowest $\chi^2$ is shown in Figure 12.23b, compared to the background-subtracted $M_C$ distribution. The statistical uncertainty of the mass measurement is determined with a toy Monte Carlo using 500 trials with signal points shifted in accordance within their statistical errors.

#### *12.4.4.3 Results*

The mass determined from the fit, $m_{\tilde{q}_R} = 1127.9$ GeV $\pm$ 5.9 GeV, is in agreement with the cross section weighted average input $m_{\tilde{q}_R} = 1123.7$ GeV. This demonstrates the possibility of a reliable reconstruction of TeV-scale right-handed squarks with a statistical uncertainty of approximately 0.5% for an integrated luminosity of 2 ab$^{-1}$.

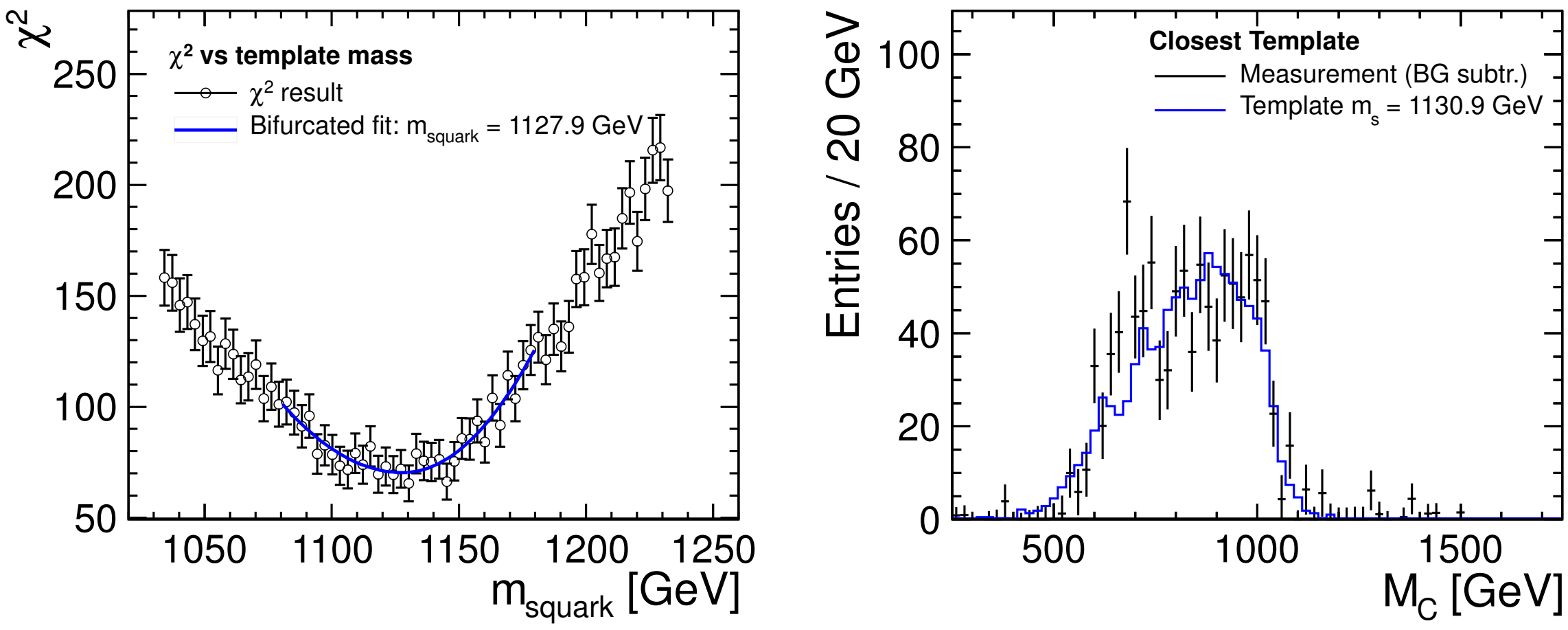

(a) (b)

Fig. 12.23: Illustration of the template fit for the squark analysis. The distribution of $\chi^2$ as a function of the squark mass for all considered templates fitted using a bifurcated parabola (left), and the $M_C$ distribution for the signal sample and the template corresponding to the lowest $\chi^2$.

Table 12.9: Summary of the results from the squark study. All numbers are obtained assuming an integrated luminosity of 2 ab$^{-1}$.

| Observable | Result | Generator value |
|---|---|---|
| Averaged right-squark mass | 1127.9 GeV $\pm$ 5.9 GeV | 1123.7 GeV |
| Combined cross section | 1.51 fb $\pm$ 0.07 fb | 1.47 fb |

Since the mass of the lightest neutralino enters the $M_C$ calculation, there is an additional uncertainty on the squark mass, depending on the uncertainty of the neutralino mass. Here, a 1 GeV uncertainty on the neutralino mass translates into a 0.54 GeV uncertainty of the squark mass. Taking the neutralino mass determination from right-handed selectrons in Section 12.4.5 with a statistical uncertainty of 3.4 GeV for an integrated luminosity of 2 ab$^{-1}$, the corresponding additional uncertainty of the squark mass measurement is 1.8 GeV.

Using the efficiency of the event selection procedure, the cross section of right-squark production is determined from the integral of the background-subtracted $M_C$ distribution. The result of 1.51 fb $\pm$ 0.07 fb, in agreement with the input value of 1.47 fb, shows that a cross section measurement with a statistical precision of 5% is possible with an integrated luminosity of 2 ab$^{-1}$. Table 12.9 summarises the results on mass and cross section. The variations of the luminosity spectrum as described in Section 12.2.2 have negligible impact on the measurements.

### 12.4.5 Slepton Searches

This section presents results of a study of the following four processes:

1. $e^+e^- \to \tilde{e}^+_R\tilde{e}^-_R \to e^+e^-\tilde{\chi}^0_1\tilde{\chi}^0_1$
2. $e^+e^- \to \tilde{\mu}^+_R\tilde{\mu}^-_R \to \mu^+\mu^-\tilde{\chi}^0_1\tilde{\chi}^0_1$

Table 12.10: Signal processes, decay modes, cross sections ($\sigma$), cross sections times branching ratio ($\sigma \times BR$) and cross sections times branching ratio into two electrons and four quarks ($\sigma \times BR$(ee4Q) ).

| Process | $\sigma$ (fb) | Decay Mode | $\sigma \times BR$ (fb) | $\sigma \times BR$ (ee4Q) (fb) |
|---|---|---|---|---|
| $e^+e^- \to \tilde{\mu}_R^+\tilde{\mu}_R^-$ | 0.72 | $\mu^+\mu^-\tilde{\chi}_1^0\tilde{\chi}_1^0$ | 0.72 | |
| $e^+e^- \to \tilde{e}_R^+\tilde{e}_R^-$ | 6.05 | $e^+e^-\tilde{\chi}_1^0\tilde{\chi}_1^0$ | 6.05 | |
| $e^+e^- \to \tilde{e}_L^+\tilde{e}_L^-$ | 3.07 | $\tilde{\chi}_1^0\ \tilde{\chi}_1^0\ e^+e^-$ (h/$Z^0$ h/$Z^0$) | 0.25 | 0.16 |
| $e^+e^- \to \tilde{\nu}_e\tilde{\nu}_e$ | 13.74 | $\tilde{\chi}_1^0\ \tilde{\chi}_1^0 e^+e^-\ W^+W^-$ | 4.30 | 1.82 |

3. $e^+e^- \to \tilde{e}_L^+\tilde{e}_L^- \to e^+e^-\tilde{\chi}_2^0\tilde{\chi}_2^0$
4. $e^+e^- \to \tilde{\nu}_e\tilde{\nu}_e \to e^+e^-\tilde{\chi}_1^+\tilde{\chi}_1^-$

All processes have been studied in the *SUSY model II* at 3 TeV using the CLIC_ILD detector framework. The branching ratio $\tilde{\ell}_R^{\pm} \to \ell^{\pm}\tilde{\chi}_1^0$ is $\sim$ 100%, and the branching ratio $\tilde{e}_L \to e^-\tilde{\chi}_1^0, \tilde{e}_L \to e^-\tilde{\chi}_2^0$, and $\tilde{\nu}_e \to e^-\tilde{\chi}_1^+$ are 16%, 29% and 56% respectively. The cross sections, the decay channels and the cross sections times the branching ratio of the signal processes under study are given in Table 12.10.

The expected accuracy on the production cross sections and on the $\tilde{e}_R$, $\tilde{\mu}_R$, $\tilde{\nu}_e$, $\tilde{\chi}_1^{\pm}$ and $\tilde{\chi}_1^0$ mass measurements are reported and the errors related to the knowledge of the luminosity spectrum are discussed. The analysis is documented in detail in [36].

#### *12.4.5.1 Event Selection*

All signal processes have two undetected $\tilde{\chi}_1^0$ in the final state. Therefore, the main characteristics of these events are missing energy, missing transverse momentum and acoplanarity. Despite this signature, the large Standard Model backgrounds given in Table 12.11, make the analysis rather challenging.

Table 12.11: Signal and Background processes considered in the slepton study. The cross sections multiplied by branching ratio, $\sigma \times BR$, for the investigated decay modes with and without pre-selection cuts are shown.

| Process | Decay mode | $\sigma \times BR$ (fb) no cuts | $\sigma \times BR$ (fb) cuts |
|---|---|---|---|
| $e^+e^- \to \mu^+\mu^-$ | $\mu^+\mu^-$ | 81.9 | 0.6 |
| $e^+e^- \to \mu^+\nu_e\mu^-\nu_e$ | $\mu^+\mu^-$ | 65.6 | 3.5 |
| $e^+e^- \to \mu^+\nu_\mu\mu^-\nu_\mu$ | $\mu^+\mu^-$ | 6.2 | 2.2 |
| $e^+e^- \to W^+\nu W^-\nu$ | $\mu^+\mu^-$ | 92.6 | 2.4 |
| $e^+e^- \to Z^0\nu Z^0\nu$ | $\mu^+\mu^-$ | 40.5 | 0.002 |
| SUSY background | $\mu^+\mu^-$ | 0.31 | 0.31 |
| $e^+e^- \to e^+e^-$ | $e^+e^-$ | 6226.1 | 77.1 |
| $e^+e^- \to e^+\nu_e e^-\nu_e$ | $e^+e^-$ | 179.3 | 91.1 |
| $e^+e^- \to W^+\nu W^-\nu$ | $e^+e^-$ | 92.6 | 2.4 |
| $e^+e^- \to Z^0\nu Z^0\nu$ | $e^+e^-$ | 40.5 | 0.002 |
| SUSY background | $e^+e^-$ | 1.04 | 1.04 |
| $e^+e^- \to W^+W^-Z^0$ | $e^+e^-W^+W^-$ | 1.35 | 0.61 |
| $e^+e^- \to Z^0Z^0Z^0$ | $e^+e^-Z^0Z^0$ | 0.045 | 0.023 |
| SUSY background | $e^+e^-$ (WW or h h or $Z^0Z^0$) | 0.77 | 0.12 |

The event reconstruction method and performances are presented in Section 12.3.2. To distinguish signal events from background events the following set of discriminating variables is used:

- di-lepton energy $\mathrm{E(L1)+E(L2)}$
- vector sum $\mathrm{p_T(L1)+p_T(L2)}$ of the two leptons
- algebraic sum $\mathrm{p_T(L1)+p_T(L2)}$ of the two leptons
- di-lepton invariant mass $\mathrm{M(L1,L2)}$
- di-lepton velocity $\beta\mathrm{(L1,L2)}$
- angle of the di-lepton missing momentum vector $\cos\theta\mathrm{(L1,L2)}$
- di-lepton acollinearity
- di-lepton energy imbalance $\Delta = |\mathrm{E(L1)-E(L2)}|/|\mathrm{E(L1)+E(L2)}|$

where L1 and L2 are the two leptons. The event selection proceeds as follows: First, the following pre-selection cuts are applied:

- $p_\mathrm{T}$(L1 and L2) $>$ 4 GeV
- $10^\circ < \theta$(L1 and L2) $< 170^\circ$
- $4^\circ < \Delta\Phi$(L1,L2) $< 176^\circ$
- $p_\mathrm{T}$(L1,L2) $>$ 10 GeV
- $M$(L1,L2) $>$ 100 GeV

Then, histograms of the discriminating variables are built for signal and background events. The events are weighted such that the data samples correspond to the same integrated luminosity. The Boosted Decision Tree classifier from the toolkit for multivariate analysis, TMVA [40], is used for the event selection. From the signal and background histograms of the discriminating variables, signal and background event PDFs are computed and combined into a total probability classifier. The signal and background samples are split into two equal data samples called “Monte Carlo” and “Data”. The Monte Carlo sample composed of signal and background events is used to train the classifier which ranks events to be signal or background-like. The method is then applied to the “Data” sample, for each event a total probability is computed and a cut is applied to separate signal from background. The cut value is chosen to optimise the significance $\mathrm{N_S}/\sqrt{\mathrm{N_S+N_B}}$, where $\mathrm{N_S}$ and $\mathrm{N_B}$ are the number of signal and background events. The selection efficiencies are 97% for the di-muon final state process and $\sim$94% for the di-electron final state processes.

#### *12.4.5.2 Slepton and Gauginos Mass Determination*

After the final selection, the slepton and gauginos masses are extracted from the position of the kinematic edges of the lepton energy distribution, a technique first proposed for squarks [51], then extensively applied to sleptons [52]:

$$m_{\tilde{\ell}^\pm} = \frac{\sqrt{s}}{2}\left(1-\frac{(E_H-E_L)^2}{(E_H+E_L)^2}\right)^{1/2} \quad \text{and} \quad m_{\tilde{\chi}_1^0} \text{ or } m_{\tilde{\chi}_1^\pm} = m_{\tilde{\ell}^\pm}\left(1-\frac{2(E_H+E_L)}{\sqrt{s}}\right)^{1/2} \tag{12.2}$$

where $E_L$ and $E_H$ are the low and high edges of the lepton energy distribution.

$$E_{L,H} = \frac{\sqrt{s}}{4}\left(1-\frac{m^2_{\tilde{\chi}_1^0}}{m^2_{\tilde{\ell}^\pm}}\right)\left(1\pm\sqrt{1-4\frac{m^2_{\tilde{\ell}^\pm}}{s}}\right) \tag{12.3}$$

The slepton, neutralino and chargino masses depend on the beam energy $\sqrt{s}/2$ and on the kinematic edges $E_{L,H}$, therefore the accuracy on the masses relies on the measurement of the shape of the luminosity

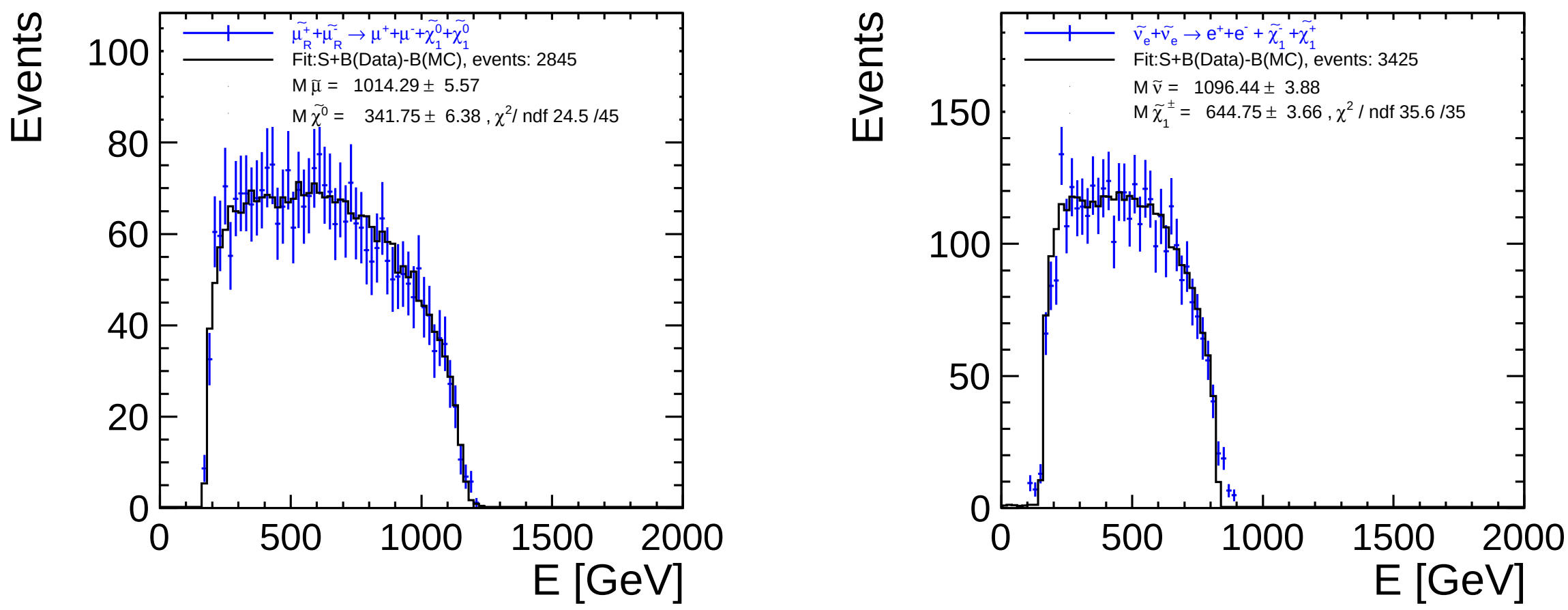

(a) $e^+e^- \to \tilde{\mu}_R^+\tilde{\mu}_R^-$ (b) $e^+e^- \to \tilde{\nu}_e\tilde{\nu}_e$

Fig. 12.24: Reconstructed lepton energy spectra for the processes $e^+e^- \to \tilde{\mu}_R^+\tilde{\mu}_R^-$ (left) and $e^+e^- \to \tilde{\nu}_e\tilde{\nu}_e$ (right). The distributions obtained for an integrated luminosity of 2 $ab^{-1}$ are compared to the fit result.

spectrum and on the lepton energy resolution. The masses are determined using a two-parameters fit to the reconstructed energy distribution with $m_{\tilde{\ell}^\pm}$ and $m_{\tilde{\chi}_1^0}$ or $m_{\tilde{\chi}_1^\pm}$ as parameters. The fit is performed with the MINUIT minimisation package [53]. The lepton energy spectrum is a uniform distribution with the end points given by Equation 12.3. For each event, a random value of $\sqrt{s}$ is generated taking into account the beamstrahlung and ISR effects, and the lepton energy resolution is included through a parametric smearing of the lepton energy. For each process the parameters of the energy resolution function are the ones obtained from the fits shown in Figure 12.5. Figure 12.24 shows, for the processes $e^+e^- \to \tilde{\mu}_R^+\tilde{\mu}_R^-$ and $e^+e^- \to \tilde{\nu}_e\tilde{\nu}_e$, the lepton energy distributions and the fit results. The fit gives also the integral of the momentum distribution, allowing to determine the process cross section. For the process $e^+e^- \to \tilde{e}_L^+\tilde{e}_L^- \to e^+e^-\tilde{\chi}_2^0\tilde{\chi}_2^0$, the cross section is determined from the fit to the boson mass distribution shown in Figure 12.11 (b). Table 12.12 shows the values of the measured slepton cross sections, slepton masses and gaugino masses assuming 2 $ab^{-1}$ of integrated luminosity.

Table 12.12: Overview of the results of the slepton study. Values for the extracted cross sections, slepton and gaugino masses are given with statistical uncertainties. All numbers are obtained assuming an integrated luminosity of 2 $ab^{-1}$.

| Process | Decay Mode | $\sigma$ (fb) | $m_{\tilde{\ell}}$ (GeV) | $m_{\tilde{\chi}_1^0}$ or $m_{\tilde{\chi}_1^\pm}$ (GeV) |
|---|---|---|---|---|
| $e^+e^- \to \tilde{\mu}_R^+\tilde{\mu}_R^-$ | $\mu^+\mu^-\tilde{\chi}_1^0\tilde{\chi}_1^0$ | $0.71 \pm 0.02$ | $1014.3 \pm 5.6$ | $341.8 \pm 6.4$ |
| $e^+e^- \to \tilde{e}_R^+\tilde{e}_R^-$ | $e^+e^-\tilde{\chi}_1^0\tilde{\chi}_1^0$ | $6.20 \pm 0.05$ | $1001.6 \pm 2.8$ | $340.6 \pm 3.4$ |
| $e^+e^- \to \tilde{e}_L^+\tilde{e}_L^-$ | $\tilde{\chi}_1^0\,\tilde{\chi}_1^0\,e^+e^-$ ($h/Z^0$ $h/Z^0$) | $2.77 \pm 0.20$ | | |
| $e^+e^- \to \tilde{\nu}_e\tilde{\nu}_e$ | $\tilde{\chi}_1^0\,\tilde{\chi}_1^0 e^+e^-\ W^+W^-$ | $13.24 \pm 0.32$ | $1096.4 \pm 3.9$ | $644.8 \pm 3.7$ |

#### *12.4.5.3 Effects of the Luminosity Spectrum Measurement Uncertainty*

In order to assess the effect of the knowledge of the luminosity spectrum on the mass measurement accuracy, a luminosity spectrum variation was introduced as described in Section 12.2.2. We compare the results of the mass fit without variation to those obtained with variation. For the process $e^+e^- \to$ $\widetilde{\mu}_R^+\widetilde{\mu}_R^- \rightarrow e^+e^-\ \tilde{\chi}_1^0\ \tilde{\chi}_1^0$, the $\tilde{\mu}_R$ and $\tilde{\chi}_1^0$ mass change by 0.2% and 0.6% respectively. For the process $e^+e^- \rightarrow \widetilde{e}_R^+\widetilde{e}_R^- \rightarrow e^+e^-\ \tilde{\chi}_1^0\ \tilde{\chi}_1^0$, the $\tilde{e}_R$ and $\tilde{\chi}_1^0$ mass change by 0.2% and 1.0% respectively. For the process $e^+e^- \rightarrow \widetilde{\nu}_e\widetilde{\nu}_e \rightarrow e^+e^-\ \tilde{\chi}_1^+\ \tilde{\chi}_1^-$, the $\tilde{\nu}_e$ and $\tilde{\chi}_1^\pm$ mass change by 0.2% and 0.4% respectively. For 2ab$^{-1}$ of integrated luminosity, the statistical errors are dominant except for the process $e^+e^- \rightarrow \widetilde{e}_R^+\widetilde{e}_R^- \rightarrow e^+e^-\ \tilde{\chi}_1^0\ \tilde{\chi}_1^0$, which has the largest cross section.

#### *12.4.5.4 Results*

Slepton cross sections, and slepton and gauginos masses can be extracted from the lepton energy distributions. With 2 ab$^{-1}$ of integrated luminosity for the process $e^+e^- \rightarrow \widetilde{\mu}_R^+\widetilde{\mu}_R^- \rightarrow \mu^+\mu^-\tilde{\chi}_1^0\tilde{\chi}_1^0$, the cross section can be determined with a relative statistical uncertainty of 2.8% and the $\widetilde{\mu}_R$ mass with an accuracy of 0.6% and the $\tilde{\chi}_1^0$ mass with an accuracy of 2.0%. For the process $e^+e^- \rightarrow \widetilde{e}_R^+\widetilde{e}_R^- \rightarrow e^+e^-\tilde{\chi}_1^0\tilde{\chi}_1^0$, the uncertainty on the cross section measurement is 0.8%, 0.3% on the $\widetilde{e}_R$ mass and 1.0% on the $\tilde{\chi}_1^0$ mass. For the processes $e^+e^- \rightarrow \widetilde{\nu}_e\widetilde{\nu}_e \rightarrow e^+e^-\tilde{\chi}_1^+\tilde{\chi}_1^-$ and $e^+e^- \rightarrow \widetilde{e}_L^+\widetilde{e}_L^- \rightarrow e^+e^-\tilde{\chi}_2^0\tilde{\chi}_2^0$ the statistical accuracy on the cross section is $\sim$ 2.4% and $\sim$ 7.0% respectively. The $\widetilde{\nu}_e$ and $\tilde{\chi}_1^\pm$ masses are determined with an accuracy of $\sim$ 0.4% and $\sim$ 0.6% respectively. A dedicated energy scan of the slepton pair production threshold can further improve the mass measurements.

### 12.4.6 Chargino and Neutralino Production at 3 TeV

The pair production of charginos and neutralinos is studied with the CLIC_SiD detector concept [54]. The main emphasis of this benchmark channel is to demonstrate the successful reconstruction of four jet final states and missing energy in the presence of machine background. The investigated signal channels are:

$$e^+e^- \rightarrow \tilde{\chi}_1^+\tilde{\chi}_1^- \rightarrow W^+W^-\tilde{\chi}_1^0\tilde{\chi}_1^0$$
$$e^+e^- \rightarrow \tilde{\chi}_2^0\tilde{\chi}_2^0 \rightarrow h(Z)h(Z)\tilde{\chi}_1^0\tilde{\chi}_1^0$$

More details on the investigated *SUSY model II* are given in Section 2.6. Cross sections for these processes as well as an overview of the physics backgrounds included in the study presented here are given in Table 12.13. All processes are fully simulated and reconstructed and include background from $\gamma\gamma \rightarrow$ hadrons.

Table 12.13: Cross sections for chargino and neutralino pair production and for SUSY and Standard Model backgrounds.

| Type | Process | Cross section (fb) | Referenced with |
|---|---|---|---|
| Signal | $\tilde{\chi}_1^+\tilde{\chi}_1^-$ | 10.6 | Chargino |
| | $\tilde{\chi}_2^0\tilde{\chi}_2^0$ | 3.3 | Neutralino |
| Background | $\tilde{\chi}_2^+\tilde{\chi}_2^-$ | 10.5 | SUSY |
| | $\tilde{\chi}_1^+\tilde{\chi}_2^-$ | 0.8 | |
| | $\tilde{\chi}_1^+\tilde{\chi}_1^-\nu\bar{\nu}$ | 1.4 | |
| | $\tilde{\chi}_2^0\tilde{\chi}_2^0\nu\bar{\nu}$ | 1.2 | |
| | $q\bar{q}q\bar{q}\,\nu\bar{\nu}$ | 95.4 | SM |
| | $q\bar{q}\,h\nu\bar{\nu}$ | 3.1 | |
| | $h\,h\,\nu\bar{\nu}$ | 0.6 | |

#### *12.4.6.1 Event Reconstruction*

Events with at least one identified electron or muon with $p_T > 20$ GeV are rejected. Jets are reconstructed from PFOs using the $k_t$ algorithm in its exclusive mode with R = 0.7 and using the E recombination scheme. The clustering is stopped when four jets are found. To reject leptonic decays of W, Z or Higgs bosons further, all jets are required to contain more than one PFO.

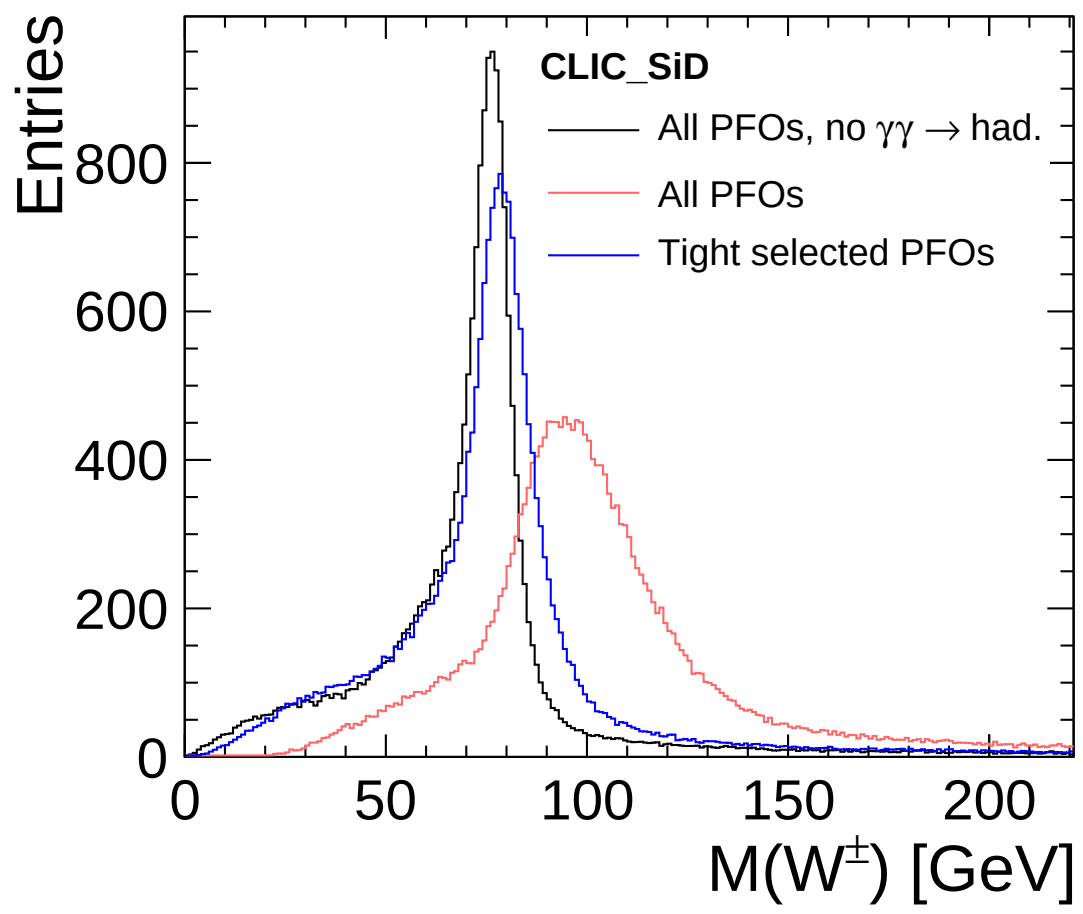

Fig. 12.25: Reconstructed mass of W candidates in $e^+e^- \to \tilde{\chi}_1^+\tilde{\chi}_1^-$ events without overlay of $\gamma\gamma \to$ hadrons (black histogram), with overlay of $\gamma\gamma \to$ hadrons (red histogram) and for tight selected PFOs (blue histogram).

Boson candidates are formed from jet pairs minimising:

$$(M_{\mathrm{jj},1} - M_{\mathrm{W,h}})^2 + (M_{\mathrm{jj},2} - M_{\mathrm{W,h}})^2, \qquad (12.4)$$

where $M_{\mathrm{jj},1}$ and $M_{\mathrm{jj},2}$ are the masses of the two reconstructed jet pairs and $M_{\mathrm{W,h}}$ is set to the world average of the W boson mass to reconstruct $\tilde{\chi}_1^{\pm}$, and to a Higgs mass of 118.52 GeV to reconstruct $\tilde{\chi}_2^0$.

As an example, the reconstruction of W bosons is illustrated in Figure 12.25. The distributions obtained with and without the overlay of $\gamma\gamma \to$ hadrons are compared. A good reconstruction of W bosons is achieved when `tight` cuts (see Section 12.1.4) are applied to select the PFOs used as input to the jet reconstruction.

#### *12.4.6.2 Event selection*

First, pre-selection cuts are applied. The following requirements are imposed:

- $40 < M_{\mathrm{jj},1} < 160$ GeV and $40 < M_{\mathrm{jj},2} < 160$ GeV
- $|\cos\theta^{\mathrm{miss}}| < 0.95$, where $\theta^{\mathrm{miss}}$ is the polar angle of the missing momentum
- Angle between the W or Higgs candidates larger than 1 radian
- $|\cos\theta^{\mathrm{jj},1}| < 0.95$ and $|\cos\theta^{\mathrm{jj},2}| < 0.95$

A selection based on BDTs using TMVA [40] is carried out. The BDTs were trained using 15 variables describing kinematic properties of the reconstructed W and Higgs candidates as well as describing the event topology. The distributions of the reconstructed W energy and mass for selected chargino signal and background events are shown in Figure 12.26. Figure 12.27 shows the same histograms for the reconstructed Higgs energy and mass observed in neutralino signal and background events. All distributions are scaled to an integrated luminosity of 2 ab$^{-1}$. The efficiencies of the selection for reconstructed chargino and neutralino signal events are 25% and 33%, respectively. The signal purities in the selected samples are 57% for the chargino and 55% for the neutralino.

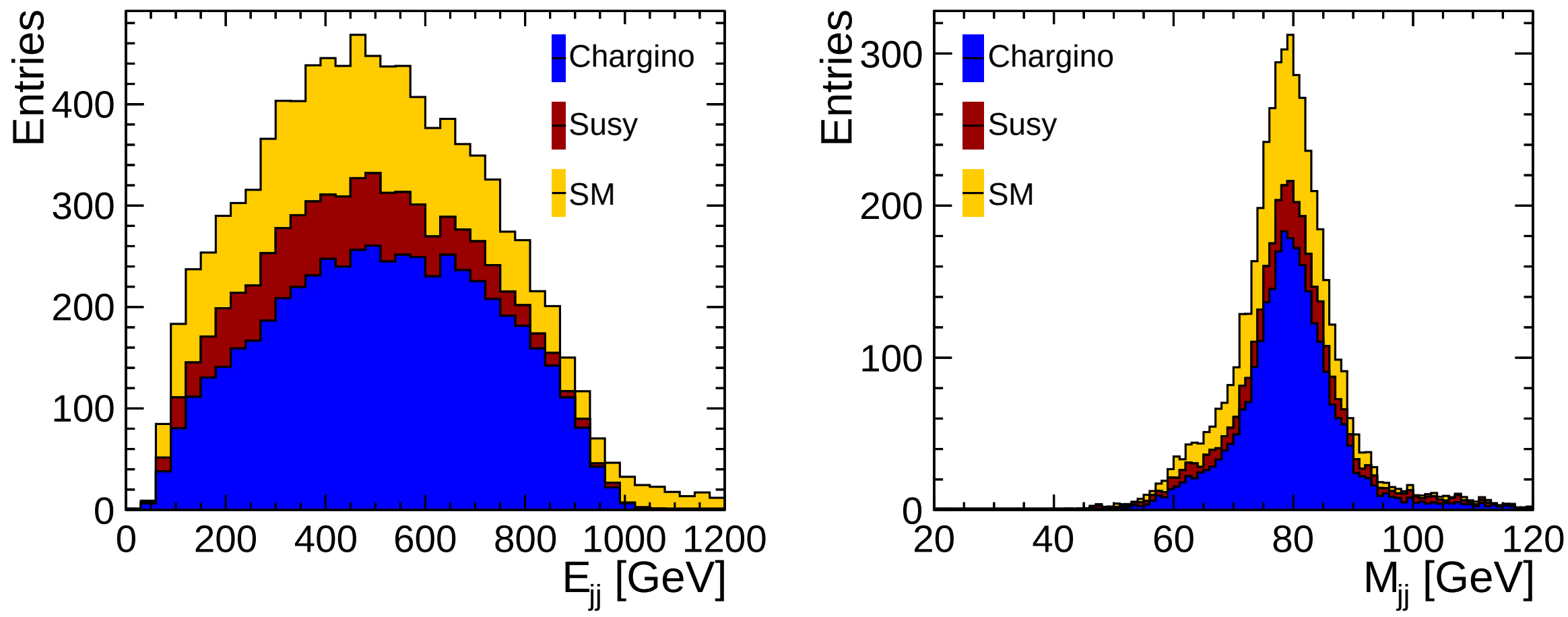

Fig. 12.26: Reconstructed energy (left) and mass (right) of W candidates for the chargino measurement. The signal is compared to the backgrounds from SM and SUSY processes. All distributions are scaled to an integrated luminosity of 2 ab$^{-1}$.

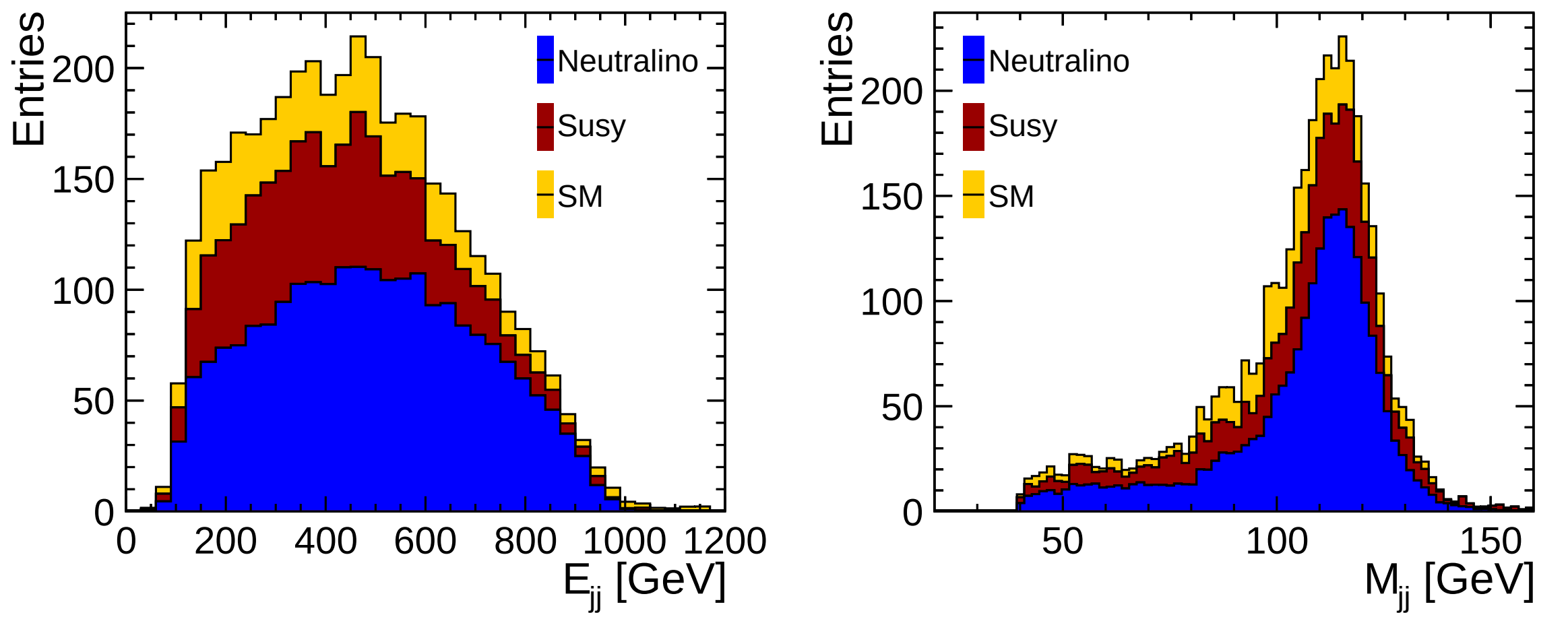

Fig. 12.27: Reconstructed energy (left) and mass (right) of Higgs candidates for the neutralino measurement. The signal is compared to the backgrounds from SM and SUSY processes. All distributions are scaled to an integrated luminosity of 2 ab$^{-1}$.

#### ***12.4.6.3 Mass and cross section measurement***

##### *12.4.6.3.1 Template Fitting*

The pair production cross sections and masses of the $\tilde{\chi}_1^{\pm}$ and $\tilde{\chi}_2^0$ particles are determined using a template method where chargino and neutralino signal Monte Carlo samples for different mass hypotheses are generated. The $\tilde{\chi}_1^0$ mass is also measured since the energy distribution of W bosons from $\tilde{\chi}_1^{\pm}$ decays is sensitive to this observable. Two-dimensional fits are performed simultaneously to the mass and production cross section for a given particle to account for the correlation between both quantities. The statistical uncertainties of the extracted masses and cross sections, determined using toy Monte Carlos, are shown in Table 12.14. All measured values are in agreement with the input values used in the Monte Carlo generation.

Table 12.14: Uncertainties of the chargino and neutralino masses and pair production cross sections obtained from two parameter template fits. An integrated luminosity of 2 ab$^{-1}$ is assumed.

| Parameter 1 | Uncertainty | Parameter 2 | Uncertainty |
|---|---|---|---|
| $M(\tilde{\chi}_1^{\pm})$ | 6.3 GeV | $\sigma(\tilde{\chi}_1^{+}\tilde{\chi}_1^{-})$ | 2.2% |
| $M(\tilde{\chi}_1^{0})$ | 3.0 GeV | $\sigma(\tilde{\chi}_1^{+}\tilde{\chi}_1^{-})$ | 1.8% |
| $M(\tilde{\chi}_2^{0})$ | 7.3 GeV | $\sigma(\tilde{\chi}_2^{0}\tilde{\chi}_2^{0})$ | 2.9% |

#### *12.4.6.3.2 Least Squares Fitting*

Linear least squares fits are performed in multiple dimensions to extract several masses simultaneously. In the histograms of the reconstructed energy of the W/Z/h, each bin is expanded linearly around the nominal masses and cross sections. The slopes are obtained by convolving a map of true to reconstructed bin contents with the true energy distributions for different chargino and neutralino masses. No fits are actually performed. Instead, the statistical uncertainties for each bin are calculated and then propagated to the fit parameter uncertainties using standard formulae for linear least squares fits.

The least squares results for two parameter fits of one gaugino mass and one cross section are summarised in Table 12.15. These are to be directly compared with the template fit results of Table 12.14. Reasonable agreement is obtained between the two techniques.

Table 12.15: Uncertainties of the chargino and neutralino masses and pair production cross sections obtained from two parameter least squares fits. The correlation coefficients, $\rho(1,2)$, between the two fit parameters are given. An integrated luminosity of 2 ab$^{-1}$ is assumed.

| Par. 1 | Uncertainty | Par. 2 | Uncertainty | $\rho(1,2)$ |
|---|---|---|---|---|
| $M(\tilde{\chi}_1^{\pm})$ | 5.7 GeV | $\sigma(\tilde{\chi}_1^{+}\tilde{\chi}_1^{-})$ | 2.0 % | 0.51 |
| $M(\tilde{\chi}_1^{0})$ | 3.3 GeV | $\sigma(\tilde{\chi}_1^{+}\tilde{\chi}_1^{-})$ | 1.8 % | 0.23 |
| $M(\tilde{\chi}_2^{0})$ | 8.5 GeV | $\sigma(\tilde{\chi}_2^{0}\tilde{\chi}_2^{0})$ | 3.0 % | 0.40 |

The two parameter fits implicitly assume that the other SUSY parameters are obtained through independent measurements, their uncertainties do not enter the fits. For example, the $\tilde{\chi}_1^0$ mass will be measured with an accuracy of $\Delta M(\tilde{\chi}_1^0) = 3$ GeV at CLIC by combining the results from the slepton analyses (see Section 12.4.5). A term constraining the $\tilde{\chi}_1^0$ mass to be within 3 GeV of the best estimate is added to a three parameter least squares fit of $M(\tilde{\chi}_1^{\pm}), M(\tilde{\chi}_1^0), \sigma(\tilde{\chi}_1^{+}\tilde{\chi}_1^{-})$ or $M(\tilde{\chi}_2^0), M(\tilde{\chi}_1^0), \sigma(\tilde{\chi}_2^0\tilde{\chi}_2^0)$ to demonstrate the expected accuracy if both measurements are combined. The results for these three parameter fits are shown in Table 12.16. The obtained mass and cross section uncertainties are somewhat larger than for the simple two parameter fits discussed above.

Table 12.16: Uncertainties of the chargino and neutralino masses and pair production cross sections obtained from three parameter least squares fits. The mass of the $\tilde{\chi}_1^0$ particle is assumed to be within 3 GeV of the value measured from the slepton analyses. The correlation coefficients, $\rho(i,j)$, between the fit parameters are shown. An integrated luminosity of 2 ab$^{-1}$ is assumed.

| Par. 1 | Uncertainty | Par. 2 | Uncertainty | Par. 3 | Uncertainty | $\rho(1,2)$ | $\rho(1,3)$ | $\rho(2,3)$ |
|---|---|---|---|---|---|---|---|---|
| $M(\tilde{\chi}_1^{\pm})$ | 7.3 GeV | $M(\tilde{\chi}_1^{0})$ | 2.9 GeV | $\sigma(\tilde{\chi}_1^{+}\tilde{\chi}_1^{-})$ | 2.4 % | 0.64 | 0.66 | 0.51 |
| $M(\tilde{\chi}_2^{0})$ | 9.9 GeV | $M(\tilde{\chi}_1^{0})$ | 3.0 GeV | $\sigma(\tilde{\chi}_2^{0}\tilde{\chi}_2^{0})$ | 3.2 % | 0.52 | 0.49 | 0.33 |

#### *12.4.6.4 Systematic uncertainties*

The measurements described in this section are sensitive to the luminosity spectrum. A variation as described in Section 12.2.2 translates to a change of the measured chargino and neutralino pair production cross sections that is similar in size as the statistical uncertainty. For the measured masses the variation of the luminosity spectrum leads to a shift that is typically half of the statistical uncertainty.

As an additional test, the normalisation of the SM background assumed in the template fits is changed by $\pm 15\%$ to evaluate the effect of the uncertainty of the Monte Carlo predictions. The impact on the fit results is found to be negligible.

#### *12.4.6.5 Conclusion and Summary*

The signals from $\tilde{\chi}_1^{\pm}$ and $\tilde{\chi}_2^0$ pair production are extracted from fully hadronic final states with four jets and missing transverse energy. Two different signal extraction procedures are in reasonable agreement. The chargino and neutralino pair production cross sections are extracted with a precision of $2-3\%$ while the masses of the $\tilde{\chi}_1^{\pm}$, $\tilde{\chi}_1^0$ and $\tilde{\chi}_2^0$ particles are determined with typical accuracies of about $1-1.5\%$.

### 12.4.7 Top Pair Production at 500 GeV

The study of top quark pair production at a 500 GeV CLIC provides the possibility for a direct comparison to the expected performance of the ILC to assess the impact of the more challenging experimental conditions at CLIC, in particular regarding the luminosity spectrum and the high bunch crossing rate, which leads to a pile-up of $\gamma\gamma \rightarrow$ hadrons background. In addition, the study tests mass reconstruction in multi-jet final states at relatively low jet energy as well as flavour tagging. The analysis is documented in detail in [55].

For the study, a slightly modified version of the CLIC_ILD detector concept is used, which is adapted to the lower collision energy. In particular, the innermost vertex detector layer is moved in by 6 mm compared to the 3 TeV detector design to a radius of 25 mm to improve flavour tagging at low momentum. Due to the lower rate of $\gamma\gamma \rightarrow$ hadrons events at 500 GeV compared to the 3 TeV case, the timing cuts in the event reconstruction are relaxed, which results in a significant improvement of the energy resolution for lower-energy jets.

Since the top quark decays predominantly into a W boson and a b quark, its decay topologies are determined by the decay of the W boson. In the analysis, the two most probable decays of the $t\bar{t}$ pair are selected, the fully-hadronic $e^+e^- \rightarrow t\bar{t} \rightarrow (q\bar{q}b)(q\bar{q}b)$ and the semi-leptonic $e^+e^- \rightarrow t\bar{t} \rightarrow (q\bar{q}b)(l\nu b)$ $(l = e, \mu)$ process, which are best suited for mass measurements since they have at most one neutrino in the final state. The mass is measured from the direct reconstruction of the invariant mass of the top decay products.

The cross section of $t\bar{t}$ pair production at a 500 GeV CLIC collider is approximately 530 fb. In the present study, a mass of 174 GeV and a width of 1.37 GeV is assumed. In addition to the signal, background processes with similar topologies, dominated by di- and tri-boson production, are considered, as summarised in Table 12.17. For both signal and background processes, all possible decay modes are simulated and consequently enter in the analysis, making the accurate selection of the desired final states a part of this study. The full simulation of events was performed with 300 bunch crossings of $\gamma\gamma \rightarrow$ hadrons events overlaid, to allow for the relaxed timing cuts at 500 GeV. The study was carried out for a data sample corresponding to an integrated luminosity of 100 $\text{fb}^{-1}$.

#### *12.4.7.1 Event Classification, Jet and Particle Identification*

Since the analysis depends on the event type (semi-leptonic or fully-hadronic), all events are classified according to the number of highly-energetic isolated charged leptons ($e^{\pm}$ or $\mu^{\pm}$). Events with two or more identified leptons are rejected, while events with zero and one lepton are classified as fully-hadronic and semi-leptonic candidates, respectively.

Table 12.17: Cross sections for the signal (inclusively, as well as separately for the investigated decay channels) and for the background processes considered in the top pair production study. All numbers are given for $\sqrt{s} = 500$ GeV using the CLIC luminosity spectrum at that energy.

| | Process | Cross section $\sigma$ (fb) |
|---|---|---|
| Signal | $e^+e^- \rightarrow t\bar{t}$ | 530 |
| | $e^+e^- \rightarrow t\bar{t} \rightarrow q\bar{q}b\, q\bar{q}b$ | 244 |
| | $e^+e^- \rightarrow t\bar{t} \rightarrow q\bar{q}b\, l\nu b$ $(l = e, \mu)$ | 159 |
| Background | $e^+e^- \rightarrow W^+W^-$ | 7100 |
| | $e^+e^- \rightarrow Z^0Z^0$ | 410 |
| | $e^+e^- \rightarrow q\bar{q}$ | 2600 |
| | $e^+e^- \rightarrow W^+W^-Z^0$ | 40 |

Based on this classification, jet finding is performed using an exclusive $k_t$ algorithm from the FASTJET package. Fully-hadronic event candidates are forced into six jets, while semi-leptonic event candidates are forced into four jets, where the identified lepton is excluded from jet finding. To mitigate the influence of $\gamma\gamma \rightarrow$ hadrons background, $\Delta\eta$, $\Delta\phi$ is used as distance measure in the jet finding, which reduces pickup of background particles in the forward region. A large jet size parameter $R = 1.3$ is used, since the jets in $t\bar{t}$ events are rather low-energetic and thus quite broad.

Efficient b-tagging is essential for the separation of $t\bar{t} \rightarrow (q\bar{q}b)(q\bar{q}b)$ and $t\bar{t} \rightarrow (q\bar{q}b)(l\nu b)$ events from multi-fermion background, and is also crucial for the assignment of jets to top candidates. Flavour tagging is performed using the LCFI package. For both samples the two jets with the highest b-tag values are classified as jets created by a b quark (b-jets). All other jets are classified as light-jets (created by u, d, s or c quarks), originating from the W decay. According to these classifications, jets are assigned to W candidates. In the semi-leptonic case, this assignment is unique, since only two light jets exist. In the fully-hadronic case, the pairing of light jets with the invariant mass closest to the W boson is chosen.

#### *12.4.7.2 Kinematic Fit and Background Rejection*

After the identification of b jets and the pairing of light jets and leptons into W bosons, the next step of the analysis is the grouping of W candidates and b jets into top quarks. This assignment is performed using a kinematic fit. Out of the two possible combinations of combining a W boson with a b jet, the one with the higher probability of the kinematic fit result is chosen. The kinematic fit itself is performed using the MarlinKinFit package [56], exploiting kinematic constraints of the signal process. In the present analysis, these are energy and momentum conservation, the W mass and the requirement of equal mass of the t and $\bar{t}$ candidates. For semi-leptonic events, a dedicated neutrino object is introduced to account for the unmeasured particle of one of the W decays. The behaviour of the fit is governed by the assumed angular and energy resolutions for the measured physics objects, which are obtained from simulation studies comparing fully reconstructed events with the generator information.

The kinematic fit fails if it is unable to satisfy the event constraints within the given accuracy of the input parameters. This rejects events with badly reconstructed jets, in particular badly reconstructed W bosons and large overlap between the two top candidates, consistent with the analysis goal to achieve measurements of the top quark properties.

Since the kinematic fit places stringent constraints on the overall event topology, it also serves as a powerful rejection of top events containing $\tau$ leptons from the W decay as well as non-$t\bar{t}$ background. Further reduction of background is achieved by means of a binned likelihood technique [57] which combines several discriminating variables into one likelihood variable. In combination with the kinematic fit, excellent rejection of background is achieved. The overall efficiency of the complete analysis chain is 35% (56%) for fully-hadronic (semi-leptonic) $t\bar{t}$ events.

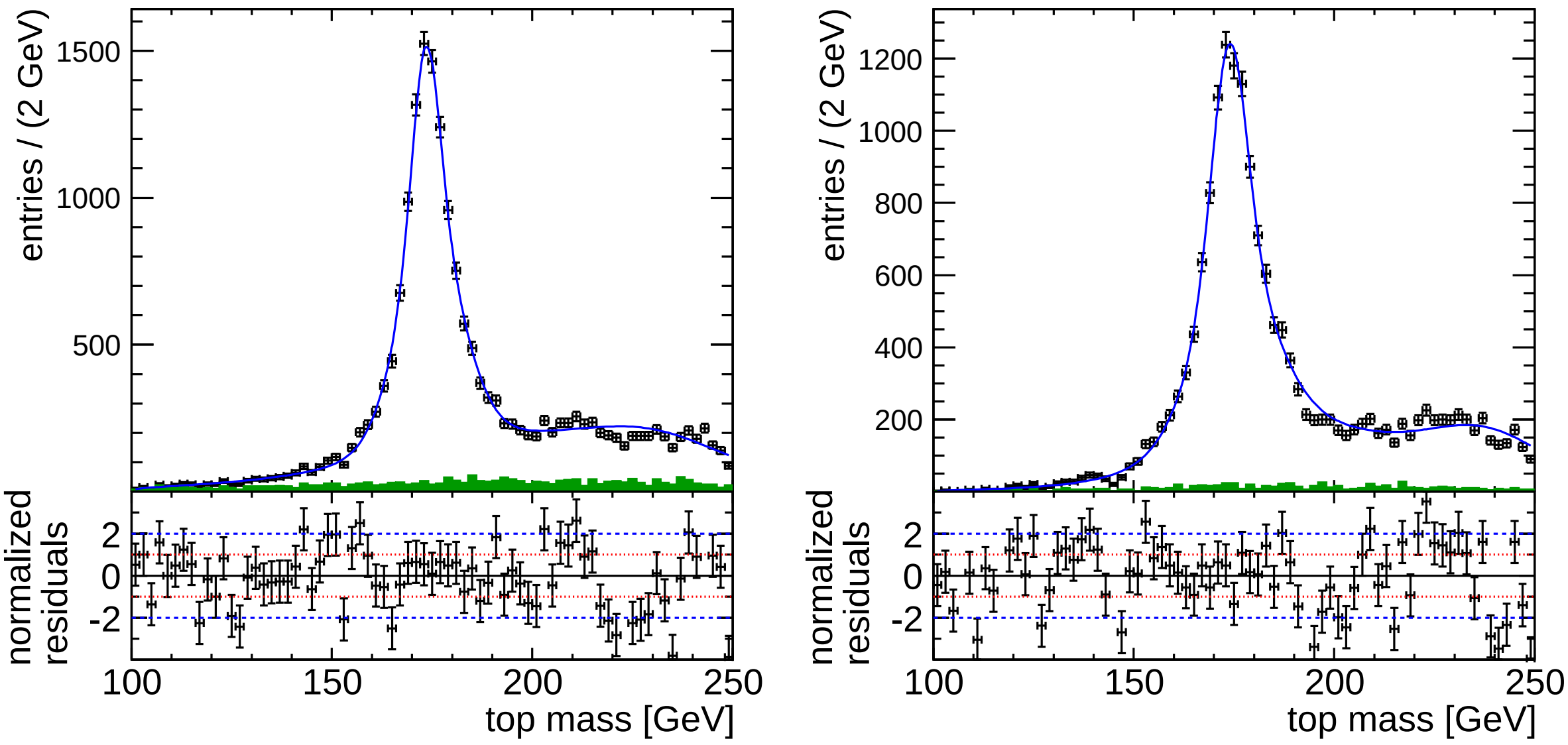

(a) hadronic

(b) semi-leptonic

Fig. 12.28: Mass distribution for 6-jet events (12.28a) and 4-jet events (12.28b). Points with error bars show simulated data classified as signal events. The green histogram illustrates the contribution of non-$t\bar{t}$ background to the distribution. The blue line shows the fit of the top mass distribution.

#### *12.4.7.3 Results: Top Mass and Width Measurement*

The top mass and width are extracted using an un-binned likelihood fit of the final top mass distribution of events (signal and background) after kinematic fitting and background rejection, as shown in Figure 12.28. The fit function consists of three components, which account for physics background, the detector resolution and the signal itself. The signal part is described by a Breit-Wigner distribution convolved with the sum of three Gaussians to account for the detector resolution. This resolution function is determined from an independent sample of fully simulated signal events by performing the fit with the parameters of the Breit-Wigner distribution fixed to the known input values. This sample has statistics corresponding to approximately 250 $\mathrm{fb}^{-1}$. Likewise, the shape of the background function is fixed in a fit to background events only. Possible biases originating from the top mass and width assumed in this training sample are excluded by tests with a sample with slightly different parameters, which yields consistent results.

In the fit of the final distribution containing signal and background events, the detector resolution contribution as well as the shape of the background distribution is fixed. The fit is performed independently for the fully-hadronic and for the semi-leptonic events. The distribution of the top mass and a fit to the data are shown in Figure 12.28.

The resulting top mass and width are summarised in Table 12.18. These results are in agreement with the generator values of 174 GeV for the top mass and 1.37 GeV for the width. This study shows that the top mass can be determined at a 500 GeV CLIC with a statistical uncertainty of 0.08 GeV and 0.09 GeV for the fully-hadronic and the semi-leptonic decay channel, respectively. These results are comparable to the ILC [18, 19].

Table 12.18: Results of the top pair production study. The fit results for fully-hadronic and semi-leptonic $t\bar{t}$ samples are shown. All numbers are obtained assuming an integrated luminosity of 100 $fb^{-1}$.

| Top decay | Top mass (GeV) | Top width (GeV) | Generator value | |
|---|---|---|---|---|
| | | | Mass (GeV) | Width (GeV) |
| Fully-hadronic | $174.07 \pm 0.08$ | $1.33 \pm 0.22$ | 174 | 1.37 |
| Semi-leptonic | $174.28 \pm 0.09$ | $1.55 \pm 0.26$ | | |

## 12.5 Summary

Several physics benchmark studies have been performed to demonstrate the performance of both the CLIC_ILD and CLIC_SiD detectors. The particular physics channels each cover individual physics observables and collectively represent a wide range of possible signatures expected in a general purpose detector at CLIC. The benchmark studies were carried out using full detector simulation and reconstruction, and they include $\gamma\gamma \rightarrow$ hadrons background overlay. The results of the benchmark studies are summarised in Table 12.19. The quoted errors represent the statistical accuracy of the results. Dependencies of the results on several systematic effects have been studied, and are reported in more detail the text of the preceding section. The impact of systematic effects is typically smaller than the statistical error, or at most equal.

The results demonstrate that the interesting physics processes at CLIC can be reconstructed with good precision despite challenging background conditions.

The top mass benchmark measurement at 500 GeV yields results that are comparable to those of similar ILC studies. This demonstrates that precision measurements are feasible with a CLIC machine operating at a few hundred GeV. One can therefore consider a scenario where CLIC is built and operated in successive energy stages if motivated by the physics landscape at the time of construction.

Table 12.19: Summary table of the CLIC benchmark analyses results. All studies at a centre-of-mass energy of 3 TeV are performed for an integrated luminosity of 2 $ab^{-1}$. The study at 500 GeV assumes an integrated luminosity of 100 $fb^{-1}$.

| $\sqrt{s}$ (TeV) | Process | Decay mode | SUSY model | Observable | Unit | Generator value | Stat. uncertainty |
|---|---|---|---|---|---|---|---|
| 3.0 | Light Higgs production | $h \to b\bar{b}$ | | $\sigma \times$ Branching ratio | fb | 285 | 0.22% |
| | | $h \to c\bar{c}$ | | | | 13 | 3.2% |
| | | $h \to \mu^+\mu^-$ | | | | 0.12 | 15.7% |
| 3.0 | Heavy Higgs production | $HA \to b\bar{b}b\bar{b}$ | I | Mass | GeV | 902.4 | 0.3% |
| | | | | Width | GeV | | 31% |
| | | | II | Mass | GeV | 742.0 | 0.2% |
| | | | | Width | GeV | | 17% |
| | | $H^+H^- \to t\bar{b}b\bar{t}$ | I | Mass | GeV | 906.3 | 0.3% |
| | | | | Width | GeV | | 27% |
| | | | II | Mass | GeV | 747.6 | 0.3% |
| | | | | Width | GeV | | 23% |
| 3.0 | Production of right-handed squarks | $\tilde{q}_R\tilde{q}_R \to q\bar{q}\tilde{\chi}_1^0\tilde{\chi}_1^0$ | I | Mass | GeV | 1123.7 | 0.52% |
| | | | | $\sigma$ | fb | 1.47 | 4.6% |
| 3.0 | Sleptons production | $\tilde{\mu}_R^+\tilde{\mu}_R^- \to \mu^+\mu^-\tilde{\chi}_1^0\tilde{\chi}_1^0$ | II | $\sigma$ | fb | 0.72 | 2.8% |
| | | | | $\tilde{\ell}$ mass | GeV | 1010.8 | 0.6% |
| | | | | $\tilde{\chi}_1^0$ mass | GeV | 340.3 | 1.9% |
| | | $\tilde{e}_R^+\tilde{e}_R^- \to e^+e^-\tilde{\chi}_1^0\tilde{\chi}_1^0$ | | $\sigma$ | fb | 6.05 | 0.8% |
| | | | | $\tilde{\ell}$ mass | GeV | 1010.8 | 0.3% |
| | | | | $\tilde{\chi}_1^0$ mass | GeV | 340.3 | 1.0% |
| | | $\tilde{e}_L^+\tilde{e}_L^- \to \tilde{\chi}_1^0\tilde{\chi}_1^0 e^+e^-hh$<br>$\tilde{e}_L^+\tilde{e}_L^- \to \tilde{\chi}_1^0\tilde{\chi}_1^0 e^+e^-Z^0Z^0$ | | $\sigma$ | fb | 3.07 | 7.2% |
| | | $\tilde{\nu}_e\tilde{\nu}_e \to \tilde{\chi}_1^0\tilde{\chi}_1^0 e^+e^-W^+W^-$ | | $\sigma$ | fb | 13.74 | 2.4% |
| | | | | $\tilde{\ell}$ mass | GeV | 1097.2 | 0.4% |
| | | | | $\tilde{\chi}_1^\pm$ mass | GeV | 643.2 | 0.6% |
| 3.0 | Chargino and neutralino production | $\tilde{\chi}_1^+\tilde{\chi}_1^- \to \tilde{\chi}_1^0\tilde{\chi}_1^0 W^+W^-$ | II | $\tilde{\chi}_1^\pm$ mass | GeV | 643.2 | 1.1% |
| | | | | $\sigma$ | fb | 10.6 | 2.4% |
| | | $\tilde{\chi}_2^0\tilde{\chi}_2^0 \to h^0/Z^0h^0/Z^0\tilde{\chi}_1^0\tilde{\chi}_1^0$ | | $\tilde{\chi}_2^0$ mass | GeV | 643.1 | 1.5% |
| | | | | $\sigma$ | fb | 3.3 | 3.2% |
| 0.5 | $t\bar{t}$ production | $t\bar{t} \to (q\bar{q}b)\,(q\bar{q}b)$ | | Mass | GeV | 174 | 0.046% |
| | | | | Width | GeV | 1.37 | 16% |
| | | $t\bar{t} \to (q\bar{q}b)\,(\ell\nu b)$, $\ell = e, \mu$ | | Mass | GeV | 174 | 0.052% |
| | | | | Width | GeV | 1.37 | 18% |

# Chapter 13

# Future Plans and R&D Prospects

## 13.1 Introduction

The main aim of this CDR is to assess the physics potential of a 3 TeV CLIC accelerator and to see whether experimental methods can be proposed to measure the physics with adequate precision. In many domains the CLIC physics and detector study has drawn on work previously carried out for lower-energy $e^+e^-$ machines under study. This holds for both detector concepts and the accompanying hardware developments, but also for the simulation software and analysis methods. A large fraction of the work for the current CDR was directed towards simulation studies to understand the issues at stake at CLIC, in particular the impact of the beam-induced background on the detectors. The detector geometries underwent a first round of optimisations to minimise the impact of beam-induced background, such as backscattering from the forward region into the vertex detector. The HCAL was made deeper and denser to reduce the effect of shower leakage, and PFA was extended to the required higher energy scale. Despite the substantial beam-induced background, good physics performance can be achieved by applying precise timing cuts and $p_T$ cuts on reconstructed physics objects and by using appropriate jet-clustering. On the engineering side, the work focused on the detector layout surrounding the incoming and outgoing beams. In particular, methods were developed by which the final focusing quadrupole can be positioned within the detector volume with unprecedented stability.

## 13.2 Activities for the next Project Phase

In the next project phase more in-depth studies will be carried out. The detector simulation studies will be pushed towards a more profound understanding and will assess areas that have already been identified as subject to improvement. The energy range, which focused on a centre-of-mass energy of 3 TeV for the CDR, will be extended to lower energies, assessing ways to gradually explore the physics with a machine built in successive stages. Most of the hardware R&D currently ongoing for linear collider detector purposes is equally valid for ILC and CLIC, however there need to be a few focused R&D efforts for CLIC, since the CLIC conditions push the demands on the detector even further than for ILC. The time line for upcoming phases of the CLIC detector study will be driven by the evolution of the physics landscape through results from LHC or elsewhere. The detector study will be linked closely to the CLIC accelerator developments, and will strongly depend on funding opportunities. In the following, an outline of activities for the next phase is presented. It is assumed that most of these activities could be carried out over a 5-year time span.

### 13.2.1 Simulation Studies and Detector Optimisation

Besides a confirmation of the strengths of the proposed detector concepts, the current studies have also revealed some weaknesses in the CLIC environment. Using the knowledge and tools acquired for the CDR, these will be addressed in the next phase. The high TPC occupancies will be studied in more detail, which may result in adaptations to either the TPC technology or the geometry. The inner tracker regions show occupancies that seem too high to be read out by current silicon strip detectors. These occupancies will be studied in more detail and, where appropriate, adapted technology choices defined. High occupancies near the lower radii of the HCAL endcap require further masking studies in the forward region and further optimisation of the readout segmentation in this region. The final focusing quadrupole (QD0) and its infrastructure for mechanical stability result in a significant reduction of the detector acceptance. Forward physics processes therefore need to be studied more extensively in the next project phase. The balance between advantages and disadvantages of having the focusing element inside the detector volume will be studied. The determination of the luminosity spectrum and the measurement of the absolute luminosity, as well as the influence of beam-beam affects on these measurements will be studied in more detail. An assessment will be made of the optimal detector aspect ratio. Detector optimisation and background mitigation studies at lower centre-of-mass energies are also foreseen. In view of the push-pull scenario, fast alignment strategies using particle tracks need to be developed. In addition, there is a broad range of simulation studies in support of detector development and beam tests. Subsequently well-validated detector readout responses for the various technologies under consideration will be included in the full simulation studies of the concepts. The main areas of work are:

- Mitigation of high occupancies in the low-angle region of the endcap calorimetry;
- Occupancies in the inner tracking regions and related technology choices;
- Origin and mitigation of high TPC occupancies;
- Location of QD0 inside or outside the detector and impact on the physics;
- Systematic effects on the measurement of the absolute luminosity, the luminosity spectrum and the beam polarisation;
- Detector optimisation and background suppression at intermediate centre-of-mass energies;
- Simulation studies in support of detector development and beam tests;
- Implementation of the response of various detector readout technologies in the full-detector simulations;
- Development of fast alignment strategies using tracks from suitable physics events.

### 13.2.2 Physics at CLIC

The physics potential of CLIC is rich for a wide variety of possible physics scenarios that are consistent with current (end 2011) data. Thus, a strong case has been made in this document for the CLIC collider. Nevertheless, results from the LHC will continue to accrue over the coming years, and they will play a critical role in the development of the more refined physics case to be made for CLIC. Staying abreast of these developments, and understanding how they impact the CLIC physics case, is a primary activity. This and the other main activities of theory are:

- Monitor the developments at the LHC and report on their implications for the physics potential of CLIC;
- Investigate the physics opportunities and challenges of a staged approach to reaching the highest energy of the CLIC machine;
- Investigate the relative merits of electron polarisation versus combined electron and positron polarisation;
- Study a supersymmetric benchmark model point in full detail to determine all the masses and mixings that can be measured, and investigate how well these measurements can lead us to answers to fundamental questions such as the verification of supersymmetry, the origin and mediation of supersymmetry breaking, the relic abundance of the lightest neutralino, and the compatibility of the model to various approaches to explaining the baryon asymmetry of the universe.

### 13.2.3 Software Development

For the studies reported in this CDR the MOKKA & MARLIN and SLIC & `org.lcsim` software frameworks, initially developed for the ILC detector concepts ILD and SiD, have been used extensively. Even though the two frameworks differ in their basic approach and programming language, they use the common LCIO data format. A number of software tools were used for the simulation studies in this CDR for both CLIC_SiD and CLIC_ILD: physics event generation using WHIZARD and PYTHIA, the ILCDIRAC grid production tool, the PANDORAPFA particle flow analysis package and the LCFI flavour tagging package. Further developments and improvements of the software for the upcoming phase of the project will build more on tools that provide common solutions for both detector concepts and are shared with other experiments. Both CLIC_ILD and CLIC_SiD reconstruction tools require improved and well-maintained tracking and flavour tagging codes. Improved tools for geometry descriptions are needed, potentially profiting from recent initiatives towards more advanced common geometry tools for GEANT4 and ROOT. In a next phase reconstruction tools will be adapted to benefit more from the granularity of hits in space and time. The main areas of software development are:

- Roadmap towards common software tools for both experiments;
- Improved and well-maintained tracking and flavour tagging codes;
- Improved software tools for geometry descriptions;
- More advanced reconstruction methods, making use of the granularity in space and time.

### 13.2.4 Vertex Detector

The vertex detector is among the largest technological challenges for a CLIC detector. A set of competing requirements has to be met in a highly integrated subdetector, reaching well beyond the current state-of-the-art. The R&D activities listed below are highly correlated and have to be pursued in parallel within a fully integrated effort:

- Developments towards a thin hybrid or integrated CMOS or multi-tier (SOI, 3D or other) pixel technology with small pixel sizes of $\mathcal{O}(20\ \mu\mathrm{m})$ and a hit time resolution of $\mathcal{O}(5\ \mathrm{ns})$;
- Development of high-density interconnect technologies towards maximum detector integration and seamless tiling;
- Thinning of wafers, ASICs or tiers and development of low-mass construction and services materials to reach $\mathcal{O}(0.2\%\ X_0)$ material per layer;
- Advanced power reduction, power delivery, power pulsing and cooling developments to reach $\mathcal{O}(0.2\%\ X_0)$ material per layer.

### 13.2.5 Silicon Tracking

In order to reach the required momentum resolution, silicon tracking systems need to be substantially thinner than in the current LHC experiments. This can be achieved through fully integrated low-mass designs, already under development for ILC. These designs are based on low-power electronics, chip-on-sensor bonding, power pulsing, air cooling and low-mass supports. In addition, silicon detectors at CLIC will include time stamping with a time window of 10 ns. Occupancies in the inner tracker regions will be high. This calls for novel solutions, either in the form of replacing the $\approx$10 cm long silicon strips by smaller cells, or by implementing multi-hit capabilities in the electronics. The main areas of activity are:

- Continued development and beam tests of low-mass silicon strip detectors with time stamping functionalities, low-power electronics, power pulsing, air cooling and low-mass supports;
- Study of technology choices to mitigate high occupancies in the inner tracking regions.

### 13.2.6 TPC-based Tracking

The ongoing TPC prototype studies and beam tests with the Large Prototype TPC are relevant for both ILC and CLIC. They assess GEM and Micromegas gas multiplication with pad readout and pixelised readout, ion backflow suppression, integration issues and cooling of the readout plane, as well as the influence of the magnetic field map on the TPC tracking performance. The main areas of development are therefore:

– Continued TPC prototype tests (GEM, Micromegas, pad, pixel, ion backflow, magnetic field effects);
– TPC endplate integration and cooling.

### 13.2.7 Calorimetry

There is a broad ongoing linear collider R&D effort in PFA-based fine-grained calorimetry, which is relevant for ILC and for CLIC. This includes extensive beam tests of large prototypes in a variety of readout technology options and engineering studies of realistic fully integrated technological prototypes. This R&D will be extended to address the timing requirements for CLIC through the study of fast signal formation in active layers and dedicated electronics developments. Beam tests with tungsten-based HCAL modules and two kinds of active layers (sensitive and insensitive to neutrons) are required to verify GEANT4 simulations to the same level as was done for steel absorbers. Further development is also needed towards a more cost-effective ECAL solution, while the forward calorimetry (BeamCal and LumiCal) requires excellent radiation hardness and new electronics approaches in view of the high rates in that area. The main areas of calorimeter development are:

– Continued beam tests of fine-grained ECAL, HCAL and forward calorimeter modules based on different active and passive layers (including tungsten for HCAL) and accompanying validation of GEANT4 modelling;
– Engineering designs and technological prototypes of ECAL, HCAL and forward calorimetry;
– Electronics developments for calorimetry at CLIC, including extensive power delivery and power pulsing tests at the system level.

### 13.2.8 Electronics and Power Delivery

The various CLIC detector subsystems call for a wide range of advanced developments in electronics. These include qualification studies of very deep sub-micron technologies and core semiconductor materials, densely packed front-end ASICs with complex functionalities, system integration and data acquisition. Across all subdetectors, there is a need for low-power designs, complemented with efficient power delivery (to reduce the material budget associated with the services) and power pulsing (to reduce heat dissipation and need for cooling). Power pulsing requires extensive testing to study any possible adverse effect on performance and system stability (e.g. following repeated mechanical stress). In view of the challenging CLIC bunch structure and beam-induced background, the front-end electronics will integrate many functions. For example, the pixel detector electronics needs to include time stamping with a time window of 10 ns, while preserving a good signal-to-noise ratio for each individual pixel of ≈20 µm size. Inner silicon strip detectors require time stamping with a time window of 10 ns, while preserving a good signal-to-noise ratio and multi-hit functionality for up to five hits in a 156 ns bunch train. The calorimeter readout requires pulse height information over a large dynamic range for each cell, time information to 1 ns precision and multi-hit functionality for up to five hits in a 156 ns bunch train. Advanced interconnect technologies and low-mass services will be an integral part of the R&D programme listed below:

– Qualification of deep sub-micron technologies for the integration of advanced functionalities in compact detector ASICs;
– Studies and prototyping of core front-end functionalities with low power consumption, in particular: pulse height and time measurements, in some cases (silicon tracking and calorimetry) combined with multi-hit functionality within the 156 ns bunch train, as well as on-chip power pulsing features;

– Development and extensive testing of power delivery and power pulsing, including development of advanced protection schemes and system tests in a 4 to 5 T magnetic field;
– Interconnect technologies for front-end electronics and low-mass services.

### 13.2.9 Magnet and Ancillary Systems

Detector magnet development will principally focus on conductor R&D and on magnet services. Modelling and measurement of new materials for conductor re-inforcement will be pursued. Extrusion tests and cold-working of conductors with large cross sections will be carried out, followed by characterisation tests. Study and tests of an appropriate winding technique for the large conductor will be pursued. The development and prototyping of a flexible high-temperature superconducting power line will profit from synergy with accelerator applications. In case the final focusing quadrupole is maintained inside the detector configuration, engineering studies and R&D towards a compact anti-solenoid will be necessary. In this case, the strong coupling forces between the solenoid and the anti-solenoid will represent a major challenge. Prototypes of safety elements, such as the water-cooled dump resistor, will be developed in a later stage. The magnet developments therefore comprise:

– Extrusion tests and characterisation of a large re-inforced superconductor;
– Material studies and tests of new conductor re-inforcement materials;
– Winding technique for a large conductor;
– Flexible high-temperature power line;
– Prototyping of safety elements, e.g. a water-cooled dump resistor;
– Development and integration of a compact anti-solenoid (only needed in case the QD0 location inside the experiment is confirmed).

### 13.2.10 Engineering and Detector Integration

Throughout the next phase, the overall detector design and integration effort will treat all aspects of the various detector components and their services in more detail. In-depth engineering studies will initially focus on new elements related to the CLIC implementation of the detector concepts. These will include small-scale prototyping for a tungsten HCAL barrel detector, for example tests of joining techniques for tungsten-tungsten and tungsten-steel connections. The studies during the CDR phase have concentrated on detectors with the last focusing quadrupole inside the detector volume. In the next phase, a solution with the quadrupole located outside of the detector will be studied with high priority. This presents additional challenges, since an even shorter detector length along the beam axis is desirable to achieve maximum luminosity. Moreover, the condition of two detectors with the same overall length remains. It will be necessary to study, in more detail, the possibility to use end-coils to achieve good field quality and small fringe fields for a short magnet. The detector movement and push-pull operations require additional studies, for example on the detector platforms and on the integration of motors, hydraulics, air-pads and cable chains for a safe and rapid movement with minimum interventions and including seismic safety. Development of detector alignment techniques will also be pursued in the upcoming phase, together with techniques for deformation measurements, which may prove beneficial to realise the ultra-low mass requirements for the tracking systems. Engineering studies of the overall mechanical stability of the central beam pipe, comprising a thin beryllium central cylinder joined to thick conical steel sections, will also be included. For this beryllium-steel junction, production techniques will be studied and prototypes will be needed.

– Design and integration of the detector concepts in gradually increasing detail;
– Construction and joining techniques with tungsten;

- Engineering and layout studies for a short detector length including end-coils;
- Detector movements and push-pull operation;
- Detector alignment techniques and deformation measurements;
- Engineering and production techniques of a beryllium with steel beam pipe.

# Summary

In this CDR, an assessment of the physics potential of a future multi-TeV $e^+e^-$ collider has been made, and detector concepts are described which are able to measure the physics with good precision. The studies presented here are mostly focused on the case of a CLIC machine operating at 3 TeV. Since the impact of background increases with energy, this is considered to be the most difficult environment.

Beam related effects that impact the experimental conditions, such as backgrounds and time-structure of the bunch trains, were simulated in order to investigate their influence on the experimental accuracy of the physics observables. Full detector simulation studies with beam background overlaid were carried out for a number of relevant detector benchmark processes. They show that the expected physics signals with mass scales from about 100 GeV up to around 1.5 TeV can be extracted with the required precision at such a 3 TeV collider and the impact of the beam related backgrounds can be effectively suppressed. Altogether, the simulation results presented in this CDR indicate that, if new physics occurs at the fore-mentioned mass scale, this physics can be studied with high precision at CLIC with the proposed detector concepts.

The CLIC physics and detector study is part of a world-wide effort towards a future linear $e^+e^-$ collider, and the work reported in this document has been carried out with broad international participation. Many of the tools and methodologies used for the CDR studies have been drawn from previous work carried out in the framework of physics and detector studies for the ILC. Specifically, the detector concepts and technologies proposed for CLIC are based on the ILD and SiD concepts and on detector hardware R&D, initially developed for lower centre-of-mass energies. These ILC detector concepts were adapted to accommodate the higher energies and the more challenging experimental environment at CLIC.

In the coming years, further in-depth studies and hardware R&D for the CLIC detectors are foreseen. The detector simulation studies will be refined and will address specific areas requiring further optimisation. These simulation studies will address the full centre-of-mass energy range that will be accessible with a future CLIC machine. Given the large overlap in performance requirements, hardware R&D will principally be carried out in common for detectors at ILC and CLIC. However, as the CLIC conditions push the demands on the detector even further, dedicated hardware R&D for CLIC will continue in a number of priority areas.

Above all, however, the evolution of LHC results at 7 TeV and subsequently at 14 TeV centre-of-mass energy will be actively followed, and the expected impact on the physics to be explored at CLIC will be studied.

# Acknowledgements

The CLIC physics and detector, described in this document, benefited from the ILC detector R&D efforts, which were supported by BMWF, Austria; MinObr, Belarus; FNRS and FWO, Belgium; NSERC, Canada; NSFC, China; MPO CR and VSC CR, Czech Republic; FP6 and FP7 European Commission, European Union; HIP, Finland; IN2P3-CNRS, CEA-DSM/IRFU, France; BMBF, DFG, HGF, MPG and AvH Foundation, Germany; DAE and DST, India; ISF, Israel; INFN, Italy; MEXT and JSPS, Japan; CRI(MST) and MOST/KOSEF, Korea; FOM, NWO, the Netherlands; NFR, Norway; MNSW, Poland; ANCS, Romania; MES of Russia and ROSATOM, Russian Federation; MON, Serbia and Montenegro; MSSR, Slovakia; MICINN and CPAN, Spain; SRC, Sweden; ETHZ and Uni-GE, Switzerland; STFC, United Kingdom; DOE and NSF, United States of America. We acknowledge the support from CERN as the host laboratory of this study. We would like to gratefully acknowledge CERN, DESY, FNAL and KEK and their accelerator staff for the reliable and efficient test beam operation.

# Appendices

# Appendix A

# Acronyms

| | |
|---|---|
| **ALEPH** | Apparatus for LEP Physics |
| **ALICE** | A Large Ion Collider Experiment |
| **ANSYS** | Package for finite-element analysis |
| **ATLAS** | A Toroidal LHC ApparatuS - one of the two general purpose LHC detectors |
| **BDS** | Beam Delivery System |
| **BDT** | Boosted Decision Tree |
| **BeamCal** | Beam Calorimeter |
| **BEBC** | Big European Bubble Chamber |
| **BPM** | Beam Position Monitor |
| **BSM** | Beyond the Standard Model |
| **CALICE** | Calorimeter for Linear Collider Experiment |
| **CDR** | Conceptual Design Report |
| **CLIC** | Compact Linear Collider |
| **CMS** | Compact Muon Solenoid - one of the two general purpose LHC detectors |
| **CMSSM** | constrained MSSM |
| **CVD** | Chemical Vapor Deposition |
| **DELPHI** | Detector with Lepton,Photon and Hadron Identification |
| **ECAL** | Electromagnetic Calorimeter |
| **ETD** | Endcap Tracking Disk |
| **EUDET** | EU Detector R&D Towards the International Linear Collider |
| **EWSB** | Electroweak Symmetry Breaking |
| **FEA** | Finite Element Analysis |
| **FSR** | Final State Radiation |
| **FTD** | Forward Tracking Disk |
| **GEM** | Gas Electron Multiplier |
| **GMSB** | Gauge Mediated Supersymmetry Brekaing |
| **GUT** | Grand Unified Theory |
| **HCAL** | Hadronic Calorimeter |

**HTS** High Temperature Superconductor

**IDAG** International Detector Advisory Group

**ILC** International Linear Collider

**ILD** International Large Detector - one of the two validated ILC detector concepts

**IP** Interaction Point

**ISR** Initial State Radiation

**LCFI** Linear Collider Flavour Identification

**L**$^*$ Distance from exit of last quadrupole (QD0) to the interaction point (IP)

**LDT** LiC Detector Toy simulation and reconstruction framework

**LEP** Large Electron Positron Collider at CERN

**LHC** Large Hadron Collider at CERN

**LSP** Lightest (R-parity odd) Supersymmetric particle

**LumiCal** Luminosity Calorimeter

**LVDS** Low Voltage Differential Signal

**MAPS** Monolithic Active Pixel Sensor

**mGMSB** Minimal Gauge Mediated Supersymmetry Breaking

**Micromegas** Micro Mesh Gaseous Structure

**MIP** Minimum Ionising Particle

**MPGD** Micro-Pattern Gas Detector

**MPPC** Multi-Pixel Photon Counter

**MSSM** Minimal Supersymmetric Standard Model

**mSUGRA** Minimal Supergravity Model

**NIEL** Non-ionizing Energy Loss

**OPAL** OmniPurpose Apparatus at LEP

**PDF** Probability Density Function

**PEEK** Poly Ether Ether Ketone

**PFA** Particle Flow Algorithms

**PFO** Particle Flow Object

**pMSSM** Phenomenological Minimal Supersymmetric Standard Model

**POISSON** Software for magnetic field calculations

**QD0** Quadrupole magnet closest to the interaction point

| | |
|---|---|
| **RF** | Radio Frequency |
| **RPC** | Resistive Plate Chambers |
| **SET** | Silicon External Tracker |
| **SiD** | Silicon Detector - one of the two validated ILC detector concepts |
| **SiPM** | Silicon Photomultiplier |
| **SIT** | Silicon Internal Tracker |
| **SLAC** | SLAC National Accelerator Laboratory |
| **SLC** | Stanford Linear Collider at SLAC |
| **SLD** | Stanford Large Detector |
| **SM** | Standard Model |
| **STAR** | Solenoidal Tracker at RHIC |
| **SUSY** | Supersymmetry |
| **T2K** | Tokai to Kamioka |
| **TID** | Total Ionising Dose |
| **TPC** | Time Projection Chamber |
| **VDSM** | Very Deep Sub-Micron |
| **VTX** | Vertex Detector |

# Appendix B

# Simulation and Reconstruction Parameters

## B.1 PFO Lists at 3 TeV

Table B.1: Cuts on the `DefaultSelectedPFO` list in the mass production

| Region | $p_T$ range | time cut |
|---|---|---|
| | Photons | |
| central | $0.75 \text{ GeV} \leq p_T < 4.0 \text{ GeV}$ | t < 2.0 ns |
| $\cos\theta \leq 0.975$ | $0 \text{ GeV} \leq p_T < 0.75 \text{ GeV}$ | t < 1.0 ns |
| forward | $0.75 \text{ GeV} \leq p_T < 4.0 \text{ GeV}$ | t < 2.0 ns |
| $\cos\theta > 0.975$ | $0 \text{ GeV} \leq p_T < 0.75 \text{ GeV}$ | t < 1.0 ns |
| | neutral hadrons | |
| central | $0.75 \text{ GeV} \leq p_T < 8.0 \text{ GeV}$ | t < 2.5 ns |
| $\cos\theta \leq 0.975$ | $0 \text{ GeV} \leq p_T < 0.75 \text{ GeV}$ | t < 1.5 ns |
| forward | $0.75 \text{ GeV} \leq p_T < 8.0 \text{ GeV}$ | t < 2.0 ns |
| $\cos\theta > 0.975$ | $0 \text{ GeV} \leq p_T < 0.75 \text{ GeV}$ | t < 1.0 ns |
| | charged particles | |
| all | $0.75 \text{ GeV} \leq p_T < 4.0 \text{ GeV}$ | t < 3.0 ns |
| | $0 \text{ GeV} \leq p_T < 0.75 \text{ GeV}$ | t < 1.5 ns |

Table B.2: Cuts on the `LooseSelectedPFO` list in the mass production

| Region | $p_T$ range | time cut |
|---|---|---|
| | Photons | |
| central | $0.75 \text{ GeV} \leq p_T < 4.0 \text{ GeV}$ | t < 2.0 ns |
| $\cos\theta \leq 0.975$ | $0 \text{ GeV} \leq p_T < 0.75 \text{ GeV}$ | t < 2.0 ns |
| forward | $0.75 \text{ GeV} \leq p_T < 4.0 \text{ GeV}$ | t < 2.0 ns |
| $\cos\theta > 0.975$ | $0 \text{ GeV} \leq p_T < 0.75 \text{ GeV}$ | t < 1.0 ns |
| | neutral hadrons | |
| central | $0.75 \text{ GeV} \leq p_T < 8.0 \text{ GeV}$ | t < 2.5 ns |
| $\cos\theta \leq 0.975$ | $0 \text{ GeV} \leq p_T < 0.75 \text{ GeV}$ | t < 1.5 ns |
| forward | $0.75 \text{ GeV} \leq p_T < 8.0 \text{ GeV}$ | t < 2.5 ns |
| $\cos\theta > 0.975$ | $0 \text{ GeV} \leq p_T < 0.75 \text{ GeV}$ | t < 1.5 ns |
| | charged particles | |
| all | $0.75 \text{ GeV} \leq p_T < 4.0 \text{ GeV}$ | t < 3.0 ns |
| | $0 \text{ GeV} \leq p_T < 0.75 \text{ GeV}$ | t < 1.5 ns |

Table B.3: Cuts on the `TightSelectedPFO` list in the mass production

| Region | $p_T$ range | time cut |
|---|---|---|
| | Photons | |
| central | 1.0 GeV $\leq p_T <$ 4.0 GeV | t < 2.0 ns |
| $\cos\theta \leq 0.95$ | 0.2 GeV $\leq p_T <$ 1.0 GeV | t < 1.0 ns |
| forward | 1.0 GeV $\leq p_T <$ 4.0 GeV | t < 2.0 ns |
| $\cos\theta > 0.95$ | 0.2 GeV $\leq p_T <$ 1.0 GeV | t < 1.0 ns |
| | neutral hadrons | |
| central | 1.0 GeV $\leq p_T <$ 8.0 GeV | t < 2.5 ns |
| $\cos\theta \leq 0.95$ | 0.5 GeV $\leq p_T <$ 1.0 GeV | t < 1.5 ns |
| forward | 1.0 GeV $\leq p_T <$ 8.0 GeV | t < 1.5 ns |
| $\cos\theta > 0.95$ | 0.5 GeV $\leq p_T <$ 1.0 GeV | t < 1.0 ns |
| | charged particles | |
| all | 1.0 GeV $\leq p_T <$ 4.0 GeV | t < 2.0 ns |
| | 0 GeV $\leq p_T <$ 1.0 GeV | t < 1.0 ns |

## B.2 PFO Lists at 500 GeV

Table B.4: Cuts on the `DefaultSelectedPFO` list at 500 GeV in the mass production

| Region | $p_T$ range | time cut |
|---|---|---|
| | Photons | |
| central | $1.0\ \text{GeV} \leq p_T < 2.0\ \text{GeV}$ | t < 5.0 ns |
| $\cos\theta \leq 0.975$ | $0\ \text{GeV} \leq p_T < 1.0\ \text{GeV}$ | t < 2.5 ns |
| forward | $0.75\ \text{GeV} \leq p_T < 4.0\ \text{GeV}$ | t < 2.0 ns |
| $\cos\theta > 0.975$ | $0\ \text{GeV} \leq p_T < 0.75\ \text{GeV}$ | t < 1.0 ns |
| | neutral hadrons | |
| central | $1.0\ \text{GeV} \leq p_T < 2.0\ \text{GeV}$ | t < 5.0 ns |
| $\cos\theta \leq 0.975$ | $0\ \text{GeV} \leq p_T < 1.0\ \text{GeV}$ | t < 2.5 ns |
| forward | $2.0\ \text{GeV} \leq p_T < 4.0\ \text{GeV}$ | t < 2.0 ns |
| $\cos\theta > 0.975$ | $0\ \text{GeV} \leq p_T < 2.0\ \text{GeV}$ | t < 1.0 ns |
| | charged particles | |
| all | $1.0\ \text{GeV} \leq p_T < 4.0\ \text{GeV}$ | t < 10.0 ns |
| | $0\ \text{GeV} \leq p_T < 1.0\ \text{GeV}$ | t < 3.0 ns |

## B.3 PYTHIA Parameters

Table B.5: PYTHIA hadronisation tuning parameters

| Parameter | label | PYTHIA default | OPAL tuning |
|---|---|---|---|
| longitudinal F | MSTJ(11) | 4 | 3 |
| source for $\Lambda$ | MSTP(3) | 2 | 1 |
| qq/q | PARJ(1) | 0.10 | 0.085 |
| s/u | PARJ(2) | 0.30 | 0.31 |
| su/du | PARJ(3) | 0.40 | 0.45 |
| S=1/S=2 diquark suppr. | PARJ(4) | 0.05 | 0.025 |
| (S=1) d,u | PARJ(11) | 0.50 | 0.60 |
| (S=1) s | PARJ(12) | 0.60 | 0.40 |
| (S=1) c,b | PARJ(13) | 0.75 | 0.72 |
| S=1, s=0 prob. | PARJ(14) | 0.0 | 0.43 |
| S=0, s=1 prob. | PARJ(15) | 0.0 | 0.08 |
| S=1, s=1 prob. | PARJ(16) | 0.0 | 0.08 |
| tensor mesons (L=1) | PARJ(17) | 0.0 | 0.17 |
| leading baryon suppr. | PARJ(19) | 1.0 | - |
| $\sigma_q$ (GeV) | PARJ(21) | 0.36 | 0.40 |
| $\eta'$ suppression | PARJ(26) | 0.40 | - |
| $a$ of LSFF | PARJ(41) | 0.30 | 0.11 |
| $b$ of LSFF (GeV$^{-2}$) | PARJ(42) | 0.58 | 0.52 |
| $\varepsilon_{charm}$ | PARJ(54) | 0.05 | -0.031 |
| $\varepsilon_{bottom}$ | PARJ(55) | 0.005 | -0.002 |
| $\Lambda_{QCD}$ (GeV) | PARJ(81) | 0.29 | 0.25 |
| PS cut-off (GeV) | PARJ(82) | 1.0 | 1.90 |

Table B.6: Masses of quarks and bosons used for the generation of Standard Model samples

| Particle | Mass (GeV/$c^2$) | Width (GeV/$c^2$) |
|---|---|---|
| up, down, strange | 0 | 0 |
| charm | 0.54 | 0 |
| bottom | 2.9 | 0 |
| top | 174 | 1.37 |
| W | 80.45 | 2.071 |
| $Z^0$ | 91.188 | 2.478 |
| higgs | 120 | 0.0036 |

# Appendix C

# Cost Methodology for a CLIC Detector

## C.1 Introduction

This appendix summarises the methodology used for estimating the tentative cost of a detector for CLIC. The actual values of the estimated detector costs will be included in a forthcoming volume of the CLIC Conceptual Design Report. The physics goals and conceptual engineering design of the detectors are described elsewhere in this volume of the CDR. In terms of costing, the present designs includes the following principal components of the experiment:

- a vertex detector and tracker;
- an electromagnetic calorimeter;
- a hadronic calorimeter;
- a magnet solenoid, return yoke and muon instrumentation;
- experiment-specific infrastructure.

The basic raw materials and principal cost drivers have been identified and include: superconducting cable, silicon sensors, steel and tungsten. For selected assumptions related to production parameters and methods, cost information is available for all of these items.

Based on experience gained from constructed detectors (e.g. for LEP, SLC, Tevatron, LHC), direct detector costs can be simulated in a rather straightforward way as a function of the required detector design and engineering parameters. Such simulation models exist for both the International Large Detector (ILD) and the Silicon Detector (SiD). Cost sensitivity analysis can be performed, and comparative cost studies can be made, once cost units and cost drivers are coherently used within these cost simulation models.

## C.2 Scope of Detector Costing

Detector-related costs which can be controlled in the process of making such estimates have been identified. These include direct construction costs related to the detector in terms of raw materials and, in a few cases, manufactured components where price information is available and has been collected. Experiment-specific infrastructure (e.g. a platform), on-detector services (gas, cooling etc.) and detector-specific safety systems are included.

Items which are not considered in the cost estimate are, e.g., generic detector R&D as well as the core personnel effort and the use of technical infrastructure in the participating institutes. Also not included are the supporting cavern infrastructure around and interfacing with the detector, such as access structures and systems, safety and security systems, technical services[1], cooling and ventilation systems etc. Moreover, the costing does not include items that could be identified as part of the Host Laboratory obligations such as access roads, surface buildings, electricity, communication, gas, water, coolants and other distribution systems to and inside the experimental caverns. In the absence of any agreement on what would be part of the Host Laboratory responsibilities, all such items are not included in the detector cost analysis presented here. (These items are, however, included in the cost estimate of the CLIC accelerator facility). Finally, off-line computing is also not included in the current study.

## C.3 Guiding Principles

In order to approach the question of CLIC detector costing in a meaningful way, and to fully benefit from the similar costing work already carried out elsewhere (e.g. by the ILC detector concepts), the following

[1] Exceptions are the services directly linked to the superconducting magnet of the detector.

underlying assumptions are made:

1. Only direct, production-related personnel and at-factory testing efforts are included in the cost estimates. Estimates of personnel efforts in participating institutes are thus excluded[2].

2. Related items not included in the above costing will be indicated for clarity, however, without providing any cost estimates.

3. Unit currency is the Swiss Franc (CHF), as applied at CERN (e.g. price information obtained by CERN).

4. Unit costs for tungsten, steel and silicon are as agreed by the CLIC detector costing Working Group in its meeting on October 21, 2010: see Table C.1, below. In this table, and reflecting approximately the situation in 2010, the US-$/CHF exchange rate is assumed to be 1.0 for simplicity, and the €/CHF is assumed to be 1.5.

5. A cost sensitivity analysis can be carried out for the detector by assuming e.g. a $\pm$ 30% variance in the above unit prices.

6. Unit costs are expressed in 2010 prices, thus not taking future inflation into account. Price information collected from earlier years are escalated to 2010 prices using official CERN materials cost variation indices (CVI);

7. No assumptions are made about future technology impact or demand fluctuations on the unit prices used.

8. Contingency is not addressed at this stage.

Table C.1: Assumed unit cost for some materials

| Unit | Agreed Unit Cost |
|---|---|
| Tungsten for HCAL | 105 CHF / kg |
| Tungsten for ECAL (tighter mechanical tolerances) | 180 CHF / kg |
| Steel for Yoke (semi-product) | 1000 CHF / t |
| Steel for Yoke (final product, including assembly supervision) | 6000 CHF / t |
| Stainless Steel for HCAL | 4500 CHF / t |
| Silicon Sensor (for tracking detectors and ECAL) | 6 CHF / cm$^2$ |

## C.4 Relative Distribution of Cost among the Main Detector Components

The main components of a CLIC detector are very different in size, weight and technical complexity. Nevertheless, it can be instructive to identify the main cost drivers. As an example, the relative distribution of costs for a typical CLIC detector is schematically illustrated in Figure C.1. Here, the option of a silicon-tungsten ECAL and an analog readout of the HCAL is assumed.

[2]Based on experience gained in the ATLAS and CMS projects, the construction and testing efforts may be assumed at 500 Full-Time-Equivalents for each year of construction for a CLIC detector.

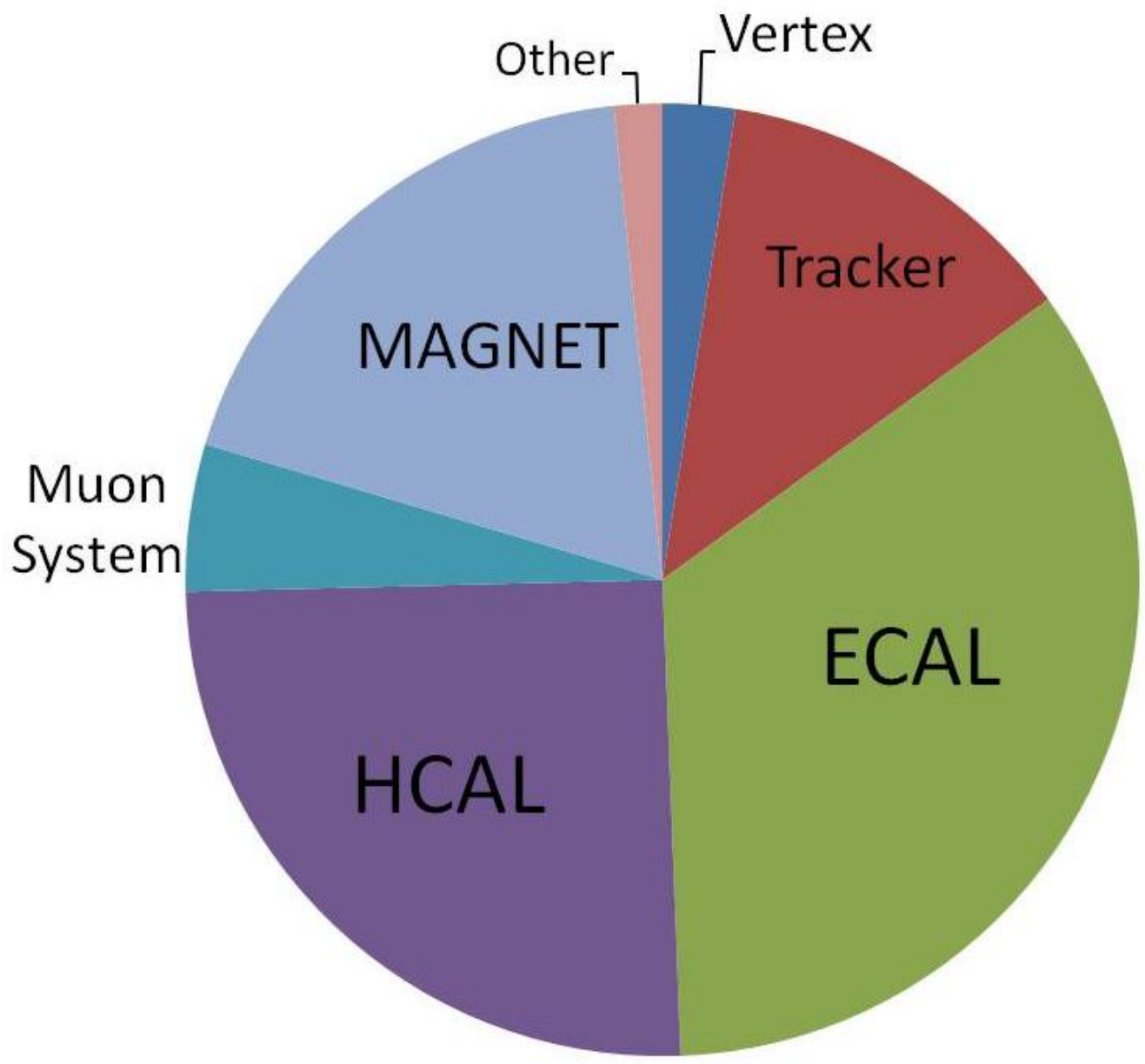

Fig. C.1: Relative contribution of subsystems to the total cost of a typical CLIC detector.

## C.5 Cost Sensitivity Analysis

The unit costs for a few materials have been agreed upon and frozen in October 2010. Evidently, it is impossible to predict the evolution of the market. Instead, we compare CLIC detector costs at the assumed unit cost with a "reduced cost" case, corresponding to a material cost that is lowered by 30%. This comparison also provides information on the relative importance of the unit cost of e.g. tungsten, steel and silicon. A few examples are given in Table C.2.

Table C.2: Sensitivity of subdetector cost and total detector cost to a reduction by 30% in the unit cost of some materials (see Table C.1)

| | Reduction in ECAL cost | Reduction in HCAL cost | Reduction in total detector cost |
|---|---|---|---|
| lower cost for silicon detector material | 22% | *n.a.* | 8% |
| lower cost for ECAL-grade tungsten | 3% | *n.a.* | 1% |
| lower cost for HCAL-grade tungsten | *n.a.* | 23% | 6% |
| lower cost for HCAL-grade stainless steel | *n.a.* | 1% | 0% |
| lower cost for magnet yoke-grade steel | *n.a.* | *n.a.* | 3% |